\documentclass[12pt,a4paper]{book}

\usepackage{ngerman}
\usepackage[T1]{fontenc}
\usepackage{fancyhdr}
\usepackage{graphicx}
\usepackage{amsmath}
\usepackage{amssymb}
\usepackage{caption}
\usepackage{mystyle}

\def\thebibliography#1{\markboth
{}{}\list
 {[\arabic{enumi}]}{\settowidth\labelwidth{[#1]}\leftmargin
\labelwidth
 \advance\leftmargin\labelsep
 \usecounter{enumi}}
 \def\newblock{\hskip .11em plus .33em minus -.07em}
 \sloppy
 \sfcode`\.=1000\relax}

\newenvironment{npb}{\interlinepenalty9999
                     \postdisplaypenalty9999
                     \interdisplaylinepenalty9999
                     \csname@beginparpenalty\endcsname9999
                     \csname@endparpenalty\endcsname9999
                     \csname@itempenalty\endcsname9999
                     \csname@secpenalty\endcsname9999}{}

\newcommand{\Int}[2]{\int\limits_{#1}^{#2}}
\newcommand{\Sum}[2]{\sum\limits_{#1}^{#2}}
\newcommand{\Prod}[2]{\prod\limits_{#1}^{#2}}

\newcommand{\D}{\displaystyle}
\newcommand{\T}{\textstyle}
\newcommand{\winkel}{<\!\!\!)}
\newcommand{\CdoT}{\!\cdot\!}
\newcommand{\Cdot}{\!\cdot}
\newcommand{\cdoT}{\cdot\!}
\newcommand{\adots}{\rotatebox[origin=c]{70}{$\ddots$}}
\newcommand{\myddots}{\rotatebox[origin=c]{-6}{$\ddots$}}
\newcommand{\Left}{\!\left}
\newcommand{\up}[2]{\raisebox{#1ex}{$\scriptscriptstyle #2$}}
\newcommand{\uP}[2]{\raisebox{#1ex}{$\scriptstyle #2$}}

\newcommand{\UP}[2]{\raisebox{#1ex}{$\displaystyle #2$}}
\newcommand{\HH}{{\text{H}}}
\newcommand{\TT}{{\!\text{T}}}
\newcommand{\Hh}{\raisebox{0.3ex}{$\scriptstyle\text{H}$}}
\newcommand{\Kk}{\raisebox{0.3ex}{$\scriptstyle *$}}
\newcommand{\Tt}{\raisebox{0.3ex}{$\scriptstyle\!\text{T}$}}
\newcommand{\hH}{\raisebox{-0.4ex}{$\scriptstyle\text{H}$}}

\newcommand{\tT}{\raisebox{-0.4ex}{$\scriptstyle\!\text{T}$}}
\newcommand{\snorm}[1]{\|#1\|_{\raisebox{-0.4ex}{$\scriptstyle\text{2}$}}}
\newcommand{\Snorm}[1]{\big\|#1\big\|_{\raisebox{-0.4ex}{$\scriptstyle\text{2}$}}}
\newcommand{\fnorm}[1]{\|#1\|_{\raisebox{-0.4ex}{$\scriptstyle\text{F}$}}}

\newcommand{\mb}[1]{{\setlength{\fboxrule}{0.3pt}\setlength{\fboxsep}{1.5pt}\fbox{#1}}}
\newcommand{\mbb}[1]{\mb{{\bf{#1}}}}
\newcommand{\zu}{\mbox{$\stackrel{\raisebox{-0.3ex}{\tiny?}}{z}_0$}}
\newcommand{\psiu}{\mbox{$\stackrel{\raisebox{-0.5ex}{\tiny?}}{\psi}_0$}}

\allowdisplaybreaks[1]

\renewcommand{\chaptermark}[1]{\markboth{\small\sf\thechapter.\ #1}{\small\sf\thechapter.\ #1}}
\renewcommand{\sectionmark}[1]{\markright{\small\sf\thesection.\ #1}}

\renewcommand{\headrulewidth}{0pt}
\newcommand{\myref}[1]{\cite{Diss}:\ref{#1}}
\newcommand{\mypageref}[1]{\cite{Diss}:\pageref{#1}}
\begin{document}
\tracingpages=1
\thispagestyle{empty}
\begin{titlepage}
\begin{center}
{\sf {\large
\vspace*{2cm}
{\LARGE \bf Ein Fenster zur gleichzeitigen Messung der "Ubertragungsfunktion
eines realen Systems und des Leistungsdichtespektrums des "uberlagerten
Rauschens am Systemausgang (Teil 2)}\\
\vspace*{8cm}
{\bf Helmut Repp}\\
\vspace*{1cm}
Erlangen - 2025
}}
\newpage
\rule{0pt}{0pt}
\vfill
Dipl.-Ing. Helmut Repp\\
Liegnitzer Stra"se 1\\
D-91058 Erlangen\\
Germany\\\vspace{6pt}
Telefon: +49-9131-3\,66\,41\\
Mobile: +49-173-57\,19\,350\\
E-mail: Helmut.Repp@gmx.de
\end{center}

\end{titlepage}
\thispagestyle{empty}
\chapter*{"Ubersicht}

Mit Hilfe des in \cite{Diss} verwendeten Rauschklirrmessverfahrens
gelingt es in einer Messung sowohl die "Ubertragungsfunktion als
auch das Rauschleistungsdichtespektrum eines gest"orten Systems zu
messen. Dort wird f"ur zeitinvariante Systeme, die von station"aren
und mittelwertfreien Prozessen erregt und gest"ort werden, gezeigt
wie sich das Rauschklirrmessverfahren um eine Fensterung erweitern
l"asst. Durch die Einf"uhrung einer Fensterung wird es unter anderem 
m"oglich, die spektralen Eigenschaften der St"orung wesentlich
frequenzselektiver und dabei dennoch leistungsrichtig zu ermitteln.

Im zweiten Teil wird nun gezeigt, wie man mit dem um die Fensterung 
erweiterten Rauschklirrmessverfahren Systeme vermessen kann, die von Prozessen 
erregt und gest"ort werden, die zeitabh"angige erste Momente aufweisen. 
Das erste und die zweiten zentralen Momente der St"orung werden dabei 
getrennt ermittelt. Zudem werden hier drei M"oglichkeiten untersucht, wie 
dieses Messverfahren zu erweitern ist, so dass man damit auch
{\setlength{\parskip}{0ex}\begin{list}{$\bullet$}{\setlength{\itemsep}{0ex}
\setlength{\labelwidth}{2em}\setlength{\labelsep}{0.5em}
\setlength{\leftmargin}{2.5em}\setlength{\parsep}{0ex}
\setlength{\topsep}{0ex}\setlength{\partopsep}{0ex}}
\item alle Korrelationen, die zwischen Ein- und Ausgangssignal bestehen,
      bei einem 
\item periodisch zeitvarianten System, das von einem
\item zyklostation"aren Rauschprozess
\end{list}
gest"ort wird, messen kann. Die dazu notwendigen Modifikationen des RKM
k"onnen dabei in beliebiger Kombination der drei m"oglichen Erweiterungen
eingesetzt werden. Wenn man alle drei Modifikationen ber"ucksichtigt,
erh"alt man ein sehr allgemeines, und daher vielseitig einsetzbares
Messverfahren f"ur komplexwertige Systeme.} 

Desweiteren wird gezeigt, wie und unter welchen Umst"anden  man Messwerte 
f"ur die periodisch zeitvariante, komplexe Impulsantwort und die beiden 
Autokorrelationsfolgen eines komplexen, zyklostation"aren 
Approximationsfehlerprozesses berechnen kann.

Desweiteren enth"alt der zweite Teil auch einige zus"atzliche 
praxisorientierte Betrachtungen zu den in \cite{Diss} und hier 
theoretisch behandelten Varianten des Messverfahrens. Es wird eine 
programmflussartige Ablaufs"ubersicht f"ur die Variante des RKM angegeben, 
mit der man das lineare Systemverhalten messen kann, das alle Korrelationen 
des Systemein- und -ausgangs beschreibt, und bei der das erste zeitabh"angige 
und die zweiten zentralen zeit"-un"-ab"-h"angigen Momente der St"orung getrennt 
gemessen werden. F"ur die anderen Varianten wird eine Aufwandsabsch"atzung 
durchgef"uhrt.

Zwei Simulationsbeispiele runden die Beschreibung des RKM ab. 
Eines besch"aftigt sich mit der Varianz und Kovarianz der Messwerte des 
zeitabh"angigen ersten Moments der St"orung. Das andere demonstriert die Messung 
eines von einem zyklostation"aren Prozess gest"orten, periodisch zeitvarianten, 
komplexwertigen Systems mit Hilfe der umfangreichsten RKM-Variante mit Fensterung. 

Auch zur Konstruktion der Fenster werden einige Erg"anzungen angegeben. 
So wird hier eine erweiterte Variante des in \cite{Diss} vorgestellten 
Algorithmus angegeben, bei der einige dort nicht genutzte Freiheitsgrade 
dazu verwendet werden k"onnen, die Eigenschaften der Fensterfolge je nach 
Applikation geeignet zu beeinflussen. Auch wird ein Algorithmus zur 
Konstruktion einer kontinuierlichen Fensterfunktion endlicher L"ange mit 
analogen Eigenschaften angegeben. Einige MATLAB Programme zeigen, wie sich 
die Fenster, deren Autokorrelationsfolgen bzw. -funktionen,  
Fourierreihenkoeffizienten und Spektren berechnen lassen. 

Es handelt sich beim zweiten Teil nicht um eine eigenst"andige
Abhandlung. Vielmehr wird hier auf der in \cite{Diss} dargestellten
Theorie aufgebaut, und es wird immer wieder auf die dort gewonnenen
Ergebnisse Bezug genommen. Die Referenzierung erfolgt dabei, indem
der entsprechenden Kapitel-, Gleichungs-, Bild-, Tabellennummer oder Seitenzahl 
die Literaturstelle \cite{Diss} vorangestellt wird 
( z.~B.\;~\ldots{}Gleichung~(\myref{2.20})~\ldots).

\tableofcontents
\newpage
\rule{0pt}{0pt}
\vfill
{\bf Denksportaufgabe:\label{sinprod}}

Zeige, dass \,\,\,${\D \Prod{\rho=1}{M-1}\sin\Big(\frac{\pi}{M}\cdot\rho\Big)\;=\;M\cdot 2^{1-M}}$\,\,\, gilt.

L"osung auf Seite \pageref{proof}.
\cleardoublepage

\pagenumbering{arabic} 
\setcounter{page}{1}
\frenchspacing
\renewcommand{\headrulewidth}{0.4pt}
\lhead[\fancyplain{\small\sf\thepage}{\small\sf\thepage}]{\fancyplain{}{{\rightmark}}}
\rhead[\fancyplain{}{{\leftmark}}]{\fancyplain{\small\sf\thepage}{\small\sf\thepage}}
\chapter{Einleitung}\label{E.Kap.1}

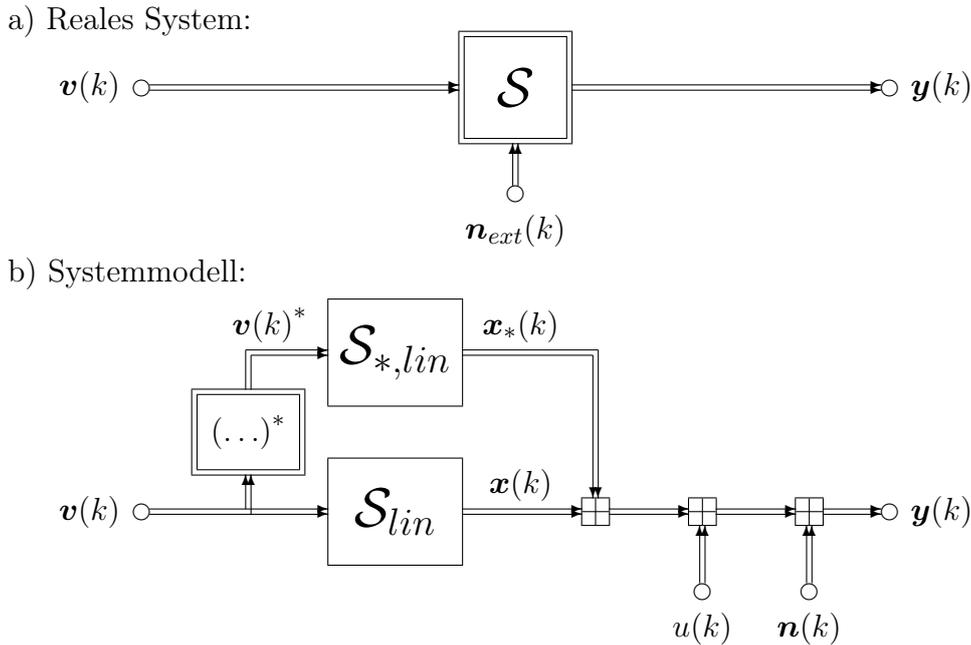
\begin{figure}[tbp]
\begin{center}
{ 
\begin{picture}(300,252)
\put(-40,245){\makebox(0,0)[l]{a) Reales System:}}
\put(10,220){\circle{6}}
\put(150,180){\circle{6}}
\put(290,220){\circle{6}}
\put(13,219){\vector(1,0){116}}
\put(13,221){\vector(1,0){116}}
\put(149,183){\vector(0,1){16}}
\put(151,183){\vector(0,1){16}}
\put(171,219){\vector(1,0){116}}
\put(171,221){\vector(1,0){116}}
\put(131,201){\framebox(38,38){\LARGE ${\cal S}$}}
\put(129,199){\framebox(42,42){}}
\put(2,220){\makebox(0,0)[r]{$\boldsymbol{v}(k)$}}
\put(150,172){\makebox(0,0)[t]{$\boldsymbol{n}_{ext}(k)$}}
\put(298,220){\makebox(0,0)[l]{$\boldsymbol{y}(k)$}}
\put(-40,150){\makebox(0,0)[l]{b) Systemmodell:}}
\put(10,60){\circle{6}}
\put(220,30){\circle{6}}
\put(260,30){\circle{6}}
\put(290,60){\circle{6}}
\put(175,60){\line(1,0){10}}
\put(180,55){\line(0,1){10}}
\put(175,55){\framebox(10,10){}}
\put(215,60){\line(1,0){10}}
\put(220,55){\line(0,1){10}}
\put(215,55){\framebox(10,10){}}
\put(255,60){\line(1,0){10}}
\put(260,55){\line(0,1){10}}
\put(255,55){\framebox(10,10){}}
\put(130,119){\line(1,0){49}}
\put(130,121){\line(1,0){51}}
\put(49,106){\line(0,1){15}}
\put(51,106){\line(0,1){13}}
\put(13,59){\vector(1,0){67}}
\put(13,61){\vector(1,0){67}}
\put(49,121){\vector(1,0){31}}
\put(51,119){\vector(1,0){29}}
\put(130,59){\vector(1,0){45}}
\put(130,61){\vector(1,0){45}}
\put(185,59){\vector(1,0){30}}
\put(185,61){\vector(1,0){30}}
\put(225,59){\vector(1,0){30}}
\put(225,61){\vector(1,0){30}}
\put(265,59){\vector(1,0){22}}
\put(265,61){\vector(1,0){22}}
\put(49,61){\vector(0,1){13}}
\put(51,59){\vector(0,1){15}}
\put(219,33){\vector(0,1){22}}
\put(221,33){\vector(0,1){22}}
\put(259,33){\vector(0,1){22}}
\put(261,33){\vector(0,1){22}}
\put(179,119){\vector(0,-1){54}}
\put(181,121){\vector(0,-1){56}}
\put(31,76){\framebox(38,28){$(\ldots)^{\Kk}$}}
\put(29,74){\framebox(42,32){}}
\put(80,40){\framebox(50,40){\LARGE ${\cal S}_{lin}$}}
\put(80,100){\framebox(50,40){\LARGE ${\cal S}_{*,lin}$}}
\put(2,60){\makebox(0,0)[r]{$\boldsymbol{v}(k)$}}
\put(58,124){\makebox(0,0)[b]{$\boldsymbol{v}(k)^{\Kk}$}}
\put(260,22){\makebox(0,0)[t]{$\boldsymbol{n}(k)$}}
\put(220,22){\makebox(0,0)[t]{$u(k)$}}
\put(152,64){\makebox(0,0)[b]{$\boldsymbol{x}(k)$}}
\put(152,124){\makebox(0,0)[b]{$\boldsymbol{x}_*(k)$}}
\put(298,60){\makebox(0,0)[l]{$\boldsymbol{y}(k)$}}
\end{picture}}
\end{center}\vspace{-20pt}
\setlength{\belowcaptionskip}{-6pt}
\caption{Erweitertes Modell eines gest"orten nichtlinearen realen Systems}
\label{E.b1h}
\rule{\textwidth}{0.5pt}\vspace{-10pt}
\end{figure}

Von vielen digitalen Systemen\footnote{F"ur die Behandlung 
zeitkontinuierlicher Systeme gilt das in \cite{Diss} auf 
Seite \mypageref{f.1} gesagte.} w"unscht man ein lineares,
zeitinvariantes oder wenigstens periodisch zeitvariantes und 
stabiles Verhalten. Reale Systeme k"onnen solches Verhalten nur 
n"aherungsweise realisieren. So ist z.~B. in der Realit"at der Bereich
der zul"assigen Aussteuerung immer begrenzt, bei digitalen Systemen kann 
nur mit endlicher Wortl"ange gearbeitet werden, und bei kontinuierlichen
Systemen zeigen nichtlineare Bauteilkennlinien ihre Wirkung. Auch 
externe St"orungen, die von au"sen in das System einstreuen und die 
im oberen Teil von Bild~\ref{E.b1h} mit \mbox{$\boldsymbol{n}_{ext}(k)$}
bezeichnet werden, verursachen Abweichungen des realen Systemverhaltens
vom gew"unschten Systemverhalten. In aller Regel sind solche Systeme nicht
nur f"ur die Erregung mit einer konkreten Signalfolge entworfen. Man wird daher
die Erregung des Systems als einen Zufallsprozess \mbox{$\boldsymbol{v}(k)$}
betrachten, von dem die wichtigsten stochastischen Eigenschaften
--- wie z.~B. die Varianz --- bekannt sind. Auch im zweiten Teil werden 
wir uns auf den Fall der Erregung mit zuf"alligen periodischen Eingangssignalen 
beschr"anken, wobei die wesentlichen stochastischen Eigenschaften dieselben sein 
sollen, wie bei dem im normalen Betrieb anliegenden Zufallsprozess. Man ist nun 
daran interessiert, den sich am Ausgang des realen Systems ${\cal S}$ ergebenden 
Prozess \mbox{$\boldsymbol{y}(k)$} des realen Systems in vier Anteile
\mbox{$\boldsymbol{x}(k)$}, \mbox{$\boldsymbol{x}_*(k)$}, \mbox{$u(k)$} und 
\mbox{$\boldsymbol{n}(k)$} aufzuspalten, wie dies in Bild~\ref{E.b1h} im unteren 
Teilbild dargestellt ist. Der eine Anteil \mbox{$\boldsymbol{x}(k)$} ist die 
Reaktion eines idealen linearen periodisch zeitvarianten Modellsystems  
${\cal S}_{lin}$, das durch seine bifrequente "Ubertragungsfunktion 
\mbox{$H(\mu_1,\mu_2)$} beschrieben wird, auf den erregenden Prozess 
\mbox{$\boldsymbol{v}(k)$}. Der zweite Anteil \mbox{$\boldsymbol{x}_*(k)$} 
ist die Reaktion eines weiteren idealen linearen periodisch zeitvarianten 
Modellsystems ${\cal S}_{*,lin}$, das durch die bifrequente "Ubertragungsfunktion 
\mbox{$H_*(\mu_1,\mu_2)$} beschrieben wird, auf denselben, nun aber konjugierten, 
erregenden  Prozess \mbox{$\boldsymbol{v}(k)^*$}. Das im allgemeinen 
zeitabh"angige erste Moment des verbleibenden Differenzprozesses 
\mbox{$\boldsymbol{y}(k)-\boldsymbol{x}(k)-\boldsymbol{x}_*(k)$} ist nicht 
zuf"allig und wird in weiteren als deterministische St"orung \mbox{$u(k)$} 
bezeichnet. Der vierte und letzte Anteil des Prozess \mbox{$\boldsymbol{y}(k)$} 
ist der Rauschprozess \mbox{$\boldsymbol{n}(k)$}, der somit mittelwertfrei 
ist. Wir wollen uns hier auf den Fall beschr"anken, dass der Prozess 
\mbox{$\boldsymbol{n}(k)$} zyklostation"ar ist.

Eine Ensemblemittelung --- d.~h. eine Mittelung "uber mehrere Messungen im 
gleichen Zeitintervall, bei denen mit unterschiedlichen Musterfolgen des 
eingangsseitigen Zufallsprozesses erregt wird --- kann zur Gewinnung der 
das System beschreibenden Gr"o"sen an einem einzelnen System i.~Allg. nicht 
vorgenommen werden. Daher muss die Messung dieser Gr"o"sen mit Hilfe einer
Zeitmittelung erfolgen. Deshalb werden bei der Messung zeitlich begrenzte 
Musterfolgen verwendet, die in unterschiedlichen Zeitintervallen das 
zu messende System erregen. Die am System auftretenden Zufallsprozesse
bed"urfen dann aber einer Interpretation, die an die Art des typischen
Betriebs angepasst ist, und die Messung muss mit dieser Betriebsart 
konform durchgef"uhrt werden. Eine Forderung an die so interpretierten
Zufallsprozesse, die sich daraus ergibt, ist die Ergodizit"at dieser
Prozesse in einem sehr weiten Sinn. Mit dieser Interpretation der 
Zufallsprozesse und mit der Ergodizit"atsforderung unter besonderer 
Ber"ucksichtigung einer deterministischen St"orung wird bei der 
Beschreibung des Systemmodells im n"achsten Kapitel begonnen.

Wie in \cite{Diss} werden wir uns bei der Beschreibung des Rauschprozesses 
\mbox{$\boldsymbol{n}(k)$} auf die Momente zweiter Ordnung\footnote{Gem"a"s der 
Definition der deterministischen St"orung \mbox{$u(k)$} ist das Moment erster 
Ordnung des Prozesses \mbox{$\boldsymbol{n}(k)$} immer null.} 
beschr"anken. Die Autokorrelationsfolge (AKF) dieses Prozesses ist der 
Teil der Momente zweiter Ordnung, der die Korrelationen der Zufallsgr"o"sen
des Prozesses zu einem Zeitpunkt mit den Zufallsgr"o"sen zu einem weiteren 
Zeitpunkt beinhaltet. Bei den hier betrachteten Fall eines zyklostation"aren  
Prozesses \mbox{$\boldsymbol{n}(k)$} ist die AKF 
\begin{equation}
\phi_{\boldsymbol{n}}(k_1,k_2)\;=
\text{E}\big\{\boldsymbol{n}(k_1)\CdoT\boldsymbol{n}(k_2)^{\Kk}\big\}
\label{E.1.1}
\end{equation}
zweidimensional. "Uber eine zweidimensionale diskrete Fouriertransformation bez"uglich 
beider Zeitvariablen erh"alt man das bifrequente Leistungsdichtespektrum~(\,LDS\,)
\begin{equation}
\Phi_{\boldsymbol{n}}(\Omega_1,\Omega_2)\;=\!
\Sum{k_1=-\infty}{\infty}\;\Sum{k_2=-\infty}{\infty}\!
\text{E}\big\{\boldsymbol{n}(k_1)\CdoT\boldsymbol{n}(k_2)^{\Kk}\big\}\cdot
e^{\!-j\cdot\Omega_1\Cdot k_1}\cdot e^{j\cdot\Omega_2\Cdot k_2},
\label{E.1.2}
\end{equation}
bei der das Vorzeichen bei der zweiten Frequenzvariable $\Omega_2$ umgedreht ist. 
Bei komplexen Rauschprozessen ist die Beschreibung der zweiten Momente 
nur vollst"andig, wenn neben der AKF auch noch die diskrete Folge der
Kovarianzen der Zufallsgr"o"sen des Prozesses zu einem Zeitpunkt
mit den konjugierten Zufallsgr"o"sen zu einem weiteren Zeitpunkt
angegeben wird. Weil bei der Definition der Kovarianz die eine daran
beteiligte Zufallsgr"o"se bereits konjugiert wird, ist also auch noch
die zweidimensionale Folge der Erwartungswerte der Produkte zweier 
{\em nicht}\/ konjugierter Zufallsgr"o"sen anzugeben.
\begin{equation}
\psi_{\boldsymbol{n}}(k_1,k_2)\;=
\text{E}\big\{\boldsymbol{n}(k_1)\CdoT \boldsymbol{n}(k_2)\big\}
\label{E.1.3}
\end{equation}
Diese sei im weiteren als Kreuzkorrelationsfolge bezeichnet, 
weil diese die kreuzweisen Korrelationen des Ein- und Ausgangs 
eines konjugierenden ---~im "ubrigen nichtlinearen~--- Systems enth"alt. 
Durch eine zweidimensionale diskrete Fouriertransformation bez"uglich beider 
Zeitvariablen ---\,wieder mit Vorzeicheninvertierung bei der zweiten Frequenzvariable\,---
gewinnt man daraus das bifrequente Kreuzleistungsdichtespektrum~(\,KLDS\,)
\begin{equation}
\Psi_{\boldsymbol{n}}(\Omega_1,\Omega_2)\;=
\Sum{k_1=-\infty}{\infty}\;\Sum{k_2=-\infty}{\infty}\!
\psi_{\boldsymbol{n}}(k_1,k_2)\cdot
e^{\!-j\cdot\Omega_1\Cdot k_1}\Cdot
e^{j\cdot\Omega_2\Cdot k_2}.
\label{E.1.4}
\end{equation}

Im Anschluss an die Interpretation der Rauschprozesse wird in Kapitel \ref{E.Kap.2.2} 
auf die Voraussetzungen eingegangen, die ein reales System erf"ullen muss,
damit eine Beschreibung durch ein Modell nach Bild \ref{E.b1h} sinnvoll ist.
Es wird gezeigt, welche "Ubertragungsfunktionen \mbox{$H(\mu_1,\mu_2)$} und 
\mbox{$H_*(\mu_1,\mu_2)$} und welche deterministische St"orung \mbox{$u(k)$}
sich im Systemmodell theoretisch ergeben, wenn das Modellsystem
das lineare Verhalten des realen Systems in dem Sinne m"oglichst gut
approximieren soll, dass das zweite Moment des Modellrauschprozesses
\mbox{$\boldsymbol{n}(k)$} m"oglichst klein wird. Die "Uberf"uhrung des
realen Systems in das Systemmodell wird somit als die L"osung einer
linearen Regression dargestellt. Der Modellrauschprozess ergibt sich
daher als der Prozess des Fehlers der Approximation des realen Systems
durch die beiden linearen periodisch zeitvarianten Systeme ${\cal S}_{lin}$ 
und ${\cal S}_{*,lin}$ und die deterministische St"orung \mbox{$u(k)$}. 
Der Rauschprozess \mbox{$\boldsymbol{n}(k)$} wird daher im weiteren auch 
oft als Approximationsfehlerprozess bezeichnet.

Da die Approximation des realen Systems durch das Systemmodell nach
Bild \ref{E.b1h} von den i.~Allg. unbekannten stochastischen Eigenschaften 
der St"orungen am realen System abh"angt, kann man die zu bestimmenden 
Folgen und Funktionen in der Regel nicht theoretisch in geschlossener 
Form angeben. Man muss daher zun"achst endlich viele Werte finden, 
mit deren Hilfe die das Systemmodell beschreibenden Folgen und Funktionen 
m"oglichst aussagekr"aftig charakterisiert werden k"onnen. Desweiteren 
wird ein Verfahren ben"otigt, das es erlaubt, mit Hilfe einer Messung 
am realen System diese endlich vielen Werte abzusch"atzen. Ein endlich 
langer Abschnitt der Folge der deterministischen St"orung \mbox{$u(k)$} 
besteht aus endlich vielen Werten, und ist daher zur Beschreibung des
Systemmodells unver"andert geeignet. Es wird gezeigt, dass bei der hier 
gew"ahlten periodischen zuf"alligen Erregung die endlich vielen Abtastwerte 
\mbox{$H(\mu_1,\mu_2)$} bzw. \mbox{$H_*(\mu_1,\mu_2)$} der beiden bifrequenten 
"Ubertragungsfunktionen die beiden linearen Systeme ${\cal S}_{lin}$ und 
${\cal S}_{*,lin}$ vollst"andig beschreiben. Analog zu \cite{Diss} wird 
hier f"ur den Fall des zyklostation"aren Approximationsfehlerprozesses 
eine dem bifrequenten LDS eng verwandte bifrequente spektrale Folge 
zur Beschreibung der zweiten Momente des Modellrauschprozesses angegeben. 
Es handelt sich dabei um die zweidimensionale Verallgemeinerung des 
Periodogramms eines gefensterten Signalausschnittes. Die dabei an die 
Fensterfolge zu stellenden Forderungen sind dabei dieselben wie in \cite{Diss}. 

In Kapitel \ref{E.Kap.3} wird dann das Rauschklirrmessverfahren in der 
Art modifiziert, dass man damit alle im Systemmodell vorkommenden Gr"o"sen, 
die in Kapitel \ref{E.Kap.2} vorgestellt und theoretisch hergeleitet worden 
sind, messtechnisch bestimmen kann. Dass diese Messwerte die entsprechenden 
theoretischen Gr"o"sen erwartungstreu und konsistent absch"atzen, wird dadurch 
gezeigt, dass die Erwartungswerte und Varianzen der Messwerte berechnet werden. 
Zus"atzlich zu den Messwertvarianzen werden auch deren Kovarianzen hergeleitet und 
es wird angegeben wie diese abgesch"atzt werden k"onnen. Damit kann man dann 
Konfidenzintervalle bzw. -gebiete f"ur die Messwerte angeben, wie dies in 
\cite{Diss} beschrieben ist. 

Aus den bis dahin bestimmten Messwerten lassen sich weitere Messwerte ableiten. 
Dies sind zum einen die beiden periodisch zeitvarianten Impulsantworten der 
beiden linearen Modellsysteme in Bild~\ref{E.b1h} und zum anderen die Messwerte 
f"ur die Auto- und Kreuzkorrelationsfolgen des zyklostation"aren 
Approximationsfehlerprozesses. Die Berechnung dieser Messwerte und die Absch"atzung 
ihrer Messwert"-(ko)"-varianzen wird in Kapitel \ref{E.Kap.4} beschrieben.

Die Spektralsch"atzung komplexer, station"arer und zyklostation"arer, 
mittelwertbehafteter Prozesse wird als ein Sonderfall des des RKM 
in Kapitel \ref{E.Kap.5} behandelt. 

In Kapitel \ref{E.Kap.6} werden mehrere Varianten zu Messung reellwertiger 
Systeme untersucht, die von Prozessen gest"ort werden, deren erstes Moment 
zeitabh"angig ist.

F"ur den Fall, dass sich ein station"arer Approximationsfehlerprozess 
\mbox{$\boldsymbol{n}(k)$} ergibt, wird eine implementierungsnahe Auf"|listung 
der bei diesem Messverfahren durchzuf"uhrenden Einzelschritte in Kapitel 
\ref{E.Kap.7} angegeben.

Kapitel \ref{E.Kap.8} enth"alt f"ur einige weitere wichtige Varianten 
des RKM bzw. der daraus abgeleiteten Spektralsch"atzverfahren 
Absch"atzungen des ben"otigten Speicherbedarfs. 

Anhand zweier am Rechner simulierter Beispielsysteme wird in Kapitel 
\ref{E.Kap.9} die Einsetzbarkeit der Fensterfolge sowie die neu 
eingef"uhrte Sch"atzung der deterministischen St"orung beim RKM 
verdeutlicht. Diese Beispiele dienen auch der Verdeutlichung 
einiger im theoretischen Teil der Abhandlung beschriebener 
Sachverhalte.

Der Rest des zweiten Teils besch"aftigt sich mit dem in Kapitel 
\myref{Algo} vorgestellten Algorithmus zur Konstruktion einer geeigneten 
Fensterfolge. Im Kapitel \ref{E.Kap.10} wird zun"achst angegeben, wie man 
die in \cite{Diss} nicht genutzten Freiheitsgrade bei der Berechnung einer 
Fensterfolge mit einbeziehen kann, um die Eigenschaften des Spektrums der 
Fensterfolge g"unstig beeinflussen zu k"onnen. Hierf"ur werden anschlie"send 
drei Beispiele angegeben. Desweiteren wird eine Methode vorgestellt, wie 
man das Fensterspektrum vor allem im Sperrbereich hochgenau berechnen kann. 
Eine weitere Methode, gibt an wie sich die Fenster AKF einzig aus den 
Fourierreihenkoeffizienten des Fensters berechnen l"asst. Wie sich der 
in Kapitel \myref{Algo} vorgestellte Algorithmus ab"andern l"asst, um eine 
kontinuierliche Fensterfunktion zu berechnen, wird ebenfalls angegeben. 
Dass dieser Algorithmus eine Grenzwertl"osung f"ur \mbox{$M\!\to\!\infty$}
darstellt, wird gezeigt. Auch wird ein Augendiagramm der AKF des 
kontinuierlichen Fensters dargestellt. In Kapitel \ref{E.Kap.11} werden 
sieben MATLAB Programme aufgelistet, die der Berechnung der Fenster,
sowie ihrer Spektren und AKFs dienen.


\chapter{Das Systemmodell}\label{E.Kap.2}

In diesem Kapitel werde ich zun"achst auf die Interpretation der im 
Systemmodell nach Bild \ref{E.b1h} vorkommenden Zufallsprozesse eingehen. 
Dann werden die beiden bifrequenten "Ubertragungsfunktionen der beiden linearen 
Modellsysteme und das deterministische St"orsignal in Abh"angigkeit der 
stochastischen Eigenschaften der Prozesse am Eingang und am Ausgang des 
realen Systems berechnet. Im letzten Teil dieses Kapitels wird die 
in \cite{Diss} angestellte "Uberlegung, statt des LDS des Rauschprozesses 
den Erwartungswert des Periodogramms des gefensterten Rauschprozesses zu 
dessen Beschreibung zu verwenden, f"ur die beiden bifrequenten 
Leistungsdichtespektren best"atigt. Es wird gezeigt, dass die in \cite{Diss} 
aufgestellten Forderungen f"ur die dabei verwendete Fensterfolge 
weiterhin g"ultig sind.

\section[Interpretation der Rauschprozesse im Systemmodell]
{Interpretation der Rauschprozesse im\\Systemmodell}\label{E.Kap.2.1}

Zun"achst eine Vorbemerkung: In der Literatur findet man zwei verschiedene 
Definitionen f"ur die Stationarit"at eines Zufallsprozesses im engen Sinne. 
Die erste Definition besagt folgendes: Greift man sich aus einem 
Zufallsprozess einen Satz von Zufallsgr"o"sen heraus, indem man f"ur 
den Parameter --- bei uns meist die diskrete Zeit $k$ --- 
des Zufallsprozesses eine beliebige Anzahl beliebiger Werte einsetzt, 
so weisen diese eine gemeinsame Verbundverteilung auf. Nimmt man 
nun statt dieser Zufallsgr"o"sen, diejenigen Zufallsgr"o"sen, bei denen 
die Werte des Parameters um einen beliebigen aber bei allen Zufallsgr"o"sen 
konstanten Wert ---  z.~B. die gleiche Zeit $\kappa$ --- verschobenen 
sind, so weisen diese ebenfalls eine gemeinsame Verbundverteilung auf. 
Ist die Verbundverteilung der Zufallsgr"o"sen ohne Parameterverschiebung 
gleich der Verbundverteilung der Zufallsgr"o"sen mit den verschobenen 
Parameterwerten, und zwar unabh"angig davon, wieviele und welche 
Parameterwerte f"ur den Satz der Zufallsgr"o"sen ausgew"ahlt wurden, 
und wie gro"s der Verschiebungswert des Parameters ist, 
so nennt man den Zufallsprozess im engen Sinne station"ar. 
Die zweite Definition ist da weniger streng. Sie besagt lediglich, 
dass alle Momente, die man aus den Verbundverteilungen der so 
herausgegriffenen Zufallsgr"o"sen berechnet, nicht von der 
Parameterverschiebung der beiden S"atze der Zufallsgr"o"sen 
abh"angen d"urfen. Da es Verteilungen gibt, bei denen keine Momente 
oder nur solche Momente bis zu einer endlichen Ordnung existieren, 
l"asst diese Definition bei solchen Zufallsprozessen keine Aussage 
"uber deren Stationarit"at im engen Sinne zu. Beispielsweise 
besitzt die Cauchy-Verteilung \cite{Fisz} kein einziges Moment. 
Einen Zufallsprozess, dessen Zufallsgr"o"sen alle unabh"angig 
sind und dieselbe Cauchy-Verteilung aufweisen, w"urde man ohne 
weiteres als station"ar bezeichnen, obwohl nach der zweiten 
Definition keine Aussage dar"uber m"oglich ist. In weiteren wird 
bei den theoretischen Herleitungen mit Stationarit"at im engen Sinne 
immer die Stationarit"at im engen Sinne gem"a"s der ersten Definition 
gemeint sein. Unter der Stationarit"at im weiten Sinne ist --- wie "ublich --- 
die Definition gem"a"s der zweiten Variante f"ur die Momente bis zur 
zweiten Ordnung gemeint. Bei realen Zufallsprozessen kann man davon 
ausgehen, dass die beiden Definitionen identisch sind, da dann der Bereich 
der m"oglichen Werte der Zufallsvariablen als begrenzt angesehen 
werden kann, so dass die Momente beliebiger Ordnung existieren. 
Des"ofteren wird von bereichsweiser Stationarit"at oder Stationarit"at 
eines Zufallsprozessausschnittes die Rede sein. Da diese Begriffe nicht 
definiert sind sei hiermit erkl"art, dass mit bereichsweiser 
Stationarit"at gemeint ist, dass bei den Definitionen der 
Stationarit"at nur solche S"atze von Zufallsgr"o"sen betrachtet 
werden, bei denen die Parameterwerte aller Zufallsgr"o"sen 
--- also sowohl der Zufallsgr"o"sen mit der urspr"unglichen, 
als auch der Zufallsgr"o"sen mit den verschobenen Parameterwerten --- 
innerhalb des angegebenen Parameterbereichs liegen. Im weiteren 
wird mit einem Zufallsprozessausschnitt ein Zufallsprozess gemeint 
sein, bei dem der Parameterbereich auf ein angegebenes Intervall 
beschr"ankt ist. Ein Ausschnitt eines Zufallsprozesses wird ab 
nun als station"ar bezeichnet, wenn der Zufallsprozess, aus dem 
er entnommen worden ist, in dem angegebenen Intervall bereichsweise 
station"ar ist. Da bei uns der Parameter immer diskret ist --- meist 
\mbox{$k\in{}\mathbb{Z}$ ---}, wird im weiteren mit einer Zuordnung 
eines Ausschnittes eines Zufallsprozesses zu einem Zufallsvektor 
der Vorgang gemeint sein, bei dem man die Zufallsgr"o"sen eines 
Zufallsprozessausschnittes als Elemente zu einem Zufallsvektor 
zusammenfasst, wobei man alle m"oglichen Werte des Parameters 
innerhalb des angegebenen Intervalls in streng monoton steigender 
Reihenfolge einsetzt. Damit kann man die Stationarit"at eines 
Zufallsvektors analog definieren, wobei eine Parameterverschiebung 
dann einer Indexverschiebung entspricht . Entsprechend definiert man 
die Zyklostationarit"at mit einer Periode, indem man bei der Definition 
die Identit"at der beiden Verbundverteilungen, bzw. die Identit"at 
der Momente nur f"ur solche Werte der Parameterverschiebung fordert, 
die sich jeweils um ganzzahlige Vielfache der Periode unterscheiden.

Jedes am realen System auftretende determinierte Signal kann 
man sich als eine unbegrenzt lange Musterfolge eines 
Zufallsprozesses vorstellen. Die Musterfolge ist unbegrenzt 
lang, weil der Parameter des Zufallsprozesses die diskrete 
Zeit $k$ ist, die ein Element der unendlichen Menge der 
ganzen Zahlen ist. Solch eine Musterfolge ist eine konkrete Realisierung 
unbegrenzter L"ange, also eine konkrete Stichprobe vom Umfang $1$ des 
Zufallsprozesses. Will man Aussagen "uber die stochastischen Eigenschaften 
des zugrundeliegenden Zufallsprozesses empirisch gewinnen, so wird 
man sich eine endliche Anzahl $L$ von endlich langen Signalausschnitten 
gleicher L"ange $F$ aus der zeitlich unbegrenzten Musterfolge in einem 
bestimmten Zeitraum --- dem gesamten Zeitraum der Messung --- 
herausschneiden, und diese in geeigneter Weise verarbeiten. 
Die daraus gewonnen Aussagen k"onnen prinzipiell nur Sch"atzwerte 
f"ur die zu bestimmenden theoretischen Gr"o"sen des Zufallsprozesses 
sein. Au"serdem kann so nur ein Teil der den Zufallsprozess bestimmenden 
Gr"o"sen abgesch"atzt werden.

Die $L$ Signalausschnitte seien mit \mbox{$\lambda=1\;(1)\;L$} 
durchnummeriert. Der Signalausschnitt mit der Nummer $\lambda$ 
stellt eine konkrete Realisierung, also eine Stichprobe vom 
Umfang $1$ des Zufallsvektors mit derselben Nummer $\lambda$ dar. 
Insgesamt hat man $L$ konkrete Realisierungen von $L$ Zufallsvektoren. 
Der Zufallsvektor mit der Nummer $\lambda$ enth"alt als Elemente 
die $F$ Zufallsgr"o"sen des Zufallsprozesses der Zeitpunkte $k$, 
die dem Zeitintervall des Signalausschnittes mit der Nummer $\lambda$ 
entsprechen. Bei jedem dieser $L$ Zufallsvektoren kann man eine 
Verbundverteilung seiner $F$ Elemente angeben. Im allgemeinen sind die 
$L$ Verbundverteilungen der $L$ Zufallsvektoren nicht identisch. 
Wenn nun die $L$ Zeitintervalle der Signalausschnitte so gew"ahlt 
worden sind, dass alle $L$ Zufallsvektoren unabh"angig sind, und dass 
alle dieselbe Verbundverteilung ihrer Elemente aufweisen, so kann 
man die $L$ Signalausschnitte als eine konkrete Stichprobe 
vom Umfang $L$ eines einzigen Zufallsvektors, der im weiteren als 
Modellzufallsvektor bezeichnet sei, mit derselben Verbundverteilung 
interpretieren. Unabh"angig sind die $L$ Zufallsvektoren dann, wenn 
sich die \mbox{$F\CdoT L$}-dimensionale Verbundverteilung von allen 
\mbox{$F\CdoT L$} Elementen aller $L$ Zufallsvektoren als das Produkt 
der $F$-dimensionalen Verbundverteilungen der jeweils $F$ Elemente 
der $L$ Zufallsvektoren schreiben l"asst. Die konkrete Stichprobe 
vom Umfang $L$, also die Gesamtheit aller $L$ aus der konkreten 
Musterfolge des Zufallsprozesses herausgeschnittenen Signalausschnitte 
der L"ange $F$, ist dann eine konkrete Realisierung einer 
mathematischen Stichprobe vom Umfang $L$ des $F$-dimensionalen 
Modellzufallsvektors. Unter einer mathematischen Stichprobe vom 
Umfang $L$ versteht man hier das \mbox{$F\CdoT L$}-dimensionale 
Zufallsexperiment, des Ziehens der konkreten Stichprobe. In unserem 
Fall ist die Zuf"alligkeit dieses Experiments dadurch gegeben, dass 
die zeitlich unbegrenzte Musterfolge, aus der alle $L$ Signalausschnitte 
entnommen worden sind, selbst eine konkrete Realisierung eines 
Zufallsprozesses ist. Mathematisch wird eine Stichprobe nur dann 
genannt, wenn alle $L$ Elemente der Stichprobe unabh"angig sind, 
und wenn die Verbundverteilung jedes Elements der Stichprobe bei 
allen Elementen der Stichprobe gleich der Verbundverteilung des 
Zufallsvektors ist, aus dem die Stichprobe entnommen wurde. Genau diese 
Forderung wurde eben an die Art der Auswahl der $L$ Signalausschnitte 
gestellt. Sinnvollerweise sollte die Auswahl der $L$ Zeitintervalle 
der Signalausschnitte noch ein weiteres Kriterium erf"ullen. 
Es ist n"amlich zu fordern, dass die interessierenden stochastischen 
Eigenschaften des Zufallsprozesses identisch sind mit den 
entsprechenden stochastischen Eigenschaften des Modellzufallsvektors, 
dessen konkrete Stichprobe die ausgew"ahlten Signalausschnitte 
der Musterfolge sind. Welche Signalausschnittl"ange man w"ahlen muss, 
h"angt dabei vor allem davon ab, welche stochastischen Eigenschaften 
man bestimmen will. Will man beispielsweise bei einem zyklostation"aren 
Prozess, dessen Momente mit $17$ periodisch sind, alle $17$ Korrelation 
zweier Signalwerte bestimmen, die um $7$ Takte auseinanderliegen, 
so ist f"ur die Signalausschnittl"ange $F$ wenigstens der Wert 
\mbox{$F\ge17\!+\!7=24$} zu w"ahlen. Anderenfalls w"aren die an einigen 
der interessierenden Korrelationen beteiligten Zufallsgr"o"sen gar nicht 
in den gew"ahlten Ausschnitten des Zufallsprozesses vorhanden. Allgemein 
empfiehlt es sich, die L"ange der Signalausschnitte schon deshalb eher 
gr"o"ser zu w"ahlen, als unbedingt erforderlich, weil dann zu erwarten ist, 
dass sich die gew"unschten stochastischen Eigenschaften bei gleichem 
Stichprobenumfang $L$ mit einer kleineren Varianz bestimmen lassen. 
Wir wollen uns im weiteren mit der Frage besch"aftigen, was man bei der 
Auswahl der $L$ Zeitintervalle der Signalausschnitte beachten sollte, damit 
die bisher aufgestellten Forderungen als erf"ullt angesehen werden k"onnen.

Die Zufallsprozesse, die an einem realen System anliegen, sind immer 
instation"ar, da das System erstens nicht ewig existiert, und zweitens 
sich nicht immer in demselben Betriebszustand befindet. So kann z.~B. 
ein und dasselbe System zur "Ubertragung verschiedener Signale dienen, 
deren stochastische Eigenschaften sich grundlegend unterscheiden, oder 
das System wird zu unterschiedlichen Zeiten in unterschiedlichen 
Umgebungen betrieben, in denen eine grundlegend andere St"orsituation 
vorliegt. Daher wird die Forderung nach Gleichheit der $L$ 
Verbundverteilungen nur erf"ullbar sein, wenn man die Wahl der 
Zeitintervalle einschr"ankt. Man wird zun"achst die unterschiedlichen 
Betriebsarten des Systems klassifizieren m"ussen. Dann wird man die 
Messung am realen System, beginnend mit der Entnahme der 
Signalausschnitte, innerhalb eines Zeitraums durchzuf"uhren, 
in dem sich das System in diesem typischen Betriebszustand befindet, 
der nicht zuletzt auch durch die typischen stochastischen Eigenschaften 
des erregenden sowie des st"orenden Prozesses festgelegt ist. Innerhalb 
des Zeitraums der Messung wird man diesen Betriebszustand nicht verlassen. 
Es sei darauf hingewiesen, dass die stochastischen Eigenschaften der 
Zufallsprozesse innerhalb dieses typischen Betriebszustandes keinesfalls 
station"ar sein m"ussen. Bei geeigneter Entnahme der Signalausschnitte 
wird man so eine Aussage "uber das System- und St"orverhalten gewinnen, 
die dann auf andere Zeitr"aume "ubertragbar ist, in denen sich das reale 
System ebenfalls in diesem typischen Betriebszustand befindet. Um die 
Verwendbarkeit der Messergebnisse sicherzustellen ist es daher wichtig, 
den Betriebszustand, in dem sich das System bei der Messung befindet, 
genau zu dokumentieren. Wie genau der Betriebszustand der Messung 
festgelegt sein sollte, h"angt davon ab, wie empfindlich das System 
auf eine "Anderung des Betriebszustands reagiert. Es ist daher immer 
im Einzelfall entweder anhand heuristischer "Uberlegungen oder durch 
Messung mit ge"andertem Betriebszustand zu pr"ufen, wie genau der bei 
der Messung vorliegende Betriebszustand beschrieben werden muss. Wir 
wollen im weiteren davon ausgehen, dass der typische Betriebszustand 
w"ahrend der Messung nicht verlassen wird, und sind uns dar"uber im Klaren, 
dass die gemessenen stochastischen Eigenschaften nicht die des f"ur alle 
Zeiten anliegenden Zufallsprozesses sind, sondern lediglich f"ur 
die Zeitr"aume g"ultig sind, innerhalb derer sich das System in 
demselben Betriebszustand wie bei der Messung befindet.

Innerhalb des Zeitraums der gesamten Messung muss es --- wie gesagt --- 
mindestens $L$ Zeitintervalle der L"ange $F$ geben, innerhalb derer die 
$L$ Verbundverteilungen der $L$ Zufallsvektoren, die durch 
Ausschneiden der Intervalle aus dem Zufallsprozess entstehen, 
gleich sind. Diese $F$-dimensionale Verbundverteilung ist die des 
Modellzufallsvektors. Ob die Forderung, dass die Verbundverteilungen 
aller $L$ Zufallsprozessausschnitte gleich sein sollen, erf"ullt 
ist, kann man letztlich nicht "uberpr"ufen, da man f"ur ein 
Zeitintervall immer nur eine konkrete Realisierung des 
zugrundeliegenden Zufallsprozesses messen kann. Bei manchen Systemen 
ist der Zugriff auf das System im realen Betrieb auf bestimmte 
Zeitintervalle begrenzt, und man kann annehmen, dass die stochastischen 
Eigenschaften aller Zufallsprozessausschnitte innerhalb der zugreifbaren 
Intervalle identisch sind, sofern der typische Betriebszustand vorliegt. 
Man kann daher beliebige zugreifbare Zeitintervalle ausw"ahlen. Bei 
anderen Systemen ist anzunehmen, dass es sich bei dem zugrundeliegenden 
Zufallsprozess um einen zyklostation"aren Prozess handelt, 
f"ur dessen Beschreibung die Kenntnis einer Periode der stochastischen 
Eigenschaften ausreicht. Bei solchen Systemen wird man die 
Zeitintervalle, die man zur Gewinnung der Signalausschnitte heranzieht, 
synchron zur Periodizit"at der Eigenschaften des Zufallsprozesses 
w"ahlen, so dass dadurch alle Zufallsprozessausschnitte dieselben 
stochastischen Eigenschaften aufweisen. Bei den meisten Systemen 
bestehen keinerlei Einschr"ankungen hinsichtlich der Zeit des Zugriffs, 
und es kann angenommen werden, dass der zugrundeliegende Zufallsprozess 
innerhalb des Zeitraums des typischen Betriebszustands station"ar ist. 
Bei solchen Systemen sind die stochastischen Eigenschaften aller 
Zufallsprozessausschnitte identisch, egal wie man die Ausschnitte w"ahlt.

Eine weitere wichtige Forderung war, dass die $L$ Zufallsvektoren, 
die als die Ausschnitte des zugrundeliegenden Zufallsprozesses 
definiert sind, die den Zeitintervallen der Signalentnahme 
entsprechen, unabh"angig sein m"ussen. Die $L$ Zeitintervalle d"urfen 
sich nicht "uberlappen, da sonst die $L$ Zufallsvektoren nicht 
unabh"angig sein k"onnten, weil unterschiedliche Zufallsvektoren 
dann gleiche Zufallsgr"o"sen als Elemente enthalten w"urden. Bei vielen 
Systemen kann man davon ausgehen, dass die Unabh"angigkeit dadurch 
erreicht werden kann, dass man die Zeitintervalle mit einem gen"ugend 
gro"sen zeitlichen Abstand w"ahlt. Wenn die Unabh"angigkeit der 
Zufallsprozessausschnitte gegeben ist, folgt aus der Faktorisierbarkeit 
der gemeinsamen Verbundverteilung aller $L$ Stichprobenelemente, dass 
das zweite {\em zentrale}\/ Moment zweier beliebiger Zufallsgr"o"sen, 
die aus unterschiedlichen Intervallen aus dem Zufallsprozesses 
entnommen worden sind, null ist. Die Autokovarianzfolge 
(\,nicht die Autokorrelationsfolge\,), die die  zeitdiskrete 
zweidimensionale Folge der zweiten {\em zentralen}\/ Momente ist 
und die von den beiden Zeitpunkten der Entnahme der Zufallsgr"o"sen 
abh"angt, ist in diesem Fall f"ur alle Zeitpunkttupel, deren Elemente 
in unterschiedlichen Zeitintervallen liegen, null. Bei vielen realen 
Systemen kann man annehmen, dass die Autokovarianzfolge null ist, 
wenn die Zeitpunkte nur weit genug auseinander liegen, so dass die 
Autokovarianzfolge auf eine festes Intervall der Differenz der 
Zeitpunkte begrenzt ist. Es gibt aber auch technisch relevante 
Zufallsprozesse, bei denen diese zeitliche Begrenzung der 
Autokovarianzfolge nicht vorliegt, oder bei denen die zeitliche Begrenzung 
der Autokovarianzfolge so gro"s ist, dass das Einf"ugen von Pausen bei 
der Signalentnahme {\it de facto}\/ nicht durchf"uhrbar ist. So kann es z.~B. 
vorkommen, dass den Signalen an einem realen System ein periodischer St"orer 
"uberlagert ist. Dieser St"orer kann entweder deterministisch oder 
stochastisch sein. Ein deterministischer periodischer St"orer 
bewirkt ein periodisch zeitabh"angiges erstes Moment, das nicht 
in die Autokovarianzfunktion eingeht. Die Unabh"angigkeit der 
Zufallsprozessausschitte kann auch hier gegeben sein (\,muss aber nicht\,). 
Ein stochastischer periodischer St"orer liegt beispielsweise vor, 
wenn die Phasenlage des periodischen St"orsignals zuf"allig ist. 
Die Autokovarianzfolge enth"alt im letztgenannten Fall einen Anteil, 
der in ihren beiden unabh"angigen Variablen --- den Zeitpunkten 
der zur Berechnung des zweiten zentralen Moments herangezogenen 
Zufallsgr"o"sen --- periodisch und somit (\,wenigstens innerhalb 
des Zeitraums der Messung\,) zeitlich unbegrenzt ist. Daher kann 
die Unabh"angigkeit der $L$ Zufallsvektoren in Falle eines additiven, 
zuf"alligen periodischen St"orers nicht dadurch erreicht werden, dass 
man die $L$ Zeitintervalle mit einem hinreichend gro"sen Abstand w"ahlt. 
Anhand der einzigen Musterfolge, die im Zeitraum der Messung vorliegt, 
kann man nicht entscheiden, ob es sich bei diesem Signalanteil 
um einen deterministischen St"orer oder um eine Musterfolge 
eines stochastischen St"orers handelt. Man muss daher mit Hilfe 
heuristischer "Uberlegungen entscheiden, wie man diese St"orung 
in der Musterfolge interpretieren will, und die Art der Festlegung 
der Zeitintervalle zur Signalentnahme eventuell entsprechend anpassen, 
um diese St"orung entweder dem ersten oder dem zweiten zentralen Moment 
zuzuordnen. Zu ber"ucksichtigen ist dabei, in welcher Art 
der St"orer im normalen Betrieb des Systems vorliegt. 

Erwartet man bei einem System, das auf einen festen Takt synchronisiert ist, 
dass im normalen Betrieb eine starre Phasenkopplung zwischen dem Systemtakt 
und dem St"orer vorliegt, so wird man diese Art von St"orung als 
deterministische St"orung und somit als zeitabh"angiges Moment erster 
Ordnung im zugrundeliegenden Zufallsprozess interpretieren. Wenn man f"ur 
die Messung der stochastischen Eigenschaften zur Signalentnahme 
nur solche Zeitintervalle ausw"ahlt, die um ganzzahlige Vielfache 
der Periode der St"orung auseinanderliegen, werden auch bei Anwesenheit 
einer deterministischen St"orung die $L$ Zufallsvektoren unabh"angig 
sein, und gleiche Verbundverteilungen aufweisen, wie dies ohne die 
deterministischen St"orung der Fall ist. Aus der so gewonnenen konkreten 
Stichprobe kann man dann Sch"atzwerte f"ur die stochastischen Eigenschaften 
des Modellzufallsvektors berechnen. 

Erwartet man jedoch, dass die Phasenlage der periodischen St"orung 
in keinem festen Zusammenhang mit der Art des Zugriffs auf das System 
steht, so k"onnen die $L$ Zufallsvektoren, die die Zufallsgr"o"sen 
des zugrundeliegenden Zufallsprozesses der $L$ festen Zeitintervalle 
enthalten, schon wegen der Periodizit"at der Autokovarianzfolge 
nicht unabh"angig sein. Dann muss die Auswahl der Zeitintervalle 
der Signalentnahme in Bezug auf die Phasenlage der Musterfolge 
des periodischen St"orers zuf"allig erfolgen. Um diesen Fall 
beschreiben zu k"onnen erzeugen wir uns zun"achst eine 
\mbox{$F\!\times\!L$} Zufallsmatrix, die wir im weiteren als 
zuf"allige Stichprobenmatrix bezeichnen wollen. Wie diese 
Matrix entsteht sei nun erl"autert. In einem ersten Schritt legen 
wir eine hinreichend gro"se Anzahl $\widetilde{L}$ (\,wenigstens $L$, 
wenn m"oglich mehr\,) von Zeitintervallen fest, die zur Signalentnahme 
geeignet sind, was hei"sen soll, dass bei diesen $\widetilde{L}$ 
Zeitintervallen die entsprechenden Zufallsprozessausschnitte 
immer die gleiche $F$-dimensionale Verbundverteilung --- 
n"amlich die des Modellzufallsvektors --- aufweisen. Die 
\mbox{$\widetilde{L}\CdoT F$} Zufallsgr"o"sen des zugrundeliegenden 
Zufallsprozesses bilden $\widetilde{L}$ Zufallsvektoren. Den 
Spaltenvektoren der zuf"alligen Stichprobenmatrix ordnet man nun 
zuf"allig maximal $L$ der $\widetilde{L}$ Zufallsvektoren zu. Wie diese 
zuf"allige Zuordnung vorgenommen wird sei zun"achst v"ollig frei, so dass 
sich die Wahrscheinlichkeiten, einem Spaltenvektor einen bestimmten 
der $\widetilde{L}$ Zufallsvektoren zuzuordnen, auch gegenseitig bedingen 
k"onnen. Beispielsweise kann man solche Zuordnungen ausschlie"sen, 
die denselben Zufallsvektor zwei unterschiedlichen Spaltenvektoren 
zuordnen. Die freie Wahl der Art der zuf"alligen Zuordnung der 
Zufallsprozessausschnitte zu den $L$ Spaltenvektoren kann dazu 
genutzt werden, gewisse stochastische Eigenschaften der zuf"alligen 
Stichprobenmatrix g"unstig zu beeinflussen. Die \mbox{$F\CdoT L$}-dimensionale 
Verbundverteilung der Elemente der zuf"alligen Stichprobenmatrix ist 
der Erwartungswert der \mbox{$F\CdoT L$}-dimensionalen Verbundverteilung 
"uber alle m"oglichen Zuordnungen. Er berechnet sich als die gewichtete 
Summe "uber alle \mbox{$F\CdoT L$}-dimensionalen Verbundverteilungen der 
Zufallsmatrizen, die sich bei allen m"oglichen konkreten Zuordnungen 
ergeben. Die Gewichtung erfolgt dabei nach der Wahrscheinlichkeit, mit 
der eine konkrete Zuordnung auftritt. Jede \mbox{$F\CdoT L$}-dimensionale 
Verbundverteilung jeder konkreten Zuordnung wird sich wegen der Anwesenheit 
des zuf"alligen periodischen St"orers nicht als das Produkt der $L$ 
identischen $F$-dimensionalen Verbundverteilungen der Spaltenvektoren 
schreiben lassen. Unabh"angig davon, wie die Wahrscheinlichkeiten der 
Zuordnungen gew"ahlt werden, sind alle $F$-dimensionalen Verbundverteilungen 
aller $L$ Spaltenvektoren der zuf"alligen Stichprobenmatrix immer 
identisch und gleich der $F$-dimensionalen Verbundverteilung des 
Modellzufallsvektors. Die \mbox{$\widetilde{L}$} Zeitintervalle wurden 
ja so gew"ahlt, dass die $F$-dimensionalen Verbundverteilungen der 
\mbox{$\widetilde{L}$} Zufallsvektoren gleich sind. Bildet man die 
$F$-dimensionale Verbundverteilung eines Spaltenvektors, indem man den 
Erwartungswert aller m"oglichen $F$-dimensionalen Verbundverteilungen 
der \mbox{$\widetilde{L}$} Zufallsvektoren --- gewichtet gem"a"s der 
Auftrittswahrscheinlichkeit der Zuordnungen --- berechnet, so kann diese 
\mbox{$F$-dimensionale} Verbundverteilung vor die Erwartungswertbildung 
gezogen werden, und der Erwartungswert ist dann nur mehr die 
Summe der Wahrscheinlichkeiten aller m"oglichen Zuordnungen, 
die immer eins ist. Die Wahrscheinlichkeiten aller m"oglichen 
Zuordnungen sollen nun m"oglichst so gew"ahlt werden, dass die 
\mbox{$F\CdoT L$}-dimensionale Verbundverteilung aller Elemente der 
zuf"alligen Stichprobenmatrix gleich dem Produkt der $L$ identischen 
\mbox{$F$-dimensionalen} Verbundverteilungen der Spaltenvektoren 
ist. Ob dies m"oglich ist, h"angt sicher von den stochastischen 
Eigenschaften des zugrundeliegenden Zufallsprozesses ab. Sollte dies 
nicht m"oglich sein, so wird man die Freiheit, die Wahrscheinlichkeiten 
aller m"oglichen Zuordnungen beliebig w"ahlen zu k"onnen, dazu 
nutzen, eine m"oglichst gute Approximation des Produkts der $L$ 
identischen $F$-dimensionalen Verbundverteilungen der Spaltenvektoren 
durch die \mbox{$F\CdoT L$}-dimensionale Verbundverteilung der 
zuf"alligen Stichprobenmatrix zu erreichen. Wie man den Fehler dieser 
Approximation sinnvollerweise definiert, kann man nicht generell 
sagen. Dies h"angt von der Wahl der zu bestimmenden stochastischen 
Eigenschaften des zugrundeliegenden Zufallsprozesses ab. 
Trotzdem kann man sicher sein, dass es wenigstens eine Wahl 
f"ur die Wahrscheinlichkeiten der Zuordnungen gibt, die das 
Approximationsziel zumindest nicht schlechter ann"ahert, als die 
beste aller bei der Erwartungswertbildung auftretenden Zuordnungen, 
also als \mbox{jede} feste Zuordnung mit der Wahrscheinlichkeit Eins. 
Hat man die Wahrscheinlichkeiten aller m"oglichen Zuordnungen 
geeignet gew"ahlt, so kann man erwarten, dass die Messwerte der 
stochastischen Eigenschaften, die man anhand der einer konkreten 
Stichprobenmatrix erh"alt, mit den zu messenden stochastischen 
Eigenschaften des Modellzufallsvektors gut "ubereinstimmen. 
Das Hauptproblem der Wahl der Wahrscheinlichkeiten der 
Zuordnungen besteht darin, dass man die \mbox{$F\CdoT L$}-dimensionalen 
Verbundverteilungen der einzelnen Zuordnungen, deren Erwartungswert 
man zu bilden hat, ebensowenig kennt, wie das zu approximierende 
Produkt der $L$ identischen $F$-dimensionalen Verbundverteilungen der 
Spaltenvektoren, deren stochastische Eigenschaften gemessen werden sollen. 
Bei den meisten Systemen, die von einer periodischen St"orung "uberlagert 
sind, kann man annehmen, dass die gleichwahrscheinliche Wahl aller 
Zuordnungen, unter Ausschluss der Zuordnungen, die 
zu identischen Spaltenvektoren f"uhren, zu dem gew"unschten Ergebnis 
f"uhrt, wenn man die Anzahl \mbox{$\widetilde{L}$} der Zeitintervalle 
gro"s genug w"ahlt und diese gleichm"a"sig "uber mehrere Perioden 
einer zu erwartenden St"orung verteilt. Gegebenfalls sollte man die 
Messung wiederholen und anhand der Abweichung der Messergebnisse 
entscheiden, ob man die Messung mit ge"anderter Stichprobenentnahme 
durchf"uhren sollte. Theoretisch k"onnte man auch die Messung mehrfach 
wiederholen, und anschlie"send die Hypothese testen, dass alle 
Messergebnisse aus ein und demselben Zufallsvektor stammen. Dieses 
Verfahren d"urfte allerdings in der Praxis wegen der \mbox{dazu} notwendigen 
immensen Messdauer kaum anwendbar seien. Heuristische "Uberlegungen 
bez"uglich der Art der zu erwartenden St"orung erscheinen hier 
angebrachter. Da bei der Messung der stochastischen Eigenschaften 
die Reihenfolge der Elemente der Stichprobe in der Regel --- wie 
auch beim RKM --- keinen Einfluss auf die Messergebnisse hat, sind 
alle Stichprobenmatrizen, die durch reine Spaltenpermutationen 
ineinander "uberf"uhrt werden k"onnen, gleich gut geeignet. 
Deshalb ist bei vielen Systemen anzunehmen, dass die Unabh"angigkeit der 
Spaltenvektoren --- also der Stichprobenelemente --- in guter N"aherung 
dadurch erreicht werden kann, dass man bei der Entnahme zweier zeitlich 
aufeinanderfolgender Stichproben eine Pause zuf"alliger L"ange einlegt, 
so dass sich die Zeitintervalle der Stichproben "uber mehrere Perioden 
der zu erwartenden St"orung etwa gleichm"a"sig verteilen. 

Die Essenz der bisher gemachten Betrachtungen besteht darin, 
dass man bei Systemen, bei denen man nicht sicher sein kann, 
dass keine St"orungen auftreten, die bei festgelegter Wahl 
der Zeitintervalle zu abh"angigen Signalausschnitten f"uhren, 
eine zuf"allige Wahl der Zeitintervalle der Stichprobenentnahme 
durchf"uhren sollte, und dass man die Messergebnisse 
"uberpr"ufen sollte, wenn man sich nicht sicher ist, dass 
die zuf"allige Entnahme zur gew"unschten Unabh"angigkeit f"uhrt. 
Es wird nun anhand eines ausf"uhrlichen realit"atsnahen Beispiels 
gezeigt, wie man bei den Ein- und Ausgangssignalen eines Systems durch 
die zuf"allige Entnahme von Signalausschnitten aus deren Musterfolgen 
innerhalb des Messzeitraums zu der gew"unschten Stichprobenmatrix kommt. 
Dabei treten typische Vertreter aller oben beschriebenen Arten von 
St"orungen auf.

\begin{figure}[tbp]
\begin{center}
{ 
\begin{picture}(430,207)
\put(0,70){\begin{picture}(430,140)
\put(100,20){\circle{6}}
\put(100,50){\circle{6}}
\put(100,70){\circle{6}}
\put(100,90){\circle{6}}
\put(100,130){\circle{6}}
\put(330,20){\circle{6}}
\put(330,50){\circle{6}}
\put(330,70){\circle{6}}
\put(330,90){\circle{6}}
\put(130,30){\circle*{2}}
\put(130,60){\circle*{2}}
\put(130,65){\circle*{2}}
\put(130,70){\circle*{2}}
\put(150,50){\circle*{3}}
\put(300,30){\circle*{2}}
\put(300,60){\circle*{2}}
\put(300,65){\circle*{2}}
\put(300,70){\circle*{2}}
\put(280,50){\circle*{3}}
\put(115,115){\circle*{2}}
\put(240,45){\line(0,1){10}}
\put(115,115){\line(0,1){15}}
\put(103,130){\line(1,0){12}}
\put(103,90){\line(1,0){17}}
\put(103,70){\line(1,0){17}}
\put(103,50){\line(1,0){17}}
\put(103,20){\line(1,0){17}}
\put(120,20){\line(1,1){10}}
\put(120,50){\line(1,1){10}}
\put(120,70){\line(2,-1){10}}
\put(120,90){\line(1,-2){10}}
\put(150,50){\line(-2,1){17}}
\put(235,50){\line(1,0){45}}
\put(280,50){\line(2,1){17}}
\put(300,30){\line(1,-1){10}}
\put(300,60){\line(1,-1){10}}
\put(300,65){\line(2,1){10}}
\put(300,70){\line(1,2){10}}
\put(310,20){\line(1,0){17}}
\put(310,50){\line(1,0){17}}
\put(310,70){\line(1,0){17}}
\put(310,90){\line(1,0){17}}
\put(333,90){\line(1,0){12}}
\put(345,90){\line(0,1){25}}
\put(130,100){\line(1,1){30}}
\put(50,115){\vector(0,-1){15}}
\put(50,115){\vector(1,0){80}}
\put(80,90){\vector(1,0){17}}
\put(150,50){\vector(1,0){20}}
\put(240,20){\vector(0,1){25}}
\put(240,75){\vector(0,-1){20}}
\put(210,50){\vector(1,0){25}}
\put(345,115){\vector(-1,0){40}}
\put(145,52.5){\dashbox{2}(0,47){}}
\put(285,52.5){\dashbox{2}(0,47){}}
\put(20,80){\framebox(60,20){Sync. Gen.}}
\put(130,100){\framebox(30,30){}}
\put(170,30){\framebox(40,40){\large$H(\Omega)$}}
\put(265,100){\framebox(40,30){Sync.}}
\put(235,45){\framebox(10,10){}}
\put(90,130){\makebox(0,0)[r]{Takt}}
\put(90,70){\makebox(0,0)[r]{Kanal 1}}
\put(90,50){\makebox(0,0)[r]{Kanal 2}}
\put(90,20){\makebox(0,0)[r]{Kanal $N$}}
\put(340,70){\makebox(0,0)[l]{Kanal 1}}
\put(340,50){\makebox(0,0)[l]{Kanal 2}}
\put(340,20){\makebox(0,0)[l]{Kanal $N$}}
\put(133,127){\makebox(0,0)[lt]{$F$}}
\put(157,103){\makebox(0,0)[rb]{1}}
\put(240,15){\makebox(0,0)[t]{$n_{\text{Netz}}(k)$}}
\put(240,80){\makebox(0,0)[b]{$n_{\text{Gau"s}}(k)$}}
\put(100,38){\makebox(0,0){$\vdots$}}
\put(130,47){\makebox(0,0){$\vdots$}}
\put(300,47){\makebox(0,0){$\vdots$}}
\put(330,38){\makebox(0,0){$\vdots$}}
\end{picture}}
\put(0,-130){\begin{picture}(430,200)
\put(10,150){\line(1,0){200}}
\put(220,150){\line(1,0){200}}
\put(10,170){\line(1,0){200}}
\put(220,170){\line(1,0){200}}
\put(22,150){\line(0,1){20}}
\put(66,150){\line(0,1){20}}
\put(110,150){\line(0,1){20}}
\put(154,150){\line(0,1){20}}
\put(198,150){\line(0,1){20}}
\put(232,150){\line(0,1){20}}
\put(276,150){\line(0,1){20}}
\put(320,150){\line(0,1){20}}
\put(364,150){\line(0,1){20}}
\put(408,150){\line(0,1){20}}
\put(390,141){\vector(1,0){30}}
\put(44,160){\makebox(0,0){$\text{Data}_{N}$}}
\put(88,160){\makebox(0,0){Sync.}}
\put(132,160){\makebox(0,0){$\text{Data}_{1}$}}
\put(176,160){\makebox(0,0){$\text{Data}_{2}$}}
\put(254,160){\makebox(0,0){$\text{Data}_{\!N\!-\!1}$}}
\put(298,160){\makebox(0,0){$\text{Data}_{N}$}}
\put(342,160){\makebox(0,0){Sync.}}
\put(386,160){\makebox(0,0){$\text{Data}_{1}$}}
\put(385,141){\makebox(0,0)[r]{$k$}}
\put(20,190){\makebox(0,0)[l]{TDMA-Zeitrahmen:}}
\put(10,160){\makebox(0,0)[r]{$\cdots$}}
\put(215,160){\makebox(0,0)[c]{$\cdots$}}
\put(420,160){\makebox(0,0)[l]{$\cdots$}}
\end{picture}}
\end{picture}}
\end{center}\vspace{-25pt}
\setlength{\belowcaptionskip}{-6pt}
\caption{TDMA-Beispielsystem und TDMA-Zeitrahmen}
\label{E.b1c}
\rule{\textwidth}{0.5pt}\vspace{-10pt}
\end{figure}
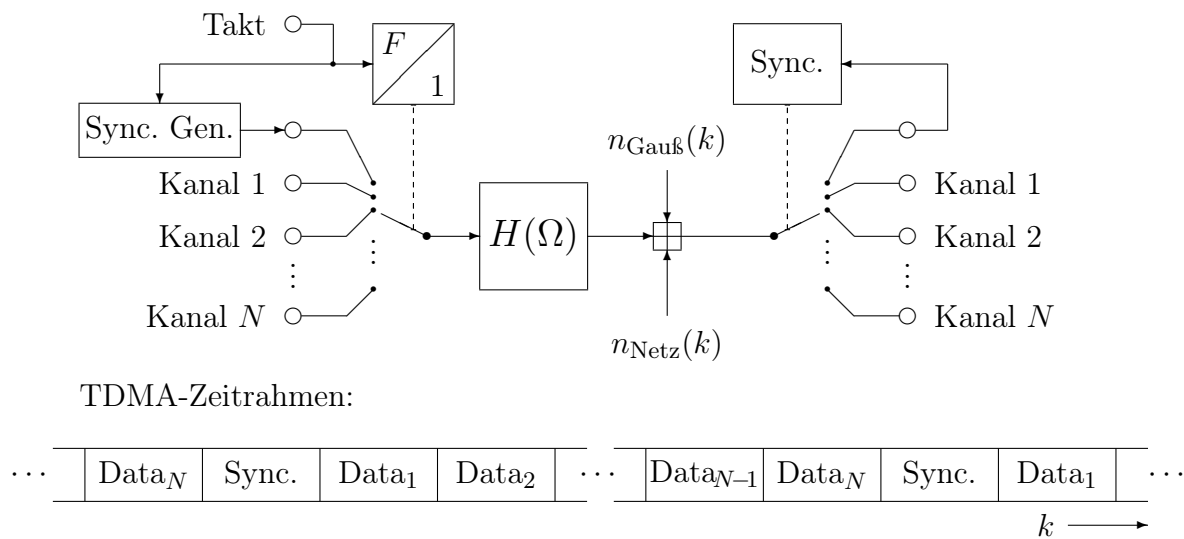

In diesem Beispiel werden verschiedene Signale "uber die $N$ logischen 
Kan"ale des in Bild \ref{E.b1c}
 dargestellten 
TDMA-Systems "uber einen linearen, zeitinvarianten, dispersiven und 
gest"orten physikalischen "Ubertragungskanal "ubertragen. Bei dem TDMA 
(\,Time-Division Multiple Access\,) Verfahren werden den logischen 
Kan"alen jeweils periodisch abwechselnd Zeitschlitze der L"ange 
$F$ zugeordnet. Nachdem jeweils ein Zeitschlitz von jedem logischen 
Kanal "ubertragen wurde, wird ein Zeitschlitz der L"ange $F$ eingef"ugt, 
in dem eine im Empf"anger bekannte Signalsequenz gesendet wird. 
Diese dient der Synchronisation des Empf"angers einerseits auf die 
Periode \mbox{$P=F\cdot(N\!+\!1)$}, mit der sich die Zeitschlitze 
aller logischen Kan"ale wiederholen und andererseits auf den Takt, 
mit dem sich die \mbox{$N\!+\!1$} Zeitschlitze abwechseln. 
In der weiteren theoretischen Interpretation der Prozesse 
am TDMA-System wird angenommen, dass die Synchronisation des 
Empf"angers auf den Sendetakt fehlerfrei erfolgt. Die sich periodisch 
wiederholende Synchronisationssequenz ist ein deterministisches Signal, 
das in Phase und Frequenz exakt festliegt, und deren Phasenlage sich 
bez"uglich der Zeitschlitze der logischen Kan"ale nicht "andert. 
Der physikalische "Ubertragungskanal, "uber den das so konstruierte
TDMA-Signal "ubertragen wird, habe eine Impulsantwort, die auf die
L"ange eines Zeitschlitzes begrenzt sei. In diesem "Ubertragungskanal
seien nun zwei St"orquellen vorhanden. Die eine St"orquelle ist ein 
additives, station"ares, wei"ses und gau"sverteiltes Rauschen, w"ahrend 
die andere St"orquelle ein durch die Energieversorgung verursachter 
netzfrequenter Sinuseintonst"orer der Kreisfrequenz \mbox{$\Omega_{Netz}$} 
ist (\,Netzbrumm\,). Der typische Betriebszustand des Systems sei unter 
anderem dadurch beschrieben, dass der Sinuseintonst"orer immer die 
gleiche Amplitude hat. Der Sinuseintonst"orer und der Takt der 
Zeitschlitze seien {\em nicht}\/ synchron, d.~h. ihre Frequenzen 
stehen nicht in einem trivialen rationalen Verh"altnis zueinander. 
Es liegt daher nahe anzunehmen, dass die Phase des netzfrequenten 
Sinuseintonst"orers in jedem Zeitpunkt innerhalb des Messzeitraums als 
im Bereich von $-\pi$ bis $\pi$ gleichverteilt angenommen werden kann. 
Jede Verbundverteilung beliebiger Abtastwerte des Sinuseintonst"orers 
"andert sich nicht, wenn man alle daran beteiligten Abtastwerte durch 
Abtastwerte ersetzt, die alle um eine beliebige aber bei allen 
Abtastwerten gleiche Zeit verschoben sind. Der Zufallsprozess des 
Sinuseintonst"orers ist daher station"ar. Die zweidimensionale 
Autokovarianzfolge zweier Abtastwerte des netzfrequenten 
Sinuseintonst"orers ist die Folge der Abtastwerte der Kosinusfunktion 
mit der Netzfrequenz und dem halben Quadrat der Amplitude der St"orung. 
Im Argument der Kosinusfunktion tritt dabei nur der zeitliche Abstand 
der beiden Abtastzeitpunkte der zuf"alligen Sinuseintonst"orung auf, 
was aufgrund der Stationarit"at dieser St"orung auch so sein muss. 
Da diese Autokovarianzfolge nicht auf eine Periode $P$ des TDMA-Rahmens 
begrenzt ist, kann die Verbundverteilung der beiden Abtastwerte des 
Sinuseintonst"orers auch dann nicht als das Produkt der beiden 
eindimensionalen Verteilungen eines Sinussignals mit gleichverteilter 
Phase geschrieben werden, wenn die Abtastwerte zeitlich um mehr als $P$ 
auseinander liegen.

In diesem Beispielsystem sollen nun die stochastischen Eigenschaften 
der Prozesse am Systemein- sowie -ausgang gemessen werden. Die 
Messung soll f"ur alle $N$ logischen Kan"ale getrennt erfolgen. 
Die Wahl der \mbox{$\widetilde{L}$} Zeitintervalle ist nicht beliebig 
m"oglich, sondern muss mit einer festen Phasenlage bez"uglich des 
$P$-fachen Systemtakts erfolgen. Alle innerhalb des Messzeitraums 
liegenden und f"ur eine Signalausschnittentnahme in Frage kommenden 
Zeitintervalle sind also immer um ganzzahlige Vielfache von $P$ 
gegeneinander verschoben. W"ahrend ein Kanal gemessen wird, werden 
auf den anderen Kan"alen modulierte Datensymbole "ubertragen, die in 
der Art moduliert sind, dass deren Anteile am Sendesignal nur in den 
Zeitschlitzen von null verschieden sind, die den jeweiligen logischen 
Kan"alen entsprechen. Von den Datensymbolen, die in den gerade nicht 
gemessenen Kan"alen "ubertragen werden, nehmen wir an, dass diese aus 
Zufallsprozessen entnommen sind, die einen Anteil am Sendesignal 
erzeugen, der mit der Periode $P$ zyklostation"ar ist, und dessen 
Erwartungswert f"ur jeden Zeitpunkt null ist. Nun erregen wir 
den zu vermessenden logischen Kanal mit einem zuf"alligen Testsignal. 
Zur Generierung des Testsignals wird f"ur die Zeitschlitze des zu 
vermessenden logischen Kanals, die innerhalb des Zeitraums der Messung 
liegen, aus einem mittelwertfreien Zufallsvektor eine konkrete Stichprobe 
entnommen. Jedes Element der Stichprobe bildet einen Signalausschnitt. 
Diese Signalausschnitte werden als Sendesignale f"ur die Zeitschlitze 
des zu vermessenden Kanals verwendet. Der Umfang der Stichprobe wird 
dabei so gro"s gew"ahlt, dass alle $\widetilde{L}$ Zeitschlitze des zu 
vermessenden Kanals, die innerhalb des Zeitraums der Messung liegen, 
mit Elementen der Stichprobe aufgef"ullt sind. Der gesamte sendeseitige 
Zufallsprozess ist daher innerhalb des Zeitraums der Messung zyklostation"ar 
mit der Periode $P$, und die auf die Zeitschlitze des zu vermessenden Kanals 
begrenzten Zufallsvektoren der Elemente der mathematischen Stichprobe sind 
unabh"angig. W"ahlt man sich aus allen $\widetilde{L}$ Zeitschlitzen 
des zu vermessenden logischen Kanals innerhalb des Zeitraums der 
Messung nun $L$ beliebige Zeitschlitze fest aus, so stellen die $L$ 
Signalausschnitte der Musterfolge des erregenden Zufallsprozesses eine 
konkrete Realisierung einer mathematischen Stichprobenmatrix dar, deren 
Spaltenvektoren alle dieselbe Verbundverteilung besitzen wie der 
Zufallsvektor, aus dem die Sendesignalsequenzen gewonnen wurden. Eine 
zuf"allige Auswahl der $L$ Zeitschlitze ist daher beim ungest"orten 
Sendesignal nicht notwendig. Eine Sch"atzung der stochastischen 
Eigenschaften des erregenden Zufallsvektors anhand einer konkreten 
Realisierung der zuf"alligen Stichprobenmatrix ist daher auch ohne 
eine zuf"allige Zuordnung der Zeitintervalle zu den Spaltenvektoren 
der Stichprobenmatrix m"oglich.

Betrachten wir nun die $\widetilde{L}$ Zufallsvektoren des {\em ersten}\/ 
logischen Kanals am Ausgang des physikalischen "Ubertragungskanals. 
Diese Zufallsvektoren enthalten die Zufallsprozessausschnitte 
der Zeitschlitze, die den $\widetilde{L}$ Synchronisationszeitschlitzen 
im Zeitbereich der Messung unmittelbar folgen. Die stochastischen Eigenschaften 
dieser $\widetilde{L}$ Zufallsvektoren werden durch die Eigenschaften der 
unterschiedlichen am Gesamtausgangsprozess beteiligten Teilprozesse bestimmt. 
Da ist zun"achst der station"are, wei"se Gau"sprozess, dessen Anteile 
bei allen $\widetilde{L}$ Zufallsvektoren unabh"angig sind. Dem "uberlagern 
sich additiv die mit der Periode $P$ zyklostation"aren Prozesse, die die 
Sendesignale der anderen gerade nicht vermessenen logischen Kan"ale erzeugen, 
und die durch den physikalischen "Ubertragungskanal linear verzerrt sind. 
Da wir angenommen hatten, dass die Impulsantwort des "Ubertragungskanals 
auf die Dauer eines Zeitschlitzes begrenzt ist, und da die Zeitschlitze 
des gerade vermessenen logischen Kanals den Synchronisationszeitschlitzen 
folgen, liefern diese Teilprozesse keinen Beitrag zu den $\widetilde{L}$ 
betrachteten Zufallsvektoren. Ein weiterer Teilprozess ist der mit der 
Periode $P$ zyklostation"are Prozess, aus dem das Testsignal gewonnen wird, 
der ebenfalls durch den physikalischen "Ubertragungskanal linear verzerrt ist. 
Von diesem Teilprozess wird nur der Anteil, der nach der linearen Verzerrung
in die Zeitschlitze des gerade vermessenen logischen Kanals f"allt, in den 
$\widetilde{L}$ Zufallsvektoren vorhanden sein. Da erstens die Impulsantwort 
des "Ubertragungskanals auf einen Zeitschlitz begrenzt ist, da zweitens die 
Zeitintervalle mit einem gr"o"seren zeitlichen Abstand entnommen werden, und 
da drittens die Stichprobenelemente des Testsignals unabh"angig sind, sind 
auch die Beitr"age des Testsignals zu den $\widetilde{L}$ Zufallsvektoren 
unabh"angig. Die Anteile des Testsignalprozesses und des Gau"sprozesses 
sind ebenfalls voneinander unabh"angig. Als weiterer Anteil ist noch 
der station"are Zufallsprozess des netzfrequenten Sinuseintonst"orers 
vorhanden, der ebenfalls von den anderen Teilprozessen unabh"angig 
ist. Da alle Signalausschnitte des Sinuseintonst"orers voneinander 
{\em abh"angig}\/ sind, sind auch deren Anteile an den $\widetilde{L}$ 
Zufallsvektoren voneinander abh"angig. Der Anteil des mit $P$ 
periodischen deterministischen Synchronisationssignals, 
der durch die lineare Verzerrung des physikalischen "Ubertragungskanals 
jeweils in die den Synchronisationszeitschlitzen folgenden, gerade 
vermessenen Zeitintervalle f"allt, ist bei allen $\widetilde{L}$ 
Zufallsvektoren gleich, und "uberlagert sich additiv. Er verursacht 
bei allen $\widetilde{L}$ Zufallsvektoren dieselbe Verschiebung ihrer 
Verbundverteilungen. Da alle Teilprozesse voneinander unabh"angig 
und station"ar bzw. mit der Periode $P$ zyklostation"ar sind, und da 
sich alle Teilprozesse additiv "uberlagern, ist die $F$-dimensionale 
Verbundverteilung aller $\widetilde{L}$ Zufallsvektoren gleich. 
Diese Verbundverteilung ist die unseres Modellzufallsvektors. Nun 
greifen wir uns aus den $\widetilde{L}$ Zufallsvektoren $L$ heraus 
und fassen diese Vektoren als Spaltenvektoren zu der zuf"alligen 
Stichprobenmatrix zusammen. Zun"achst erfolgt das Herausgreifen 
der $L$ Zufallsvektoren --- also die Zuordnung eines Teils der 
$\widetilde{L}$ Zufallsvektoren zu den $L$ Spaltenvektoren der 
Stichprobenmatrix --- zwar fast beliebig aber {\em nicht}\/ zuf"allig. 
Wir wollen nur solche Zuordnungen zulassen, bei denen die $L$ 
Spaltenvektoren der zuf"alligen Stichprobenmatrix aus $L$ 
unterschiedlichen Zeitschlitzen entnommen werden, so dass es nicht 
vorkommen kann, dass irgendeiner der $\widetilde{L}$ Zufallsvektoren 
mehrfach als Spaltenvektor auftritt. Bei dem Element der zuf"alligen 
Stichprobenmatrix in der Zeile $\kappa$ und in der Spalte $\lambda$ 
erzeugt der Teilprozess des Sinuseintonst"orers den Anteil 
\mbox{$\sin(\Omega_{Netz}\CdoT\kappa+\boldsymbol{\phi}_{\lambda})$}, wobei 
\mbox{$\boldsymbol{\phi}_{\lambda}$} eine zuf"allige Phase ist, die bei 
allen Elementen eines Spaltenvektors gleich ist, und die sich aus 
zwei Anteilen zusammensetzt. Der erste Anteil ist die zuf"allige 
Phase $\boldsymbol{\phi}$ der im Zeitraum der Messung vorliegenden 
Sinusst"orung. Sie ist bei allen Elementen der zuf"alligen 
Stichprobenmatrix gleich. Der zweite Anteil ist die Phasenverschiebung 
\mbox{$\Omega_{Netz}\CdoT P\CdoT\Tilde{\lambda}_{\lambda}$}, die bei allen 
Elementen eines Spaltenvektors jeweils gleich ist. $\Tilde{\lambda}_{\lambda}$ 
ist dabei die Nummer des Zeitschlitzes, dessen Zufallsprozessausschnitt 
bei der gew"ahlten Zuordnung dem $\lambda$-ten Spaltenvektor zugeordnet wird. 
Die zuf"alligen Abtastwerte der Sinusst"orung, die in allen Elementen 
der zuf"alligen Stichprobenmatrix additiv vorhanden sind, sind also 
unterschiedlich phasenverschobene nichtlineare Funktionen von ein 
und derselben Zufallsgr"o"se und somit nicht unabh"angig. Die 
\mbox{$F\CdoT L$}-dimensionale Verbundverteilung der zuf"alligen 
Abtastwerte der Sinusst"orung ist daher nicht faktorisierbar. Weil 
die an der zuf"alligen Stichprobenmatrix beteiligten Teilprozesse alle 
voneinander unabh"angig sind, ergibt sich die \mbox{$F\CdoT L$}-dimensionale 
Verbundverteilung der Elemente der zuf"alligen Stichprobenmatrix als 
die mehrfache \mbox{$F\CdoT L$}-dimensionale Faltung\footnote{ 
Begonnen wird mit einer Verbundverteilung eines Teilprozesses. 
Dann wird wiederholt die Faltung der Verbundverteilung eines noch 
nicht ber"ucksichtigten Teilprozesses mit der bisher berechneten 
Verbundverteilung durchgef"uhrt, bis alle Teilprozesse verarbeitet 
wurden. Bei der \mbox{$F\CdoT L$}-dimensionale Faltung handelt es 
sich um eine Integration "uber alle \mbox{$F\CdoT L$} Dimensionen 
des Raumes. In welchem Sinne die Integration hier definiert ist 
(\,Riemann, Stieltjes oder im Sinne der Distributionentheorie\,) 
sei dahingestellt. Integriert wird "uber das Produkt der 
\mbox{$F\CdoT L$}-dimensionalen Verbundverteilungsdichte des einen 
Teilprozesses und der \mbox{$F\CdoT L$}-dimensionalen Verbundverteilung 
des anderen Teilprozesses (\,Beim Stieltjes-Integral muss es hei"sen: 
Integral der Verbundverteilungsdichte bez"uglich der Verbundverteilung\,). 
Bei der \mbox{$F\CdoT L$}-dimensionalen Verbundverteilung treten 
dabei die Ver"anderlichen der \mbox{$F\CdoT L$}-dimensionalen 
Verbundverteilungsdichte gespiegelt und verschoben auf. "Uber die 
\mbox{$F\CdoT L$} Ver"anderlichen der \mbox{$F\CdoT L$}-dimensionalen 
Verbundverteilungsdichte wird integriert, w"ahrend die \mbox{$F\CdoT L$} 
Parameter der dabei auftretenden Verschiebung die Ver"anderlichen des 
Ergebnisses der Faltung sind. Festgehalten sei auch noch, dass sich beide 
Verbundverteilungen faktorisieren lassen, wenn die Spaltenvektoren der an 
der Faltung beteiligten Teilprozessausschnitte unabh"angig sind. Dann kann die 
\mbox{$F\CdoT L$}-fache Integration als Produkt von $L$ Mehrfachintegralen 
geschrieben werden, wobei jedes Mehrfachintegral nur "uber einem 
$F$-dimensionalen Raum zu berechnen ist. Daher ist die sich ergebende 
\mbox{$F\CdoT L$}-dimensionale Verbundverteilung ebenfalls wieder 
faktorisierbar.} der \mbox{$F\CdoT L$}-dimensionalen Verbundverteilungen 
der einzelnen beteiligten Teilprozesse. Weil wir identische Spaltenvektoren 
in der zuf"alligen Stichprobenmatrix ausgeschlossen haben, ergibt sich 
{\em ohne}\/ den Zufallsprozess des netzfrequenten Sinuseintonst"orers bei 
der Faltung eine \mbox{$F\CdoT L$}-dimensionale Verbundverteilung, die 
als Produkt der $L$ identischen $F$-dimensionalen Verbundverteilungen 
der Spaltenvektoren geschrieben werden kann. Bei der 
\mbox{$F\CdoT L$}-dimensionalen Faltung mit der 
\mbox{$F\CdoT L$}-dimensionalen Verbundverteilung des zuf"alligen 
Sinuseintonst"orers ergibt sich keine faktorisierbare 
Verbundverteilung, weil die \mbox{$F\CdoT L$}-dimensionale 
Verbundverteilung  des zuf"alligen Sinuseintonst"orers nicht 
faktorisierbar ist. Es besteht nun die Hoffnung, dass sich bei einer 
zuf"alligen Zuordnung eine faktorisierbare \mbox{$F\CdoT L$}-dimensionale 
Verbundverteilung ergibt. Wie oben hergeleitet, ergibt sich diese 
durch eine Erwartungswertbildung aus allen nicht faktorisierbaren 
Verbundverteilungen aller m"oglichen Zuordnungen. Wir haben 
also den Erwartungswert eines \mbox{$F\CdoT L$}-dimensionale 
Faltungsintegrals zu bilden, wobei an diesem Faltungsintegral 
einerseits die {\em nicht}\/ faktorisierbare 
\mbox{$F\CdoT L$}-dimensionale Verbundverteilung des 
Sinuseintonst"orers und andererseits die faktorisierbare 
\mbox{$F\CdoT L$}-dimensionale Verbundverteilung aller anderen Anteile 
beteiligt sind. Die Erwartungswertbildung kann nun in das 
\mbox{$F\CdoT L$}-dimensionale Faltungsintegral gezogen werden. 
{\em Ohne}\/ Ber"ucksichtigung des Sinuseintonst"orers ist die 
\mbox{$F\CdoT L$}-dimensionale Verbundverteilung das Produkt der 
$L$ \mbox{identischen} $F$-dimensionalen Verbundverteilungen der 
Spaltenvektoren, das bei allen m"oglichen Zuordnungen gleich ist. 
Diese \mbox{$F\CdoT L$}-dimensionale Verbundverteilung kann daher 
vor die Erwartungswertbildung gezogen werden. Daher ergibt sich 
bei einer zuf"alligen Zuordnung die \mbox{$F\CdoT L$}-dimensionale 
Verbundverteilung der Stichprobenmatrix als die 
\mbox{$F\CdoT L$}-dimensionale Faltung des Produkts der $L$ 
identischen $F$-dimensionalen Verbundverteilungen der Spaltenvektoren 
ohne Ber"ucksichtigung des Sinuseintonst"orers mit dem Erwartungswert 
der \mbox{$F\CdoT L$}-dimensionalen Verbundverteilung des zuf"alligen 
Sinuseintonst"orers. Dieser Erwartungswert sollte durch die Wahl 
der Wahrscheinlichkeiten der m"oglichen Zuordnungen faktorisierbar 
gemacht werden. Die Zuordnung $\Tilde{\boldsymbol{\lambda}}_{\lambda}$, 
die dem $\lambda$-ten Spaltenvektor die Nummer des Zeitschlitzes zuordnet, 
ist nun selbst eine Zufallsgr"o"se, die von der zuf"alligen Phase 
$\boldsymbol{\phi}$ der im Zeitraum der Messung vorliegenden Sinusst"orung 
unabh"angig ist. Die zuf"alligen Phasen \mbox{$\boldsymbol{\phi}_{\lambda}$}, 
setzen sich daher jeweils aus zwei unabh"angigen Zufallsgr"o"sen 
additiv zusammen, sind aber immer noch bei allen Elementen {\em eines}\/ 
Spaltenvektors jeweils gleich. Weil alle zuf"alligen Abtastwerte eines 
Spaltenvektors jeweils Funktionen von ein und derselben Zufallsgr"o"se 
\mbox{$\boldsymbol{\phi}_{\lambda}$} sind, gen"ugt es, die Unabh"angigkeit 
der $L$ Phasen \mbox{$\boldsymbol{\phi}_{\lambda}$} zu zeigen, um die 
Unabh"angigkeit der Spaltenvektoren der Zufallsmatrix der Abtastwerte 
des Sinuseintonst"orers zu zeigen. Nun wollen wir annehmen, dass 
die Messdauer gro"s genug gew"ahlt worden sei, so dass die $\widetilde{L}$ 
Ausschnitte des netzfrequenten Sinuseintonst"orers Phasenlagen 
\mbox{$\Omega_{Netz}\CdoT P\CdoT\Tilde{\boldsymbol{\lambda}}_{\lambda}$} 
besitzen, die im gesamten Zeitraum der Messung in einen Bereich von mehreren
Vielfachen von $2\pi$ fallen. W"ahlt man nun alle m"oglichen Zuordnungen, 
mit Ausnahme der Zuordnungen, die zu identischen Spaltenvektoren f"uhren, 
gleich wahrscheinlich, so ist die Phasendifferenz zweier Stichprobenelemente 
des Sinuseintonst"orers bei allen \mbox{$\widetilde{L}\!-\!1$} m"oglichen 
diskreten Phasendifferenzen gleichwahrscheinlich, und die m"oglichen 
Phasendifferenzen liegen --- modulo $2\pi$ berechnet --- etwa gleichm"a"sig 
im Bereich von $-\pi$ bis $\pi$ verstreut. Die $L$-dimensionale 
Verbundverteilungsdichte der Phasen \mbox{$\boldsymbol{\phi}_{\lambda}$} 
ist als Ableitung der Verbundverteilung nach allen Ver"anderlichen 
definiert, und daher eine Funktion, die nur au"serhalb eines 
eindimensionalen nichtlinearen Raumes (\,ein kontinuierlicher 
Freiheitsgrad der zuf"alligen Phase $\boldsymbol{\phi}$ des St"orers\,) 
existiert, und dort null ist. Weil die Phase $\boldsymbol{\phi}$ in 
alle Zufallsphasen \mbox{$\boldsymbol{\phi}_{\lambda}$} additiv eingeht, 
besteht der eindimensionale nichtlineare Raum aus 
\mbox{$\widetilde{L}!/(\widetilde{L}\!-\!L)!$} Geradenst"ucken, 
die alle parallel zu dem Vektor liegen, der nur Einsen enth"alt. 

\begin{figure}[btp]
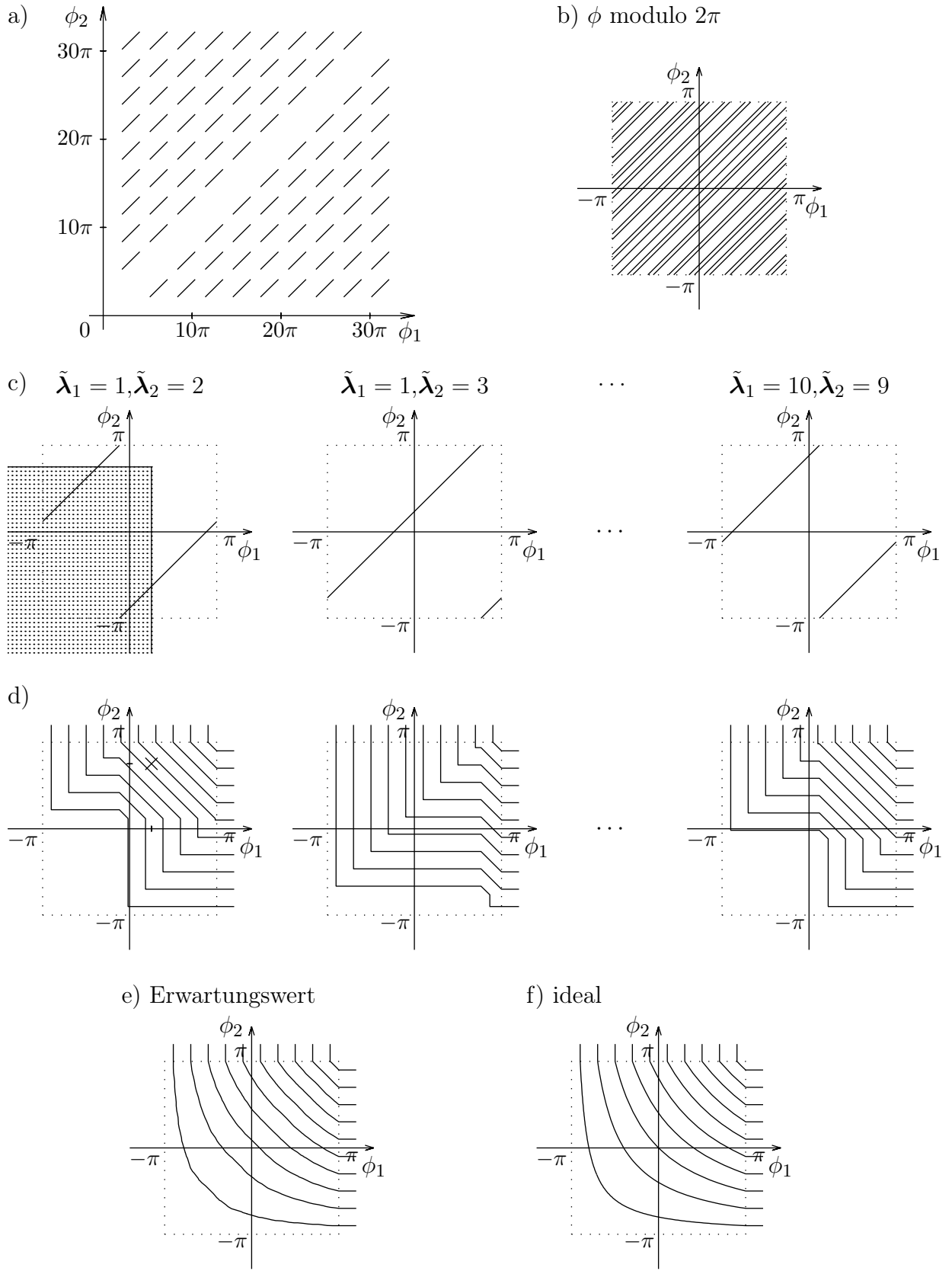

\begin{center}
{ 
\begin{picture}(454,626)

\put(0,620){\makebox(0,0)[l]{a)}}
\input{mbild1d1}
\put(38,465){\makebox(0,0){\small$0$}}
\put(91,465){\makebox(0,0){\small$10\pi$}}
\put(134,465){\makebox(0,0){\small$20\pi$}}
\put(178,465){\makebox(0,0){\small$30\pi$}}
\put(42,516){\makebox(0,0)[r]{\small$10\pi$}}
\put(42,560){\makebox(0,0)[r]{\small$20\pi$}}
\put(42,603){\makebox(0,0)[r]{\small$30\pi$}}
\put(198,464){\makebox(0,0){$\phi_1$}}
\put(40,620){\makebox(0,0)[r]{$\phi_2$}}

\put(270,620){\makebox(0,0)[l]{b) $\phi$ modulo $2\pi$}}
\input{mbild1d2}
\put(338,490){\makebox(0,0)[rt]{\small$-\pi$}}
\put(338,580){\makebox(0,0)[rb]{\small$\pi$}}
\put(295,533){\makebox(0,0)[rt]{\small$-\pi$}}
\put(386,532){\makebox(0,0)[lt]{\small$\pi$}}
\put(398,531){\makebox(0,0)[t]{$\phi_1$}}
\put(336,592){\makebox(0,0)[r]{$\phi_2$}}

\put(0,438){\makebox(0,0)[l]{c)}}
\put(60,438){\makebox(0,0){$\tilde{\boldsymbol{\lambda}}_1=1$,$\tilde{\boldsymbol{\lambda}}_2=2$}}
\put(200,438){\makebox(0,0){$\tilde{\boldsymbol{\lambda}}_1=1$,$\tilde{\boldsymbol{\lambda}}_2=3$}}
\put(297,438){\makebox(0,0){$\ldots$}}
\put(394,438){\makebox(0,0){$\tilde{\boldsymbol{\lambda}}_1=10$,$\tilde{\boldsymbol{\lambda}}_2=9$}}
\input{mbild1d3}
\put(58,322){\makebox(0,0)[rt]{\small$-\pi$}}
\put(58,410){\makebox(0,0)[rb]{\small$\pi$}}
\put(15,363){\makebox(0,0)[rt]{\small$-\pi$}}
\put(106,363){\makebox(0,0)[lt]{\small$\pi$}}
\put(119,361){\makebox(0,0)[t]{$\phi_1$}}
\put(56,422){\makebox(0,0)[r]{$\phi_2$}}
\input{mbild1d4}
\put(198,322){\makebox(0,0)[rt]{\small$-\pi$}}
\put(198,410){\makebox(0,0)[rb]{\small$\pi$}}
\put(155,363){\makebox(0,0)[rt]{\small$-\pi$}}
\put(246,363){\makebox(0,0)[lt]{\small$\pi$}}
\put(259,361){\makebox(0,0)[t]{$\phi_1$}}
\put(196,422){\makebox(0,0)[r]{$\phi_2$}}
\put(297,365){\makebox(0,0)[c]{$\cdots$}}
\input{mbild1d5}
\put(392,322){\makebox(0,0)[rt]{\small$-\pi$}}
\put(392,410){\makebox(0,0)[rb]{\small$\pi$}}
\put(349,363){\makebox(0,0)[rt]{\small$-\pi$}}
\put(440,363){\makebox(0,0)[lt]{\small$\pi$}}
\put(453,361){\makebox(0,0)[t]{$\phi_1$}}
\put(390,422){\makebox(0,0)[r]{$\phi_2$}}

\put(0,283){\makebox(0,0)[l]{d)}}
\input{mbild1d6}
\put(58,175){\makebox(0,0)[rt]{\small$-\pi$}}
\put(58,263){\makebox(0,0)[rb]{\small$\pi$}}
\put(15,216){\makebox(0,0)[rt]{\small$-\pi$}}
\put(106,216){\makebox(0,0)[lt]{\small$\pi$}}
\put(120,214){\makebox(0,0)[t]{$\phi_1$}}
\put(56,277){\makebox(0,0)[r]{$\phi_2$}}
\input{mbild1d7}
\put(198,175){\makebox(0,0)[rt]{\small$-\pi$}}
\put(198,263){\makebox(0,0)[rb]{\small$\pi$}}
\put(155,216){\makebox(0,0)[rt]{\small$-\pi$}}
\put(246,216){\makebox(0,0)[lt]{\small$\pi$}}
\put(260,214){\makebox(0,0)[t]{$\phi_1$}}
\put(196,277){\makebox(0,0)[r]{$\phi_2$}}
\put(297,218){\makebox(0,0)[c]{$\cdots$}}
\input{mbild1d8}
\put(392,175){\makebox(0,0)[rt]{\small$-\pi$}}
\put(392,263){\makebox(0,0)[rb]{\small$\pi$}}
\put(349,216){\makebox(0,0)[rt]{\small$-\pi$}}
\put(440,216){\makebox(0,0)[lt]{\small$\pi$}}
\put(454,214){\makebox(0,0)[t]{$\phi_1$}}
\put(390,277){\makebox(0,0)[r]{$\phi_2$}}

\put(56,135){\makebox(0,0)[l]{e) Erwartungswert}}
\input{mbild1d9}
\put(118,17){\makebox(0,0)[rt]{\small$-\pi$}}
\put(118,105){\makebox(0,0)[rb]{\small$\pi$}}
\put(75,58){\makebox(0,0)[rt]{\small$-\pi$}}
\put(166,58){\makebox(0,0)[lt]{\small$\pi$}}
\put(180,56){\makebox(0,0)[t]{$\phi_1$}}
\put(116,120){\makebox(0,0)[r]{$\phi_2$}}

\put(255,135){\makebox(0,0)[l]{f) ideal}}
\input{mbild1d10}
\put(318,17){\makebox(0,0)[rt]{\small$-\pi$}}
\put(318,105){\makebox(0,0)[rb]{\small$\pi$}}
\put(275,58){\makebox(0,0)[rt]{\small$-\pi$}}
\put(366,58){\makebox(0,0)[lt]{\small$\pi$}}
\put(380,56){\makebox(0,0)[t]{$\phi_1$}}
\put(316,120){\makebox(0,0)[r]{$\phi_2$}}
\end{picture}}
\end{center}\vspace{-13pt}
\caption{Zur Verbundverteilung der Phase des Sinusst"orers}
\label{E.b1d}
\end{figure}

Im Teilbild a) des Bildes \ref{E.b1d}
 ist dieser 
eindimensionale Raum, in dem die Verbundverteilungsdichte der Phasen 
\mbox{$\boldsymbol{\phi}_{\lambda}$} Impulslinien aufweist, 
f"ur \mbox{$L=2$} und \mbox{$\widetilde{L}=10$} dargestellt. 
Jedes Geradenst"uck entspricht dabei einer konkreten Zuordnung. 
Das Geradenst"uck einer konkreten Zuordnung ist um einen Vektor 
verschoben, der die bei der Zuordnung auftretenden \mbox{$L=2$} 
Phasenvers"atze \mbox{$\Omega_{Netz}\CdoT P\CdoT\Tilde{\lambda}_1$} und 
\mbox{$\Omega_{Netz}\CdoT P\CdoT\Tilde{\lambda}_2$} beider Spaltenvektoren 
enth"alt. Da die Elemente der Spaltenvektoren der Zufallsmatrix der 
Abtastwerte des Sinuseintonst"orers phasenverschobene Sinusfunktionen 
von \mbox{$\boldsymbol{\phi}_{\lambda}$} sind, und somit mit $2\pi$ 
periodisch sind, kann man ebensogut die modulo $2\pi$ berechneten 
Zufallsgr"o"sen betrachten. Wie man in Teilbild b) sieht, werden 
durch die modulo $2\pi$ Reduktion die verschobenen Geradenst"ucke 
zerteilt und durch Parallelverschiebung um Vielfache von $2\pi$ 
auf den Bereich des Quadrats mit der Kantenl"ange $2\pi$ um den 
Koordinatenursprung abgebildet. Diese zerteilten Geradenst"ucke 
der ersten beiden und der letzten der insgesamt $90$ m"oglichen 
konkreten Zuordnungen sind in Teilbild c) des Bildes \ref{E.b1d} 
dargestellt. Innerhalb jedes zerteilten Geradenst"uckes einer konkreten 
Zuordnung, in dem die zweidimensionale Verbundverteilungsdichte nicht 
definiert ist, ist die eindimensionale bedingte\footnote{unter der 
Bedingung der konkreten Zuordnung} Verteilungsdichte der zuf"alligen 
Phase $\boldsymbol{\phi}$ eine Gleichverteilung. Will man daraus f"ur 
eine konkrete Zuordnung die $L$-dimensionale bedingte Verbundverteilung 
berechnen, so hat man allgemein das uneigentliche $L$-dimensionale 
Volumenintegral "uber die $L$-dimensionale bedingte Verbundverteilungsdichte 
zu berechnen, wobei die oberen Integrationsgrenzen die freien 
Ver"anderlichen der $L$-dimensionalen Verbundverteilung sind. 
Im ersten Bild des Teilbildes c), das die Lage der Impulslinien 
der zweidimensionalen Verteilungsdichte bei der ersten m"oglichen 
Zuordnung darstellt, ist das Gebiet hervorgehoben, "uber das zu 
integrieren ist, um einen bestimmten Wert der zweidimensionalen 
Verbundverteilung zu erhalten. In unserem Fall des Sinuseintonst"orers 
sind ein oder zwei eindimensionale Integrale "uber den konstanten 
Wert \mbox{$1/(2\pi)$} zu berechnen, wobei die Integrationsgrenzen 
von den beiden freien Ver"anderlichen der zweidimensionalen 
Verbundverteilung abh"angen. Bei der hier vorliegenden Gleichverteilung 
innerhalb der Geradenst"ucke ist dieses Integral proportional zur 
L"ange der Anteile der Geradenst"ucke, die in das hervorgehobene 
Gebiet fallen. Die sich bei der Integration ergebenden zweidimensionalen 
bedingten Verbundverteilungen der drei f"ur die Graphik ausgew"ahlten 
Zuordnungen sind jeweils unter den in Teilbild d) dargestellten 
Verbundverteilungsdichten als "`H"ohenliniengraphik"' dargestellt. 
Das Phasentupel, f"ur das sich der Wert des Integrals "uber das 
hervorgehobene Gebiet ergibt, ist im linken Teilbild der ersten 
m"oglichen Zuordnung als "`$\times$"' markiert. Die 
Erwartungswertbildung "uber die zweidimensionalen bedingten 
Verbundverteilungen aller m"oglichen Zuordnungen liefert die zweidimensionale 
Verbundverteilung der zuf"alligen modulo $2\pi$ reduzierten Phasen 
\mbox{$\boldsymbol{\phi}_{\lambda}$}. Sie ist f"ur unser einfaches Beispiel
in Teilbild e) des Bildes \ref{E.b1d} dargestellt. Wie man in Teilbild b) 
erkennt, sind die zerteilten Geradenst"ucke dadurch, dass die Phasenvers"atze 
\mbox{$\Omega_{Netz}\CdoT P\CdoT\Tilde{\boldsymbol{\lambda}}_{\lambda}$} im 
Bereich von $-\pi$ bis $\pi$ etwa gleichm"a"sig verstreut sind, ebenfalls 
etwa gleichm"a"sig auf den gesamten Bereich des Quadrats verstreut. 
Es wird sich --- nicht nur in unserem Beispiel mit \mbox{$L=2$} --- bei der 
Berechnung des Integrals etwa dieselbe $L$-dimensionalen Verbundverteilung 
ergeben, wie bei der $L$-fachen Integration "uber eine $L$-dimensionale 
Gleichverteilung \mbox{${\D(2\pi)^{\!-L}}$}, die sich problemlos 
faktorisieren l"asst. Zum Vergleich ist die sich bei der Integration 
"uber die zweidimensionale Gleichverteilung ergebende Verbundverteilung 
in Teilbild f) zu sehen. Man kann daher annehmen, dass die modulo $2\pi$ 
reduzierten Phasen \mbox{$\boldsymbol{\phi}_{\lambda}$} bei einer 
zuf"alligen Zeitschlitzzuordnung in wesentlich besserer N"aherung 
als unabh"angig angesehen werden k"onnen, als im Fall einer festen 
Zuordnung der Zeitschlitze, bei der sich die zweidimensionale 
Verbundverteilung als das Integral "uber ein einziges zerteiltes 
Geradenst"uck ergibt, wie dies in den Teilbildern d) zu sehen ist. 
Daher kann man auch die Spaltenvektoren der Zufallsmatrix der 
Abtastwerte des Sinuseintonst"orers als n"aherungsweise unabh"angig 
betrachten. 

\enlargethispage*{2pt}Wie wir oben gesehen haben wird somit auch die 
\mbox{$F\CdoT L$}-dimensionale Verbundverteilung des zuf"alligen 
Sinuseintonst"orers n"aherungsweise faktorisierbar sein. Da diese bei 
der Berechnung der \mbox{$F\CdoT L$}-dimensionalen Verbundverteilung 
aller Anteile der zuf"alligen Stichprobenmatrix mit der 
faktorisierbaren \mbox{$F\CdoT L$}-dimensionalen Verbundverteilung 
der restlichen Anteile gefaltet wird, ist anzunehmen, dass sich auch 
die \mbox{$F\CdoT L$}-dimensionale Verbundverteilung aller Anteile 
der zuf"alligen Stichprobenmatrix am Ausgang des TDMA-Systems 
n"aherungsweise faktorisieren l"asst. Somit sind die Spaltenvektoren der 
zuf"alligen Stichprobenmatrix n"aherungsweise unabh"angig. Es sei nochmal 
darauf hingewiesen, dass die $F$-dimensionale Verbundverteilung jedes 
Spaltenvektors von der zuf"alligen Zuordnung nicht beeinflusst 
wird und bei allen Spaltenvektoren gleich ist. Insgesamt bilden 
die Spaltenvektoren der zuf"alligen Stichprobenmatrix daher 
n"aherungsweise eine mathematische Stichprobe des Modellzufallsvektors, 
der dieselbe $F$-dimensionale Verbundverteilung aufweist, wie jeder 
beliebige der $\widetilde{L}$ Zufallsvektoren. Eine konkrete Stichprobe 
vom Umfang $L$ \mbox{--- also} eine konkrete Realisierung der zuf"alligen 
Stichprobenmatrix~--- erh"alt man, indem man aus der konkreten 
Musterfolge des Ausgangssignals aus den $\widetilde{L}$ 
konkreten Signalausschnitten zuf"allig einen Satz von $L$ 
Signalausschnitten ausw"ahlt. Wenn man anhand dieser konkreten 
Stichprobe die gesuchten stochastischen Eigenschaften empirisch bestimmt, 
so wird man erwarten k"onnen, dass diese die stochastischen Eigenschaften 
jedes beliebigen der $\widetilde{L}$ Zufallsvektoren und somit 
auch die stochastischen Eigenschaften des Ausgangsprozesses 
des TDMA-Systems innerhalb des Zeitschlitzes des ersten logischen 
Kanals ad"aquat absch"atzen.

Abschlie"send wird noch untersucht, was sich bei den eben 
durchgef"uhrten Betrachtungen "andert, wenn man die Messung 
an einem anderen als dem logischen Kanal durchf"uhrt, der dem 
Zeitschlitz entspricht, der dem Synchronisationszeitschlitz folgt. 
In allen anderen logischen Kan"alen ist die deterministische St"orung 
durch das Synchronisationssignal aufgrund der begrenzt angenommen 
Dauer der Systemimpulsantwort in dem analog gebildeten 
Modellzufallsvektor nicht mehr vorhanden. Stattdessen wird der 
mit der Periode $P$ zyklostation"are Prozess, der durch die 
lineare Verzerrung des "Ubertragungskanals aus dem Prozess entsteht, 
aus dem die Sendesignale des der Messung vorangehenden Kanals erzeugt 
werden, als St"orer mit zeitabh"angigen stochastischen Eigenschaften 
in unserem Modellzufallsvektor auftreten. Diese Art der St"orung ist 
auch diejenige, die man beim normalen Betrieb des TDMA-Systems in einem 
Kanal, der nicht dem Synchronisationszeitschlitz folgt, erwartet. 
Je nachdem ob die Zufallsvektoren, aus denen die Datensignale entstammen, 
die in dem vor dem zu messenden logischen Kanal entsprechenden Zeitschlitz 
"ubertragen werden, unabh"angig sind oder nicht, sind deren 
Anteile an den Spaltenvektoren der zuf"alligen Stichprobenmatrix 
bei einer festen Zuordnung ebenfalls unabh"angig oder nicht. Wenn 
diese unabh"angig sind, wird sich deren Unabh"angigkeit nicht "andern, 
wenn man die feste Zuordnung durch eine davon unabh"angige zuf"allige 
Zuordnung ersetzt. Sind die Spaltenvektoranteile jedoch abh"angig, 
so kann man auch hier erwarten, dass die Qualit"at\footnote{Abweichung 
der Approximation des Produkts der $L$ identischen $F$-dimensionalen 
Randverbundverteilungen der Spaltenvektoren durch die 
\mbox{$F\CdoT L$}-dimensionale Verbundverteilung der Matrix} der 
Abh"angigkeit der Spaltenvektoren der zuf"alligen Stichprobenmatrix 
sich deutlich verringert, wenn man eine zuf"allige Zuordnung der 
Zeitschlitze zu den Spaltenvektoren der zuf"alligen Stichprobenmatrix 
einf"uhrt. Eine Messung der stochastischen Eigenschaften der 
Zufallsprozessausschnitte der Zeitschlitze anhand einer konkreten 
Stichprobenmatrix wird dann auch bei diesen logischen Kan"alen erst 
durch die zuf"allige Wahl der Zuordnung erm"oglicht, selbst wenn
kein netzfrequenter St"orer vorhanden ist.

Will man stochastische Eigenschaften bestimmen, in die sowohl 
Zufallsgr"o"sen am Eingang wie auch am Ausgang unseres 
TDMA-Beispielsystems eingehen --- um daraus z.~B. die 
"Ubertragungsfunktion des physikalischen "Ubertragungskanals zu 
berechnen ---, so muss man weiterhin fordern, dass auch alle $\widetilde{L}$ 
Zufallsvektoren, die sich aus den $\widetilde{L}$ Zufallsvektoren 
des Ein- und Ausgangsprozesses zusammensetzen, eine Verbundverteilung 
besitzen, die bei allen $\widetilde{L}$ zusammengesetzten Zufallsvektoren 
identisch ist. Man muss dann beim Eingangssignal dieselbe Zuordnung 
verwenden, wie beim Ausgangssignal, auch wenn \mbox{--- wie} bei unserem 
Beispielsystem --- eine zuf"allige Zuordnung beim Signal am Systemeingang 
nicht notwendig ist, wenn man nur die stochastischen Eigenschaften 
des Eingangssignals bestimmt. Wenn man dann die ggf. zuf"allige 
Zuordnung der Spaltenvektoren der zuf"alligen Stichprobenmatrizen 
am Systemein- und ausgang so vornimmt, dass auch noch sichergestellt 
ist, dass die Spaltenvektoren der einen Stichprobenmatrix von den 
Spaltenvektoren der anderen unabh"angig sind, wenn sie einen 
unterschiedlichen Spaltenindex aufweisen, bilden die Paare der 
Spaltenvektoren mit gleichen Spaltenindex der beiden Stichprobenmatrizen 
des Ein- und Ausgangs die Elemente einer mathematischen Stichprobe 
des Zufallsvektors, der sich aus den beiden Zufallsvektoren des Ein- 
und Ausgangs zusammensetzt.

Damit m"ochte ich das TDMA-Beispiel zur Interpretation der Zufallsprozesse 
im Systemmodell nach Bild \ref{E.b1h} abschlie"sen. Es sollte damit vor 
allem gezeigt werden, dass man immer darauf achten sollte, ob und wie 
eine geeignete Stichprobenentnahme an einem realen System m"oglich ist. 
Im weiteren werden unter den im Systemmodell \mbox{auftretenden} 
Zufallsvektoren nicht mehr die Ausschnitte der zeitverschobenen 
Zeitintervalle des tats"achlich am System anliegenden Zufallsprozesses 
gemeint sein, sondern immer die entsprechenden zeitverschiebungsunabh"angigen 
Modellzufallsvektoren, die wegen der geforderten Wahl der Zeitintervalle 
dieselbe Verbundverteilung aufweisen, wie alle entsprechenden 
Zufallsprozessausschnitte. Es wird dabei angenommen, dass die Entnahme 
der Signalausschnitte \mbox{--- wenn} notwendig --- zuf"allig vorgenommen wurde, 
so dass die oben beschriebene Betrachtungsweise der $L$ Signalausschnitte 
als eine konkrete Stichprobe einer mathematischen Stichprobe vom Umfang $L$ 
--- also als ein Ensemble --- des Modellzufallsvektors ebenso zutrifft, 
wie die zuletzt beschriebene Interpretation des Paares der beiden 
Stichprobenmatrizen des Ein- und des Ausgangs des Systems. Die diskrete
Zeitvariable $k$ wird im weiteren sowohl f"ur die tats"achliche physikalisch
unaufhaltsam verrinnende Zeit --- etwa beim Zugriff auf eine bestimmte
Zufallsgr"o"se eines tats"achlich am System anliegenden Zufallsprozesses ---
als auch als Lauf"|index der Elemente des Modellzufallsvektors verwendet, ohne 
dass die sich dabei ergebende ggf. zuf"allige Zeitverschiebung ber"ucksichtigt 
wird. Dadurch werden mehrere konkrete Realisierungen des Modellzufallsvektors 
m"oglich, ohne dass jedesmal eine andere Zeitverschiebung zu ber"ucksichtigen 
ist oder eine neue Variable f"ur die Elemente des Modellzufallsvektors 
eingef"uhrt werden muss. Ob der Zufallsprozess mit der physikalischen Zeit 
oder die Zufallsgr"o"sen des Modellzufallsvektors mit der verschobenen Zeit 
gemeint sind, wird aus dem Zusammenhang ersichtlich. Die Forderung, 
dass die Verbundverteilung des Modellzufallsvektors gleich der 
Verbundverteilung aller Zufallsvektoren derselben L"ange ist, die aus 
dem realen Zufallsprozess innerhalb der Zeitintervalle entnommen werden, 
die alle in einem Zeitraum liegen, in dem sich das System in einem typischen 
Betriebszustand befindet, ist eine der Ergodizit"at analoge Forderung f"ur 
Prozesse, die nicht station"ar oder zyklostation"ar sein m"ussen.

Die bisher durchgef"uhrte Betrachtung bez"uglich der Gewinnung einer 
Stichprobenmatrix l"asst es zu, dass man beliebige stochastische 
Eigenschaften anhand einer konkreten Stichprobenmatrix bestimmen 
kann. Manchmal kann sich jedoch der Fall ergeben, dass die geeignete 
Gewinnung einer Stichprobenmatrix nicht m"oglich oder praktisch 
nicht durchf"uhrbar ist. Wenn man jedoch nur ganz bestimmte 
stochastische Eigenschaften absch"atzen will --- beim RKM 
sind dies die ersten und zweiten zentralen Momente ---, kann es 
gen"ugen, dass man die Stichprobenmatrix in einer Art erzeugt, 
die bez"uglich der zu messenden Merkmale als zuf"allig angesehen 
werden kann. Das bedeutet, dass lediglich die Sch"atzwerte der zu 
bestimmenden Merkmale keine systematischen oder methodischen Fehler 
aufweisen, die durch die Art der Stichprobenerhebung verursacht werden. 
Wie solch eine Stichprobe zu erheben ist, kann allgemein nicht gesagt 
werden, sondern ist immer anhand des konkret zu vermessenden Systems zu 
entscheiden. Die im weiteren hergeleitete Theorie geht immer davon aus, 
dass die Art der Erhebung der Stichprobe so gew"ahlt wurde, dass sich 
eine mathematische Stichprobenmatrix ergibt.

\section[Theoretische Werte der "Ubertragungsfunktionen und der 
deterministischen St"orung]{Theoretische Werte der "Ubertragungsfunktionen\\ 
und der deterministischen St"orung}\label{E.Kap.2.2}

\begin{figure}[tbp]
\begin{center}
{ 
\begin{picture}(400,330)
\put(-25,320){\makebox(0,0)[lb]{a) Erweitertes, nichtlineares Modellsystem:}}
\put(-25,125){\makebox(0,0)[lb]{b) "Aquivalentes Modellsystem, das zwei lineare komplexwertige Modellsysteme enth"alt:}}
\put(60,30){\circle{6}}
\put(60,235){\circle{6}}
\put(320,30){\circle{6}}
\put(320,235){\circle{6}}
\put(120,30){\circle*{4}}
\put(105,175){\circle*{3}}
\put(105,295){\circle*{3}}
\put(220,99){\line(1,0){59}}
\put(220,101){\line(1,0){61}}
\put(275,30){\line(1,0){10}}
\put(285,29){\line(1,0){32}}
\put(285,31){\line(1,0){32}}
\put(63,234){\line(1,0){17}}
\put(63,236){\line(1,0){17}}
\put(220,215){\line(1,0){40}}
\put(220,255){\line(1,0){40}}
\put(255,175){\line(1,0){45}}
\put(255,295){\line(1,0){45}}
\put(300,234){\line(1,0){17}}
\put(300,236){\line(1,0){17}}
\put(119,80){\line(0,1){21}}
\put(121,80){\line(0,1){19}}
\put(280,25){\line(0,1){10}}
\put(80,175){\line(0,1){59}}
\put(80,236){\line(0,1){59}}
\put(105,175){\line(0,1){40}}
\put(105,255){\line(0,1){40}}
\put(260,170){\line(0,1){10}}
\put(260,290){\line(0,1){10}}
\put(300,175){\line(0,1){59}}
\put(300,236){\line(0,1){59}}
\put(105,215){\line(1,1){40}}
\put(105,255){\line(1,-1){40}}
\put(63,29){\vector(1,0){97}}
\put(63,31){\vector(1,0){97}}
\put(121,99){\vector(1,0){39}}
\put(119,101){\vector(1,0){41}}
\put(220,29){\vector(1,0){55}}
\put(220,31){\vector(1,0){55}}
\put(80,175){\vector(1,0){80}}
\put(80,295){\vector(1,0){80}}
\put(145,255){\vector(1,0){15}}
\put(145,215){\vector(1,0){15}}
\put(220,295){\vector(1,0){35}}
\put(220,175){\vector(1,0){35}}
\put(279,99){\vector(0,-1){64}}
\put(281,101){\vector(0,-1){66}}
\put(260,215){\vector(0,-1){35}}
\put(119,31){\vector(0,1){19}}
\put(121,29){\vector(0,1){21}}
\put(260,255){\vector(0,1){35}}
\put(275,25){\framebox(10,10){}}
\put(255,170){\framebox(10,10){}}
\put(255,290){\framebox(10,10){}}
\put(160,15){\framebox(60,30){$h_{\kappa}(k)$}}
\put(160,85){\framebox(60,30){$h_{*,\kappa}(k)$}}
\put(160,280){\framebox(60,30){$h_{\alpha,\kappa}(k)$}}
\put(160,240){\framebox(60,30){$h_{\beta,\kappa}(k)$}}
\put(160,200){\framebox(60,30){$h_{\gamma,\kappa}(k)$}}
\put(160,160){\framebox(60,30){$h_{\delta,\kappa}(k)$}}
\put(100,50){\framebox(40,30){$(\ldots)^{\Kk}$}}
\put(50,30){\makebox(0,0)[r]{$\boldsymbol{v}(k)$}}
\put(50,235){\makebox(0,0)[r]{$\boldsymbol{v}(k)$}}
\put(115,100){\makebox(0,0)[r]{$\boldsymbol{v}(k)^{\Kk}$}}
\put(75,175){\makebox(0,0)[r]{$\Im\{\boldsymbol{v}(k)\}$}}
\put(75,295){\makebox(0,0)[r]{$\Re\{\boldsymbol{v}(k)\}$}}
\put(305,175){\makebox(0,0)[l]{$\Im\{\boldsymbol{x}(k)\!+\!\boldsymbol{x}_*(k)\}$}}
\put(305,295){\makebox(0,0)[l]{$\Re\{\boldsymbol{x}(k)\!+\!\boldsymbol{x}_*(k)\}$}}
\put(232,33){\makebox(0,0)[lb]{$\boldsymbol{x}(k)$}}
\put(242,103){\makebox(0,0)[lb]{$\boldsymbol{x}_*(k)$}}
\put(330,30){\makebox(0,0)[l]{$\boldsymbol{x}(k)\!+\!\boldsymbol{x}_*(k)$}}
\put(330,235){\makebox(0,0)[l]{$\boldsymbol{x}(k)\!+\!\boldsymbol{x}_*(k)$}}
\put(265,27){\makebox(0,0)[t]{$+$}}
\put(267,45){\makebox(0,0)[l]{$+$}}
\put(245,170){\makebox(0,0)[t]{$+$}}
\put(265,190){\makebox(0,0)[l]{$+$}}
\put(245,300){\makebox(0,0)[b]{$+$}}
\put(265,280){\makebox(0,0)[l]{$+$}}
\end{picture}}
\end{center}\vspace{-20pt}
\setlength{\belowcaptionskip}{-6pt}
\caption{Erweitertes Modellsystem, das alle Ein- und Ausgangskovarianzen ber"ucksichtigt}
\label{E.b3a}
\rule{\textwidth}{0.5pt}\vspace{-10pt}
\end{figure}
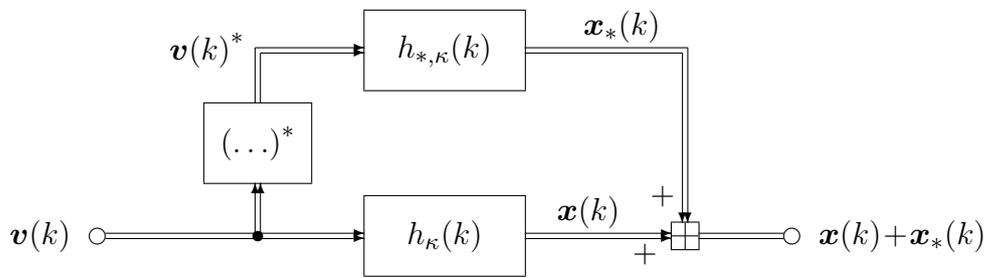

In \cite{Diss} haben wir nur {\em ein}\/ lineares Modellsystem angesetzt. 
Ein solches Systemmodell liefert jedoch nur eine unvollst"andige Aussage 
"uber die Korrelationen, die zwischen den Real- und Imagin"arteilen 
des Ein- und Ausgangssignals am realen System vorhanden sein k"onnen. 
Das erste Teilbild in  Bild~\ref{E.b3a}
 zeigt ein 
erweitertes Modellsystem, bei dem alle vier m"oglichen linearen Verkn"upfungen 
der Real- und Imagin"arteile von Ein- und Ausgang vorhanden sind. 
Es besteht aus den vier reellwertigen linearen Teilsystemen, die durch 
ihre zeitvarianten Impulsantworten \mbox{$h_{\alpha,\kappa}(k)$}, 
\mbox{$h_{\beta,\kappa}(k)$}, \mbox{$h_{\gamma,\kappa}(k)$} und 
\mbox{$h_{\delta,\kappa}(k)$} beschrieben werden. Wie ein Vergleich
der Ein- und Ausgangssignale zeigt, l"asst sich dieses i.~Allg. 
nichtlineare, erweiterte Modellsystem in die beiden im zweiten 
Teilbild dargestellten linearen komplexwertigen Modellsysteme "uberf"uhren,
wenn man f"ur deren zeitvariante Impulsantworten\vspace{-6pt}
\begin{subequations}\label{E.2.1}
\begin{flalign}
&&h_{\kappa}(k)&\;=\;
      \frac{\;h_{\alpha,\kappa}(k)+h_{\delta,\kappa}(k)\,}{2}\;+\;
j\CdoT\frac{\;h_{\gamma,\kappa}(k)-h_{\beta,\kappa}(k)\,}{2}&&
\label{E.2.1.a}\\*[10pt]
\text{und}&&
h_{*,\kappa}(k)&\;=\;
      \frac{\;h_{\alpha,\kappa}(k)-h_{\delta,\kappa}(k)\,}{2}\;+\;
j\CdoT\frac{\;h_{\gamma,\kappa}(k)+h_{\beta,\kappa}(k)\,}{2}&&
\label{E.2.1.b}
\end{flalign}
\end{subequations}
w"ahlt. Somit enth"alt das Systemmodell in Bild \ref{E.b1h} nun zwei lineare, 
stabile, zeitdiskrete und i.~allg. komplexwertige Modellsysteme ${\cal S}_{lin}$ 
und ${\cal S}_{*,lin}$, deren Antworten \mbox{$h_{\kappa}(k)$} und 
\mbox{$h_{*,\kappa}(k)$} auf einen Impuls  \mbox{$v(k)=\gamma_0(k\!-\!\kappa)$} 
zum Zeitpunkt $\kappa$  wir zun"achst als {\em abh"angig}\/ von dem Zeitpunkt $\kappa$ 
ansetzen. Da reale Systeme immer kausal sind, ist es ausreichend Modellsysteme 
anzusetzen, deren Impulsantworten f"ur \mbox{$k\!<\!\kappa$} null sind. 
Desweiteren kann man bei einem realen stabilen System davon ausgehen, dass dessen 
Impulsantwort nach einer hinreichend gro"s gew"ahlten Einschwingzeit $E$ soweit 
abgeklungen ist, dass die Impulsantworten beider Modellsysteme in guter N"aherung 
als zeitlich begrenzt anzusehen sind. Da die Modellsysteme von dem Zufallsprozess 
\mbox{$\boldsymbol{v}(k)$}, bzw. dem konjugierten Zufallsprozess 
\mbox{$\boldsymbol{v}(k)^{\Kk}$} erregt werden, erhalten wir an den Ausg"angen der 
Modellsysteme die Zufallsprozesse
\begin{subequations}\label{E.2.2}
\begin{align}
\boldsymbol{x}(k)\;&=
\Sum{\kappa=k-E}{k}\boldsymbol{v}(\kappa)\CdoT h_{\kappa}(k)
&&\text{und}\label{E.2.2.a}\\[4pt]
\boldsymbol{x}_*(k)\;&=
\Sum{\kappa=k-E}{k}\boldsymbol{v}(\kappa)^{\Kk}\CdoT h_{*,\kappa}(k)
&\qquad&\forall\qquad k\in\mathbb{Z}
\label{E.2.2.b}
\end{align}
\end{subequations}
Der Prozess des Fehlers der Approximation des realen Systems durch 
die Modellsysteme und das deterministische St"orsignal wird mit 
\mbox{$\boldsymbol{n}(k)$} bezeichnet. Er berechnet sich zu
\begin{equation}
\boldsymbol{n}(k)\,=\,\boldsymbol{y}(k)-\boldsymbol{x}(k)-\boldsymbol{x}_*(k)-u(k)
\label{E.2.3}
\end{equation}
Bei der Approximation des realen Systems durch die Modellsysteme 
und das deterministische St"orsignal w"ahlt man die Impulsantworten 
\mbox{$h_{\kappa}(k)$}, \mbox{$h_{*,\kappa}(k)$} und das Signal \mbox{$u(k)$} so, 
dass das zweite Moment\footnote{Also nicht die Varianz, die das zweite 
{\em zentrale}\/ Moment ist} des Approximationsfehlerprozesses 
\mbox{$\boldsymbol{n}(k)$} innerhalb des Beobachtungszeitraums 
\mbox{$0\!\le\!k\!<\!F$} m"oglichst klein wird. Bei realen Systemen 
kann man davon ausgehen, dass dieses Moment immer existiert, da 
der Zufallsprozess \mbox{$\boldsymbol{n}(k)$} in der Amplitude 
begrenzt ist, so dass dessen Verteilungsfunktion unter- bzw. oberhalb 
bestimmter Werte immer $0$ bzw. $1$ ist. Das bei der Berechnung 
des zweiten Moments "uber \mbox{$n(k)^2$} gebildete uneigentliche 
Integral\footnote{Gegebenenfalls ist dieses Integration im Stieltjesschen 
Sinne durchzuf"uhren.} existiert daher immer. F"ur jeden Zeitpunkt $k$ 
erhalten wir eine Minimierungsaufgabe f"ur das zweite Moment des 
Prozesses \mbox{$\boldsymbol{n}(k)$}:
\begin{gather}
\text{E}\big\{|\boldsymbol{n}(k)|^2\big\}\,=\,
\text{E}\Bigg\{\bigg|\,\boldsymbol{y}(k)-\!\!\!\!
\Sum{\kappa=k-E}{k}\!\!\boldsymbol{v}(\kappa)\CdoT h_{\kappa}(k)-\!\!\!
\Sum{\kappa=k-E}{k}\!\!\boldsymbol{v}(\kappa)^{\Kk}\!\CdoT h_{*,\kappa}(k)-
u(k)\bigg|^2\Bigg\}\stackrel{!}{=}\,\text{minimal}
\notag\\*[4pt]
{}\qquad\qquad\qquad\forall\qquad0\!\le\!k\!<\!F\qquad\wedge\qquad k\in\mathbb{Z}.
\label{E.2.4}
\end{gather}
Um die L"osung dieser Minimierungsaufgabe f"ur einen Zeitpunkt $k$ zu erhalten, 
leitet man den zu minimierenden Term nach den zu bestimmenden Gr"o"sen --- den 
Werten der beiden Impulsantworten und des deterministischen St"orsignals ---
partiell ab\footnote{Der zu minimierende Term ist reell. Diesen Term leitet 
man zun"achst nach allen Realteilen, und dann nach allen Imagin"arteilen beider 
Impulsantworten bzw. des deterministischen St"orsignals partiell ab. Man berechnet 
also ganz konventionell die partiellen Ableitungen einer reellen Funktion mehrerer 
reeller Variablen. Man erh"alt so zwei Gleichungen f"ur jeden Wert jeder Impulsantwort
und zwei Gleichungen f"ur jeden Wert des deterministischen St"orsignals.
Die Gleichung, die man bei der partiellen Ableitung nach dem Imagin"arteil
erh"alt, multipliziert man anschie"send mit $j$ und fasst sie mit der Gleichung, 
die man bei der partiellen Ableitung nach dem Realteil erh"alt, zu einer komplexen 
Gleichung zusammen.}. Indem man nun alle partiellen Ableitungen zugleich zu null 
setzt, erh"alt man ein Gleichungssystem, mit dem man die gesuchten Werte der beiden 
Impulsantworten und des deterministischen St"orsignals bestimmen kann. Beim partiellen 
Ableiten nach \mbox{$u(k)$} erh"alt man die Gleichung
\begin{equation}
u(k)\;=\; 
\text{E}\big\{\boldsymbol{y}(k)\big\}-\!\!
\Sum{\kappa=k-E}{k}\!\!\text{E}\big\{\boldsymbol{v}(\kappa)\big\}\CdoT h_{\kappa}(k)-\!\!
\Sum{\kappa=k-E}{k}\!\!\text{E}\big\{\boldsymbol{v}(\kappa)\big\}^{\!\Kk}\!\CdoT h_{*,\kappa}(k).
\label{E.2.5}
\end{equation}
Diese Gleichung besagt, dass \mbox{$u(k)$} gerade so zu w"ahlen ist, dass der 
verbleibende Approximationsfehlerprozess \mbox{$\boldsymbol{n}(k)$} mittelwertfrei ist:
\begin{equation}
\text{E}\big\{\boldsymbol{n}(k)\big\}\;=\;0
\qquad\qquad\qquad\forall\qquad k\in\mathbb{Z}.
\label{E.2.6}
\end{equation}
Dies zeigt man, indem man nach und nach die Gleichungen (\ref{E.2.3}), (\ref{E.2.2}) 
und (\ref{E.2.5}) einsetzt, und dann den Erwartungswert der Summe als die Summe der 
Erwartungswerte der einzelnen Summanden berechnet. Die optimalen Werte \mbox{$u(k)$}
nach Gleichung (\ref{E.2.5}) kann man in alle anderen Gleichungen einsetzen.
\vadjust{\penalty-100}Man erh"alt dadurch f"ur jeden Zeitpunkt $k$ ein lineares Gleichungssystem
zur Bestimmung der Werte der Impulsantworten, das sich in Matrixschreibweise
folgenderma"sen darstellen l"asst:
\begin{equation}
\Big[\;\Vec{h}\;\;\;\Vec{h}_*\;\Big]\cdot\begin{bmatrix}
{\D\,\underline{C}_{\boldsymbol{v},\boldsymbol{v}}}&
{\D\,\underline{C}_{\boldsymbol{v},\boldsymbol{v}^*}}\\
{\D\,\underline{C}_{\boldsymbol{v}^*,\boldsymbol{v}}}&
{\D\,\underline{C}_{\boldsymbol{v}^*,\boldsymbol{v}^*}}
\end{bmatrix}\;=\;
\Big[\;\Vec{C}_{\boldsymbol{y},\boldsymbol{v}}\;\;\;
\Vec{C}_{\boldsymbol{y},\boldsymbol{v}^*}\;\Big].
\label{E.2.7}
\end{equation}
Dabei treten die vier \mbox{$(E\!+\!1)\times(E\!+\!1)$} Kovarianzmatrizen 
$\underline{C}_{\boldsymbol{v},\boldsymbol{v}}$, 
$\underline{C}_{\boldsymbol{v},\boldsymbol{v}^*}$, 
$\underline{C}_{\boldsymbol{v}^*,\boldsymbol{v}}$ und 
$\underline{C}_{\boldsymbol{v}^*,\boldsymbol{v}^*}$, 
sowie die zwei \mbox{$1\times(E\!+\!1)$} Kovarianzvektoren 
$\Vec{C}_{\boldsymbol{y},\boldsymbol{v}}$ und 
$\Vec{C}_{\boldsymbol{y},\boldsymbol{v}^*}$ und die zwei 
\mbox{$1\times(E\!+\!1)$} Zeilenvektoren $\Vec{h}$ und $\Vec{h}_*$ 
der Werte der Impulsantworten auf. Wie deren Elemente definiert sind, 
zeigt die nachfolgende Tabelle:
\[\begin{array}{||c|c|c|l||}
\hline
\hline
&\!\text{Zeile}\!&\!\text{Spalte}\!&\text{Wert}\\
\hline
\rule[-7pt]{0pt}{25pt}\underline{C}_{\boldsymbol{v},\boldsymbol{v}}&i&j&
\text{E}\Big\{\!\big(\boldsymbol{v}(k\!+\!1\!-\!i)\!-\!\text{E}\{\boldsymbol{v}(k\!+\!1\!-\!i)\}\big)
\CdoT\big(\boldsymbol{v}(k\!+\!1\!-\!j)\!-\!\text{E}\{\boldsymbol{v}(k\!+\!1\!-\!j)\}\big)^{\!\Kk}\Big\}\!\\
\rule[-7pt]{0pt}{25pt}\underline{C}_{\boldsymbol{v},\boldsymbol{v}^*}&i&j&
\text{E}\Big\{\!\big(\boldsymbol{v}(k\!+\!1\!-\!i)\!-\!\text{E}\{\boldsymbol{v}(k\!+\!1\!-\!i)\}\big)
\CdoT\big(\boldsymbol{v}(k\!+\!1\!-\!j)\!-\!\text{E}\{\boldsymbol{v}(k\!+\!1\!-\!j)\}\big)\Big\}\\
\rule[-7pt]{0pt}{25pt}\underline{C}_{\boldsymbol{v}^*,\boldsymbol{v}}&i&j&
\text{E}\Big\{\!\big(\boldsymbol{v}(k\!+\!1\!-\!i)\!-\!\text{E}\{\boldsymbol{v}(k\!+\!1\!-\!i)\}\big)^{\!\Kk}
\!\CdoT\big(\boldsymbol{v}(k\!+\!1\!-\!j)\!-\!\text{E}\{\boldsymbol{v}(k\!+\!1\!-\!j)\}\big)^{\!\Kk}\Big\}\!\!\\
\rule[-12pt]{0pt}{30pt}\underline{C}_{\boldsymbol{v}^*,\boldsymbol{v}^*}\!&i&j&
\text{E}\Big\{\!\big(\boldsymbol{v}(k\!+\!1\!-\!i)\!-\!\text{E}\{\boldsymbol{v}(k\!+\!1\!-\!i)\}\big)^{\!\Kk}
\!\CdoT\big(\boldsymbol{v}(k\!+\!1\!-\!j)\!-\!\text{E}\{\boldsymbol{v}(k\!+\!1\!-\!j)\}\big)\Big\}\\
\hline
\rule[-7pt]{0pt}{25pt}\Vec{C}_{\boldsymbol{y},\boldsymbol{v}}&1&j&
\text{E}\Big\{\big(\boldsymbol{y}(k)-\text{E}\{\boldsymbol{y}(k)\}\big)
\cdot\big(\boldsymbol{v}(k\!+\!1\!-\!j)-\text{E}\{\boldsymbol{v}(k\!+\!1\!-\!j)\}\big)^{\!\Kk}\Big\}\\
\rule[-12pt]{0pt}{30pt}\Vec{C}_{\boldsymbol{y},\boldsymbol{v}^*}&1&j&
\text{E}\Big\{\big(\boldsymbol{y}(k)-\text{E}\{\boldsymbol{y}(k)\}\big)
\cdot\big(\boldsymbol{v}(k\!+\!1\!-\!j)-\text{E}\{\boldsymbol{v}(k\!+\!1\!-\!j)\}\big)\Big\}\\
\hline
\rule[-7pt]{0pt}{21pt}\Vec{h}&1&j&h_{k+1-j}(k)\\
\rule[-7pt]{0pt}{21pt}\Vec{h}_*&1&j&h_{*,k+1-j}(k)\\
\hline
\hline
\end{array}\]
Da das lineare Gleichungssystem (\ref{E.2.7}) durch partielles Ableiten eines 
stets positiven Terms entstanden ist, und da die Minimierungsparameter in diesem 
Term nur linear oder quadratisch auftreten, existiert f"ur jeden einzelnen Zeitpunkt $k$ 
wenigstens eine L"osung des Gleichungssystems, selbst wenn die Matrix singul"ar 
ist. Gegebenenfalls kann sich auch ein ein- oder mehrdimensionaler L"osungsraum 
ergeben. Nur wenn es eine L"osung gibt, die alle Gleichungssysteme f"ur alle 
Zeitpunkte $k$ zugleich l"ost, macht es Sinn, zeitinvariante Modellsysteme anzusetzen. 
Im Fall eines im weiten Sinne zyklostation"aren Verbundprozesses\footnote{Ein 
station"arer Prozess ist hier der Sonderfall eines zyklostation"aren Prozesses mit
der Periodizit"at Eins.} aus \mbox{$\boldsymbol{v}(k)$} und  \mbox{$\boldsymbol{y}(k)$} 
"andert sich sowohl die Matrix des Gleichungssystems als auch der Vektor auf der 
rechten Seite nicht, wenn man den Zeitpunkt $k$ der Minimierung um ein ganzzahliges 
Vielfaches der Periode der Zyklostationarit"at ver"andert. Es existiert dann immer eine 
periodisch zeitvariante L"osung f"ur die Impulsantworten der Modellsysteme. Da nur 
zweite zentrale Momente in die Matrix des Gleichungssystems und den Vektor auf der 
rechten Seite eingehen, gibt es selbst dann eine periodisch zeitvariante L"osung 
f"ur die Impulsantworten der Modellsysteme, wenn sich damit in Gleichung 
(\ref{E.2.5}) eine nicht periodische, zeitabh"angige, deterministische St"orung 
\mbox{$u(k)$} ergibt. Auch falls der Verbundprozess aus \mbox{$\boldsymbol{v}(k)$} 
und  \mbox{$\boldsymbol{y}(k)$} nicht im weiten Sinn zyklostation"ar ist, kann es 
sein, dass es eine periodisch zeitvariante L"osung f"ur die Impulsantworten der 
Modellsysteme gibt. Dies ist dann jedoch anhand heuristischer "Uberlegungen zu 
zeigen, wenn man das RKM zur Vermessung solcher Systeme verwenden will. Im weiteren 
gehen wir davon aus, dass nur solche Systeme untersucht werden, f"ur die eine 
periodisch zeitvariante L"osung existiert. Nur dann ist es sinnvoll, diese mit Hilfe 
des RKM messtechnisch abzusch"atzen. Als periodisch zeitvariant wird die L"osung 
f"ur die Impulsantworten der Modellsysteme dann bezeichnet, wenn eine Verschiebung 
des Zeitpunktes $k$, f"ur den die Minimierungsaufgabe zu l"osen ist, um eine feste 
Zeitdifferenz $K_H$ lediglich zu einer zeitlichen Verschiebung der L"osung um 
dieselbe Zeitdifferenz f"uhrt, so dass sich die L"osungen mit $K_H$ periodisch 
wiederholen. Die L"osungen selbst sind jedoch i.~Allg. keine periodischen Folgen. 
F"ur die mit dieser Variante des RKM untersuchbaren Systeme muss also 
\begin{subequations}\label{E.2.8}
\begin{flalign}
&&h_{\kappa+K_H}(k\!+\!K_H)&\;=\;h_{\kappa}(k)&&&&
\label{E.2.8.a}\\*[4pt]
\text{und}&&h_{*,\kappa+K_H}(k\!+\!K_H)&\;=\;h_{*,\kappa}(k)&
\qquad\qquad&\forall\qquad k,\kappa\in\mathbb{Z}&&
\label{E.2.8.b}
\end{flalign}
\end{subequations}
gelten. Da wir uns auf kausale Systeme mit zeitlich begrenzter Impulsantwort 
beschr"ankt hatten, gilt weiterhin: 
\begin{equation}
h_{\kappa}(k)\;=\;h_{*,\kappa}(k)\;=\;0
\qquad\qquad\qquad\forall\qquad k\!-\!\kappa\notin[0;E].
\label{E.2.9}
\end{equation}
Bei solchen Systemen gen"ugt es ---\,wie im Fall des
zeitinvarianten Systems\,--- die Erregung im Zeitintervall
\mbox{$-E\!\le\!k\!<\!F$} an das zu vermessende System anzulegen,
da die Werte der Erregung au"serhalb dieses Intervalls nicht in die
Minimierung des zweiten Moments des Approximationsfehlers nach 
Gleichung~(\ref{E.2.4}) eingehen. 

Wenn man bei einem realen System, bei dem der Ausgangsprozess mit dem konjugierten 
Eingangsprozess in der Art korreliert ist, dass sich bei der eben dargestellten 
theoretischen L"osung eine von null verschiedene Impulsantwort f"ur das 
Modellsystem ${\cal S}_{*,lin}$ ergibt, lediglich das eine Modellsystem ${\cal S}_{lin}$ 
ansetzt, so erh"alt man ebenfalls eine L"osung minimaler Approximationsfehlerleistung.
Diese L"osung wird in aller Regel aber von der L"osung des vollst"andigen Systems 
abweichen und eine h"ohere Approximationsfehlerleistung aufweisen. Auch wenn man 
das deterministische St"orsignal wegl"asst, obwohl beim vollst"andigen Systemmodell 
\mbox{$u(k)$} eine von null verschiedene L"osung annehmen w"urde, wird sich zwar 
eine L"osung minimaler, aber dennoch gr"o"serer Approximationsfehlerleistung ergeben, 
die von der L"osung bei vollst"andigem Systemmodell normalerweise abweicht. Da das 
RKM bestenfalls die theoretische L"osung erwartungstreu absch"atzen kann, ist es 
auch f"ur die Messung wichtig, das zum realen System passende Systemmodell zu w"ahlen. 
Wenn man sich nicht sicher ist, ob \mbox{$u(k)$} oder das Modellsystem ${\cal S}_{*,lin}$ 
weggelassen werden kann, sollte man sich im Zweifel f"ur das vollst"andige Systemmodell 
entscheiden. Die Messergebnisse werden dann zeigen, ob eine einfachere Modellierung auch 
m"oglich w"are.

Wie beim zeitinvarianten Modellsystem in \cite{Diss} wird auch hier 
wieder eine gem"a"s Gleichung (\myref{2.8}) mit $M$ periodische Erregung 
verwendet. Die Periode $M$ wird dabei als ganzzahliges Vielfaches von 
$K_H$ gew"ahlt, so dass \mbox{$M/K_H\in{}\mathbb{N}$} gilt. Wenn die Erregung 
im gesamten Zeitintervall \mbox{$-E\!\le\!k\!<\!F$} anliegt, kann man 
f"ur die Ausgangssignale der periodischen zeitvarianten Modellsysteme
\begin{subequations}\label{E.2.10}
\begin{gather}
\boldsymbol{x}(k)\;=
\Sum{\kappa=-\infty}{\infty}\!
\boldsymbol{v}(\kappa)\CdoT h_{\kappa}(k)\;=\,
\Sum{\kappa=0}{M-1}\;\Sum{\Tilde{\kappa}=-\infty}{\infty}\!
\boldsymbol{v}(\kappa\!+\!\Tilde{\kappa}\CdoT M)\CdoT
h_{\kappa+\Tilde{\kappa}\cdot M}(k)\;=
\label{E.2.10.a}\\[2pt]
=\;\Sum{\kappa=0}{M-1}\boldsymbol{v}(\kappa)\cdoT\!\!
\Sum{\Tilde{\kappa}=-\infty}{\infty}\!
h_{\kappa}(k\!-\!\Tilde{\kappa}\CdoT M)\;=
\Sum{\kappa=0}{M-1}
\boldsymbol{v}(\kappa)\CdoT\Tilde{h}_{\kappa}(k)\;=
\notag\\[6pt]
=\;\frac{1}{M}\CdoT\Sum{\mu=0}{M-1}\;\Sum{\Hat{\mu}=0}{K_H-1}\!
H\!\big({\T \mu,\mu\!+\!\Hat{\mu}\CdoT\frac{M}{K_H}}\big)\cdoT
\Sum{\kappa=0}{M-1}\boldsymbol{v}(\kappa)\cdot
e^{j\cdot\frac{2\pi}{M}\cdot\mu\cdot(k-\kappa)}\cdot
e^{\!-j\cdot\frac{2\pi}{K_H}\cdot\Hat{\mu}\cdot\kappa}\;=
\notag\\[4pt]
=\;\frac{1}{M}\CdoT\Sum{\mu=0}{M-1}\;\Sum{\Hat{\mu}=0}{K_H-1}
H\!\big({\T \mu,\mu\!+\!\Hat{\mu}\CdoT\frac{M}{K_H}}\big)\CdoT
\boldsymbol{V}\!\big({\T\mu\!+\!\Hat{\mu}\CdoT\frac{M}{K_H}}\big)\cdot
e^{j\cdot\frac{2\pi}{M}\cdot\mu\cdot k}\notag
\end{gather}
und analog 
\begin{gather}
\boldsymbol{x}_*(k)\;=\!
\Sum{\kappa=-\infty}{\infty}\!\!
\boldsymbol{v}(\kappa)^{\Kk}\!\CdoT h_{*,\kappa}(k)\;=\;
\frac{1}{M}\CdoT\Sum{\mu=0}{M-1}\;\Sum{\Hat{\mu}=0}{K_H-1}\!\!
H_*\!\big({\T \mu,\mu\!+\!\Hat{\mu}\CdoT\frac{M}{K_H}}\big)\CdoT
\boldsymbol{V}\!\big({\T-\mu\!-\!\Hat{\mu}\CdoT\frac{M}{K_H}}\big)^{\!\Kk}\!\Cdot
e^{j\cdot\frac{2\pi}{M}\cdot\mu\cdot k}\notag\\
\forall\qquad 0\le k< F\label{E.2.10.b}
\end{gather}
\end{subequations}
schreiben. Dabei treten die aus den Impulsantworten durch periodische 
Fortsetzung gewonnenen periodischen Folgen
\begin{subequations}\label{E.2.11}
\begin{gather}
\Tilde{h}_{\kappa}(k)\;=\!
\Sum{\Tilde{\kappa}=-\infty}{\infty}\!
h_{\kappa}(k\!-\!\Tilde{\kappa}\CdoT M)\;=
\label{E.2.11.a}\\
=\;\frac{1}{M}\cdoT\Sum{\mu=0}{M-1}\;\Sum{\Hat{\mu}=0}{K_H-1}\!
H\!\big({\T \mu,\mu\!+\!\Hat{\mu}\CdoT\frac{M}{K_H}}\big)\cdot
e^{ j\cdot\frac{2\pi}{M}\cdot\mu\cdot(k-\kappa)}\cdot
e^{\!-j\cdot\frac{2\pi}{K_H}\cdot\Hat{\mu}\cdot\kappa}
\notag
\end{gather}
und\vspace{-12pt}
\begin{gather}
\Tilde{h}_{*,\kappa}(k)\;=\!
\Sum{\Tilde{\kappa}=-\infty}{\infty}\!
h_{*,\kappa}(k\!-\!\Tilde{\kappa}\CdoT M)\;=
\label{E.2.11.b}\\
=\;\frac{1}{M}\cdoT\Sum{\mu=0}{M-1}\;\Sum{\Hat{\mu}=0}{K_H-1}\!
H_*\!\big({\T \mu,\mu\!+\!\Hat{\mu}\CdoT\frac{M}{K_H}}\big)\cdot
e^{ j\cdot\frac{2\pi}{M}\cdot\mu\cdot(k-\kappa)}\cdot
e^{\!-j\cdot\frac{2\pi}{K_H}\cdot\Hat{\mu}\cdot\kappa}
\notag
\end{gather}
\end{subequations}
sowie die daraus durch diskrete Fouriertransformation bez"uglich $k$ 
und inverse diskrete Fouriertransformation bez"uglich $\kappa$ 
gewonnenen diskreten Werte der bifrequenten "Ubertragungsfunktionen 
\begin{subequations}\label{E.2.12}
\begin{equation}
H(\mu,\mu_1)\,=\,\frac{1}{M}\cdoT\Sum{\kappa=0}{M-1}\;\Sum{k=0}{M-1}\,
\Tilde{h}_{\kappa}(k)\cdot
e^{\!-j\cdot\frac{2\pi}{M}\cdot\mu\cdot k}\cdot
e^{ j\cdot\frac{2\pi}{M}\cdot\mu_1\cdot\kappa}
\label{E.2.12.a}
\end{equation}
und 
\begin{equation}
H_*(\mu,\mu_1)\,=\,\frac{1}{M}\cdoT\Sum{\kappa=0}{M-1}\;\Sum{k=0}{M-1}\,
\Tilde{h}_{*,\kappa}(k)\cdot
e^{\!-j\cdot\frac{2\pi}{M}\cdot\mu\cdot k}\cdot
e^{ j\cdot\frac{2\pi}{M}\cdot\mu_1\cdot\kappa}
\label{E.2.12.b}
\end{equation}
\end{subequations}
auf. Setzen wir in diese Gleichungen 
\mbox{$\mu_1=\mu\!+\!\Tilde{\mu}\!+\!\Hat{\mu}\CdoT M/K_H$} ein, so 
erhalten wir, wenn wir die Periodizit"at gem"a"s der Gleichungen~(\ref{E.2.8})
ber"ucksichtigen, f"ur die Werte der diskreten bifrequenten
"Ubertragungsfunktionen die Aussagen
\begin{subequations}\label{E.2.13}
\begin{gather}
H\!\big({\T \mu,\mu\!+\!\Tilde{\mu}\!+\!\Hat{\mu}\CdoT\frac{M}{K_H}}\big)\,=\,
\begin{cases}
{\D\;\frac{1}{K_H}\cdoT\Sum{k=0}{M-1}\;\Sum{\kappa=0}{K_H-1}\!
\Tilde{h}_{\kappa}(k\!+\!\kappa)\CdoT
e^{ j\cdot\frac{2\pi}{K_H}\cdot\Hat{\mu}\cdot\kappa}\CdoT
e^{\!-j\cdot\frac{2\pi}{M}\cdot\mu\cdot k}}&
\text{ f"ur }\;\;\Tilde{\mu}=0\\
\;0&\text{ f"ur }0 \!<\!\Tilde{\mu}\!<\!\frac{M}{K_H}
\end{cases}\notag\\*[4pt]
\forall\qquad\mu,\Hat{\mu},\Tilde{\mu}\in\mathbb{Z},
\label{E.2.13.a}
\end{gather}
und 
\begin{gather}
H_*\!\big({\T \mu,\mu\!+\!\Tilde{\mu}\!+\!\Hat{\mu}\CdoT\frac{M}{K_H}}\big)\,=\,
\begin{cases}
{\D\,\frac{1}{K_H}\cdoT\!\Sum{k=0}{M-1}\;\Sum{\kappa=0}{K_H-1}\!
\Tilde{h}_{*,\kappa}(k\!+\!\kappa)\CdoT
e^{ j\cdot\frac{2\pi}{K_H}\cdot\Hat{\mu}\cdot\kappa}\CdoT
e^{\!-j\cdot\frac{2\pi}{M}\cdot\mu\cdot k}}&
\text{ f"ur }\;\;\Tilde{\mu}=0\\
\,0&\text{f"ur }0\!<\!\Tilde{\mu}\!<\!\frac{M}{K_H}
\end{cases}\notag\\*[4pt]
\forall\qquad\mu,\Hat{\mu},\Tilde{\mu}\in\mathbb{Z},
\label{E.2.13.b}
\end{gather}
\end{subequations}
die besagen, dass nur die Werte von null verschieden sein k"onnen, deren 
Argumente $\mu$ und $\mu_1$ eine Differenz aufweisen, die ein ganzzahliges 
Vielfaches von \mbox{$M/K_H$} ist. F"ur einen festen Wert von $\mu$ 
werden die Werte der bifrequenten "Ubertragungsfunktionen mit 
\mbox{$\mu_1=\mu\!+\!\Hat{\mu}\CdoT M/K_H$}, die von null verschieden sein 
k"onnen, zu einem Zeilenvektor zusammengefasst: 
\begin{equation}
\Vec{H}(\mu)\;=\;
\begin{bmatrix}
H(\mu,\mu)\\[2pt]
H\!\big({\T\mu,\mu\!+\!\frac{M}{K_H}}\big)\\[-2pt]
\vdots\\[-2pt]
H\!\big({\T\mu,\mu\!+\!(K_H\!-\!1)\CdoT\frac{M}{K_H}}\big)\\[2pt]
H_*(\mu,\mu)\\[2pt]
H_*\big({\T\mu,\mu\!+\!\frac{M}{K_H}}\big)\\[-2pt]
\vdots\\[-2pt]
H_*\big({\T\mu,\mu\!+\!(K_H\!-\!1)\CdoT\frac{M}{K_H}}\big)
\end{bmatrix}^{\Tt}
\qquad\qquad\forall\qquad\mu=0\;(1)\;M\!-\!1.
\label{E.2.14}
\end{equation}
Die Definition der bifrequenten "Ubertragungsfunktionen wurde so gew"ahlt, 
dass f"ur den Fall eines zeitinvarianten Systems mit \mbox{$K_H\!=\!1$} die 
Hauptdiagonalenelemente der bifrequenten "Ubertragungsfunktionen gerade 
die Werte \mbox{$H(\mu\CdoT 2\pi/M)$} bzw. \mbox{$H_*(\mu\CdoT 2\pi/M)$}  
der "Ubertragungsfunktionen ohne einen weiteren von eins verschiedenen 
Faktor sind.  Es sei wieder darauf hingewiesen, dass "ahnlich wie beim 
Fall eines zeitinvarianten Modellsystems aufgrund des Abtasttheorems 
durch die f"ur \mbox{$H(\mu,\mu\!+\!\Hat{\mu}\CdoT M/K_H)$} und 
\mbox{$H_*(\mu,\mu\!+\!\Hat{\mu}\CdoT M/K_H)$} approximierten Gr"o"sen 
nur die Werte der periodisch fortgesetzten Impulsantworten 
\mbox{$\Tilde{h}_{\kappa}(k)$} und \mbox{$\Tilde{h}_{*,\kappa}(k)$} festgelegt 
sind. Nur wenn heuristische "Uberlegungen vermuten lassen, dass alle \mbox{$2\CdoT K_H$} 
Impulsantworten der beiden realen Systeme auf ein Intervall der L"ange $M$
beschr"ankt sind, k"onnen auch die Impulsantworten \mbox{$h_{\kappa}(k)$}
und \mbox{$h_{*,\kappa}(k)$} eindeutig aus den $2\CdoT M\CdoT K_H$ 
Optimalwerten der "Ubertragungsfunktionen bestimmt werden.

In den Gleichungen~(\ref{E.2.10}) kommt auch noch das diskrete Spektrum der
periodischen Erregung, das nach Gleichung~(\myref{2.7}) definiert ist, vor.
Auch einige Zufallsgr"o"sen dieses Spektrums lassen sich zu einen
Spaltenvektor
\begin{equation}
\Tilde{\Vec{\boldsymbol{V}}}(\mu)\;=\;
\begin{bmatrix}
\boldsymbol{V}(\mu)\\[2pt]
\boldsymbol{V}\big({\T\mu\!+\!\frac{M}{K_H}}\big)\\[-2pt]
\vdots\\[-2pt]
\boldsymbol{V}\big({\T\mu\!+\!(K_H\!-\!1)\CdoT\frac{M}{K_H}}\big)\\[2pt]
\boldsymbol{V}(\!-\mu)^{\Kk}\\[2pt]
\boldsymbol{V}\big(\!{\T-\mu\!-\!\frac{M}{K_H}}\big)^{\!\Kk}\\[-2pt]
\vdots\\[-2pt]
\boldsymbol{V}\big(\!{\T-\mu\!-\!(K_H\!-\!1)\CdoT\frac{M}{K_H}}\big)^{\!\Kk}
\end{bmatrix}
\qquad\qquad\forall\qquad\mu=0\;(1)\;M\!-\!1.
\label{E.2.15}
\end{equation}
zusammenfassen. Mit den Vektoren \mbox{$\Tilde{\Vec{\boldsymbol{V}}}(\mu)$}
und \mbox{$\Vec{H}(\mu)$} erhalten wir f"ur den Summenprozess am Ausgang der
beiden periodisch zeitvarianten Modellsysteme die kompaktere Schreibweise\vspace{-8pt}
\begin{equation}
\boldsymbol{x}(k)+\boldsymbol{x}_*(k)\;=\;
\frac{1}{M}\cdoT\Sum{\mu=0}{M-1}\Vec{H}(\mu)\CdoT
\Tilde{\Vec{\boldsymbol{V}}}(\mu)\cdot
e^{j\cdot\frac{2\pi}{M}\cdot\mu\cdot k}
\qquad\qquad\forall\qquad 0\le k< F.
\label{E.2.16}
\end{equation}
Setzt man \mbox{$\boldsymbol{x}(k)+\boldsymbol{x}_*(k)$} in den Approximationsfehler 
nach Gleichung~(\ref{E.2.3}), so erh"alt man zun"achst\vspace{-8pt}
\begin{equation}
\boldsymbol{n}(k)\,=\,
\boldsymbol{y}(k)-
\frac{1}{M}\cdoT\Sum{\mu=0}{M-1}\Vec{H}(\mu)\CdoT
\Tilde{\Vec{\boldsymbol{V}}}(\mu)\cdot
e^{j\cdot\frac{2\pi}{M}\cdot\mu\cdot k}-u(k),
\label{E.2.17}
\end{equation}
und damit im Zeitbereich der Messung die modifizierte Minimierungsaufgabe
\begin{gather}
\text{E}\big\{\,|\boldsymbol{n}(k)|^2\big\}\;=\;
\text{E}\Bigg\{\,\bigg|\,\boldsymbol{y}(k)\,-
\frac{1}{M}\cdoT\Sum{\mu=0}{M-1}
\Vec{H}(\mu)\CdoT\Tilde{\Vec{\boldsymbol{V}}}(\mu)\CdoT
e^{j\cdot\frac{2\pi}{M}\cdot\mu\cdot k}\,-\,u(k)\,\bigg|^2\Bigg\}\,
\stackrel{!}{=}\,\text{minimal}\notag\\*[4pt]
\forall\qquad k=0\;(1)\;F\!-\!1.
\label{E.2.18}
\end{gather}
Um die Optimall"osung f"ur \mbox{$u(k)$} zu bestimmen, leitet man diese 
Gleichung f"ur alle Werte von $k$ in der oben beschriebenen Art nach 
\mbox{$u(k)$} ab. Die optimale L"osung f"ur die Regressionskoeffizienten 
\mbox{$u(k)$} ergibt sich auch hier, wenn der Erwartungswert von 
\mbox{$\boldsymbol{n}(k)$} f"ur alle Zeitpunkte $k$ null wird:\vspace{-8pt}
\begin{gather}
u(k)\;=\;\text{E}\{\boldsymbol{y}(k)\}-
\frac{1}{M}\cdoT\Sum{\mu=0}{M-1}
\Vec{H}(\mu)\CdoT\text{E}\big\{\Tilde{\Vec{\boldsymbol{V}}}(\mu)\big\}\CdoT
e^{j\cdot\frac{2\pi}{M}\cdot\mu\cdot k}\notag\\*[4pt]
\forall\qquad k=0\;(1)\;F\!-\!1.
\label{E.2.19}
\end{gather}
Dieselbe L"osung erh"alt man, wenn man in die allgemeinere L"osung~(\ref{E.2.5}) 
den im Zeitintervall \mbox{$-E\!\le\!k\!<\!F$} gem"a"s (\myref{2.8}) periodischen 
Eingangsprozess einsetzt, und die periodische Zeitvarianz des Systems ber"ucksichtigt, 
indem man nach und nach die Gleichungen~(\ref{E.2.8}) bis (\ref{E.2.15}) verwendet.

Um die L"osungen f"ur die  beiden bifrequenten "Ubertragungsfunktionen 
zu erhalten, setzen wir zun"achst die optimale L"osung f"ur \mbox{$u(k)$} in 
den zu minimierenden, reellen Term~(\ref{E.2.18}) ein. Dann leiten wir diesen 
Term partiell nach Real- und Imagin"arteil der \mbox{$M\CdoT K_H$} 
Werte der bifrequenten "Ubertragungsfunktion
\mbox{$H(\Tilde{\mu},\Tilde{\mu}\!+\!\Hat{\Tilde{\mu}}\CdoT M/K_H)$} 
ab, uns setzen die Ableitungen mit null gleich. Wir erhalten so insgesamt 
\mbox{$2\CdoT M\CdoT K_H$} Gleichungen f"ur jeden der $F$ Zeitpunkte $k$.
Jeweils zwei dieser Gleichungen fassen wir danach zu einer komplexen
Gleichung zusammen. Wir erhalten so die \mbox{$F\CdoT M\CdoT K_H$}
komplexen Gleichungen:\vspace{4pt}
\begin{subequations}\label{E.2.20}
\begin{gather}
\begin{aligned}
\frac{1}{M}\cdoT\Sum{\mu=0}{M-1}\;\Sum{\Hat{\mu}=0}{K_H-1}\!
\Bigg(H\!\big({\T\mu,\mu\!+\!\Hat{\mu}\CdoT\frac{M}{K_H}}\big)\cdot
\text{E}\bigg\{&
\Big(\boldsymbol{V}\!\big({\T\mu\!+\!\Hat{\mu}\CdoT\frac{M}{K_H}}\big)\!-\!
\text{E}\big\{\boldsymbol{V}\!\big({\T\mu\!+\!\Hat{\mu}\CdoT\frac{M}{K_H}}\big)
\big\}\Big)\Cdot\\*
{}\cdot\,&\Big(\boldsymbol{V}\!\big({\T\Tilde{\mu}\!+\!\Hat{\Tilde{\mu}}\CdoT\frac{M}{K_H}}\big)\!-\!
\text{E}\big\{\boldsymbol{V}\!\big({\T\Tilde{\mu}\!+\!\Hat{\Tilde{\mu}}\CdoT\frac{M}{K_H}}\big)
\big\}\Big)^{\!\!\Kk}\bigg\}\;+{}
\notag\\[2pt]
H_*\!\big({\T\mu,\mu\!+\!\Hat{\mu}\CdoT\frac{M}{K_H}}\big)\cdot
\text{E}\bigg\{&
\Big(\boldsymbol{V}\!\big({\T-\mu\!-\!\Hat{\mu}\CdoT\frac{M}{K_H}}\big)\!-\!
\text{E}\big\{\boldsymbol{V}\!\big({\T-\mu\!-\!\Hat{\mu}\CdoT\frac{M}{K_H}}\big)
\big\}\Big)\Cdot\\*
{}\cdot\,&\Big(\boldsymbol{V}\!\big({\T\Tilde{\mu}\!+\!\Hat{\Tilde{\mu}}\CdoT\frac{M}{K_H}}\big)\!-\!
\text{E}\big\{\boldsymbol{V}\!\big({\T\Tilde{\mu}\!+\!\Hat{\Tilde{\mu}}\CdoT\frac{M}{K_H}}\big)
\big\}\!\Big)\!\bigg\}^{\!\!\Kk}\Bigg)\cdot
e^{j\cdot\frac{2\pi}{M}\cdot\mu\cdot k}\;=
\end{aligned}\notag\\[10pt]
=\;\text{E}\bigg\{\Big(\boldsymbol{y}(k)\!-\!
\text{E}\{\boldsymbol{y}(k)\}\Big)\cdot
\Big(\boldsymbol{V}\!
\big({\T\Tilde{\mu}\!+\!\Hat{\Tilde{\mu}}\CdoT\frac{M}{K_H}}\big)\!-\!
\text{E}\big\{\boldsymbol{V}\!
\big({\T\Tilde{\mu}\!+\!\Hat{\Tilde{\mu}}\CdoT\frac{M}{K_H}}\big)
\big\}\Big)^{\!\!\Kk}\bigg\}\notag\\*[10pt]
\forall\qquad k=0\;(1)\;F\!-\!1,\quad\Tilde{\mu}=0\;(1)\;M\!-\!1
\quad\text{ und }\quad\Hat{\Tilde{\mu}}=0\;(1)\;K_H\!-\!1.
\label{E.2.20.a}
\end{gather}
F"ur die partiellen Ableitungen nach Real- und Imagin"arteil der 
\mbox{$M\CdoT K_H$} Werte der anderen bifrequenten "Ubertragungsfunktion  
\mbox{$H_*(\Tilde{\mu},\Tilde{\mu}\!+\!\Hat{\Tilde{\mu}}\CdoT M/K_H)$}
erhalten wir analog die \mbox{$F\CdoT M\CdoT K_H$}
komplexen Gleichungen:
\begin{gather}
\begin{aligned}
\frac{1}{M}\cdoT\Sum{\mu=0}{M-1}\;\Sum{\Hat{\mu}=0}{K_H-1}\!
\Bigg(H\!\big({\T\mu,\mu\!+\!\Hat{\mu}\CdoT\frac{M}{K_H}}\big)\cdot
\text{E}\bigg\{&
\Big(\boldsymbol{V}\!\big({\T\mu\!+\!\Hat{\mu}\CdoT\frac{M}{K_H}}\big)\!-\!
\text{E}\big\{\boldsymbol{V}\!\big({\T\mu\!+\!\Hat{\mu}\CdoT\frac{M}{K_H}}\big)
\big\}\Big)\Cdot\\*
{}\cdot\,&\Big(\boldsymbol{V}\!\big({\T\Tilde{\mu}\!+\!\Hat{\Tilde{\mu}}\CdoT\frac{M}{K_H}}\big)\!-\!
\text{E}\big\{\boldsymbol{V}\!\big({\T\Tilde{\mu}\!+\!\Hat{\Tilde{\mu}}\CdoT\frac{M}{K_H}}\big)
\big\}\Big)\bigg\}\;+{}
\notag\\[2pt]
H_*\!\big({\T\mu,\mu\!+\!\Hat{\mu}\CdoT\frac{M}{K_H}}\big)\cdot
\text{E}\bigg\{&
\Big(\boldsymbol{V}\!\big({\T-\mu\!-\!\Hat{\mu}\CdoT\frac{M}{K_H}}\big)\!-\!
\text{E}\big\{\boldsymbol{V}\!\big({\T-\mu\!-\!\Hat{\mu}\CdoT\frac{M}{K_H}}\big)
\big\}\Big)^{\!\!\Kk}\Cdot\\*
{}\cdot\,&\Big(\boldsymbol{V}\!\big({\T\Tilde{\mu}\!+\!\Hat{\Tilde{\mu}}\CdoT\frac{M}{K_H}}\big)\!-\!
\text{E}\big\{\boldsymbol{V}\!\big({\T\Tilde{\mu}\!+\!\Hat{\Tilde{\mu}}\CdoT\frac{M}{K_H}}\big)
\big\}\Big)\!\bigg\}\!\Bigg)\cdot
e^{j\cdot\frac{2\pi}{M}\cdot\mu\cdot k}\;=
\end{aligned}\notag\\[8pt]
=\;\text{E}\bigg\{\Big(\boldsymbol{y}(k)\!-\!
\text{E}\{\boldsymbol{y}(k)\}\Big)\cdot
\Big(\boldsymbol{V}\!
\big({\T\Tilde{\mu}\!+\!\Hat{\Tilde{\mu}}\CdoT\frac{M}{K_H}}\big)\!-\!
\text{E}\big\{\boldsymbol{V}\!
\big({\T\Tilde{\mu}\!+\!\Hat{\Tilde{\mu}}\CdoT\frac{M}{K_H}}\big)
\big\}\Big)\bigg\}\notag\\*[6pt]
\forall\qquad k=0\;(1)\;F\!-\!1,\quad\Tilde{\mu}=0\;(1)\;M\!-\!1
\quad\text{ und }\quad\Hat{\Tilde{\mu}}=0\;(1)\;K_H\!-\!1.
\label{E.2.20.b}
\end{gather}
\end{subequations}
Bei den Gleichungen aller partiellen Ableitungen stellen wir fest, dass eine 
Erh"ohung der diskreten Frequenzvariable \mbox{$\Tilde{\mu}$} um ein ganzzahliges 
Vielfaches $n$ von \mbox{$M/K_H$} zu derselben Gleichung f"uhrt, wie eine 
Erh"ohung der diskreten Variable \mbox{$\Hat{\Tilde{\mu}}$} um dieselbe ganze 
Zahl $n$. Dies liegt daran, dass in dem gesamten Gleichungssystem nur die Summe
\mbox{$\Tilde{\mu}\!+\!\Hat{\Tilde{\mu}}\CdoT M/K_H$} vorkommt, und niemals eine 
der Variablen allein. Wir k"onnen daher alle mehrfach aufgef"uhrten Gleichungen 
eliminieren, indem wir nur die Gleichungen mit \mbox{$\Hat{\Tilde{\mu}}=0$} 
verwenden. Mit den Vektoren \mbox{$\Vec{H}(\mu)$} nach Gleichung~(\ref{E.2.14}) 
und \mbox{$\Tilde{\Vec{\boldsymbol{V}}}(\mu)$} nach Gleichung~(\ref{E.2.15}) ergeben 
sich die \mbox{$M\CdoT F$} Gleichungssysteme
\begin{gather}
\frac{1}{M}\cdoT\Sum{\mu=0}{M-1}\Vec{H}(\mu)\cdot
\text{E}\Big\{
\big(\Tilde{\Vec{\boldsymbol{V}}}(\mu)\!-\!
\text{E}\{\Tilde{\Vec{\boldsymbol{V}}}(\mu)\}\big)\CdoT
\big(\Tilde{\Vec{\boldsymbol{V}}}(\Tilde{\mu})\!-\!
\text{E}\{\Tilde{\Vec{\boldsymbol{V}}}(\Tilde{\mu})\}\big)^{\HH}\Big\}\cdot
e^{j\cdot\frac{2\pi}{M}\cdot\mu\cdot k}\;={}\notag\\*
{}=\;\text{E}\Big\{
\big(\boldsymbol{y}(k)\!-\!
\text{E}\{\boldsymbol{y}(k)\}\big)\CdoT
\big(\Tilde{\Vec{\boldsymbol{V}}}(\Tilde{\mu})\!-\!
\text{E}\{\Tilde{\Vec{\boldsymbol{V}}}(\Tilde{\mu})\}\big)^{\HH}\Big\}
\notag\\*[6pt]
\forall\qquad k=0\;(1)\;F\!-\!1\quad\text{ und }
\quad\Tilde{\mu}=0\;(1)\;M\!-\!1,
\label{E.2.21}
\end{gather}
die jeweils auf beiden Seiten des Gleichheitszeichens einen Zeilenvektor 
der Dimension \mbox{$1\times(2\CdoT K_H)$} aufweisen. 

Die L"osung f"ur diese Gleichungssysteme ist bei der Wahl einer Fensterfolge, 
deren Spektrum der Bedingung~(\myref{2.27}) gen"ugt, identisch mit der L"osung 
der folgenden $M$ Gleichungssysteme, bei denen auf beiden Seiten ebenfalls ein 
Zeilenvektor der Dimension \mbox{$1\times(2\CdoT K_H)$} steht.
\begin{gather}
\Vec{H}(\mu)\CdoT\text{E}\Big\{
\big(\Tilde{\Vec{\boldsymbol{V}}}(\mu)\!-\!
\text{E}\{\Tilde{\Vec{\boldsymbol{V}}}(\mu)\}\big)\CdoT
\big(\Tilde{\Vec{\boldsymbol{V}}}(\mu)\!-\!
\text{E}\{\Tilde{\Vec{\boldsymbol{V}}}(\mu)\}\big)^{\HH}\Big\}\;=\;
\text{E}\Big\{\boldsymbol{Y}_{\!\!\!f}(\mu)\CdoT
\big(\Tilde{\Vec{\boldsymbol{V}}}(\mu)\!-\!
\text{E}\{\Tilde{\Vec{\boldsymbol{V}}}(\mu)\}\big)^{\HH}\Big\}\notag\\*[6pt]
\forall\qquad\mu=0\;(1)\;M\!-\!1
\label{E.2.22}
\end{gather}
Im Anhang~\ref{E.Kap.A.1} wird die Identit"at der L"osungsr"aume der Gleichungssysteme 
(\ref{E.2.21}) und (\ref{E.2.22}) hergeleitet, wobei dort f"ur den Parameter der 
Anzahl der Elemente des Zufallsvektors \mbox{$\Tilde{\Vec{\boldsymbol{V}}}(\mu)$} 
der Wert \mbox{$R=2\CdoT K_H$} einzusetzten ist. Das Gleichungssystem~(\ref{E.2.22}) 
hat nun mit \mbox{$2\CdoT M\CdoT K_H$} Gleichungen genauso viele Gleichungen 
wie Unbekannte. Die darin auftretenden zuf"alligen Spektralwerte 
\mbox{$\boldsymbol{Y}_{\!\!\!f}(\mu)$} berechnen sich wieder durch Fensterung 
und anschlie"sende diskrete Fouriertransformation gem"a"s Gleichung~(\myref{2.25}) 
aus dem Zufallsvektor des Prozesses am Ausgang des realen Systems. Durch die
Verwendung einer Fensterfolge, deren Spektrum die in Gleichung~(\myref{2.27})
angegebene Nullstelleneigenschaft erf"ullt, ist es auch hier gelungen ein 
Gleichungssystem zu erhalten, das in $M$ unabh"angige Gleichungssysteme 
(\,f"ur jeden Wert von $\mu$\,) zu je \mbox{$2\CdoT K_H$} Gleichungen mit je 
\mbox{$2\CdoT K_H$} Unbekannten zerf"allt. Diese Unbekannten sind diejenigen 
\mbox{$2\CdoT K_H$} Werte der beiden bifrequenten "Ubertragungsfunktion f"ur einen
festen Wert $\mu$, die gem"a"s der Gleichungen~(\ref{E.2.13}) von null verschieden 
sein k"onnen.

Unter Verwendung der \mbox{$(2\CdoT K_H)\times(2\CdoT K_H)$} Autokovarianzmatrix
\begin{equation}
\underline{C}_{\Tilde{\Vec{\boldsymbol{V}}}(\mu),\Tilde{\Vec{\boldsymbol{V}}}(\mu)}\;=\;
\text{E}\Big\{
\big(\Tilde{\Vec{\boldsymbol{V}}}(\mu)\!-\!
\text{E}\{\Tilde{\Vec{\boldsymbol{V}}}(\mu)\}\big)\CdoT
\big(\Tilde{\Vec{\boldsymbol{V}}}(\mu)\!-\!
\text{E}\{\Tilde{\Vec{\boldsymbol{V}}}(\mu)\}\big)^{\HH}\Big\}
\label{E.2.23}
\end{equation}
des Zufallsvektors \mbox{$\Tilde{\Vec{\boldsymbol{V}}}(\mu)$} und des 
\mbox{$1\times(2\CdoT K_H)$} Kreuzkovarianzvektors 
\begin{equation}
\underline{C}_{\boldsymbol{Y}_{\!\!\!f}(\mu),\Tilde{\Vec{\boldsymbol{V}}}(\mu)}\;=\;
\text{E}\Big\{
\big(\boldsymbol{Y}_{\!\!\!f}(\mu)\!-\!
\text{E}\{\boldsymbol{Y}_{\!\!\!f}(\mu)\}\big)\CdoT
\big(\Tilde{\Vec{\boldsymbol{V}}}(\mu)\!-\!
\text{E}\{\Tilde{\Vec{\boldsymbol{V}}}(\mu)\}\big)^{\HH}\Big\}
\label{E.2.24}
\end{equation}
eines Spektralwertes des Signals am Systemausgang mit einigen der Spektralwerte
des Signals am Systemeingang l"asst sich das Gleichungssystem (\ref{E.2.22}) sehr 
kompakt als 
\begin{equation}
\Vec{H}(\mu)\cdot
\underline{C}_{\Tilde{\Vec{\boldsymbol{V}}}(\mu),\Tilde{\Vec{\boldsymbol{V}}}(\mu)}\;=\;
\underline{C}_{\boldsymbol{Y}_{\!\!\!f}(\mu),\Tilde{\Vec{\boldsymbol{V}}}(\mu)}
\qquad\qquad\forall\qquad\mu=0\;(1)\;M\!-\!1
\label{E.2.25}
\end{equation}
schreiben. Im weiteren wollen wir uns auf den Fall beschr"anken, 
dass der periodische Prozess am Eingang des realen Systems 
in der Art gew"ahlt wurde, dass die Autokovarianzmatrix 
\raisebox{0.4ex}{$\underline{C}_{\Tilde{\Vec{\boldsymbol{V}}}(\mu),\Tilde{\Vec{\boldsymbol{V}}}(\mu)}$}
regul"ar ist. Dann l"asst sich diese invertieren, und wir erhalten die L"osung f"ur die 
\mbox{$2\CdoT K_H$} Werte der beiden bifrequenten "Ubertragungsfunktionen f"ur einen
festen Wert $\mu$, indem wir beide Seiten des Gleichungssystems (\ref{E.2.25}) mit 
der inversen Matrix von links multiplizieren. 

Auch hier wollen wir uns nun wieder der Frage widmen, welche Optimall"osungen man 
erh"alt, wenn man nicht das zweite Moment des Approximationsfehlerprozesses 
\mbox{$\boldsymbol{n}(k)$} minimiert, sondern stattdessen die $M$ zweiten Momente 
der Spektralwerte des gefensterten Approximationsfehlerprozesses. 
Dazu formen wir die nach Gleichung (\myref{2.16}) definierten Spektralwerte des 
gefensterten Approximationsfehlerprozesses zun"achst um:
\begin{gather}
\boldsymbol{N}_{\!\!f}(\mu)\;=\;
\Sum{k=-\infty}{\infty}\!\boldsymbol{n}(k)\CdoT
f(k)\cdot e^{\!-j\cdot\frac{2\pi}{M}\cdot\mu\cdot k}\;=
\notag
\\[6pt]
=\;\Sum{k=-\infty}{\infty}\!
\big(\,\boldsymbol{y}(k)-\boldsymbol{x}(k)-\boldsymbol{x}_*(k)-u(k)\,\big)\cdot f(k)\cdot
e^{\!-j\cdot\frac{2\pi}{M}\cdot\mu\cdot k}\;=
\notag
\\[6pt]
=\;\boldsymbol{Y}_{\!\!\!f}(\mu)-\frac{1}{M}\cdoT\!
\Sum{\Breve{\mu}=0}{M-1}\!\Vec{H}(\Breve{\mu})\CdoT
\Tilde{\Vec{\boldsymbol{V}}}(\Breve{\mu})\cdot
\Sum{k=-\infty}{\infty}\!f(k)\cdot e^{j\cdot\frac{2\pi}{M}\cdot(\Breve{\mu}-\mu)\cdot k}
-U_{\!f}(\mu)\;=
\notag
\\[6pt]
=\;\boldsymbol{Y}_{\!\!\!f}(\mu)-\frac{1}{M}\cdoT\!
\Sum{\Breve{\mu}=0}{M-1}\!\Vec{H}(\Breve{\mu})\CdoT
\Tilde{\Vec{\boldsymbol{V}}}(\Breve{\mu})\cdot
F\big({\T(\mu\!-\!\Breve{\mu})\CdoT\frac{2\pi}{M}}\big)-
U_{\!f}(\mu)\;=\;
\boldsymbol{Y}_{\!\!\!f}(\mu)-
\Vec{H}(\mu)\CdoT\Tilde{\Vec{\boldsymbol{V}}}(\mu)-U_{\!f}(\mu)
\notag\\[2pt]
\forall\qquad \mu=0\;(1)\;M\!-\!1.
\label{E.2.26}
\end{gather}
Hier wurde nach und nach der Approximationsfehlerprozess nach Gleichung (\ref{E.2.3}), 
der Ausgangsprozess der beiden Modellsysteme nach Gleichung (\ref{E.2.16}), das Spektrum 
des gefensterten Ausgangsprozesses des realen Systems nach Gleichung (\myref{2.25}) und 
das analog definierte Spektrum der gefensterten deterministischen St"orung gem"a"s
\begin{equation}
U_{\!f}(\mu)\;=\,
\Sum{k=-\infty}{\infty}\!u(k)\CdoT f(k)\cdot
e^{\!-j\cdot\frac{2\pi}{M}\cdot\mu\cdot k}
\qquad\qquad\forall\qquad \mu=0\;(1)\;M\!-\!1
\label{E.2.27}
\end{equation}
eingesetzt. Zuletzt wurde ber"ucksichtigt, dass bei Verwendung einer Fensterfolge, 
deren Spektrum die in Gleichung~(\myref{2.27}) angegebene Nullstelleneigenschaft erf"ullt, 
in der Summe lediglich ein Summand verbleibt. 
Die neue Aufgabe der Minimierung der $M$ zweiten Momente 
der Spektralwerte des gefensterten Approximationsfehlerprozesses lautet nun:
\begin{gather}
\text{E}\big\{\,|\boldsymbol{N}_{\!\!f}(\mu)|^2\big\}\;=\;
\text{E}\Bigg\{
\bigg|\,\boldsymbol{Y}_{\!\!\!f}(\mu)-
\Vec{H}(\mu)\CdoT\Tilde{\Vec{\boldsymbol{V}}}(\mu)-
U_{\!f}(\mu)\,\bigg|^2\Bigg\}\;\stackrel{!}{=}\;\text{minimal}
\notag\\[2pt]
\forall\qquad \mu=0\;(1)\;M\!-\!1.
\label{E.2.28}
\end{gather}
Im Gegensatz zur urspr"unglichen 
Minimierungsaufgabe (\ref{E.2.18}) treten in (\ref{E.2.28}) nicht mehr die $F$ Abtastwerte 
\mbox{$u(k)$} der deterministischen St"orung, sondern nur mehr die $M$ Spektralwerte 
\mbox{$U_{\!f}(\mu)$} als Parameter der Optimierung auf. Um die optimalen 
Approximationsparameter \mbox{$U_{\!f}(\mu)$} zu berechnen, leitet man 
den Term in der Gleichung (\ref{E.2.28}) --- wieder nach Real- und Imagin"arteil 
von \mbox{$U_{\!f}(\mu)$} getrennt --- partiell ab. Die dabei entstehenden 
Gleichungen lassen sich wieder zu komplexen Gleichungen zusammenfassen. 
Man erh"alt so die Optimall"osungen
\begin{equation}
U_{\!f}(\mu)\;=\;
\text{E}\big\{\boldsymbol{Y}_{\!\!\!f}(\mu)\big\}-
\Vec{H}(\mu)\CdoT\text{E}\big\{\Tilde{\Vec{\boldsymbol{V}}}(\mu)\big\}
\qquad\qquad\forall\qquad \mu=0\;(1)\;M\!-\!1.
\label{E.2.29}
\end{equation}
Dieselben  $M$ Spektralwerte \mbox{$U_{\!f}(\mu)$} erh"alt man, wenn man 
die $F$ optimalen Werte \mbox{$u(k)$} der urspr"unglichen Minimierungsaufgabe 
(\ref{E.2.18}) mit einer Fensterfolge fenstert, deren Spektrum der 
Gleichung~(\myref{2.27}) gen"ugt, und anschlie"send fouriertransformiert.
Wie man durch Einsetzen der Gleichungen (\ref{E.2.26}) und (\ref{E.2.29})
verifizieren kann, ergeben sich die optimalen Werte \mbox{$U_{\!f}(\mu)$} in der 
Art, dass alle $M$ Spektralwerte des gefensterten Approximationsfehlerprozesses 
mittelwertfrei sind:
\begin{gather}
\text{E}\big\{\boldsymbol{N}_{\!\!f}(\mu)\big\}\;=\;
\text{E}\Big\{\boldsymbol{Y}_{\!\!\!f}(\mu)-
\Vec{H}(\mu)\CdoT\Tilde{\Vec{\boldsymbol{V}}}(\mu)-
U_{\!f}(\mu)\Big\}\;=
\notag
\\[6pt]
=\;\text{E}\bigg\{\boldsymbol{Y}_{\!\!\!f}(\mu)-
\Vec{H}(\mu)\CdoT\Tilde{\Vec{\boldsymbol{V}}}(\mu)-
\text{E}\big\{\boldsymbol{Y}_{\!\!\!f}(\mu)\big\}+
\Vec{H}(\mu)\CdoT\text{E}\big\{\Tilde{\Vec{\boldsymbol{V}}}(\mu)\big\}
\bigg\}\;=
\notag
\\[6pt]
=\;\text{E}\big\{\boldsymbol{Y}_{\!\!\!f}(\mu)\big\}-
\Vec{H}(\mu)\CdoT\text{E}\big\{\Tilde{\Vec{\boldsymbol{V}}}(\mu)\big\}-
\text{E}\big\{\boldsymbol{Y}_{\!\!\!f}(\mu)\big\}+
\Vec{H}(\mu)\CdoT\text{E}\big\{\Tilde{\Vec{\boldsymbol{V}}}(\mu)\big\}\;=\;0
\notag\\[6pt]
\forall\qquad \mu=0\;(1)\;M\!-\!1.
\label{E.2.30}
\end{gather}
Die Optimall"osungen \mbox{$U_{\!f}(\mu)$} setzen wir wieder in die Minimierungsaufgabe (\ref{E.2.28}) 
ein und erhalten so f"ur jeden festen Wert $\mu$ eine Minimierungsaufgabe zur Bestimmung der 
\mbox{$2\CdoT K_H$} Werte der beiden bifrequenten "Ubertragungsfunktionen f"ur diesen Wert $\mu$:
\begin{gather}
\text{E}\Bigg\{
\bigg|\,\boldsymbol{Y}_{\!\!\!f}(\mu)-\text{E}\big\{\boldsymbol{Y}_{\!\!\!f}(\mu)\big\}-
\Vec{H}(\mu)\CdoT\Big(\Tilde{\Vec{\boldsymbol{V}}}(\mu)-
\text{E}\big\{\Tilde{\Vec{\boldsymbol{V}}}(\mu)\big\}\Big)
\,\bigg|^2\Bigg\}\;\stackrel{!}{=}\;\text{minimal}
\notag\\[2pt]
\forall\qquad \mu=0\;(1)\;M\!-\!1.
\label{E.2.31}
\end{gather}
Indem wir wieder die partiellen Ableitungen berechnen und diese zu null setzten, 
erhalten wir exakt dieselben $M$ Gleichungssysteme (\ref{E.2.22}), und somit dieselben 
L"osungen f"ur die optimalen Werte der beiden bifrequenten "Ubertragungsfunktionen, 
wie bei der urspr"unglichen Minimierungsaufgabe (\ref{E.2.18}). Dabei spielt es keine Rolle, 
ob der Approximationsfehlerprozess station"ar, zyklostation"ar oder gar instation"ar ist, und 
ob die deterministische St"orung konstant, periodisch oder beliebig zeitvariant ist. Wenn man eine 
andere Fensterfolge verwenden w"urde, deren Spektrum die in Gleichung~(\myref{2.27}) angegebene 
Nullstelleneigenschaft {\em nicht}\/ erf"ullt, w"urden sich in aller Regel abweichende 
Gleichungssysteme zur Bestimmung der Werte der beiden bifrequenten "Ubertragungsfunktionen 
ergeben, und somit abweichende Optimall"osungen, die dann nicht die urspr"ungliche 
Minimierungsaufgabe (\ref{E.2.18}) l"osen w"urden. Somit w"urde auch ein Messverfahren 
wie das RKM versagen, bei dem die empirischen Werte der beiden bifrequenten 
"Ubertragungsfunktionen in der Art bestimmt werden, dass sie die $M$ empirischen zweiten 
Momente der Spektralwerte des gefensterten Approximationsfehlerprozesses minimieren.

Durch Einsetzen der Gleichungen (\ref{E.2.26}), (\ref{E.2.29}) und (\ref{E.2.22}) 
l"asst sich zeigen, dass die L"osung der theoretischen Regression dem 
Orthogonalit"atsprinzip
\begin{equation}
\qquad\qquad\text{E}\Big\{\boldsymbol{N}_{\!\!f}(\mu)\CdoT
\Tilde{\Vec{\boldsymbol{V}}}(\mu)^{\Hh}\Big\}\;=\;\Vec{0}
\label{E.2.32}
\end{equation}
folgt.  Hier ist nun allerdings die Orthogonalit"at zu allen konjugierten, 
{\em nicht}\/ konjugierten und teilweise verschobenen Spektralwerten der 
Erregung enthalten, die in dem Vektor \mbox{$\Tilde{\Vec{\boldsymbol{V}}}(\mu)$} 
zusammengefasst sind.

\section{Leistungsdichtespektren und Fensterung}\label{E.Kap.2.3}

Wir haben gesehen, dass wir nur solche Systeme geeignet modellieren 
k"onnen, f"ur die die Minimierung des Terms in Gleichung (\ref{E.2.4}) 
eine periodisch zeitvariante L"osung f"ur die Werte der beiden 
Impulsantworten \mbox{$h_{*,\kappa}(k)$} und \mbox{$h_{*,\kappa}(k)$} 
liefert. I.~Allg. besagt dies jedoch {\em nicht}, dass auch die verbleibende 
Restdispersion, also der Wert \mbox{$\text{E}\big\{\,|\boldsymbol{n}(k)|^2\big\}$} 
des zu minimierenden Ausdrucks in Gleichung (\ref{E.2.18}), den man durch 
Einsetzen der optimalen Regressionskoeffizienten gem"a"s der Gleichung (\ref{E.2.19}) 
und der L"osung des Gleichungssystems (\ref{E.2.25}) erhalten, eine Folge ist, die 
sich abh"angig vom Zeitpunkt $k$ der Minimierung periodisch wiederholt. 
Der Prozess des Approximationsfehlers \mbox{$\boldsymbol{n}(k)$} des realen 
Systems durch die beiden linearen, periodisch zeitvarianten Modellsysteme und 
die deterministische St"orung ist daher i.~Allg. instation"ar. Bei einem 
instation"aren Prozess ist es prinzipiell nicht m"oglich, die zweidimensionale 
Autokorrelationsfolge \mbox{$\phi_{\boldsymbol{n}}(k_1,k_2)$} nach Gleichung 
(\ref{E.1.1}) und die ebenfalls zweidimensionale Kreuzkorrelationsfolge 
\mbox{$\psi_{\boldsymbol{n}}(k_1,k_2)$} nach Gleichung (\ref{E.1.3}) mit einer 
Messung endlicher Dauer zu bestimmen. Man m"usste dazu f"ur jedes der unbegrenzt 
vielen Paare $k_1$, $k_2$ ein Ensemble der beiden daran beteiligten Zufallsgr"o"sen 
\mbox{$\boldsymbol{n}(k_1)$} und \mbox{$\boldsymbol{n}(k_2)$} zur Verf"ugung 
haben, um daraus deren Korrelationen empirisch ermitteln zu k"onnen. An einem realen 
System wird man jedoch immer nur jeweils {\em eine}\/ konkrete Realisierung 
f"ur jeden Zeitpunkt zur Verf"ugung haben. Daher wollen wir uns im weiteren auf 
die Untersuchung solcher Systeme beschr"anken, bei denen der Prozess des 
Approximationsfehlers \mbox{$\boldsymbol{n}(k)$} zyklostation"ar ist. Allenfalls 
kann die Zyklostationarit"at des Fehlers durch die in Kapitel \ref{E.Kap.2.1} 
beschriebene zuf"allige Auswahl der Intervalle des Zugriffs auf das reale System 
erreicht werden. 

Aufgrund der Zyklostationarit"at gilt f"ur das zweite Moment des
mittelwertfreien Approximationsfehlerprozesses zu den beiden diskreten Zeiten
$k_1$ und $k_2$ ---\,also f"ur die von zwei Zeitvariablen abh"angige
und daher zweidimensionale AKF\,---
\begin{equation}
\text{E}\big\{\boldsymbol{n}(k_1)\CdoT\boldsymbol{n}(k_2)^{\Kk}\big\}\;=\;
\text{E}\big\{\boldsymbol{n}(k_1\!+\!K_{\Phi})\CdoT
              \boldsymbol{n}(k_2\!+\!K_{\Phi})^{\Kk}\big\}.
\label{E.2.33}
\end{equation}
Dabei ist \mbox{$K_{\Phi}$} die Periode der Momente des Prozesses 
\mbox{$\boldsymbol{n}(k)$}. Es sei hier darauf hingewiesen, dass 
\mbox{$K_{\Phi}$} nicht unbedingt mit der Periode \mbox{$K_H$} der 
Zeitvarianz der beiden Modellsysteme "ubereinstimmen muss. 
Das aus der zweidimensionalen AKF durch zweidimensionale diskrete 
Fouriertransformation\footnote{Bei der Frequenzvariable \mbox{$\Omega_2$} 
wird dabei das Vorzeichen invertiert.} gewonnene zweidimensionale 
(\,bifrequente\,) LDS l"asst sich als "Uberlagerung von zur Gerade 
\mbox{$\Omega_2\!=\!\Omega_1$} parallelen Impulslinien im Abstand 
von \mbox{$2\pi/K_{\Phi}$} schreiben.
\begin{gather}
\Phi_{\boldsymbol{n}}(\Omega_1,\Omega_2)\;=\!
\Sum{k_1=-\infty}{\infty}\;\Sum{k_2=-\infty}{\infty}\!
\text{E}\big\{\boldsymbol{n}(k_1)\CdoT
              \boldsymbol{n}(k_2)^{\Kk}\big\}\cdot
e^{\!-j\cdot\Omega_1\Cdot k_1}\cdot e^{j\cdot\Omega_2\Cdot k_2}=
\notag\\[8pt]
=\Sum{\Bar{\mu}=-\infty}{\infty}
\frac{1}{K_{\Phi}}\cdoT\Sum{k=0}{K_{\Phi}-1}
\Acute{\Phi}_{\boldsymbol{n}}(\Omega_1,k)\cdot
 e^{j\cdot\frac{2\pi}{K_{\Phi}}\cdot\Bar{\mu}\cdot k}\cdot
2\pi\CdoT\delta_0\big({\T\Omega_2\!-\!\Omega_1\!-\!
\Bar{\mu}\CdoT\frac{2\pi}{K_{\Phi}}}\big)\;=\notag\\[6pt]
=\Sum{\Bar{\mu}=-\infty}{\infty}\!
\Grave{\Phi}_{\boldsymbol{n}}(\Omega_1,\Bar{\mu})\cdot
2\pi\CdoT\delta_0\big({\T\Omega_2\!-\!\Omega_1\!-\!
\Bar{\mu}\CdoT\frac{2\pi}{K_{\Phi}}}\big)
\label{E.2.34}\\[-6pt]\intertext{Dabei sind\vspace{-6pt}}
\Acute{\Phi}_{\boldsymbol{n}}(\Omega,k)\;=\;
\Acute{\Phi}_{\boldsymbol{n}}(\Omega,k+K_{\Phi})\;=
\Sum{\kappa=-\infty}{\infty}\!
\text{E}\big\{\boldsymbol{n}(k\!+\!\kappa)\CdoT
              \boldsymbol{n}(k)^{\Kk}\big\}\cdot
e^{\!-j\cdot\Omega\cdot\kappa}
\label{E.2.35}
\end{gather}
die von der Zeit $k$ abh"angigen und in $\Omega$ kontinuierlichen diskreten
Fouriertransformierten der Autokorrelationsfolgen zu den Zeitpunkten $k$,
die in $\Omega$ mit $2\pi$ und in $k$ mit $K_{\Phi}$ periodisch sind.
Die diskret invers Fouriertransformierten
\begin{equation}
\Grave{\Phi}_{\boldsymbol{n}}(\Omega,\Bar{\mu})\;=\;
\Grave{\Phi}_{\boldsymbol{n}}(\Omega,\Bar{\mu}+K_{\Phi})\;=\;
\frac{1}{K_{\Phi}}\cdoT\Sum{k=0}{K_{\Phi}-1}
\Acute{\Phi}_{\boldsymbol{n}}(\Omega,k)\cdot
 e^{j\cdot\frac{2\pi}{K_{\Phi}}\cdot\Bar{\mu}\cdot k}
\label{E.2.36}
\end{equation}
einer Periode in $k$ der Leistungsdichtespektren 
\mbox{$\Acute{\Phi}_{\boldsymbol{n}}(\Omega,k)$}
bei den Frequenzen \mbox{$\Bar{\mu}\CdoT2\pi/K_{\Phi}$} sind in
$\Bar{\mu}$ mit $K_{\Phi}$ periodisch und treten in dem bifrequenten
LDS ---\,abgesehen von dem in Frequenzbereich typischerweise auftretenden
Vorfaktor $2\pi$\,--- als St"arken der Impulslinien auf, die in der
\mbox{$\Omega_1,\Omega_2$}-Ebene in einem Abstand von Vielfachen
von \mbox{$2\pi/K_{\Phi}$} parallel zur Winkelhalbierenden
\mbox{$\Omega_2\!=\!\Omega_1$} verlaufen. Diese Linien sind in Bild~\ref{E.b3b1}
f"ur \mbox{$K_{\Phi}\!=\!4$} in der \mbox{$\Omega_1,\Omega_2$}-Ebene
fettgedruckt dargestellt. Da jeder beliebige
Abtastwert ---\,sofern man bei Distributionen "uberhaupt von 
Abtastwerten reden mag\,--- des bifrequenten LDS entweder null\footnote{
Im distributionstheoretischen Sinn ist eine Distribution au"serhalb
eines abgeschlossenen Intervalls null, wenn die Distribution jeder Funktion,
die innerhalb des Intervalls null ist, den Wert Null zuordnet.},
oder nicht definiert ist, kann man das bifrequente LDS nicht
durch endlich viele Abtastwert beschreiben. Sinnvoller erscheint
es schon f"ur jeden der $K_{\Phi}$ Zeitpunkte $k$ endlich viele
Abtastwerte der zeitabh"angigen Leistungsdichtespektren
\mbox{$\Acute{\Phi}_{\boldsymbol{n}}(\Omega,k)$}, oder f"ur jede der
$K_{\Phi}$ Impulslinien endlich viele Abtastwerte der vor den
Impulslinien auftretenden von $\Bar{\mu}$ abh"angigen Funktionen 
\mbox{$\Grave{\Phi}_{\boldsymbol{n}}(\Omega,\Bar{\mu})$}
anzugeben.
\begin{figure}[btp]
\begin{center}
{ 
\begin{picture}(400,430)

\input{mbild3b1}
\put(32,20){\makebox(0,0)[lb]{\small$0$}}
\put(110,17){\makebox(0,0)[b]{\small$\pi/2$}}
\put(190,20){\makebox(0,0)[b]{\small$\pi$}}
\put(270,17){\makebox(0,0)[b]{\small$3\pi/2$}}
\put(352,19){\makebox(0,0)[b]{\small$2\pi$}}
\put(25,32){\makebox(0,0)[b]{\small$0$}}
\put(25,110){\makebox(0,0)[r]{\small$\pi/2$}}
\put(25,190){\makebox(0,0)[r]{\small$\pi$}}
\put(25,270){\makebox(0,0)[r]{\small$3\pi/2$}}
\put(25,350){\makebox(0,0)[r]{\small$2\pi$}}
\put(400,25){\makebox(0,0)[tr]{$\Omega_1$}}
\put(35,392){\makebox(0,0)[tl]{$\Omega_2$}}

\end{picture}}
\end{center}\vspace{4pt}
\caption[Zur n"aherungsweisen Beschreibung des bifrequenten LDS]{
Zur n"aherungsweisen Beschreibung des bifrequenten LDS:\\
H"ohenlinien des Betrags des Produktes
\mbox{$F\big({\T\mu\CdoT\frac{2\pi}{M}\!-\!\Omega_1}\big)\CdoT
F\big({\T\mu\CdoT\frac{2\pi}{M}\!+\!
\Tilde{\mu}\CdoT\frac{2\pi}{K_{\Phi}}\!-\!\Omega_2}\big)^{\!\Kk}$}
der beiden verschobenen Fensterspektren im Integranden der Gleichung
(\ref{E.2.39}) f"ur drei verschiedene Fensterfolgen am 
Beispiel mit $M\!=\!16$, $K_{\Phi}\!=\!4$ und\\
$\mu\!=\!7$, $\protect\Tilde{\mu}\!=\!2$ beim si-Fensters,\\
$\mu\!=\!13$, $\protect\Tilde{\mu}\!=\!-2$ beim Rechteckfenster mit
$F\!=\!M$ und\\
$\mu\!=\!4$, $\protect\Tilde{\mu}\!=\!2$ beim Fensters nach
Kapitel \myref{Algo} mit $F\!=\!4\CdoT M$\vspace{30pt}\\\mbox{}.}
\label{E.b3b1}
\end{figure}
\begin{figure}[btp]
{ 
\begin{picture}(454,599)

\input{mbild3b2}
\put(15,540){\makebox(0,0)[r]{\small$50$}}
\put(15,440){\makebox(0,0)[r]{\small$0$}}
\put(22,410){\makebox(0,0)[lb]{\small$0$}}
\put(110,407){\makebox(0,0)[b]{\small$\pi/2$}}
\put(200,410){\makebox(0,0)[b]{\small$\pi$}}
\put(220,415){\makebox(0,0)[tr]{$\Omega$}}
\put(23,600){\makebox(0,0)[tl]{
$10\CdoT\log\Big(\Big|F(\Omega)\CdoT F\big({\T\Omega\!-\!
(\Bar{\mu}\!-\!\Tilde{\mu})\CdoT\frac{2\pi}{K_{\Phi}}}\big)^{\!\Kk}\Big|\Big)$}}
\put(120,540){\makebox(0,0)[b]{$\Bar{\mu}=\Tilde{\mu}$}}

\input{mbild3b3}
\put(245,540){\makebox(0,0)[r]{\small$50$}}
\put(245,440){\makebox(0,0)[r]{\small$0$}}
\put(252,410){\makebox(0,0)[lb]{\small$0$}}
\put(340,407){\makebox(0,0)[b]{\small$\pi/2$}}
\put(430,410){\makebox(0,0)[b]{\small$\pi$}}
\put(450,415){\makebox(0,0)[tr]{$\Omega$}}
\put(253,600){\makebox(0,0)[tl]{
$10\CdoT\log\Big(\Big|F(\Omega)\CdoT F\big({\T\Omega\!-\!
(\Bar{\mu}\!-\!\Tilde{\mu})\CdoT\frac{2\pi}{K_{\Phi}}}\big)^{\!\Kk}\Big|\Big)$}}
\put(350,540){\makebox(0,0)[b]{$\Bar{\mu}=\Tilde{\mu}\!+\!1$}}

\input{mbild3b4}
\put(15,340){\makebox(0,0)[r]{\small$50$}}
\put(15,240){\makebox(0,0)[r]{\small$0$}}
\put(22,210){\makebox(0,0)[lb]{\small$0$}}
\put(110,207){\makebox(0,0)[b]{\small$\pi/2$}}
\put(200,210){\makebox(0,0)[b]{\small$\pi$}}
\put(220,215){\makebox(0,0)[tr]{$\Omega$}}
\put(23,400){\makebox(0,0)[tl]{
$10\CdoT\log\Big(\Big|F(\Omega)\CdoT F\big({\T\Omega\!-\!
(\Bar{\mu}\!-\!\Tilde{\mu})\CdoT\frac{2\pi}{K_{\Phi}}}\big)^{\!\Kk}\Big|\Big)$}}
\put(120,340){\makebox(0,0)[b]{$\Bar{\mu}=\Tilde{\mu}\!+\!2$}}

\input{mbild3b5}
\put(245,340){\makebox(0,0)[r]{\small$50$}}
\put(245,240){\makebox(0,0)[r]{\small$0$}}
\put(252,210){\makebox(0,0)[lb]{\small$0$}}
\put(340,207){\makebox(0,0)[b]{\small$\pi/2$}}
\put(430,210){\makebox(0,0)[b]{\small$\pi$}}
\put(450,215){\makebox(0,0)[tr]{$\Omega$}}
\put(253,400){\makebox(0,0)[tl]{
$10\CdoT\log\Big(\Big|F(\Omega)\CdoT F\big({\T\Omega\!-\!
(\Bar{\mu}\!-\!\Tilde{\mu})\CdoT\frac{2\pi}{K_{\Phi}}}\big)^{\!\Kk}\Big|\Big)$}}
\put(350,340){\makebox(0,0)[b]{$\Bar{\mu}=\Tilde{\mu}\!+\!5$}}

\input{mbild3b10}
\put(45,180){\makebox(0,0)[r]{\small$60$}}
\put(45,140){\makebox(0,0)[r]{\small$50$}}
\put(45,100){\makebox(0,0)[r]{\small$40$}}
\put(45,60){\makebox(0,0)[r]{\small$30$}}
\put(52,7){\makebox(0,0)[lb]{\small$0$}}
\put(100,7){\makebox(0,0)[b]{\small$1$}}
\put(150,7){\makebox(0,0)[b]{\small$2$}}
\put(200,7){\makebox(0,0)[b]{\small$3$}}
\put(250,7){\makebox(0,0)[b]{\small$4$}}
\put(300,7){\makebox(0,0)[b]{\small$5$}}
\put(350,7){\makebox(0,0)[b]{\small$6$}}
\put(400,7){\makebox(0,0)[b]{\small$7$}}
\put(450,15){\makebox(0,0)[tr]{${\T\Omega\CdoT\frac{M}{2\pi}}$}}
\put(53,200){\makebox(0,0)[tl]{
$10\CdoT\log\Big(\Big|F(\Omega)\CdoT F\big({\T\Omega\!-\!
(\Bar{\mu}\!-\!\Tilde{\mu})\CdoT\frac{2\pi}{K_{\Phi}}}\big)^{\!\Kk}\Big|\Big)$}}
\put(300,180){\makebox(0,0)[l]{$\Bar{\mu}=\Tilde{\mu}$}}
\put(300,165){\makebox(0,0)[l]{$\Bar{\mu}=\Tilde{\mu}\!+\!1$}}
\put(300,150){\makebox(0,0)[l]{$\Bar{\mu}=\Tilde{\mu}\!+\!2$}}
\put(300,135){\makebox(0,0)[l]{$\Bar{\mu}=\Tilde{\mu}\!+\!5$}}

\end{picture}}
\caption[D"ampfung der Haupt- und Nebenlinien des bifrequenten LDS
mit einem Rechteckfenster]{D"ampfung der Haupt- und Nebenlinien
des bifrequenten LDS am Beispiel des Rechteckfensters mit der
Fensterl"ange \mbox{$F\!=\!M\!=\!1024$} und mit \mbox{$M\!=\!K_{\Phi}$}.}
\label{E.b3b2}
\end{figure}
\begin{figure}[btp]
{ 
\begin{picture}(454,599)

\input{mbild3b6}
\put(15,570){\makebox(0,0)[r]{\small$50$}}
\put(15,540){\makebox(0,0)[r]{\small$0$}}
\put(15,510){\makebox(0,0)[r]{\small$-50$}}
\put(15,480){\makebox(0,0)[r]{\small$-100$}}
\put(15,450){\makebox(0,0)[r]{\small$-150$}}
\put(22,410){\makebox(0,0)[lb]{\small$0$}}
\put(110,407){\makebox(0,0)[b]{\small$\pi/2$}}
\put(190,410){\makebox(0,0)[b]{\small$\pi$}}
\put(220,415){\makebox(0,0)[tr]{$\Omega$}}
\put(23,600){\makebox(0,0)[tl]{
$10\CdoT\log\Big(\Big|F(\Omega)\CdoT F\big({\T\Omega\!-\!
(\Bar{\mu}\!-\!\Tilde{\mu})\CdoT\frac{2\pi}{K_{\Phi}}}\big)^{\!\Kk}\Big|\Big)$}}
\put(120,540){\makebox(0,0)[b]{$\Bar{\mu}=\Tilde{\mu}$}}

\input{mbild3b7}
\put(245,570){\makebox(0,0)[r]{\small$50$}}
\put(245,540){\makebox(0,0)[r]{\small$0$}}
\put(245,510){\makebox(0,0)[r]{\small$-50$}}
\put(245,480){\makebox(0,0)[r]{\small$-100$}}
\put(245,450){\makebox(0,0)[r]{\small$-150$}}
\put(252,410){\makebox(0,0)[lb]{\small$0$}}
\put(340,407){\makebox(0,0)[b]{\small$\pi/2$}}
\put(420,410){\makebox(0,0)[b]{\small$\pi$}}
\put(450,415){\makebox(0,0)[tr]{$\Omega$}}
\put(253,600){\makebox(0,0)[tl]{
$10\CdoT\log\Big(\Big|F(\Omega)\CdoT F\big({\T\Omega\!-\!
(\Bar{\mu}\!-\!\Tilde{\mu})\CdoT\frac{2\pi}{K_{\Phi}}}\big)^{\!\Kk}\Big|\Big)$}}
\put(350,540){\makebox(0,0)[b]{$\Bar{\mu}=\Tilde{\mu}\!+\!1$}}

\input{mbild3b8}
\put(15,370){\makebox(0,0)[r]{\small$50$}}
\put(15,340){\makebox(0,0)[r]{\small$0$}}
\put(15,310){\makebox(0,0)[r]{\small$-50$}}
\put(15,280){\makebox(0,0)[r]{\small$-100$}}
\put(15,250){\makebox(0,0)[r]{\small$-150$}}
\put(22,210){\makebox(0,0)[lb]{\small$0$}}
\put(110,207){\makebox(0,0)[b]{\small$\pi/2$}}
\put(190,210){\makebox(0,0)[b]{\small$\pi$}}
\put(220,215){\makebox(0,0)[tr]{$\Omega$}}
\put(23,400){\makebox(0,0)[tl]{
$10\CdoT\log\Big(\Big|F(\Omega)\CdoT F\big({\T\Omega\!-\!
(\Bar{\mu}\!-\!\Tilde{\mu})\CdoT\frac{2\pi}{K_{\Phi}}}\big)^{\!\Kk}\Big|\Big)$}}
\put(120,340){\makebox(0,0)[b]{$\Bar{\mu}=\Tilde{\mu}\!+\!2$}}

\input{mbild3b9}
\put(245,370){\makebox(0,0)[r]{\small$50$}}
\put(245,340){\makebox(0,0)[r]{\small$0$}}
\put(245,310){\makebox(0,0)[r]{\small$-50$}}
\put(245,280){\makebox(0,0)[r]{\small$-100$}}
\put(245,250){\makebox(0,0)[r]{\small$-150$}}
\put(252,210){\makebox(0,0)[lb]{\small$0$}}
\put(340,207){\makebox(0,0)[b]{\small$\pi/2$}}
\put(420,210){\makebox(0,0)[b]{\small$\pi$}}
\put(450,215){\makebox(0,0)[tr]{$\Omega$}}
\put(253,400){\makebox(0,0)[tl]{
$10\CdoT\log\Big(\Big|F(\Omega)\CdoT F\big({\T\Omega\!-\!
(\Bar{\mu}\!-\!\Tilde{\mu})\CdoT\frac{2\pi}{K_{\Phi}}}\big)^{\!\Kk}\Big|\Big)$}}
\put(350,340){\makebox(0,0)[b]{$\Bar{\mu}=\Tilde{\mu}\!+\!5$}}

\input{mbild3b11}
\put(45,170){\makebox(0,0)[r]{\small$50$}}
\put(45,95){\makebox(0,0)[r]{\small$0$}}
\put(52,7){\makebox(0,0)[lb]{\small$0$}}
\put(100,7){\makebox(0,0)[b]{\small$1$}}
\put(150,7){\makebox(0,0)[b]{\small$2$}}
\put(200,7){\makebox(0,0)[b]{\small$3$}}
\put(250,7){\makebox(0,0)[b]{\small$4$}}
\put(300,7){\makebox(0,0)[b]{\small$5$}}
\put(350,7){\makebox(0,0)[b]{\small$6$}}
\put(400,7){\makebox(0,0)[b]{\small$7$}}
\put(450,15){\makebox(0,0)[tr]{${\T\Omega\CdoT\frac{M}{2\pi}}$}}
\put(53,201){\makebox(0,0)[tl]{
$10\CdoT\log\Big(\Big|F(\Omega)\CdoT F\big({\T\Omega\!-\!
(\Bar{\mu}\!-\!\Tilde{\mu})\CdoT\frac{2\pi}{K_{\Phi}}}\big)^{\!\Kk}\Big|\Big)$}}
\put(300,175){\makebox(0,0)[l]{$\Bar{\mu}=\Tilde{\mu}$}}
\put(300,160){\makebox(0,0)[l]{$\Bar{\mu}=\Tilde{\mu}\!+\!1$}}
\put(300,145){\makebox(0,0)[l]{$\Bar{\mu}=\Tilde{\mu}\!+\!2$}}
\put(300,130){\makebox(0,0)[l]{$\Bar{\mu}=\Tilde{\mu}\!+\!5$}}

\end{picture}}
\caption[D"ampfung der Haupt- und Nebenlinien des bifrequenten LDS mit einem
Fenster nach \myref{Algo}]{ D"ampfung der Haupt- und Nebenlinien
des  bifrequenten LDS am Beispiel des Fensters nach Kapitel
\myref{Algo} mit der Fensterl"ange \mbox{$F=4\CdoT M$} und mit
\mbox{$M\!=\!K_{\Phi}\!=\!1024$}.}
\label{E.b3b3}
\end{figure}


Da jedoch f"ur die Abtastwerte des zeitabh"angigen LDS
\mbox{$\Acute{\Phi}_{\boldsymbol{n}}(\mu\CdoT2\pi/M,k)$} dasselbe gilt, wie
f"ur die Abtastwerte des zeitunabh"angigen LDS im Fall des station"aren
Approximationsfehlerprozesses, verwenden wir stattdessen zun"achst die
zeitabh"angigen Werte einer fl"achengleichen Stufenapproximation, die
analog zu Gleichung~(\myref{2.13}) als
\begin{gather}
\Check{\Phi}_{\boldsymbol{n}}(\mu,k)\;=\;
\frac{M}{2\pi}\;\cdoT\!\!\Int{\mu\CdoT\frac{2\pi}{M}-\frac{\pi}{M}}
{\mu\CdoT\frac{2\pi}{M}+\frac{\pi}{M}}\!\!\!\!
\Acute{\Phi}_{\boldsymbol{n}}(\Omega,k)\cdot d\Omega\;=\;
\frac{M}{2\pi}\cdoT\!\Int{-\frac{\pi}{M}}{\frac{\pi}{M}}\!\!
\Acute{\Phi}_{\boldsymbol{n}}\big({\T\mu\CdoT\frac{2\pi}{M}\!-\!\Omega,k}\big)
\cdot d\Omega\notag\\*[8pt]
\forall\qquad k=0\;(1)\;K_{\Phi}\!-\!1
\quad\text{ und }\quad\mu=0\;(1)\;M\!-\!1.
\label{E.2.37}
\end{gather}
definiert werden. Um die Analogie zum bifrequenten LDS zu wahren,
und um die auf $2\pi$ normierten\footnote{Der Faktor $2\pi$ wird
hier aus zwei Gr"unden als Vorfaktor des Dirac-Impulses betrachtet.
Zum einen entsteht bei der Fouriertransformation eines periodischen
Signals immer ein Impulslinienspektrum, dessen Impulsst"arken gegen"uber
den Fourierreihenkoeffizienten um den Faktor $2\pi$ gr"o"ser sind, und
zum anderen ergeben sich so f"ur $K_{\Phi}\!=\!1$ exakt dieselben Werte f"ur
die Stufenapproximation wie im Fall eines station"aren Prozesses.
Alternativ ---\,aber eher un"ublich\,--- kann man auch gleich das
bifrequente LDS als auf $2\pi$ normiert definieren, wie dies beispielsweise
in \cite{Heinle} gemacht wurde, und erh"alt dann unmittelbar die
Stufenapproximationswerte als Vorfaktoren der Impulslinien.}
St"arken \mbox{$\Grave{\Phi}_{\boldsymbol{n}}(\Omega,\Bar{\mu})$}
der Impulslinien absch"atzen zu k"onnen, berechnen wir daraus die
bez"uglich $k$ invers diskret Fouriertransformierten einer Periode von
\mbox{$\Bar{\Phi}_{\boldsymbol{n}}(\mu,k)$}: 
\begin{gather}
\Bar{\Phi}_{\boldsymbol{n}}
\big({\T\mu,\mu\!+\!\Tilde{\mu}\CdoT\frac{M}{K_{\Phi}}}\big)\;=\;
\frac{1}{K_{\Phi}}\cdoT\Sum{k=0}{K_{\Phi}-1}
\Check{\Phi}_{\boldsymbol{n}}(\mu,k)\cdot
e^{j\cdot\frac{2\pi}{K_{\Phi}}\cdot\Tilde{\mu}\cdot k}\notag\\*[4pt]
\forall\qquad\mu=0\;(1)\;M\!-\!1
\quad\text{ und }\quad\Tilde{\mu}=0\;(1)\;K_{\Phi}\!-\!1.
\label{E.2.38}
\end{gather}
Die bereichsweise Integration in Gleichung~(\ref{E.2.37}) kann man
wieder als eine \mbox{Integration} interpretieren, die man f"ur alle
\mbox{$-\pi\le\Omega<\pi$} durchf"uhrt, bei der man aber zuvor
aus dem zeitabh"angigen LDS mit Hilfe einer verschobenen Rechteckfunktionen
den Bereich ausblendet, der den Integrationsgrenzen entspricht. Da sich
ein rechteckf"ormiger Verlauf des Spektrums nicht mit einer
endlich langen Zeitfolge realisieren l"asst, kann man wie im Fall des
station"aren Approximationsfehlerprozesses die \mbox{$M\CdoT K_{\Phi}$} Werte
\mbox{$\Bar{\Phi}_{\boldsymbol{n}}(\mu,\mu\!+\!\Tilde{\mu}\CdoT M/K_{\Phi})$}
nicht als Funktion endlich vieler Werte der zweidimensionalen
Autokorrelationsfolge angeben. Daher muss man sich auch hier damit
begn"ugen, die N"aherungswerte
\begin{gather}
\Tilde{\Phi}_{\boldsymbol{n}}
\big({\T\mu,\mu\!+\!\Tilde{\mu}\CdoT\frac{M}{K_{\Phi}}}\big)\;=\;
\frac{1}{M}\cdot\text{E}\Big\{\boldsymbol{N}_{\!\!f}(\mu)\CdoT
\boldsymbol{N}_{\!\!f}
\big({\T\mu\!+\!\Tilde{\mu}\CdoT\frac{M}{K_{\Phi}}}\big)^{\!\Kk}
\Big\}\;=\notag\\[4pt]
=\;\frac{1}{(2\pi)^2\CdoT M}\cdoT\Int{-\pi}{\pi}\Int{-\pi}{\pi}
F\big({\T\mu\CdoT\frac{2\pi}{M}\!-\!\Omega_1}\big)\CdoT
F\big({\T\mu\CdoT\frac{2\pi}{M}\!+\!
\Tilde{\mu}\CdoT\frac{2\pi}{K_{\Phi}}\!-\!\Omega_2}\big)^{\!\Kk}\Cdot
\Phi_{\boldsymbol{n}}(\Omega_1,\Omega_2)\cdot
d\Omega_1\Cdot d\Omega_2\;=\notag\\[2pt]
=\Sum{\Bar{\mu}=0}{K_{\Phi}-1}
\frac{1}{2\pi\CdoT M}\cdoT\Int{-\pi}{\pi}
F(\Omega)\CdoT F\big({\T\Omega\!-\!
(\Bar{\mu}\!-\!\Tilde{\mu})\CdoT\frac{2\pi}{K_{\Phi}}}\big)^{\!\Kk}\Cdot
\Grave{\Phi}_{\boldsymbol{n}}
\big({\T\mu\CdoT\frac{2\pi}{M}\!-\!\Omega,\Bar{\mu}}\big)\cdot d\Omega\qquad
\notag\\*[2pt]
\qquad\forall\qquad\mu=0\;(1)\;M\!-\!1\quad\text{ und }
\quad\Tilde{\mu}=0\;(1)\;K_{\Phi}\!-\!1,
\label{E.2.39}
\end{gather}
die man durch Erwartungswertbildung aus den zuf"alligen Spektralwerten
des gefensterten Approximationsfehlerprozesses gewinnt, f"ur
\mbox{$\Bar{\Phi}_{\boldsymbol{n}}(\mu,\mu\!+\!\Tilde{\mu}\CdoT M/K_{\Phi})$}
angeben zu k"onnen. Sinnvollerweise wird man $M$ als ganzzahliges
Vielfaches von $K_{\Phi}$ w"ahlen, da sich nur in diesem Fall die
zuf"alligen Spektralwerte
\mbox{$\boldsymbol{N}_{\!\!f}(\mu\!+\!\Tilde{\mu}\CdoT M/K_{\Phi})$}
durch eine blockweise "Uberlagerung und eine anschlie"sende DFT aus
den gefensterten Zufallsgr"o"sen des Prozesses \mbox{$\boldsymbol{n}(k)$}
in der in Kapitel~\myref{W} geschilderten Art berechnen lassen.
Wie wir sp"ater noch sehen werden, ben"otigt man zur Berechnung der
Messwertkovarianzen auch die Spektralwerte des gefensterten
Approximationsfehlerprozesses bei den Frequenzen
\mbox{$\mu=-\Tilde{\mu}\CdoT M/(2\CdoT K_{\Phi}\!)$} mit
\mbox{$\Tilde{\mu}\in\mathbb{Z}$}. Daher ist $M$ als gerades
Vielfaches von $K_{\Phi}$ zu w"ahlen, so dass
\mbox{$M/(2\CdoT K_{\Phi}\!)\in\mathbb{N}$} gelten soll.

Setzt man in die letzte Gleichung einerseits das bifrequente LDS nach
Gleichung~(\ref{E.2.34}) und andererseits eine Fensterfolge ein, deren Spektrum
einen rechteckf"ormigen Betragsfrequenzgang mit einer Amplitude von $M$ und
einer Breite von \mbox{$2\pi/M$} sowie einen beliebigen Phasenfrequenzgang
aufweist, so wird durch die beiden Fensterspektren vom bifrequenten LDS
nur die Impulslinie mit \mbox{$\Bar{\mu}\!=\!\Tilde{\mu}$} ausgeblendet, wenn
man \mbox{$M\!\ge\!K_{\Phi}$} w"ahlt. Die Integration "uber $\Omega_2$ liefert
im Sinne der Distributionentheorie den Wert der zu integrierenden Funktion
bei der Frequenz \mbox{$\Omega_2=\Omega_1\!+\!\Bar{\mu}\CdoT2\pi/K_{\Phi}=
\Omega_1\!+\!\Tilde{\mu}\CdoT2\pi/K_{\Phi}$}. Bei dem vom $\Omega_2$
abh"angigen Fensterspektrum erh"alt man den Wert des Spektrums
bei der Frequenz \mbox{$\mu\CdoT2\pi/M\!-\!\Omega_1$}. Dadurch
entsteht innerhalb des Integrals "uber $\Omega_1$ der Faktor
\mbox{$\big|F(\mu\CdoT2\pi/M\!-\!\Omega_1)\big|^2=M^2$},
der vom Phasenfrequenzgang der Fensterfolge {\em nicht}\/ abh"angt.
Ein Vergleich mit der Definition von
\mbox{$\Bar{\Phi}_{\boldsymbol{n}}(\mu,\mu\!+\!\Tilde{\mu}\CdoT M/K_{\Phi})$}
nach Gleichung~(\ref{E.2.38}) in Verbindung mit Gleichung~(\ref{E.2.37}) zeigt,
dass man mit einer Fensterfolge mit rechteckf"ormigem Betragsfrequenzgang
gerade diese Werte f"ur \mbox{$
\Tilde{\Phi}_{\boldsymbol{n}}(\mu,\mu\!+\!\Tilde{\mu}\CdoT M/K_{\Phi})$}
erh"alt. Da der Phasenfrequenzgang des Fensters mit dem rechteckf"ormigen
Betragsfrequenzgang bei dem zyklostation"aren Prozess f"ur
\mbox{$M\ge K_{\Phi}$} bedeutungslos ist, kann man ihn auch konstant
zu null setzten ---\,also eine abgetastete si-Funktion als Fensterfolge 
verwenden\,---, und es ergibt sich, wenn man das rechteckige Spektrum
des Fensters in den Integrationsgrenzen ber"ucksichtigt, f"ur die diskret
Fouriertransformierte der zeitabh"angigen Stufenapproximation folgendes:
\begin{gather}
\Bar{\Phi}_{\boldsymbol{n}}
\big({\T\mu,\mu\!+\!\Tilde{\mu}\CdoT\frac{M}{K_{\Phi}}}\big)\;=\notag\\[8pt]
=\;\frac{1}{M}\cdot\Big(\frac{M}{2\pi}\Big)^{\!2}\cdoT
\Int{\mu\CdoT\frac{2\pi}{M}-\frac{\pi}{M}}
{\mu\CdoT\frac{2\pi}{M}+\frac{\pi}{M}}\;\;\,\Int
{\mu\CdoT\frac{2\pi}{M}+\Tilde{\mu}\CdoT\frac{2\pi}{K_{\Phi}}-\frac{\pi}{M}}
{\mu\CdoT\frac{2\pi}{M}+\Tilde{\mu}\CdoT\frac{2\pi}{K_{\Phi}}+\frac{\pi}{M}}
\!\!\!\Phi_{\boldsymbol{n}}(\Omega_1,\Omega_2)\cdot
d\Omega_2\Cdot d\Omega_1\;=\;
\frac{M}{2\pi}\cdoT\!\Int{-\frac{\pi}{M}}{\frac{\pi}{M}}
\Grave{\Phi}_{\boldsymbol{n}}(\Omega,\Tilde{\mu})\cdot d\Omega\notag\\*[6pt]
\forall\qquad\mu=0\;(1)\;M\!-\!1,
\quad\Tilde{\mu}=0\;(1)\;K_{\Phi}\!-\!1
\quad\text{ und }\quad K_{\Phi}\le M
\label{E.2.40}
\end{gather}
Man erh"alt also das zweidimensionale Integral "uber das 
bifrequente LDS \mbox{$\Phi_{\boldsymbol{n}}(\Omega_1,\Omega_2)$} in 
einem quadratischen Bereich der \mbox{$\Omega_1,\Omega_2$}-Ebene, 
der um \mbox{$\mu\CdoT2\pi/M$} in $\Omega_1$-Richtung und um 
\mbox{$\mu\CdoT2\pi/M\!+\!\Tilde{\mu}\CdoT2\pi/K_{\Phi}$} in 
$\Omega_2$-Richtung verschoben ist. Das zweidimensionale Integral ist 
einerseits auf den Fl"acheninhalt \mbox{${\D(2\pi/M)^2}$} des Quadrats, 
und andererseits auf $M$ normiert. Der Bereich, "uber den integriert 
wird, ist in Bild~\ref{E.b3b1} am Beispiel mit \mbox{$M\!=\!16$}, 
\mbox{$K_{\Phi}\!=\!4$}, \mbox{$\mu\!=\!7$} und \mbox{$\Tilde{\mu}\!=\!2$} 
als Quadrat der Breite \mbox{$2\pi/M$} eingezeichnet. Der Wert 
\mbox{$\Bar{\Phi}_{\boldsymbol{n}}(\mu,\mu\!+\!2\CdoT M/K_{\Phi})$} 
ist das Integral "uber den Vorfaktor 
\mbox{$\Grave{\Phi}_{\boldsymbol{n}}(\Omega,2)$} des Teils 
der Impulslinie, der die Diagonale des Quadrats bildet, wobei das 
Integral noch auf die Seitenl"ange des Quadrats normiert ist. 
Warum es wichtig ist, dass bei dem zweidimensionalen Integral die 
zus"atzliche Normierung auf $M$, und nicht nur die Normierung auf den 
Fl"acheninhalt des Quadrats durchgef"uhrt wird, sei kurz erl"autert. 
Betrachten wir dazu einen zyklostation"aren Prozess, bei dem die 
St"arke der Impulslinien f"ur hinreichend gro"se Werte von $M$ 
---\,also hinreichend schmale Integrationsbereiche\,---, als im Bereich 
der Integration konstant anzusehen ist. Da bei einem zyklostation"aren 
Prozess sich die gesamte Energie des LDS auf die zur Winkelhalbierenden 
der \mbox{$\Omega_1,\Omega_2$}-Ebene parallelen Geraden konzentriert, 
wird bei einer Ver"anderung von $M$ der Wert des Integrals vor der 
Normierung proportional zur L"ange der Diagonale des Quadrats, 
"uber das integriert wird, ---\,also indirekt proportional zu $M$\,--- sein. 
Es ist also {\em keine}\/ Proportionalit"at zur Fl"ache "uber die 
integriert wird ---\,also keine Proportionalit"at zu $M^{-2}$\,--- vorhanden, 
wie dies bei einem LDS eines instation"aren Prozesses der Fall w"are, bei 
dem das LDS sich auf die gesamte \mbox{$\Omega_1,\Omega_2$}-Ebene verteilt. 
Nach der Normierung auf die Gr"o"se der Fl"ache, "uber die integriert
wurde, ergibt sich somit eine Proportionalit"at zu $M$, 
was f"ur steigende Werte von $M$ den Impulsliniencharakter des
bifrequenten LDS widerspiegelt. Eine Stufenapproximation der Impulslinien,
die von der Wahl von $M$ abh"angt, ist jedoch nicht das, was
wir zur Beschreibung des LDS mittels endlich vieler Werte w"unschen.
Damit wir das erhalten, was wir eigentlich bestimmen wollen,
n"amlich N"aherungen f"ur die auf $2\pi$ normierten
St"arken \mbox{$\Grave{\Phi}_{\boldsymbol{n}}(\Omega,\Bar{\mu})$}
der Impulslinien, muss man zus"atzlich auf $M$ normieren,
um so zu erreichen, dass die Werte
\mbox{$\Bar{\Phi}_{\boldsymbol{n}}(\mu,\mu\!+\!\Tilde{\mu}\CdoT M/K_{\Phi})$}
f"ur steigende Werte von $M$ gegen diese Werte konvergieren, und nicht
divergieren.

\enlargethispage{2pt}Da eine Fensterfolge mit einem rechteckf"ormigen Betragsquadratspektrum 
zeitlich unbegrenzt ist, wird man versuchen, eine zeitlich begrenzte 
Fensterfolge zu verwenden, bei der das Betragsquadrat des Spektrums 
den rechteckigen Wunschverlauf m"oglichst gut approximiert. Da jedoch 
keine zeitlich begrenzte Fensterfolge abrupt vom Durchlassbereich 
\mbox{(\,$|\Omega|\le\pi/M$\,)}, in dem der Betrag des Spektrums nahe 
bei $M$ liegt, in den Sperrbereich \mbox{(\,$2\pi/M\le|\Omega|\le\pi$\,)} 
mit hoher D"ampfung wechselt, empfiehlt es sich $M$ wenigstens doppelt so 
gro"s wie $K_{\Phi}$ ---\,wenn m"oglich nat"urlich noch gr"o"ser\,--- 
zu w"ahlen, so dass au"ser dem Bereich der einen Impulslinie, der aus 
dem bifrequenten LDS durch das Betragsquadratspektrum der Fensterfolge 
herausgeschnitten werden soll, keine weitere Impulslinie bei den beiden 
verschobenen Spektren der Fensterfolge gleichzeitig durch den 
"Ubergangsbereich \mbox{(\,$\pi/M\le|\Omega|\le2\pi/M$\,)} der 
D"ampfung verl"auft. So kann eine zus"atzliche Verf"alschung der Werte 
\mbox{$\Tilde{\Phi}_{\boldsymbol{n}}(\mu,\mu\!+\!\Tilde{\mu}\CdoT M/K_{\Phi})$}
weitgehend vermieden werden. Weil das LDS des zyklostation"aren 
Approximationsfehlerprozesses aus einzelnen parallelen 
Impulslinien besteht, kommt dann bei Verwendung einer hoch 
frequenzselektiven Fensterfolge bei der Berechnung von 
\mbox{$\Tilde{\Phi}_{\boldsymbol{n}}(\mu,\mu\!+\!\Tilde{\mu}\CdoT M/K_{\Phi})$}
nach Gleichung~\ref{E.2.39} in guter N"aherung nur mehr die eine 
Impulslinie mit \mbox{$\Bar{\mu}\!=\!\Tilde{\mu}$} zum tragen, so 
dass nur das Betragsquadrat des Spektrums der Fensterfolge im Integral 
steht, und daher der Phasenfrequenzgang der Fensterfolge beliebig gew"ahlt 
werden kann. In Bild~\ref{E.b3b1} sind f"ur zwei verschiedene Fensterfolgen 
die H"ohenlinien des Betrags des Produktes der Fensterspektren 
im zweidimensionalen Integral in Gleichung~\ref{E.2.39} am Beispiel 
mit \mbox{$M\!=\!16$} und  \mbox{$K_{\Phi}\!=\!4$} eingetragen. Das eine 
Fenster ist das Rechteckfenster mit einer L"ange von $M$ und einer H"ohe 
von eins. Hier wurde \mbox{$\mu\!=\!13$}, \mbox{$\Tilde{\mu}\!=\!-2$} und 
f"ur die beiden H"ohenlinien die Werte \mbox{$M^2/2$} und \mbox{$M^2/100$} 
gew"ahlt. Da die H"ohenlinien jeweils geschlossene Kurven bilden, 
die die Maxima des Betrags des Produktes der Fensterspektren umschlie"sen, 
erkennt man an der H"ohenlinie mit \mbox{$M^2/2$} die ungef"ahre Lage 
des Hauptmaximums und an der H"ohenlinie mit \mbox{$M^2/100$} 
die ungef"ahre Lage der Nebenmaxima. Desweiteren erkennt man, dass 
der Betrag des Produktes der Fensterspektren in weiten Gebieten der 
\mbox{$\Omega_1,\Omega_2$}-Ebene nicht unter \mbox{$M^2/100$} absinkt, 
so dass bei dem Rechteckfenster zu erwarten ist, dass man mit 
\mbox{$\Tilde{\Phi}_{\boldsymbol{n}}(\mu,\mu\!+\!\Tilde{\mu}\CdoT M/K_{\Phi})$} 
nur eine schlechte N"aherung f"ur 
\mbox{$\Bar{\Phi}_{\boldsymbol{n}}(\mu,\mu\!+\!\Tilde{\mu}\CdoT M/K_{\Phi})$} 
erh"alt, wenn einerseits der zu n"ahernde Wert klein, und andererseits 
die durch die Ausblendung zu unterdr"uckenden Anteile gro"s sind. 
Das zweite Fenster ist das mit dem Algorithmus nach Kapitel~\myref{Algo} 
konstruierte Fenster, wobei die Fensterl"ange auf \mbox{$4\CdoT M$} 
festgelegt wurde. Bei diesem Fenster wurde \mbox{$\mu\!=\!4$}, 
\mbox{$\Tilde{\mu}\!=\!2$} und f"ur die beiden H"ohenlinien die Werte 
\mbox{$M^2/2$} und \mbox{$M^2/10000$} gew"ahlt. Die zweite H"ohenlinie 
ist hier also um den Faktor $100$ kleiner als bei dem Rechteckfenster. 
Man erkennt dass die Nebenmaxima wesentlich schneller abklingen, und 
au"serdem wesentlich niedriger liegen. Daher ist zu erwarten, dass 
durch die bessere Unterdr"uckung der Nebenlinien des LDS, also der Linien, 
die f"ur den eingestellten Wert von $\Tilde{\mu}$ gerade nicht in 
\mbox{$\Tilde{\Phi}_{\boldsymbol{n}}(\mu,\mu\!+\!\Tilde{\mu}\CdoT M/K_{\Phi})$} 
eingehen sollen, eine deutlich bessere N"aherung f"ur 
\mbox{$\Bar{\Phi}_{\boldsymbol{n}}(\mu,\mu\!+\!\Tilde{\mu}\CdoT M/K_{\Phi})$} 
erhalten werden kann. Um zu demonstrieren, warum man nicht \mbox{$M\!=\!K_{\Phi}$} 
w"ahlen sollte, wurde in den Bildern~\ref{E.b3b2} und \ref{E.b3b3} bei dieser 
Einstellung mit \mbox{$M\!=\!K_{\Phi}\!=\!1024$} der Betrag des Produktes der 
Fensterspektren im eindimensionalen Integral in Gleichung~(\ref{E.2.39}) einerseits 
f"ur den gew"unschten Summanden mit \mbox{$\Bar{\mu}\!=\!\Tilde{\mu}$} 
und andererseits f"ur die unerw"unschten Summanden mit 
\mbox{$\Bar{\mu}=\Tilde{\mu}\!+\!1$}, \mbox{$\Bar{\mu}=\Tilde{\mu}\!+\!2$} 
und \mbox{$\Bar{\mu}=\Tilde{\mu}\!+\!5$} als Funktionen "uber $\Omega$ 
aufgetragen. Wieder wurden dieselben beiden Fensterfolgen, allerdings mit dem 
Wert $M\!=\!1024$ statt $M\!=\!16$ verwendet. Da sich die wesentlichen Anteile 
des Fensterspektrums im Bereich kleiner Frequenzen befinden, ist dieser nochmals 
vergr"o"sert dargestellt. Bei der Darstellung des gesamten Frequenzbereichs 
erkennt man, einerseits den deutlich schnelleren und st"arkeren Abfall des 
Spektrums des nach Kapitel~\myref{Algo} berechneten Fensters zu hohen 
Frequenzen hin, so dass man davon ausgehen kann, dass selbst extrem starke 
Impulse auf den Nebenlinien gut unterdr"uckt werden k"onnen, die gegen"uber 
der gerade vermessenen Nebenlinie weit genug entfernt liegen. 
Andererseits sieht man, dass bei \mbox{$M\!=\!1024$} die Nullstellen der 
Fensterspektren so nahe beieinander liegen, dass man in der graphischen 
Darstellung unterhalb einer Linie, die die Nebenmaxima verbindet, nur 
mehr eine schwarze Fl"ache erh"alt. In der Darstellung des Bereichs 
niedriger Frequenzen ist bei beiden Fensterfolgen zu sehen, dass der 
Anteil der Nebenlinie mit \mbox{$\Bar{\mu}=\Tilde{\mu}\!+\!1$} f"ur 
die Frequenz \mbox{$\Omega\!=\!\pi/M$} nur mit etwa dem Faktor $0.5$ 
unterdr"uckt wird. Dies kann auch nicht anders sein, da diese 
Frequenz bei dem Spektrum der Fensterfolge gerade die Grenze 
des gew"unschten Durchlassbereichs darstellt. W"urde man einen 
schm"aleren Durchlassbereich w"ahlen, so w"are es praktisch unm"oglich, 
die Bedingung~(\myref{2.20}) zu erf"ullen, nach der die "Uberlagerung 
aller um Vielfache von \mbox{$2\pi/M$} verschobenen Betragsquadrate 
des Spektrums der Fensterfolge eine Konstante sein soll. Daher sollte 
man unbedingt vermeiden $M$ und $K_{\Phi}$ gleich zu w"ahlen. Bei dem 
nach Kapitel~\myref{Algo} berechneten Fenster ergibt sich in 
Bild~\ref{E.b3b3} bereits bei der zweiten Nebenlinie des LDS mit 
\mbox{$\Bar{\mu}=\Tilde{\mu}\!+\!2$} eine ganz brauchbare Unterdr"uckung 
von mehr als $30$dB im Maximum, und bei \mbox{$\Bar{\mu}=\Tilde{\mu}\!+\!5$} 
ist die Nebenlinie des LDS schon wenigstens um mehr als $50$dB abgeschw"acht. 
Wie Bild~\ref{E.b3b2} zeigt ist diese Unterdr"uckung bei dem Rechteckfenster 
bei weitem nicht so gut, und nimmt zu weiter entfernten Nebenlinien auch 
nicht so rasch ab.

Im weiteren wird angenommen, dass wir $M$ gro"s genug gew"ahlt haben und eine
hoch frequenzselektive Fensterfolge verwenden, so dass die Erwartungswerte
\begin{gather}
\text{E}\big\{\boldsymbol{N}_{\!\!f}(\mu)\CdoT
\boldsymbol{N}_{\!\!f}(\Hat{\mu})^{\Kk}\big\}\;=
\notag\\*[2pt]
=\;\frac{1}{(2\pi)^2}\cdoT\Int{-\pi}{\pi}\Int{-\pi}{\pi}
F\big({\T\mu\CdoT\frac{2\pi}{M}\!-\!\Omega_1}\big)\CdoT
F\big({\T\Hat{\mu}\CdoT\frac{2\pi}{M}\!-\!\Omega_2}\big)^{\!\Kk}\Cdot
\Phi_{\boldsymbol{n}}(\Omega_1,\Omega_2)\cdot
d\Omega_1\Cdot d\Omega_2\;\approx\;0
\notag\\*[-4pt]
{\T\forall\qquad\Hat{\mu}\neq\mu\!+\!\Tilde{\mu}\CdoT\frac{M}{K_{\Phi}}}
\label{E.2.41}
\end{gather}
f"ur alle Kombinationen von $\mu$ und $\Hat{\mu}$, die nicht bei der Berechnung der N"aherungswerte 
\mbox{$\Tilde{\Phi}_{\boldsymbol{n}}(\mu,\mu\!+\!\Tilde{\mu}\CdoT M/K_{\Phi})$}
auftreten, in guter N"aherung null sind. Diese Vernachl"assigung gilt vor allem 
dann, wenn diese Erwartungswerte in Summen auftreten, die au"serdem noch die 
N"aherungswerte des LDS enthalten, die vom Hauptmaximum des Spektrums der Fensterfolge 
aus dem LDS herausgeschnitten wurden. Es sei noch erw"ahnt, dass f"ur die Werte 
\mbox{$\Tilde{\Phi}_{\boldsymbol{n}}(\mu,\mu\!+\!\Tilde{\mu}\CdoT M/K_{\Phi})$} 
erstens die Symmetrie
\begin{gather}
\Tilde{\Phi}_{\boldsymbol{n}}
\big({\T\mu\!+\!\Tilde{\mu}\CdoT\frac{M}{K_{\Phi}},\mu}\big)\;=\;
\frac{1}{M}\cdot\text{E}\Big\{
\boldsymbol{N}_{\!\!f}
\big({\T\mu\!+\!\Tilde{\mu}\CdoT\frac{M}{K_{\Phi}}}\big)
\CdoT\boldsymbol{N}_{\!\!f}(\mu)^{\Kk}\Big\}\;=\notag\\[4pt]
=\;\frac{1}{M}\cdot\text{E}\Big\{\boldsymbol{N}_{\!\!f}(\mu)\CdoT
\boldsymbol{N}_{\!\!f}
\big({\T\mu\!+\!\Tilde{\mu}\CdoT\frac{M}{K_{\Phi}}}\big)^{\!\Kk}
\Big\}^{\!\!*}\,=\;
\Tilde{\Phi}_{\boldsymbol{n}}
\big({\T\mu,\mu\!+\!\Tilde{\mu}\CdoT\frac{M}{K_{\Phi}}}\big)^{\!\Kk}\!,
\label{E.2.42}
\end{gather}
und zweitens die Ungleichung
\begin{equation}
\Big|\Tilde{\Phi}_{\boldsymbol{n}}
\big({\T\mu,\mu\!+\!\Tilde{\mu}\CdoT\frac{M}{K_{\Phi}}}\big)\Big|^2
\;\le\;\Tilde{\Phi}_{\boldsymbol{n}}(\mu,\mu)\cdot
\Tilde{\Phi}_{\boldsymbol{n}}
\big({\T\mu\!+\!\Tilde{\mu}\CdoT\frac{M}{K_{\Phi}},
\mu\!+\!\Tilde{\mu}\CdoT\frac{M}{K_{\Phi}}}\big).
\label{E.2.43}
\end{equation}
gilt. Letzteres zeigt man indem man in Gleichung~(\myref{A.2.5}) des
Anhangs~\myref{Cauchy} die Substitutionen
\mbox{${\boldsymbol{X}} = \boldsymbol{N}_{\!\!f}(\mu)$} und
\mbox{$\boldsymbol{Y}\!= 
\boldsymbol{N}_{\!\!f}(\mu\!+\!\Tilde{\mu}\CdoT M/K_{\Phi})^{\Kk}$}
vornimmt.

Die Beschreibung der zweiten Momente des zyklostation"aren, mittelwertfreien
Approximationsfehlerprozesses \mbox{$\boldsymbol{n}(k)$} ist nur dann
vollst"andig, wenn man auch noch die Korrelationsfolge
\mbox{$\text{E}\big\{\boldsymbol{n}(k_1)\CdoT \boldsymbol{n}(k_2)\big\}$}
angibt, die wegen der Zyklostationarit"at die Periodizit"at
\begin{equation}
\text{E}\big\{\boldsymbol{n}(k_1)\CdoT \boldsymbol{n}(k_2)\big\}\;=\;
\text{E}\big\{\,\boldsymbol{n}(k_1\!+\!K_{\Phi})\cdot
\boldsymbol{n}(k_2\!+\!K_{\Phi})\,\big\}.
\label{E.2.44}
\end{equation}
aufweist. Durch zweidimensionale diskrete Fouriertransformation
---\,wieder mit Vorzeicheninvertierung bei der zweiten Frequenzvariable\,---
gewinnen wir daraus die bifrequente kontinuierliche und in
beiden Frequenzvariablen mit \mbox{$2\pi$} periodische Funktion
\begin{gather}
\Psi_{\boldsymbol{n}}(\Omega_1,\Omega_2)\;=
\Sum{k_1=-\infty}{\infty}\;\Sum{k_2=-\infty}{\infty}\!
\text{E}\big\{\boldsymbol{n}(k_1)\CdoT \boldsymbol{n}(k_2)\big\}\cdot
e^{\!-j\cdot\Omega_1\Cdot k_1}\Cdot
e^{j\cdot\Omega_2\Cdot k_2}\;=\notag\\[4pt]
=\Sum{\Bar{\mu}=-\infty}{\infty}
\frac{1}{K_{\Phi}}\cdoT\Sum{k=0}{K_{\Phi}-1}
\Acute{\Psi}_{\boldsymbol{n}}(\Omega_1,k)\cdot
e^{j\cdot\frac{2\pi}{K_{\Phi}}\cdot\Bar{\mu}\cdot k}\Cdot
2\pi\CdoT\delta_0\big({\T\Omega_2\!-\!\Omega_1\!-\!
\Bar{\mu}\CdoT\frac{2\pi}{K_{\Phi}}}\big)\;=\notag\\[4pt]
=\Sum{\Bar{\mu}=-\infty}{\infty}
\Grave{\Psi}_{\boldsymbol{n}}(\Omega_1,\Bar{\mu})\cdot
2\pi\CdoT\delta_0\big({\T\Omega_2\!-\!\Omega_1\!-\!
\Bar{\mu}\CdoT\frac{2\pi}{K_{\Phi}}}\big),
\label{E.2.45}
\end{gather}
die sich ebenfalls als "Uberlagerung von zur Gerade
\mbox{$\Omega_2=\Omega_1$} parallelen Impulslinien im Abstand
von \mbox{$2\pi/K_{\Phi}$} schreiben l"asst. Dabei sind
\begin{equation}
\Acute{\Psi}_{\boldsymbol{n}}(\Omega,k)\;=\;
\Acute{\Psi}_{\boldsymbol{n}}(\Omega,k+K_{\Phi})\;=
\Sum{\kappa=-\infty}{\infty}\!
\text{E}\big\{\boldsymbol{n}(k\!+\!\kappa)\CdoT
              \boldsymbol{n}(k)\big\}\cdot
e^{\!-j\cdot\Omega\cdot\kappa}
\label{E.2.46}
\end{equation}
die von der Zeit $k$ abh"angigen und in $\Omega$ kontinuierlichen diskreten
Fouriertransformierten der Kreuzkorrelationsfolgen zu den Zeitpunkten $k$,
die in $\Omega$ mit $2\pi$ und in $k$ mit $K_{\Phi}$ periodisch sind.
Die diskret invers Fouriertransformierten
\begin{equation}
\Grave{\Psi}_{\boldsymbol{n}}(\Omega,\Bar{\mu})\;=\;
\Grave{\Psi}_{\boldsymbol{n}}(\Omega,\Bar{\mu}+K_{\Phi})\;=\;
\frac{1}{K_{\Phi}}\cdoT\Sum{k=0}{K_{\Phi}-1}
\Acute{\Psi}_{\boldsymbol{n}}(\Omega,k)\cdot
 e^{j\cdot\frac{2\pi}{K_{\Phi}}\cdot\Bar{\mu}\cdot k}
\label{E.2.47}
\end{equation}
einer Periode in $k$ der Kreuzleistungsdichtespektren 
\mbox{$\Acute{\Psi}_{\boldsymbol{n}}(\Omega,k)$}
bei den Frequenzen \mbox{$\Bar{\mu}\CdoT\frac{2\pi}{K_{\Phi}}$} sind in
$\Bar{\mu}$ mit $K_{\Phi}$ periodisch und treten in dem bifrequenten
KLDS ---\,abgesehen von dem in Frequenzbereich typischerweise auftretenden
Vorfaktor $2\pi$\,--- als St"arken der Impulslinien auf, die in der
\mbox{$\Omega_1,\Omega_2$}-Ebene in einem Abstand von Vielfachen
von \mbox{$2\pi/K_{\Phi}$} parallel zur Winkelhalbierenden
\mbox{$\Omega_2\!=\!\Omega_1$} verlaufen. Um eine sinnvolle Aussage "uber diese
kontinuierlichen Funktionen machen zu k"onnen, geben wir zun"achst
wieder die endlich vielen Werte
\begin{gather}
\Check{\Psi}_{\boldsymbol{n}}(\mu,k)\;=\;
\frac{M}{2\pi}\cdoT\!\Int{-\frac{\pi}{M}}{\frac{\pi}{M}}\!
\Acute{\Psi}_{\boldsymbol{n}}\big({\T\mu\CdoT\frac{2\pi}{M}\!-\!\Omega,k}\big)
\cdot d\Omega\notag\\*[2pt]
\forall\qquad k=0\;(1)\;K_{\Phi}\!-\!1
\quad\text{ und }\quad\mu=0\;(1)\;M\!-\!1
\label{E.2.48}
\end{gather}
der fl"achengleichen Stufenapproximation an. Die daraus durch
inverse diskrete Fouriertransformation einer Periode gewonnenen
\mbox{$M\Cdot K_{\Phi}$} Werte
\begin{gather}
\Bar{\Psi}_{\boldsymbol{n}}
\big({\T\mu,\mu\!+\!\Tilde{\mu}\CdoT\frac{M}{K_{\Phi}}}\big)\;=\;
\frac{1}{K_{\Phi}}\cdoT\Sum{k=0}{K_{\Phi}-1}
\Check{\Psi}_{\boldsymbol{n}}(\mu,k)\cdot
e^{j\cdot\frac{2\pi}{K_{\Phi}}\cdot\Tilde{\mu}\cdot k}\notag\\*[3pt]
\forall\qquad\mu=0\;(1)\;M\!-\!1
\quad\text{ und }\quad\Tilde{\mu}=0\;(1)\;K_{\Phi}\!-\!1
\label{E.2.49}
\end{gather}
werden durch deren N"aherungswerte
\begin{gather}
\Tilde{\Psi}_{\boldsymbol{n}}
\big({\T\mu,\mu\!+\!\Tilde{\mu}\CdoT\frac{M}{K_{\Phi}}}\big)\;=\;
\frac{1}{M}\cdot\text{E}\Big\{\boldsymbol{N}_{\!\!f}(\mu)\CdoT
\boldsymbol{N}_{\!\!f}
\big(\!{\T-\mu\!-\!\Tilde{\mu}\CdoT\frac{M}{K_{\Phi}}}\big)
\Big\}\;=\notag\\[6pt]
=\;\frac{1}{(2\pi)^2\CdoT M}\cdoT\Int{-\pi}{\pi}\;\Int{-\pi}{\pi}
F\big({\T\mu\CdoT\frac{2\pi}{M}\!-\!\Omega_1}\big)\CdoT
F\big({\T\Omega_2\!-\!\mu\CdoT\frac{2\pi}{M}\!-\!
\Tilde{\mu}\CdoT\frac{2\pi}{K_{\Phi}}}\big)\Cdot
\Psi_{\boldsymbol{n}}(\Omega_1,\Omega_2)\cdot
d\Omega_1\Cdot d\Omega_2\;=
\notag\\[4pt]
=\;\Sum{\Bar{\mu}=0}{K_{\Phi}-1}
\frac{1}{2\pi\CdoT M}\cdoT\Int{-\pi}{\pi}
F(\Omega)\CdoT F\big({\T(\Bar{\mu}\!-\!\Tilde{\mu})\CdoT
\frac{2\pi}{K_{\Phi}}\!-\!\Omega}\big)\Cdot
\Grave{\Psi}_{\boldsymbol{n}}
\big({\T\mu\CdoT\frac{2\pi}{M}\!-\!\Omega,\Bar{\mu}}\big)\cdot d\Omega
\notag\\*[2pt]
\forall\qquad\mu=0\;(1)\;M\!-\!1\quad\text{ und }
\quad\Tilde{\mu}=0\;(1)\;K_{\Phi}\!-\!1,
\label{E.2.50}
\end{gather}
die man durch Erwartungswertbildung aus den zuf"alligen Spektralwerten
des gefensterten Approximationsfehlerprozesses gewinnen kann, ersetzt.
Es sei wieder angemerkt, das f"ur eine reelle Fensterfolge
dieselben Fensterspektren im Integral stehen wie in Gleichung~(\ref{E.2.39}). 
Auch hier ist festzustellen, dass der Erwartungswert
\begin{gather}
\text{E}\Big\{\boldsymbol{N}_{\!\!f}(\mu)\CdoT
\boldsymbol{N}_{\!\!f}(\!-\Hat{\mu})\;\Big\}\;=
\notag\\[4pt]
=\;\frac{1}{(2\pi)^2}\cdoT\Int{-\pi}{\pi}\Int{-\pi}{\pi}
F\big({\T\mu\CdoT\frac{2\pi}{M}\!-\!\Omega_1}\big)\CdoT
F\big({\T\Omega_2\!-\!\Hat{\mu}\CdoT\frac{2\pi}{M}}\big)\Cdot
\Psi_{\boldsymbol{n}}(\Omega_1,\Omega_2)\cdot
d\Omega_1\Cdot d\Omega_2\;\approx\;0
\notag\\*
\forall\qquad{\T\Hat{\mu}\neq\mu\!+\!\Tilde{\mu}\CdoT\frac{M}{K_{\Phi}}}
\label{E.2.51}
\end{gather}
f"ur alle Kombinationen von $\mu$ und $\Hat{\mu}$, die nicht bei der Berechnung von 
\mbox{$\Tilde{\Psi}_{\boldsymbol{n}}\big(\mu,\mu\!+\!\Tilde{\mu}\CdoT\frac{M}{K_{\Phi}}\big)$}
auftreten, in guter N"aherung null ist, wenn man $M$ gro"s genug 
w"ahlt, und eine hoch frequenzselektive Fensterfolge verwendet. 
\mbox{$\Tilde{\Psi}_{\boldsymbol{n}}\big(\mu,\mu\!+\!\Tilde{\mu}\CdoT\frac{M}{K_{\Phi}}\big)$}
weist die Symmetrie
\begin{gather}
\Tilde{\Psi}_{\boldsymbol{n}}
\big({\T\mu\!+\!\Tilde{\mu}\CdoT\frac{M}{K_{\Phi}},\mu}\big)\;=\;
\frac{1}{M}\cdot\text{E}\Big\{
\boldsymbol{N}_{\!\!f}
\big({\T\mu\!+\!\Tilde{\mu}\CdoT\frac{M}{K_{\Phi}}}\big)
\CdoT\boldsymbol{N}_{\!\!f}(\!-\mu)\Big\}\;=
\notag\\[3pt]
=\;\frac{1}{M}\cdot\text{E}\Big\{\boldsymbol{N}_{\!\!f}(\!-\mu)\CdoT
\boldsymbol{N}_{\!\!f}
\big({\T\mu\!+\!\Tilde{\mu}\CdoT\frac{M}{K_{\Phi}}}\big)
\Big\}\;=\;
\Tilde{\Psi}_{\boldsymbol{n}}
\big(\!{\T-\mu,-\mu\!-\!\Tilde{\mu}\CdoT\frac{M}{K_{\Phi}}}\big)
\label{E.2.52}
\end{gather}
auf, und erf"ullt die Ungleichung
\begin{equation}
\Big|\Tilde{\Psi}_{\boldsymbol{n}}
\big({\T\mu,\mu\!+\!\Tilde{\mu}\CdoT\frac{M}{K_{\Phi}}}\big)\Big|^2
\;\le\;\Tilde{\Phi}_{\boldsymbol{n}}(\mu,\mu)\cdot
\Tilde{\Phi}_{\boldsymbol{n}}
\big(\!{\T-\mu\!-\!\Tilde{\mu}\CdoT\frac{M}{K_{\Phi}},
-\mu\!-\!\Tilde{\mu}\CdoT\frac{M}{K_{\Phi}}}\big),
\label{E.2.53}
\end{equation}
deren G"ultigkeit man mit
\mbox{${\boldsymbol{X}} = \boldsymbol{N}_{\!\!f}(\mu)$} und
\mbox{$\boldsymbol{Y}\!=
\boldsymbol{N}_{\!\!f}(\!-\mu\!-\!\Tilde{\mu}\CdoT M/K_{\Phi})$}
in Gleichung~(\myref{A.2.5}) des Anhangs~\myref{Cauchy} zeigt.

Wenn man ber"ucksichtigt, dass sich die zweidimensionale
Autokorrelationsfolge nach Gleichung~(\ref{E.1.1}) durch die inverse
zweidimensionale Fourierr"ucktransformation ---\,wieder mit
Vorzeicheninvertierung bei $\Omega_2$\,--- aus dem bifrequenten LDS gem"a"s
\begin{equation}
\text{E}\big\{\boldsymbol{n}(k_1)\CdoT \boldsymbol{n}(k_2)^{\Kk}\big\}\;=\;
\frac{1}{(2\pi)^2}\cdoT
\Int{-\pi}{\pi}\;\Int{-\pi}{\pi}
\Phi_{\boldsymbol{n}}(\Omega_1,\Omega_2)\cdot
e^{j\cdot(\Omega_1\Cdot k_1-\Omega_2\Cdot k_2)}\Cdot
d\Omega_1\Cdot d\Omega_2
\label{E.2.54}
\end{equation}
berechnen l"asst, kann man die zeitabh"angige Varianz des
Approximationsfehlers, die man mit \mbox{$k_1\!=\!k_2\!=\!k$} erh"alt,
als endliche Summe
\begin{gather}
\text{E}\big\{|\boldsymbol{n}(k)|^2\,\big\}\;=\;
\frac{1}{(2\pi)^2}\cdoT
\Int{-\pi}{\pi}\;\Int{-\pi}{\pi}\Phi_{\boldsymbol{n}}(\Omega_1,\Omega_2)\CdoT
e^{j\cdot(\Omega_1-\Omega_2)\cdot k}
\cdot\:d\Omega_1\cdot\,d\Omega_2\;=\notag\\[6pt]
=\;\frac{1}{M}\cdoT\Sum{\mu=0}{M-1}\;\Sum{\Tilde{\mu}=0}{K_{\Phi}-1}
\Bar{\Phi}_{\boldsymbol{n}}
\big({\T\mu,\mu\!+\!\Tilde{\mu}\CdoT\frac{M}{K_{\Phi}}}\big)\cdot
e^{\!-j\cdot\frac{2\pi}{K_{\Phi}}\cdot\Tilde{\mu}\cdot k}
\label{E.2.55}
\end{gather}
angeben. Wenn man die gleiche Summe mit den N"aherungswerten 
\mbox{$\Tilde{\Phi}_{\boldsymbol{n}}(\mu,\mu\!+\!\Tilde{\mu}\CdoT M/K_{\Phi})$} 
bildet, und wenn man $M$ hinreichend gro"s gew"ahlt hat, so dass man vom 
bifrequenten LDS die Impulslinien mit \mbox{$\Bar{\mu}\neq\Tilde{\mu}$} 
vernachl"assigen kann, erh"alt man in guter N"aherung
\begin{gather}
\frac{1}{M}\cdoT
\Sum{\mu=0}{M-1}\;
\Sum{\Tilde{\mu}=0}{K_{\Phi}-1}
\Tilde{\Phi}_{\boldsymbol{n}}
\big({\T\mu,\mu\!+\!\Tilde{\mu}\CdoT\frac{M}{K_{\Phi}}}\big)\cdot
e^{\!-j\cdot\frac{2\pi}{K_{\Phi}}\cdot\Tilde{\mu}\cdot k}\;=
\notag\\[6pt]
=\!\frac{1}{2\pi\CdoT M^2}\cdoT\!
\Sum{\mu=0}{M-1}\;
\Sum{\Tilde{\mu}=0}{K_{\Phi}-1}\;
\Sum{\Bar{\mu}=0}{K_{\Phi}-1}\,
\Int{-\pi}{\pi}\!\!
F(\Omega)\CdoT F\big({\T\Omega\!-\!
(\Bar{\mu}\!-\!\Tilde{\mu})\CdoT\frac{2\pi}{K_{\Phi}}}\big)^{\!\Kk}\!\CdoT
\Grave{\Phi}_{\boldsymbol{n}}
\big({\T\mu\CdoT\frac{2\pi}{M}\!-\!\Omega,\Bar{\mu}}\big)\CdoT d\Omega\CdoT
e^{\!\!-j\cdot\frac{2\pi}{K_{\Phi}}\cdot\Tilde{\mu}\cdot k}\!\approx\!
\notag\\[4pt]
\approx\;\frac{1}{2\pi\CdoT M^2}\cdoT
\Sum{\mu=0}{M-1}\;
\Sum{\Tilde{\mu}=0}{K_{\Phi}-1}\;
\Int{-\pi}{\pi}
\big|F(\Omega)\big|^2\Cdot
\Grave{\Phi}_{\boldsymbol{n}}
\big({\T\mu\CdoT\frac{2\pi}{M}\!-\!\Omega,\Tilde{\mu}}\big)\cdot d\Omega\cdot
e^{\!-j\cdot\frac{2\pi}{K_{\Phi}}\cdot\Tilde{\mu}\cdot k}\;=
\notag\\[4pt]
=\;\frac{1}{2\pi\CdoT M^2}\cdoT
\Sum{\mu=0}{M-1}\;
\Int{-\pi}{\pi}\big|F(\Omega)\big|^2\cdoT
\Sum{\Tilde{\mu}=0}{K_{\Phi}-1}
\Grave{\Phi}_{\boldsymbol{n}}
\big({\T\mu\CdoT\frac{2\pi}{M}\!-\!\Omega,\Tilde{\mu}}\big)\cdot
e^{\!-j\cdot\frac{2\pi}{K_{\Phi}}\cdot\Tilde{\mu}\cdot k}\Cdot d\Omega\;=
\notag\displaybreak[2]\\[4pt]
=\;\frac{1}{2\pi\CdoT M^2}\cdoT
\Sum{\mu=0}{M-1}\;
\Int{-\pi}{\pi}\big|F(\Omega)\big|^2\Cdot
\Acute{\Phi}_{\boldsymbol{n}}
\big({\T\mu\CdoT\frac{2\pi}{M}\!-\!\Omega,k}\big)\cdot d\Omega\;=
\notag\\[4pt]
=\;\frac{1}{2\pi\CdoT M^2}\cdoT
\Sum{\mu=0}{M-1}\;
\Int{-\pi}{\pi}
\big|F(\Omega)\big|^2\CdoT\!
\Sum{\kappa=-\infty}{\infty}\!
\text{E}\big\{\boldsymbol{n}(k\!+\!\kappa)\CdoT
             \boldsymbol{n}(k)^{\Kk}\big\}\Cdot
e^{\!-j\cdot\big(\mu\cdot\frac{2\pi}{M}-\Omega\big)\cdot\kappa}\Cdot
d\Omega\;=
\notag\\[6pt]
=\;\frac{1}{2\pi\CdoT M^2}\cdoT
\Sum{\mu=0}{M-1}\;
\Int{-\pi}{\pi}\big|
F\big({\T\mu\CdoT\frac{2\pi}{M}\!-\!\Tilde{\Omega}}\big)\big|^2\CdoT\!
\Sum{\kappa=-\infty}{\infty}\!
\text{E}\big\{\boldsymbol{n}(k\!+\!\kappa)\CdoT
              \boldsymbol{n}(k)^{\Kk}\big\}\cdot
e^{\!-j\cdot\Tilde{\Omega}\cdot\kappa}\Cdot d\Tilde{\Omega}\;=
\notag\\[6pt]
=\;\frac{1}{2\pi\CdoT M^2}\cdoT
\Sum{\kappa=-\infty}{\infty}\;\;
\Int{-\pi}{\pi}\;
\underbrace{
\Sum{\mu=0}{M-1}\big|
F\big({\T\mu\CdoT\frac{2\pi}{M}\!-\!\Tilde{\Omega}}\big)\big|^2}_
{=\;M^2\text{ nach (\myref{2.20})}}\Cdot\,
\text{E}\big\{\boldsymbol{n}(k\!+\!\kappa)\CdoT
              \boldsymbol{n}(k)^{\Kk}\big\}\cdot
e^{\!-j\cdot\Tilde{\Omega}\cdot\kappa}\Cdot d\Tilde{\Omega}\;=
\notag\\
=\Sum{\kappa=-\infty}{\infty}\!
\text{E}\big\{\boldsymbol{n}(k\!+\!\kappa)\CdoT
              \boldsymbol{n}(k)^{\Kk}\big\}\cdot
\frac{1}{2\pi}\cdoT\!\Int{-\pi}{\pi}
e^{\!-j\cdot\Tilde{\Omega}\cdot\kappa}\Cdot d\Tilde{\Omega}\;=
\notag\\[2pt]
=\Sum{\kappa=-\infty}{\infty}\!
\text{E}\big\{\boldsymbol{n}(k\!+\!\kappa)\CdoT
              \boldsymbol{n}(k)^{\Kk}\big\}\cdot\gamma_0(\kappa)\;=\;
\text{E}\big\{|\boldsymbol{n}(k)|^2\big\},
\label{E.2.56}
\end{gather}
wobei vorausgesetzt wurde, dass man eine Fensterfolge verwendet, deren 
Spektrum der Bedingung (\myref{2.20}) gen"ugt.

\enlargethispage{2pt}{\small Anmerkung: Wenn man eine Fensterfolge verwendet, die mit dem 
in Kapitel~\myref{Algo} vorgestellten Algorithmus berechnet wird, 
und wenn man die Fensterl"ange gro"s genug w"ahlt, erh"alt man bereits 
f"ur \mbox{$M=2\CdoT K_{\Phi}$} eine so hohe Sperrd"ampfung, dass die eben 
gemachte N"aherung zu Fehlern bei der Berechnung der zeitabh"angigen 
theoretischen Varianz f"uhrt, die um Gr"o"senordnungen kleiner sind als 
die zuf"alligen Abweichungen der empirischen Varianz, die man mit dem 
RKM selbst f"ur extrem gro"se Mittelungsanzahlen $L$ misst. Es gibt 
jedoch wenigstens zwei M"oglichkeiten, die N"aherung bei der Berechnung 
der theoretischen Varianz zu umgehen. Die eine M"oglichkeit besteht 
darin, den zweidimensionalen Approximationsfehlerprozess 
\mbox{$\boldsymbol{n}(k_1)\CdoT\boldsymbol{n}(k_2)$} zu bilden, und 
diesen mit einer zweidimensionalen Fensterfolge zu multiplizieren, deren 
zweidimensionales Spektrum in der \mbox{$\Omega_1,\Omega_2$}-Ebene dort 
Nullinien aufweist, wo das zweidimensionale LDS seine Impulslinien hat 
(\,mit Ausnahme der Linie \mbox{$\Omega_1\!=\!\Omega_2$}\,) und die auch 
noch einige andere Bedingungen erf"ullen muss. Eine solche zweidimensionale 
Fensterfolge kann man z.~B. dadurch erhalten, dass man aus der mit dem in 
Kapitel~\myref{Algo} vorgestellten Algorithmus berechneten eindimensionalen 
Fensterfolge \mbox{$f(k)$} die zweidimensionale Fensterfolge 
\mbox{$f(k_1)\CdoT f(k_2)$} bildet und diese mit der zweidimensionalen 
Folge faltet, die nur f"ur \mbox{$0\!\le\!k_1\!=\!k_2\!<\!K_{\Phi}$} 
eins ist und sonst null. Die so entstandene zweidimensionale 
Fensterfolge erf"ullt alle Forderungen, die f"ur die Anwendung beim 
RKM notwendig sind, und erm"oglicht ebenfalls eine erwartungstreue Messung 
der zeitabh"angigen Varianz. Da der sich durch die Verwendung des 
zweidimensional gefensterten Approximationsfehlerprozesses beim RKM 
ergebende Mehraufwand nicht unerheblich ist, ist diese Modifikation 
jedoch nicht zu rechtfertigen. Daher wird auf die ausf"uhrliche 
Darstellung dieser Modifikationen verzichtet. Eine weitere 
M"oglichkeit besteht darin, die Messung mit dem RKM f"ur die 
Korrelationen aller \mbox{$K_{\Phi}$} Polyphasenkomponenten \cite{Fliege} 
des Eingangs- und des Ausgangssignals getrennt durchzuf"uhren, 
und so die einzelnen Polyphasenkomponenten mit einer entsprechend 
um den Faktor \mbox{$K_{\Phi}$} gespreizten Fensterfolge zu fenstern, 
wie dies in \cite{Heinle} f"ur die Messung von Multiratensystemen 
durchgef"uhrt wird. Die dabei auftretenden Prozesse der einzelnen 
Polyphasenkomponenten sind dann alle station"ar, so dass 
Nebenlinien im bifrequenten LDS der Polyphasenkomponenten 
dann nicht mehr auftreten.}

Analog erh"alt man, wenn man $M$ hinreichend gro"s gew"ahlt hat, 
so dass man bei \mbox{$\Psi_{\boldsymbol{n}}(\Omega_1,\Omega_2)$} 
nach Gleichung~(\ref{E.2.25}) die Impulslinien mit 
\mbox{$\Bar{\mu}\neq-\Tilde{\mu}$} vernachl"assigen kann, 
f"ur die Summe "uber alle mit denselben Drehfaktoren wie 
in Gleichung~(\ref{E.2.56}) multiplizierten Werte 
\mbox{$\Tilde{\Psi}_{\boldsymbol{n}}(\mu,\mu\!+\!\Tilde{\mu}\CdoT M/K_{\Phi})$} 
ein gute N"aherung f"ur den zeitabh"angigen Erwartungswert 
\mbox{$\text{E}\big\{\!\boldsymbol{n}(k)^2\big\}$}.
\begin{gather}
\frac{1}{M}\cdoT
\Sum{\mu=0}{M-1}\;
\Sum{\Tilde{\mu}=0}{K_{\Phi}-1}
\Tilde{\Psi}_{\boldsymbol{n}}
\big({\T\mu,\mu\!+\!\Tilde{\mu}\CdoT\frac{M}{K_{\Phi}}}\big)\cdot
e^{\!-j\cdot\frac{2\pi}{K_{\Phi}}\cdot\Tilde{\mu}\cdot k}\;=
\notag\\[6pt]
=\frac{1}{2\pi\CdoT M^2}\cdoT\!
\Sum{\mu=0}{M-1}\;
\Sum{\Tilde{\mu}=0}{K_{\Phi}-1}\;
\Sum{\Bar{\mu}=0}{K_{\Phi}-1}\,
\Int{-\pi}{\pi}\!\!F(\Omega)\CdoT
F\big({\T(\Bar{\mu}\!-\!\Tilde{\mu})\CdoT\frac{2\pi}{K_{\Phi}}\!-\!\Omega}\big)\CdoT
\Grave{\Psi}_{\boldsymbol{n}}
\big({\T\mu\CdoT\frac{2\pi}{M}\!-\!\Omega,\Bar{\mu}}\big)\CdoT d\Omega\CdoT
e^{\!-j\cdot\frac{2\pi}{K_{\Phi}}\cdot\Tilde{\mu}\cdot k}\approx\!
\notag\\[4pt]
\approx\;\frac{1}{2\pi\CdoT M^2}\cdoT
\Sum{\mu=0}{M-1}\;
\Sum{\Tilde{\mu}=0}{K_{\Phi}-1}\;
\Int{-\pi}{\pi}\!
F(\Omega)\CdoT F(\!-\Omega)\cdot
\Grave{\Psi}_{\boldsymbol{n}}
\big({\T\mu\CdoT\frac{2\pi}{M}\!-\!\Omega,\Tilde{\mu}}\big)\cdot d\Omega\cdot
e^{\!-j\cdot\frac{2\pi}{K_{\Phi}}\cdot\Tilde{\mu}\cdot k}\;=
\notag\\[4pt]
=\;\frac{1}{2\pi\CdoT M^2}\cdoT
\Sum{\mu=0}{M-1}\;
\Int{-\pi}{\pi}\!F(\Omega)\CdoT F(\!-\Omega)\cdoT
\Sum{\Tilde{\mu}=0}{K_{\Phi}-1}
\Grave{\Psi}_{\boldsymbol{n}}
\big({\T\mu\CdoT\frac{2\pi}{M}\!-\!\Omega,\Tilde{\mu}}\big)\cdot
e^{\!-j\cdot\frac{2\pi}{K_{\Phi}}\cdot\Tilde{\mu}\cdot k}\Cdot
d\Omega\;=
\notag\\[4pt]
=\;\frac{1}{2\pi\CdoT M^2}\cdoT
\Sum{\mu=0}{M-1}\;
\Int{-\pi}{\pi}\!F(\Omega)\CdoT F(\!-\Omega)\cdot
\Acute{\Psi}_{\boldsymbol{n}}
\big({\T\mu\CdoT\frac{2\pi}{M}\!-\!\Omega,k}\big)\cdot
d\Omega\;=
\notag\\[4pt]
=\;\frac{1}{2\pi\CdoT M^2}\cdoT
\Sum{\mu=0}{M-1}\;
\Int{-\pi}{\pi}
\big|F(\Omega)\big|^2\CdoT\!
\Sum{\kappa=-\infty}{\infty}\!
\text{E}\big\{\boldsymbol{n}(k\!+\!\kappa)\CdoT
              \boldsymbol{n}(k)\big\}\Cdot
e^{\!-j\cdot\big(\mu\cdot\frac{2\pi}{M}-\Omega\big)\cdot\kappa}\Cdot
d\Omega\;=
\notag\\[6pt]
=\;\frac{1}{2\pi\CdoT M^2}\cdoT
\Sum{\mu=0}{M-1}\;
\Int{-\pi}{\pi}
\big|F\big({\T\mu\CdoT\frac{2\pi}{M}\!-\!\Tilde{\Omega}}\big)\big|^2\CdoT\!
\Sum{\kappa=-\infty}{\infty}\!
\text{E}\big\{\boldsymbol{n}(k\!+\!\kappa)\CdoT
              \boldsymbol{n}(k)\big\}\cdot
e^{\!-j\cdot\Tilde{\Omega}\cdot\kappa}\Cdot d\Tilde{\Omega}\;=
\notag\\[6pt]
=\;\frac{1}{2\pi\CdoT M^2}\cdoT
\Sum{\kappa=-\infty}{\infty}\;\;
\Int{-\pi}{\pi}\;
\underbrace{
\Sum{\mu=0}{M-1}\big|
F\big({\T\mu\CdoT\frac{2\pi}{M}\!-\!\Tilde{\Omega}}\big)\big|^2}_
{=\;M^2\text{ nach (\myref{2.20})}}\Cdot\,
\text{E}\big\{\boldsymbol{n}(k\!+\!\kappa)\CdoT
              \boldsymbol{n}(k)\big\}\cdot
e^{\!-j\cdot\Tilde{\Omega}\cdot\kappa}\Cdot d\Tilde{\Omega}\;=
\notag\displaybreak[2]\\
=\Sum{\kappa=-\infty}{\infty}\!
\text{E}\big\{\boldsymbol{n}(k\!+\!\kappa)\CdoT
              \boldsymbol{n}(k)\big\}\cdot
\frac{1}{2\pi}\cdoT\!\Int{-\pi}{\pi}
e^{\!-j\cdot\Tilde{\Omega}\cdot\kappa}\Cdot d\Tilde{\Omega}\;=
\notag\\[2pt]
=\Sum{\kappa=-\infty}{\infty}\!
\text{E}\big\{\boldsymbol{n}(k\!+\!\kappa)\CdoT
              \boldsymbol{n}(k)\big\}\cdot\gamma_0(\kappa)\;=\;
\text{E}\big\{\boldsymbol{n}(k)^2\big\}
\label{E.2.57}
\end{gather}
Dabei wurde vorausgesetzt, dass man eine reelle Fensterfolge verwendet,
deren Spektrum der Bedingung~(\myref{2.20}) gen"ugt.

In diesem Kapitel wurde gezeigt, wie sich ein reales gest"ortes 
System durch das Systemmodell nach Bild \ref{E.b1h} modellieren l"asst.
Die sich ergebenden {\em theoretisch} optimalen Werte der deterministischen 
St"orung und der beiden bifrequenten "Ubertragungsfunktionen wurden in der 
Art bestimmt, dass der verbleibende Approximationsfehlerprozess eine 
minimale Varianz aufweist. Die theoretischen Optimall"osungen berechnen 
sich gem"a"s der Gleichungen (\ref{E.2.25}) und (\ref{E.2.19}) bzw. 
(\ref{E.2.29}), falls folgende Voraussetzungen erf"ullt sind:
{\setlength{\parskip}{0ex}\begin{itemize}
\item Das System wird mit einem bereichsweise periodischen
      Zufallssignal erregt.
\item Die Autokovarianzmatrix 
      \raisebox{0.4ex}{$\underline{C}_{\Tilde{\Vec{\boldsymbol{V}}}(\mu),\Tilde{\Vec{\boldsymbol{V}}}(\mu)}$}
      der Spektralwerte des erregenden Zufallssignals ist f"ur alle $M$ diskreten Frequenzen 
      \mbox{$\mu\CdoT\frac{2\pi}{M}$} regul"ar.
\item Das System l"asst sich durch ein lineares periodisch zeitvariantes System
      approximieren, d.~h. die Approximationsl"osung wiederholt sich periodisch 
      mit dem Zeitpunkt der Approximation. 
\item Bei der Berechnung der Optimall"osungen  wird eine
      Fensterfolge verwendet, deren Spektralwerte bei den diskreten
      Frequenzen \mbox{$\mu\CdoT\frac{2\pi}{M}$} die Bedingung (\myref{2.27}) erf"ullen.
\end{itemize}
Es wurde zur Beschreibung des bifrequenten LDS und des KLDS des Fehlerprozesses 
in den Gleichungen (\ref{E.2.39}) und (\ref{E.2.50}) jeweils eine zweidimensionale 
Spektralfolge endlich vieler Werte angegeben, die sich mit Hilfe einer Fensterung, 
einer DFT und einer Erwartungswertbildung aus den Ein- und Ausgangsprozessen berechnen 
l"asst. Dabei stellen die folgenden Voraussetzungen sicher, dass die beiden Spektralfolgen 
in der Lage sind das LDS und das KLDS des Approximationsfehlerprozesses aussagekr"aftig 
zu beschreiben.
\begin{itemize}
\item Der bei der Approximation verbleibende Fehlerprozess ist zyklostation"ar.
\item Es wird eine Fensterfolge verwendet, deren Spektrum die Bedingung (\myref{2.20})
      erf"ullt, und das eine hohe Sperrd"ampfung aufweist. 
\end{itemize}
Sollten diese Voraussetzungen vom realen System nicht a priori erf"ullt sein, kann eine
zuf"allige Verschiebung des Zugriffszeitpunktes auf das reale System, wie sie im 
ersten Unterkapitel beschrieben ist, daf"ur sorgen, dass die so entstehenden 
Modellzufallsvektoren die Voraussetzungen erf"ullen.}


\chapter{Das Rauschklirrmessverfahren mit Fensterung}\label{E.Kap.3}

Nun wollen wir uns der Frage widmen, wie man die mit Hilfe
einer Messung Sch"atzwerte
\begin{align*}
\Hat{H}\big({\T\mu,\mu\!+\!\Tilde{\mu}\CdoT\frac{M}{K_H}}\big)&
\qquad\text{ mit }\quad\mu=0\;(1)\;M\!-\!1\quad\text{ und }\quad\Tilde{\mu}=0\;(1)\;K_H\!-\!1,\\
\Hat{H}_*\big({\T\mu,\mu\!+\!\Tilde{\mu}\CdoT\frac{M}{K_H}}\big)&
\qquad\text{ mit }\quad\mu=0\;(1)\;M\!-\!1\quad\text{ und }\quad\Tilde{\mu}=0\;(1)\;K_H\!-\!1,\\
\Hat{u}(k)&
\qquad\text{ mit }\quad k=0\;(1)\;F\!-\!1,\\
\Hat{U}_{\!f}(\mu)&
\qquad\text{ mit }\quad\mu=0\;(1)\;M\!-\!1,\\
\Hat{\Phi}_{\boldsymbol{n}}\big({\T\mu,\mu\!+\!\Tilde{\mu}\CdoT\frac{M}{K_{\Phi}}}\big)&
\qquad\text{ mit }\quad\mu=0\;(1)\;M\!-\!1\quad\text{ und }\quad\Tilde{\mu}=0\;(1)\;K_{\Phi}\!-\!1,\\
\Hat{\Psi}_{\boldsymbol{n}}\big({\T\mu,\mu\!+\!\Tilde{\mu}\CdoT\frac{M}{K_{\Phi}}}\big)&
\qquad\text{ mit }\quad\mu=0\;(1)\;M\!-\!1\quad\text{ und }\quad\Tilde{\mu}=0\;(1)\;K_{\Phi}\!-\!1\\
\intertext{f"ur die Optimall"osungen}
H\!\big({\T \mu,\mu\!+\!\Tilde{\mu}\CdoT\frac{M}{K_H}}\big)&
\qquad\text{ mit }\quad\mu=0\;(1)\;M\!-\!1\quad\text{ und }\quad\Tilde{\mu}=0\;(1)\;K_H\!-\!1,\\
H_*\!\big({\T \mu,\mu\!+\!\Tilde{\mu}\CdoT\frac{M}{K_H}}\big)&
\qquad\text{ mit }\quad\mu=0\;(1)\;M\!-\!1\quad\text{ und }\quad\Tilde{\mu}=0\;(1)\;K_H\!-\!1,\\
u(k)&
\qquad\text{ mit }\quad k=0\;(1)\;F\!-\!1\quad\text{ und }\\
U_{\!f}(\mu)&
\qquad\text{ mit }\quad\mu=0\;(1)\;M\!-\!1,\\
\intertext{sowie f"ur die N"aherungen}
\Tilde{\Phi}_{\boldsymbol{n}}\big({\T\mu,\mu\!+\!\Tilde{\mu}\CdoT\frac{M}{K_{\Phi}}}\big)&
\qquad\text{ mit }\quad\mu=0\;(1)\;M\!-\!1\quad\text{ und }\quad\Tilde{\mu}=0\;(1)\;K_{\Phi}\!-\!1,\\
\Tilde{\Psi}_{\boldsymbol{n}}\big({\T\mu,\mu\!+\!\Tilde{\mu}\CdoT\frac{M}{K_{\Phi}}}\big)&
\qquad\text{ mit }\quad\mu=0\;(1)\;M\!-\!1\quad\text{ und }\quad\Tilde{\mu}=0\;(1)\;K_{\Phi}\!-\!1
\end{align*}
der Stufenapproximationen des LDS und des KLDS des Approximationsfehlers
gewinnen kann. Wir gehen dazu im wesentlichen so vor, wie dies bei
der empirischen Bestimmung der Regressionskoeffizienten "ublich ist.
Die mit einer Messung f"ur diese Gr"o"sen gewonnenen Sch"atzwerte
werden im weiteren meist als Messwerte bezeichnet.

\mbox{ }

\section[Messung der "Ubertragungsfunktionen und der deterministischen 
St"orung]{Messung der "Ubertragungsfunktionen und der\\
deterministischen St"orung}\label{E.Kap.3.1}

Um die Sch"atzwerte mit Hilfe einer Messung bestimmen zu k"onnen, 
erregen wir das reale System in $L$ Einzelmessungen mit einer Stichprobe 
vom Umfang $L$ des Zufallsvektors \mbox{$\Vec{\boldsymbol{v}}$} wie dies in 
Kapitel \myref{RKM} beschrieben ist. Am Ausgang des Systems messen wir 
die $L$ Ausgangssignalfolgen \mbox{$y_{\lambda}(k)$} mit 
\mbox{$\lambda=1\;(1)\;L$} und berechnen daraus durch Fensterung und 
Fouriertransformation wie in \cite{Diss} die \mbox{$M\CdoT L$} Spektralwerte 
\mbox{$Y_{\!f,\lambda}(\mu)$}. 

Nun werden die Sch"atzwerte f"ur die 
Regressionskoeffizienten dadurch gewonnen, dass wir anhand dieser Stichproben 
vom Umfang $L$ die Sch"atzwerte f"ur die Regressionskoeffizienten in der Art 
w"ahlen, dass sich der kleinste quadratische Fehler bei der Approximation der 
Stichprobe des Spektrums des gefensterten  Ausgangssignals des realen Systems 
durch die Stichproben der Spektren der Ausgangssignale der beiden Modellsysteme 
und die deterministische Modellst"orung ergibt. Wir minimieren also nicht die 
theoretische Varianz des Spektrums \mbox{$\boldsymbol{N}_{\!\!f}(\mu)$} des 
gefensterten Approximationsfehlers, sondern das Betragsquadrat der L"ange des 
Vektors, der sich als die Differenz des Stichprobenvektors 
\mbox{$\Vec{Y}_{\!f}(\mu)$} des Spektrums des Ausgangssignals des 
realen Systems nach Gleichung~(\myref{3.12}) und der Summe der Stichproben der 
Spektren der Ausgangssignale der beiden Modellsysteme und des Spektrums der 
gefensterten deterministischen St"orung ergibt. Wir suchen daher eine Ausgleichsl"osung
f"ur das gegen"uber Gleichung~(\myref{3.3}) modifizierte Gleichungssystem
\begin{flalign}
\frac{1}{M}\cdoT\Sum{\Tilde{\mu}=0}{M-1}\;\Sum{\Bar{\mu}=0}{K_H-1}
&\bigg(\Hat{H}\big({\T\Tilde{\mu},\Tilde{\mu}\!+\!\Bar{\mu}\CdoT\frac{M}{K_H}}\big)
\CdoT V_{\lambda}\big({\T\Tilde{\mu}\!+\!\Bar{\mu}\CdoT\frac{M}{K_H}}\big)+
\Hat{H}_*\big({\T\Tilde{\mu},\Tilde{\mu}\!+\!\Bar{\mu}\CdoT\frac{M}{K_H}}\big)
\CdoT V_{\lambda}\big({\T-\Tilde{\mu}\!-\!\Bar{\mu}\CdoT\frac{M}{K_H}}\big)^{\!\Kk}\bigg)\cdot\hspace*{-\textwidth}{}&&
\notag\\*
&&&{}\cdot F\big({\T(\mu\!-\!\Tilde{\mu})\CdoT\frac{2\pi}{M}}\big)\,+\,
\Hat{U}_{\!f}(\mu)\;=\;Y_{\!f,\lambda}(\mu)\notag\\*[10pt]
&\forall\qquad\mu=0\;(1)\;M\!-\!1\quad\text{ und }
\quad\lambda=1\;(1)\;L.\hspace*{-\textwidth}{}&&
\label{E.3.1}
\end{flalign}
Wenn man eine Fensterfunktion verwendet, die der Bedingung~(\myref{2.27})
gen"ugt, bleibt von der Summe "uber $\Tilde{\mu}$ nur der Summand 
mit \mbox{$\Tilde{\mu}\!=\!\mu$} "ubrig, und es entstehen in Matrixschreibweise 
die $M$ Gleichungssysteme
\begin{equation}
\Hat{\Vec{H}}(\mu)\cdot\Tilde{\underline{V}}(\mu)\;+\;
\Hat{U}_{\!f}(\mu)\cdot\Vec{1} \;=\; \Vec{Y}_{\!f}(\mu)
\qquad\qquad\forall\qquad \mu=0\;(1)\;M\!-\!1,
\label{E.3.2}
\end{equation}
bei denen auf beiden Seiten jeweils ein \mbox{$1\times L$} Zeilenvektor
steht. Diese $M$ Gleichungssysteme haben paarweise verschiedene S"atze von 
Unbekannten. Es tritt jeweils ein Spektralwert \mbox{$\Hat{U}_{\!f}(\mu)$} der 
gefensterten deterministischen St"orung, sowie ein Satz von Sch"atzwerten f"ur die 
beiden bifrequenten "Ubertragungsfunktionen, die zu dem \mbox{$1\times2\CdoT K_H$}
Zeilenvektor 
\begin{equation}
\Hat{\Vec{H}}(\mu)\;=\;
\begin{bmatrix}
\Hat{H}(\mu,\mu)\\[2pt]
\Hat{H}\big({\T\mu,\mu\!+\!\frac{M}{K_H}}\big)\\[-2pt]
\vdots\\[-2pt]
\Hat{H}\big({\T\mu,\mu\!+\!(K_H\!-\!1)\CdoT\frac{M}{K_H}}\big)\\[2pt]
\Hat{H}_*(\mu,\mu)\\[2pt]
\Hat{H}_*\big({\T\mu,\mu\!+\!\frac{M}{K_H}}\big)\\[-2pt]
\vdots\\[-2pt]
\Hat{H}_*\big({\T\mu,\mu\!+\!(K_H\!-\!1)\CdoT\frac{M}{K_H}}\big)
\end{bmatrix}^{\Tt}
\qquad\qquad\forall\qquad\mu=0\;(1)\;M\!-\!1
\label{E.3.3}
\end{equation}
zusammengefasst sind, auf. Der Spektralwert \mbox{$\Hat{U}_{\!f}(\mu)$} tritt 
in allen $L$ Gleichungen (\ref{E.3.1}) f"ur eine diskrete Frequenz $\mu$ mit 
dem Koeffizienten Eins auf. Daher tritt in Gleichung (\ref{E.3.2}) der 
Zeilenvektor \mbox{$\Vec{1}$} auf, bei dem alle $L$ Elemente eins sind. 
Die gesuchten Sch"atzwerte f"ur die beiden bifrequenten 
"Ubertragungsfunktionen werden in den $L$ Gleichungen (\ref{E.3.1}) f"ur eine 
diskrete Frequenz $\mu$ mit den unterschiedlichen Spektralwerten der Musterfolgen 
der Erregung gewichtet. In der \mbox{$2\CdoT K_H\times L$} Matrix 
\mbox{$\Tilde{\underline{V}}(\mu)$} sind diese zusammengefasst:
\begin{equation}
\Tilde{\underline{V}}(\mu)\;=\;
\begin{bmatrix}
\Vec{V}\!(\mu)\\[2pt]
\Vec{V}\!\big({\T\mu\!+\!\frac{M}{K_H}}\big)\\[-2pt]
\vdots\\[-2pt]
\Vec{V}\!\big({\T\mu\!+\!(K_H\!-\!1)\CdoT\frac{M}{K_H}}\big)\\[2pt]
\Vec{V}\!(\!-\mu)^{\Kk}\\[2pt]
\Vec{V}\!\big(\!{\T-\mu\!-\!\frac{M}{K_H}}\big)^{\!\Kk}\\[-2pt]
\vdots\\[-2pt]
\Vec{V}\!\big(\!{\T-\mu\!-\!(K_H\!-\!1)\CdoT\frac{M}{K_H}}\big)^{\!\Kk}
\end{bmatrix}
\qquad\qquad\forall\qquad\mu=0\;(1)\;M\!-\!1.
\label{E.3.4}
\end{equation}
Dabei sind die Zeilen dieser Matrix die Stichprobenvektoren der Spektralwerte 
der Erregung, die nach Gleichung (\myref{3.13}) definiert sind. Die Matrix selbst 
ist somit eine konkrete Stichprobe vom Umfang $L$ des nach Gleichung (\ref{E.2.15}) 
definierten Zufallsvektors \mbox{$\Tilde{\Vec{\boldsymbol{V}}}(\mu)$}. 

Die Berechnung der Ausgleichsl"osung jedes dieser Gleichungssysteme
(\ref{E.3.2}) f"uhren wir in zwei Schritten durch. Zun"achst wird jedes
Gleichungssystem jeweils nach dem Anteil aufgel"ost, der den gesuchten
Messwert des Spektrums der deterministischen St"orung enth"alt, indem man 
\mbox{$\Hat{\Vec{H}}(\mu)\cdot\Tilde{\underline{V}}(\mu)$} auf beiden Seiten
subtrahiert. Die Ausgleichsl"osung f"ur die Messwerte \mbox{$\Hat{U}_{\!f}(\mu)$}
erhalten wir, indem wir zun"achst jedes Gleichungssystem mit dem konjugiert
transponierten des Vektors von rechts multiplizieren, der mit der gesuchten
Gr"o"se \mbox{$\Hat{U}_{\!f}(\mu)$} von rechts multipliziert wird, und
anschlie"send beide Seiten durch das Quadrat der euklidischen Norm
dieses Vektors dividieren. In unserem Fall handelt es sich dabei um
den Einservektor \mbox{$\Vec{1}$}, dessen euklidische Norm $\sqrt{\!L\,}$ ist.
\begin{equation}
\Hat{U}_{\!f}(\mu)\;=\;\frac{1}{L}\cdot
\Big(\,\Vec{Y}_{\!f}(\mu)-
\Hat{\Vec{H}}(\mu)\cdot\Tilde{\underline{V}}(\mu)\,\Big)\cdot\Vec{1}^{\,\Hh}
\qquad\qquad\forall\qquad\mu=0\;(1)\;M\!-\!1
\label{E.3.5}
\end{equation}
Dabei ist
\begin{equation}
\frac{1}{L}\cdot\Vec{Y}_{\!f}(\mu)\cdot\Vec{1}^{\,\Hh}\;=\;
\frac{1}{L}\cdoT\Sum{\lambda=1}{L}Y_{\!f,\lambda}(\mu)
\qquad\qquad\forall\qquad\mu=0\;(1)\;M\!-\!1\quad{}
\label{E.3.6}
\end{equation}
der empirische Mittelwert der Zufallsgr"o"se \mbox{$\boldsymbol{Y}_{\!\!\!f}(\mu)$}. 
Man kann diesen berechnen, indem man zu einem Akkumulator, den man zu null initialisiert, 
nach und nach bei jeder der $L$ Einzelmessungen den Wert \mbox{$Y_{\!f,\lambda}(\mu)$} 
addiert, den man durch Fensterung und anschlie"sende DFT des bei der Einzelmessung
$\lambda$ am Systemausgang gemessenen Signals \mbox{$y_{\lambda}(k)$} berechnet. 
Analog --- allerdings ohne Fensterung --- l"asst sich auch der empirische Mittelwert
\begin{equation}
\frac{1}{L}\cdot\Tilde{\underline{V}}(\mu)\cdot\Vec{1}^{\,\Hh}\;=\;
\frac{1}{L}\cdoT\Sum{\lambda=1}{L}\Tilde{\Vec{V}}_{\!\!\lambda}(\mu)
\qquad\qquad\forall\qquad\mu=0\;(1)\;M\!-\!1\quad{}
\label{E.3.7}
\end{equation}
des Zufallsvektors \mbox{$\Tilde{\Vec{\boldsymbol{V}}}(\mu)$} durch Akkumulation der 
bei den Einzelmessungen verwendeten Eingangsspektren berechnen. Setzt man nun Gleichung 
(\ref{E.3.5}) in die Gleichungssysteme (\ref{E.3.2}) ein, so erh"alt man die Messwerte 
f"ur die beiden "Ubertragungsfunktionen:
\begin{gather}
\Hat{\Vec{H}}(\mu)\;=\;
\Vec{Y}_{\!f}(\mu)\cdot
\Big(\,\underline{E}-\frac{1}{L}\cdot\Vec{1}^{\,\Hh}\Cdot\Vec{1}\:\Big)\cdot
\Tilde{\underline{V}}(\mu)^{\Hh}\cdot
\bigg(\,\Tilde{\underline{V}}(\mu)\cdot
      \Big(\,\underline{E}-\frac{1}{L}\cdot\Vec{1}^{\,\Hh}\Cdot\Vec{1}\:\Big)
      \cdot\Tilde{\underline{V}}(\mu)^{\Hh}\,\bigg)^{\!\!-1}=
\notag\\*[6pt]
=\;\frac{1}{L\!-\!1}\cdot
\Vec{Y}_{\!f}(\mu)\cdot\underline{1}_{\bot}\Cdot
\Tilde{\underline{V}}(\mu)^{\Hh}\Cdot
\Hat{\underline{C}}_{\Tilde{\Vec{\boldsymbol{V}}}(\mu),\Tilde{\Vec{\boldsymbol{V}}}(\mu)}^{\uP{0.4}{\!-1}}\;=\;
\Hat{\Vec{C}}_{\boldsymbol{Y}_{\!\!\!f}(\mu),\Tilde{\Vec{\boldsymbol{V}}}(\mu)}\Cdot
\Hat{\underline{C}}_{\Tilde{\Vec{\boldsymbol{V}}}(\mu),\Tilde{\Vec{\boldsymbol{V}}}(\mu)}^{\uP{0.4}{\!-1}}
\notag\\*[6pt]
\qquad\forall\qquad\mu=0\;(1)\;M\!-\!1.
\label{E.3.8}
\end{gather}
Die $L$ Zeilenvektoren der Matrix\vspace{-8pt}
\begin{equation}
\qquad\qquad\underline{1}_{\bot}\;=\;
\underline{E}-\frac{1}{L}\cdot\Vec{1}^{\,\Hh}\Cdot\Vec{1}
\label{E.3.9}
\end{equation}
spannen den \mbox{$L\!-\!1$}-dimensionalen Nullraum des Einservektors
\mbox{$\Vec{1}$} in der Art auf, dass das Produkt eines beliebigen
Vektors $\Vec{x}$ mit dieser Matrix den auf den Nullraum projizierten Vektor
liefert. Der Anteil des Vektors $\Vec{x}$ in Richtung des Einservektors
\mbox{$=\Vec{x}\cdot\Vec{1}^{\,\Hh}\Cdot\Vec{1}\,/\,L$}
wird n"amlich von dem Vektor selbst abgezogen. Da das Produkt
\mbox{$\Vec{1}\cdot\underline{1}_{\bot}$} den Nullvektor ergibt, ist der
Einservektor ein Eigenvektor des Eigenwertes Null. Da jede unit"are
Basis --- \mbox{$L\!-\!1$} orthonormale Vektoren der Dimension
\mbox{$1\!\times\!L$} --- des Nullraums des Einservektors 
auf sich selbst abgebildet wird, sind die Basisvektoren
Eigenvektoren zum \mbox{$L\!-\!1$}-fachen Eigenwert Eins. Die Spur dieser
Matrix, die gleich der Summe ihrer Eigenwerte ist, ist daher gleich
\mbox{$L\!-\!1$}. Diese Matrix \mbox{$\underline{1}_{\bot}$} ist abgesehen
davon, dass sie hermitesch ist, auch noch idempotent. Sie kann also
beliebig oft mit sich selbst multipliziert werden ohne sich dadurch zu
"andern.

Das Produkt\vspace{-6pt}
\begin{gather}
\Hat{\underline{C}}_{\Tilde{\Vec{\boldsymbol{V}}}(\mu_1),\Tilde{\Vec{\boldsymbol{V}}}(\mu_2)}\;=\;
\frac{1}{L\!-\!1}\cdot
\Tilde{\underline{V}}(\mu_1)\cdot
\underline{1}_{\bot}\Cdot
\Tilde{\underline{V}}(\mu_2)^{\Hh}\;=
\label{E.3.10}\\*[4pt]
=\;\frac{1}{L\!-\!1}\cdot
\Tilde{\underline{V}}(\mu_1)\cdot
\Big(\,\underline{E}-\frac{1}{L}\cdot\Vec{1}^{\,\Hh}\Cdot\Vec{1}\:\Big)\cdot
\Tilde{\underline{V}}(\mu_2)^{\Hh}\;=
\notag\\[4pt]
=\;\frac{1}{L\!-\!1}\cdot\Big(\,
\Tilde{\underline{V}}(\mu_1)\cdot\Tilde{\underline{V}}(\mu_2)^{\Hh}-
\frac{1}{L}\cdot
\Tilde{\underline{V}}(\mu_1)\cdot\Vec{1}^{\,\Hh}\cdot
\Vec{1}\cdot\Tilde{\underline{V}}(\mu_2)^{\Hh}\,\Big)\;=
\notag\\[6pt]
=\;\frac{1}{L\!-\!1}\cdot\bigg(\,\Sum{\lambda=1}{L}\,
\Tilde{\Vec{V}}_{\!\!\lambda}(\mu_1)\cdot\Tilde{\Vec{V}}_{\!\!\lambda}(\mu_2)^{\Hh}-\,
\frac{1}{L}\cdot
\Sum{\lambda=1}{L}\Tilde{\Vec{V}}_{\!\!\lambda}(\mu_1)\cdot
\Sum{\lambda=1}{L}\Tilde{\Vec{V}}_{\!\!\lambda}(\mu_2)^{\Hh}\bigg)
\notag\\*[4pt]
\forall\qquad\mu_1=0\;(1)\;M\!-\!1\quad\text{und}
\quad\mu_2=0\;(1)\;M\!-\!1,\notag
\end{gather}
das mit \mbox{$\mu_1=\mu_2=\mu$} in Gleichung (\ref{E.3.8}) auftritt,
ist die empirische Kovarianzmatrix der beiden Zufallsvektoren
\mbox{$\Tilde{\Vec{\boldsymbol{V}}}(\mu_1)$} und \mbox{$\Tilde{\Vec{\boldsymbol{V}}}(\mu_2)$}.
Man beachte, dass bei der Berechnung der empirischen Kovarianzmatrix
das Konjugieren und Transponieren des zweiten beteiligten Stichprobenzeilenvektors
in dieser Definition bereits mit eingeschlossen ist.
Dies erfolgt konform mit der Definition der Kovarianzmatrix zweier
komplexer Zufallsvektoren, bei der ebenfalls der zweite der
daran beteiligten Zufallsvektoren konjugiert wird (\,siehe Liste der 
Formelzeichen in \cite{Diss}\,). Wie die letzte Umformung zeigt, l"asst sich 
auch die empirische Kovarianzmatrix berechnen, ohne dass dazu die Spektralwerte 
aller Einzelmessungen abgespeichert werden m"ussen. Um die erste Summe zu 
berechnen, addiert man zu einem Akkumulatorfeld, das man zu null initialisiert, 
nach und nach bei jeder der $L$ Einzelmessungen das dyadische Vektorprodukt 
\mbox{$\Tilde{\Vec{V}}_{\!\!\lambda}(\mu_1)\cdot\Tilde{\Vec{V}}_{\!\!\lambda}(\mu_2)^{\Hh}$}.
Die beiden anderen Summen wurden bereits bei der Berechnung der empirischen 
Mittelwerte mit Gleichung (\ref{E.3.7}) analog berechnet.
Nach  Gleichung (\ref{E.3.8}) ben"otigt man zur Berechnung
der Messwerte der beiden "Ubertragungsfunktionen auch noch die Kreuzkovarianzvektoren
\begin{gather}
\Hat{\Vec{C}}_{\boldsymbol{Y}_{\!\!\!f}(\mu_1),\Tilde{\Vec{\boldsymbol{V}}}(\mu_2)}\;=\;
\frac{1}{L\!-\!1}\cdot
\Vec{Y}_{\!f}(\mu_1)\cdot\underline{1}_{\bot}\Cdot
\Tilde{\underline{V}}(\mu_2)^{\Hh}\;=
\label{E.3.11}\\[4pt]
=\;\frac{1}{L\!-\!1}\cdot
\Vec{Y}_{\!f}(\mu_1)\cdot
\Big(\,\underline{E}-\frac{1}{L}\cdot\Vec{1}^{\,\Hh}\Cdot\Vec{1}\:\Big)\cdot
\Tilde{\underline{V}}(\mu_2)^{\Hh}\;=
\notag\\[6pt]
=\;\frac{1}{L-\!1}\cdot\Big(\,
\Vec{Y}_{\!f}(\mu_1)\cdot\Tilde{\underline{V}}(\mu_2)^{\Hh}-
\frac{1}{L}\cdot
\Vec{Y}_{\!f}(\mu_1)\cdot\Vec{1}^{\,\Hh}\cdot
\Vec{1}\cdot\Tilde{\underline{V}}(\mu_2)^{\Hh}\,\Big)\;=
\notag\\[6pt]
=\;\frac{1}{L\!-\!1}\cdot\bigg(\,\Sum{\lambda=1}{L}\,
Y_{\!f,\lambda}(\mu_1)\cdot\Tilde{\Vec{V}}_{\!\!\lambda}(\mu_2)^{\Hh}-\,
\frac{1}{L}\cdot
\Sum{\lambda=1}{L}Y_{\!f,\lambda}(\mu_1)\cdot
\Sum{\lambda=1}{L}\Tilde{\Vec{V}}_{\!\!\lambda}(\mu_2)^{\Hh}\bigg)
\notag\\*[4pt]
\forall\qquad\mu_1=0\;(1)\;M\!-\!1\quad\text{und}
\quad\mu_2=0\;(1)\;M\!-\!1,\notag
\end{gather}
der Spektralwerte der Signale am Ein- und Ausgang des Systems
f"ur \mbox{$\mu_1=\mu_2=\mu$}, die sich ebenfalls durch Akkumulation
aus den bei den Einzelmessungen $\lambda$ gemessenen Signalen
berechnen lassen.

Da sich die L"osungen (\ref{E.3.8}) f"ur die beiden bifrequenten "Ubertragungsfunktionen 
nur dann numerisch gut berechnen lassen, wenn die empirischen Kovarianzmatrizen 
\mbox{$\Hat{\underline{C}}_{\Tilde{\Vec{\boldsymbol{V}}}(\mu),\Tilde{\Vec{\boldsymbol{V}}}(\mu)}$}
gut konditioniert sind, muss man zur Messung der beiden "Ubertragungsfunktionen einen 
Zufallsvektor $\Vec{\boldsymbol{V}}$ verwenden, bei dem die nach Gleichung~(\ref{E.2.23}) 
definierten theoretischen Kovarianzmatrizen gut konditioniert sind, so dass auch die 
empirischen Kovarianzmatrizen nach Gleichung~(\ref{E.3.10}) mit hoher Wahrscheinlichkeit 
gut konditioniert sind (\,siehe Anhang~\ref{E.Kap.A.2}\,).
Auch wenn der Fall einer singul"aren empirischen Kovarianzmatrix dann
so extrem unwahrscheinlich\footnote{F"ur den Falle einer \mbox{$1\times1$} Kovarianzmatrix 
wird dies im Anhang \ref{E.Kap.A.3} diskutiert} wird, dass er f"ur die praktische Anwendung
keinerlei Bedeutung hat, ist es f"ur die theoretische Berechnung der
Erwartungswerte und der Varianzen der Messwerte notwendig, die F"alle, bei
denen die Stichprobe \mbox{$\Tilde{\underline{V}}(\mu)$} zu einer singul"aren
empirischen Kovarianzmatrix f"uhrt, in einer Weise zu handhaben, dass der
Erwartungswert und die Varianz des Messwertes existiert. Dazu transformieren
wir die empirische Kovarianzmatrix mit Hilfe der unit"aren Transformationsmatrix 
\mbox{$\underline{U}$} kongruent auf ihre Diagonalform: 
\begin{equation}
\underline{U}\cdot
\Hat{\underline{C}}_{\Tilde{\Vec{\boldsymbol{V}}}(\mu),\Tilde{\Vec{\boldsymbol{V}}}(\mu)}\cdot
\underline{U}^{\Hh}\;=\;
\frac{1}{L\!-\!1}\cdot\Big(\,\underline{U}\cdot\Tilde{\underline{V}}(\mu)\cdot
\underline{1}_{\bot}\Big)\cdot\Big(\,\underline{1}_{\bot}\Cdot
\Tilde{\underline{V}}(\mu)^{\Hh}\Cdot\underline{U}^{\Hh}\,\Big)\;=\;
\underline{D}.
\label{E.3.12}
\end{equation}
Die Diagonalmatrix \mbox{$\underline{D}$} enth"alt nur reelle, nichtnegative
Diagonalelemente, wovon im singul"aren Fall ein oder mehrere null sind.
Alle Diagonalelemente, die unterhalb einer beliebig kleinen positiven
Schranke liegen, werden auf diese begrenzt. Um wirklich nur die singul"aren 
F"alle zu modifizieren, sollte die Schranke wenigstens so klein sein, dass 
wirklich nur diejenigen Diagonalelemente  begrenzt werden, die exakt null sind. 
Wir erhalten so die Diagonalmatrix \mbox{$\underline{D}_{\text{Limit}}$}. 
Diese l"asst sich dann invertieren. Mit der oben verwendeten unit"aren 
kongruenten Transformation erhalten wir daraus eine Matrix
\mbox{$\underline{U}^{\Hh}\Cdot\underline{D}_{\text{Limit}}^{-1}
\Cdot\underline{U}$}, die auch im singul"aren Fall existiert,
und die im nichtsingul"aren Fall gleich der inversen der empirischen
Kovarianzmatrix ist. Mit dieser Matrix berechnen wir die Messwerte
\mbox{$\Hat{\Vec{H}}(\mu)$} nach Gleichung~(\ref{E.3.8}).
\begin{equation}
\Hat{\Vec{H}}(\mu)\;=\;
\frac{1}{L\!-\!1}\cdot
\Vec{Y}_{\!f}(\mu)\cdot\Big(\,\underline{1}_{\bot}\Cdot
\Tilde{\underline{V}}(\mu)^{\Hh}\Cdot\underline{U}^{\Hh}\,\Big)\cdot
\underline{D}_{\text{Limit}}^{-1}\Cdot\underline{U}
\label{E.3.13}
\end{equation}
Im singul"aren Fall weist die Matrix \mbox{$\underline{1}_{\bot}\Cdot
\Tilde{\underline{V}}(\mu)^{\Hh}\Cdot\underline{U}^{\Hh}$} in den 
Spalten Nullvektoren auf, die den Diagonalelementen Null entsprechen. 

Diese Nullvektoren werden durch die Multiplikation mit den inversen
Diagonalelementen der limitierten Diagonalmatrix multipliziert, und bleiben
daher Nullvektoren. Der Fehler, der durch die Limitierung der Singul"arwerte
der empirischen Kovarianzmatrix in den Messwerten der beiden
"Ubertragungsfunktionen entsteht, ist begrenzt, da die Elemente des
Stichprobenvektors \mbox{$\Vec{Y}_{\!f}(\mu)$} begrenzt sind.
Dieser Fehler wird bei der Berechnung des Erwartungswertes und der Varianz
des Messwertes mit der Auftrittswahrscheinlichkeit des singul"aren Falls,
der zu diesem fehlerhaften Messwert gef"uhrt hat, multipliziert.
Wenn die theoretische Kovarianzmatrix gut konditioniert ist, werden die 
Auftrittswahrscheinlichkeiten aller singul"aren F"alle so klein,
dass deren Beitrag zum Erwartungswert und zur Varianz so gering wird,
dass er gegen"uber der Berechnung unter Ausschluss aller singul"aren
F"alle vernachl"assigt werden kann.

Indem man die Messwerte \mbox{$\Hat{\Vec{H}}(\mu)$}, die man mit 
Gleichung~(\ref{E.3.8}) berechnet, in Gleichung~(\ref{E.3.5}) einsetzt, 
erh"alt man die Messwerte \mbox{$\Hat{U}_{\!f}(\mu)$} f"ur das Spektrum der 
gefensterten deterministischen St"orung \mbox{$u(k)$}. Dabei wurde der Anteil 
des mittleren Spektrums \mbox{$\Tilde{\underline{V}}(\mu)\cdot\Vec{1}^{\,\Hh}/L$} 
der Erregung mit den Messwerten der "Ubertragungsfunktionen multipliziert, und 
dieses Produkt vom mittleren Spektrum \mbox{$\Vec{Y}_{\!f}(\mu)\cdot\Vec{1}^{\,\Hh}/L$} 
des gefensterten Ausgangssignals abgezogen. Analog dazu erh"alt man die Messwerte 
\mbox{$\Hat{u}(k)$} f"ur die deterministische St"orung \mbox{$u(k)$} dadurch,
dass man von dem mittleren Ausgangssignal \mbox{$\Vec{y}(k)\cdot\Vec{1}^{\,\Hh}/L$} 
die mit den gesch"atzten Modellsystemen linear verzerrte mittlere Erregung subtrahiert.
\begin{gather}
\Hat{u}(k)=\frac{1}{L}\cdoT\!\Sum{\lambda=1}{L}
y_{\lambda}(k)\!-\!x_{\lambda}(k)\!-\!x_{*,\lambda}(k)\,=\,
\frac{1}{L}\CdoT\Vec{y}(k)\cdoT\Vec{1}^{\,\Hh}\!-
\frac{1}{M\CdoT L}\cdoT\!\Sum{\mu=0}{M-1}\Hat{\Vec{H}}(\mu)\CdoT
\Tilde{\underline{V}}(\mu)\CdoT\Vec{1}^{\,\Hh}\!\Cdot
e^{j\cdot\frac{2\pi}{M}\cdot\mu\cdot k}
\notag\\*[6pt]
\forall\qquad k=0\;(1)\;F\!-\!1.
\label{E.3.14}
\end{gather}
Die $F$ dabei auftretenden \mbox{$1\!\times\!L$} Zeilenvektoren
\begin{equation}
\Vec{y}(k)\;=\;\big[\,y_{1}(k),\,\ldots\,,
y_{\lambda}(k),\,\ldots\,,y_{L}(k)\,\big]
\qquad\qquad\forall\qquad k=0\;(1)\;F\!-\!1\quad{}
\label{E.3.15}
\end{equation}
setzten sich jeweils aus den $L$ Elementen \mbox{$y_{\lambda}(k)$} zusammen.
Jeder dieser Vektoren ist also eine Stichprobe vom Umfang $L$ der 
Zufallsgr"o"se \mbox{$\boldsymbol{y}_(k)$} f"ur einen Zeitpunkt $k$. 
Die empirischen Mittelwerte \mbox{$\Vec{y}(k)\cdot\Vec{1}^{\,\Hh}/L$} 
lassen sich f"ur alle Zeitpunkte $k$ auch hier wieder berechnen, 
indem man zu einem Akkumulatorvektor, den man zu null initialisiert, 
nach und nach bei jeder der $L$ Einzelmessungen die Werte \mbox{$y_{\lambda}(k)$} 
des bei der Einzelmessung am Systemausgang gemessenen Signals f"ur 
alle Zeitpunkte $k$ addiert.

\newsavebox{\TitleTmp}
\sbox{\TitleTmp}{\small\sf\thesection.\ Erwartungstreue der Messwerte \mbox{$H(\mu_1,\mu_2)$}, 
\mbox{$H_*(\mu_1,\mu_2)$}, \mbox{$\Hat{U}_{\!f}(\mu)$} und \mbox{$u(k)$}}
\markright{\usebox{\TitleTmp}}
\section{Erwartungstreue der Messwerte der "Ubertragungsfunktionen 
und der deterministischen St"orung}\label{E.Kap.3.2}
\markright{\usebox{\TitleTmp}}

Die in \cite{Diss} angestellte Vor"uberlegung zur Berechnung der 
Erwartungswerte der Messwerte der "Ubertragungsfunktion ist hier nun zu 
modifizieren, um das erweiterte Systemmodell mit den beiden periodisch 
zeitvarianten Modellsystemen und der deterministischen St"orung behandeln zu k"onnen. 
Nun ist es Gleichung (\ref{E.2.26}) die zeigt, wie sich 
die $M$ Zufallsgr"o"sen  \mbox{$\boldsymbol{N}_{\!\!f}(\mu)$} des Spektrums 
des gefensterten Approximationsfehlers aus den $M$ Zufallsgr"o"sen 
\mbox{$\boldsymbol{V}(\mu)$} am Eingang und den $M$ Zufallsgr"o"sen 
\mbox{$\boldsymbol{Y}_{\!\!\!f}(\mu)$} am Ausgang des realen Systems 
ergeben. Im Gegensatz zu \cite{Diss} treten nun bei jeder diskreten Frequenz 
$\mu$ mehrere zuf"allige Spektralwerte der Erregung auf, die jeweils in dem 
Vektor \mbox{$\Tilde{\Vec{\boldsymbol{V}}}(\mu)$} nach Gleichung (\ref{E.2.15})
zusammengefasst sind. Die weiterhin nach Gleichung (\myref{3.17}) definierten 
$M$ konkreten Stichprobenvektoren \mbox{$\Vec{N}_{\!f}(\mu)$} vom Umfang $L$ 
der $M$ Zufallsgr"o"sen \mbox{$\boldsymbol{N}_{\!\!f}(\mu)$} erhalten wir, 
wenn in Gleichung (\ref{E.2.26}) statt der Zufallsvektoren \mbox{$\Tilde{\Vec{\boldsymbol{V}}}(\mu)$} 
und  der Zufallsgr"o"sen \mbox{$\boldsymbol{Y}_{\!\!\!f}(\mu)$} deren konkrete Realisierungen 
\mbox{$\Tilde{\underline{V}}(\mu)$} und \mbox{$\Vec{Y}_{\!f}(\mu)$} einsetzen. 
\begin{equation}
\Vec{N}_{\!f}(\mu)\;=\;\Vec{Y}_{\!f}(\mu)-
\Vec{H}(\mu)\CdoT\Tilde{\underline{V}}(\mu)-
U_{\!f}(\mu)\CdoT\Vec{1}
\qquad\qquad\forall\qquad \mu=0\;(1)\;M\!-\!1\quad{}
\label{E.3.16}
\end{equation}
Hier treten mit \mbox{$U_{\!f}(\mu)$} und mit den Elementen der Vektoren 
\mbox{$\Vec{H}(\mu)$} die unbekannten Optimall"osungen der theoretischen 
Regression auf, die durch die Messung abgesch"atzt werden sollen. 
Die letzte Gleichung l"asst sich nach \mbox{$\Vec{Y}_{\!f}(\mu)$} 
auf"|l"osen und in die Ausgleichsl"osung (\ref{E.3.8}) einsetzen und 
wir erhalten die Messwerte der "Ubertragungsfunktionen als 
Funktion der Stichprobe des Spektrums der Erregung, der 
Stichprobe des Approximationsfehlerspektrums und der theoretisch 
optimalen Regressionskoeffizienten:
\begin{gather}
\Hat{\Vec{H}}(\mu)\;=\;
\frac{1}{L\!-\!1}\cdot
\Big(\Vec{H}(\mu)\CdoT\Tilde{\underline{V}}(\mu)+U_{\!f}(\mu)\CdoT\Vec{1}+\Vec{N}_{\!f}(\mu)\Big)\cdot
\underline{1}_{\bot}\Cdot\Tilde{\underline{V}}(\mu)^{\Hh}\Cdot
\Hat{\underline{C}}_{\Tilde{\Vec{\boldsymbol{V}}}(\mu),\Tilde{\Vec{\boldsymbol{V}}}(\mu)}^{\uP{0.4}{\!-1}}\;={}
\label{E.3.17}\\*[10pt]
{}=\;\Vec{H}(\mu)\cdot\underbrace{
\frac{1}{L\!-\!1}\cdot\Tilde{\underline{V}}(\mu)\cdot\underline{1}_{\bot}\Cdot\Tilde{\underline{V}}(\mu)^{\Hh}\Cdot
\Hat{\underline{C}}_{\Tilde{\Vec{\boldsymbol{V}}}(\mu),\Tilde{\Vec{\boldsymbol{V}}}(\mu)}^{\uP{0.4}{\!-1}}
}_{\scriptstyle=\underline{E}}\;+{}
\notag\\*[3pt]
{}+\;\frac{1}{L\!-\!1}\cdot U_{\!f}(\mu)\cdot\underbrace{\Vec{1}\cdot\underline{1}_{\bot}
}_{\scriptstyle=\Vec{0}}\Cdot\,\Tilde{\underline{V}}(\mu)^{\Hh}\Cdot
\Hat{\underline{C}}_{\Tilde{\Vec{\boldsymbol{V}}}(\mu),\Tilde{\Vec{\boldsymbol{V}}}(\mu)}^{\uP{0.4}{\!-1}}\;+{}
\notag\\*[3pt]
{}+\;\frac{1}{L\!-\!1}\cdot\Vec{N}_{\!f}(\mu)\cdot\underline{1}_{\bot}\Cdot\Tilde{\underline{V}}(\mu)^{\Hh}\Cdot
\Hat{\underline{C}}_{\Tilde{\Vec{\boldsymbol{V}}}(\mu),\Tilde{\Vec{\boldsymbol{V}}}(\mu)}^{\uP{0.4}{\!-1}}\;={}
\notag\\*[18pt]
{}=\;\Vec{H}(\mu)\;+\;\frac{1}{L\!-\!1}\cdot\Vec{N}_{\!f}(\mu)\cdot\underline{1}_{\bot}\Cdot\Tilde{\underline{V}}(\mu)^{\Hh}\Cdot
\Hat{\underline{C}}_{\Tilde{\Vec{\boldsymbol{V}}}(\mu),\Tilde{\Vec{\boldsymbol{V}}}(\mu)}^{\uP{0.4}{\!-1}}
\qquad\qquad\forall\qquad \mu=0\;(1)\;M\!-\!1.
\notag
\end{gather}

Nun wollen wir die Erwartungswerte der Messwerte der beiden 
"Ubertragungsfunktionen bestimmen. Dazu betrachten wir diese 
Messwerte jeweils als eine konkrete Realisierung --- also eine 
Stichprobe vom Umfang Eins --- der Zufallsgr"o"sen 
\mbox{$\Hat{\Vec{\boldsymbol{H}}}(\mu)$}. Diese Zufallsgr"o"sen 
erh"alt man, wenn man statt der konkreten Stichprobenmatrizen 
\mbox{$\Tilde{\underline{V}}(\mu)$} und der konkreten Stichprobenvektoren 
\mbox{$\Vec{N}_{\!f}(\mu)$} die mathematischen Stichprobenmatrizen 
\mbox{$\Tilde{\underline{\boldsymbol{V}}}(\mu)$} und die mathematischen 
Stichprobenvektoren \mbox{$\Vec{\boldsymbol{N}}_{\!\!f}(\mu)$}, 
die aufgrund der zuf"alligen Stichprobenentnahme selbst Zufallsvektoren sind,
in die letzte Gleichung einsetzt. Wir bilden also den Erwartungswert
"uber alle m"oglichen Messungen, die sich jeweils aus $L$
Einzelmessungen zusammensetzen.

Die Erwartungswerte der Messwerte der beiden "Ubertragungsfunktionen lassen 
sich nur berechnen, wenn man voraussetzt, dass einerseits alle Stichproben 
voneinander unabh"angig gewonnen wurden, und dass andererseits f"ur jede diskrete 
Frequenz $\mu$ der zuf"allige Spektralwert \mbox{$\boldsymbol{N}_{\!\!f}(\mu)$} 
von dem Zufallsvektor \mbox{$\Tilde{\Vec{\boldsymbol{V}}}(\mu)$} bei derselben 
Frequenz unabh"angig ist. Die Unabh"angigkeit der Zufallsgr"o"sen, die in dem  
Zufallsvektor \mbox{$\Tilde{\Vec{\boldsymbol{V}}}(\mu)$} nach Gleichung (\ref{E.2.15}) 
jeweils zusammengefasst worden sind, muss jedoch {\em nicht}\/ gefordert werden. 
Wenn wir nun den Erwartungswert des Produkts einer beliebigen Funktion der 
Zufallsmatrix \mbox{$\Tilde{\underline{\boldsymbol{V}}}(\mu)$}  und einer anderen 
beliebigen Funktion des Zufallsvektors \mbox{$\Vec{\boldsymbol{N}}_{\!\!f}(\mu)$} 
bilden, berechnet sich dieser als das Produkt der beiden Erwartungswerte der 
einzelnen zuf"alligen Faktoren. Dies gilt auch f"ur nichtlineare Funktionen, 
wie z.~B. f"ur die Inverse der empirischen Kovarianzmatrix.

Mit Gleichung (\ref{E.3.17}) berechnen wir nun die gesuchten Erwartungswerte:
\begin{gather}
\text{E}\big\{\Hat{\Vec{\boldsymbol{H}}}(\mu)\big\}\;=\;
\Vec{H}(\mu)\,+\,
\frac{1}{L\!-\!1}\cdot
\text{E}\Big\{\Vec{\boldsymbol{N}}_{\!\!f}(\mu)\cdot
\underline{1}_{\bot}\Cdot\Tilde{\underline{\boldsymbol{V}}}(\mu)^{\Hh}\Cdot
\Hat{\underline{\boldsymbol{C}}}_{\Tilde{\Vec{\boldsymbol{V}}}(\mu),\Tilde{\Vec{\boldsymbol{V}}}(\mu)}^{\uP{0.4}{\!-1}}\Big\}\;={}
\label{E.3.18}\\*[8pt]
{}=\;\Vec{H}(\mu)\,+\,
\frac{1}{L\!-\!1}\cdot\underbrace{
\text{E}\big\{\boldsymbol{N}_{\!\!f}(\mu)\big\}}_{\scriptstyle=0}\cdot
\text{E}\Big\{\underbrace{\Vec{1}\cdot\underline{1}_{\bot}
}_{\scriptstyle=\Vec{0}}\Cdot\,\Tilde{\underline{\boldsymbol{V}}}(\mu)^{\Hh}\Cdot
\Hat{\underline{\boldsymbol{C}}}_{\Tilde{\Vec{\boldsymbol{V}}}(\mu),\Tilde{\Vec{\boldsymbol{V}}}(\mu)}^{\uP{0.4}{\!-1}}\Big\}\;=\;\Vec{H}(\mu)
\notag
\end{gather}

Bis hierher hatten wir bei der Berechnung des Erwartungswertes den Fall, dass die 
empirische Kovarianzmatrix singul"ar ist, nicht ber"ucksichtigt. Wenn wir diesen 
Fall in der Art behandeln, indem wir die Singul"arwerte der Kovarianzmatrix nach 
unten limitieren, wie dies im letzten Unterkapitel beschrieben ist, hat dies zwei 
Auswirkungen auf Erwartungswertberechnung. Zum einen existiert der in der letzten 
Gleichung angegebene von der Erregung abh"angige Erwartungswert nur dann, wenn man 
die Kovarianzmatrix entsprechend modifiziert, so dass sie sich invertieren l"asst. 
Zum anderen ergibt sich im singul"aren Fall der Messwertvektor \mbox{$\Hat{\Vec{H}}(\mu)=\Vec{0}$}, 
dessen Erwartungswert ebenfalls der Nullvektor ist. Damit berechnet sich der 
Erwartungswert zu\vspace{4pt}
\begin{gather}
\text{E}\big\{\Hat{\Vec{\boldsymbol{H}}}(\mu)\big\}\;=\;
\text{E}\Big\{\Hat{\Vec{\boldsymbol{H}}}(\mu)\,\pmb{\big|}\,
\det\big(\Hat{\underline{C}}_{\Tilde{\Vec{\boldsymbol{V}}}(\mu),\Tilde{\Vec{\boldsymbol{V}}}(\mu)}\big)\!=\!0\Big\}\cdot
P\Big(\det\big(\Hat{\underline{C}}_{\Tilde{\Vec{\boldsymbol{V}}}(\mu),\Tilde{\Vec{\boldsymbol{V}}}(\mu)}\big)\!=\!0\Big)\;+{}
\notag\\*[4pt]
{}+\;\text{E}\Big\{\Hat{\Vec{\boldsymbol{H}}}(\mu)\,\pmb{\big|}\,
\det\big(\Hat{\underline{C}}_{\Tilde{\Vec{\boldsymbol{V}}}(\mu),\Tilde{\Vec{\boldsymbol{V}}}(\mu)}\big)\!\neq\!0\Big\}\cdot
P\Big(\det\big(\Hat{\underline{C}}_{\Tilde{\Vec{\boldsymbol{V}}}(\mu),\Tilde{\Vec{\boldsymbol{V}}}(\mu)}\big)\!\neq\!0\Big)\;={}
\notag\\*[16pt]
{}=\;\Vec{0}\cdot
P\Big(\det\big(\Hat{\underline{C}}_{\Tilde{\Vec{\boldsymbol{V}}}(\mu),\Tilde{\Vec{\boldsymbol{V}}}(\mu)}\big)\!=\!0\Big)+
\Vec{H}(\mu)\cdot
P\Big(\det\big(\Hat{\underline{C}}_{\Tilde{\Vec{\boldsymbol{V}}}(\mu),\Tilde{\Vec{\boldsymbol{V}}}(\mu)}\big)\!\neq\!0\Big)\;={}
\notag\\*[12pt]
{}=\;\Vec{H}(\mu)\cdot
P\Big(\det\big(\Hat{\underline{C}}_{\Tilde{\Vec{\boldsymbol{V}}}(\mu),\Tilde{\Vec{\boldsymbol{V}}}(\mu)}\big)\!\neq\!0\Big)
\label{E.3.19}
\end{gather}
In der Praxis wird man zur Erregung des Systems immer Zufallsprozesse verwenden, 
bei denen die Wahrscheinlichkeit, eine singul"are empirische Kovarianzmatrix zu erhalten,
so extrem klein ist, dass die Wahrscheinlichkeit des gegenteiligen Falles als eins anzusehen 
ist, und somit sind die Messwerte der beiden bifrequenten "Ubertragungsfunktionen 
praktisch erwartungstreu.

Bevor wir mit der Berechnung der Erwartungswerte \mbox{$\Hat{\boldsymbol{U}}_{\!f}(\mu)$} 
der Messwerte des Spektrums der gefensterten deterministischen St"orung beginnen, 
wollen wir zun"achst diese nach Gleichung~(\ref{E.3.5}) definierten Messwerte 
anders darstellen, indem wir zuerst die Messwerte \mbox{$\Hat{\Vec{H}}(\mu)$} 
der beiden "Ubertragungsfunktionen nach Gleichung~(\ref{E.3.8}) einsetzen. Dann 
l"osen wir Gleichung~(\ref{E.3.16}) nach den $M$ konkreten Stichprobenvektoren 
\mbox{$\Vec{Y}_{\!f}(\mu)$} der Spektralwerte des gefensterten Signals am 
Systemausgang auf und setzen diese ebenfalls ein. Wir erhalten so:
\begin{gather}
\Hat{U}_{\!f}(\mu)\;=\;\frac{1}{L}\cdot
\Vec{Y}_{\!f}(\mu)\cdot
\Big(\,\underline{E}-
\frac{\underline{1}_{\bot}}{L\!-\!1}\cdot
\Tilde{\underline{V}}(\mu)^{\Hh}\Cdot
\Hat{\underline{C}}_{\Tilde{\Vec{\boldsymbol{V}}}(\mu),\Tilde{\Vec{\boldsymbol{V}}}(\mu)}^{\uP{0.4}{\!-1}}\Cdot
\Tilde{\underline{V}}(\mu)\,\Big)\cdot\Vec{1}^{\,\Hh}\;={}
\notag\\*[8pt]
{}=\,\frac{1}{L}\CdoT
\Big(\Vec{H}(\mu)\CdoT\Tilde{\underline{V}}(\mu)+U_{\!f}(\mu)\CdoT\Vec{1}+\Vec{N}_{\!f}(\mu)\Big)\CdoT
\Big(\underline{E}-
\frac{\underline{1}_{\bot}}{L\!-\!1}\CdoT
\Tilde{\underline{V}}(\mu)^{\Hh}\!\CdoT
\Hat{\underline{C}}_{\Tilde{\Vec{\boldsymbol{V}}}(\mu),\Tilde{\Vec{\boldsymbol{V}}}(\mu)}^{\uP{0.4}{\!-1}}\CdoT
\Tilde{\underline{V}}(\mu)\Big)\CdoT\Vec{1}^{\,\Hh}\,={}
\notag\\*[8pt]
{}=\;U_{\!f}(\mu)+\frac{1}{L}\cdot\Vec{N}_{\!f}(\mu)\cdot
\Big(\,\underline{E}-
\frac{\underline{1}_{\bot}}{L\!-\!1}\cdot
\Tilde{\underline{V}}(\mu)^{\Hh}\Cdot
\Hat{\underline{C}}_{\Tilde{\Vec{\boldsymbol{V}}}(\mu),\Tilde{\Vec{\boldsymbol{V}}}(\mu)}^{\uP{0.4}{\!-1}}\Cdot
\Tilde{\underline{V}}(\mu)\,\Big)\cdot\Vec{1}^{\,\Hh}
\notag\\*[8pt]
\forall\qquad\mu=0\;(1)\;M\!-\!1.
\label{E.3.20}
\end{gather}
Nun k"onnen wir die Erwartungswerte der Messwerte des Spektrums der deterministischen 
St"orung berechnen:
\begin{gather}
\text{E}\big\{\Hat{\boldsymbol{U}}_{\!\!f}(\mu)\big\}\;=\;
U_{\!f}(\mu)+\frac{1}{L}\cdot\text{E}\Big\{\Vec{\boldsymbol{N}}_{\!\!f}(\mu)\CdoT
\Big(\,\underline{E}-
\frac{\underline{1}_{\bot}}{L\!-\!1}\cdot
\Tilde{\underline{\boldsymbol{V}}}(\mu)^{\Hh}\Cdot
\Hat{\underline{\boldsymbol{C}}}_{\Tilde{\Vec{\boldsymbol{V}}}(\mu),\Tilde{\Vec{\boldsymbol{V}}}(\mu)}^{\uP{0.4}{\!-1}}\Cdot
\Tilde{\underline{\boldsymbol{V}}}(\mu)\Big)\CdoT\Vec{1}^{\,\Hh}\Big\}\;={}
\notag\\*[8pt]
{}=\;U_{\!f}(\mu)+\frac{1}{L}\cdot\underbrace{
\text{E}\big\{\boldsymbol{N}_{\!\!f}(\mu)\big\}}_{\scriptstyle=0}\cdot\;
\text{E}\Big\{\Vec{1}\CdoT\Vec{1}^{\,\Hh}\!-\frac{1}{L\!-\!1}\cdot
\underbrace{\Vec{1}\cdot\underline{1}_{\bot}
}_{\scriptstyle=\Vec{0}}\Cdot
\Tilde{\underline{\boldsymbol{V}}}(\mu)^{\Hh}\Cdot
\Hat{\underline{\boldsymbol{C}}}_{\Tilde{\Vec{\boldsymbol{V}}}(\mu),\Tilde{\Vec{\boldsymbol{V}}}(\mu)}^{\uP{0.4}{\!-1}}\Cdot
\Tilde{\underline{\boldsymbol{V}}}(\mu)\cdoT\Vec{1}^{\,\Hh}\Big\}\;={}
\notag\\*[8pt]
{}=\;U_{\!f}(\mu)\qquad\qquad\forall\qquad\mu=0\;(1)\;M\!-\!1.
\label{E.3.21}
\end{gather}
Auch diese Messwerte sind also erwartungstreu.

Gleichung~(\ref{E.2.17}) zeigt, wie sich die $F$ Zufallsgr"o"sen  
\mbox{$\boldsymbol{n}(k)$} des gefensterten Approximationsfehlers 
aus den $M$ Zufallsgr"o"sen \mbox{$\boldsymbol{V}(\mu)$} des Spektrums am Eingang des realen Systems 
und den $F$ Zufallsgr"o"sen \mbox{$\boldsymbol{y}(k)$} am Ausgang des realen Systems 
ergeben. Die $F$ konkreten Stichprobenvektoren \mbox{$\Vec{n}(k)$} vom Umfang $L$ 
der $F$ Zufallsgr"o"sen \mbox{$\boldsymbol{n}(k)$} erhalten wir, 
wenn in Gleichung~(\ref{E.2.17}) statt der Zufallsvektoren \mbox{$\Tilde{\Vec{\boldsymbol{V}}}(\mu)$} 
und  der Zufallsgr"o"sen \mbox{$\boldsymbol{y}(k)$} deren konkrete Realisierungen 
\mbox{$\Tilde{\underline{V}}(\mu)$} und \mbox{$\Vec{y}(k)$} nach Gleichung~(\ref{E.3.15}) 
einsetzen. Nach \mbox{$\Vec{y}(k)$} aufgel"ost ergibt sich:
\begin{equation}
\Vec{y}(k)\,=\,
\Vec{n}(k)+
\frac{1}{M}\cdoT\Sum{\mu=0}{M-1}\Vec{H}(\mu)\CdoT
\Tilde{\underline{V}}(\mu)\cdot
e^{j\cdot\frac{2\pi}{M}\cdot\mu\cdot k}+
u(k)\cdoT\Vec{1}.
\label{E.3.22}
\end{equation}
Zur Berechnung der Erwartungswerte \mbox{$\Hat{\boldsymbol{u}}(k)$} 
der Messwerte der gefensterten deterministischen St"orung ersetzen wir 
in Gleichung~(\ref{E.3.14}) den Vektor \mbox{$\Vec{y}(k)$} mit dieser Gleichung, 
die Messwerte der beiden "Ubertragungsfunktionen mit Gleichung~(\ref{E.3.8}) 
und anschlie"send den Stichprobenvektor \mbox{$\Vec{Y}_{\!f}(\mu)$} mit 
Gleichung~(\ref{E.3.16}). So erhalten wir:
\begin{gather}
\Hat{u}(k)\;={}
\notag\\*[4pt]
{}=\;\frac{1}{L}\CdoT\Vec{y}(k)\cdoT\Vec{1}^{\,\Hh}\!-
\frac{1}{M\CdoT L}\cdoT\!\Sum{\mu=0}{M-1}
\frac{1}{L\!-\!1}\cdot
\Vec{Y}_{\!f}(\mu)\cdot\underline{1}_{\bot}\Cdot
\Tilde{\underline{V}}(\mu)^{\Hh}\Cdot
\Hat{\underline{C}}_{\Tilde{\Vec{\boldsymbol{V}}}(\mu),\Tilde{\Vec{\boldsymbol{V}}}(\mu)}^{\uP{0.4}{\!-1}}\cdot
\Tilde{\underline{V}}(\mu)\CdoT\Vec{1}^{\,\Hh}\!\Cdot
e^{j\cdot\frac{2\pi}{M}\cdot\mu\cdot k}\;={}
\notag\\[8pt]
{}=\;\frac{1}{L}\CdoT\bigg(
\Vec{n}(k)+
\frac{1}{M}\cdoT\Sum{\mu=0}{M-1}\Vec{H}(\mu)\CdoT
\Tilde{\underline{V}}(\mu)\cdot
e^{j\cdot\frac{2\pi}{M}\cdot\mu\cdot k}+
u(k)\cdoT\Vec{1}\bigg)\cdoT\Vec{1}^{\,\Hh}\;-{}
\notag\\*[3pt]
{}-\;\frac{1}{M\CdoT L}\cdoT\!\Sum{\mu=0}{M-1}
\frac{1}{L\!-\!1}\cdot
\Big(\Vec{H}(\mu)\CdoT\Tilde{\underline{V}}(\mu)+U_{\!f}(\mu)\CdoT\Vec{1}+\Vec{N}_{\!f}(\mu)\Big)\cdot{}
\notag\\*[3pt]
{}\cdot\underline{1}_{\bot}\Cdot\Tilde{\underline{V}}(\mu)^{\Hh}\Cdot
\Hat{\underline{C}}_{\Tilde{\Vec{\boldsymbol{V}}}(\mu),\Tilde{\Vec{\boldsymbol{V}}}(\mu)}^{\uP{0.4}{\!-1}}\cdot
\Tilde{\underline{V}}(\mu)\CdoT\Vec{1}^{\,\Hh}\!\Cdot
e^{j\cdot\frac{2\pi}{M}\cdot\mu\cdot k}\;={}
\notag\\[8pt]
{}=\;u(k)+\frac{1}{L}\CdoT\Vec{n}(k)\cdoT\Vec{1}^{\,\Hh}\!-
\frac{1}{M\CdoT L}\cdoT\!\Sum{\mu=0}{M-1}
\frac{\Vec{N}_{\!f}(\mu)\cdot\underline{1}_{\bot}\Cdot\Tilde{\underline{V}}(\mu)^{\Hh}}{L\!-\!1}\cdot
\Hat{\underline{C}}_{\Tilde{\Vec{\boldsymbol{V}}}(\mu),\Tilde{\Vec{\boldsymbol{V}}}(\mu)}^{\uP{0.4}{\!-1}}\cdot
\Tilde{\underline{V}}(\mu)\CdoT\Vec{1}^{\,\Hh}\!\Cdot
e^{j\cdot\frac{2\pi}{M}\cdot\mu\cdot k}
\notag\\*[8pt]
\forall\qquad\mu=0\;(1)\;M\!-\!1.
\label{E.3.23}
\end{gather}
Nun k"onnen wir wieder die Erwartungswerte der Messwerte der deterministischen 
St"orung berechnen:\vspace{-6pt}
\begin{gather}
\text{E}\big\{\Hat{\boldsymbol{u}}(k)\big\}\;=\;
u(k)+\underbrace{\text{E}\big\{\boldsymbol{n}(k)\big\}}_{\scriptstyle=0}-{}
\notag\\*[0pt]
{}-\frac{1}{M\CdoT L}\cdoT\!\Sum{\mu=0}{M-1}
\frac{1}{L\!-\!1}\cdot\underbrace{
\text{E}\big\{\boldsymbol{N}_{\!\!f}(\mu)\big\}}_{\scriptstyle=0}\cdot\;
\text{E}\Big\{\underbrace{\Vec{1}\cdot\underline{1}_{\bot}
}_{\scriptstyle=\Vec{0}}\Cdot
\Tilde{\underline{\boldsymbol{V}}}(\mu)^{\Hh}\Cdot
\Hat{\underline{\boldsymbol{C}}}_{\Tilde{\Vec{\boldsymbol{V}}}(\mu),\Tilde{\Vec{\boldsymbol{V}}}(\mu)}^{\uP{0.4}{\!-1}}\Cdot
\Tilde{\underline{\boldsymbol{V}}}(\mu)\cdoT\Vec{1}^{\,\Hh}\Big\}\cdot
e^{j\cdot\frac{2\pi}{M}\cdot\mu\cdot k}\;={}
\notag\\*[8pt]
{}=\;u(k)\qquad\qquad\forall\qquad 0\le k< F.
\label{E.3.24}
\end{gather}
Auch diese Messwerte sind erwartungstreu. Der Leser m"oge sich selbst davon "uberzeugen, 
dass die Auswirkungen der Behandlung des singul"aren Falls auf die Erwartungswerte der 
deterministischen St"orung und ihrer Spektralwerte in der Praxis bedeutungslos sind.

\section{Messung der beiden Leistungsdichtespektren}\label{E.Kap.3.3}

Da man auch hier durch eine Messung keine Stichprobe \mbox{$\Vec{N}_{\!f}(\mu)$} 
des wahren Spektrums des gefensterten Approximationsfehlerprozesses 
\mbox{$\boldsymbol{n}(k)$} gewinnen kann, weil man weder die wahren Parameter 
der beiden Modellsysteme, noch das wahre Spektrum der gefensterten 
deterministischen St"orung kennt, ben"otigt man wieder aus den gemessenen 
Spektren abgeleitete Zufallsgr"o"sen, deren Erwartungswerte gleich den zu 
messenden Gr"o"sen sind. Dazu verwenden wir wieder den nach 
Gleichung~(\myref{3.12}) definierten gemessenen Stichprobenvektor
\mbox{$\Vec{Y}_{\!f}(\mu)$}, den wir wieder mit einer Matrix
\mbox{$\underline{V}_{\bot}\!(\mu)$} wie in Gleichung ~(\myref{3.25})
linear abbilden, um so den Vektor \mbox{$\Hat{\Vec{N}}_{\!f}(\mu)$} 
zu erhalten.  Mit dem Stichprobenvektor \mbox{$\Vec{N}_{\!f}(\mu)$} nach 
Gleichung (\ref{3.16}) erhalten wir:
\begin{gather}
\Hat{\Vec{N}}_{\!f}(\mu)\:=\:
\Vec{N}_{\!f}(\mu)\CdoT\underline{V}_{\bot}\!(\mu)\:=\:
\Big(\Vec{Y}_{\!f}(\mu)-
\Vec{H}(\mu)\CdoT\Tilde{\underline{V}}(\mu)-
U_{\!f}(\mu)\CdoT\Vec{1}\Big)
\cdot\underline{V}_{\bot}\!(\mu)\:=\:
\Vec{Y}_{\!f}(\mu)\CdoT\underline{V}_{\bot}\!(\mu)
\notag\\*[3pt]
\forall\qquad\mu=0\;(1)\;M\!-\!1.
\label{E.3.25}
\end{gather}
Damit diese Gleichung gilt, muss die Matrix \mbox{$\underline{V}_{\bot}\!(\mu)$}
nun au"ser dem Stichprobenvektor \mbox{$\Vec{V}(\mu)$}, den wir in \cite{Diss} 
verwendet haben, auch noch alle weiteren Zeilenvektoren der Stichprobenmatrix 
\mbox{$\Tilde{\underline{V}}(\mu)$}, die wir bei der Berechnung der Messwerte 
der "Ubertragungsfunktion und der deterministischen St"orung verwendet haben,
sowie den Einservektor $\Vec{1}$ als Eigenvektoren zum Eigenwert Null aufweisen. 
\begin{equation}
\Tilde{\underline{V}}(\mu)\cdot\underline{V}_{\bot}\!(\mu)\;=\;\underline{0}
\qquad\wedge\qquad
\Vec{1}\cdot\underline{V}_{\bot}\!(\mu)\;=\;\Vec{0}
\label{E.3.26}
\end{equation}
$\underline{0}$ ist hier eine \mbox{$(2\CdoT K_H)\times L$} Matrix, 
deren Elemente ebenso null sind, wie die $L$ Elemente des 
Zeilenvektors $\Vec{0}$. Der Vektor \mbox{$\Hat{\Vec{N}}_{\!f}(\mu)$} 
ist dann orthogonal zu den Zeilenvektoren der Stichprobenmatrix 
\mbox{$\Tilde{\underline{V}}(\mu)$} und zum Einservektor $\Vec{1}$ und daher ist
er unabh"angig von den beiden bifrequenten "Ubertragungsfunktionen 
und dem Spektralwert \mbox{$U_{\!f}(\mu)$} der gefensterten St"orung. 
F"ur jede Frequenz $\mu$ kann man eine Matrix \mbox{$\underline{V}_{\bot}\!(\mu)$}, 
die die Bedingung~(\ref{E.3.26}) auf jeden Fall erf"ullt, analog zu Gleichung~(\myref{3.45}), 
nun aber unter Ber"ucksichtigung der Modellierung des Spektrums der deterministischen 
St"orung, konstruieren. Dazu ben"otigen wir eine Matrix \mbox{$\Breve{\underline{V}}(\mu)$},
deren Zeilenvektoren voneinander unabh"angig sind, so dass sie vollen Rang hat. 
Die Zeilenvektoren dieser Matrix m"ussen so gew"ahlt werden, dass alle Zeilenvektoren 
der Matrix \mbox{$\Tilde{\underline{V}}(\mu)$} in dem Raum liegen, der durch die 
Zeilenvektoren der Matrix \mbox{$\Breve{\underline{V}}(\mu)$} aufgespannt wird, 
w"ahrend der Einservektor $\Vec{1}$ nicht in diesem Raum liegen darf.
Die Elemente der so konstruierten Matrix \mbox{$\Breve{\underline{V}}(\mu)$} k"onnen 
zuf"allig sein, m"ussen aber von den zuf"alligen Spektralwerten des gefensterten 
Approximationsfehlers unabh"angig ein. F"ur jede Frequenz $\mu$ kann man mit 
dieser Matrix \mbox{$\Breve{\underline{V}}(\mu)$} wie in Gleichung~(\ref{E.3.10}) 
eine hermitesche, empirische Kovarianzmatrix 
\begin{gather}
\Hat{\underline{C}}_{\Breve{\Vec{\boldsymbol{V}}}(\mu),\Breve{\Vec{\boldsymbol{V}}}(\mu)}\;=\;
\Hat{\underline{C}}_{\Breve{\Vec{\boldsymbol{V}}}(\mu),\Breve{\Vec{\boldsymbol{V}}}(\mu)}^{\Hh}\;=\;
\frac{1}{L\!-\!1}\cdot
\Breve{\underline{V}}(\mu)\cdot
\underline{1}_{\bot}\Cdot
\Breve{\underline{V}}(\mu)^{\Hh}\;=
\label{E.3.27}\\*[4pt]
=\;\frac{1}{L\!-\!1}\cdot
\Breve{\underline{V}}(\mu)\cdot
\Big(\,\underline{E}-\frac{1}{L}\cdot\Vec{1}^{\,\Hh}\Cdot\Vec{1}\:\Big)\cdot
\Breve{\underline{V}}(\mu)^{\Hh}\;=
\notag\\[4pt]
=\;\frac{1}{L\!-\!1}\cdot\Big(\,
\Breve{\underline{V}}(\mu)\cdot\Breve{\underline{V}}(\mu)^{\Hh}-
\frac{1}{L}\cdot
\Breve{\underline{V}}(\mu)\cdot\Vec{1}^{\,\Hh}\cdot
\Vec{1}\cdot\Breve{\underline{V}}(\mu)^{\Hh}\,\Big)\;=
\notag\displaybreak[2]\\[6pt]
=\;\frac{1}{L\!-\!1}\cdot\bigg(\,\Sum{\lambda=1}{L}\,
\Breve{\Vec{V}}_{\!\!\lambda}(\mu)\cdot\Breve{\Vec{V}}_{\!\!\lambda}(\mu)^{\Hh}-\,
\frac{1}{L}\cdot
\Sum{\lambda=1}{L}\Breve{\Vec{V}}_{\!\!\lambda}(\mu)\cdot
\Sum{\lambda=1}{L}\Breve{\Vec{V}}_{\!\!\lambda}(\mu)^{\Hh}\bigg)
\notag\\*[4pt]
\forall\qquad\mu=0\;(1)\;M\!-\!1\notag
\end{gather}
berechnen, die auf Grund der Konstruktion der Matrix \mbox{$\Breve{\underline{V}}(\mu)$} regul"ar ist. 
Mit dieser empirischen Kovarianzmatrix und der Matrix \mbox{$\Breve{\underline{V}}(\mu)$} wird dann die 
idempotente und hermitesche Matrix 
\begin{gather}
\underline{V}_{\bot}\!(\mu)\;=\;
\underline{V}_{\bot}\!(\mu)^{n}\;=\;
\underline{V}_{\bot}\!(\mu)^{\Hh}\;=\;
\underline{1}_{\bot}-\frac{1}{L\!-\!1}\cdot
\underline{1}_{\bot}\!\CdoT\Breve{\underline{V}}(\mu)^{\Hh}\!\Cdot
\Hat{\underline{C}}_{\Breve{\Vec{\boldsymbol{V}}}(\mu),\Breve{\Vec{\boldsymbol{V}}}(\mu)}^{\uP{0.4}{\!-1}}\CdoT
\Breve{\underline{V}}(\mu)\CdoT\underline{1}_{\bot}
\notag\\*[2pt]
\forall\qquad\mu=0\;(1)\;M\!-\!1\quad\wedge\quad n\in\mathbb{N},
\label{E.3.28}
\end{gather}
die die Bedingung~(\ref{E.3.26}) erf"ullt, f"ur jede Frequenz $\mu$ gebildet. 
Der Rangdefekt der Matrix \mbox{$\underline{V}_{\bot}\!(\mu)$} ist um eins gr"o"ser als 
die Zeilenanzahl der Matrix \mbox{$\Breve{\underline{V}}(\mu)$}, da sowohl der Einservektor 
als auch alle Zeilenvektoren der Matrix \mbox{$\Breve{\underline{V}}(\mu)$} Eigenvektoren 
zum Eigenwert Null sind. Da alle dazu orthogonalen Vektoren mit der Matrix \mbox{$\underline{V}_{\bot}\!(\mu)$} 
auf sich selbst abgebildet werden, sind sie Eigenvektoren zum Eigenwert Eins. Da die Matrix 
\mbox{$\underline{V}_{\bot}\!(\mu)$} keine anderen Eigenwerte als Null und Eins aufweist, 
ist die Spur der Matrix gleich dem Rang der Matrix. 

Die einfachste M"oglichkeit die Matrix \mbox{$\Breve{\underline{V}}(\mu)$} 
zu konstruieren besteht darin, mit dem Einservektor $\Vec{1}$ zu beginnen und 
aus der Matrix \mbox{$\Tilde{\underline{V}}(\mu)$} nach und nach Zeilenvektoren 
hinzuzuf"ugen, die von allen bisher ausgew"ahlten Zeilenvektoren linear 
unabh"angig sind. Dies wiederholt man, bis sich keine weiteren linear unabh"angigen 
Vektoren mehr in der Matrix \mbox{$\Tilde{\underline{V}}(\mu)$} finden lassen.
Da die Matrix \mbox{$\underline{V}_{\bot}\!(\mu)$} nicht unbedingt den Nullraum 
der bis dahin ausgew"ahlten Zeilenvektoren der Matrix \mbox{$\Tilde{\underline{V}}(\mu)$} 
vollst"andig aufspannen muss, kann man weitere linear unabh"angige Zeilenvektoren
hinzuf"ugen. Somit kann man, auch wenn im Systemmodell kein Modellsystem, das mit
dem konjugierten Eingangssignal erregt wird, vorgesehen worden ist, zur 
Konstruktion der Matrix \mbox{$\underline{V}_{\bot}\!(\mu)$} die Matrix 
\mbox{$\Breve{\underline{V}}(\mu)$} verwenden, die sich bei einem 
vollst"andigen Systemmodell ergeben w"urde. Abschlie"send wird die Zeile der 
Matrix \mbox{$\Breve{\underline{V}}(\mu)$}, die den Einservektor $\Vec{1}$ enth"alt, 
wieder entfernt, und mit der so konstruierten Matrix \mbox{$\Breve{\underline{V}}(\mu)$}
wird die Matrix \mbox{$\underline{V}_{\bot}\!(\mu)$} mit den Gleichungen~(\ref{E.3.27}) 
und (\ref{E.3.28}) berechnet. Im weiteren soll die Matrix 
\mbox{$\underline{V}_{\bot}\!(\mu)$} immer in der Art konstruiert sein, dass der Rangdefekt  
der Matrix \mbox{$\underline{V}_{\bot}\!(\mu)$} auf eine von $L$ unabh"angige Konstante 
begrenzt ist, d.~h. die Zahl der Zeilenvektoren der Matrix \mbox{$\Breve{\underline{V}}(\mu)$} 
darf mit steigender Mittelungsanzahl $L$ nicht "uber eine von $L$ unabh"angige Konstante steigen. 
Am besten w"ahlt man die Anzahl der Zeilen der Matrix \mbox{$\Breve{\underline{V}}(\mu)$} konstant. 

Prinzipiell muss die Matrix \mbox{$\underline{V}_{\bot}\!(\mu)$} auch hier 
nicht idempotent und hermitesch sein, sofern sie der Bedingung~(\ref{E.3.26})
gen"ugt, und ihre Elemente von den zuf"alligen Spektralwerten des
gefensterten Approximationsfehlers unabh"angig sind. Im weiteren 
werde ich mich aber auf die Verwendung dieser idempotenten und 
hermiteschen Matrizen beschr"anken, die einen Rangdefekt aufweisen, 
der von $L$ unabh"angig nach oben begrenzt ist. Dadurch ist die G"ultigkeit der 
im weiteren hergeleiteten Ergebnisse sichergestellt. Bei Verwendung 
anderer Matrizen w"are dies ggf. im Einzelfall zu "uberpr"ufen. 

Da \mbox{$\underline{V}_{\bot}\!(\mu)$} die Bedingung~(\ref{E.3.26})
erf"ullt, ist nach Gleichung~(\ref{E.3.25}) der durch die Abbildung 
mit der Matrix \mbox{$\underline{V}_{\bot}\!(\mu)$} aus dem des 
Stichprobenvektor \mbox{$\Vec{Y}_{\!f}(\mu)$} entstandene und daher 
bekannte Stichprobenvektor \mbox{$\Hat{\Vec{N}}_{\!f}(\mu)$} gleich 
dem Bildvektor des unbekannten Stichprobenvektors \mbox{$\Vec{N}_{\!f}(\mu)$}, 
der bei der Abbildung mit derselben Matrix entsteht. Daher verwenden wir 
in Analogie zu den Messwerten nach Gleichung~(\myref{3.34}) und (\myref{3.35}) 
im Falle eines station"aren Approximationsfehlerprozesses die Messwerte
\begin{subequations}\label{E.3.29}
\begin{gather}
\Hat{\Phi}_{\boldsymbol{n}}
\big({\T\mu,\mu\!+\!\Tilde{\mu}\CdoT\frac{M}{K_{\Phi}}}\big)\;=\;
\frac{\Hat{\Vec{N}}_{\!f}(\mu)\CdoT\Hat{\Vec{N}}_{\!\!f}
\big({\T\mu\!+\!\Tilde{\mu}\CdoT\frac{M}{K_{\Phi}}}\big)^{\HH}}
{M\CdoT\text{spur}\Big(\underline{V}_{\bot}\!(\mu)\cdot
\underline{V}_{\bot}\!
\big({\T\mu\!+\!\Tilde{\mu}\CdoT\frac{M}{K_{\Phi}}}\big)^{\HH}\Big)}\;=
\label{E.3.29.a}\\[6pt]
=\;\frac{\Vec{Y}_{\!f}(\mu)\cdot
\underline{V}_{\bot}\!(\mu)\cdot
\underline{V}_{\bot}\!
\big({\T\mu\!+\!\Tilde{\mu}\CdoT\frac{M}{K_{\Phi}}}\big)^{\HH}\Cdot
\Vec{Y}_{\!f}
\big({\T\mu\!+\!\Tilde{\mu}\CdoT\frac{M}{K_{\Phi}}}\big)^{\HH}}
{M\CdoT\text{spur}\Big(\underline{V}_{\bot}\!(\mu)\cdot
\underline{V}_{\bot}\!
\big({\T\mu\!+\!\Tilde{\mu}\CdoT\frac{M}{K_{\Phi}}}\big)^{\HH}\Big)}\;=
\notag\\[6pt]
=\;\frac{\Vec{N}_{\!f}(\mu)\cdot
\underline{V}_{\bot}\!(\mu)\cdot
\underline{V}_{\bot}\!
\big({\T\mu\!+\!\Tilde{\mu}\CdoT\frac{M}{K_{\Phi}}}\big)^{\HH}\Cdot
\Vec{N}_{\!f}
\big({\T\mu\!+\!\Tilde{\mu}\CdoT\frac{M}{K_{\Phi}}}\big)^{\HH}}
{M\CdoT\text{spur}\Big(\underline{V}_{\bot}\!(\mu)\cdot
\underline{V}_{\bot}\!
\big({\T\mu\!+\!\Tilde{\mu}\CdoT\frac{M}{K_{\Phi}}}\big)^{\HH}\Big)}
\notag\\[4pt]
\forall\qquad\mu=0\;(1)\;M\!-\!1\quad\text{ und }
\quad\Tilde{\mu}=0\;(1)\;K_{\Phi}\!-\!1
\notag\\[-20pt]\intertext{und\vspace{-20pt}}
\Hat{\Psi}_{\boldsymbol{n}}
\big({\T\mu,\mu\!+\!\Tilde{\mu}\CdoT\frac{M}{K_{\Phi}}}\big)\;=\;
\frac{\Hat{\Vec{N}}_{\!f}(\mu)\CdoT\Hat{\Vec{N}}_{\!\!f}
\big(\!{\T-\mu\!-\!\Tilde{\mu}\CdoT\frac{M}{K_{\Phi}}}\big)^{\TT}}
{M\CdoT\text{spur}\Big(\underline{V}_{\bot}\!(\mu)\cdot
\underline{V}_{\bot}\!
\big(\!{\T-\mu\!-\!\Tilde{\mu}\CdoT\frac{M}{K_{\Phi}}}\big)^{\TT}\Big)}\;=
\label{E.3.29.b}\\[6pt]
=\;\frac{\Vec{Y}_{\!f}(\mu)\cdot
\underline{V}_{\bot}\!(\mu)\cdot
\underline{V}_{\bot}\!
\big(\!{\T-\mu\!-\!\Tilde{\mu}\CdoT\frac{M}{K_{\Phi}}}\big)^{\TT}\Cdot
\Vec{Y}_{\!f}
\big(\!{\T-\mu\!-\!\Tilde{\mu}\CdoT\frac{M}{K_{\Phi}}}\big)^{\TT}}
{M\CdoT\text{spur}\Big(\underline{V}_{\bot}\!(\mu)\cdot
\underline{V}_{\bot}\!
\big(\!{\T-\mu\!-\!\Tilde{\mu}\CdoT\frac{M}{K_{\Phi}}}\big)^{\TT}\Big)}
\;=\notag\\[6pt]
=\;\frac{\Vec{N}_{\!f}(\mu)\cdot
\underline{V}_{\bot}\!(\mu)\cdot
\underline{V}_{\bot}\!
\big(\!{\T-\mu\!-\!\Tilde{\mu}\CdoT\frac{M}{K_{\Phi}}}\big)^{\TT}\Cdot
\Vec{N}_{\!f}
\big(\!{\T-\mu\!-\!\Tilde{\mu}\CdoT\frac{M}{K_{\Phi}}}\big)^{\TT}}
{M\CdoT\text{spur}\Big(\underline{V}_{\bot}\!(\mu)\cdot
\underline{V}_{\bot}\!
\big(\!{\T-\mu\!-\!\Tilde{\mu}\CdoT\frac{M}{K_{\Phi}}}\big)^{\TT}\Big)}
\notag\\*[4pt]
\forall\qquad\mu=0\;(1)\;M\!-\!1\quad\text{ und }
\quad\Tilde{\mu}=0\;(1)\;K_{\Phi}\!-\!1.
\notag
\end{gather}
\end{subequations}
Diese Messwerte weisen die gleichen Symmetrien auf wie die
entsprechenden theoretischen Gr"o"sen:\vspace{-14pt}
\begin{subequations}\label{E.3.30}
\begin{flalign}
&&\Hat{\Phi}_{\boldsymbol{n}}
\big({\T\mu\!+\!\Tilde{\mu}\CdoT\frac{M}{K_{\Phi}},\mu}\big)&\;=\;
\Hat{\Phi}_{\boldsymbol{n}}
\big({\T\mu,\mu\!+\!\Tilde{\mu}\CdoT\frac{M}{K_{\Phi}}}\big)^{\!\Kk}&&
\label{E.3.30.a}\\[6pt]
\text{und}&&\Hat{\Psi}_{\boldsymbol{n}}
\big({\T\mu\!+\!\Tilde{\mu}\CdoT\frac{M}{K_{\Phi}},\mu}\big)&\;=\;
\Hat{\Psi}_{\boldsymbol{n}}
\big(\!{\T-\mu,-\mu\!-\!\Tilde{\mu}\CdoT\frac{M}{K_{\Phi}}}\big).&&
\label{E.3.30.b}
\end{flalign}
\end{subequations}
Man wird diese Messwerte {\em nicht}\/ dadurch berechnen, dass man 
zun"achst die Vektoren \mbox{$\Vec{Y}_{\!f}(\mu)$} und die Matrizen 
\mbox{$\underline{V}_{\bot}\!(\mu)$} bestimmt, und diese dann 
---\,wie in den Gleichungen~(\ref{E.3.29}) angegeben\,--- multipliziert. 
Dies w"urde n"amlich einen mit der Mittelungsanzahl $L$ quadratisch 
anwachsenden Speicher erfordern. Im Anhang~\ref{E.Kap.A.6} wird eine 
geeignetere Vorgehensweise zur Berechnung der Messwerte 
\mbox{$\Hat{\Phi}_{\boldsymbol{n}}\big({\T\mu,\mu\!+\!\Tilde{\mu}\CdoT\frac{M}{K_{\Phi}}}\big)$} und 
\mbox{$\Hat{\Psi}_{\boldsymbol{n}}\big({\T\mu,\mu\!+\!\Tilde{\mu}\CdoT\frac{M}{K_{\Phi}}}\big)$}
angegeben, bei der der Speicherbedarf von $L$ unabh"angig ist.

Am sinnvollsten erscheint es mir, die Matrizen \mbox{$\Breve{\underline{V}}(\mu)$} 
f"ur die unterschiedlichen Frequenzen $\mu$ folgenderma"sen zu konstruieren. 
Zun"achst berechnet man sich aus der Periode $K_H$ der Zeitvarianz des Systems 
und der Periode $K_{\Phi}$ der Zyklostationarit"at der St"orung deren kleinstes 
gemeinsames Vielfaches\vspace{-6pt}
\begin{equation}
\qquad K_S = \text{kgv}(K_H,\,K_{\Phi}).
\label{E.3.31}
\end{equation}
Jede der beiden Frequenzverschiebungen \mbox{$\Tilde{\mu}\CdoT M/K_{\Phi}$} 
und \mbox{$\Hat{\mu}\CdoT M/K_H$} ist dann eine Vielfaches von \mbox{$M/K_S$}. 
Danach greifen wir uns f"ur jede Frequenz $\mu$ aus den Zufallsgr"o"sen 
\mbox{$\boldsymbol{V}\!(\mu\!+\!\Breve{\mu}\CdoT M/K_S)$} und 
\mbox{$\boldsymbol{V}\!(\!-\mu\!-\!\Breve{\mu}\CdoT M/K_S)^{\Kk}$} 
mehrere Zufallsgr"o"sen heraus, und fassen diese in dem Zufallsvektor 
\mbox{$\Breve{\Vec{\boldsymbol{V}}}(\mu)$} zusammen. Die analog zu Gleichung~(\ref{E.2.23}) 
definierte theoretische Kovarianzmatrix der herausgegriffenen Zufallsgr"o"sen 
darf dabei keinen Rangdefekt aufweisen, und muss demselben Rang besitzen, wie 
die theoretische Kovarianzmatrix aller Zufallsgr"o"sen f"ur alle Werte von 
$\Breve{\mu}$ aus denen die Zufallsgr"o"sen herausgegriffen wurden. Es sind 
also bei jeder Frequenz $\mu$ soviele Zufallsgr"o"sen herauszugreifen, dass 
alle linearen Abh"angigkeiten innerhalb aller Zufallsgr"o"sen f"ur alle Werte 
von $\Breve{\mu}$ eliminiert sind, und trotzdem alle Freiheitsgrade 
ber"ucksichtigt sind. Die Anzahl der bei der Frequenz $\mu$ 
herausgegriffenen Zufallsgr"o"sen sei mit \mbox{$K(\mu)$} bezeichnet. 
Da bei jeder Frequenz maximal \mbox{$2\CdoT\!K_S$} Zufallsgr"o"sen 
herausgegriffen werden k"onnen, gilt die Ungleichung 
\begin{equation}
K(\mu)\,\le\,2\cdoT K_S.
\label{E.3.32}
\end{equation}
Die Zeilenvektoren der Stichproben der herausgegriffenen Zufallsgr"o"sen 
fassen wir zu der Matrix \mbox{$\Breve{\underline{V}}(\mu)$} zusammen.
Auch wenn die theoretische Kovarianzmatrix der bei einer Frequenz $\mu$ 
herausgegriffenen Zufallsgr"o"sen gut konditioniert ist, kann doch 
der Rang der Matrix \mbox{$\Breve{\underline{V}}(\mu)$} kleiner 
als deren Zeilendimension, und somit die empirische Kovarianzmatrix 
singul"ar sein. Verwendet man Zufallgr"o"sen 
\mbox{$\boldsymbol{V}\!(\mu\!+\!\Breve{\mu}\CdoT M/K_S)$}, 
bei denen alle theoretischen Kovarianzmatrizen der jeweils herausgegriffenen 
Zufallgr"o"sen gut konditioniert sind, so wird dieser Fall f"ur hinreichend 
gro"se Werte von $L$ mit einer Wahrscheinlichkeit auftreten, die so extrem 
klein ist, dass der Fall einer singul"aren empirischen Kovarianzmatrix 
praktisch bedeutungslos ist. F"ur die theoretischen Herleitungen wird 
angenommen, dass bei der Matrix \mbox{$\Breve{\underline{V}}(\mu)$} 
ggf. die linear abh"angigen Zeilenvektoren durch linear unabh"angige 
Zufallsvektoren ersetzt werden. Die Konstruktion dieser Zufallsvektoren muss 
dabei in einer Weise erfolgen, dass deren Elemente --- unter Ber"ucksichtigung der 
Zuf"alligkeit der Messung und der Matrizen \mbox{$\Breve{\underline{\boldsymbol{V}}}(\mu)$} 
und der Vektoren \mbox{$\Vec{\boldsymbol{N}}_{\!\!f}(\mu)$} --- von den Zufallsgr"o"sentupeln 
\mbox{$\big[\boldsymbol{N}_{\!\!f}(\mu),
\boldsymbol{N}_{\!\!f}(\mu\!+\!\Tilde{\mu}\CdoT M/K_{\Phi})\big]^{\Tt}$} und 
\mbox{$\big[\boldsymbol{N}_{\!\!f}(\mu),
\boldsymbol{N}_{\!\!f}(\!-\mu\!-\!\Tilde{\mu}\CdoT M/K_{\Phi})\big]^{\Tt}$}
unabh"angig sind, um sicherzustellen, dass die im weiteren durchgef"uhrten Berechnungen der 
Erwartungswerte der Messwerte auch im singul"aren Fall ihre G"ultigkeit behalten. 

Mit den so konstruierten Matrizen \mbox{$\Breve{\underline{V}}(\mu)$} berechnen wir uns
mit Gleichung~(\ref{E.3.28}) die Matrizen \mbox{$\underline{V}_{\bot}\!(\mu)$}. Diese 
haben dann immer die Spur \mbox{$L\!-\!1\!-\!K(\mu)$}. Dies zeigt man ganz analog zu der 
Herleitung ab Gleichung~(\myref{3.48}). Es sei auch angemerkt, dass der Rangdefekt 
dieser Matrizen gem"a"s der Ungleichung~(\ref{E.3.32}) f"ur hinreichend gro"se 
Werte von $L$ auf eine von $L$ unabh"angige obere Schranke begrenzt ist. 

Betrachten wir nun zwei Matrizen \mbox{$\underline{V}_{\bot}\!(\mu)$} 
f"ur zwei Frequenzen $\mu$, die sich um ein ganzzahliges Vielfaches von 
\mbox{$M/K_S$} unterscheiden. Die bei beiden Frequenzen herausgegriffenen 
Zufallgr"o"sen entstammen dem gleichen Satz der Zufallsgr"o"sen 
\mbox{$\boldsymbol{V}\!(\mu\!+\!\Breve{\mu}\CdoT M/K_S)$} und 
\mbox{$\boldsymbol{V}\!(\!-\mu\!-\!\Breve{\mu}\CdoT M/K_S)^{\Kk}$}. Da bei 
beiden Frequenzen $\mu$ die herausgegriffenen Zufallgr"o"sen eine regul"are 
theoretische Kovarianzmatrix gr"o"stm"oglichen Ranges (\,$=$~Dimension\,) 
aufweisen, lassen sich die herausgegriffenen Zufallgr"o"sen bei der einen 
Frequenz als regul"are Linearkombinationen der herausgegriffenen 
Zufallgr"o"sen bei der anderen Frequenz darstellen. Damit l"asst sich 
mit Hilfe der Gleichungen~(\ref{E.3.27}) und (\ref{E.3.28}) 
zeigen, dass bei beiden Frequenzen die nach dem eben beschriebenen Verfahren 
berechneten idempotenten Matrizen \mbox{$\underline{V}_{\bot}\!(\mu)$} 
identisch sind. Da bei der Berechnung der Matrizen 
\mbox{$\underline{V}_{\bot}\!(\mu)$} neben den Zufallsgr"o"sen 
\mbox{$\boldsymbol{V}\!(\mu\!+\!\Breve{\mu}\CdoT M/K_S)$} 
auch die konjugierten Zufallsgr"o"sen 
\mbox{$\boldsymbol{V}\!(\!-\mu\!-\!\Breve{\mu}\CdoT M/K_S)^{\Kk}$} 
bei der negativen Frequenz ber"ucksichtigt worden sind, 
erf"ullen die Matrizen \mbox{$\underline{V}_{\bot}\!(\mu)$} 
auch die Bedingung~(\myref{3.39}). Zusammenfassend gilt daher 
\begin{subequations}\label{E.3.33}
\begin{gather}
\label{E.3.33.a}
\underline{V}_{\bot}\!(\mu)\,=\,
\underline{V}_{\bot}\!(\mu)^{\Hh}\!=\,
\underline{V}_{\bot}\!
\big({\T\mu\!+\!\Bar{\mu}\CdoT\frac{M}{K_S}}\big)\,=\,
\underline{V}_{\bot}\!
\big(\!{\T-\mu\!-\!\Bar{\mu}\CdoT\frac{M}{K_S}}\big)^{\Tt}\!=\,
\underline{V}_{\bot}\!(\!-\mu)^{\Tt}\!=\,
\underline{V}_{\bot}\!(\!-\mu)^{\Kk}\\*[12pt]
\label{E.3.33.b}\begin{flalign}
\text{und}&&
\text{spur}\big(\underline{V}_{\bot}\!(\mu)\big)&\;=\;
L\!-\!1\!-\!K(\mu)&&
\end{flalign}\notag
\end{gather}
\end{subequations}
Es gen"ugt daher, wenn man die Matrizen \mbox{$\underline{V}_{\bot}\!(\mu)$} 
f"ur \mbox{$\mu=0\;(1)\;(M\!-\!K_S)/(2\cdoT K_S)$} berechnet, da man die 
Matrizen aller anderen Frequenzen ggf. durch Transponieren daraus erh"alt.

Ein Teil der Zufallsgr"o"sen aller Werte von $\Breve{\mu}$ ist bei 
jeder Frequenz $\mu$ an der theoretischen Kovarianzmatrix beteiligt, 
die bei der Berechnung der theoretischen Werte der "Ubertragungsfunktionen 
\mbox{$H(\mu,\mu\!+\!\Hat{\mu}\CdoT M/K_H)$} und 
\mbox{$H_*(\mu,\mu\!+\!\Hat{\mu}\CdoT M/K_H)$} 
auftritt. Dort wurde gefordert, dass der Teil der Zufallsgr"o"sen 
\mbox{$\boldsymbol{V}\!(\mu\!+\!\Breve{\mu}\CdoT M/K_S)$}, 
der dort verwendet wird und der den Vektor \mbox{$\Tilde{\Vec{\boldsymbol{V}}}(\mu)$}
bildet, so zu w"ahlen ist, dass deren theoretische 
Kovarianzmatrix bei der Berechnung der theoretischen Werte der 
"Ubertragungsfunktionen gut konditioniert und somit regul"ar ist. 
Jede der Zufallsgr"o"sen des Zufallsvektors \mbox{$\Tilde{\Vec{\boldsymbol{V}}}(\mu)$}
 l"asst sich daher als Linearkombination der f"ur \mbox{$\Breve{\Vec{\boldsymbol{V}}}(\mu)$} 
herausgegriffenen Zufallgr"o"sen schreiben, da andernfalls die 
theoretische Kovarianzmatrix des Zufallvektors \mbox{$\Breve{\Vec{\boldsymbol{V}}}(\mu)$}
nicht die maximal m"ogliche Dimension \mbox{$K(\mu)\!\times\!K(\mu)$} 
h"atte, bei der die theoretische Kovarianzmatrix noch regul"ar ist. 
Genau dieselben Linearkombinationen, durch die sich die Zufallsgr"o"sen 
des Zufallsvektors \mbox{$\Tilde{\Vec{\boldsymbol{V}}}(\mu)$}
durch die f"ur \mbox{$\Breve{\Vec{\boldsymbol{V}}}(\mu)$} herausgegriffenen 
Zufallsgr"o"sen ausdr"ucken lassen, gelten f"ur die entsprechenden 
Stichprobenvektoren dieser Zufallsgr"o"sen. Da alle Stichprobenvektoren 
der f"ur \mbox{$\Breve{\Vec{\boldsymbol{V}}}(\mu)$} herausgegriffenen 
Zufallgr"o"sen Eigenvektoren von \mbox{$\underline{V}_{\bot}\!(\mu)$}
zum Eigenwert Null sind, sind auch die Stichprobenvektoren der Zufallsgr"o"sen
des Zufallsvektors \mbox{$\Tilde{\Vec{\boldsymbol{V}}}(\mu)$} Eigenvektoren 
zum gleichen Eigenwert. Die mit der Stichprobenmatrix \mbox{$\Breve{\underline{V}}(\mu)$} 
konstruierten Matrizen \mbox{$\underline{V}_{\bot}\!(\mu)$} erf"ullen daher
die Bedingung~(\ref{E.3.26}) wie gefordert.

Wenn man \mbox{$\underline{V}_{\bot}\!(\mu)$} nach dem oben
angegebenen Verfahren konstruiert, so dass die Gleichungen~(\ref{E.3.33})
erf"ullt sind, erh"alt man die einfacheren Messwerte
\begin{subequations}\label{E.3.34}
\begin{gather}
\Hat{\Phi}_{\boldsymbol{n}}\big({\T\mu,\mu\!+\!\Tilde{\mu}\CdoT\frac{M}{K_{\Phi}}}\big)\;=\;
\frac{\;\Hat{\Vec{N}}_{\!f}(\mu)\CdoT
\Hat{\Vec{N}}_{\!f}\big({\T\mu\!+\!\Tilde{\mu}\CdoT\frac{M}{K_{\Phi}}}\big)^{\HH}\;}
{M\CdoT\big(L\!-\!1\!-\!K(\mu)\big)}\;=
\label{E.3.34.a}\\[5pt]
=\;\frac{\;\Vec{Y}_{\!f}(\mu)\cdot
\underline{V}_{\bot}\!(\mu)\cdot
\Vec{Y}_{\!f}\big({\T\mu\!+\!\Tilde{\mu}\CdoT\frac{M}{K_{\Phi}}}\big)^{\HH}\;}
{M\CdoT\big(L\!-\!1\!-\!K(\mu)\big)}\;=\;
\frac{\;\Vec{N}_{\!f}(\mu)\cdot
\underline{V}_{\bot}\!(\mu)\cdot
\Vec{N}_{\!f}\big({\T\mu\!+\!\Tilde{\mu}\CdoT\frac{M}{K_{\Phi}}}\big)^{\HH}\;}
{M\CdoT\big(L\!-\!1\!-\!K(\mu)\big)}\;=
\notag\\[3pt]
=\;\frac{1}{M}\cdot\frac{L\!-\!1}{L\!-\!1\!-\!K(\mu)}\cdot\Big(
\Hat{C}_{\boldsymbol{Y}_{\!\!\!f}(\mu),\boldsymbol{Y}_{\!\!\!f}(\mu+\Tilde{\mu}\cdot\frac{M}{K_{\Phi}})}-
\Hat{\Vec{C}}_{\boldsymbol{Y}_{\!\!\!f}(\mu),\Breve{\Vec{\boldsymbol{V}}}(\mu)}\Cdot
\Hat{\underline{C}}_{\Breve{\Vec{\boldsymbol{V}}}(\mu),\Breve{\Vec{\boldsymbol{V}}}(\mu)}^{\uP{0.4}{\!-1}}\Cdot
\Hat{\Vec{C}}_{\Breve{\Vec{\boldsymbol{V}}}(\mu),\boldsymbol{Y}_{\!\!\!f}(\mu+\Tilde{\mu}\cdot\frac{M}{K_{\Phi}})}\Big)
\notag\\[3pt]
\forall\quad\mu=0\;(1)\;M\!-\!1\quad\text{ und }
\quad\Tilde{\mu}=0\;(1)\;K_{\Phi}\!-\!1
\notag\\[-10pt]\intertext{und\vspace{-10pt}}
\Hat{\Psi}_{\boldsymbol{n}}\big({\T\mu,\mu\!+\!\Tilde{\mu}\CdoT\frac{M}{K_{\Phi}}}\big)\;=\;
\frac{\;\Hat{\Vec{N}}_{\!f}(\mu)\CdoT
\Hat{\Vec{N}}_{\!f}\big(\!{\T-\mu\!-\!\Tilde{\mu}\CdoT\frac{M}{K_{\Phi}}}\big)^{\TT}\;}
{M\CdoT\big(L\!-\!1\!-\!K(\mu)\big)}\;=
\label{E.3.34.b}\\[5pt]
=\;\frac{\;\Vec{Y}_{\!f}(\mu)\cdot
\underline{V}_{\bot}\!(\mu)\cdot
\Vec{Y}_{\!f}\big(\!{\T-\mu\!-\!\Tilde{\mu}\CdoT\frac{M}{K_{\Phi}}}\big)^{\TT}\;}
{M\CdoT\big(L\!-\!1\!-\!K(\mu)\big)}\;=\;
\frac{\;\Vec{N}_{\!f}(\mu)\cdot
\underline{V}_{\bot}\!(\mu)\cdot
\Vec{N}_{\!f}\big(\!{\T-\mu\!-\!\Tilde{\mu}\CdoT\frac{M}{K_{\Phi}}}\big)^{\TT}\;}
{M\CdoT\big(L\!-\!1\!-\!K(\mu)\big)}\;=
\notag\\[3pt]
=\frac{1}{M}\cdot\frac{L\!-\!1}{L\!-\!1\!-\!K(\mu)}\cdot\Big(
\Hat{C}_{\boldsymbol{Y}_{\!\!\!f}(\mu),\boldsymbol{Y}_{\!\!\!f}(-\mu-\Tilde{\mu}\cdot\frac{M}{K_{\Phi}})^{\Kk}}-
\Hat{\Vec{C}}_{\boldsymbol{Y}_{\!\!\!f}(\mu),\Breve{\Vec{\boldsymbol{V}}}(\mu)}\!\Cdot
\Hat{\underline{C}}_{\Breve{\Vec{\boldsymbol{V}}}(\mu),\Breve{\Vec{\boldsymbol{V}}}(\mu)}^{\uP{0.4}{\!-1}}\!\Cdot
\Hat{\Vec{C}}_{\Breve{\Vec{\boldsymbol{V}}}(\mu),\boldsymbol{Y}_{\!\!\!f}(-\mu-\Tilde{\mu}\cdot\frac{M}{K_{\Phi}})^{\Kk}}\!\Big)\!{}
\notag\\[3pt]
\forall\qquad\mu=0\;(1)\;M\!-\!1\quad\text{ und }
\quad\Tilde{\mu}=0\;(1)\;K_{\Phi}\!-\!1,
\notag
\end{gather}
\end{subequations}
Dabei wurden die empirischen Kovarianzen
\begin{subequations}\label{E.3.35}
\begin{gather}
\Hat{C}_{\boldsymbol{Y}_{\!\!\!f}(\mu_1),\boldsymbol{Y}_{\!\!\!f}(\mu_2)}\;=\;
\frac{1}{L\!-\!1}\cdot
\Vec{Y}_{\!f}(\mu_1)\cdot\underline{1}_{\bot}\Cdot\Vec{Y}_{\!f}(\mu_2)^{\Hh}\;=
\label{E.3.35.a}\\*[4pt]
=\;\frac{1}{L\!-\!1}\cdot\bigg(\,\Sum{\lambda=1}{L}\,
Y_{\!f,\lambda}(\mu_1)\CdoT Y_{\!f,\lambda}(\mu_2)^{\Kk}-\,
\frac{1}{L}\cdot
\Sum{\lambda=1}{L}Y_{\!f,\lambda}(\mu_1)\cdot
\Sum{\lambda=1}{L}Y_{\!f,\lambda}(\mu_2)^{\Kk}\bigg)
\notag\\*[4pt]
\forall\qquad\mu_1=0\;(1)\;M\!-\!1\quad\text{und}
\quad\mu_2=0\;(1)\;M\!-\!1,\notag
\end{gather}
und 
\begin{gather}
\Hat{C}_{\boldsymbol{Y}_{\!\!\!f}(\mu_1),\boldsymbol{Y}_{\!\!\!f}(\mu_2)^{\Kk}}\;=\;
\frac{1}{L\!-\!1}\cdot
\Vec{Y}_{\!f}(\mu_1)\cdot\underline{1}_{\bot}\Cdot\Vec{Y}_{\!f}(\mu_2)^{\Tt}\;=
\label{E.3.35.b}\\*[4pt]
=\;\frac{1}{L\!-\!1}\cdot\bigg(\,\Sum{\lambda=1}{L}\,
Y_{\!f,\lambda}(\mu_1)\CdoT Y_{\!f,\lambda}(\mu_2)-\,
\frac{1}{L}\cdot
\Sum{\lambda=1}{L}Y_{\!f,\lambda}(\mu_1)\cdot
\Sum{\lambda=1}{L}Y_{\!f,\lambda}(\mu_2)\bigg)
\notag\\*[4pt]
\forall\qquad\mu_1=0\;(1)\;M\!-\!1\quad\text{und}
\quad\mu_2=0\;(1)\;M\!-\!1,\notag
\end{gather}
sowie die Kovarianzvektoren
\begin{gather}
\Hat{\Vec{C}}_{\Breve{\Vec{\boldsymbol{V}}}(\mu_1),\boldsymbol{Y}_{\!\!\!f}(\mu_2)}\;=\;
\Hat{\Vec{C}}_{\boldsymbol{Y}_{\!\!\!f}(\mu_2),\Breve{\Vec{\boldsymbol{V}}}(\mu_1)}^{\;\HH}\;=\;
\frac{1}{L\!-\!1}\cdot
\Breve{\underline{V}}(\mu_1)\cdot\underline{1}_{\bot}\Cdot\Vec{Y}_{\!f}(\mu_2)^{\Hh}\;=
\label{E.3.35.c}\\*[4pt]
=\;\frac{1}{L\!-\!1}\cdot\bigg(\,\Sum{\lambda=1}{L}\,
\Breve{\Vec{V}}_{\!\!\lambda}(\mu_1)\CdoT Y_{\!f,\lambda}(\mu_2)^{\Kk}-\,
\frac{1}{L}\cdot
\Sum{\lambda=1}{L}\Breve{\Vec{V}}_{\!\!\lambda}(\mu_1)\cdot
\Sum{\lambda=1}{L}Y_{\!f,\lambda}(\mu_2)^{\Kk}\bigg)
\notag\\*[4pt]
\forall\qquad\mu_1=0\;(1)\;M\!-\!1\quad\text{und}
\quad\mu_2=0\;(1)\;M\!-\!1,\notag
\end{gather}
und
\begin{gather}
\Hat{\Vec{C}}_{\Breve{\Vec{\boldsymbol{V}}}(\mu_1),\boldsymbol{Y}_{\!\!\!f}(\mu_2)^{\Kk}}\;=\;
\frac{1}{L\!-\!1}\cdot
\Breve{\underline{V}}(\mu_1)\cdot\underline{1}_{\bot}\Cdot\Vec{Y}_{\!f}(\mu_2)^{\Tt}\;=
\label{E.3.35.d}\\*[4pt]
=\;\frac{1}{L\!-\!1}\cdot\bigg(\,\Sum{\lambda=1}{L}\,
\Breve{\Vec{V}}_{\!\!\lambda}(\mu_1)\CdoT Y_{\!f,\lambda}(\mu_2)-\,
\frac{1}{L}\cdot
\Sum{\lambda=1}{L}\Breve{\Vec{V}}_{\!\!\lambda}(\mu_1)\cdot
\Sum{\lambda=1}{L}Y_{\!f,\lambda}(\mu_2)\bigg)
\notag\\*[4pt]
\forall\qquad\mu_1=0\;(1)\;M\!-\!1\quad\text{und}
\quad\mu_2=0\;(1)\;M\!-\!1\notag
\end{gather}
\end{subequations}
verwendet. 

Dass die mit den Gleichungen~(\ref{E.3.34}) berechneten Messwerte zus"atzlich die Ungleichungen
\begin{subequations}\label{E.3.36}
\begin{flalign}
&&\Big|\Hat{\Phi}_{\boldsymbol{n}}\big({\T\mu,\mu\!+\!\Tilde{\mu}\CdoT\frac{M}{K_{\Phi}}}\big)\Big|^2&
\;\le\;\Hat{\Phi}_{\boldsymbol{n}}(\mu,\mu)\cdot
\Hat{\Phi}_{\boldsymbol{n}}\big({\T\mu\!+\!\Tilde{\mu}\CdoT\frac{M}{K_{\Phi}},\mu\!+\!\Tilde{\mu}\CdoT\frac{M}{K_{\Phi}}}\big)&&
\label{E.3.36.a}\\*[8pt]
\text{und}&&
\Big|\Hat{\Psi}_{\boldsymbol{n}}\big({\T\mu,\mu\!+\!\Tilde{\mu}\CdoT\frac{M}{K_{\Phi}}}\big)\Big|^2
&\;\le\;\Hat{\Phi}_{\boldsymbol{n}}(\mu,\mu)\cdot
\Hat{\Phi}_{\boldsymbol{n}}\big(\!{\T-\mu\!-\!\Tilde{\mu}\CdoT\frac{M}{K_{\Phi}},-\mu\!-\!\Tilde{\mu}\CdoT\frac{M}{K_{\Phi}}}\big)&&
\label{E.3.36.b}
\end{flalign}
\end{subequations}
erf"ullen, die den Ungleichungen~(\ref{E.2.43}) und (\ref{E.2.53})
der zu messenden theoretischen Gr"o"sen entsprechen, ergibt sich
einerseits aus der Cauchy-Schwarzschen Ungleichung angewandt auf die
Vektorenpaare \mbox{$\Hat{\Vec{N}}_{\!f}(\mu)$} und
\mbox{$\Hat{\Vec{N}}_{\!f}(\mu\!+\!\Tilde{\mu}\CdoT M/K_{\Phi})$}
bzw. \mbox{$\Hat{\Vec{N}}_{\!f}(\mu)$} und \mbox{$
\Hat{\Vec{N}}_{\!f}(\!-\mu\!-\!\Tilde{\mu}\CdoT M/K_{\Phi})^{\Kk}\!\!$},
und andererseits aus der Tatsache, dass bei allen in diesen
Ungleichungen auftretenden Messwerten der gleiche Vorfaktor auftritt.
Wie wir noch sehen werden, erm"oglicht es uns die Erf"ullung der
letzten beiden Ungleichungen wieder, Konfidenzgebiete f"ur alle
Messwerte anzugeben.

Wenn wir die Messwerte wieder als zuf"allig betrachten, und wenn der 
Zufallsvektor \mbox{$\Breve{\Vec{\boldsymbol{V}}}(\mu)$}, der aus den 
Zufallgr"o"sen \mbox{$\boldsymbol{V}\!(\mu\!+\!\Breve{\mu}\CdoT M/K_S)$} und
\mbox{$\boldsymbol{V}\!(\!-\mu\!-\!\Breve{\mu}\CdoT M/K_S)^{\Kk}$} des Spektrums
der Erregung gebildet ist, die in die Elemente der dann ebenfalls zuf"alligen
Matrizen \mbox{$\underline{\boldsymbol{V}}_{\bot}\!(\mu)$} eingehen,
unabh"angig ist von dem Zufallsgr"o"senpaar der Spektralwerte
\mbox{$\big[\boldsymbol{N}_{\!\!f}(\mu),
\boldsymbol{N}_{\!\!f}(\mu\!+\!\Tilde{\mu}\CdoT\frac{M}{K_{\Phi}})\big]^{\Tt}\!{}$} bzw.
\mbox{$\big[\boldsymbol{N}_{\!\!f}(\mu),
\boldsymbol{N}_{\!\!f}(\!-\mu\!-\!\Tilde{\mu}\CdoT\frac{M}{K_{\Phi}})\big]^{\Tt}$}
der gefensterten St"orung des realen Systems, kann man wieder Erwartungswerte,
Varianzen und Kovarianzen f"ur die Messwerte berechnen, und erwartungstreue
Sch"atzwerte f"ur die Messwert"-(ko)"-varianzen angeben.

Die Erwartungstreue der Messwerte gem"a"s der Gleichungen~(\ref{E.3.29})
und somit auch der Messwerte gem"a"s der Gleichungen~(\ref{E.3.34}),
die einen Spezialfall der erstgenannten darstellen, zeigt man wie im 
Fall des station"aren Approximationsfehlerprozesses in \cite{Diss} mit 
den in Anhang \ref{E.Kap.A.5} dargestellten Umformungen.
Auch hier gehen nur die Hauptdiagonalelemente der Matrizen der bilinearen
Formen additiv in die Erwartungswerte ein, da die Nebendiagonalelemente
dieser Matrizen bei der Berechnung der bilinearen Form mit Produkten
von Stichprobenelementen der Zufallsgr"o"sen
\mbox{$\boldsymbol{N}_{\!\!f}(\mu)$} und
\mbox{$\boldsymbol{N}_{\!\!f}(\mu\!+\!\Tilde{\mu}\CdoT M/K_{\Phi})$} bzw.
\mbox{$\boldsymbol{N}_{\!\!f}(\!-\mu\!-\!\Tilde{\mu}\CdoT M/K_{\Phi})$}
verkn"upft werden, die aus unterschiedlichen und unabh"angigen
Einzelmessungen stammen, und deren Erwartungswerte (\mbox{$=\!0$}\,) daher 
faktorisierbar sind. In dem zu bildenden Erwartungswert
der nur von den Spektralwerten der Erregung abh"angt, l"asst sich
daher jeweils die Spur der Matrix, auf die der jeweilige Messwert
normiert ist, k"urzen. Bei der Berechnung des Erwartungswertes des Messwertes 
\mbox{$\Hat{\boldsymbol{\Phi}}_{\!\boldsymbol{n}}(\mu,\mu\!+\!\Tilde{\mu}\CdoT M/K_{\Phi})$}
verbleibt der Erwartungswert
\begin{subequations}\label{E.3.37}
\begin{align}
\frac{1}{M}\cdot\text{E}\Big\{\boldsymbol{N}_{\!\!f}(\mu)\CdoT
\boldsymbol{N}_{\!\!f}
\big({\T\mu\!+\!\Tilde{\mu}\CdoT\frac{M}{K_{\Phi}}}\big)^{\!\Kk}\Big\}&-
\text{E}\big\{\boldsymbol{N}_{\!\!f}(\mu)\big\}\cdot
\text{E}\Big\{\boldsymbol{N}_{\!\!f}
\big({\T\mu\!+\!\Tilde{\mu}\CdoT\frac{M}{K_{\Phi}}}\big)\Big\}^{\!*}
\label{E.3.37.a}\\\intertext{w"ahrend sich bei dem Messwert
\mbox{$\Hat{\boldsymbol{\Psi}}_{\!\boldsymbol{n}}(\mu,\mu\!+\!\Tilde{\mu}\CdoT M/K_{\Phi})$}
der Erwartungswert}
\frac{1}{M}\cdot\text{E}\Big\{\boldsymbol{N}_{\!\!f}(\mu)\CdoT
\boldsymbol{N}_{\!\!f}
\big(\!{\T-\mu\!-\!\Tilde{\mu}\CdoT\frac{M}{K_{\Phi}}}\big)\Big\}&-
\text{E}\big\{\boldsymbol{N}_{\!\!f}(\mu)\big\}\cdot
\text{E}\Big\{\boldsymbol{N}_{\!\!f}
\big(\!{\T-\mu\!-\!\Tilde{\mu}\CdoT\frac{M}{K_{\Phi}}}\big)\Big\}
\label{E.3.37.b}
\end{align}
\end{subequations}
ergibt. Da die aus dem Approximationsfehlerprozess durch Fensterung und DFT
gewonnenen Zufallsgr"o"sen \mbox{$\boldsymbol{N}_{\!\!f}(\mu)$} mittelwertfrei 
sind, stimmen diese Erwartungswerte mit den gew"unschten Werten nach
Gleichung~(\ref{E.2.39}) bzw. (\ref{E.2.50}) "uberein.

\section{Varianzen und Kovarianzen der Messwerte}\label{E.Kap.3.4}

Mit dem Einheitszeilenvektor\vspace{-2pt}
\begin{equation}
\Vec{E}_n=[0,\ldots,1,\ldots,0],
\label{E.3.38}
\end{equation}
dessen $n$-tes Element eins ist, w"ahrend alle anderen Elemente
null sind, und der hier \mbox{$2\CdoT K_H$} Elemente enth"alt, 
kann man aus dem Zeilenvektor \mbox{$\Hat{\Vec{H}}(\mu)$}, der die Messwerte 
der beiden bifrequenten "Ubertragungsfunktionen enth"alt, den $n$-ten 
Messwert herausgreifen, indem man diesen Vektor mit dem transponierten 
Einheitsvektor multipliziert:
\begin{equation}
\Hat{H}_n(\mu)\;=\;
\Hat{\Vec{H}}(\mu)\CdoT\Vec{E}_n^{\,\Hh}.
\label{E.3.39}
\end{equation}
Wieder kann man den $n$-ten Messwert \mbox{$\Hat{\boldsymbol{H}}_{\!n}(\mu)$}
als eine Zufallsgr"o"se betrachten, die aus einer mathematische Stichprobe
der Zufallsprozesse am Systemein- und -ausgang entstanden ist.
Die Varianz des Messwertes \mbox{$\Hat{\boldsymbol{H}}_{\!n}(\mu)$} berechnet
sich mit Gleichung~(\ref{E.3.17}) als Erwartungswert einer quadratischen Form. 
Die Berechnung erfolgt nach dem im Anhang \ref{E.Kap.A.5} beschriebenen Verfahren. 
Wir erhalten:
\begin{gather}
C_{\Hat{\boldsymbol{H}}_n(\mu),\Hat{\boldsymbol{H}}_n(\mu)}\;=\;
\text{E}\bigg\{\Big|\Hat{\boldsymbol{H}}_n(\mu)\!-\!
\text{E}\big\{\Hat{\boldsymbol{H}}_n(\mu)\big\}\Big|^2\bigg\}\;=\;
\text{E}\bigg\{\Big|\big(\Hat{\Vec{\boldsymbol{H}}}(\mu)\!-\!
\Vec{H}(\mu)\big)\CdoT\Vec{E}_n^{\,\Hh}\Big|^2\bigg\}\;=
\notag\\[4pt]
=\text{E}\bigg\{\frac{1}{(L\!-\!1)^2}\cdot
\Vec{\boldsymbol{N}}_{\!\!f}(\mu)\CdoT
\underline{1}_{\bot}\!\CdoT
\Tilde{\underline{\boldsymbol{V}}}(\mu)^{\Hh}\!\Cdot
\Hat{\underline{\boldsymbol{C}}}_{\Tilde{\Vec{\boldsymbol{V}}}(\mu),\Tilde{\Vec{\boldsymbol{V}}}(\mu)}^{\uP{0.4}{\!-1}}\CdoT
\Vec{E}_n^{\,\Hh}\CdoT\Vec{E}_n\Cdot
\Hat{\underline{\boldsymbol{C}}}_{\Tilde{\Vec{\boldsymbol{V}}}(\mu),\Tilde{\Vec{\boldsymbol{V}}}(\mu)}^{\uP{0.4}{\!-1}}\CdoT
\Tilde{\underline{\boldsymbol{V}}}(\mu)\CdoT\underline{1}_{\bot}\!\CdoT
\Vec{\boldsymbol{N}}_{\!\!f}(\mu)^{\Hh}\bigg\}=
\notag\\[5pt]\begin{flalign*}
{}=&\;\text{E}\bigg\{\text{spur}\bigg(\frac{1}{(L\!-\!1)^2}\cdot
\underline{1}_{\bot}\!\CdoT
\Tilde{\underline{\boldsymbol{V}}}(\mu)^{\Hh}\!\Cdot
\Hat{\underline{\boldsymbol{C}}}_{\Tilde{\Vec{\boldsymbol{V}}}(\mu),\Tilde{\Vec{\boldsymbol{V}}}(\mu)}^{\uP{0.4}{\!-1}}\CdoT
\Vec{E}_n^{\,\Hh}\CdoT\Vec{E}_n\Cdot
\Hat{\underline{\boldsymbol{C}}}_{\Tilde{\Vec{\boldsymbol{V}}}(\mu),\Tilde{\Vec{\boldsymbol{V}}}(\mu)}^{\uP{0.4}{\!-1}}\CdoT
\Tilde{\underline{\boldsymbol{V}}}(\mu)\CdoT\underline{1}_{\bot}\!\bigg)\bigg\}\cdot{}&&
\end{flalign*}\notag\\*\begin{flalign*}
&&{}\cdot\Big(\text{E}\big\{|\boldsymbol{N}_{\!\!f}(\mu)|^2\big\}\!-\!
\big|\text{E}\{\boldsymbol{N}_{\!\!f}(\mu)\}\big|^2\Big)&=\!{}
\end{flalign*}\notag\\[5pt]\begin{flalign*}
{}=&\;\text{E}\bigg\{\text{spur}\bigg(\frac{1}{(L\!-\!1)^2}\cdot\Vec{E}_n\Cdot
\Hat{\underline{\boldsymbol{C}}}_{\Tilde{\Vec{\boldsymbol{V}}}(\mu),\Tilde{\Vec{\boldsymbol{V}}}(\mu)}^{\uP{0.4}{\!-1}}\CdoT
\Tilde{\underline{\boldsymbol{V}}}(\mu)\CdoT\underline{1}_{\bot}\!\CdoT
\underline{1}_{\bot}\!\CdoT\Tilde{\underline{\boldsymbol{V}}}(\mu)^{\Hh}\!\Cdot
\Hat{\underline{\boldsymbol{C}}}_{\Tilde{\Vec{\boldsymbol{V}}}(\mu),\Tilde{\Vec{\boldsymbol{V}}}(\mu)}^{\uP{0.4}{\!-1}}\CdoT
\Vec{E}_n^{\,\Hh}\bigg)\!\bigg\}\cdot{}&&
\end{flalign*}\notag\\*\begin{flalign*}
&&{}\cdot\text{E}\big\{|\boldsymbol{N}_{\!\!f}(\mu)|^2\big\}&=\!{}
\end{flalign*}\notag\\[5pt]
{}=\;\frac{1}{(L\!-\!1)^2}\cdot\text{E}\Big\{
\Hat{\Vec{\boldsymbol{C}}}_n(\mu)\CdoT
\Hat{\Vec{\boldsymbol{C}}}_n(\mu)^{\Hh}\Big\}\cdot
\text{E}\big\{|\boldsymbol{N}_{\!\!f}(\mu)|^2\big\}\;={}
\notag\\[5pt]
{}=\;\frac{M}{L\!-\!1}\cdot\text{E}\big\{\Vec{E}_n\CdoT
\Hat{\underline{\boldsymbol{C}}}_{\Tilde{\Vec{\boldsymbol{V}}}(\mu),\Tilde{\Vec{\boldsymbol{V}}}(\mu)}^{\uP{0.4}{\!-1}}\CdoT
\Vec{E}_n^{\,\Hh}\,\big\}\CdoT\Tilde{\Phi}_{\boldsymbol{n}}(\mu,\mu).
\label{E.3.40}
\end{gather}
In einem Zwischenschritt wurde dabei der Zeilenvektor
\begin{equation}
\Hat{\Vec{\boldsymbol{C}}}_n(\mu)\;=\;
\Vec{E}_n\Cdot
\Hat{\underline{\boldsymbol{C}}}_{\Tilde{\Vec{\boldsymbol{V}}}(\mu),\Tilde{\Vec{\boldsymbol{V}}}(\mu)}^{\uP{0.4}{\!-1}}\CdoT
\Tilde{\underline{\boldsymbol{V}}}(\mu)\CdoT\underline{1}_{\bot}
\label{E.3.41}
\end{equation}
eingef"uhrt, der durch eine lineare Abbildung des Einheitsvektors
\mbox{$\Vec{E}_n$} entsteht, und der sp"ater noch von Bedeutung
sein wird. Ein Faktor der Messwertvarianz ist der Erwartungswert 
des $n$-ten Hauptdiagonalelements der inversen Kovarianzmatrix. 
Im weiteren setzen wir die oben beschriebene Behandlung des extrem 
unwahrscheinlichen, aber dennoch m"oglichen Falls einer singul"aren 
empirischen Kovarianzmatrix voraus. In Anhang~\ref{E.Kap.A.4} wird gezeigt, 
dass die Hauptdiagonalelemente immer reell und gr"o"ser als das Inverse 
des gr"o"sten Singul"arwertes der empirischen Kovarianzmatrix sind und 
kleiner als das Inverse des kleinsten ---\,nach unten begrenzten\,--- 
Singul"arwertes der empirischen Kovarianzmatrix sind. Somit ist die 
Existenz des Erwartungswertes des Hauptdiagonalelements der inversen 
Kovarianzmatrix gesichert. Der von der Erregung unabh"angige Vorfaktor
\mbox{$1/(L\!-\!1)$} sorgt daf"ur, dass die Messwertvarianz mit steigender 
Mittelungsanzahl $L$ gegen null strebt, so dass die Messwerte konsistent 
sind. Durch die Wahl eines Zufallsvektors $\Vec{\boldsymbol{V}}$, bei dem 
die nach Gleichung~(\ref{E.3.10}) definierten theoretischen Kovarianzmatrizen 
gut konditioniert sind, kann erreicht werden, dass die Wahrscheinlichkeit 
eine schlecht konditionierte empirische Kovarianzmatrix zu erhalten extrem 
klein wird (\,siehe Kapitel~\ref{E.Kap.A.2}\,). Daher wird auch der Erwartungswert
des $n$-ten Hauptdiagonalelements der inversen Kovarianzmatrix 
---\,und somit die Messwertvarianz\,--- klein, wenn man einerseits
die Varianzen der Zufallsgr"o"sen des Vektors 
\mbox{$\Tilde{\Vec{\boldsymbol{V}}}(\mu)$}  m"oglichst gro"s w"ahlt, 
und andererseits darauf achtet, dass die theoretische Kovarianzmatrix 
dieser Zufallsgr"o"sen gut konditioniert ist.

F"ur die Kovarianz des $n$-ten Messwertes \mbox{$\Hat{\boldsymbol{H}}_{\!n}(\mu)$}
des Messwertvektors \mbox{$\Hat{\Vec{\boldsymbol{H}}}(\mu)$}
der beiden "Ubertragungsfunktionen ergibt sich analog:
\begin{gather}
C_{ \Hat{\boldsymbol{H}}_n(\mu),\Hat{\boldsymbol{H}}_n(\mu)^{\Kk}}\;=\;
\text{E}\bigg\{\Big(\Hat{\boldsymbol{H}}_n(\mu)\!-\!
\text{E}\big\{\Hat{\boldsymbol{H}}_n(\mu)\big\}\Big)^2\bigg\}\;=\;
\text{E}\bigg\{\Big(\big(\Hat{\Vec{\boldsymbol{H}}}(\mu)\!-\!
\Vec{H}(\mu)\big)\CdoT\Vec{E}_n^{\,\Hh}\Big)^2\bigg\}\;=
\notag\\[4pt]
=\text{E}\bigg\{\!\frac{1}{(L\!-\!1)^2}\CdoT
\Vec{\boldsymbol{N}}_{\!\!f}(\mu)\CdoT
\underline{1}_{\bot}\!\CdoT
\Tilde{\underline{\boldsymbol{V}}}(\mu)^{\Hh}\!\Cdot
\Hat{\underline{\boldsymbol{C}}}_{\Tilde{\Vec{\boldsymbol{V}}}(\mu),\Tilde{\Vec{\boldsymbol{V}}}(\mu)}^{\uP{0.4}{\!-1}}\CdoT
\Vec{E}_n^{\,\Hh}\CdoT\Vec{E}_n\CdoT
\big(\Hat{\underline{\boldsymbol{C}}}_{\Tilde{\Vec{\boldsymbol{V}}}(\mu),\Tilde{\Vec{\boldsymbol{V}}}(\mu)}^{\uP{0.4}{\!-1}}\big)^{\!\Kk}\!\CdoT
\Tilde{\underline{\boldsymbol{V}}}(\mu)^{\Kk}\!\CdoT\underline{1}_{\bot}\!\CdoT
\Vec{\boldsymbol{N}}_{\!\!f}(\mu)^{\Tt}\!\bigg\}=
\notag\\[5pt]\begin{flalign*}
{}=&\;\text{E}\bigg\{\text{spur}\bigg(\frac{1}{(L\!-\!1)^2}\cdot
\underline{1}_{\bot}\!\CdoT
\Tilde{\underline{\boldsymbol{V}}}(\mu)^{\Hh}\!\Cdot
\Hat{\underline{\boldsymbol{C}}}_{\Tilde{\Vec{\boldsymbol{V}}}(\mu),\Tilde{\Vec{\boldsymbol{V}}}(\mu)}^{\uP{0.4}{\!-1}}\CdoT
\Vec{E}_n^{\,\Hh}\CdoT\Vec{E}_n\Cdot
\big(\Hat{\underline{\boldsymbol{C}}}_{\Tilde{\Vec{\boldsymbol{V}}}(\mu),\Tilde{\Vec{\boldsymbol{V}}}(\mu)}^{\uP{0.4}{\!-1}}\big)^{\!\Kk}\!\CdoT
\Tilde{\underline{\boldsymbol{V}}}(\mu)^{\Kk}\!\CdoT\underline{1}_{\bot}\!\bigg)\bigg\}\cdot{}&&
\end{flalign*}\notag\\*\begin{flalign*}
&&{}\cdot\Big(\text{E}\big\{\boldsymbol{N}_{\!\!f}(\mu)^2\big\}-
\text{E}\{\boldsymbol{N}_{\!\!f}(\mu)\}^2\Big)&={}
\end{flalign*}\notag\\[5pt]\begin{flalign*}
{}=&\;\text{E}\bigg\{\text{spur}\bigg(\frac{1}{(L\!-\!1)^2}\cdot\Vec{E}_n\Cdot
\big(\Hat{\underline{\boldsymbol{C}}}_{\Tilde{\Vec{\boldsymbol{V}}}(\mu),\Tilde{\Vec{\boldsymbol{V}}}(\mu)}^{\uP{0.4}{\!-1}}\big)^{\!\Kk}\!\CdoT
\Tilde{\underline{\boldsymbol{V}}}(\mu)^{\Kk}\!\CdoT\underline{1}_{\bot}\!\CdoT
\underline{1}_{\bot}\!\CdoT\Tilde{\underline{\boldsymbol{V}}}(\mu)^{\Hh}\!\Cdot
\Hat{\underline{\boldsymbol{C}}}_{\Tilde{\Vec{\boldsymbol{V}}}(\mu),\Tilde{\Vec{\boldsymbol{V}}}(\mu)}^{\uP{0.4}{\!-1}}\CdoT
\Vec{E}_n^{\,\Hh}\bigg)\!\bigg\}\cdot{}&&
\end{flalign*}\notag\\*\begin{flalign*}
&&{}\cdot\text{E}\big\{\boldsymbol{N}_{\!\!f}(\mu)^2\big\}&={}
\end{flalign*}\notag\\[5pt]
{}=\;\frac{1}{(L\!-\!1)^2}\cdot\text{E}\Big\{
\Hat{\Vec{\boldsymbol{C}}}_n(\mu)\CdoT
\Hat{\Vec{\boldsymbol{C}}}_n(\mu)^{\Tt}\Big\}^{\!*}\cdot
\text{E}\big\{\boldsymbol{N}_{\!\!f}(\mu)^2\big\}\;={}
\notag\\[5pt]
{}=\;\frac{1}{L\!-\!1}\cdot\text{E}\Big\{\Vec{E}_n\Cdot
\big(\Hat{\underline{\boldsymbol{C}}}_{\Tilde{\Vec{\boldsymbol{V}}}(\mu),\Tilde{\Vec{\boldsymbol{V}}}(\mu)}^{\uP{0.4}{\!-1}}\big)^{\!\Kk}\!\CdoT
\Hat{\underline{\boldsymbol{C}}}_{\Tilde{\Vec{\boldsymbol{V}}}(\mu)^{\Kk},\Tilde{\Vec{\boldsymbol{V}}}(\mu)}\CdoT
\Hat{\underline{\boldsymbol{C}}}_{\Tilde{\Vec{\boldsymbol{V}}}(\mu),\Tilde{\Vec{\boldsymbol{V}}}(\mu)}^{\uP{0.4}{\!-1}}\CdoT
\Vec{E}_n^{\,\Hh}\,\Big\}\cdot
\text{E}\big\{\boldsymbol{N}_{\!\!f}(\mu)^2\big\}.
\label{E.3.42}
\end{gather}
Bei der Kovarianz des Messwertes \mbox{$\Hat{\boldsymbol{H}}_{\!n}(\mu)$}
ist zu beachten, dass hier der Erwartungswert \mbox{$\text{E}\{\boldsymbol{N}_{\!\!f}(\mu)^2\}$}
nicht mehr nur f"ur \mbox{$\mu\!=\!0$} und \mbox{$\mu\!=\!M/2$} nennenswert
von null verschieden ist, da er sich nicht mehr nach Gleichung~(\myref{3.61})
durch einfache Integration des KLDS ergibt, das mit dem auf zwei 
unterschiedliche Weisen verschobenen Fensterspektrum multipliziert wird. 
Stattdessen findet hier das entsprechende Doppelintegral nach 
Gleichung~(\ref{E.2.51}) mit \mbox{$\Hat{\mu}\!=\!-\mu$} Anwendung, 
bei dem im Integranden ebenfalls das Fensterspektrum mit zwei 
unterschiedlichen Frequenzverschiebungen auftritt. In diesem Integral
steht nun aber statt des eindimensionalen das bifrequente KLDS
\mbox{$\Psi_{\boldsymbol{n}}(\Omega_1,\Omega_2)$}, das nicht nur f"ur
\mbox{$\Omega_1\!=\!\Omega_2$} eine Impulslinie aufweist, wie
das bei einem station"aren Approximationsfehlerprozess der
Fall ist. Deshalb erhalten wir selbst bei Verwendung einer hoch 
frequenzselektiven Fensterfolge f"ur den Erwartungswert 
\mbox{$\text{E}\big\{\boldsymbol{N}_{\!\!f}(\mu)^2\big\}$} nur dann 
n"aherungsweise null, wenn $\mu$ kein ganzzahliges Vielfaches von 
\mbox{$M/2/K_{\Phi}$} ist. Die Kovarianz des Messwertes 
\mbox{$\Hat{\boldsymbol{H}}_{\!n}(\mu)$} kann also nur von null 
verschieden sein, wenn $\mu$ ein ganzzahliges Vielfaches von 
\mbox{$M/2/K_{\Phi}$} ist. Die von null verschiedenen Werte 
der Messwertkovarianz erhalten wir mit Gleichung~(\ref{E.2.50}):
\begin{gather}
C_{ \Hat{\boldsymbol{H}}_n(\mu),\Hat{\boldsymbol{H}}_n(\mu)^{\Kk}}\;=\;
\frac{M}{L\!-\!1}\cdot\text{E}\Big\{\Vec{E}_n\Cdot
\big(\Hat{\underline{\boldsymbol{C}}}_{\Tilde{\Vec{\boldsymbol{V}}}(\mu),\Tilde{\Vec{\boldsymbol{V}}}(\mu)}^{\uP{0.4}{\!-1}}\big)^{\!\Kk}\!\CdoT
\Hat{\underline{\boldsymbol{C}}}_{\Tilde{\Vec{\boldsymbol{V}}}(\mu)^{\Kk},\Tilde{\Vec{\boldsymbol{V}}}(\mu)}\CdoT
\Hat{\underline{\boldsymbol{C}}}_{\Tilde{\Vec{\boldsymbol{V}}}(\mu),\Tilde{\Vec{\boldsymbol{V}}}(\mu)}^{\uP{0.4}{\!-1}}\CdoT
\Vec{E}_n^{\,\Hh}\,\Big\}\CdoT\Tilde{\Psi}_{\boldsymbol{n}}(\mu,\!-\mu)
\notag\\[5pt]
{\T\qquad\forall\qquad\mu=\Breve{\mu}\CdoT\frac{M}{2\cdot K_{\Phi}}\quad\wedge\quad
\Breve{\mu}\in\mathbb{Z}}
\label{E.3.43}
\end{gather}
Die wahren Messwert"-(ko)"-varianzen lassen sich mit
\begin{subequations}\label{E.3.44}
\begin{gather*}
\label{E.3.44.a}\begin{flalign}
&\Hat{C}_{\Hat{\boldsymbol{H}}_n(\mu),\Hat{\boldsymbol{H}}_n(\mu)}\;=\;
\frac{M}{(L\!-\!1)^2}\cdot\Hat{\Vec{C}}_n(\mu)\cdot
\Hat{\Vec{C}}_n(\mu)^{\Hh}\Cdot\Hat{\Phi}_{\boldsymbol{n}}(\mu,\mu)\;=&&
\end{flalign}\\*\begin{flalign*}
&&\;=\frac{M}{L\!-\!1}\cdot\Vec{E}_n\cdot
\Hat{\underline{C}}_{\Tilde{\Vec{\boldsymbol{V}}}(\mu),\Tilde{\Vec{\boldsymbol{V}}}(\mu)}^{\uP{0.4}{\!-1}}\cdot
\Vec{E}_n^{\,\Hh}\cdot
\Hat{\Phi}_{\boldsymbol{n}}(\mu,\mu)&
\end{flalign*}\\*[6pt]
\forall\qquad \mu=0\;(1)\;M\!-\!1
\\[-16pt]\intertext{und\vspace{-6pt}}\label{E.3.44.b}\begin{flalign}
&\Hat{C}_{ \Hat{\boldsymbol{H}}_n(\mu),\Hat{\boldsymbol{H}}_n(\mu)^{\Kk}}\;=\;
\frac{M}{(L\!-\!1)^2}\cdot\Big(\Hat{\Vec{C}}_n(\mu)\cdot
\Hat{\Vec{C}}_n(\mu)^{\Tt}\Big)^{\!\Kk}\cdot
\Hat{\Psi}_{\boldsymbol{n}}(\mu,-\mu)\;=&&
\end{flalign}\\*[6pt]\begin{flalign*}
&&=\;\frac{M}{L\!-\!1}\cdot
\Vec{E}_n\cdot
\big(\Hat{\underline{C}}_{\Tilde{\Vec{\boldsymbol{V}}}(\mu),\Tilde{\Vec{\boldsymbol{V}}}(\mu)}^{\uP{0.4}{\!-1}}\big)^{\!\Kk}\!\CdoT
\Hat{\underline{C}}_{\Tilde{\Vec{\boldsymbol{V}}}(\mu)^{\Kk},\Tilde{\Vec{\boldsymbol{V}}}(\mu)}\CdoT
\Hat{\underline{C}}_{\Tilde{\Vec{\boldsymbol{V}}}(\mu),\Tilde{\Vec{\boldsymbol{V}}}(\mu)}^{\uP{0.4}{\!-1}}\cdot
\Vec{E}_n^{\,\Hh}\cdot
\Hat{\Psi}_{\boldsymbol{n}}(\mu,-\mu)&
\end{flalign*}\\*[6pt]
{\T\qquad\forall\qquad\mu=\Breve{\mu}\CdoT\frac{M}{2\cdot K_{\Phi}}\quad\wedge\quad
\Breve{\mu}\in\mathbb{Z}}
\end{gather*}
\end{subequations}
absch"atzen. Setzt man hier
die Messwerte \mbox{$\Hat{\Phi}_{\boldsymbol{n}}(\mu,\mu)$}
und \mbox{$\Hat{\Psi}_{\boldsymbol{n}}(\mu,-\mu)$} gem"a"s der
Gleichungen~(\ref{E.3.29}) in der Form ein, die die Vektoren
\mbox{$\Vec{N}_{\!f}(\ldots)$} enth"alt, betrachtet man die
Sch"atzwerte wieder als zuf"allig, und bildet man deren Erwartungswerte mit Anhang \ref{E.Kap.A.5},
gehen wieder nur die Hauptdiagonalelemente der quadratischen bzw.
bilinearen Formen additiv in die Erwartungswerte ein. In dem zu
bildenden Erwartungswert, der nur von den Spektralwerten der
Erregung abh"angt, l"asst sich daher jeweils die Spur der Matrix,
auf die der jeweilige Messwert normiert ist, k"urzen, so dass
derselbe vom Spektrum der Erregung abh"angige Erwartungswert
verbleibt, wie bei den theoretischen Messwert"-(ko)"-varianzen. 
Der bei der Berechnung der Erwartungswerte der Sch"atzwerte auftretende,
von dem Spektrum des gefensterten Approximationsfehlerprozesses
abh"angige Erwartungswert ist identisch mit dem Erwartungswert,
der in den theoretischen Messwert"-(ko)"-varianzen zu finden ist.
Die Sch"atzwerte der Messwert"-(ko)"-varianzen sind daher erwartungstreu. 
Wenn man Messwerte \mbox{$\Hat{\Phi}_{\boldsymbol{n}}(\mu,\mu)$}
und \mbox{$\Hat{\Psi}_{\boldsymbol{n}}(\mu,-\mu)$} verwendet, die mit
Matrizen \mbox{$\underline{V}_{\bot}\!(\mu)$} gewonnen wurden,
die die Bedingungen~(\ref{E.3.33}) erf"ullen, l"asst sich zeigen, 
dass die konkreten Varianzsch"atzwerte niemals kleiner als die 
Betr"age der entsprechenden konkreten Kovarianzsch"atzwerte sind. 
Dabei ber"ucksichtigt man, dass zum einen die Messwerte 
\mbox{$\Hat{\Phi}_{\boldsymbol{n}}(\mu,\mu)$} und 
\mbox{$\Hat{\Psi}_{\boldsymbol{n}}(\mu,-\mu)$} 
die Ungleichung~(\ref{E.3.36.b}) erf"ullen (\,man setzt dort 
\mbox{$\mu=\Breve{\mu}\CdoT M/2/K_{\Phi}$} und 
\mbox{$\Tilde{\mu}\!=\!-\Breve{\mu}$} ein\,), und dass zum anderen der Betrag 
\mbox{$\Big|\Hat{\Vec{C}}_n(\mu)\cdot\Hat{\Vec{C}}_n(\mu)^{\Tt}\Big|$}
eines Skalarprodukts niemals gr"o"ser ist, als das Produkt der
euklidischen Normen der beiden daran beteiligten Vektoren
(\,Cauchy-Schwarzsche Ungleichung\,),
das in diesem Fall gleich dem Skalarprodukt
\mbox{$\Hat{\Vec{C}}_n(\mu)\cdot\Hat{\Vec{C}}_n(\mu)^{\Hh}$}
ist.

Auch die Messwerte \mbox{$\Hat{\boldsymbol{U}}_{\!\!f}(\mu)$} betrachten wir
nun wieder als Zufallsgr"o"sen, die aus mathematischen Stichproben 
der Zufallsprozesse am Systemein- und -ausgang entstanden sind.
Die Varianzen und die Kovarianzen der Messwerte \mbox{$\Hat{\boldsymbol{U}}_{\!\!f}(\mu)$} 
berechnen sich mit Gleichung~(\ref{E.3.20}) als Erwartungswerte quadratischer bzw. bilinearer 
Formen, die man mit der in Anhang \ref{E.Kap.A.5} beschriebenen Methode umformt. 
In weiser Voraussicht berechnen wir nicht nur die Varianzen und
Kovarianzen der Messwerte, sondern auch gleich alle Kovarianzen zweier 
Messwerte unterschiedlicher Frequenzen $\mu_1$ und $\mu_2$ als Erwartungswerte 
bilinearer Formen. Die Messwert"-(ko)"-varianzen ergeben sich dann mit 
\mbox{$\mu\!=\!\mu_1\!=\!\mu_2$}. Die Berechnung erfolgt in "ahnlicher Weise 
wie in den Gleichungen \myref{3.30}. Wir erhalten:
\begin{subequations}\label{E.3.45}
\begin{gather}
C_{\Hat{\boldsymbol{U}}_{\!\!f}(\mu_1),\Hat{\boldsymbol{U}}_{\!\!f}(\mu_2)}\;=\;
\text{E}\bigg\{\Big(\Hat{\boldsymbol{U}}_{\!\!f}(\mu_1)\!-\!
\text{E}\big\{\Hat{\boldsymbol{U}}_{\!\!f}(\mu_1)\big\}\Big)\CdoT
\Big(\Hat{\boldsymbol{U}}_{\!\!f}(\mu_2)\!-\!
\text{E}\big\{\Hat{\boldsymbol{U}}_{\!\!f}(\mu_2)\big\}\Big)^{\HH}\bigg\}\;={}
\label{E.3.45.a}\\[8pt]
{}=\;\text{E}\bigg\{\Big(\Hat{\boldsymbol{U}}_{\!\!f}(\mu_1)\!-\!
U_{\!f}(\mu_1)\Big)\CdoT\Big(\Hat{\boldsymbol{U}}_{\!\!f}(\mu_2)\!-\!
U_{\!f}(\mu_2)\Big)^{\HH}\bigg\}\;={}
\notag\\[8pt]\begin{flalign*}
{}=\;{}&\text{E}\Bigg\{\Vec{\boldsymbol{N}}_{\!\!f}(\mu_1)\cdot
\frac{\D(L\!-\!1)\CdoT\underline{E}-
\underline{1}_{\bot}\!\CdoT
\Tilde{\underline{\boldsymbol{V}}}(\mu_1)^{\Hh}\!\Cdot
\Hat{\underline{\boldsymbol{C}}}_{\Tilde{\Vec{\boldsymbol{V}}}(\mu_1),\Tilde{\Vec{\boldsymbol{V}}}(\mu_1)}^{\uP{0.4}{\!-1}}\CdoT
\Tilde{\underline{\boldsymbol{V}}}(\mu_1)}
{\D L\cdot(\,L-1\,)}\cdot\Vec{1}^{\,\Hh}\cdot{}&&
\end{flalign*}\notag\\*[4pt]\begin{flalign*}
&&{}\cdot\Vec{1}\cdot\frac{\D(L\!-\!1)\CdoT\underline{E}-
\Tilde{\underline{\boldsymbol{V}}}(\mu_2)^{\Hh}\!\Cdot
\Hat{\underline{\boldsymbol{C}}}_{\Tilde{\Vec{\boldsymbol{V}}}(\mu_2),\Tilde{\Vec{\boldsymbol{V}}}(\mu_2)}^{\uP{0.4}{\!-1}}\CdoT
\Tilde{\underline{\boldsymbol{V}}}(\mu_2)\CdoT\underline{1}_{\bot}}
{\D L\cdot(\,L-1\,)}\cdot\Vec{\boldsymbol{N}}_{\!\!f}(\mu_2)^{\Hh}\Bigg\}\;&={}
\end{flalign*}\notag\\[8pt]\begin{flalign*}
{}=\;\text{E}\bigg\{\text{spur}\bigg(\frac{1}{L^2\CdoT(L\!-\!1)^2}\cdot{}&
\Big((L\!-\!1)\CdoT\underline{E}-
\underline{1}_{\bot}\Cdot\Tilde{\underline{\boldsymbol{V}}}(\mu_1)^{\Hh}\Cdot
\Hat{\underline{\boldsymbol{C}}}_{\Tilde{\Vec{\boldsymbol{V}}}(\mu_1),\Tilde{\Vec{\boldsymbol{V}}}(\mu_1)}^{\uP{0.4}{\!-1}}\CdoT
\Tilde{\underline{\boldsymbol{V}}}(\mu_1)\Big)
\cdot\Vec{1}^{\,\Hh}\Cdot{}&&\\*[0pt]
{}\cdot\Vec{1}\cdot{}&\Big((L\!-\!1)\CdoT\underline{E}-
\Tilde{\underline{\boldsymbol{V}}}(\mu_2)^{\Hh}\!\Cdot
\Hat{\underline{\boldsymbol{C}}}_{\Tilde{\Vec{\boldsymbol{V}}}(\mu_2),\Tilde{\Vec{\boldsymbol{V}}}(\mu_2)}^{\uP{0.4}{\!-1}}\CdoT
\Tilde{\underline{\boldsymbol{V}}}(\mu_2)\cdot
\underline{1}_{\bot}\Big)\bigg)\bigg\}\cdot{}&&
\end{flalign*}\notag\\*[2pt]\begin{flalign*}
&&{}\cdot\Big(\text{E}\big\{\boldsymbol{N}_{\!\!f}(\mu_1)\CdoT
\boldsymbol{N}_{\!\!f}(\mu_2)^{\Kk}\big\}\!-\!
\text{E}\big\{\boldsymbol{N}_{\!\!f}(\mu_1)\big\}\CdoT
\text{E}\big\{\boldsymbol{N}_{\!\!f}(\mu_2)^{\Kk}\big\}\Big)+
\text{E}\big\{\boldsymbol{N}_{\!\!f}(\mu_1)\big\}\CdoT
\text{E}\big\{\boldsymbol{N}_{\!\!f}(\mu_2)^{\Kk}\big\}\;&={}
\end{flalign*}\notag\\[10pt]
=\;\frac{\text{E}\Big\{\Hat{\Vec{\boldsymbol{C}}}_U(\mu_2)\CdoT
\Hat{\Vec{\boldsymbol{C}}}_U(\mu_1)^{\Hh}\Big\}}
{L^2\CdoT(L\!-\!1)^2}\cdot
\text{E}\big\{\boldsymbol{N}_{\!\!f}(\mu_1)\CdoT
\boldsymbol{N}_{\!\!f}(\mu_2)^{\Kk}\big\}\;=
\notag\\[10pt]\begin{flalign*}
{}=\;{}&\bigg(\,\frac{1}{L}+\frac{1}{L^2\CdoT(L\!-\!1)^2}\cdot
\text{E}\Big\{\,\Vec{1}\CdoT
\Tilde{\underline{\boldsymbol{V}}}(\mu_2)^{\Hh}\!\Cdot
\Hat{\underline{\boldsymbol{C}}}_{\Tilde{\Vec{\boldsymbol{V}}}(\mu_2),\Tilde{\Vec{\boldsymbol{V}}}(\mu_2)}^{\uP{0.4}{\!-1}}\CdoT
\Tilde{\underline{\boldsymbol{V}}}(\mu_2)\CdoT
\underline{1}_{\bot}\CdoT{}&&
\end{flalign*}\notag\\*[-2pt]\begin{flalign*}
&&{}\cdot\underline{1}_{\bot}\!\CdoT
\Tilde{\underline{\boldsymbol{V}}}(\mu_1)^{\Hh}\!\Cdot
\Hat{\underline{\boldsymbol{C}}}_{\Tilde{\Vec{\boldsymbol{V}}}(\mu_1),\Tilde{\Vec{\boldsymbol{V}}}(\mu_1)}^{\uP{0.4}{\!-1}}\CdoT
\Tilde{\underline{\boldsymbol{V}}}(\mu_1)\cdoT
\Vec{1}^{\,\Hh}\Big\}\bigg)\cdot
\text{E}\Big\{\boldsymbol{N}_{\!\!f}(\mu_1)\CdoT
\boldsymbol{N}_{\!\!f}(\mu_2)^{\Kk}\Big\}&
\end{flalign*}\notag\\[-6pt]\intertext{und\vspace{-6pt}}
C_{\Hat{\boldsymbol{U}}_{\!\!f}(\mu_1),\Hat{\boldsymbol{U}}_{\!\!f}(\mu_2)^{\Kk}}\;=\;
\text{E}\bigg\{\Big(\Hat{\boldsymbol{U}}_{\!\!f}(\mu_1)\!-\!
\text{E}\big\{\Hat{\boldsymbol{U}}_{\!\!f}(\mu_1)\big\}\Big)\CdoT
\Big(\Hat{\boldsymbol{U}}_{\!\!f}(\mu_2)\!-\!
\text{E}\big\{\Hat{\boldsymbol{U}}_{\!\!f}(\mu_2)\big\}\Big)\bigg\}\;={}
\label{E.3.45.b}\\[8pt]
{}=\;\text{E}\bigg\{\Big(\Hat{\boldsymbol{U}}_{\!\!f}(\mu_1)\!-\!
U_{\!f}(\mu_1)\Big)\CdoT\Big(\Hat{\boldsymbol{U}}_{\!\!f}(\mu_2)\!-\!
U_{\!f}(\mu_2)\Big)\bigg\}\;={}
\notag\\[8pt]\begin{flalign*}
{}=\;{}&\text{E}\Bigg\{\Vec{\boldsymbol{N}}_{\!\!f}(\mu_1)\cdot
\frac{\D(L\!-\!1)\CdoT\underline{E}-
\underline{1}_{\bot}\!\CdoT
\Tilde{\underline{\boldsymbol{V}}}(\mu_1)^{\Hh}\!\Cdot
\Hat{\underline{\boldsymbol{C}}}_{\Tilde{\Vec{\boldsymbol{V}}}(\mu_1),\Tilde{\Vec{\boldsymbol{V}}}(\mu_1)}^{\uP{0.4}{\!-1}}\CdoT
\Tilde{\underline{\boldsymbol{V}}}(\mu_1)}
{\D L\cdot(\,L-1\,)}\cdot\Vec{1}^{\,\Hh}\cdot{}&&
\end{flalign*}\notag\\*[4pt]\begin{flalign*}
&&{}\cdot\Vec{1}\cdot\frac{\D(L\!-\!1)\CdoT\underline{E}-
\Tilde{\underline{\boldsymbol{V}}}(\mu_2)^{\Tt}\!\Cdot
\big(\Hat{\underline{\boldsymbol{C}}}_{\Tilde{\Vec{\boldsymbol{V}}}(\mu_2),\Tilde{\Vec{\boldsymbol{V}}}(\mu_2)}^{\uP{0.4}{\!-1}}\big)^{\!\Kk}\!\CdoT
\Tilde{\underline{\boldsymbol{V}}}(\mu_2)^{\Kk}\CdoT\underline{1}_{\bot}}
{\D L\cdot(\,L-1\,)}\cdot\Vec{\boldsymbol{N}}_{\!\!f}(\mu_2)^{\Tt}\Bigg\}\;&={}
\end{flalign*}\notag\\[8pt]\begin{flalign*}
{}=\;\text{E}\bigg\{\text{spur}\bigg(\frac{1}{L^2\CdoT(L\!-\!1)^2}\cdot{}&
\Big((L\!-\!1)\CdoT\underline{E}-
\underline{1}_{\bot}\Cdot\Tilde{\underline{\boldsymbol{V}}}(\mu_1)^{\Hh}\Cdot
\Hat{\underline{\boldsymbol{C}}}_{\Tilde{\Vec{\boldsymbol{V}}}(\mu_1),\Tilde{\Vec{\boldsymbol{V}}}(\mu_1)}^{\uP{0.4}{\!-1}}\CdoT
\Tilde{\underline{\boldsymbol{V}}}(\mu_1)\Big)
\cdot\Vec{1}^{\,\Hh}\Cdot{}&&\\*[0pt]
{}\cdot\Vec{1}\cdot{}&\Big((L\!-\!1)\CdoT\underline{E}-
\Tilde{\underline{\boldsymbol{V}}}(\mu_2)^{\Tt}\!\Cdot
\big(\Hat{\underline{\boldsymbol{C}}}_{\Tilde{\Vec{\boldsymbol{V}}}(\mu_2),\Tilde{\Vec{\boldsymbol{V}}}(\mu_2)}^{\uP{0.4}{\!-1}}\big)^{\!\Kk}\!\CdoT
\Tilde{\underline{\boldsymbol{V}}}(\mu_2)^{\Kk}\cdot
\underline{1}_{\bot}\Big)\bigg)\bigg\}\cdot{}&&
\end{flalign*}\notag\\*[4pt]\begin{flalign*}
&&{}\cdot\Big(\text{E}\big\{\boldsymbol{N}_{\!\!f}(\mu_1)\CdoT
\boldsymbol{N}_{\!\!f}(\mu_2)\big\}\!-\!
\text{E}\big\{\boldsymbol{N}_{\!\!f}(\mu_1)\big\}\CdoT
\text{E}\big\{\boldsymbol{N}_{\!\!f}(\mu_2)\big\}\Big)+
\text{E}\big\{\boldsymbol{N}_{\!\!f}(\mu_1)\big\}\CdoT
\text{E}\big\{\boldsymbol{N}_{\!\!f}(\mu_2)\big\}\;&={}
\end{flalign*}\notag\\[10pt]
=\;\frac{\text{E}\Big\{\Hat{\Vec{\boldsymbol{C}}}_U(\mu_2)\CdoT
\Hat{\Vec{\boldsymbol{C}}}_U(\mu_1)^{\Tt}\Big\}^{\!\Kk}}
{(L\!-\!1)^2}\cdot
\text{E}\big\{\boldsymbol{N}_{\!\!f}(\mu_1)\CdoT
\boldsymbol{N}_{\!\!f}(\mu_2)\big\}\;=
\notag\\[10pt]\begin{flalign*}
{}=\;{}&\bigg(\,\frac{1}{L}+\frac{1}{L^2\CdoT(L\!-\!1)^2}\cdot
\text{E}\Big\{\,\Vec{1}\CdoT
\Tilde{\underline{\boldsymbol{V}}}(\mu_2)^{\Tt}\!\Cdot
\big(\Hat{\underline{\boldsymbol{C}}}_{\Tilde{\Vec{\boldsymbol{V}}}(\mu_2),\Tilde{\Vec{\boldsymbol{V}}}(\mu_2)}^{\uP{0.4}{\!-1}}\big)^{\!\Kk}\!\CdoT
\Tilde{\underline{\boldsymbol{V}}}(\mu_2)^{\Kk}\CdoT
\underline{1}_{\bot}\CdoT{}&&
\end{flalign*}\notag\\*[0pt]\begin{flalign*}
&&{}\cdot\underline{1}_{\bot}\!\CdoT
\Tilde{\underline{\boldsymbol{V}}}(\mu_1)^{\Hh}\!\Cdot
\Hat{\underline{\boldsymbol{C}}}_{\Tilde{\Vec{\boldsymbol{V}}}(\mu_1),\Tilde{\Vec{\boldsymbol{V}}}(\mu_1)}^{\uP{0.4}{\!-1}}\CdoT
\Tilde{\underline{\boldsymbol{V}}}(\mu_1)\cdoT
\Vec{1}^{\,\Hh}\Big\}\bigg)\cdot
\text{E}\Big\{\boldsymbol{N}_{\!\!f}(\mu_1)\CdoT
\boldsymbol{N}_{\!\!f}(\mu_2)\Big\}.&
\end{flalign*}
\notag
\end{gather}
\end{subequations}
Dabei wurde jeweils ber"ucksichtigt, dass die optimale theoretische
Anpassung des Spektrums der deterministischen St"orung nach Gleichung~(\ref{E.2.29})
dazu f"uhrte, dass der Erwartungswert der Zufallsgr"o"se
\mbox{$\boldsymbol{N}_{\!\!f}(\mu)$} nach Gleichung~(\ref{E.2.30}) null ist. 
Wieder wurde in einem Zwischenschritt ein Zeilenvektor
\begin{equation}
\Hat{\Vec{\boldsymbol{C}}}_U(\mu)\;=\;
\,\frac{L\!-\!1}{L}\cdot\Vec{1}\;-\;
\frac{\Vec{1}\cdot\Tilde{\underline{\boldsymbol{V}}}(\mu)^{\Hh}}{L}\cdot
\Hat{\underline{\boldsymbol{C}}}_{\Tilde{\Vec{\boldsymbol{V}}}(\mu),\Tilde{\Vec{\boldsymbol{V}}}(\mu)}^{\uP{0.4}{\!-1}}\cdot
\Tilde{\underline{\boldsymbol{V}}}(\mu)\cdot\underline{1}_{\bot}
\label{E.3.46}
\end{equation}
eingef"uhrt, der zur Absch"atzung der Betr"age der empirischen Varianz und 
Kovarianz des Messwertes \mbox{$\Hat{\boldsymbol{U}}_{\!\!f}(\mu)$} gebraucht wird. 
Wenn man eine hoch frequenzselektive Fensterfolge verwendet, sind die Kovarianzen 
\mbox{$\text{E}\big\{\boldsymbol{N}_{\!\!f}(\mu_1)\CdoT\boldsymbol{N}_{\!\!f}(\mu_2)^{\Kk}\big\}$}, 
die in Gleichung~(\ref{E.3.45.a}) auftreten, gem"a"s Gleichung~(\ref{E.2.41}) f"ur die Frequenzen 
mit \mbox{$\mu_2\!\neq\!\mu_1\!+\!\Tilde{\mu}\CdoT M/K_{\Phi}$} gegen"uber den Kovarianzen mit 
\mbox{$\mu_2\!=\!\mu_1\!+\!\Tilde{\mu}\CdoT M/K_{\Phi}$} vernachl"assigbar klein. 
Entsprechendes gilt nach Gleichung~(\ref{E.2.51}) f"ur die Kovarianzen 
\mbox{$\text{E}\big\{\boldsymbol{N}_{\!\!f}(\mu_1)\CdoT\boldsymbol{N}_{\!\!f}(\mu_2)\big\}$}
in Gleichung~(\ref{E.3.45.b}) f"ur die Frequenzen mit 
\mbox{$\mu_2\neq-\mu_1\!-\!\Tilde{\mu}\CdoT M/K_{\Phi}$}. F"ur die
nennenswert von null verschiedenen Messwert"-(ko)"-varianzen erhalten wir:
\begin{subequations}\label{E.3.47}
\begin{gather}
C_{\Hat{\boldsymbol{U}}_{\!\!f}(\mu),\Hat{\boldsymbol{U}}_{\!\!f}(\mu+\Tilde{\mu}\cdot\frac{M}{K_{\Phi}})}\;=\;
M\cdot\frac{\text{E}\Big\{
\Hat{\Vec{\boldsymbol{C}}}_U\big({\T\mu\!+\!\Tilde{\mu}\CdoT\frac{M}{K_{\Phi}}}\big)\CdoT
\Hat{\Vec{\boldsymbol{C}}}_U(\mu)^{\Hh}\Big\}}{(L\!-\!1)^2}\cdot
\Tilde{\Phi}_{\boldsymbol{n}}
\big({\T\mu,\mu\!+\!\Tilde{\mu}\CdoT\frac{M}{K_{\Phi}}}\big)\;=
\notag\\[4pt]\label{E.3.47.a}\begin{flalign}
{}=\;{}&\frac{M}{L}\cdot\Bigg(1+\frac{L}{L\!-\!1}\cdot\text{E}\bigg\{
\frac{\Vec{1}\CdoT\Tilde{\underline{\boldsymbol{V}}}\big({\T\mu\!+\!\Tilde{\mu}\CdoT\frac{M}{K_{\Phi}}}\big)^{\HH}\!}{L}\CdoT
\Hat{\underline{\boldsymbol{C}}}_{\Tilde{\Vec{\boldsymbol{V}}}(\mu+\Tilde{\mu}\cdot\frac{M}{K_{\Phi}}),\Tilde{\Vec{\boldsymbol{V}}}(\mu+\Tilde{\mu}\cdot\frac{M}{K_{\Phi}})}^{\uP{0.4}{\!-1}}\cdot{}&&
\end{flalign}\notag\\*[2pt]\begin{flalign*}
&&{}\Cdot\Hat{\underline{\boldsymbol{C}}}_{\Tilde{\Vec{\boldsymbol{V}}}(\mu+\Tilde{\mu}\cdot\frac{M}{K_{\Phi}}),\Tilde{\Vec{\boldsymbol{V}}}(\mu)}\CdoT
\Hat{\underline{\boldsymbol{C}}}_{\Tilde{\Vec{\boldsymbol{V}}}(\mu),\Tilde{\Vec{\boldsymbol{V}}}(\mu)}^{\uP{0.4}{\!-1}}\CdoT
\frac{\Tilde{\underline{\boldsymbol{V}}}(\mu)\cdoT\Vec{1}^{\,\Hh}}{L}\bigg\}\!\Bigg)\cdot
\Tilde{\Phi}_{\boldsymbol{n}}\big({\T\mu,\mu\!+\!\Tilde{\mu}\CdoT\frac{M}{K_{\Phi}}}\big)&
\end{flalign*}\notag\\[-10pt]\intertext{und\vspace{-6pt}}
C_{\Hat{\boldsymbol{U}}_{\!\!f}(\mu),\Hat{\boldsymbol{U}}_{\!\!f}(-\mu-\Tilde{\mu}\cdot\frac{M}{K_{\Phi}})^{\Kk}}\,=\;
M\cdot\frac{\text{E}\Big\{
\Hat{\Vec{\boldsymbol{C}}}_U\big(\!{\T-\mu\!-\!\Tilde{\mu}\CdoT\frac{M}{K_{\Phi}}}\big)\CdoT
\Hat{\Vec{\boldsymbol{C}}}_U(\mu)^{\Tt}\Big\}^{\!\Kk}\!}
{L^2\Cdot(L\!-\!1)^2}\cdot
\Tilde{\Psi}_{\boldsymbol{n}}\big({\T\mu,\mu\!+\!\Tilde{\mu}\CdoT\frac{M}{K_{\Phi}}}\big)\,=
\notag\\[4pt]\label{E.3.47.b}\begin{flalign}
{}=\;{}&\frac{M}{L}\cdot\Bigg(1+\frac{L}{L\!-\!1}\cdot\text{E}\bigg\{
\frac{\Vec{1}\CdoT\Tilde{\underline{\boldsymbol{V}}}\big({\T\!-\mu\!-\!\Tilde{\mu}\CdoT\frac{M}{K_{\Phi}}}\big)^{\TT}\!}{L}\CdoT
\big(\Hat{\underline{\boldsymbol{C}}}_{\Tilde{\Vec{\boldsymbol{V}}}(-\mu-\Tilde{\mu}\cdot\frac{M}{K_{\Phi}}),\Tilde{\Vec{\boldsymbol{V}}}(-\mu-\Tilde{\mu}\cdot\frac{M}{K_{\Phi}})}^{\uP{0.4}{\!-1}}\big)^{\!\Kk}\Cdot{}&&
\end{flalign}\notag\\*[2pt]\begin{flalign*}
&&{}\Cdot\Hat{\underline{\boldsymbol{C}}}_{\Tilde{\Vec{\boldsymbol{V}}}(-\mu-\Tilde{\mu}\cdot\frac{M}{K_{\Phi}})^{\Kk},\Tilde{\Vec{\boldsymbol{V}}}(\mu)}\CdoT
\Hat{\underline{\boldsymbol{C}}}_{\Tilde{\Vec{\boldsymbol{V}}}(\mu),\Tilde{\Vec{\boldsymbol{V}}}(\mu)}^{\uP{0.4}{\!-1}}\CdoT
\frac{\Tilde{\underline{\boldsymbol{V}}}(\mu)\cdoT\Vec{1}^{\,\Hh}}{L}\bigg\}\!\Bigg)\cdot
\Tilde{\Psi}_{\boldsymbol{n}}\big({\T\mu,\mu\!+\!\Tilde{\mu}\CdoT\frac{M}{K_{\Phi}}}\big)&
\end{flalign*}\notag
\end{gather}
\end{subequations}
Wie man sieht ist die Messwertkovarianz \mbox{$C_{\Hat{\boldsymbol{U}}_{\!\!f}(\mu),\Hat{\boldsymbol{U}}_{\!\!f}(\mu)^{\Kk}}$}
nur dann nennenswert von null verschieden, wenn $\mu$ ein ganzzahliges Vielfaches von
\mbox{$\frac{M}{2\cdot K_{\Phi}}$} ist, weil es nur dann ein ganzzahliges $\!\Tilde{\mu}$ gibt, 
bei dem \mbox{$\mu=-\mu\!-\!\Tilde{\mu}\cdot\frac{M}{K_{\Phi}}$} gilt.
Wenn man den Fall ausschlie"st, dass die empirischen Kovarianzmatrizen des Spektrums der 
Erregung singul"ar werden\footnote{Siehe Messwertvarianz der "Ubertragungsfunktionen}, 
existieren die von der Erregung abh"angigen Erwartungswerte, die in den letzten Gleichungen  
als Vorfaktoren auftreten, da es sich dabei um zwei Vektoren von empirischen Mittelwerten 
handelt, die gemeinsam mit einem Produkt empirischer, teils inverser,  Kovarianzmatrizen 
eine bilineare Form bilden. Von diesen Vorfaktoren kann man erwarten, dass sie f"ur steigende 
Werte von $L$ gegen einen festen endlichen Wert gehen. Die Varianz der Messwerte sinkt dann 
asymptotisch mit \mbox{$1/L$}, so dass die Messwerte konsistent sind. Die letzten Gleichungen 
zeigen auch, dass sich die Messwert"-(ko)"-varianzen hier aus jeweils zwei Anteilen zusammensetzen. 
Der eine Anteil ist der Summand mit dem konstanten Vorfaktor $M/L$, der durch den 
Approximationsfehlers selbst verursacht wird, w"ahrend der zweite Anteil, dessen Vorfaktor 
der von der Erregung abh"angige Erwartungswert ist, durch die verrauschte Messung der empirischen 
Mittelwerte der Erregung, die mit den beiden linearen Modellsystemen gefiltert werden, verursacht wird. 
Mit \mbox{$\Tilde{\mu}\!=\!0$} erh"alt man in Gleichung~(\ref{E.3.47.a}) die Messwertvarianz. 
Da in diesem Fall in dem von der Erregung abh"angigen Erwartungswert eine positiv semidefinite 
quadratische Form steht, wird die Messwertvarianz minimal, wenn man einen erregenden Prozess f"ur 
die Messung verwendet, die keinen zeitabh"angigen Mittelwert aufweist. Einen mittelwertfreien 
erregenden Prozess f"ur die Messung zu verwenden ist nat"urlich nur dann m"oglich, wenn man sicher 
sein kann, dass sich dadurch die f"ur den Betriebszustand typischen, optimalen, theoretischen 
Regressionskoeffizienten nicht "andern.

Die theoretischen Messwertkovarianzen lassen sich mit
\begin{subequations}\label{E.3.48}
\begin{gather}
\Hat{C}_{\Hat{\boldsymbol{U}}_{\!\!f}(\mu),\Hat{\boldsymbol{U}}_{\!\!f}(\mu+\Tilde{\mu}\cdot\frac{M}{K_{\Phi}})}\;=\;
\frac{M}{(L\!-\!1)^2}\cdot
\Hat{\Vec{C}}_U\big({\T\mu\!+\!\Tilde{\mu}\CdoT\frac{M}{K_{\Phi}}}\big)\CdoT
\Hat{\Vec{C}}_U(\mu)^{\Hh}\Cdot
\Hat{\Phi}_{\boldsymbol{n}}\big({\T\mu,\mu\!+\!\Tilde{\mu}\CdoT\frac{M}{K_{\Phi}}}\big)\;=
\notag\\[7pt]\label{E.3.48.a}\begin{flalign}
{}=\;{}&\frac{M}{L}\cdot\Bigg(1+\frac{L}{L\!-\!1}\cdot
\frac{\Vec{1}\CdoT\Tilde{\underline{V}}\big({\T\mu\!+\!\Tilde{\mu}\CdoT\frac{M}{K_{\Phi}}}\big)^{\HH}\!}{L}\CdoT
\Hat{\underline{C}}_{\Tilde{\Vec{\boldsymbol{V}}}(\mu+\Tilde{\mu}\cdot\frac{M}{K_{\Phi}}),\Tilde{\Vec{\boldsymbol{V}}}(\mu+\Tilde{\mu}\cdot\frac{M}{K_{\Phi}})}^{\uP{0.4}{\!-1}}\cdoT{}&&
\end{flalign}\notag\\*[0pt]\begin{flalign*}
&&{}\Cdot\Hat{\underline{C}}_{\Tilde{\Vec{\boldsymbol{V}}}(\mu+\Tilde{\mu}\cdot\frac{M}{K_{\Phi}}),\Tilde{\Vec{\boldsymbol{V}}}(\mu)}\CdoT
\Hat{\underline{C}}_{\Tilde{\Vec{\boldsymbol{V}}}(\mu),\Tilde{\Vec{\boldsymbol{V}}}(\mu)}^{\uP{0.4}{\!-1}}\CdoT
\frac{\Tilde{\underline{V}}(\mu)\cdoT\Vec{1}^{\,\Hh}}{L}\!\Bigg)\cdot
\Hat{\Phi}_{\boldsymbol{n}}\big({\T\mu,\mu\!+\!\Tilde{\mu}\CdoT\frac{M}{K_{\Phi}}}\big)&
\end{flalign*}\notag\\*[4pt]
\forall\qquad\qquad\mu=0\;(1)\;M\!-\!1
\qquad\text{ und }\qquad\Tilde{\mu}=0\;(1)\;K_{\Phi}\!-\!1\qquad{}
\notag\\[-10pt]\intertext{und\vspace{-6pt}}
\Hat{C}_{\Hat{\boldsymbol{U}}_{\!\!f}(\mu),\Hat{\boldsymbol{U}}_{\!\!f}(-\mu-\Tilde{\mu}\cdot\frac{M}{K_{\Phi}})^{\Kk}}\,=\;
\frac{M}{(L\!-\!1)^2}\cdot
\Hat{\Vec{C}}_U\big({\T\!-\mu\!-\!\Tilde{\mu}\CdoT\frac{M}{K_{\Phi}}}\big)^{\!\Kk}\!\CdoT
\Hat{\Vec{C}}_U(\mu)^{\Hh}\!\CdoT
\Hat{\Psi}_{\boldsymbol{n}}\big({\T\mu,\mu\!+\!\Tilde{\mu}\CdoT\frac{M}{K_{\Phi}}}\big)\,=
\notag\\[7pt]\label{E.3.48.b}\begin{flalign}
{}=\;{}&\frac{M}{L}\cdot\Bigg(1+\frac{L}{L\!-\!1}\cdot
\frac{\Vec{1}\CdoT\Tilde{\underline{V}}\big({\T\!-\mu\!-\!\Tilde{\mu}\CdoT\frac{M}{K_{\Phi}}}\big)^{\TT}\!}{L}\CdoT
\big(\Hat{\underline{C}}_{\Tilde{\Vec{\boldsymbol{V}}}(-\mu-\Tilde{\mu}\cdot\frac{M}{K_{\Phi}}),\Tilde{\Vec{\boldsymbol{V}}}(-\mu-\Tilde{\mu}\cdot\frac{M}{K_{\Phi}})}^{\uP{0.4}{\!-1}}\big)^{\!\Kk}\CdoT{}&&
\end{flalign}\notag\\*[0pt]\begin{flalign*}
&&{}\Cdot\Hat{\underline{C}}_{\Tilde{\Vec{\boldsymbol{V}}}(-\mu-\Tilde{\mu}\cdot\frac{M}{K_{\Phi}})^{\Kk},\Tilde{\Vec{\boldsymbol{V}}}(\mu)}\CdoT
\Hat{\underline{C}}_{\Tilde{\Vec{\boldsymbol{V}}}(\mu),\Tilde{\Vec{\boldsymbol{V}}}(\mu)}^{\uP{0.4}{\!-1}}\CdoT
\frac{\Tilde{\underline{V}}(\mu)\cdoT\Vec{1}^{\,\Hh}}{L}\!\Bigg)\cdot
\Hat{\Psi}_{\boldsymbol{n}}\big({\T\mu,\mu\!+\!\Tilde{\mu}\CdoT\frac{M}{K_{\Phi}}}\big)&
\end{flalign*}\notag\\*[4pt]
\forall\qquad\qquad\mu=0\;(1)\;M\!-\!1
\qquad\text{ und }\qquad\Tilde{\mu}=0\;(1)\;K_{\Phi}\!-\!1\qquad{}
\notag
\end{gather}
\end{subequations}
absch"atzen. Dass diese Sch"atzwerte erwartungstreu sind, zeigt man genauso
wie bei den Sch"atzwerten der Varianzen und Kovarianzen der Messwerte der
"Ubertragungsfunktion mit der im Anhang \ref{E.Kap.A.5} beschriebenen Methode.

Wie gesagt, erhalten wir mit \mbox{$\mu_1\!=\!\mu_2$} die
ben"otigten Messwert"-(ko)"-varianzen. Mit den neuen
Frequenzvariablen $\mu$ und $\Tilde{\mu}$ bedeutet das bei den
Messwertvarianzen, dass man \mbox{$\Tilde{\mu}\!=\!0$} einsetzt,
und dass man f"ur alle \mbox{$\mu=0\;(1)\;M\!-\!1$} von null verschiedene
Messwertvarianzen erhalten kann. Bei den Messwertkovarianzen kann man
lediglich f"ur \mbox{$\mu=\Breve{\mu}\CdoT M/(2\CdoT K_{\Phi})$} mit
\mbox{$\Breve{\mu}=0\;(1)\;2\CdoT K_{\Phi}\!-\!1$} nennenswert von null
verschiedene Werte erhalten, und in der letzten Gleichung ist 
\mbox{$\Tilde{\mu}\!=\!-\Breve{\mu}$} einzusetzen. 

Wenn man Messwerte
\mbox{$\Hat{\Phi}_{\boldsymbol{n}}(\mu,\mu)$} und
\mbox{$\Hat{\Psi}_{\boldsymbol{n}}(\mu,-\mu)$} verwendet, die mit
Matrizen \mbox{$\underline{V}_{\bot}\!(\mu)$} gewonnen wurden,
die den Bedingungen~(\ref{E.3.33}) gen"ugen, l"asst sich analog zu
den Sch"atzwerten der Varianzen und Kovarianzen der Messwerte der
"Ubertragungsfunktion zeigen, dass die konkreten Varianzsch"atzwerte
niemals kleiner als die Betr"age der entsprechenden konkreten
Kovarianzsch"atzwerte sind.  Dabei greift man auf die
Ungleichung~(\ref{E.3.36.b}) und auf die nach Gleichung~(\ref{E.3.46})
definierten Vektoren \mbox{$\Hat{\Vec{C}}_U(\ldots)$} zur"uck.

Auch die Messwerte \mbox{$\Hat{\boldsymbol{u}}(k)$} betrachten wir
nun wieder als Zufallsgr"o"sen, die aus mathematischen Stichproben 
der Zufallsprozesse am Systemein- und -ausgang entstanden sind.
Die Varianzen der Messwerte \mbox{$\Hat{\boldsymbol{u}}(k)$}
berechnen sich mit Gleichung~(\ref{E.3.23}) zu:
\begin{gather}
C_{\Hat{\boldsymbol{u}}(k),\Hat{\boldsymbol{u}}(k)}\;=\;
\text{E}\bigg\{\Big|\Hat{\boldsymbol{u}}(k)\!-\!
\text{E}\big\{\Hat{\boldsymbol{u}}(k)\big\}\Big|^2\bigg\}\;=\;
\text{E}\Big\{\big|\Hat{\boldsymbol{u}}(k)\!-\!u(k)\big|^2\Big\}\;={}
\label{E.3.49}\\[10pt]
{}=\,\text{E}\Bigg\{\Bigg|
\frac{1}{L}\CdoT\Vec{\boldsymbol{n}}(k)\cdoT\Vec{1}^{\,\Hh}\!-
\frac{1}{M\CdoT L}\cdoT\!\Sum{\mu=0}{M-1}
\frac{\Vec{\boldsymbol{N}}_{\!\!f}(\mu)\cdot
\underline{1}_{\bot}\Cdot\Tilde{\underline{\boldsymbol{V}}}(\mu)^{\Hh}}{L\!-\!1}\cdot
\Hat{\underline{\boldsymbol{C}}}_{\Tilde{\Vec{\boldsymbol{V}}}(\mu),\Tilde{\Vec{\boldsymbol{V}}}(\mu)}^{\uP{0.4}{\!-1}}\cdot
\Tilde{\underline{\boldsymbol{V}}}(\mu)\CdoT\Vec{1}^{\,\Hh}\!\Cdot
e^{j\cdot\frac{2\pi}{M}\cdot\mu\cdot k}\Bigg|^2\Bigg\}.
\notag
\end{gather}
Als Abk"urzung f"uhren wir den Zufallsvektor
\begin{equation}
\Hat{\Vec{\boldsymbol{C}}}_u(\mu)\;=\;
\frac{\Vec{1}\cdot\Tilde{\underline{\boldsymbol{V}}}(\mu)^{\Hh}}{L}\cdot
\Hat{\underline{\boldsymbol{C}}}_{\Tilde{\Vec{\boldsymbol{V}}}(\mu),\Tilde{\Vec{\boldsymbol{V}}}(\mu)}^{\uP{0.4}{\!-1}}\cdot
\Tilde{\underline{\boldsymbol{V}}}(\mu)\cdot\underline{1}_{\bot}
\label{E.3.50}
\end{equation}
ein und erhalten mit 
\begin{gather}
C_{\Hat{\boldsymbol{u}}(k),\Hat{\boldsymbol{u}}(k)}\;=\;
\text{E}\Bigg\{\Bigg|
\frac{1}{L}\CdoT\Vec{\boldsymbol{n}}(k)\cdoT\Vec{1}^{\,\Hh}\!-
\frac{1}{M\CdoT(L\!-\!1)}\cdoT\!\Sum{\mu=0}{M-1}
\Vec{\boldsymbol{N}}_{\!\!f}(\mu)\cdot
\Hat{\Vec{\boldsymbol{C}}}_u(\mu)^{\Hh}\Cdot
e^{j\cdot\frac{2\pi}{M}\cdot\mu\cdot k}
\Bigg|^2\Bigg\}\;={}
\label{E.3.51}\\[10pt]\begin{flalign*}
{}=\;{}&\frac{1}{L^2}\CdoT\text{E}\Big\{\Vec{\boldsymbol{n}}(k)\cdoT\Vec{1}^{\,\Hh}\Cdot\Vec{1}\CdoT\Vec{\boldsymbol{n}}(k)^{\Hh}\Big\}-{}&&
\end{flalign*}\notag\\*[4pt]
{}-\frac{1}{M\CdoT L\CdoT(L\!-\!1)}\cdoT\!\Sum{\mu=0}{M-1}
\text{E}\Big\{\Vec{\boldsymbol{n}}(k)\cdoT\Vec{1}^{\,\Hh}\Cdot
\Hat{\Vec{\boldsymbol{C}}}_u(\mu)\cdot
\Vec{\boldsymbol{N}}_{\!\!f}(\mu)^{\Hh}\Big\}\cdot
e^{\!-j\cdot\frac{2\pi}{M}\cdot\mu\cdot k}-{}
\notag\\*[4pt]
{}-\frac{1}{M\CdoT L\CdoT(L\!-\!1)}\cdoT\!\Sum{\mu=0}{M-1}
\text{E}\Big\{\Vec{\boldsymbol{N}}_{\!\!f}(\mu)\cdot
\Hat{\Vec{\boldsymbol{C}}}_u(\mu)^{\Hh}\Cdot
\Vec{1}\CdoT\Vec{\boldsymbol{n}}(k)^{\Hh}\Big\}\cdot
e^{j\cdot\frac{2\pi}{M}\cdot\mu\cdot k}+{}
\notag\\*[4pt]\begin{flalign*}
&&&{}+\frac{1}{M^2\CdoT(L\!-\!1)^2}\cdoT\!\Sum{\mu_1=0}{M-1}\Sum{\mu_2=0}{M-1}
\text{E}\Big\{\Vec{\boldsymbol{N}}_{\!\!f}(\mu_1)\cdot
\Hat{\Vec{\boldsymbol{C}}}_u(\mu_1)^{\Hh}\Cdot
\Hat{\Vec{\boldsymbol{C}}}_u(\mu_2)\cdot
\Vec{\boldsymbol{N}}_{\!\!f}(\mu_2)^{\Hh}\Big\}\cdot
e^{j\cdot\frac{2\pi}{M}\cdot(\mu_1-\mu_2)\cdot k}
\end{flalign*}\notag
\end{gather}
eine Summe von Erwartungswerten von bilinearen Formen. Die Erwartungswerte der bilinearen Formen 
lassen sich mit Anhang \ref{E.Kap.A.5} berechnen. Dabei erkennt man dass die beiden Einfachsummen in 
der letzten Gleichung keinen Beitrag liefern. Als n"achstes ber"ucksichtigen wir, dass die 
Zufallsgr"o"sen \mbox{$\boldsymbol{n}(k)$} und \mbox{$\boldsymbol{N}_{\!\!f}(\mu)$} 
mittelwertfrei sind, und dass die Kovarianzen
\mbox{$\text{E}\big\{\boldsymbol{N}_{\!\!f}(\mu_1)\CdoT
\boldsymbol{N}_{\!\!f}(\mu_2)^{\Kk}\big\}$} nach Gleichung~(\ref{E.2.41}) 
f"ur die Frequenzen mit 
\mbox{$\mu_2\!\neq\!\mu_1\!+\!\Tilde{\mu}\CdoT M/K_{\Phi}$}
gegen"uber den Kovarianzen mit 
\mbox{$\mu_2\!=\!\mu_1\!+\!\Tilde{\mu}\CdoT M/K_{\Phi}$} vernachl"assigbar
klein sind, wenn wir eine hoch frequenzselektive Fensterfolge verwenden. 
Damit ergibt sich:
\begin{subequations}\label{E.3.52}
\begin{gather}
\begin{flalign}
&C_{\Hat{\boldsymbol{u}}(k),\Hat{\boldsymbol{u}}(k)}\;=\;
\frac{1}{L}\cdot\text{E}\big\{|\boldsymbol{n}(k)|^2\big\}+{}&&
\end{flalign}\label{E.3.52.a}\\*[2pt]\begin{flalign*}
&&{}\!\!\!\!{}+\frac{1}{M^2\CdoT(L\!-\!1)^2}\cdoT
\Sum{\mu_1=0}{M-1}\;\Sum{\mu_2=0}{M-1}
\text{E}\Big\{\Hat{\Vec{\boldsymbol{C}}}_u(\mu_2)\CdoT
\Hat{\Vec{\boldsymbol{C}}}_u(\mu_1)^{\Hh}\Big\}\CdoT
\text{E}\Big\{\boldsymbol{N}_{\!\!f}(\mu_1)\CdoT
\boldsymbol{N}_{\!\!f}(\mu_2)^{\Kk}\Big\}\CdoT
e^{j\cdot\frac{2\pi}{M}\cdot(\mu_1-\mu_2)\cdot k}\approx{}&
\end{flalign*}\notag\\[12pt]\begin{flalign*}
&{}\approx\frac{1}{M\CdoT L}\cdoT
\Sum{\mu=0}{M-1}\,\Sum{\Tilde{\mu}=0}{K_{\Phi}-1}
\Tilde{\Phi}_{\boldsymbol{n}}
\big({\T\mu,\mu\!+\!\Tilde{\mu}\CdoT\frac{M}{K_{\Phi}}}\big)\CdoT
e^{\!-j\cdot\frac{2\pi}{K_{\Phi}}\cdot\Tilde{\mu}\cdot k}+{}&&
\end{flalign*}\notag\\*[2pt]\begin{flalign*}
&&{}\!\!\!\!{}+\frac{1}{M\CdoT(L\!-\!1)^2}\cdoT
\Sum{\mu=0}{M-1}\,\Sum{\Tilde{\mu}=0}{K_{\Phi}-1}
\text{E}\Big\{\Hat{\Vec{\boldsymbol{C}}}_u\big({\T\mu\!+\!\Tilde{\mu}\CdoT\frac{M}{K_{\Phi}}}\big)\CdoT
\Hat{\Vec{\boldsymbol{C}}}_u(\mu)^{\Hh}\Big\}\CdoT
\Tilde{\Phi}_{\boldsymbol{n}}\big({\T\mu,\mu\!+\!\Tilde{\mu}\CdoT\frac{M}{K_{\Phi}}}\big)\CdoT
e^{\!-j\cdot\frac{2\pi}{K_{\Phi}}\cdot\Tilde{\mu}\cdot k}={}&
\end{flalign*}\notag\\[16pt]\begin{flalign*}
&{}=\frac{1}{M\CdoT(L\!-\!1)^2}\cdoT
\Sum{\mu=0}{M-1}\,\Sum{\Tilde{\mu}=0}{K_{\Phi}-1}
\bigg(\frac{L\!-\!1}{L}\cdot\Vec{1}\cdot\Vec{1}^{\,\Hh}\Cdot\frac{L\!-\!1}{L}+
\text{E}\Big\{\Hat{\Vec{\boldsymbol{C}}}_u\big({\T\mu\!+\!\Tilde{\mu}\CdoT\frac{M}{K_{\Phi}}}\big)\CdoT
\Hat{\Vec{\boldsymbol{C}}}_u(\mu)^{\Hh}\Big\}\bigg)\cdot{}&&
\end{flalign*}\notag\\*[2pt]\begin{flalign*}
&&{}\cdot\Tilde{\Phi}_{\boldsymbol{n}}\big({\T\mu,\mu\!+\!\Tilde{\mu}\CdoT\frac{M}{K_{\Phi}}}\big)\cdot
e^{\!-j\cdot\frac{2\pi}{K_{\Phi}}\cdot\Tilde{\mu}\cdot k}={}&
\end{flalign*}\notag\\[12pt]\begin{flalign*}
&{}=\frac{1}{M\CdoT(L\!-\!1)^2}\cdoT
\Sum{\mu=0}{M-1}\;\Sum{\Tilde{\mu}=0}{K_{\Phi}-1}
\text{E}\bigg\{\Big(\frac{L\!-\!1}{L}\cdot\Vec{1}-
\Hat{\Vec{\boldsymbol{C}}}_u\big({\T\mu\!+\!\Tilde{\mu}\CdoT\frac{M}{K_{\Phi}}}\big)\Big)\cdot
\Big(\Vec{1}^{\,\Hh}\Cdot\frac{L\!-\!1}{L}-
\Hat{\Vec{\boldsymbol{C}}}_u(\mu)^{\Hh}\Big)\bigg\}\cdot{}&&
\end{flalign*}\notag\\*[0pt]\begin{flalign*}
&&{}\cdot\Tilde{\Phi}_{\boldsymbol{n}}\big({\T\mu,\mu\!+\!\Tilde{\mu}\CdoT\frac{M}{K_{\Phi}}}\big)\CdoT
e^{\!-j\cdot\frac{2\pi}{K_{\Phi}}\cdot\Tilde{\mu}\cdot k}={}&
\end{flalign*}\notag\\[18pt]
=\frac{1}{M\CdoT(L\!-\!1)^2}\cdoT
\Sum{\mu=0}{M-1}\,\Sum{\Tilde{\mu}=0}{K_{\Phi}-1}\text{E}\Big\{
\Hat{\Vec{\boldsymbol{C}}}_U\big({\T\mu\!+\!\Tilde{\mu}\CdoT\frac{M}{K_{\Phi}}}\big)\CdoT
\Hat{\Vec{\boldsymbol{C}}}_U(\mu)^{\Hh}\Big\}\CdoT
\Tilde{\Phi}_{\boldsymbol{n}}
\big({\T\mu,\mu\!+\!\Tilde{\mu}\CdoT\frac{M}{K_{\Phi}}}\big)\CdoT
e^{\!-j\cdot\frac{2\pi}{K_{\Phi}}\cdot\Tilde{\mu}\cdot k}=\!\!
\notag\\[12pt]
=\;\frac{1}{M^2}\cdoT
\Sum{\mu=0}{M-1}\;\Sum{\Tilde{\mu}=0}{K_{\Phi}-1}
C_{\Hat{\boldsymbol{U}}_{\!\!f}(\mu),\Hat{\boldsymbol{U}}_{\!\!f}(\mu+\Tilde{\mu}\cdot\frac{M}{K_{\Phi}})}\cdot 
e^{\!-j\cdot\frac{2\pi}{K_{\Phi}}\cdot\Tilde{\mu}\cdot k}\qquad
\notag\\*[4pt]
\qquad\qquad\qquad\forall\qquad k=0\;(1)\;F\!-\!1,
\notag
\end{gather}
wobei nach und nach die Gleichungen~(\ref{E.2.55}), (\ref{E.3.46}) und 
(\ref{E.3.47.a}) Verwendung fanden.

Ganz analog erhalten wir aus den Messwertkovarianzen nach
Gleichung~(\ref{E.3.47.b}) f"ur die Kovarianzen der Messwerte
\mbox{$\Hat{u}(k)$} die N"aherungen
\begin{gather}
C_{\Hat{\boldsymbol{u}}(k),\Hat{\boldsymbol{u}}(k)^{\Kk}}\;\approx\;
\frac{1}{M^2}\cdot
\Sum{\mu=0}{M-1}\;\Sum{\Tilde{\mu}=0}{K_{\Phi}-1}
C_{\Hat{\boldsymbol{U}}_{\!\!f}(\mu),\Hat{\boldsymbol{U}}_{\!\!f}(-\mu-\Tilde{\mu}\cdot\frac{M}{K_{\Phi}})^{\Kk}}
\cdot e^{\!-j\cdot\frac{2\pi}{K_{\Phi}}\cdot\Tilde{\mu}\cdot k}\qquad
\label{E.3.52.b}\\*[4pt]
\forall\qquad k=0\;(1)\;F\!-\!1.
\notag
\end{gather}
\end{subequations}
Diese Messwert"-(ko)"-varianzen sch"atzen wir dadurch ab, dass wir die Werte der
wahren Kovarianzen der Messwerte \mbox{$\Hat{\boldsymbol{U}}_{\!\!f}(\mu)$}
durch deren Sch"atzwerte ersetzen:
\begin{subequations}\label{E.3.53}
\begin{align}
\Hat{C}_{\Hat{\boldsymbol{u}}(k),\Hat{\boldsymbol{u}}(k)}&\;=\;
\frac{1}{M^2}\cdot
\Sum{\mu=0}{M-1}\;\Sum{\Tilde{\mu}=0}{K_{\Phi}-1}
\Hat{C}_{\Hat{\boldsymbol{U}}_{\!\!f}(\mu),\Hat{\boldsymbol{U}}_{\!\!f}(\mu+\Tilde{\mu}\cdot\frac{M}{K_{\Phi}})}\cdot
 e^{\!-j\cdot\frac{2\pi}{K_{\Phi}}\cdot\Tilde{\mu}\cdot k}
\label{E.3.53.a}\\[10pt]
\Hat{C}_{\Hat{\boldsymbol{u}}(k),\Hat{\boldsymbol{u}}(k)^{\Kk}}&\;=\;
\frac{1}{M^2}\cdot
\Sum{\mu=0}{M-1}\;\Sum{\Tilde{\mu}=0}{K_{\Phi}-1}
\Hat{C}_{\Hat{\boldsymbol{U}}_{\!\!f}(\mu),\Hat{\boldsymbol{U}}_{\!\!f}(-\mu-\Tilde{\mu}\cdot\frac{M}{K_{\Phi}})^{\Kk}}\cdot
e^{\!-j\cdot\frac{2\pi}{K_{\Phi}}\cdot\Tilde{\mu}\cdot k}
\label{E.3.53.b}\\*[4pt]
\forall\qquad&\qquad k=0\;(1)\;F\!-\!1
\notag
\end{align}
\end{subequations}
Da sich diese Sch"atzwerte als eine endliche "Uberlagerung erwartungstreuer 
und konsistenter Sch"atzwerte berechnen, sch"atzen sie die N"aherung 
der Messwert"-(ko)"-varianzen der deterministischen St"orung ebenfalls 
erwartungstreu und konsistent ab. Wenn wir zur Berechnung der Sch"atzwerte
\mbox{$\Hat{C}_{\Hat{\boldsymbol{U}}_{\!\!f}(\mu),\Hat{\boldsymbol{U}}_{\!\!f}(\mu+\Tilde{\mu}\cdot M/K_{\Phi})}$}
und \mbox{$\Hat{C}_{\Hat{\boldsymbol{U}}_{\!\!f}(\mu),\Hat{\boldsymbol{U}}_{\!\!f}(-\mu-\Tilde{\mu}\cdot M/K_{\Phi})^{\Kk}}$}
die Messwerte \mbox{$\Hat{\Phi}_{\boldsymbol{n}}(\ldots,\ldots)$}
und \mbox{$\Hat{\Psi}_{\boldsymbol{n}}(\ldots,-\ldots)$} verwenden, die mit
Matrizen \mbox{$\underline{V}_{\bot}\!(\ldots)$} gewonnen wurden,
die die Bedingungen~(\ref{E.3.33}) erf"ullen, l"asst sich auch f"ur die
Sch"atzwerte der Messwert"-(ko)"-varianzen zeigen, dass die konkreten
Varianzsch"atzwerte niemals kleiner als die Betr"age der
entsprechenden konkreten Kovarianzsch"atzwerte sind.
Im Unterkapitel~\ref{E.Kap.A.9.2} der Anhangs wird diese Aussage bewiesen.

Nun bestimmen wir die Varianzen und Kovarianzen der Messwerte
\mbox{$\Hat{\boldsymbol{\Phi}}_{\!\boldsymbol{n}}(\mu,\mu\!+\!\Tilde{\mu}\CdoT M/K_{\Phi})$}
und \mbox{$\Hat{\boldsymbol{\Psi}}_{\!\boldsymbol{n}}(\mu,\mu\!+\!\Tilde{\mu}\CdoT M/K_{\Phi})$}
indem wir mit den Gleichungen~(\ref{E.3.29}) die vier zweiten zentralen Momente
\begin{subequations}\label{E.3.54}
\begin{gather*}
\begin{flalign*}
&C_{\Hat{\boldsymbol{\Phi}}_{\!\boldsymbol{n}}(\mu,\mu+\Tilde{\mu}\cdot\frac{M}{K_{\Phi}}),\Hat{\boldsymbol{\Phi}}_{\!\boldsymbol{n}}(\mu,\mu+\Tilde{\mu}\cdot\frac{M}{K_{\Phi}})}\;=\;
\text{E}\bigg\{\Big|\Hat{\boldsymbol{\Phi}}_{\!\boldsymbol{n}}\big({\T\mu,\mu\!+\!\Tilde{\mu}\CdoT\frac{M}{K_{\Phi}}}\big)\!-\!
\Tilde{\Phi}_{\boldsymbol{n}}\big({\T\mu,\mu\!+\!\Tilde{\mu}\CdoT\frac{M}{K_{\Phi}}}\big)\Big|^2\bigg\}\;={}&&
\end{flalign*}\\*[10pt]\label{E.3.54.a}\begin{flalign}
&=\;\text{E}\bigg\{\bigg|\Vec{\boldsymbol{N}}_{\!\!f}(\mu)\cdot
\frac{\underline{\boldsymbol{V}}_{\bot}\!(\mu)\CdoT
\underline{\boldsymbol{V}}_{\bot}\!\big({\T\mu\!+\!\Tilde{\mu}\CdoT\frac{M}{K_{\Phi}}}\big)^{\HH}}
{M\CdoT\text{spur}\Big(\underline{\boldsymbol{V}}_{\bot}\!(\mu)\CdoT
\underline{\boldsymbol{V}}_{\bot}\!\big({\T\mu\!+\!\Tilde{\mu}\CdoT\frac{M}{K_{\Phi}}}\big)^{\HH}\Big)}\cdot
\Vec{\boldsymbol{N}}_{\!\!f}\big({\T\mu\!+\!\Tilde{\mu}\CdoT\frac{M}{K_{\Phi}}}\big)^{\HH}\bigg|^2\bigg\}\;-{}&&
\end{flalign}\\*\begin{flalign*}
&&{}-\;\frac{1}{M^2}\cdot\Big|\text{E}\Big\{\boldsymbol{N}_{\!\!f}(\mu)\CdoT
\boldsymbol{N}_{\!\!f}\big({\T\mu\!+\!\Tilde{\mu}\CdoT\frac{M}{K_{\Phi}}}\big)^{\!\Kk}\Big\}\Big|^2\!\!,&
\end{flalign*}\\[16pt]\begin{flalign*}
&C_{\Hat{\boldsymbol{\Phi}}_{\!\boldsymbol{n}}(\mu,\mu+\Tilde{\mu}\cdot\frac{M}{K_{\Phi}}),\Hat{\boldsymbol{\Phi}}_{\!\boldsymbol{n}}(\mu,\mu+\Tilde{\mu}\cdot\frac{M}{K_{\Phi}})^{\Kk}}\;=\;
\text{E}\bigg\{\Big(\Hat{\boldsymbol{\Phi}}_{\!\boldsymbol{n}}\big({\T\mu,\mu\!+\!\Tilde{\mu}\CdoT\frac{M}{K_{\Phi}}}\big)\!-\!
\Tilde{\Phi}_{\boldsymbol{n}}\big({\T\mu,\mu\!+\!\Tilde{\mu}\CdoT\frac{M}{K_{\Phi}}}\big)\Big)^2\bigg\}\;={}&&
\end{flalign*}\\*[10pt]\label{E.3.54.b}\begin{flalign}
&=\;\text{E}\bigg\{\bigg(\Vec{\boldsymbol{N}}_{\!\!f}(\mu)\cdot
\frac{\underline{\boldsymbol{V}}_{\bot}\!(\mu)\CdoT
\underline{\boldsymbol{V}}_{\bot}\!\big({\T\mu\!+\!\Tilde{\mu}\CdoT\frac{M}{K_{\Phi}}}\big)^{\HH}}
{M\CdoT\text{spur}\Big(\underline{\boldsymbol{V}}_{\bot}\!(\mu)\CdoT
\underline{\boldsymbol{V}}_{\bot}\!\big({\T\mu\!+\!\Tilde{\mu}\CdoT\frac{M}{K_{\Phi}}}\big)^{\HH}\Big)}\cdot
\Vec{\boldsymbol{N}}_{\!\!f}\big({\T\mu\!+\!\Tilde{\mu}\CdoT\frac{M}{K_{\Phi}}}\big)^{\HH}\bigg)^{\!\!2}\bigg\}\;-{}&&
\end{flalign}\\*\begin{flalign*}
&&{}-\;\frac{1}{M^2}\cdot\text{E}\Big\{\boldsymbol{N}_{\!\!f}(\mu)\CdoT
\boldsymbol{N}_{\!\!f}\big({\T\mu\!+\!\Tilde{\mu}\CdoT\frac{M}{K_{\Phi}}}\big)^{\!\Kk}\Big\}^{\!2}\!,&
\end{flalign*}\displaybreak[2]\\[6pt]\begin{flalign*}
&C_{\Hat{\boldsymbol{\Psi}}_{\!\boldsymbol{n}}(\mu,\mu+\Tilde{\mu}\cdot\frac{M}{K_{\Phi}}),\Hat{\boldsymbol{\Psi}}_{\!\boldsymbol{n}}(\mu,\mu+\Tilde{\mu}\cdot\frac{M}{K_{\Phi}})}\;=\;
\text{E}\bigg\{\Big|\Hat{\boldsymbol{\Psi}}_{\!\boldsymbol{n}}\big({\T\mu,\mu\!+\!\Tilde{\mu}\CdoT\frac{M}{K_{\Phi}}}\big)\!-\!
\Tilde{\Psi}_{\boldsymbol{n}}\big({\T\mu,\mu\!+\!\Tilde{\mu}\CdoT\frac{M}{K_{\Phi}}}\big)\Big|^2\bigg\}\;={}&&
\end{flalign*}\\*[14pt]\label{E.3.54.c}\begin{flalign}
&=\;\text{E}\bigg\{\bigg|\Vec{\boldsymbol{N}}_{\!\!f}(\mu)\cdot
\frac{\underline{\boldsymbol{V}}_{\bot}\!(\mu)\CdoT
\underline{V}_{\bot}\!\big(\!{\T-\mu\!-\!\Tilde{\mu}\CdoT\frac{M}{K_{\Phi}}}\big)^{\TT}}
{M\CdoT\text{spur}\Big(\underline{\boldsymbol{V}}_{\bot}\!(\mu)\CdoT
\underline{\boldsymbol{V}}_{\bot}\!\big(\!{\T-\mu\!-\!\Tilde{\mu}\CdoT\frac{M}{K_{\Phi}}}\big)^{\TT}\Big)}\cdot
\Vec{\boldsymbol{N}}_{\!\!f}\big(\!{\T-\mu\!-\!\Tilde{\mu}\CdoT\frac{M}{K_{\Phi}}}\big)^{\TT}\bigg|^2\bigg\}\;-{}&&
\end{flalign}\\*\begin{flalign*}
&&{}-\;\frac{1}{M^2}\cdot\Big|\text{E}\Big\{\boldsymbol{N}_{\!\!f}(\mu)\CdoT
\boldsymbol{N}_{\!\!f}\big(\!{\T-\mu\!-\!\Tilde{\mu}\CdoT\frac{M}{K_{\Phi}}}\big)\Big\}\Big|^2&
\end{flalign*}\\[-16pt]\intertext{und\vspace{-4pt}}\begin{flalign*}
&C_{\Hat{\boldsymbol{\Psi}}_{\!\boldsymbol{n}}(\mu,\mu+\Tilde{\mu}\cdot\frac{M}{K_{\Phi}}),\Hat{\boldsymbol{\Psi}}_{\!\boldsymbol{n}}(\mu,\mu+\Tilde{\mu}\cdot\frac{M}{K_{\Phi}})^{\Kk}}\;=\;
\text{E}\bigg\{\Big(\Hat{\boldsymbol{\Psi}}_{\!\boldsymbol{n}}\big({\T\mu,\mu\!+\!\Tilde{\mu}\CdoT\frac{M}{K_{\Phi}}}\big)\!-\!
\Tilde{\Psi}_{\boldsymbol{n}}\big({\T\mu,\mu\!+\!\Tilde{\mu}\CdoT\frac{M}{K_{\Phi}}}\big)\Big)^2\bigg\}\;={}&&
\end{flalign*}\\*[14pt]\label{E.3.54.d}\begin{flalign}
&=\;\text{E}\bigg\{\bigg(\Vec{\boldsymbol{N}}_{\!\!f}(\mu)\cdot
\frac{\underline{\boldsymbol{V}}_{\bot}\!(\mu)\CdoT
\underline{\boldsymbol{V}}_{\bot}\!\big(\!{\T-\mu\!-\!\Tilde{\mu}\CdoT\frac{M}{K_{\Phi}}}\big)^{\TT}}
{M\CdoT\text{spur}\Big(\underline{\boldsymbol{V}}_{\bot}\!(\mu)\CdoT
\underline{\boldsymbol{V}}_{\bot}\!\big(\!{\T-\mu\!-\!\Tilde{\mu}\CdoT\frac{M}{K_{\Phi}}}\big)^{\TT}\Big)}\cdot
\Vec{\boldsymbol{N}}_{\!\!f}\big(\!{\T-\mu\!-\!\Tilde{\mu}\CdoT\frac{M}{K_{\Phi}}}\big)^{\TT}\bigg)^{\uP{-0.2}{\!\!2}}\bigg\}\;-{}&&
\end{flalign}\\*\begin{flalign*}
&&{}-\;\frac{1}{M^2}\cdot\text{E}\Big\{\boldsymbol{N}_{\!\!f}(\mu)\CdoT
\boldsymbol{N}_{\!\!f}\big(\!{\T-\mu\!-\!\Tilde{\mu}\CdoT\frac{M}{K_{\Phi}}}\big)\Big\}^{\!2}&
\end{flalign*}\\*[6pt]
\forall\qquad\mu=0\;(1)\;M\!-\!1\quad\text{ und }
\quad\Tilde{\mu}=0\;(1)\;K_{\Phi}\!-\!1
\end{gather*}
\end{subequations}
berechnen. Dazu unterscheiden wir zwei F"alle.

Im ersten Fall ist $\mu$ ein ganzzahliges Vielfaches von
\mbox{$M/(2\CdoT K_{\Phi})$}.
\begin{subequations}\label{E.3.55}
\begin{equation}
\mu \;=\; \Hat{\mu}\cdot\frac{M}{2\CdoT K_{\Phi}}
\qquad\text{ mit }\quad \Hat{\mu}\in\mathbb{Z}
\label{E.3.55.a}
\end{equation}
In diesem Fall ist nicht nur die diskrete Frequenz
\mbox{$\mu\!+\!\Tilde{\mu}\CdoT M/K_{\Phi}$} um ein
ganzzahliges Vielfaches von \mbox{$M/K_{\Phi}$} gegen"uber
$\mu$ verschoben. Auch die negativen diskreten Frequenzen $-\mu$ und
\mbox{$-\mu\!-\!\Tilde{\mu}\CdoT M/K_{\Phi}$} liegen in demselben
Frequenzraster:
\begin{gather}
-\mu\;=\;-\Hat{\mu}\CdoT\frac{M}{2\CdoT K_{\Phi}}\;=\;
\Hat{\mu}\CdoT\frac{M}{2\CdoT K_{\Phi}}-
\Hat{\mu}\CdoT\frac{M}{K_{\Phi}}\;=\;
\mu-\Hat{\mu}\CdoT\frac{M}{K_{\Phi}}
\label{E.3.55.b}\\*[10pt]
-\mu-\Tilde{\mu}\CdoT\frac{M}{K_{\Phi}}\;=\;
-\Hat{\mu}\CdoT\frac{M}{2\CdoT K_{\Phi}}-
\Tilde{\mu}\CdoT\frac{M}{K_{\Phi}}\;=\;
\Hat{\mu}\CdoT\frac{M}{2\CdoT K_{\Phi}}+
\Breve{\mu}\CdoT\frac{M}{K_{\Phi}}\;=\;
\mu+\Breve{\mu}\CdoT\frac{M}{K_{\Phi}}
\label{E.3.55.c}\\*[4pt]
\text{ mit }\quad \Breve{\mu}=-\Hat{\mu}\!-\!\Tilde{\mu}\in\mathbb{Z}.
\notag
\end{gather}
\end{subequations}
Daher k"onnen nach den Gleichungen~(\ref{E.2.41}) und (\ref{E.2.51}) 
in  diesem Fall alle m"oglichen Kovarianzen zweier Zufallsgr"o"sen, 
die man sich aus aus den vier Zufallsgr"o"sen \mbox{$\boldsymbol{N}_{\!\!f}(\mu)$}, 
\mbox{$\boldsymbol{N}_{\!\!f}\big({\T\mu\!+\!\Tilde{\mu}\CdoT\frac{M}{K_{\Phi}}}\big)$}, 
\mbox{$\boldsymbol{N}_{\!\!f}(\!-\mu)$}, 
\mbox{$\boldsymbol{N}_{\!\!f}\big(\!{\T-\mu\!-\!\Tilde{\mu}\CdoT\frac{M}{K_{\Phi}}}\big)$} 
und den dazu konjugierten Zufallsgr"o"sen herausgreift, von null verschieden 
sein. Ebenso finden sich Sch"atzwerte f"ur das $M$-fache all dieser Kovarianzen 
unter den in den Gleichungen~(\ref{E.3.29}) angegebenen Messwerten.

Nun k"onnen wir jeweils mit der ersten Zeile des Vektorgleichungssystems~(\ref{E.A.41})
des Anhangs~\ref{E.Kap.A.8} die vier gesuchten theoretischen Messwertvarianzen und
-kovarianzen berechnen, indem wir in den Gleichungen~(\ref{E.A.40})
die in Tabelle~\ref{T.3.1} aufgelisteten Substitutionen vornehmen,
und die Ergebnisse auf $M$ normieren. 
\begin{table}[!b]
\rule{0pt}{10pt}\\
\rule{\textwidth}{0.5pt}\vspace{5pt}
\[
{\renewcommand{\arraystretch}{1.4}
\begin{array}{||c||c|c|c|c||}
\hline
\hline
\rule{0pt}{18pt}\text{Berechnung von}&
\Vec{\boldsymbol{N}}_{\!1}&
\Vec{\boldsymbol{N}}_{\!2}&
\Vec{\boldsymbol{N}}_{\!3}&
\Vec{\boldsymbol{N}}_{\!4}\\
\hline
\rule{0pt}{18pt}C_{\Hat{\boldsymbol{\Phi}}_{\!\boldsymbol{n}}(\mu,\mu+\Tilde{\mu}\cdot\frac{M}{K_{\Phi}}),\Hat{\boldsymbol{\Phi}}_{\!\boldsymbol{n}}(\mu,\mu+\Tilde{\mu}\cdot\frac{M}{K_{\Phi}})}&
\Vec{\boldsymbol{N}}_{\!\!f}(\mu)&
\Vec{\boldsymbol{N}}_{\!\!f}\big({\T\mu\!+\!\Tilde{\mu}\CdoT\frac{M}{K_{\Phi}}}\big)&
\Vec{\boldsymbol{N}}_{\!\!f}(\mu)^{\Kk}&
\Vec{\boldsymbol{N}}_{\!\!f}\big({\T\mu\!+\!\Tilde{\mu}\CdoT\frac{M}{K_{\Phi}}}\big)^{\!\Kk}\\
C_{\Hat{\boldsymbol{\Phi}}_{\!\boldsymbol{n}}(\mu,\mu+\Tilde{\mu}\cdot\frac{M}{K_{\Phi}}),\Hat{\boldsymbol{\Phi}}_{\!\boldsymbol{n}}(\mu,\mu+\Tilde{\mu}\cdot\frac{M}{K_{\Phi}})^{\Kk}}&
\Vec{\boldsymbol{N}}_{\!\!f}(\mu)&
\Vec{\boldsymbol{N}}_{\!\!f}\big({\T\mu\!+\!\Tilde{\mu}\CdoT\frac{M}{K_{\Phi}}}\big)&
\Vec{\boldsymbol{N}}_{\!\!f}(\mu)&
\Vec{\boldsymbol{N}}_{\!\!f}\big({\T\mu\!+\!\Tilde{\mu}\CdoT\frac{M}{K_{\Phi}}}\big)\\
C_{\Hat{\boldsymbol{\Psi}}_{\!\boldsymbol{n}}(\mu,\mu+\Tilde{\mu}\cdot\frac{M}{K_{\Phi}}),\Hat{\boldsymbol{\Psi}}_{\!\boldsymbol{n}}(\mu,\mu+\Tilde{\mu}\cdot\frac{M}{K_{\Phi}})}&
\Vec{\boldsymbol{N}}_{\!\!f}(\mu)&
\Vec{\boldsymbol{N}}_{\!\!f}\big(\!{\T-\mu\!-\!\Tilde{\mu}\CdoT\frac{M}{K_{\Phi}}}\big)^{\!\Kk}&
\Vec{\boldsymbol{N}}_{\!\!f}(\mu)^{\Kk}&
\Vec{\boldsymbol{N}}_{\!\!f}\big(\!{\T-\mu\!-\!\Tilde{\mu}\CdoT\frac{M}{K_{\Phi}}}\big)\\
\rule[-12pt]{0pt}{12pt}C_{\Hat{\boldsymbol{\Psi}}_{\!\boldsymbol{n}}(\mu,\mu+\Tilde{\mu}\cdot\frac{M}{K_{\Phi}}),\Hat{\boldsymbol{\Psi}}_{\!\boldsymbol{n}}(\mu,\mu+\Tilde{\mu}\cdot\frac{M}{K_{\Phi}})^{\Kk}}&
\Vec{\boldsymbol{N}}_{\!\!f}(\mu)&
\Vec{\boldsymbol{N}}_{\!\!f}\big(\!{\T-\mu\!-\!\Tilde{\mu}\CdoT\frac{M}{K_{\Phi}}}\big)^{\!\Kk}&
\Vec{\boldsymbol{N}}_{\!\!f}(\mu)&
\Vec{\boldsymbol{N}}_{\!\!f}\big(\!{\T-\mu\!-\!\Tilde{\mu}\CdoT\frac{M}{K_{\Phi}}}\big)^{\!\Kk}\\
\hline
\hline
\text{Berechnung von}&
\boldsymbol{N}_{\!1}&
\boldsymbol{N}_{\!2}&
\boldsymbol{N}_{\!3}&
\boldsymbol{N}_{\!4}\\
\hline
C_{\Hat{\boldsymbol{\Phi}}_{\!\boldsymbol{n}}(\mu,\mu+\Tilde{\mu}\cdot\frac{M}{K_{\Phi}}),\Hat{\boldsymbol{\Phi}}_{\!\boldsymbol{n}}(\mu,\mu+\Tilde{\mu}\cdot\frac{M}{K_{\Phi}})}&
\boldsymbol{N}_{\!\!f}(\mu)&
\boldsymbol{N}_{\!\!f}\big({\T\mu\!+\!\Tilde{\mu}\CdoT\frac{M}{K_{\Phi}}}\big)&
\boldsymbol{N}_{\!\!f}(\mu)^{\Kk}&
\boldsymbol{N}_{\!\!f}\big({\T\mu\!+\!\Tilde{\mu}\CdoT\frac{M}{K_{\Phi}}}\big)^{\!\Kk}\\
C_{\Hat{\boldsymbol{\Phi}}_{\!\boldsymbol{n}}(\mu,\mu+\Tilde{\mu}\cdot\frac{M}{K_{\Phi}}),\Hat{\boldsymbol{\Phi}}_{\!\boldsymbol{n}}(\mu,\mu+\Tilde{\mu}\cdot\frac{M}{K_{\Phi}})^{\Kk}}&
\boldsymbol{N}_{\!\!f}(\mu)&
\boldsymbol{N}_{\!\!f}\big({\T\mu\!+\!\Tilde{\mu}\CdoT\frac{M}{K_{\Phi}}}\big)&
\boldsymbol{N}_{\!\!f}(\mu)&
\boldsymbol{N}_{\!\!f}\big({\T\mu\!+\!\Tilde{\mu}\CdoT\frac{M}{K_{\Phi}}}\big)\\
C_{\Hat{\boldsymbol{\Psi}}_{\!\boldsymbol{n}}(\mu,\mu+\Tilde{\mu}\cdot\frac{M}{K_{\Phi}}),\Hat{\boldsymbol{\Psi}}_{\!\boldsymbol{n}}(\mu,\mu+\Tilde{\mu}\cdot\frac{M}{K_{\Phi}})}&
\boldsymbol{N}_{\!\!f}(\mu)&
\boldsymbol{N}_{\!\!f}\big(\!{\T-\mu\!-\!\Tilde{\mu}\CdoT\frac{M}{K_{\Phi}}}\big)^{\!\Kk}&
\boldsymbol{N}_{\!\!f}(\mu)^{\Kk}&
\boldsymbol{N}_{\!\!f}\big(\!{\T-\mu\!-\!\Tilde{\mu}\CdoT\frac{M}{K_{\Phi}}}\big)\\
\rule[-12pt]{0pt}{12pt}C_{\Hat{\boldsymbol{\Psi}}_{\!\boldsymbol{n}}(\mu,\mu+\Tilde{\mu}\cdot\frac{M}{K_{\Phi}}),\Hat{\boldsymbol{\Psi}}_{\!\boldsymbol{n}}(\mu,\mu+\Tilde{\mu}\cdot\frac{M}{K_{\Phi}})^{\Kk}}&
\boldsymbol{N}_{\!\!f}(\mu)&
\boldsymbol{N}_{\!\!f}\big(\!{\T-\mu\!-\!\Tilde{\mu}\CdoT\frac{M}{K_{\Phi}}}\big)^{\!\Kk}&
\boldsymbol{N}_{\!\!f}(\mu)&
\boldsymbol{N}_{\!\!f}\big(\!{\T-\mu\!-\!\Tilde{\mu}\CdoT\frac{M}{K_{\Phi}}}\big)^{\!\Kk}\\
\hline
\hline
\text{Berechnung von}&
\underline{\boldsymbol{V}}_1&
\underline{\boldsymbol{V}}_2&
\underline{\boldsymbol{V}}_3&
\underline{\boldsymbol{V}}_4\\
\hline
C_{\Hat{\boldsymbol{\Phi}}_{\!\boldsymbol{n}}(\mu,\mu+\Tilde{\mu}\cdot\frac{M}{K_{\Phi}}),\Hat{\boldsymbol{\Phi}}_{\!\boldsymbol{n}}(\mu,\mu+\Tilde{\mu}\cdot\frac{M}{K_{\Phi}})}&
\underline{\boldsymbol{V}}_{\bot}\!(\mu)&
\underline{\boldsymbol{V}}_{\bot}\!\big({\T\mu\!+\!\Tilde{\mu}\CdoT\frac{M}{K_{\Phi}}}\big)&
\underline{\boldsymbol{V}}_{\bot}\!(\mu)^{\Kk}&
\underline{\boldsymbol{V}}_{\bot}\!\big({\T\mu\!+\!\Tilde{\mu}\CdoT\frac{M}{K_{\Phi}}}\big)^{\!\Kk}\\
C_{\Hat{\boldsymbol{\Phi}}_{\!\boldsymbol{n}}(\mu,\mu+\Tilde{\mu}\cdot\frac{M}{K_{\Phi}}),\Hat{\boldsymbol{\Phi}}_{\!\boldsymbol{n}}(\mu,\mu+\Tilde{\mu}\cdot\frac{M}{K_{\Phi}})^{\Kk}}&
\underline{\boldsymbol{V}}_{\bot}\!(\mu)&
\underline{\boldsymbol{V}}_{\bot}\!\big({\T\mu\!+\!\Tilde{\mu}\CdoT\frac{M}{K_{\Phi}}}\big)&
\underline{\boldsymbol{V}}_{\bot}\!(\mu)&
\underline{\boldsymbol{V}}_{\bot}\!\big({\T\mu\!+\!\Tilde{\mu}\CdoT\frac{M}{K_{\Phi}}}\big)\\
C_{\Hat{\boldsymbol{\Psi}}_{\!\boldsymbol{n}}(\mu,\mu+\Tilde{\mu}\cdot\frac{M}{K_{\Phi}}),\Hat{\boldsymbol{\Psi}}_{\!\boldsymbol{n}}(\mu,\mu+\Tilde{\mu}\cdot\frac{M}{K_{\Phi}})}&
\underline{\boldsymbol{V}}_{\bot}\!(\mu)&
\underline{\boldsymbol{V}}_{\bot}\!\big(\!{\T-\mu\!-\!\Tilde{\mu}\CdoT\frac{M}{K_{\Phi}}}\big)^{\!\Kk}&
\underline{\boldsymbol{V}}_{\bot}\!(\mu)^{\Kk}&
\underline{\boldsymbol{V}}_{\bot}\!\big(\!{\T-\mu\!-\!\Tilde{\mu}\CdoT\frac{M}{K_{\Phi}}}\big)\\
\rule[-12pt]{0pt}{12pt}C_{\Hat{\boldsymbol{\Psi}}_{\!\boldsymbol{n}}(\mu,\mu+\Tilde{\mu}\cdot\frac{M}{K_{\Phi}}),\Hat{\boldsymbol{\Psi}}_{\!\boldsymbol{n}}(\mu,\mu+\Tilde{\mu}\cdot\frac{M}{K_{\Phi}})^{\Kk}}&
\underline{\boldsymbol{V}}_{\bot}\!(\mu)&
\underline{\boldsymbol{V}}_{\bot}\!\big(\!{\T-\mu\!-\!\Tilde{\mu}\CdoT\frac{M}{K_{\Phi}}}\big)^{\!\Kk}&
\underline{\boldsymbol{V}}_{\bot}\!(\mu)&
\underline{\boldsymbol{V}}_{\bot}\!\big(\!{\T-\mu\!-\!\Tilde{\mu}\CdoT\frac{M}{K_{\Phi}}}\big)^{\!\Kk}\\
\hline
\hline
\end{array}}
\]
\setlength{\belowcaptionskip}{2pt}
\caption{Substitutionen in den Gleichungen~(\ref{E.A.40}) bzw. (\ref{E.A.42}).
Die Variablen $\protect\boldsymbol{c}_1$,
$\protect\boldsymbol{c}_2$ und $\protect\boldsymbol{c}_3$
berechnen sich gem"a"s des unteren Teils der Tabelle~\ref{T.A.1} mit den
hier angegebenen Substitutionen f"ur $\underline{\boldsymbol{V}}_1$ bis
$\underline{\boldsymbol{V}}_4$.}
\label{T.3.1}
\end{table}
Wir erhalten:\vfill

\begin{subequations}\label{E.3.56}
\begin{gather*}\label{E.3.56.a}
\begin{flalign}
&C_{\Hat{\boldsymbol{\Phi}}_{\!\boldsymbol{n}}(\mu,\mu+\Tilde{\mu}\cdot\frac{M}{K_{\Phi}}),\Hat{\boldsymbol{\Phi}}_{\!\boldsymbol{n}}(\mu,\mu+\Tilde{\mu}\cdot\frac{M}{K_{\Phi}})}\;={}&&
\end{flalign}\\*[4pt]\begin{flalign*}
&&{}\!\!\!\!\!\!\!\!\!\!{}=\;E\big\{\,\mbb{h}\CdoT\mbb{d}^{-2}\big\}\cdot
\big|\Tilde{\Psi}_{\boldsymbol{n}}\big({\T\mu,-\mu\!-\!\Tilde{\mu}\CdoT\frac{M}{K_{\Phi}}}\big)\big|^2+
E\big\{\,\mbb{d}^{-1}\big\}\cdot
\Tilde{\Phi}_{\boldsymbol{n}}(\mu,\mu)\CdoT
\Tilde{\Phi}_{\boldsymbol{n}}\big({\T\mu\!+\!\Tilde{\mu}\CdoT\frac{M}{K_{\Phi}},\mu\!+\!\Tilde{\mu}\CdoT\frac{M}{K_{\Phi}}}\big)&
\end{flalign*}\displaybreak[2]\\[15pt]\label{E.3.56.b}\begin{flalign}
&C_{\Hat{\boldsymbol{\Phi}}_{\!\boldsymbol{n}}(\mu,\mu+\Tilde{\mu}\cdot\frac{M}{K_{\Phi}}),\Hat{\boldsymbol{\Phi}}_{\!\boldsymbol{n}}(\mu,\mu+\Tilde{\mu}\cdot\frac{M}{K_{\Phi}})^{\Kk}}\;={}&&
\end{flalign}\\*[4pt]\begin{flalign*}
&&{}\!\!\!\!\!\!\!\!\!\!{}=\;E\big\{\,\mbb{i}\CdoT\mbb{d}^{-2}\big\}\cdot
\Tilde{\Phi}_{\boldsymbol{n}}\big({\T\mu,\mu\!+\!\Tilde{\mu}\CdoT\frac{M}{K_{\Phi}}}\big)^{\!2}+
E\big\{\,\mbb{j}\CdoT\mbb{d}^{-2}\big\}\cdot
\Tilde{\Psi}_{\boldsymbol{n}}(\mu,-\mu)\CdoT
\Tilde{\Psi}_{\boldsymbol{n}}\big({\T\mu\!+\!\Tilde{\mu}\CdoT\frac{M}{K_{\Phi}},-\mu\!-\!\Tilde{\mu}\CdoT\frac{M}{K_{\Phi}}}\big)^{\!\Kk}&
\end{flalign*}\\[15pt]\label{E.3.56.c}\begin{flalign}
&C_{\Hat{\boldsymbol{\Psi}}_{\!\boldsymbol{n}}(\mu,\mu+\Tilde{\mu}\cdot\frac{M}{K_{\Phi}}),\Hat{\boldsymbol{\Psi}}_{\!\boldsymbol{n}}(\mu,\mu+\Tilde{\mu}\cdot\frac{M}{K_{\Phi}})}\;={}&&
\end{flalign}\\*[4pt]\begin{flalign*}
&&{}\!\!\!\!\!\!\!\!\!\!{}=\;E\big\{\,\mbb{k}\CdoT\mbb{g}^{-2}\big\}\cdot
\big|\Tilde{\Phi}_{\boldsymbol{n}}\big({\T\mu,-\mu\!-\!\Tilde{\mu}\CdoT\frac{M}{K_{\Phi}}}\big)\big|^2+
E\big\{\,\mbb{g}^{-1}\big\}\cdot
\Tilde{\Phi}_{\boldsymbol{n}}(\mu,\mu)\CdoT
\Tilde{\Phi}_{\boldsymbol{n}}\big(\!{\T-\mu\!-\!\Tilde{\mu}\CdoT\frac{M}{K_{\Phi}},-\mu\!-\!\Tilde{\mu}\CdoT\frac{M}{K_{\Phi}}}\big)&
\end{flalign*}\\[15pt]\label{E.3.56.d}\begin{flalign}
&C_{\Hat{\boldsymbol{\Psi}}_{\!\boldsymbol{n}}(\mu,\mu+\Tilde{\mu}\cdot\frac{M}{K_{\Phi}}),\Hat{\boldsymbol{\Psi}}_{\!\boldsymbol{n}}(\mu,\mu+\Tilde{\mu}\cdot\frac{M}{K_{\Phi}})^{\Kk}}\;={}&&
\end{flalign}\\*[4pt]\begin{flalign*}
&&{}\!\!\!\!\!\!\!\!\!\!{}=\;E\big\{\,\mbb{l}\CdoT\mbb{g}^{-2}\big\}\cdot
\Tilde{\Psi}_{\boldsymbol{n}}\big({\T\mu,\mu\!+\!\Tilde{\mu}\CdoT\frac{M}{K_{\Phi}}}\big)^2+
E\big\{\mbb{m}\CdoT\mbb{g}^{-2}\big\}\cdot
\Tilde{\Psi}_{\boldsymbol{n}}(\mu,-\mu)\CdoT
\Tilde{\Psi}_{\boldsymbol{n}}\big(\!{\T-\mu\!-\!\Tilde{\mu}\CdoT\frac{M}{K_{\Phi}},\mu\!+\!\Tilde{\mu}\CdoT\frac{M}{K_{\Phi}}}\big)&
\end{flalign*}\\[10pt]
{\T\forall\qquad\mu=0\;\big(\frac{M}{2\cdot K_{\Phi}}\big)\;M\!-\!1
\quad\qquad\text{ und }\qquad\Tilde{\mu}=0\;(1)\;K_{\Phi}\!-\!1.}
\end{gather*}
\end{subequations}
Die bei der Berechnung der Messwert"-(ko)"-varianzen auftretenden
Terme \mbb{d}, \mbb{g}, \mbb{h}, \mbb{i}, \mbb{j}, \mbb{k}, \mbb{l}
und \mbb{m} lassen sich mit Hilfe der Tabelle\footnote{In
der Tabelle~\ref{T.3.2} sind die aus den konkreten Stichproben
berechneten nicht zuf"alligen, und daher nicht fettgedruckten Werte
eingetragen, die bei der Berechnung der konkreten Sch"atzwerte der
Messwert"-(ko)"-varianzen ben"otigt werden. Bei der hier vorliegenden
Substitution sind jedoch die aus den mathematischen Stichproben abgeleiteten
zuf"alligen Werte einzusetzen. Da sich diese in derselben Art berechnen,
wie die konkreten Werte aus den konkreten Stichproben, wurde darauf
verzichtet, dieselbe Tabelle noch einmal in Fettdruck abzudrucken.}~\ref{T.3.2}
aus den bei den unterschiedlichen Frequenzen verwendeten Matrizen
\mbox{$\underline{\boldsymbol{V}}_{\bot}\!(\mu)$} berechnen.
\begin{table}[btp]
\[
{\renewcommand{\arraystretch}{1.8}
\begin{array}{||r@{\;\;=\;\;}l||}
\hline
\hline
\mb{a}&\text{spur}\Big(
\underline{V}_{\bot}\!(\mu)\Big)\\
\mb{b}&\text{spur}\Big(
\underline{V}_{\bot}\big({\T\mu\!+\!\Tilde{\mu}\CdoT\frac{M}{K_{\Phi}}}\big)\Big)\\
\rule[-12pt]{0pt}{12pt}\mb{c}&\text{spur}\Big(
\underline{V}_{\bot}\big(\!{\T-\mu\!-\!\Tilde{\mu}\CdoT\frac{M}{K_{\Phi}}}\big)\Big)\\
\hline
\mb{d}&\text{spur}\Big(
\underline{V}_{\bot}\!(\mu)\cdot
\underline{V}_{\bot}\big({\T\mu\!+\!\Tilde{\mu}\CdoT\frac{M}{K_{\Phi}}}\big)\Big)\\
\mb{e}&\text{spur}\Big(
\underline{V}_{\bot}\!(\mu)\cdot
\underline{V}_{\bot}\big({\T\mu\!+\!\Tilde{\mu}\CdoT\frac{M}{K_{\Phi}}}\big)^{\!\Kk}\Big)\\
\mb{f}&\text{spur}\Big(
\underline{V}_{\bot}\!(\mu)\cdot
\underline{V}_{\bot}\big(\!{\T-\mu\!-\!\Tilde{\mu}\CdoT\frac{M}{K_{\Phi}}}\big)\Big)\\
\rule[-12pt]{0pt}{12pt}\mb{g}&\text{spur}\Big(
\underline{V}_{\bot}\!(\mu)\cdot
\underline{V}_{\bot}\big(\!{\T-\mu\!-\!\Tilde{\mu}\CdoT\frac{M}{K_{\Phi}}}\big)^{\!\Kk}\Big)\\
\hline
\mb{h}&\text{spur}\Big(
\underline{V}_{\bot}\!(\mu)\cdot
\underline{V}_{\bot}\big({\T\mu\!+\!\Tilde{\mu}\CdoT\frac{M}{K_{\Phi}}}\big)^{\!\Kk}\Cdot
\underline{V}_{\bot}\!(\mu)^{\Kk}\Cdot
\underline{V}_{\bot}\big({\T\mu\!+\!\Tilde{\mu}\CdoT\frac{M}{K_{\Phi}}}\big)\Big)\\
\mb{i}&\text{spur}\Big(
\underline{V}_{\bot}\!(\mu)\cdot
\underline{V}_{\bot}\big({\T\mu\!+\!\Tilde{\mu}\CdoT\frac{M}{K_{\Phi}}}\big)\cdot
\underline{V}_{\bot}\!(\mu)\cdot
\underline{V}_{\bot}\big({\T\mu\!+\!\Tilde{\mu}\CdoT\frac{M}{K_{\Phi}}}\big)\Big)\\
\mb{j}&\text{spur}\Big(
\underline{V}_{\bot}\!(\mu)\cdot
\underline{V}_{\bot}\big({\T\mu\!+\!\Tilde{\mu}\CdoT\frac{M}{K_{\Phi}}}\big)\cdot
\underline{V}_{\bot}\big({\T\mu\!+\!\Tilde{\mu}\CdoT\frac{M}{K_{\Phi}}}\big)^{\!\Kk}\Cdot
\underline{V}_{\bot}\!(\mu)^{\Kk}\Big)\\
\mb{k}&\text{spur}\Big(
\underline{V}_{\bot}\!(\mu)\cdot
\underline{V}_{\bot}\big(\!{\T-\mu\!-\!\Tilde{\mu}\CdoT\frac{M}{K_{\Phi}}}\big)\cdot
\underline{V}_{\bot}\!(\mu)^{\Kk}\Cdot
\underline{V}_{\bot}\big(\!{\T-\mu\!-\!\Tilde{\mu}\CdoT\frac{M}{K_{\Phi}}}\big)^{\!\Kk}\Big)\\
\mb{l}&\text{spur}\Big(
\underline{V}_{\bot}\!(\mu)^{\Kk}\Cdot
\underline{V}_{\bot}\big(\!{\T-\mu\!-\!\Tilde{\mu}\CdoT\frac{M}{K_{\Phi}}}\big)\cdot
\underline{V}_{\bot}\!(\mu)^{\Kk}\Cdot
\underline{V}_{\bot}\big(\!{\T-\mu\!-\!\Tilde{\mu}\CdoT\frac{M}{K_{\Phi}}}\big)\Big)\\
\rule[-12pt]{0pt}{12pt}\mb{m}&\text{spur}\Big(
\underline{V}_{\bot}\!(\mu)\cdot
\underline{V}_{\bot}\big(\!{\T-\mu\!-\!\Tilde{\mu}\CdoT\frac{M}{K_{\Phi}}}\big)^{\!\Kk}\Cdot
\underline{V}_{\bot}\big(\!{\T-\mu\!-\!\Tilde{\mu}\CdoT\frac{M}{K_{\Phi}}}\big)\cdot
\underline{V}_{\bot}\!(\mu)^{\Kk}\Big)\;\\
\hline
\mb{n}&\text{spur}\Big(
\underline{V}_{\bot}\!(\mu)\cdot
\underline{V}_{\bot}\!(\mu)^{\Kk}\Big)\\
\mb{o}&\text{spur}\Big(
\underline{V}_{\bot}\big({\T\mu\!+\!\Tilde{\mu}\CdoT\frac{M}{K_{\Phi}}}\big)\cdot
\underline{V}_{\bot}\big({\T\mu\!+\!\Tilde{\mu}\CdoT\frac{M}{K_{\Phi}}}\big)^{\!\Kk}\Big)\\
\rule[-12pt]{0pt}{12pt}\mb{p}&\text{spur}\Big(
\underline{V}_{\bot}\big(\!{\T-\mu\!-\!\Tilde{\mu}\CdoT\frac{M}{K_{\Phi}}}\big)\cdot
\underline{V}_{\bot}\big(\!{\T-\mu\!-\!\Tilde{\mu}\CdoT\frac{M}{K_{\Phi}}}\big)^{\!\Kk}\Big)\\
\hline
\hline
\end{array}}
\]
\caption{Substitutionen in den Gleichungen~(\ref{E.3.56}), (\ref{E.3.60})
und der Tabelle~(\ref{T.3.3}). Im Anhang~\ref{E.Kap.A.6} wird gezeigt, wie
sich alle in dieser Tabelle aufgelisteten Matrixspuren aus
den bereits bei der Berechnung der Messwerte des LDS und KLDS
verwendeten empirischen Kovarianzmatrizen berechnen lassen, wenn man
die Matrizen \mbox{$\protect\underline{V}_{\bot}\!(\mu)$} mit Hilfe der 
Gleichungen~(\ref{E.3.27}) und (\ref{E.3.28}) berechnet.}
\label{T.3.2}
\end{table}
In Anhang~\ref{E.Kap.A.7} wird gezeigt, dass alle in Tabelle~\ref{T.3.2} aufgelisteten 
Terme asymptotisch mit der Mittelungsanzahl $L$ proportional steigen. Alle Erwartungswerte 
in den Gleichungen~(\ref{E.3.56}), und somit auch die Messwertvarianzen, fallen daher 
asymptotisch indirekt proportional mit $L$. Somit sind die Messwerte 
\mbox{$\Hat{\boldsymbol{\Phi}}_{\!\boldsymbol{n}}(\mu,\mu\!+\!\Tilde{\mu}\CdoT M/K_{\Phi})$}
und \mbox{$\Hat{\boldsymbol{\Psi}}_{\!\boldsymbol{n}}(\mu,\mu\!+\!\Tilde{\mu}\CdoT M/K_{\Phi})$}
konsistent. Falls man die nach dem oben beschriebenen Verfahren konstruierten Matrizen
verwendet, die die Gleichungen~(\ref{E.3.33}) erf"ullen, so kann man mit den 
Gleichungen~(\ref{E.3.55}) die Gleichheit der Matrizen
\begin{gather}
\underline{\boldsymbol{V}}_{\bot}\!\big({\T\mu\!+\!\Tilde{\mu}\CdoT\frac{M}{K_{\Phi}}}\big)^{\!\Kk}=\;
\underline{\boldsymbol{V}}_{\bot}\!(\mu)^{\Kk}=\;
\underline{\boldsymbol{V}}_{\bot}\!\big(\!{\T-\mu\!+\!\Hat{\mu}\cdot\frac{M}{K_{\Phi}}}\big)^{\!\Kk}=\;
\underline{\boldsymbol{V}}_{\bot}\!\big(\!{\T-\mu\!+\!\Hat{\mu}\cdot\frac{M}{K_{\Phi}}}\big)^{\TT}=
\notag\\*[4pt]
=\;\underline{\boldsymbol{V}}_{\bot}\!\big({\T\mu\!-\!\Hat{\mu}\cdot\frac{M}{K_{\Phi}}}\big)\;=\;
\underline{\boldsymbol{V}}_{\bot}\!(\mu)\;=\;
\underline{\boldsymbol{V}}_{\bot}\!\big({\T\mu\!+\!\Tilde{\mu}\CdoT\frac{M}{K_{\Phi}}}\big).
\label{E.3.57}
\end{gather}
zeigen. In diesem Fall sind alle in Tabelle~\ref{T.3.2} auftretenden Matrizen gleich der Matrix
\mbox{$\underline{\boldsymbol{V}}_{\bot}\!(\mu)$}, alle Matrixprodukte lassen sich aufgrund der 
Idempotenz als eine Matrix schreiben, und alle Matrixspuren sind nicht zuf"allig und gleich dem 
Wert \mbox{$L\!-\!1\!-\!K(\mu)$}. Die in den 
Gleichungen~(\ref{E.3.56}) vor den Produkten, Quadraten und Betragsquadraten 
der theoretischen Werte des LDS bzw. KLDS als Vorfaktoren auftretenden 
Erwartungswerte sind alle gleich \mbox{$\big(L\!-\!1\!-\!K(\mu)\big)^{\!-1}$}
und somit von der Erregung unabh"angig.
\begin{table}[btp]
{\small\rule{0pt}{0pt}\vspace{-16pt}
\[
{\renewcommand{\arraystretch}{2.1}
\begin{array}{||r@{\;\;\;=\;\;}c@{\;\;=\;\;}l||}
\hline
\hline
\multicolumn{1}{||c|}{\raisebox{0ex}[0pt][0pt]{$\;\underline{\boldsymbol{V}}_{\bot}\!(\mu)$}}&
\multicolumn{1}{c|}{\raisebox{0ex}[0pt][0pt]{idempotent und hermitesch}}&
\multicolumn{1}{c||}{\raisebox{0ex}[0pt][0pt]{erf"ulltGleichungen~(\ref{E.3.33})}}\\[0pt]
\hline
\mb{A}&\rule[-0pt]{0pt}{0pt}{\D
\frac{2\CdoT\mb{d}\CdoT\mb{e}\CdoT\mb{h}\!-\!
\mb{a}\CdoT\mb{b}\CdoT\mb{h}^2\!-\!
\mb{d}^2\!\CdoT\mb{e}^2}
{\mb{a}\CdoT\mb{b}\CdoT
\mb{d}^2\!\CdoT\mb{e}^2\!+\!
2\CdoT\mb{d}\CdoT\mb{e}\CdoT\mb{h}\!-\!
2\CdoT\mb{d}^2\!\CdoT\mb{e}^2\!-\!
\mb{a}\CdoT\mb{b}\CdoT\mb{h}^2}}&
{\D\frac{-2}
{\big(L\!-\!2\!-\!K(\mu)\big)\CdoT\big(L\!+\!1\!-\!K(\mu)\big)}}\;\\
\mb{B}&\rule[-0pt]{0pt}{0pt}{\D
\frac{\mb{a}\CdoT\mb{b}\CdoT
\mb{e}^2\!\CdoT\mb{h}\!-\!
\mb{d}\CdoT\mb{e}^3}
{\mb{a}\CdoT\mb{b}\CdoT
\mb{d}^2\!\CdoT\mb{e}^2\!+\!
2\CdoT\mb{d}\CdoT\mb{e}\CdoT\mb{h}\!-\!
2\CdoT\mb{d}^2\!\CdoT\mb{e}^2\!-\!
\mb{a}\CdoT\mb{b}\CdoT\mb{h}^2}}&
{\D\frac{L\!-\!1\!-\!K(\mu)}
{\big(L\!-\!2\!-\!K(\mu)\big)\CdoT\big(L\!+\!1\!-\!K(\mu)\big)}}\\
\mb{C}&\rule[-0pt]{0pt}{0pt}{\D
\frac{\mb{a}\CdoT\mb{b}
\CdoT\mb{d}\CdoT\mb{e}^2\!-\!
\mb{a}\CdoT\mb{b}\CdoT\mb{e}\CdoT\mb{h}}
{\mb{a}\CdoT\mb{b}\CdoT
\mb{d}^2\!\CdoT\mb{e}^2\!+\!
2\CdoT\mb{d}\CdoT\mb{e}\CdoT\mb{h}\!-\!
2\CdoT\mb{d}^2\!\CdoT\mb{e}^2\!-\!
\mb{a}\CdoT\mb{b}\CdoT\mb{h}^2}}&
{\D\frac{L\!-\!1\!-\!K(\mu)}
{\big(L\!-\!2\!-\!K(\mu)\big)\CdoT\big(L\!+\!1\!-\!K(\mu)\big)}}\\
\mb{D}&\rule[-0pt]{0pt}{0pt}{\D
\frac{\mb{i}\CdoT\mb{n}\CdoT\mb{o}\!-\!
2\CdoT|\mb{j}|^2}
{\mb{d}^2\!\CdoT\mb{n}\CdoT\mb{o}\!+\!
\mb{i}\CdoT\mb{n}\CdoT\mb{o}\!-\!
2\CdoT|\mb{j}|^2}}&
{\D\frac{L\!-\!3\!-\!K(\mu)}
{\big(L\!-\!2\!-\!K(\mu)\big)\CdoT\big(L\!+\!1\!-\!K(\mu)\big)}}\\
\mb{E}&\rule[-0pt]{0pt}{0pt}{\D
\frac{\mb{j}\CdoT\mb{n}\CdoT\mb{o}}
{\mb{d}^2\!\CdoT\mb{n}\CdoT\mb{o}\!+\!
\mb{i}\CdoT\mb{n}\CdoT\mb{o}\!-\!
2\CdoT|\mb{j}|^2}}&
{\D\frac{L\!-\!1\!-\!K(\mu)}
{\big(L\!-\!2\!-\!K(\mu)\big)\CdoT\big(L\!+\!1\!-\!K(\mu)\big)}}\\
\mb{F}&\rule[-0pt]{0pt}{0pt}{\D
\frac{2\CdoT\mb{g}\CdoT\mb{f}\CdoT\mb{k}\!-\!
\mb{a}\CdoT\mb{c}\CdoT\mb{k}^2\!-\!
\mb{g}^2\!\CdoT\mb{f}^2}
{\mb{a}\CdoT\mb{c}\CdoT
\mb{g}^2\!\CdoT\mb{f}^2\!+\!
2\CdoT\mb{g}\CdoT\mb{f}\CdoT\mb{k}\!-\!
2\CdoT\mb{g}^2\!\CdoT\mb{f}^2\!-\!
\mb{a}\CdoT\mb{c}\CdoT\mb{k}^2}}&
{\D\frac{-2}
{\big(L\!-\!2\!-\!K(\mu)\big)\CdoT\big(L\!+\!1\!-\!K(\mu)\big)}}\\
\mb{G}&\rule[-0pt]{0pt}{0pt}{\D
\frac{\mb{a}\CdoT\mb{c}\CdoT
\mb{f}^2\!\CdoT\mb{k}\!-\!
\mb{g}\CdoT\mb{f}^3}
{\mb{a}\CdoT\mb{c}\CdoT
\mb{g}^2\!\CdoT\mb{f}^2\!+\!
2\CdoT\mb{g}\CdoT\mb{f}\CdoT\mb{k}\!-\!
2\CdoT\mb{g}^2\!\CdoT\mb{f}^2\!-\!
\mb{a}\CdoT\mb{c}\CdoT\mb{k}^2}}&
{\D\frac{L\!-\!1\!-\!K(\mu)}
{\big(L\!-\!2\!-\!K(\mu)\big)\CdoT\big(L\!+\!1\!-\!K(\mu)\big)}}\\
\mb{H}&\rule[-0pt]{0pt}{0pt}{\D
\frac{\mb{a}\CdoT\mb{c}\CdoT
\mb{g}\CdoT\mb{f}^2\!-\!
\mb{a}\CdoT\mb{c}\CdoT\mb{f}\CdoT\mb{k}}
{\mb{a}\CdoT\mb{c}\CdoT
\mb{g}^2\!\CdoT\mb{f}^2\!+\!
2\CdoT\mb{g}\CdoT\mb{f}\CdoT\mb{k}\!-\!
2\CdoT\mb{g}^2\!\CdoT\mb{f}^2\!-\!
\mb{a}\CdoT\mb{c}\CdoT\mb{k}^2}}&
{\D\frac{L\!-\!1\!-\!K(\mu)}
{\big(L\!-\!2\!-\!K(\mu)\big)\CdoT\big(L\!+\!1\!-\!K(\mu)\big)}}\\
\mb{I}&\rule[-0pt]{0pt}{0pt}{\D
\frac{\mb{l}\CdoT\mb{n}\CdoT\mb{p}\!-\!
2\CdoT|\mb{m}|^2}
{\mb{g}^2\!\CdoT\mb{n}\CdoT\mb{p}\!+\!
\mb{l}\CdoT\mb{n}\CdoT\mb{p}\!-\!
2\CdoT|\mb{m}|^2}}&
{\D\frac{L\!-\!3\!-\!K(\mu)}
{\big(L\!-\!2\!-\!K(\mu)\big)\CdoT\big(L\!+\!1\!-\!K(\mu)\big)}}\\
\mb{J}&\rule[-16pt]{0pt}{16pt}{\D
\frac{\mb{m}\CdoT\mb{n}\CdoT\mb{p}}
{\mb{g}^2\!\CdoT\mb{n}\CdoT\mb{p}\!+\!
\mb{l}\CdoT\mb{n}\CdoT\mb{p}\!-\!
2\CdoT|\mb{m}|^2}}&
{\D\frac{L\!-\!1\!-\!K(\mu)}
{\big(L\!-\!2\!-\!K(\mu)\big)\CdoT\big(L\!+\!1\!-\!K(\mu)\big)}}\\
\hline
\mb{K}&\rule[-0pt]{0pt}{0pt}{\D
\frac{1}{\;1\!-\!\mb{a}\CdoT\mb{b}\;}}&
{\D\frac{-1}
{\big(L\!-\!2\!-\!K(\mu)\big)\CdoT\big(L\!-\!K(\mu)\big)}}\\
\mb{L}&\rule[-0pt]{0pt}{0pt}{\D
\frac{\mb{a}\CdoT\mb{b}}{\;\mb{a}\CdoT
\mb{b}\CdoT\mb{d}\!-\!\mb{d}\;}}&
{\D\frac{L\!-\!1\!-\!K(\mu)}
{\big(L\!-\!2\!-\!K(\mu)\big)\CdoT\big(L\!-\!K(\mu)\big)}}\\
\mb{M}&\rule[-0pt]{0pt}{0pt}{\D
\frac{\mb{i}}{\;\mb{d}^2\!+\!\mb{i}\;}}&
{\D\frac{1}{L\!-\!K(\mu)}}\\
\mb{N}&\rule[-0pt]{0pt}{0pt}{\D
\frac{1}{\;1\!-\!\mb{a}\CdoT\mb{c}\;}}&
{\D\frac{-1}
{\big(L\!-\!2\!-\!K(\mu)\big)\CdoT\big(L\!-\!K(\mu)\big)}}\\
\mb{O}&\rule[-0pt]{0pt}{0pt}{\D
\frac{\mb{a}\CdoT\mb{c}}{\;\mb{a}\CdoT
\mb{c}\CdoT\mb{g}\!-\!\mb{g}\;}}&
{\D\frac{L\!-\!1\!-\!K(\mu)}
{\big(L\!-\!2\!-\!K(\mu)\big)\CdoT\big(L\!-\!K(\mu)\big)}}\\
\mb{P}&\rule[-14pt]{0pt}{14pt}{\D
\frac{\mb{l}}{\;\mb{g}^2\!+\!\mb{l}\;}}&
{\D\frac{1}{L\!-\!K(\mu)}}\\
\hline
\hline
\end{array}}
\]}\vspace{-16pt}
\caption{Substitutionen in den Gleichungen~(\ref{E.3.58}) und (\ref{E.3.61}).
\mbb{a}\,--\,\mbb{p} siehe Tabelle~\ref{T.3.2}.}
\label{T.3.3}
\end{table}

Im Anhang~\ref{E.Kap.A.8} ist ebenfall angegeben, wie man die Varianzen und Kovarianzen der Messwerte 
\mbox{$\Hat{\boldsymbol{\Phi}}_{\!\boldsymbol{n}}(\mu,\mu\!+\!\Tilde{\mu}\CdoT M/K_{\Phi})$} und 
\mbox{$\Hat{\boldsymbol{\Psi}}_{\!\boldsymbol{n}}(\mu,\mu\!+\!\Tilde{\mu}\CdoT M/K_{\Phi})$} 
absch"atzen kann. Dazu setzt man die in Tabelle~\ref{T.3.1} angegebenen Substitutionen in die
Gleichungen~(\ref{E.A.40}) ein. Die Zufallsgr"o"sen, die jeweils die ersten Elemente der 
Zufallsvektoren~(\ref{E.A.47}) sind, sch"atzen die theoretischen Kovarianzen der dort 
auftretenden empirischen Kovarianzsch"atzwerte erwartungstreu ab. Diese empirischen
Kovarianzsch"atzwerte sind mit den angegebenen Substitutionen gerade
die $M$-fachen LDS bzw. KLDS-Messwerte, und somit sind die auf $M$
normierten konkreten Realisierungen der jeweils ersten Zufallsgr"o"sen
der Zufallsvektoren~(\ref{E.A.47}) die gesuchten konkreten Sch"atzwerte
f"ur die Messwert"-(ko)"-varianzen.\vspace{4pt}
\begin{subequations}\label{E.3.58}
\begin{align}
\Hat{C}_{\Hat{\boldsymbol{\Phi}}_{\!\boldsymbol{n}}(\mu,\mu+\Tilde{\mu}\cdot\frac{M}{K_{\Phi}}),\Hat{\boldsymbol{\Phi}}_{\!\boldsymbol{n}}(\mu,\mu+\Tilde{\mu}\cdot\frac{M}{K_{\Phi}})}&\;=\;
\UP{0.2}{\big[}\,\mb{A}\,,\,\mb{B}\,,\,\mb{C}\,\UP{0.2}{\big]}\cdot
\begin{bmatrix}
\big|\Hat{\Phi}_{\boldsymbol{n}}\big({\T\mu,\mu\!+\!\Tilde{\mu}\CdoT\frac{M}{K_{\Phi}}}\big)\big|^2\\[4pt]
\big|\Hat{\Psi}_{\boldsymbol{n}}\big({\T\mu,-\mu\!-\!\Tilde{\mu}\CdoT\frac{M}{K_{\Phi}}}\big)\big|^2\\[4pt]
\Hat{\Phi}_{\boldsymbol{n}}(\mu,\mu)\CdoT
\Hat{\Phi}_{\boldsymbol{n}}\big({\T\mu\!+\!\Tilde{\mu}\CdoT\frac{M}{K_{\Phi}},\mu\!+\!\Tilde{\mu}\CdoT\frac{M}{K_{\Phi}}}\big)
\end{bmatrix}\label{E.3.58.a}\\[12pt]
\Hat{C}_{\Hat{\boldsymbol{\Phi}}_{\!\boldsymbol{n}}(\mu,\mu+\Tilde{\mu}\cdot\frac{M}{K_{\Phi}}),\Hat{\boldsymbol{\Phi}}_{\!\boldsymbol{n}}(\mu,\mu+\Tilde{\mu}\cdot\frac{M}{K_{\Phi}})^{\Kk}}&\;=\;
\UP{0.2}{\big[}\,\mb{D}\,,\,\mb{E}\,\UP{0.2}{\big]}\cdot
\begin{bmatrix}
\Hat{\Phi}_{\boldsymbol{n}}\big({\T\mu,\mu\!+\!\Tilde{\mu}\CdoT\frac{M}{K_{\Phi}}}\big)^2\\[4pt]
\Hat{\Psi}_{\boldsymbol{n}}(\mu,-\mu)\CdoT
\Hat{\Psi}_{\boldsymbol{n}}\big({\T\mu\!+\!\Tilde{\mu}\CdoT\frac{M}{K_{\Phi}},-\mu\!-\!\Tilde{\mu}\CdoT\frac{M}{K_{\Phi}}}\big)^{\!\Kk}
\end{bmatrix}\label{E.3.58.b}\\[12pt]
\Hat{C}_{\Hat{\boldsymbol{\Psi}}_{\!\boldsymbol{n}}(\mu,\mu+\Tilde{\mu}\cdot\frac{M}{K_{\Phi}}),\Hat{\boldsymbol{\Psi}}_{\!\boldsymbol{n}}(\mu,\mu+\Tilde{\mu}\cdot\frac{M}{K_{\Phi}})}&\;=\;
\UP{0.2}{\big[}\,\mb{F}\,,\,\mb{G}\,,\,\mb{H}\,\UP{0.2}{\big]}\CdoT
\begin{bmatrix}
\big|\Hat{\Psi}_{\boldsymbol{n}}\big({\T\mu,\mu\!+\!\Tilde{\mu}\CdoT\frac{M}{K_{\Phi}}}\big)\big|^2\\[4pt]
\big|\Hat{\Phi}_{\boldsymbol{n}}\big({\T\mu,-\mu\!-\!\Tilde{\mu}\CdoT\frac{M}{K_{\Phi}}}\big)\big|^2\\[4pt]
\Hat{\Phi}_{\boldsymbol{n}}(\mu,\mu)\CdoT
\Hat{\Phi}_{\boldsymbol{n}}\big(\!{\T-\mu\!-\!\Tilde{\mu}\CdoT\frac{M}{K_{\Phi}},-\mu\!-\!\Tilde{\mu}\CdoT\frac{M}{K_{\Phi}}}\big)
\end{bmatrix}\label{E.3.58.c}\\[12pt]
\Hat{C}_{\Hat{\boldsymbol{\Psi}}_{\!\boldsymbol{n}}(\mu,\mu+\Tilde{\mu}\cdot\frac{M}{K_{\Phi}}),\Hat{\boldsymbol{\Psi}}_{\!\boldsymbol{n}}(\mu,\mu+\Tilde{\mu}\cdot\frac{M}{K_{\Phi}})^{\Kk}}&\;=\;
\UP{0.2}{\big[}\,\mb{I}\,,\,\mb{J}\,\UP{0.2}{\big]}\cdot
\begin{bmatrix}
\Hat{\Psi}_{\boldsymbol{n}}\big({\T\mu,\mu\!+\!\Tilde{\mu}\CdoT\frac{M}{K_{\Phi}}}\big)^2\\[4pt]
\Hat{\Psi}_{\boldsymbol{n}}(\mu,-\mu)\CdoT
\Hat{\Psi}_{\boldsymbol{n}}\big(\!{\T-\mu\!-\!\Tilde{\mu}\CdoT\frac{M}{K_{\Phi}},\mu\!+\!\Tilde{\mu}\CdoT\frac{M}{K_{\Phi}}}\big)
\end{bmatrix}\label{E.3.58.d}\\*[8pt]
{\T\forall\qquad\mu=0\;\big(\frac{M}{2\cdot K_{\Phi}}\big)\;M\!-\!1}&
\qquad\qquad\text{ und }\qquad\Tilde{\mu}=0\;(1)\;K_{\Phi}\!-\!1
\notag
\end{align}
\end{subequations}
Die hier auftretenden Terme \mb{A} bis \mb{J} sind im oberen Teil der Tabelle~\ref{T.3.3} 
zusammengestellt. Die dabei verwendeten Abk"urzungen \mb{a} bis \mb{p} lassen sich mit 
Tabelle~\ref{T.3.2} aus den bei den unterschiedlichen Frequenzen verwendeten Matrizen 
\mbox{$\underline{V}_{\bot}\!(\mu)$} berechnen. Falls man die 
nach dem oben beschriebenen Verfahren konstruierten Matrizen verwendet,
die die Gleichungen~(\ref{E.3.33}) erf"ullen, sind die in den Termen
\mb{A} bis \mb{J} auftretenden Matrizen alle gleich. Es ergeben sich 
dann f"ur die Abk"urzungen \mb{a} bis \mb{p} immer dieselben Matrixspuren 
\mbox{$L\!-\!1\!-\!K(\mu)$} und damit f"ur die Terme \mb{A} bis \mb{J} 
die in der rechten Spalte der Tabelle~\ref{T.3.3} angegebenen Quotienten. 
Von den mit diesen Quotienten berechneten Sch"atzwerten f"ur die 
Messwert"-(ko)"-varianzen kann man zeigen, dass der konkrete Sch"atzwert 
der Messwertvarianz jeweils gr"o"ser oder gleich dem Betrag des konkreten 
Sch"atzwertes der Messwertkovarianz ist. Dazu setzt man in die vier
Sch"atzwerte der Messwert"-(ko)"-varianzen in den Gleichungen~(\ref{E.3.58}) 
die Messwerte gem"a"s der Gleichungen~(\ref{E.3.34}) jeweils in der Form ein, 
die die Vektoren \mbox{$\Hat{\Vec{N}}_{\!f}(\ldots)$} enth"alt. Mit der f"ur
\mbox{$L\ge 3\!+\!K(\mu)$} stets positiven Konstante 
\mbox{$\alpha = M^2\cdot\big(L\!-\!1\!-\!K(\mu)\big)^2\cdot
\big(L\!-\!2\!-\!K(\mu)\big)\cdot\big(L\!+\!1\!-\!K(\mu)\big)$}
ist dann zu zeigen, dass
\begin{subequations}\label{E.3.59}
\begin{gather}
\alpha\cdot
\Hat{C}_{\Hat{\boldsymbol{\Phi}}_{\!\boldsymbol{n}}(\mu,\mu+\Tilde{\mu}\cdot\frac{M}{K_{\Phi}}),\Hat{\boldsymbol{\Phi}}_{\!\boldsymbol{n}}(\mu,\mu+\Tilde{\mu}\cdot\frac{M}{K_{\Phi}})}\;=
\label{E.3.59.a}\\[8pt]\begin{flalign*}
&=\;-2\cdot
\Hat{\Vec{N}}_{\!f}(\mu)\cdot
\Hat{\Vec{N}}_{\!f}\big({\T\mu\!+\!\Tilde{\mu}\CdoT\frac{M}{K_{\Phi}}}\big)^{\HH}\Cdot
\Hat{\Vec{N}}_{\!f}\big({\T\mu\!+\!\Tilde{\mu}\CdoT\frac{M}{K_{\Phi}}}\big)\cdot
\Hat{\Vec{N}}_{\!f}(\mu)^{\Hh}+{}&&
\end{flalign*}\notag\\*
{}+\;\big(L\!-\!1\!-\!K(\mu)\big)\cdot
\Hat{\Vec{N}}_{\!f}(\mu)\cdot
\Hat{\Vec{N}}_{\!f}\big({\T\mu\!+\!\Tilde{\mu}\CdoT\frac{M}{K_{\Phi}}}\big)^{\TT}\Cdot
\Hat{\Vec{N}}_{\!f}\big({\T\mu\!+\!\Tilde{\mu}\CdoT\frac{M}{K_{\Phi}}}\big)^{\!\Kk}\Cdot
\Hat{\Vec{N}}_{\!f}(\mu)^{\Hh}+{}
\notag\\*\begin{flalign*}
&&{}+\;\big(L\!-\!1\!-\!K(\mu)\big)\cdot
\Hat{\Vec{N}}_{\!f}(\mu)\cdot
\Hat{\Vec{N}}_{\!f}(\mu)^{\Hh}\Cdot
\Hat{\Vec{N}}_{\!f}\big({\T\mu\!+\!\Tilde{\mu}\CdoT\frac{M}{K_{\Phi}}}\big)\cdot
\Hat{\Vec{N}}_{\!f}\big({\T\mu\!+\!\Tilde{\mu}\CdoT\frac{M}{K_{\Phi}}}\big)^{\HH}\ge{}&
\end{flalign*}\notag\\[10pt]\begin{flalign*}
&{}\ge\;\Big|\big(L\!-\!3\!-\!K(\mu)\big)\cdot
\Hat{\Vec{N}}_{\!f}(\mu)\cdot
\Hat{\Vec{N}}_{\!f}\big({\T\mu\!+\!\Tilde{\mu}\CdoT\frac{M}{K_{\Phi}}}\big)^{\HH}\Cdot
\Hat{\Vec{N}}_{\!f}\big({\T\mu\!+\!\Tilde{\mu}\CdoT\frac{M}{K_{\Phi}}}\big)^{\!\Kk}\Cdot
\Hat{\Vec{N}}_{\!f}(\mu)^{\Tt}+{}&&
\end{flalign*}\notag\\*\begin{flalign*}
&&{}+\;\big(L\!-\!1\!-\!K(\mu)\big)\cdot
\Hat{\Vec{N}}_{\!f}(\mu)\cdot
\Hat{\Vec{N}}_{\!f}(\mu)^{\Tt}\Cdot
\Hat{\Vec{N}}_{\!f}\big({\T\mu\!+\!\Tilde{\mu}\CdoT\frac{M}{K_{\Phi}}}\big)^{\!\Kk}\Cdot
\Hat{\Vec{N}}_{\!f}\big({\T\mu\!+\!\Tilde{\mu}\CdoT\frac{M}{K_{\Phi}}}\big)^{\HH}\Big|\;={}&
\end{flalign*}\notag\\[8pt]
=\;\alpha\cdot\big|
\Hat{C}_{\Hat{\boldsymbol{\Phi}}_{\!\boldsymbol{n}}(\mu,\mu+\Tilde{\mu}\cdot\frac{M}{K_{\Phi}}),\Hat{\boldsymbol{\Phi}}_{\!\boldsymbol{n}}(\mu,\mu+\Tilde{\mu}\cdot\frac{M}{K_{\Phi}})^{\Kk}}\big|
\notag\\[10pt]\label{E.3.59.b}\begin{flalign}
\text{und}&&
\alpha\cdot
\Hat{C}_{\Hat{\boldsymbol{\Psi}}_{\!\boldsymbol{n}}(\mu,\mu+\Tilde{\mu}\cdot\frac{M}{K_{\Phi}}),\Hat{\boldsymbol{\Psi}}_{\!\boldsymbol{n}}(\mu,\mu+\Tilde{\mu}\cdot\frac{M}{K_{\Phi}})}\;={}&&&
\end{flalign}\notag\\[8pt]\begin{flalign*}
&{}=\;-2\cdot
\Hat{\Vec{N}}_{\!f}(\mu)\cdot
\Hat{\Vec{N}}_{\!f}\big(\!{\T-\mu\!-\!\Tilde{\mu}\CdoT\frac{M}{K_{\Phi}}}\big)^{\TT}\Cdot
\Hat{\Vec{N}}_{\!f}\big(\!{\T-\mu\!-\!\Tilde{\mu}\CdoT\frac{M}{K_{\Phi}}}\big)^{\!\Kk}\Cdot
\Hat{\Vec{N}}_{\!f}(\mu)^{\Hh}+{}&&
\end{flalign*}\notag\\*
{}+\;\big(L\!-\!1\!-\!K(\mu)\big)\cdot
\Hat{\Vec{N}}_{\!f}(\mu)\cdot
\Hat{\Vec{N}}_{\!f}\big(\!{\T-\mu\!-\!\Tilde{\mu}\CdoT\frac{M}{K_{\Phi}}}\big)^{\HH}\Cdot
\Hat{\Vec{N}}_{\!f}\big(\!{\T-\mu\!-\!\Tilde{\mu}\CdoT\frac{M}{K_{\Phi}}}\big)\cdot
\Hat{\Vec{N}}_{\!f}(\mu)^{\Hh}+{}
\notag\\*\begin{flalign*}
&&{}+\;\big(L\!-\!1\!-\!K(\mu)\big)\cdot
\Hat{\Vec{N}}_{\!f}(\mu)\cdot
\Hat{\Vec{N}}_{\!f}(\mu)^{\Hh}\Cdot
\Hat{\Vec{N}}_{\!f}\big(\!{\T-\mu\!-\!\Tilde{\mu}\CdoT\frac{M}{K_{\Phi}}}\big)\cdot
\Hat{\Vec{N}}_{\!f}\big(\!{\T-\mu\!-\!\Tilde{\mu}\CdoT\frac{M}{K_{\Phi}}}\big)^{\HH}\ge{}&
\end{flalign*}\notag\\[10pt]\begin{flalign*}
&{}\ge\;\Big|\big(L\!-\!3\!-\!K(\mu)\big)\cdot
\Hat{\Vec{N}}_{\!f}(\mu)\cdot
\Hat{\Vec{N}}_{\!f}\big(\!{\T-\mu\!-\!\Tilde{\mu}\CdoT\frac{M}{K_{\Phi}}}\big)^{\TT}\Cdot
\Hat{\Vec{N}}_{\!f}\big(\!{\T-\mu\!-\!\Tilde{\mu}\CdoT\frac{M}{K_{\Phi}}}\big)\cdot
\Hat{\Vec{N}}_{\!f}(\mu)^{\Tt}+{}&&
\end{flalign*}\notag\\*\begin{flalign*}
&&{}+\;\big(L\!-\!1\!-\!K(\mu)\big)\cdot
\Hat{\Vec{N}}_{\!f}(\mu)\cdot
\Hat{\Vec{N}}_{\!f}(\mu)^{\Tt}\Cdot
\Hat{\Vec{N}}_{\!f}\big(\!{\T-\mu\!-\!\Tilde{\mu}\CdoT\frac{M}{K_{\Phi}}}\big)\cdot
\Hat{\Vec{N}}_{\!f}\big(\!{\T-\mu\!-\!\Tilde{\mu}\CdoT\frac{M}{K_{\Phi}}}\big)^{\TT}\Big|\;={}&
\end{flalign*}\notag\\[8pt]
{}=\;\alpha\cdot\big|
\Hat{C}_{\Hat{\boldsymbol{\Psi}}_{\!\boldsymbol{n}}(\mu,\mu+\Tilde{\mu}\cdot\frac{M}{K_{\Phi}}),\Hat{\boldsymbol{\Psi}}_{\!\boldsymbol{n}}(\mu,\mu+\Tilde{\mu}\cdot\frac{M}{K_{\Phi}})^{\Kk}}\big|\notag\\[6pt]
{\T\forall\qquad\mu=0\;\big(\frac{M}{2\cdot K_{\Phi}}\big)\;M\!-\!1
\qquad\qquad\text{ und }\qquad\Tilde{\mu}=0\;(1)\;K_{\Phi}\!-\!1}
\notag
\end{gather}
\end{subequations}
gilt. Im Anhang~\ref{E.Kap.A.9.1} wird gezeigt, dass diese
Ungleichungen erf"ullt sind. Dazu ist in Ungleichung~(\ref{E.A.54})
des Anhangs \mbox{$a=L\!-\!1\!-\!K(\mu)$},
\mbox{$\Vec{X}=\Hat{\Vec{N}}_{\!f}(\mu)$}
und \mbox{$\Vec{Y}=
\Hat{\Vec{N}}_{\!f}(\mu\!+\!\Tilde{\mu}\CdoT M/K_{\Phi})^{\Kk}$} bzw.
\mbox{$\Vec{Y}=
\Hat{\Vec{N}}_{\!f}(\!-\mu\!-\!\Tilde{\mu}\CdoT M/K_{\Phi})$} einzusetzen.
Die Gleichung im Anhang gilt nur f"ur \mbox{$a\ge 2$}, was hier bedeutet,
dass die Mittelungsanzahl $L$ mindestens \mbox{$3\!+\!K(\mu)$} sein muss.

Nun wollen wir den Fall behandeln, dass die negativen diskreten Frequenzen 
$-\mu$ und \mbox{$-\mu\!-\!\Tilde{\mu}\CdoT M/K_{\Phi}$} {\em nicht}\/ um 
ein ganzzahliges Vielfaches von \mbox{$M/K_{\Phi}$} gegen"uber $\mu$ verschoben 
sind, und somit die Bedingung~(\ref{E.3.55.a}) nicht erf"ullt ist. In diesem 
Fall sind nach den Gleichungen~(\ref{E.2.41}) und (\ref{E.2.51}) viele der 
m"oglichen Kovarianzen zweier Zufallsgr"o"sen, die man sich aus aus den Zufallsgr"o"sen
\mbox{$\boldsymbol{N}_{\!\!f}(\mu)$}, 
\mbox{$\boldsymbol{N}_{\!\!f}\big({\T\mu\!+\!\Tilde{\mu}\CdoT\frac{M}{K_{\Phi}}}\big)$}, 
\mbox{$\boldsymbol{N}_{\!\!f}(\!-\mu)$}, 
\mbox{$\boldsymbol{N}_{\!\!f}\big(\!{\T-\mu\!-\!\Tilde{\mu}\CdoT\frac{M}{K_{\Phi}}}\big)$} 
und den dazu konjugierten Zufallsgr"o"sen herausgreift, bei Verwendung einer hoch 
frequenzselektiven Fensterfolge in guter N"aherung null. F"ur diese Kovarianzen 
haben wir daher auch keine Messwerte berechnet. Daher verwenden wir nun die in 
Tabelle~\ref{T.3.1} aufgelisteten Substitutionen in den Gleichungen~(\ref{E.A.42}) 
---\,und nicht in den Gleichungen~(\ref{E.A.40}) wie im Fall~(\ref{E.3.55.a})\,---, 
um so eine reduzierte Vektorgleichung~(\ref{E.A.41}) zur Berechnung der vier 
gesuchten theoretischen Messwert"-(ko)"-varianzen und -kovarianzen zu erhalten. 
Jeweils die auf $M$ normierte erste Zeile dieser reduzierten Vektorgleichung 
liefert uns die theoretischen Messwert"-(ko)"-varianzen:\vspace{-10pt}
\begin{subequations}\label{E.3.60}
\begin{align}
C_{\Hat{\boldsymbol{\Phi}}_{\!\boldsymbol{n}}(\mu,\mu+\Tilde{\mu}\cdot\frac{M}{K_{\Phi}}),\Hat{\boldsymbol{\Phi}}_{\!\boldsymbol{n}}(\mu,\mu+\Tilde{\mu}\cdot\frac{M}{K_{\Phi}})}\,&=\,
E\big\{\,\mbb{d}^{-1}\big\}\cdot
\Tilde{\Phi}_{\boldsymbol{n}}(\mu,\mu)\CdoT
\Tilde{\Phi}_{\boldsymbol{n}}\big({\T\mu\!+\!\Tilde{\mu}\CdoT\frac{M}{K_{\Phi}},\mu\!+\!\Tilde{\mu}\CdoT\frac{M}{K_{\Phi}}}\big)
\label{E.3.60.a}\\[6pt]
\!\!C_{\Hat{\boldsymbol{\Phi}}_{\!\boldsymbol{n}}(\mu,\mu+\Tilde{\mu}\cdot\frac{M}{K_{\Phi}}),\Hat{\boldsymbol{\Phi}}_{\!\boldsymbol{n}}(\mu,\mu+\Tilde{\mu}\cdot\frac{M}{K_{\Phi}})^{\Kk}}&=\,
E\big\{\,\mbb{i}\CdoT\mbb{d}^{-2}\big\}\cdot
\Tilde{\Phi}_{\boldsymbol{n}}\big({\T\mu,\mu\!+\!\Tilde{\mu}\CdoT\frac{M}{K_{\Phi}}}\big)^2
\label{E.3.60.b}\\[6pt]
C_{\Hat{\boldsymbol{\Psi}}_{\!\boldsymbol{n}}(\mu,\mu+\Tilde{\mu}\cdot\frac{M}{K_{\Phi}}),\Hat{\boldsymbol{\Psi}}_{\!\boldsymbol{n}}(\mu,\mu+\Tilde{\mu}\cdot\frac{M}{K_{\Phi}})}\,&=\,
E\big\{\mbb{g}^{-1}\big\}\CdoT
\Tilde{\Phi}_{\boldsymbol{n}}(\mu,\mu)\CdoT
\Tilde{\Phi}_{\boldsymbol{n}}\big(\!{\T-\mu\!-\!\Tilde{\mu}\CdoT\frac{M}{K_{\Phi}},-\mu\!-\!\Tilde{\mu}\CdoT\frac{M}{K_{\Phi}}}\big)\!\!
\label{E.3.60.c}\\[6pt]
\!\!C_{\Hat{\boldsymbol{\Psi}}_{\!\boldsymbol{n}}(\mu,\mu+\Tilde{\mu}\cdot\frac{M}{K_{\Phi}}),\Hat{\boldsymbol{\Psi}}_{\!\boldsymbol{n}}(\mu,\mu+\Tilde{\mu}\cdot\frac{M}{K_{\Phi}})^{\Kk}}&=\,
E\big\{\,\mbb{l}\CdoT\mbb{g}^{-2}\big\}\cdot
\Tilde{\Psi}_{\boldsymbol{n}}\big({\T\mu,\mu\!+\!\Tilde{\mu}\CdoT\frac{M}{K_{\Phi}}}\big)^2
\label{E.3.60.d}\\*[6pt]
\forall\qquad\Tilde{\mu}=0\;(1)\;K_{\Phi}\!-\!1\qquad&
{\T\text{ und }\qquad\mu=1\;(1)\;M\!-\!1
\quad\text{ ohne }\quad\mu=0\;\big(\frac{M}{2\cdot K_{\Phi}}\big)\;M\!-\!1.}
\notag
\end{align}
\end{subequations}
Die bei der Berechnung der Messwert"-(ko)"-varianzen auftretenden Terme  \mbb{d}, \mbb{g}, \mbb{i} 
und \mbb{l} sind als Spuren der bei den unterschiedlichen Frequenzen verwendeten Matrizen
\mbox{$\underline{\boldsymbol{V}}_{\bot}\!(\mu)$} zu berechnen, wie dies in 
Tabelle~\ref{T.3.2} angegeben ist. Alle Erwartungswerte in den Gleichungen~(\ref{E.3.60}), 
und somit auch die Messwertvarianzen, fallen asymptotisch indirekt proportional mit $L$
(\,siehe Anhang~\ref{E.Kap.A.7}\,). Somit sind die Messwerte 
\mbox{$\Hat{\boldsymbol{\Phi}}_{\!\boldsymbol{n}}(\mu,\mu\!+\!\Tilde{\mu}\CdoT M/K_{\Phi})$}
und \mbox{$\Hat{\boldsymbol{\Psi}}_{\!\boldsymbol{n}}(\mu,\mu\!+\!\Tilde{\mu}\CdoT M/K_{\Phi})$}
konsistent. Falls man die nach dem oben beschriebenen Verfahren konstruierten Matrizen 
\mbox{$\underline{\boldsymbol{V}}_{\bot}\!(\mu)$} verwendet, die die Gleichungen~(\ref{E.3.33}) 
erf"ullen, sind die alle Erwartungswerte, die als Faktoren vor den Produkten, Quadraten und 
Betragsquadraten der theoretischen Werte des LDS bzw. KLDS auftreten, auch hier gleich 
\mbox{$\big(L\!-\!1\!-\!K(\mu)\big)^{-1}$} und somit von der Erregung unabh"angig. 
Jeweils eine auf $M$ normierte konkrete Realisierung des ersten Elementes des erwartungstreuen 
Zufallsvektors~(\ref{E.A.47}) liefert uns die gesuchten konkreten Sch"atzwerte f"ur die 
Messwert"-(ko)"-varianzen:\vspace{-4pt}
\begin{subequations}\label{E.3.61}
\begin{align}
\Hat{C}_{\Hat{\boldsymbol{\Phi}}_{\!\boldsymbol{n}}(\mu,\mu+\Tilde{\mu}\cdot\frac{M}{K_{\Phi}}),\Hat{\boldsymbol{\Phi}}(\mu,\mu+\Tilde{\mu}\cdot\frac{M}{K_{\Phi}})}\,&=\,
\UP{0.2}{\big[}\,\mb{K}\,,\,\mb{L}\,\UP{0.2}{\big]}\CdoT
\begin{bmatrix}
\big|\Hat{\Phi}_{\boldsymbol{n}}\big({\T\mu,\mu\!+\!\Tilde{\mu}\CdoT\frac{M}{K_{\Phi}}}\big)\big|^2\\[4pt]
\Hat{\Phi}_{\boldsymbol{n}}(\mu,\mu)\CdoT
\Hat{\Phi}_{\boldsymbol{n}}\big({\T\mu\!+\!\Tilde{\mu}\CdoT\frac{M}{K_{\Phi}},\mu\!+\!\Tilde{\mu}\CdoT\frac{M}{K_{\Phi}}}\big)
\end{bmatrix}\!\!\label{E.3.61.a}\\[6pt]
\!\!\Hat{C}_{\Hat{\boldsymbol{\Phi}}_{\!\boldsymbol{n}}(\mu,\mu+\Tilde{\mu}\cdot\frac{M}{K_{\Phi}}),\Hat{\boldsymbol{\Phi}}_{\!\boldsymbol{n}}(\mu,\mu+\Tilde{\mu}\cdot\frac{M}{K_{\Phi}})^{\Kk}}&=\;
\mb{M}\cdot
\Hat{\Phi}_{\boldsymbol{n}}\big({\T\mu,\mu\!+\!\Tilde{\mu}\CdoT\frac{M}{K_{\Phi}}}\big)^2
\label{E.3.61.b}\displaybreak[2]\\[6pt]
\Hat{C}_{\Hat{\boldsymbol{\Psi}}_{\!\boldsymbol{n}}(\mu,\mu+\Tilde{\mu}\cdot\frac{M}{K_{\Phi}}),\Hat{\boldsymbol{\Psi}}_{\!\boldsymbol{n}}(\mu,\mu+\Tilde{\mu}\cdot\frac{M}{K_{\Phi}})}\,&=\,
\UP{0.2}{\big[}\,\mb{N}\,,\,\mb{O}\,\UP{0.2}{\big]}\CdoT
\begin{bmatrix}
\big|\Hat{\Psi}_{\boldsymbol{n}}\big({\T\mu,\mu\!+\!\Tilde{\mu}\CdoT\frac{M}{K_{\Phi}}}\big)\big|^2\\[4pt]
\Hat{\Phi}_{\boldsymbol{n}}(\mu,\mu)\CdoT
\Hat{\Phi}_{\boldsymbol{n}}\big(\!{\T-\mu\!-\!\Tilde{\mu}\CdoT\frac{M}{K_{\Phi}},-\mu\!-\!\Tilde{\mu}\CdoT\frac{M}{K_{\Phi}}}\big)
\end{bmatrix}\!\!\label{E.3.61.c}\\[6pt]
\!\!\Hat{C}_{\Hat{\boldsymbol{\Psi}}_{\!\boldsymbol{n}}(\mu,\mu+\Tilde{\mu}\cdot\frac{M}{K_{\Phi}}),\Hat{\boldsymbol{\Psi}}_{\!\boldsymbol{n}}(\mu,\mu+\Tilde{\mu}\cdot\frac{M}{K_{\Phi}})^{\Kk}}&=\;
\mb{P}\cdot
\Hat{\Psi}_{\boldsymbol{n}}\big({\T\mu,\mu\!+\!\Tilde{\mu}\CdoT\frac{M}{K_{\Phi}}}\big)^2
\label{E.3.61.d}\\*[4pt]
\forall\qquad\Tilde{\mu}=0\;(1)\;K_{\Phi}\!-\!1\qquad&
{\T\text{ und }\qquad\mu=1\;(1)\;M\!-\!1
\quad\text{ ohne }\quad\mu=0\;\big(\frac{M}{2\cdot K_{\Phi}}\big)\;M\!-\!1}.
\notag
\end{align}
\end{subequations}
Die bei der Berechnung der Sch"atzwerte der Messwert"-(ko)"-varianzen 
auftretenden Terme \mb{K} bis \mb{P} entnimmt man dem unteren Teil 
der Tabelle~\ref{T.3.3}. Dabei lassen sich die dort auftretenden 
Abk"urzungen \mb{a}, \mb{b}, \mb{c}, \mb{d}, \mb{g}, \mb{i} und 
\mb{l} mit Hilfe der Tabelle~\ref{T.3.2} aus den bei den 
unterschiedlichen Frequenzen verwendeten Matrizen 
\mbox{$\underline{V}_{\bot}\!(\mu)$} berechnen. Falls man die nach 
dem oben beschriebenen Verfahren konstruierten Matrizen verwendet, 
sind die in den Termen \mb{K} bis \mb{P} auftretenden Matrixspuren 
gem"a"s der Gleichungen~(\ref{E.3.33.b}) alle gleich \mbox{$L\!-\!1\!-\!K(\mu)$}. 
Es ergeben sich dann f"ur diese Terme die in der rechten Spalte 
der Tabelle~\ref{T.3.3} angegebenen Quotienten. Bei den mit 
diesen Quotienten berechneten Sch"atzwerten kann man zeigen, 
dass jeweils der konkrete Sch"atzwert der Messwertvarianz gr"o"ser 
oder gleich dem Betrag des konkreten Sch"atzwertes der Messwertkovarianz 
ist, falls die Mittelungsanzahl $L$ mindestens \mbox{$3\!+\!K(\mu)$} ist. 
Wir beginnen mit den Ungleichungen~(\ref{E.3.36}) und formen diese nach 
und nach um, bis wir die gew"unschten Ungleichungen f"ur die Sch"atzwerte
der Messwert"-(ko)"-varianzen erhalten:
\begin{subequations}\label{E.3.62}
\begin{gather}
\Hat{\Phi}_{\boldsymbol{n}}(\mu,\mu)\CdoT
\Hat{\Phi}_{\boldsymbol{n}}
\big({\T\mu\!+\!\Tilde{\mu}\CdoT\frac{M}{K_{\Phi}},
\mu\!+\!\Tilde{\mu}\CdoT\frac{M}{K_{\Phi}}}\big)
\;\ge\;
\big|\Hat{\Phi}_{\boldsymbol{n}}
\big({\T\mu,\mu\!+\!\Tilde{\mu}\CdoT\frac{M}{K_{\Phi}}}\big)\big|^2
\notag\\[8pt]
\big(L\!-\!1\!-\!K(\mu)\big)\CdoT
\Hat{\Phi}_{\boldsymbol{n}}(\mu,\mu)\CdoT
\Hat{\Phi}_{\boldsymbol{n}}
\big({\T\mu\!+\!\Tilde{\mu}\CdoT\frac{M}{K_{\Phi}},
\mu\!+\!\Tilde{\mu}\CdoT\frac{M}{K_{\Phi}}}\big)
\;\ge\;\big(L\!-\!1\!-\!K(\mu)\big)\CdoT
\big|\Hat{\Phi}_{\boldsymbol{n}}
\big({\T\mu,\mu\!+\!\Tilde{\mu}\CdoT\frac{M}{K_{\Phi}}}\big)\big|^2
\notag\\[12pt]\begin{flalign*}
&\big(L\!-\!1\!-\!K(\mu)\big)\CdoT
\Hat{\Phi}_{\boldsymbol{n}}(\mu,\mu)\CdoT
\Hat{\Phi}_{\boldsymbol{n}}
\big({\T\mu\!+\!\Tilde{\mu}\CdoT\frac{M}{K_{\Phi}},
\mu\!+\!\Tilde{\mu}\CdoT\frac{M}{K_{\Phi}}}\big)-
\big|\Hat{\Phi}_{\boldsymbol{n}}
\big({\T\mu,\mu\!+\!\Tilde{\mu}\CdoT\frac{M}{K_{\Phi}}}\big)\big|^2\ge{}&&
\end{flalign*}\notag\\*[2pt]\begin{flalign*}
&&{}\ge\;\big(L\!-\!1\!-\!K(\mu)\big)\CdoT
\big|\Hat{\Phi}_{\boldsymbol{n}}
\big({\T\mu,\mu\!+\!\Tilde{\mu}\CdoT\frac{M}{K_{\Phi}}}\big)\big|^2\!-
\big|\Hat{\Phi}_{\boldsymbol{n}}
\big({\T\mu,\mu\!+\!\Tilde{\mu}\CdoT\frac{M}{K_{\Phi}}}\big)\big|^2&
\end{flalign*}\notag\\[12pt]\begin{flalign*}
&\big(L\!-\!1\!-\!K(\mu)\big)\CdoT
\Hat{\Phi}_{\boldsymbol{n}}(\mu,\mu)\CdoT
\Hat{\Phi}_{\boldsymbol{n}}
\big({\T\mu\!+\!\Tilde{\mu}\CdoT\frac{M}{K_{\Phi}},
\mu\!+\!\Tilde{\mu}\CdoT\frac{M}{K_{\Phi}}}\big)-
\big|\Hat{\Phi}_{\boldsymbol{n}}
\big({\T\mu,\mu\!+\!\Tilde{\mu}\CdoT\frac{M}{K_{\Phi}}}\big)\big|^2\ge{}&&
\end{flalign*}\notag\\*\begin{flalign*}
&&{}\ge\;\big(L\!-\!2\!-\!K(\mu)\big)\CdoT
\big|\Hat{\Phi}_{\boldsymbol{n}}
\big({\T\mu,\mu\!+\!\Tilde{\mu}\CdoT\frac{M}{K_{\Phi}}}\big)\big|^2&
\end{flalign*}\notag\\[12pt]\begin{flalign*}
&\frac{\big[\;-1\;,\,\,L\!-\!1\!-\!K(\mu)\;\big]}
{\big(L\!-\!2\!-\!K(\mu)\big)\CdoT\big(L\!-\!K(\mu)\big)}\cdot
\begin{bmatrix}
{\D\big|\Hat{\Phi}_{\boldsymbol{n}}
\big({\T\mu,\mu\!+\!\Tilde{\mu}\CdoT\frac{M}{K_{\Phi}}}\big)\big|^2}\\[4pt]
\Hat{\Phi}_{\boldsymbol{n}}(\mu,\mu)\CdoT
\Hat{\Phi}_{\boldsymbol{n}}
\big({\T\mu\!+\!\Tilde{\mu}\CdoT\frac{M}{K_{\Phi}},
\mu\!+\!\Tilde{\mu}\CdoT\frac{M}{K_{\Phi}}}\big)
\end{bmatrix}\ge{}&&
\end{flalign*}\notag\\*[2pt]\begin{flalign*}
&&{}\ge\;\frac{L\!-\!2\!-\!K(\mu)}
{\big(L\!-\!2\!-\!K(\mu)\big)\CdoT\big(L\!-\!K(\mu)\big)}\cdot
\big|\Hat{\Phi}_{\boldsymbol{n}}
\big({\T\mu,\mu\!+\!\Tilde{\mu}\CdoT\frac{M}{K_{\Phi}}}\big)\big|^2&
\end{flalign*}\notag\\[8pt]
\Hat{C}_{\Hat{\boldsymbol{\Phi}}_{\!\boldsymbol{n}}(\mu,\mu+\Tilde{\mu}\cdot\frac{M}{K_{\Phi}}),\Hat{\boldsymbol{\Phi}}_{\!\boldsymbol{n}}(\mu,\mu+\Tilde{\mu}\cdot\frac{M}{K_{\Phi}})}\;\ge\;
\big|\Hat{C}_{\Hat{\boldsymbol{\Phi}}_{\!\boldsymbol{n}}(\mu,\mu+\Tilde{\mu}\cdot\frac{M}{K_{\Phi}}),\Hat{\boldsymbol{\Phi}}_{\!\boldsymbol{n}}(\mu,\mu+\Tilde{\mu}\cdot\frac{M}{K_{\Phi}})^{\Kk}}\big|
\label{E.3.62.a}\\[-4pt]\intertext{und\vspace{-12pt}}\begin{flalign*}
&&\Hat{\Phi}_{\boldsymbol{n}}(\mu,\mu)\CdoT
\Hat{\Phi}_{\boldsymbol{n}}
\big(\!{\T-\mu\!-\!\Tilde{\mu}\CdoT\frac{M}{K_{\Phi}},
-\mu\!-\!\Tilde{\mu}\CdoT\frac{M}{K_{\Phi}}}\big)&\;\ge\;
\big|\Hat{\Psi}_{\boldsymbol{n}}
\big({\T\mu,\mu\!+\!\Tilde{\mu}\CdoT\frac{M}{K_{\Phi}}}\big)\big|^2&&
\end{flalign*}\notag\\[8pt]
\big(L\!-\!1\!-\!K(\mu)\big)\CdoT
\Hat{\Phi}_{\boldsymbol{n}}(\mu,\mu)\CdoT
\Hat{\Phi}_{\boldsymbol{n}}
\big({\T\!-\mu\!-\!\Tilde{\mu}\CdoT\frac{M}{K_{\Phi}},\!
-\mu\!-\!\Tilde{\mu}\CdoT\frac{M}{K_{\Phi}}}\big)\!\ge\!
\big(L\!-\!1\!-\!K(\mu)\big)\CdoT
\big|\Hat{\Psi}_{\boldsymbol{n}}
\big({\T\mu,\mu\!+\!\Tilde{\mu}\CdoT\frac{M}{K_{\Phi}}}\big)\big|^2
\notag\\[12pt]\begin{flalign*}
&\big(L\!-\!1\!-\!K(\mu)\big)\CdoT
\Hat{\Phi}_{\boldsymbol{n}}(\mu,\mu)\CdoT
\Hat{\Phi}_{\boldsymbol{n}}
\big(\!{\T-\mu\!-\!\Tilde{\mu}\CdoT\frac{M}{K_{\Phi}},
-\mu\!-\!\Tilde{\mu}\CdoT\frac{M}{K_{\Phi}}}\big)-
\big|\Hat{\Psi}_{\boldsymbol{n}}
\big({\T\mu,\mu\!+\!\Tilde{\mu}\CdoT\frac{M}{K_{\Phi}}}\big)\big|^2\ge{}&&
\end{flalign*}\notag\\*[2pt]\begin{flalign*}
&&{}\ge\;\big(L\!-\!1\!-\!K(\mu)\big)\CdoT
\big|\Hat{\Psi}_{\boldsymbol{n}}
\big({\T\mu,\mu\!+\!\Tilde{\mu}\CdoT\frac{M}{K_{\Phi}}}\big)\big|^2\!-
\big|\Hat{\Psi}_{\boldsymbol{n}}
\big({\T\mu,\mu\!+\!\Tilde{\mu}\CdoT\frac{M}{K_{\Phi}}}\big)\big|^2&
\end{flalign*}\notag\\[12pt]\begin{flalign*}
&\big(L\!-\!1\!-\!K(\mu)\big)\CdoT
\Hat{\Phi}_{\boldsymbol{n}}(\mu,\mu)\CdoT
\Hat{\Phi}_{\boldsymbol{n}}
\big(\!{\T-\mu\!-\!\Tilde{\mu}\CdoT\frac{M}{K_{\Phi}},
-\mu\!-\!\Tilde{\mu}\CdoT\frac{M}{K_{\Phi}}}\big)-
\big|\Hat{\Psi}_{\boldsymbol{n}}
\big({\T\mu,\mu\!+\!\Tilde{\mu}\CdoT\frac{M}{K_{\Phi}}}\big)\big|^2\ge{}&&
\end{flalign*}\notag\\*\begin{flalign*}
&&{}\ge\;\big(L\!-\!2\!-\!K(\mu)\big)\CdoT
\big|\Hat{\Psi}_{\boldsymbol{n}}
\big({\T\mu,\mu\!+\!\Tilde{\mu}\CdoT\frac{M}{K_{\Phi}}}\big)\big|^2&
\end{flalign*}\notag\\[12pt]\begin{flalign*}
&\frac{\big[\;-1\;,\,\,L\!-\!1\!-\!K(\mu)\;\big]}
{\big(L\!-\!2\!-\!K(\mu)\big)\CdoT\big(L\!-\!K(\mu)\big)}\cdot
\begin{bmatrix}
{\D\big|\Hat{\Psi}_{\boldsymbol{n}}
\big({\T\mu,\mu\!+\!\Tilde{\mu}\CdoT\frac{M}{K_{\Phi}}}\big)\big|^2}\\[4pt]
\Hat{\Phi}_{\boldsymbol{n}}(\mu,\mu)\CdoT
\Hat{\Phi}_{\boldsymbol{n}}
\big(\!{\T-\mu\!-\!\Tilde{\mu}\CdoT\frac{M}{K_{\Phi}},
-\mu\!-\!\Tilde{\mu}\CdoT\frac{M}{K_{\Phi}}}\big)
\end{bmatrix}\ge{}&&
\end{flalign*}\notag\\*[2pt]\begin{flalign*}
&&{}\ge\;\frac{L\!-\!2\!-\!K(\mu)}
{\big(L\!-\!2\!-\!K(\mu)\big)\CdoT\big(L\!-\!K(\mu)\big)}\cdot
\big|\Hat{\Psi}_{\boldsymbol{n}}
\big({\T\mu,\mu\!+\!\Tilde{\mu}\CdoT\frac{M}{K_{\Phi}}}\big)\big|^2&
\end{flalign*}\notag\\[8pt]
\Hat{C}_{\Hat{\boldsymbol{\Psi}}_{\!\boldsymbol{n}}(\mu,\mu+\Tilde{\mu}\cdot\frac{M}{K_{\Phi}}),\Hat{\boldsymbol{\Psi}}_{\!\boldsymbol{n}}(\mu,\mu+\Tilde{\mu}\cdot\frac{M}{K_{\Phi}})}\;\ge\;
\big|\Hat{C}_{\Hat{\boldsymbol{\Psi}}_{\!\boldsymbol{n}}(\mu,\mu+\Tilde{\mu}\cdot\frac{M}{K_{\Phi}}),\Hat{\boldsymbol{\Psi}}_{\!\boldsymbol{n}}(\mu,\mu+\Tilde{\mu}\cdot\frac{M}{K_{\Phi}})^{\Kk}}\big|
\label{E.3.62.b}\\[8pt]
{\T\forall\qquad\Tilde{\mu}=0\;(1)\;K_{\Phi}\!-\!1
\qquad\text{ und}\qquad\mu=1\;(1)\;M\!-\!1
\quad\text{ ohne}\quad\mu=0\;\big(\frac{M}{2\cdot K_{\Phi}}\big)\;M\!-\!1}
\notag
\end{gather}
\end{subequations}

Damit k"onnen wir  f"ur alle Messwerte Sch"atzwerte f"ur deren 
Varianzen und Kovarianzen angeben. Auch hier war lediglich bei 
den Messwerten des LDS und des KLDS eine Verbundnormalverteilung 
angenommen worden, um die dort auftretenden vierten Momente auf 
deren bereits gemessenen zweiten Momente zur"uckf"uhren zu k"onnen. 
Alle anderen Messwert"-(ko)"-varianzen konnten ohne die Kenntnis 
ihrer Verbundverteilung berechnet werden. Die Messwerte 
\mbox{$\Hat{\Phi}_{\boldsymbol{n}}(\mu_1,\mu_2)$} des bifrequenten LDS 
mit gleichen Frequenzen \mbox{$\mu_1\!=\!\mu_2$} sind die einzigen 
Messwerte die immer reell sind. Deren Kovarianzsch"atzwerte sind 
immer gleich ihren Varianzsch"atzwerten. Bei diesen Messwerten 
kann man, wenn man annimmt, dass sie normalverteilt sind, 
wieder mit Hilfe der komplement"aren Fehlerfunktion 
Konfidenzintervalle nach Gleichung~(\myref{3.73}) 
absch"atzen, indem man in Gleichung~(\myref{3.72}) die Sch"atzwerte 
\mbox{$\Hat{C}_{\Hat{\boldsymbol{\Phi}}_{\!\boldsymbol{n}}(\mu,\mu),\Hat{\boldsymbol{\Phi}}_{\!\boldsymbol{n}}(\mu,\mu)}$} 
der Messwertvarianzen einsetzt. Alle anderen Messwerte sind echt komplex. 
Wenn man von deren Real- und Imagin"arteilen annimmt, dass sie 
verbundnormalverteilt sind, ergeben sich wieder Konfidenzellipsen, 
deren Halbachsen man mit Gleichung~(\myref{3.80}) absch"atzen kann, 
wenn man dort die entsprechenden Messwertvarianz- und 
-kovarianzsch"atzwerte einsetzt.

\chapter{Weitere Messwerte}\label{E.Kap.4}

\section{Messwerte der zeitvarianten Impulsantworten}\label{E.Kap.4.1}

Aus den Abtastwerten \mbox{$H(\mu,\mu\!+\!\Hat{\mu}\CdoT M/K_H)$}
der sich bei der L"osung der Regression theoretisch ergebenden
bifrequenten "Ubertragungsfunktion kann man durch eine DFT der
L"ange $K_H$ bez"uglich $\Hat{\mu}$ und eine inverse DFT der
L"ange $M$ bez"uglich $\mu$ die mit $M$ periodisch fortgesetzte,
zeitvariante Impulsantwort \mbox{$\Tilde{h}_{\kappa}(k\!+\!\kappa)$}
berechnen, die die zeitvariante Impulsantwort
\mbox{$h_{\kappa}(k\!+\!\kappa)$} nur dann vollst"andig beschreibt,
wenn diese zeitlich auf das Intervall
\mbox{$k\!-\!\kappa\in[0,M\!-\!1]$} begrenzt ist.
Der Fall einer zeitinvarianten Impulsantwort
\mbox{$h_{\kappa}(k\!+\!\kappa)=h(k)\;\;\forall\,\;k\!\in\![0,E]$}
ergibt sich im weiteren mit \mbox{$K_H\!=\!1$} und wird daher
nicht separat betrachtet. Wenn wir auch das von dem konjugierten
Eingangssignal erregte Modellsystem verwenden, berechnet sich die
mit $M$ periodisch fortgesetzte, zeitvariante Impulsantwort
\mbox{$\Tilde{h}_{*,\kappa}(k\!+\!\kappa)$} analog aus den
Abtastwerten der theoretischen "Ubertragungsfunktion
\mbox{$H_*(\mu,\mu\!+\!\Hat{\mu}\CdoT M/K_H)$}.
Aus den Messwerten \mbox{$\Hat{H}(\mu,\mu+\Hat{\mu}\CdoT M/K_H)$}
und ggf. \mbox{$\Hat{H}_*(\mu,\mu+\Hat{\mu}\CdoT M/K_H)$}
lassen sich in derselben Art Messwerte
\mbox{$\Hat{\Tilde{h}}_{\kappa}(k\!+\!\kappa)$} bzw.
\mbox{$\Hat{\Tilde{h}}_{*,\kappa}(k\!+\!\kappa)$}
f"ur die mit $M$ periodisch fortgesetzten,
zeitvarianten Impulsantworten berechnen.
\begin{subequations}\label{E.4.1}
\begin{gather}
\Hat{\Tilde{h}}_{\kappa}(k\!+\!\kappa)\;=\;
\frac{1}{M}\cdoT\Sum{\mu=0}{M-1}\;\Sum{\Hat{\mu}=0}{K_H-1}
\Hat{H}\big({\T \mu,\mu\!+\!\Hat{\mu}\CdoT\frac{M}{K_H}}\big)\cdot
e^{\!-j\cdot\frac{2\pi}{K_H}\cdot\Hat{\mu}\cdot\kappa}\cdot
e^{j\cdot\frac{2\pi}{M}\cdot\mu\cdot k}
\label{E.4.1.a}\\*[8pt]
\Hat{\Tilde{h}}_{*,\kappa}(k\!+\!\kappa)\;=\;
\frac{1}{M}\cdoT\Sum{\mu=0}{M-1}\;\Sum{\Hat{\mu}=0}{K_H-1}
\Hat{H}_*\big({\T \mu,\mu\!+\!\Hat{\mu}\CdoT\frac{M}{K_H}}\big)\cdot
e^{\!-j\cdot\frac{2\pi}{K_H}\cdot\Hat{\mu}\cdot\kappa}\cdot
e^{j\cdot\frac{2\pi}{M}\cdot\mu\cdot k}
\label{E.4.1.b}\\*[4pt]
\forall\qquad\qquad k\!+\!\kappa=0\;(1)\;M\!-\!1
\qquad\text{ und }\qquad\kappa=0\;(1)\;K_H\!-\!1.
\notag
\end{gather}
\end{subequations}
Diese Messwerte sind erwartungstreu, da einerseits die Messwerte der
"Ubertragungsfunktionen erwartungstreu sind, und andererseits
die zweidimensionale DFT eine Linearkombination mit konstanten
Koeffizienten ist.

Die theoretischen Messwertvarianzen und -kovarianzen erh"alt man wieder,
indem man jeweils den Erwartungswert des Betragsquadrats und des Quadrats
der Messwertabweichung berechnet. Die Messwertabweichung
l"asst sich als zweidimensionale DFT der in Gleichung~(\ref{E.3.18})
angegebenen Messwertfehler der "Ubertragungsfunktionen berechnen. Wenn
man eine hoch frequenzselektive Fensterfolge verwendet, kann man in den
so berechneten Vierfachsummen der theoretischen Messwert"-(ko)"-varianzen
wieder diejenigen Summanden vernachl"assigen, die eine der in den
Gleichungen~(\ref{E.2.41}) und (\ref{E.2.51}) genannten Kovarianzen
\mbox{$\text{E}\big\{\boldsymbol{N}_{\!\!f}(\mu)\CdoT\boldsymbol{N}_{\!\!f}(\Hat{\mu})^{\Kk}\big\}$} oder
\mbox{$\text{E}\big\{\boldsymbol{N}_{\!\!f}(\mu)\CdoT\boldsymbol{N}_{\!\!f}(\!-\Hat{\mu})\big\}$}
der Spektralwerte des Approximationsfehlerprozesses
als Faktor enthalten. Bei den verbleibenden Summanden kann man
die Spektralwertkovarianzen des Approximationsfehlers durch die
$M$-fachen theoretischen Werte
\mbox{$\Tilde{\Phi}_{\boldsymbol{n}}\big({\T\mu,\mu\!+\!\Tilde{\mu}\CdoT\frac{M}{K_{\Phi}}}\big)$} und
\mbox{$\Tilde{\Psi}_{\boldsymbol{n}}\big({\T\mu,\mu\!+\!\Tilde{\mu}\CdoT\frac{M}{K_{\Phi}}}\big)$}
ersetzen. Neben den Spektralwertkovarianzen des Approximationsfehlers
treten in den Summanden auch noch die Erwartungswerte der Elemente
des Produkts dreier Matrizen als Faktoren auf. Diese Matrizen sind
zwei inverse empirische Kovarianzmatrizen, die von links und rechts
an eine weitere Kovarianzmatrix heranmultipliziert werden. Die Elemente
dieser Matrizen h"angen nur von den Spektralwerten der Erregung ab.
Verzichtet man bei diesen Faktoren auf die Erwartungswertbildung,
indem man stattdessen wieder die konkreten Realisierungen
dieser Faktoren verwendet, und sch"atzt man noch die theoretischen 
Werte des bifrequenten LDS und KLDS durch ihre Messwerte 
\mbox{$\Hat{\Phi}_{\boldsymbol{n}}\big({\T\mu,\mu\!+\!\Tilde{\mu}\CdoT\frac{M}{K_{\Phi}}}\big)$} und 
\mbox{$\Hat{\Psi}_{\boldsymbol{n}}\big({\T\mu,\mu\!+\!\Tilde{\mu}\CdoT\frac{M}{K_{\Phi}}}\big)$}
ab, so erh"alt man die Sch"atzwerte
\begin{subequations}\label{E.4.2}
\begin{gather}
\begin{flalign*}
&\Hat{C}_{\Hat{\Tilde{\boldsymbol{h}}}_{\kappa}(k+\kappa),\Hat{\Tilde{\boldsymbol{h}}}_{\kappa}(k+\kappa)}\;=\;
\frac{1}{(L\!-\!1)\CdoT M}\cdot{}&&
\end{flalign*}\notag\\*
{}\!\CdoT\!\!\Sum{\mu=0}{\frac{M}{K_H}-1}\;\Sum{\Tilde{\mu}=0}{K_{\Phi}-1}
\Vec{w}_{K_H,\kappa}\cdot
\Hat{\underline{C}}_{\Tilde{\Vec{\boldsymbol{V}}}(\mu+\Tilde{\mu}\cdot\frac{M}{K_{\Phi}}),\Tilde{\Vec{\boldsymbol{V}}}(\mu+\Tilde{\mu}\cdot\frac{M}{K_{\Phi}})}^{\uP{0.4}{\!-1}}\CdoT
\Hat{\underline{C}}_{\Tilde{\Vec{\boldsymbol{V}}}(\mu+\Tilde{\mu}\cdot\frac{M}{K_{\Phi}}),\Tilde{\Vec{\boldsymbol{V}}}(\mu)}\CdoT
\Hat{\underline{C}}_{\Tilde{\Vec{\boldsymbol{V}}}(\mu),\Tilde{\Vec{\boldsymbol{V}}}(\mu)}^{\uP{0.4}{\!-1}}\CdoT
\Vec{w}_{K_H,\kappa}^{\Hh}\Cdot{}
\notag\\*[-4pt]\begin{flalign*}
&&{}\cdoT\Sum{\Hat{\mu}=0}{K_H-1}
\Hat{\Phi}_{\boldsymbol{n}}\big({\T\mu\!+\!\Hat{\mu}\CdoT\frac{M}{K_H},\mu\!+\!\Hat{\mu}\CdoT\frac{M}{K_H}\!+\!\Tilde{\mu}\CdoT\frac{M}{K_{\Phi}}}\big)\cdot 
e^{\!-j\cdot\frac{2\pi}{K_{\Phi}}\cdot\Tilde{\mu}\cdot k}&
\end{flalign*}\notag\\*[4pt]
\forall\qquad\qquad k\!+\!\kappa=0\;(1)\;M\!-\!1
\qquad\text{ und }\qquad\kappa=0\;(1)\;K_H\!-\!1
\label{E.4.2.a}
\end{gather}
f"ur die Messwertvarianzen und die Sch"atzwerte
\begin{gather}\begin{flalign*}
&\Hat{C}_{\Hat{\Tilde{\boldsymbol{h}}}_{\kappa}(k+\kappa),\Hat{\Tilde{\boldsymbol{h}}}_{\kappa}(k+\kappa)^{\Kk}}\;=\;
\frac{1}{(L\!-\!1)\CdoT M}\cdoT{}&&
\end{flalign*}\notag\\*
{}\!\CdoT\!\!\Sum{\mu=0}{\frac{M}{K_H}-1}\,\Sum{\Tilde{\mu}=0}{K_{\Phi}-1}\!
\Vec{w}_{K_H,\kappa}^{\Kk}\Cdot
\big(\Hat{\underline{C}}_{\Tilde{\Vec{\boldsymbol{V}}}(-\mu-\Tilde{\mu}\cdot\frac{M}{K_{\Phi}}),\Tilde{\Vec{\boldsymbol{V}}}(-\mu-\Tilde{\mu}\cdot\frac{M}{K_{\Phi}})}^{\uP{0.4}{\!-1}}\big)^{\!\Kk}\!\CdoT
\Hat{\underline{C}}_{\Tilde{\Vec{\boldsymbol{V}}}(-\mu-\Tilde{\mu}\cdot\frac{M}{K_{\Phi}})^{\Kk},\Tilde{\Vec{\boldsymbol{V}}}(\mu)}\CdoT
\Hat{\underline{C}}_{\Tilde{\Vec{\boldsymbol{V}}}(\mu),\Tilde{\Vec{\boldsymbol{V}}}(\mu)}^{\uP{0.4}{\!-1}}\CdoT
\Vec{w}_{K_H,\kappa}^{\Hh}\Cdot{}
\notag\\*[-4pt]\begin{flalign*}
&&{}\cdoT\Sum{\Hat{\mu}=0}{K_H-1}
\Hat{\Psi}_{\boldsymbol{n}}\big({\T\mu\!+\!\Hat{\mu}\CdoT\frac{M}{K_H},\mu\!+\!\Hat{\mu}\CdoT\frac{M}{K_H}\!+\!\Tilde{\mu}\CdoT\frac{M}{K_{\Phi}}}\big)\cdot 
e^{\!-j\cdot\frac{2\pi}{K_{\Phi}}\cdot\Tilde{\mu}\cdot k}&
\end{flalign*}\notag\\*[4pt]
\forall\qquad\qquad k\!+\!\kappa=0\;(1)\;M\!-\!1
\qquad\text{ und }\qquad\kappa=0\;(1)\;K_H\!-\!1.
\label{E.4.2.b}
\end{gather}
\end{subequations}
f"ur die Messwertkovarianzen der Impulsantwort des von \mbox{$v(k)$}
erregten Systems. Die Erwartungstreue dieser Sch"atzwerte l"asst sich
f"ur die einzelnen Summanden ---\,und somit f"ur die gesamte Summe\,---
analog zum Fall eines station"aren Approximationsfehlers zeigen.

Der Vektor 
\begin{subequations}\label{E.4.3}
\begin{equation}
\Vec{w}_{K_H,\kappa}\;=\;\Big[\;1\;,\,\,
e^{ j\cdot\frac{2\pi}{K_H}\cdot\kappa},\;\ldots,\;
e^{\!-j\cdot\frac{2\pi}{K_H}\cdot(K_H-1)\cdot\kappa},\,\,\Vec{0}\;\Big],
\label{E.4.3.a}
\end{equation}
enth"alt vorne die Drehfaktoren der inversen DFT der L"ange $K_H$, sowie 
am Ende den Nullzeilenvektor $\Vec{0}$ mit $K_H$ Elementen. Damit werden 
aus dem zwischen den Vektoren liegenden Matrixprodukt der linke, 
obere Block herausgeschnitten. Die Messwert"-(ko)"-varianzen 
der Impulsantwort des von \mbox{$v(k)^{\Kk}$} erregten Systems
erh"alt  man ebenfalls mit den Gleichungen~(\ref{E.4.2})
mit denselben Kovarianzmatrizen, allerdings ist hier der Vektor der 
DFT-Drehfaktoren {\em am Anfang}\/ um den Nullzeilenvektor $\Vec{0}$ mit $K_H$ 
Elementen zu verl"angern 
\begin{equation}
\Vec{w}_{K_H,\kappa}\;=\;\Big[\;\Vec{0}\;,\,\,1\;,\,\,
e^{ j\cdot\frac{2\pi}{K_H}\cdot\kappa},\,\ldots,\;
e^{\!-j\cdot\frac{2\pi}{K_H}\cdot(K_H-1)\cdot\kappa}\;\Big],
\label{E.4.3.b}
\end{equation}
\end{subequations}
wodurch aus dem dazwischenliegenden Matrixprodukt der
rechte, untere Block herausgeschnitten wird.

Es ist bei Verwendung von Messwerten
\mbox{$\Hat{\Phi}_{\boldsymbol{n}}(\ldots,\ldots)$}
und \mbox{$\Hat{\Psi}_{\boldsymbol{n}}(\ldots,\ldots)$}, die mit
Matrizen \mbox{$\underline{V}_{\bot}\!(\mu)$} gewonnen wurden,
die die Bedingungen~(\ref{E.3.33}) erf"ullen, zu vermuten, dass
auch bei diesen Messwerten die Sch"atzwerte der Kovarianzen betraglich
nicht gr"o"ser werden als die Sch"atzwerte der Varianzen.
Ein Beweis daf"ur wurden nicht erbracht. Er d"urfte "ahnlich ablaufen,
wie bei den Messwert"-(ko)"-varianzen der deterministischen St"orung
in Kapitel~\ref{E.Kap.A.9.2} des Anhangs.

\section{Messwerte der Auto- und Kreuzkorrelationsfolgen}\label{E.Kap.4.2}

Unter der Voraussetzung, dass sowohl die Autokovarianzfolge
\mbox{$\text{E}\big\{\boldsymbol{n}(k\!+\!\kappa)\CdoT\boldsymbol{n}(k)^{\Kk}\big\}$} 
als auch die Kreuzkovarianzfolge
\mbox{$\text{E}\big\{\boldsymbol{n}(k\!+\!\kappa)\CdoT\boldsymbol{n}(k)\big\}$} 
auf das Intervall \mbox{$\kappa\in(-M/2;M/2)$} beschr"ankt ist, und
dass man eine reelle Fensterfolge verwendet, bei der die 
zweidimensionale Fensterautokorrelationsfolge
\begin{equation}
d_{k}(\kappa)\;=\;
\frac{K_{\Phi}}{M}\cdoT\!\Sum{\Tilde{k}=-\infty}{\infty}\!
f(k+\Tilde{k}\CdoT K_{\Phi})^{\Kk}\Cdot f(k+\kappa+\Tilde{k}\CdoT K_{\Phi})
\label{E.4.4}
\end{equation}
f"ur \mbox{$k=0\;(1)\;K_{\Phi}\!-\!1$} und \mbox{$\kappa=1\!-\!M/2\;(1)\;M/2\!-\!1$} 
keine Nullstellen aufweist, kann man aus den Messwerten
\mbox{$\Hat{\Phi}_{\boldsymbol{n}}(\mu,\mu\!+\!\Tilde{\mu}\CdoT M/K_{\Phi})$} und
\mbox{$\Hat{\Psi}_{\boldsymbol{n}}(\mu,\mu\!+\!\Tilde{\mu}\CdoT M/K_{\Phi})$}
erwartungstreue Sch"atz\-werte f"ur die beiden Kovarianzfolgen berechnen,
indem man die Messwerte zun"achst einer DFT der L"ange $K_{\Phi}$ bez"uglich
$\Tilde{\mu}$ und dann einer inversen DFT der L"ange $M$ bez"uglich $\mu$
unterwirft. Anschlie"send wird das Ergebnis der zweidimensionalen DFT
durch die Fensterautokorrelationsfolge \mbox{$d_{k}(\kappa)$} dividiert.

\begin{gather}
\Hat{\phi}_{\boldsymbol{n}}(k\!+\!\kappa,k)\;=\;\text{ Sch"atzwert f"ur }
\text{E}\Big\{\boldsymbol{n}(k\!+\!\kappa)\CdoT
              \boldsymbol{n}(k)^{\Kk}\Big\}\;=\notag\\[5pt]
\frac{1}{M\CdoT d_{k}(\kappa)}\CdoT
\Sum{\mu=0}{M-1}\;\Sum{\Tilde{\mu}=0}{K_{\Phi}-1}
\Hat{\Phi}_{\boldsymbol{n}}(\mu,\mu\!+\!\Tilde{\mu}\CdoT M/K_{\Phi})\cdot
e^{\!-j\cdot\frac{2\pi}{K_{\Phi}}\cdot\Tilde{\mu}\cdot k}\cdot
e^{j\cdot\frac{2\pi}{M}\cdot\mu\cdot\kappa}\notag\\*[3pt]
\label{E.4.5}
{\T\forall\qquad\qquad \kappa=1\!-\!\frac{M}{2}\;(1)\;\frac{M}{2}\!-\!1
\qquad\text{ und }\qquad k=0\;(1)\;K_H\!-\!1.}
\end{gather}
Die Berechnung der Messwerte \mbox{$\Hat{\psi}_{\boldsymbol{n}}(k\!+\!\kappa,k)$} f"ur
\mbox{$\text{E}\big\{\boldsymbol{n}(k\!+\!\kappa)\cdoT\boldsymbol{n}(k)\big\}$} 
erfolgt analog. Es ist in der letzten Gleichung lediglich
\mbox{$\Hat{\Phi}_{\boldsymbol{n}}\big({\T\mu,\mu\!+\!\Tilde{\mu}\CdoT\frac{M}{K_{\Phi}}}\big)$} durch
\mbox{$\Hat{\Psi}_{\boldsymbol{n}}\big({\T\mu,\mu\!+\!\Tilde{\mu}\CdoT\frac{M}{K_{\Phi}}}\big)$}
zu ersetzen. Die Erwartungstreue dieser Messwerte ergibt sich u.~a. aus der
Tatsache, dass die Multiplikation der Kovarianzfolgen
\mbox{$\text{E}\big\{\boldsymbol{n}(k\!+\!\kappa)\CdoT\boldsymbol{n}(k)^{\Kk}\big\}$} und
\mbox{$\text{E}\big\{\boldsymbol{n}(k\!+\!\kappa)\CdoT\boldsymbol{n}(k)\big\}$} 
mit der Fensterautokorrelationsfolge \mbox{$d_{k}(\kappa)$} gerade der in den 
Gleichungen~(\ref{E.2.39}) und (\ref{E.2.50}) dargestellten zweidimensionalen Faltung 
der beiden Spektren der Fensterfolge mit dem zweidimensionalen LDS bzw. KLDS entspricht. 
Man erh"alt also die theoretischen  Kovarianzfolgen durch dieselbe 
zweidimensionale DFT wie die gemessenen Folgen, indem man in die letzte Gleichung
\mbox{$\Tilde{\Phi}_{\boldsymbol{n}}(\mu,\mu\!+\!\Tilde{\mu}\CdoT M/K_{\Phi})$} bzw.
\mbox{$\Tilde{\Psi}_{\boldsymbol{n}}(\mu,\mu\!+\!\Tilde{\mu}\CdoT M/K_{\Phi})$}
f"ur die entsprechenden, erwartungstreuen Messwerte einsetzt.

Auch f"ur die Messwerte der beiden Korrelationsfolgen kann man
Sch"atzwerte f"ur deren \mbox{Varianz} und Kovarianz aus den bisher berechneten
Messwerten und Matrixspuren der Tabelle~\ref{T.3.2} gewinnen, wenn
man wieder davon ausgeht, dass die Spektralwerte
\mbox{$\boldsymbol{N}_{\!\!f}(\mu)$} unterschiedlicher Frequenzen
jeweils verbundnormalverteilte Quadrupel bilden. Auf eine ausf"uhrliche
Darstellung der Herleitung dieser Sch"atzwerte wird verzichtet. Es soll
nun lediglich die prinzipielle Vorgehensweise f"ur deren Berechnung
am Beispiel der Varianzen der Messwerte der Autokovarianzfolge
\mbox{$\text{E}\big\{\boldsymbol{n}(k\!+\!\kappa)\CdoT\boldsymbol{n}(k)^{\Kk}\big\}$} 
verbal erl"autert werden. Die Messwertvarianzen berechnen sich prinzipiell als die
Differenz der Erwartungswerte der Betragsquadrate der Messwerte
und der Betragsquadrate der Erwartungswerte der Messwerte. Die
Erwartungswerte der Messwerte stimmen mit den theoretischen Werten
"uberein. Wenn man die Messwerte nach Gleichung~(\ref{E.3.29}) in
die Messwerte der Autokovarianzfolge nach Gleichung~(\ref{E.4.5})
einsetzt, und von diesen anschlie"send das Betragsquadrat bildet,
erh"alt man eine Vierfachsumme, bei der jeder Summand das Produkt zweier
bilinearer Formen ist. Der Erwartungswert dieser Vierfachsumme
ist die Summe der Erwartungswerte der Summanden. Um die Erwartungswerte
der Summanden berechnen zu k"onnen, bedarf es wieder einer
Fallunterscheidung bez"uglich der Indizes, die als Frequenzen in den
Stichprobenvektoren des gefensterten Approximationsfehlerprozesses
auftreten. Man muss nun vier F"alle unterscheiden. Im ersten Fall
liegen alle vier diskreten Frequenzen im Raster
\mbox{$0\;\big(\frac{M}{K_{\Phi}}\big)\;M\!-\!1$}, oder alle vier im Raster
\mbox{$\frac{M}{2\cdot K_{\Phi}}\;\big(\frac{M}{K_{\Phi}}\big)\;M\!-\!1$},
und es k"onnen somit alle m"oglichen Kovarianzen zweier zuf"alliger
Fehlerspektralwerte der vier an dem Produkt der bilinearen Formen
beteiligten Spektralwerte von null verschieden sein. Man kann
bei diesen Summanden deren Erwartungswerte, sowie Sch"atzwerte
f"ur diese, mit Hilfe der Gleichungen~(\ref{E.A.41}) und (\ref{E.A.47})
des Anhangs~\ref{E.Kap.A.8} erhalten, indem man dort die Matrizen und
Vektoren nach Gleichung~(\ref{E.A.40}) einsetzt, wobei man diese wiederum
durch die Matrizen \mbox{$\underline{\boldsymbol{V}}_{\bot}\!(\mu)$}
und Vektoren \mbox{$\Vec{\boldsymbol{N}}_{\!\!f}(\mu)$} 
sowie die Zufallsgr"o"sen \mbox{$\boldsymbol{N}_{\!\!f}(\mu)$}
in derselben Art wie bei der Berechnung der Messwertvarianzen des LDS
mit Tabelle~\ref{T.3.1} substituiert. Man erh"alt so f"ur den Erwartungswert
eines Summanden jeweils die Summe dreier Produkte theoretischer
Spektralwertkovarianzen. Im zweiten Fall liegen alle vier 
diskreten Frequenzen in einem Frequenzraster, das gegen"uber dem
Frequenzraster des ersten Falls um {\em kein}\/ ganzzahliges Vielfaches
von \mbox{$M/2/K_{\Phi}$} verschoben ist. Dann sind einige der
m"oglichen theoretischen Kovarianzen bei Verwendung einer hoch
frequenzselektiven Fensterfolge in guter N"aherung null, und
in den Gleichungen~(\ref{E.A.41}) und (\ref{E.A.47}) sind die Matrizen
und Vektoren nach Gleichung~(\ref{E.A.42}) einzusetzen, so dass sich
reduzierte Gleichungssysteme~(\ref{E.A.39}) ergeben. In dritten Fall
liegen die beiden zuf"alligen Stichprobenvektoren der zuf"alligen
Spektralwerte des Approximationsfehlerprozesses bei der ersten
bilinearen Form in einem Frequenzraster des zweiten Falls, und die beiden
Stichprobenvektoren der zweiten bilinearen Form in einem
Frequenzraster, das gegen"uber dem anderen Frequenzraster gerade die
negativen Frequenzen enth"alt. Die Substitution der Matrizen und Vektoren
gem"a"s der Gleichungen~(\ref{E.A.43}) in den Gleichungen~(\ref{E.A.41})
und (\ref{E.A.47}) liefert dann f"ur den Erwartungswert des Summanden
die Summe zweier Produkte theoretischer Spektralwertkovarianzen.
Im vierten Fall, wenn die beiden
Stichprobenvektoren der ersten bilinearen Form ebenso wie die beiden
Stichprobenvektoren der zweiten bilinearen Form in einem Frequenzraster
mit dem Frequenzabstand \mbox{$M/K_{\Phi}$} liegen, wobei die
beiden Frequenzraster aber keine gemeinsamen Frequenzen enthalten,
k"onnen nur sehr wenige der m"oglichen Spektralwertkovarianzen
nennenswert von null verschieden sein. Es sind dann die Matrizen
und Vektoren nach Gleichung~(\ref{E.A.44}) in die Gleichungen~(\ref{E.A.41})
und (\ref{E.A.47}) einzusetzen. Summanden, die keinem der vier
Falle zuzuordnen sind, kommen in der Vierfachsumme nicht vor,
da bei der DFT der L"ange $K_{\Phi}$ bei der Berechnung der Messwerte
der Autokorrelationsfolge nur solche bilineare Formen auftreten, bei
denen die beiden Stichprobenvektoren in einem Frequenzraster mit dem
Frequenzabstand \mbox{$M/K_{\Phi}$} liegen. Bei allen Summanden
der Vierfachsumme treten nur solche Messwerte f"ur das LDS und KLDS
auf, die bereits gemessen worden sind, und auch nur die Matrixspuren 
nach Tabelle~\ref{T.3.2}, die man zur Absch"atzung der Messwert"-(ko)"-varianzen 
sowieso schon berechnet hat. Ob die so berechneten Sch"atzwerte 
f"ur die Kovarianzen der Messwerte der Korrelationsfolgen 
\mbox{$\text{E}\big\{\boldsymbol{n}(k\!+\!\kappa)\CdoT\boldsymbol{n}(k)^{\Kk}\big\}$} und
\mbox{$\text{E}\big\{\boldsymbol{n}(k\!+\!\kappa)\CdoT\boldsymbol{n}(k)\big\}$} 
betraglich niemals gr"o"ser sind als die entsprechenden Varianzsch"atzwerte, 
wurde nicht untersucht.

\chapter{Spektralsch"atzung mit zeitabh"angigem ersten Moment}\label{E.Kap.5}

Wir wollen nun die Spektralsch"atzung des Zufallsprozesses
\mbox{$\boldsymbol{y}(k)$} als einen Sonderfall des RKM interpretieren.
Diesen erhalten wir, wenn wir bei einem System ohne Eingang nur die
spektralen Eigenschaften des Ausgangsprozesses vermessen wollen.
Im System nach Bild~\ref{E.b1h} setzen wir folglich die Erregung zu null,
und die Optimierungsparameter der Werte der beiden Impulsantworten treten 
nicht mehr in den zu minimierenden Termen nach  Gleichung (\ref{2.4}) auf. 
Unser Modellsystem besteht dann nur mehr aus der deterministischen St"orung
\mbox{$u(k)$} und der zuf"alligen St"orung \mbox{$\boldsymbol{n}(k)$}. 
Wir beschr"anken uns in diesem Kapitel auf den Fall, dass sich bei der Minimierung 
ein station"arer Approximationsfehlerprozess ergibt, f"ur den wir wie in \cite{Diss} 
dessen eindimensionales LDS und dessen ebenfalls eindimensionales KLDS 
durch die je $M$ Werte \mbox{$\Tilde{\Phi}_{\boldsymbol{n}}(\mu)$} und 
\mbox{$\Tilde{\Psi}_{\boldsymbol{n}}(\mu)$} beschreiben wollen.

\section{Spektralsch"atzung komplexer Prozesse}\label{E.Kap.5.1}

Die Minimierungsaufgabe lautet nun: 
\begin{equation}
\text{E}\big\{|\boldsymbol{n}(k)|^2\big\}\,=\,
\text{E}\Big\{\big|\boldsymbol{y}(k)\!-\!u(k)\big|^2\Big\}\;\stackrel{!}{=}\;\text{minimal}
\qquad\qquad\forall\qquad k=0\;(1)\;F\!-\!1.\quad
\label{E.5.1}
\end{equation}
Wenn wir diesen Ausdruck f"ur jeden Zeitpunkt $k$ jeweils nach dem Real-
und Imagin"arteil von \mbox{$u(k)$} partiell ableiten, und die beiden sich
dann ergebenden reellen Gleichungen wieder zu einer komplexen Gleichung
zusammenfassen, erhalten wir mit
\begin{equation}
u(k)\;=\;\text{E}\{\boldsymbol{y}(k)\}
\label{E.5.2}
\end{equation}
das zeitabh"angige erste Moment von \mbox{$\boldsymbol{y}(k)$} als die $F$
optimalen Regressionskoeffizienten. Die Werte
\mbox{$\Tilde{\Phi}_{\boldsymbol{n}}(\mu)$} sind nur mehr von der Wahl der
Werte der deterministischen St"orung abh"angig.\vspace{-4pt}
\begin{gather}
\Tilde{\Phi}_{\boldsymbol{n}}(\mu)\;=\;
\frac{1}{M}\cdot\text{E}\big\{|\boldsymbol{N}_{\!\!f}(\mu)|^2\big\}\;=\;
\frac{1}{M}\cdot\text{E}\bigg\{\,
\bigg|\Sum{k=-\infty}{\infty}\!\!\boldsymbol{n}(k)\CdoT
f(k)\cdot e^{\!-j\cdot\frac{2\pi}{M}\cdot\mu\cdot k}
\bigg|^2\,\bigg\}\;=\notag\\[6pt]
=\;\frac{1}{M}\cdot\text{E}\bigg\{\,
\bigg|\Sum{k=-\infty}{\infty}\!\!
\big(\boldsymbol{y}(k)\!-\!u(k)\big)\CdoT f(k)\CdoT
e^{\!-j\cdot\frac{2\pi}{M}\cdot\mu\cdot k}\bigg|^2\,\bigg\}\;=\;
\frac{1}{M}\cdot\text{E}\Big\{\big|\boldsymbol{Y}_{\!\!\!f}(\mu)\!-\!
U_{\!f}(\mu)\big|^2\Big\}
\notag\\
\qquad\qquad\forall\qquad\mu=0\;(1)\;M\!-\!1
\label{E.5.3}
\end{gather}
Die Minimierung dieser Werte legt die $M$ Optimierungsparameter 
\mbox{$U_{\!f}(\mu)$} fest. Dieselben Werte \mbox{$U_{\!f}(\mu)$} 
erh"alt man auch durch Fensterung und anschlie"sende DFT aus den $F$ 
optimalen Werten von \mbox{$u(k)$}, die das zweite Moment des 
Approximationsfehlers \mbox{$\boldsymbol{n}(k)$} minimieren. 
Die Wahl der optimalen Parameter f"uhrt auch hier dazu, dass die
ersten Momente \mbox{$\text{E}\{\boldsymbol{n}(k)\}$} und
\mbox{$\text{E}\{\boldsymbol{N}_{\!\!f}(\mu)\}$} null werden.

Bei der Spektralsch"atzung muss die Nullstellenbedingung (\myref{2.27})
f"ur das Spektrum der Fensterfolge nicht mehr erf"ullt werden. Es empfiehlt 
sich jedoch auch hier eine Fensterfolge zu verwenden, die der Bedingung~(\myref{2.20}) 
gen"ugt, um auch im Spektralbereich die richtige Varianz als Mittelwert aller $M$ Werte
\mbox{$\Tilde{\Phi}_{\boldsymbol{n}}(\mu)$} zu erhalten. Da diese Forderung von der im 
Kapitel \myref{Algo} vorgestellten Fensterfolge ebenso erf"ullt wird, wie auch die bei 
der Spektralsch"atzung gew"unschte hohe Frequenzselektivit"at, ist die Verwendung 
der damit berechneten Fensterfolge immer dann zu empfehlen, wenn das zu messende
LDS "uber der Frequenz stark schwankt.

Der Messwert \mbox{$\Hat{u}(k)$} der gefensterten deterministischen 
St"orung zum Zeitpunkt $k$ ist die Ausgleichsl"osung der Approximation 
der Stichprobe vom Umfang $L$ des Ausgangssignals des realen Systems 
zum Zeitpunkt $k$ durch den bez"uglich der Einzelmessungen $\lambda$ 
konstanten Wert \mbox{$\Hat{u}(k)$}. Es ergibt sich der empirische 
Mittelwert 
\begin{equation}
\Hat{u}(k)\;=\;\frac{1}{L}\cdoT\Sum{\lambda=1}{L}\,
y_{\lambda}(k)\;=\;\frac{1}{L}\CdoT\Vec{y}(k)\cdoT\Vec{1}^{\,\Hh}
\qquad\qquad\forall\qquad k=0\;(1)\;F\!-\!1
\label{E.5.4}
\end{equation}
als Ausgleichsl"osung. Die Messwerte \mbox{$\Hat{U}_{\!f}(\mu)$} f"ur das
Spektrum der gefensterten deterministischen St"orung erhalten wir wieder
dadurch, dass wir den quadratischen Fehler bei der Approximation der
Stichprobe des Spektrums des gefensterten Ausgangssignals des realen Systems durch die 
Werte der empirischen Regressionskoeffizienten \mbox{$\Hat{U}_{\!f}(\mu)$}
minimieren. Als Ausgleichsl"osung erhalten wir die empirischen
Mittelwerte der Spektralwerte des gefensterten Systemausgangsprozesses: 
\begin{equation}
\Hat{U}_{\!f}(\mu)\;=\;
\frac{1}{L}\cdoT\Sum{\lambda=1}{L}Y_{\!f,\lambda}(\mu)\;=\;
\frac{1}{L}\cdot\Vec{Y}_{\!f}(\mu)\CdoT\Vec{1}^{\Hh}
\qquad\qquad\forall\qquad\mu=0\;(1)\;M\!-\!1.
\label{E.5.5}
\end{equation}
Die Messwerte \mbox{$\Hat{\Phi}_{\boldsymbol{n}}(\mu)$} und
\mbox{$\Hat{\Psi}_{\boldsymbol{n}}(\mu)$} erhalten wir durch Projektion des
Stichprobenvektors \mbox{$\Vec{Y}_{\!f}(\mu)$} in einen
Unterraum, der zum Vektor \mbox{$\Vec{1}$} orthogonal ist.
Sinnvollerweise erfolgt diese Projektion mit Hilfe der Matrix
\mbox{$\underline{1}_{\bot}$} nach Gleichung (\ref{E.3.9}), deren
Spur \mbox{$L\!-\!1$} ist. Als erwartungstreue Messwerte erhalten wir
die empirischen Varianzen
\begin{subequations}\label{E.5.6}
\begin{gather}
\Hat{\Phi}_{\boldsymbol{n}}(\mu)\;=\;
\frac{\Hat{\Vec{N}}_{\!f}(\mu)\CdoT\Hat{\Vec{N}}_{\!f}(\mu)^{\Hh}\!}
{M\CdoT(L\!-\!1)}\;=\;
\frac{\Vec{N}_{\!f}(\mu)\CdoT\underline{1}_{\bot}\!\CdoT
\Vec{N}_{\!f}(\mu)^{\Hh}\!}{M\CdoT(L\!-\!1)}\;=\;
\frac{\Vec{Y}_{\!f}(\mu)\CdoT\underline{1}_{\bot}\!\CdoT
\Vec{Y}_{\!f}(\mu)^{\Hh}\!}{M\CdoT(L\!-\!1)}\;=\;
\frac{1}{M}\cdot\Hat{C}_{\boldsymbol{Y}_{\!\!\!f}(\mu),\boldsymbol{Y}_{\!\!\!f}(\mu)}
\notag\\[4pt]
\label{E.5.6.a}
\forall\qquad\qquad\mu=0\;(1)\;M\!-\!1\\[-24pt]\notag
\end{gather}
und Kovarianzen
\begin{gather}
\Hat{\Psi}_{\boldsymbol{n}}(\mu)=
\frac{\Hat{\Vec{N}}_{\!f}(\mu)\CdoT\Hat{\Vec{N}}_{\!f}(\!-\mu)^{\Tt}\!}
{M\CdoT(L\!-\!1)}=
\frac{\Vec{N}_{\!f}(\mu)\CdoT\underline{1}_{\bot}\!\CdoT
\Vec{N}_{\!f}(\!-\mu)^{\Tt}\!}{M\CdoT(L\!-\!1)}=
\frac{\Vec{Y}_{\!f}(\mu)\CdoT\underline{1}_{\bot}\!\CdoT
\Vec{Y}_{\!f}(\!-\mu)^{\Tt}\!}{M\CdoT(L\!-\!1)}=
\frac{1}{M}\CdoT\Hat{C}_{\boldsymbol{Y}_{\!\!\!f}(\mu),\boldsymbol{Y}_{\!\!\!f}(-\mu)^{\Kk}}
\notag\\[4pt]
\label{E.5.6.b}
\forall\qquad\qquad\mu=0\;(1)\;M\!-\!1
\end{gather}
\end{subequations}
der Spektralwerte des gefensterten Prozesses \mbox{$\boldsymbol{y}(k)$},
die die Ungleichung (\myref{3.38}) immer erf"ullen, und die ohne
Zwischenspeicherung der Spektralwerte aller Einzelmessungen berechnet
werden k"onnen.

Die Varianz und Kovarianz der Messwerte
\mbox{$\Hat{\boldsymbol{U}}_{\!f}(\mu)$} ergibt sich zu
\begin{subequations}\label{E.5.7}
\begin{gather}
C_{\Hat{\boldsymbol{U}}_{\!\!f}(\mu),\Hat{\boldsymbol{U}}_{\!\!f}(\mu)}=
\text{E}\Big\{\big|\Hat{\boldsymbol{U}}_{\!\!f}(\mu)\!-\!
\text{E}\{\Hat{\boldsymbol{U}}_{\!\!f}(\mu)\}\big|^2\Big\}=
\text{E}\Big\{\big|\Hat{\boldsymbol{U}}_{\!\!f}(\mu)\!-\!U_{\!f}(\mu)\big|^2\Big\}=
\text{E}\bigg\{\bigg|\Vec{\boldsymbol{N}}_{\!\!f}(\mu)\CdoT
\frac{\Vec{1}^{\Hh}}{L}\bigg|^2\bigg\}=
\notag\\*[4pt]
=\;\text{E}\bigg\{\text{spur}\bigg(\,\frac{\Vec{1}^{\Hh}\Cdot\Vec{1}}{\D L^2}\,\bigg)\bigg\}\CdoT
\Big(\text{E}\big\{|\boldsymbol{N}_{\!\!f}(\mu)|^2\big\}\!-\!
\big|\text{E}\{\boldsymbol{N}_{\!\!f}(\mu)\}\big|^2\Big)-
\big|\text{E}\{\boldsymbol{N}_{\!\!f}(\mu)\}\big|^2\;=\notag\\[4pt]
=\;\frac{1}{L}\cdot\text{E}\big\{|\boldsymbol{N}_{\!\!f}(\mu)|^2\big\}\;=\;
\frac{M}{L}\cdot\Tilde{\Phi}_{\boldsymbol{n}}(\mu)
\label{E.5.7.a}\\[-16pt]
\intertext{und\vspace{-8pt}}
C_{\Hat{\boldsymbol{U}}_{\!\!f}(\mu),\Hat{\boldsymbol{U}}_{\!\!f}(\mu)^{\Kk}}=
\text{E}\Big\{\!\big(\Hat{\boldsymbol{U}}_{\!\!f}(\mu)\!-\!
\text{E}\{\Hat{\boldsymbol{U}}_{\!\!f}(\mu)\}\big)^{\!2}\Big\}=
\text{E}\Big\{\!\big(\Hat{\boldsymbol{U}}_{\!\!f}(\mu)\!-\!
U_{\!f}(\mu)\big)^{\!2}\Big\}=
\text{E}\bigg\{\!\bigg(\!\Vec{\boldsymbol{N}}_{\!\!f}(\mu)\CdoT
\frac{\Vec{1}^{\Hh}}{L}\bigg)^{\!\!2}\bigg\}=
\notag\\[4pt]
=\;\text{E}\bigg\{\text{spur}\bigg(\,\frac{\Vec{1}^{\Hh}\Cdot\Vec{1}}{L^2}\,\bigg)\bigg\}\CdoT
\Big(\text{E}\big\{\boldsymbol{N}_{\!\!f}(\mu)^2\big\}\!-\!
\text{E}\{\boldsymbol{N}_{\!\!f}(\mu)\}^2\Big)-
\text{E}\{\boldsymbol{N}_{\!\!f}(\mu)\}^2\;=
\notag\\*[6pt]
=\;\frac{1}{L}\cdot\text{E}\big\{\boldsymbol{N}_{\!\!f}(\mu)^2\big\}\;
\begin{cases}
{\D\;=\;\frac{M}{L}\cdot\Tilde{\Psi}_{\boldsymbol{n}}(\mu)}
&{\T\text{ f"ur }\mu\in\big\{0;\frac{M}{2}\big\}}\\[4pt]
\;\approx\;0&\text{ sonst,}
\end{cases}
\label{E.5.7.b}
\end{gather}
\end{subequations}
und kann mit\vspace{-4pt}
\begin{subequations}\label{E.5.8}
\begin{gather}
\Hat{C}_{\Hat{\boldsymbol{U}}_{\!\!f}(\mu),\Hat{\boldsymbol{U}}_{\!\!f}(\mu)}\;=\;
\frac{M}{L}\cdot\Hat{\Phi}_{\boldsymbol{n}}(\mu)
\qquad\qquad\forall\qquad\mu=0\;(1)\;M\!-\!1
\label{E.5.8.a}\\[-16pt]
\intertext{und\vspace{-8pt}}
\Hat{C}_{\Hat{\boldsymbol{U}}_{\!\!f}(\mu),\Hat{\boldsymbol{U}}_{\!\!f}(\mu)^{\Kk}}\;=\;
\frac{M}{L}\cdot\Hat{\Psi}_{\boldsymbol{n}}(\mu)
{\T\qquad\qquad\text{ f"ur }\qquad\mu\in\big\{0;\frac{M}{2}\big\}}
\label{E.5.8.b}
\end{gather}
\end{subequations}
erwartungstreu abgesch"atzt werden, wobei die Kovarianzsch"atzwerte
betraglich niemals gr"o"ser als die Varianzsch"atzwerte sind.
Daher kann man mit diesen Varianz- und Kovarianzsch"atzwerten die Sch"atzwerte
der Halbachsen der Konfidenzellipsen nach Gleichung (\myref{3.80}) f"ur ein 
gew"unschtes Konfidenzniveau angeben, wobei man dort die eben berechneten Sch"atzwerte 
f"ur die Messwert"-(ko)"-varianzen statt der Messwert"-(ko)"-varianz-sch"atz"-werte
\mbox{$\Hat{C}_{\Hat{\boldsymbol{H}}(\mu),\Hat{\boldsymbol{H}}(\mu)}$} und
\mbox{$\Hat{C}_{\Hat{\boldsymbol{H}}(\mu),\Hat{\boldsymbol{H}}(\mu)^{\Kk}}$}
einsetzt.

Bei der Berechnung der Varianzen und Kovarianzen der Messwerte
\mbox{$\Hat{u}(k)$} der gefensterten deterministischen St"orung
ist nun keine N"aherung mehr n"otig. Die Messwertabweichung
\begin{gather}
\Hat{u}(k)\!-\!u(k)\;=\;
\frac{1}{L}\cdoT\Sum{\lambda=1}{L}y_{\lambda}(k)\!-\!u(k)\;=\;
\frac{1}{L}\cdoT\Sum{\lambda=1}{L}\big(n_{\lambda}(k)\!+\!u(k)\big)\!-\!u(k)\;=\;
\frac{1}{L}\cdoT\Sum{\lambda=1}{L}n_{\lambda}(k)
\notag
\\*[4pt]
\forall\qquad k=0\;(1)\;F\!-\!1
\label{E.5.9}
\end{gather}
enth"alt nun in Gegensatz zu Gleichung (\ref{E.3.23}) keine Terme mehr, die 
von den verrauschten Messwerten der beiden "Ubertragungsfunktionen abh"angen. 
Daher ergibt sich bei einem station"aren Approximationsfehlerprozess
bei Verwendung einer Fensterfolge, die der Bedingung (\myref{2.20})
gen"ugt, die zeitunabh"angige Messwertvarianz
\begin{subequations}\label{E.5.10}
\begin{gather}
C_{\Hat{\boldsymbol{u}}(k),\Hat{\boldsymbol{u}}(k)}\,=\;
\text{E}\Big\{\big|\Hat{\boldsymbol{u}}(k)\!-\!u(k)\big|^2\Big\}\;=\;
\frac{1}{L}\cdot\text{E}\big\{|\boldsymbol{n}(k)|^2\big\}\;=\;
\frac{1}{L\CdoT M}\cdoT\Sum{\mu=0}{M-1}\Tilde{\Phi}_{\boldsymbol{n}}(\mu)
\label{E.5.10.a}\\[0pt]
\intertext{und die ebenfalls zeitunabh"angige Messwertkovarianz}
C_{\Hat{\boldsymbol{u}}(k),\Hat{\boldsymbol{u}}(k)^{\Kk}}\;=\;
\text{E}\Big\{\big(\Hat{\boldsymbol{u}}(k)\!-\!u(k)\big)^2\Big\}\;=\;
\frac{1}{L}\cdot\text{E}\big\{\boldsymbol{n}(k)^2\big\}\;=\;
\frac{1}{L\CdoT M}\cdoT\Sum{\mu=0}{M-1}\Tilde{\Psi}_{\boldsymbol{n}}(\mu).
\label{E.5.10.b}
\end{gather}
\end{subequations}
Als Summe erwartungstreuer Messwerte sind die Sch"atzwerte
\begin{subequations}\label{E.5.11}
\begin{flalign}
&&\Hat{C}_{\Hat{\boldsymbol{u}}(k),\Hat{\boldsymbol{u}}(k)}\;=\;&
\frac{1}{L\CdoT M}\cdoT\Sum{\mu=0}{M-1}\Hat{\Phi}_{\boldsymbol{n}}(\mu)&&
\label{E.5.11.a}\\[6pt]
\text{und}&&
\Hat{C}_{\Hat{\boldsymbol{u}}(k),\Hat{\boldsymbol{u}}(k)^{\Kk}}\;=\;&
\frac{1}{L\CdoT M}\cdoT\Sum{\mu=0}{M-1}\Hat{\Psi}_{\boldsymbol{n}}(\mu)&&
\label{E.5.11.b}
\end{flalign}
\end{subequations}
ebenfalls erwartungstreu. Dass der Sch"atzwert der Kovarianz nie betraglich
gr"o"ser ist als der Sch"atzwert der Varianz, zeigt man indem man die Summe bei
der Berechnung der Sch"atzwerte in Gleichung (\ref{E.5.11.a}) durch die Summe
der arithmetischen Mittel der Messwerte \mbox{$\Hat{\Phi}_{\boldsymbol{n}}(\!-\mu)$}
und \mbox{$\Hat{\Phi}_{\boldsymbol{n}}(\mu)$} ersetzt. Diese Summe ist immer 
gr"o"ser als die Summe der entsprechenden geometrischen Mittel, deren Summanden 
nach Gleichung~(\myref{3.38}) gr"o"ser als die Betr"age von 
\mbox{$\Hat{\Psi}_{\boldsymbol{n}}(\mu)$} sind. Da der Betrag einer Summe kleiner
oder gleich der Summe der Betr"age der Summanden ist, ist der Betrag des
Kovarianzsch"atzwertes nie gr"o"ser als der Varianzsch"atzwert.
Die Sch"atzwerte der zeitunabh"angigen Halbachsen der Konfidenzellipse
erh"alt man mit dem gew"unschten Konfidenzniveau und den eben berechneten 
Messwert"-(ko)"-varianz"-sch"atz"-werten, indem man diese statt der Messwert"-(ko)"-varianz\-sch"atz\-werte
\mbox{$\Hat{C}_{\Hat{\boldsymbol{H}}(\mu),\Hat{\boldsymbol{H}}(\mu)}$} und
\mbox{$\Hat{C}_{\Hat{\boldsymbol{H}}(\mu),\Hat{\boldsymbol{H}}(\mu)^{\Kk}}$}
in Gleichung (\myref{3.80}) einsetzt.

Bei der Berechnung der Varianzen und Kovarianzen der Messwerte
\mbox{$\Hat{\boldsymbol{\Phi}}_{\boldsymbol{n}}(\mu)$} und
\mbox{$\Hat{\boldsymbol{\Psi}}_{\boldsymbol{n}}(\mu)$}
ist nun ein Rangdefekt von eins einzusetzen, weil nun die Matrix
\mbox{$\underline{1}_{\bot}$} statt der Matrix 
\mbox{$\underline{V}_{\bot}\!(\mu)$}, zur Berechnung der 
Messwerte verwendet wird. Da bei der Berechnung der 
Messwert"-(ko)"-varianzen in den Gleichungen (\myref{3.69}) und (\myref{3.70}) 
ein Rangdefekt von Zwei eingesetzt wurde, ist dort lediglich
$L$ durch \mbox{$L\!+\!1$} zu substituieren. Wir erhalten daher
die theoretischen Messwert"-(ko)"-varianzen
\begin{subequations}\label{E.5.12}
\begin{align}
C_{\Hat{\boldsymbol{\Phi}}_{\!\boldsymbol{n}}(\mu),\Hat{\boldsymbol{\Phi}}_{\!\boldsymbol{n}}(\mu)}\;&=\;
\frac{1}{L\!-\!1}\cdot\big|\Tilde{\Psi}_{\boldsymbol{n}}(\mu)\big|^2+
\frac{1}{L\!-\!1}\cdot\Tilde{\Phi}_{\boldsymbol{n}}(\mu)^{\uP{0.4}{\!2}}\!,
\notag\\*[5pt]
C_{\Hat{\boldsymbol{\Psi}}_{\!\boldsymbol{n}}(\mu),\Hat{\boldsymbol{\Psi}}_{\!\boldsymbol{n}}(\mu)}\;&=\;
\frac{2}{L\!-\!1}\cdot\Tilde{\Phi}_{\boldsymbol{n}}(\mu)^{\uP{0.4}{\!2}}
\notag\\*[5pt]
\text{und\quad}
C_{\Hat{\boldsymbol{\Psi}}_{\!\boldsymbol{n}}(\mu),\Hat{\boldsymbol{\Psi}}_{\!\boldsymbol{n}}(\mu)^{\Kk}}&=\;
\frac{2}{L\!-\!1}\cdot\Tilde{\Psi}_{\boldsymbol{n}}(\mu)^{\uP{0.4}{\!2}}
\notag\\*[5pt]
\text{f"ur}\quad&{\T\qquad\mu\in\big\{0;\frac{M}{2}\big\}}
\label{E.5.12.a}\\[-18pt]\intertext{bzw.\vspace{-14pt}}
C_{\Hat{\boldsymbol{\Phi}}_{\!\boldsymbol{n}}(\mu),\Hat{\boldsymbol{\Phi}}_{\!\boldsymbol{n}}(\mu)}\;&=\;
\frac{1}{L\!-\!1}\cdot\Tilde{\Phi}_{\boldsymbol{n}}(\mu)^{\uP{0.4}{\!2}}\!,
\notag\\*[5pt]
C_{\Hat{\boldsymbol{\Psi}}_{\!\boldsymbol{n}}(\mu),\Hat{\boldsymbol{\Psi}}_{\!\boldsymbol{n}}(\mu)}\;&=\;
\frac{1}{L\!-\!1}\cdot
\Tilde{\Phi}_{\boldsymbol{n}}(\!-\mu)\cdot
\Tilde{\Phi}_{\boldsymbol{n}}(\mu)
\notag\\*[5pt]
\text{und\quad}
C_{\Hat{\boldsymbol{\Psi}}_{\!\boldsymbol{n}}(\mu),\Hat{\boldsymbol{\Psi}}_{\!\boldsymbol{n}}(\mu)^{\Kk}}&=\;
\frac{1}{L\!-\!1}\cdot\Tilde{\Psi}_{\boldsymbol{n}}(\mu)^{\uP{0.4}{\!2}}
\notag\\*[5pt]
\text{f"ur}\quad&{\T\qquad\mu=1\;(1)\;M\!-\!1\quad\wedge\quad\mu\neq\frac{M}{2},}
\label{E.5.12.b}\end{align}
\end{subequations}
die wir mit
\begin{subequations}\label{E.5.13}
\begin{align}
\Hat{C}_{\Hat{\boldsymbol{\Phi}}_{\!\boldsymbol{n}}(\mu),\Hat{\boldsymbol{\Phi}}_{\!\boldsymbol{n}}(\mu)}\;&=\;
\frac{L\!-\!3}{(L\!+\!1)\CdoT(L\!-\!2)}\cdot
\Hat{\Phi}_{\boldsymbol{n}}(\mu)^{\uP{0.4}{\!2}}+\,
\frac{L\!-\!1}{(L\!+\!1)\CdoT(L\!-\!2)}\cdot
\big|\Hat{\Psi}_{\boldsymbol{n}}(\mu)\big|^2\!,
\notag\\*[5pt]
\Hat{C}_{\Hat{\boldsymbol{\Psi}}_{\!\boldsymbol{n}}(\mu),\Hat{\boldsymbol{\Psi}}_{\!\boldsymbol{n}}(\mu)}\;&=\;
\frac{2\CdoT(L\!-\!1)}{(L\!+\!1)\CdoT(L\!-\!2)}\cdot
\Hat{\Phi}_{\boldsymbol{n}}(\mu)^{\uP{0.4}{\!2}}-\,
\frac{2}{(L\!+\!1)\CdoT(L\!-\!2)}\cdot
\big|\Hat{\Psi}_{\boldsymbol{n}}(\mu)\big|^2
\notag\\*[5pt]
\text{und\quad}
\Hat{C}_{\Hat{\boldsymbol{\Psi}}_{\!\boldsymbol{n}}(\mu),\Hat{\boldsymbol{\Psi}}_{\!\boldsymbol{n}}(\mu)^{\Kk}}&=\;
\frac{2}{L\!+\!1}\cdot\Hat{\Psi}_{\boldsymbol{n}}(\mu)^{\uP{0.4}{\!2}}
\notag\\*[5pt]
\text{f"ur}\quad&{\T\qquad\mu\in\big\{0;\frac{M}{2}\big\}}
\label{E.5.13.a}\\[-16pt]\intertext{bzw.\vspace{-12pt}}
\Hat{C}_{\Hat{\boldsymbol{\Phi}}_{\!\boldsymbol{n}}(\mu),\Hat{\boldsymbol{\Phi}}_{\!\boldsymbol{n}}(\mu)}\;&=\;
\frac{1}{L}\cdot\Hat{\Phi}_{\boldsymbol{n}}(\mu)^{\uP{0.4}{\!2}}\!,
\notag\\*[5pt]
\Hat{C}_{\Hat{\boldsymbol{\Psi}}_{\!\boldsymbol{n}}(\mu),\Hat{\boldsymbol{\Psi}}_{\!\boldsymbol{n}}(\mu)}\;&=\;
\frac{L\!-\!1}{L\CdoT(L\!-\!2)}\cdot
\Hat{\Phi}_{\boldsymbol{n}}(\!-\mu)\cdot\Hat{\Phi}_{\boldsymbol{n}}(\mu)-
\frac{1}{L\CdoT(L\!-\!2)}\cdot
\big|\Hat{\Psi}_{\boldsymbol{n}}(\mu)\big|^2
\notag\\*[5pt]
\text{und\quad}
\Hat{C}_{\Hat{\boldsymbol{\Psi}}_{\!\boldsymbol{n}}(\mu),\Hat{\boldsymbol{\Psi}}_{\!\boldsymbol{n}}(\mu)^{\Kk}}&=\;
\frac{1}{L}\cdot\Hat{\Psi}_{\boldsymbol{n}}(\mu)^{\uP{0.4}{\!2}}
\notag\\*[5pt]
\text{f"ur}\quad&{\T\qquad\mu=1\;(1)\;M\!-\!1\quad\wedge\quad\mu\neq\frac{M}{2}}
\label{E.5.13.b}
\end{align}
\end{subequations}
erwartungstreu absch"atzen. Auch hier sind die Sch"atzwerte der Kovarianz
\mbox{$\Hat{C}_{\Hat{\boldsymbol{\Psi}}_{\!\boldsymbol{n}}(\mu),\Hat{\boldsymbol{\Psi}}_{\!\boldsymbol{n}}(\mu)^{\Kk}}$}
nie betraglich gr"o"ser als die Sch"atzwerte der Varianz
\mbox{$\Hat{C}_{\Hat{\boldsymbol{\Psi}}_{\!\boldsymbol{n}}(\mu),\Hat{\boldsymbol{\Psi}}_{\!\boldsymbol{n}}(\mu)}$}.
F"ur die Konfidenz"-intervalle (\myref{3.73}) der reellen Messwerte
\mbox{$\Hat{\boldsymbol{\Phi}}_{\boldsymbol{n}}(\mu)$} setzt man die
Sch"atzwerte der halben Intervallbreite nach Gleichung (\myref{3.72}) mit
den eben berechneten Messwertvarianzsch"atzwerten ein. Die Sch"atzwerte 
der Halbachsen der Konfidenzellipsen der Messwerte
\mbox{$\Hat{\boldsymbol{\Psi}}_{\boldsymbol{n}}(\mu)$}
erh"alt man mit dem gew"unschten Konfidenzniveau
und den eben berechneten Messwert"-(ko)"-varianz"-sch"atz"-werten,
indem man diese statt der Messwert"-(ko)"-varianz\-sch"atz\-werte
\mbox{$\Hat{C}_{\Hat{\boldsymbol{H}}(\mu),\Hat{\boldsymbol{H}}(\mu)}$} und
\mbox{$\Hat{C}_{\Hat{\boldsymbol{H}}(\mu),\Hat{\boldsymbol{H}}(\mu)^{\Kk}}$}
in die Gleichungen (\myref{3.80}) einsetzt.

\section{Spektralsch"atzung reeller Prozesse}\label{E.Kap.5.2}

Die Spektralsch"atzung reeller Prozesse l"auft im wesentlichen genau 
wie eben beschrieben ab. Auf einige Besonderheiten, die sich bei reellen 
Prozessen ergeben, soll nun hier eingegangen werden. 

Bei einem reellen Zufallsprozess am Systemausgang erhalten wir
als Optimall"osung der Minimierung (\ref{E.5.1}) eine reelle deterministische
Modellst"orung, die sich wie im Fall eines komplexwertigen Systems
nach Gleichung (\ref{E.5.2}) als das erste Moment
\mbox{$\text{E}\{\boldsymbol{y}(k)\}$} des zu messenden Prozesses
ergibt.  Somit erhalten wir einen reellen mittelwertfreien
Modellzufallsvektor f"ur den Approximationsfehlerprozess
\mbox{$\boldsymbol{n}(k)=\boldsymbol{y}(k)\!-\!u(k)$}.
Da bei einem  reellen Modellzufallsvektor die
Autokorrelationsfolge \mbox{$\text{E}\big\{\boldsymbol{n}(k)^{\Kk}\!\CdoT
\boldsymbol{n}(k\!+\!\kappa)\big\}$} und die Korrelationsfolge
\mbox{$\text{E}\big\{\boldsymbol{n}(k)\CdoT
\boldsymbol{n}(k\!+\!\kappa)\big\}$} identisch, reell und
geradesymmetrisch sind, ist das KLDS
\mbox{$\Psi_{\boldsymbol{n}}(\Omega)$} reell und geradesymmetrisch und
identisch mit dem LDS
\mbox{$\Phi_{\boldsymbol{n}}(\Omega)$}. Auch die N"aherungen
\begin{gather}
\Tilde{\Phi}_{\boldsymbol{n}}(\mu)\;=\;
\frac{1}{M}\cdot\text{E}\big\{|\boldsymbol{N}_{\!\!f}(\mu)|^2\big\}\;=\;
\frac{1}{M}\cdot\text{E}\big\{|\boldsymbol{N}_{\!\!f}(\!-\mu)|^2\big\}\;=\;
\Tilde{\Phi}_{\boldsymbol{n}}(\!-\mu)\;=
\notag\\[6pt]
=\;\frac{1}{M}\cdot\text{E}\big\{
\boldsymbol{N}_{\!\!f}(\!-\mu)\CdoT\boldsymbol{N}_{\!\!f}(\mu)\big\}\;=\;
\Tilde{\Psi}_{\boldsymbol{n}}(\mu)\;=\;
\frac{1}{M}\cdot\text{E}\bigg\{
\bigg|\Sum{k=-\infty}{\infty}\!\!\boldsymbol{n}(k)\CdoT
f(k)\cdot e^{\!-j\cdot\frac{2\pi}{M}\cdot\mu\cdot k}\bigg|^2\bigg\}\;=
\notag\\[6pt]
=\;\frac{1}{M}\cdot\text{E}\bigg\{
\bigg|\Sum{k=-\infty}{\infty}\!\!\big(\boldsymbol{y}(k)\!-\!u(k)\big)\CdoT
f(k)\cdot
e^{\!-j\cdot\frac{2\pi}{M}\cdot\mu\cdot k}\bigg|^2\bigg\}\;=\;
\frac{1}{M}\cdot\text{E}\Big\{\big|\boldsymbol{Y}_{\!\!\!f}(\mu)\!-\!
U_{\!f}(\mu)\big|^2\Big\}\notag\\*[6pt]
{\T\forall\qquad\qquad\mu=0\;(1)\;\frac{M}{2}}
\label{E.5.14}
\end{gather}
der entsprechenden Stufenapproximationen
\mbox{$\Bar{\Phi}_{\boldsymbol{n}}(\mu)$} sind bei Verwendung einer
reellen Fensterfolge identisch, reell und geradesymmetrisch, und nur
von der Wahl der $M$ Optimierungsparameter \mbox{$U_{\!f}(\mu)$} abh"angig.
Diese ergeben sich mit
\begin{equation}
U_{\!f}(\mu)\;=\;U_{\!f}(\!-\mu)^{\Kk}\,=\;
\text{E}\big\{\boldsymbol{Y}_{\!\!\!f}(\mu)\big\}
{\T\qquad\qquad\forall\qquad\mu=0\;(1)\;\frac{M}{2}}
\label{E.5.15}
\end{equation}
als die Mittelwerte der Spektralwerte des gemessenen und gefensterten
Zufallsvektors am Systemausgang, und sind somit dieselben Werte,
die man auch durch Fensterung und anschlie"sende DFT aus den $F$
optimalen Werten von \mbox{$u(k)$}, die das zweite Moment des
Approximationsfehlers \mbox{$\boldsymbol{n}(k)$} minimieren, erh"alt.
Daher sind auch die Spektralwerte des gefensterten Approximationsfehlers 
mittelwertfrei.

Auch bei der Spektralsch"atzung reeller Prozesse muss die
Nullstellenbedingung (\myref{2.27}) f"ur das Spektrum der Fensterfolge
nicht mehr erf"ullt werden. Es empfiehlt sich jedoch auch hier eine
reelle Fensterfolge zu verwenden, die der Bedingung (\myref{2.20}) gen"ugt,
um auch im Spektralbereich die richtige Varianz als Mittelwert aller
$M$ Werte \mbox{$\Tilde{\Phi}_{\boldsymbol{n}}(\mu)$} zu erhalten.
Da diese Forderung von der im Kapitel \ref{Algo} vorgestellten 
Fensterfolge ebenso erf"ullt wird, wie auch die
bei der Spektralsch"atzung gew"unschte hohe Frequenzselektivit"at erzielt
wird, ist die Verwendung der damit berechneten Fensterfolge immer dann zu
empfehlen, wenn das zu messende LDS "uber der Frequenz stark schwankt.

Die Messwerte \mbox{$\Hat{u}(k)$} berechnen sich wie bei der
Spektralsch"atzung eines komplexen Prozesses nach Gleichung (\ref{E.5.4})
als die empirischen Mittelwerte des Approximationsfehlerprozesses, sind
jedoch hier immer reell. Die Messwerte f"ur das Spektrum der gefensterten
deterministischen St"orung nach Gleichung (\ref{E.5.5}) sind dementsprechend
die empirischen Mittelwerte des Spektrums des gefensterten Prozesses 
\mbox{$\boldsymbol{y}(k)$}. Sie sind symmetrisch 
\mbox{(\,$\Hat{U}_{\!f}(\mu)=\Hat{U}_{\!f}(\!-\mu)^{\Kk}$\,)}
und minimieren den quadratischen Fehler der Ausgleichsl"osung der
Gleichungssysteme
\begin{equation}
\Hat{U}_{\!f}(\mu)\CdoT\Vec{1}\;=\;\Vec{Y}_{\!f}(\mu)
\qquad\qquad\forall\qquad\mu=0\;(1)\;M\!-\!1.
\label{E.5.16}
\end{equation}
Die Messwerte \mbox{$\Hat{\Phi}_{\boldsymbol{n}}(\mu)$}
erhalten wir aus dem Vektor \mbox{$\Hat{\Vec{N}}_{\!f}(\mu)$}, der durch
Projektion des Stichprobenvektors \mbox{$\Vec{Y}_{\!f}(\mu)$}
in einen Unterraum, der zum Vektor \mbox{$\Vec{1}$} orthogonal ist, entsteht.
Sinnvollerweise erfolgt die Projektion mit der idempotenten Matrix
\mbox{$\underline{1}_{\bot}$} nach Gleichung~(\ref{E.3.9}), deren Spur
mit \mbox{$L\!-\!1$} maximal ist. Wir erhalten damit die erwartungstreuen
Messwerte
\mbox{$\Hat{\Phi}_{\boldsymbol{n}}(\mu)=\Hat{\Phi}_{\boldsymbol{n}}(\!-\mu)=
\Hat{\Psi}_{\boldsymbol{n}}(\mu)$} gem"a"s Gleichung (\ref{E.5.6}),
die ohne Zwischenspeicherung der Spektralwerte aller
Einzelmessungen berechnet werden k"onnen.

Die Varianz der Messwerte
\mbox{$\Hat{\boldsymbol{U}}_{\!\!f}(\mu)$} ergibt sich wie bei
einem komplexwertigen System gem"a"s Gleichung (\ref{E.5.7.a})
und kann mit (\ref{E.5.8.a}) abgesch"atzt werden. Die Messwerte
\mbox{$\Hat{\boldsymbol{U}}_{\!\!f}(\mu)$} sind f"ur die
beiden diskreten Frequenzen \mbox{$\mu\!=\!0$} und \mbox{$\mu\!=\!M/2$}
immer reell. Daher ist f"ur diese beiden Frequenzen die
Messwertkovarianz gleich der Messwertvarianz, was sich
auch in Gleichung~(\ref{E.5.7.b}) mit
\mbox{$\Hat{\Phi}_{\boldsymbol{n}}(\mu)\!=\!\Hat{\Psi}_{\boldsymbol{n}}(\mu)$}
ergibt. Man gibt daher f"ur diese beiden Messwerte Konfidenzintervalle nach Gleichung~(\myref{3.73}) 
an, deren halbe Intervallbreiten sich nach Gleichung~(\myref{3.72}) mit dem 
eben angegebenen Sch"atzwerten der Messwertvarianzen absch"atzen lassen. 
Bei allen anderen Frequenzen ist die Messwertkovarianz wegen der
geforderten Stationarit"at des Prozesses \mbox{$\boldsymbol{n}(k)$}
in guter N"aherung null. Die Radien der sich ergebenden Konfidenzkreise 
der komplexen Messwerte sch"atzt man mit dem gew"unschten Konfidenzniveau 
und den Sch"atzwerten der Messwertvarianzen ab, indem man in Gleichung
(\myref{3.80.c}) die Messwertvarianzen
\mbox{$\Hat{C}_{\Hat{\boldsymbol{H}}(\mu),\Hat{\boldsymbol{H}}(\mu)}$}
durch die eben berechneten Messwertvarianzen
\mbox{$\Hat{C}_{\Hat{\boldsymbol{U}}_{\!\!f}(\mu),\Hat{\boldsymbol{U}}_{\!\!f}(\mu)}$}
ersetzt.

Bei der Berechnung der Varianzen und Kovarianzen der Messwerte
\mbox{$\Hat{\boldsymbol{u}}(k)$} der gefensterten deterministischen St"orung
ist auch beim reellwertigen System keine N"aherung mehr erforderlich.
Die Messwertabweichung ist nun reell, berechnet aber sich
wie im Fall einer komplexen deterministischen St"orung
nach Gleichung~(\ref{E.5.9}). Bei einem station"aren 
Approximationsfehlerprozess und bei Verwendung einer Fensterfolge, 
die der Bedingung~(\myref{2.20}) gen"ugt, ergibt sich dieselbe zeitunabh"angige
Messwertvarianz nach Gleichung~(\ref{E.5.10.a}) wie im komplexen Fall.
Als Summe erwartungstreuer Messwerte ist der Sch"atzwert der 
zeitunabh"angigen Messwertvarianz, die sich wieder nach Gleichung~(\ref{E.5.11.a}) 
berechnet, ebenfalls erwartungstreu. Somit kann man hier
ein zeitunabh"angiges Konfidenzintervall nach Gleichung~(\myref{3.73})
absch"atzen, wobei dort die halbe Intervallbreite nach Gleichung~(\myref{3.72}) 
mit dem gew"unschten Konfidenzniveau und mit der nach 
Gleichung~(\ref{E.5.11.a}) berechneten Messwertvarianz statt der Messwertvarianz
\mbox{$\Hat{C}_{\Hat{\boldsymbol{\Phi}}_{\!\boldsymbol{n}}(\mu),\Hat{\boldsymbol{\Phi}}_{\!\boldsymbol{n}}(\mu)}$}
abgesch"atzt wird.

Bei der Berechnung der Varianzen der Messwerte
\mbox{$\Hat{\boldsymbol{\Phi}}_{\boldsymbol{n}}(\mu)$} ist nun in den 
Gleichungen~(\ref{E.5.12}) die Gleichheit  
\mbox{$\Tilde{\Phi}_{\boldsymbol{n}}(\mu)=\Tilde{\Psi}_{\boldsymbol{n}}(\mu)$} 
einzusetzen:
\begin{equation}
C_{\Hat{\boldsymbol{\Phi}}_{\!\boldsymbol{n}}(\mu),\Hat{\boldsymbol{\Phi}}_{\!\boldsymbol{n}}(\mu)}\;=\;
\begin{cases}
{\D\;\frac{2}{L\!-\!1}\cdot\Tilde{\Phi}_{\boldsymbol{n}}(\mu)^{\uP{0.4}{\!2}}}&
\text{ f"ur}\qquad\mu\in\big\{0;\frac{M}{2}\big\}\\[8pt]
{\D\;\frac{1}{L\!-\!1}\cdot\Tilde{\Phi}_{\boldsymbol{n}}(\mu)^{\uP{0.4}{\!2}}}&
\text{ f"ur}\qquad\mu=1\;(1)\;\frac{M-1}{2}.
\end{cases}
\label{E.5.17}
\end{equation}
Da auch f"ur die Messwerte \mbox{$\Hat{\Phi}_{\boldsymbol{n}}(\mu)=\Hat{\Psi}_{\boldsymbol{n}}(\mu)$}
gilt, sch"atzt man die Messwertvarianz mit den Gleichungen~(\ref{E.5.13}) als
\begin{equation}
\Hat{C}_{\Hat{\boldsymbol{\Phi}}_{\!\boldsymbol{n}}(\mu),\Hat{\boldsymbol{\Phi}}_{\!\boldsymbol{n}}(\mu)}\;=\;
\begin{cases}
{\D\;\frac{2}{L+1}\cdot\Hat{\Phi}_{\boldsymbol{n}}(\mu)^{\uP{0.4}{\!2}}}&
\text{ f"ur}\qquad\mu\in\big\{0;\frac{M}{2}\big\}\\[8pt]
{\D\;\frac{1}{L}\cdot\Hat{\Phi}_{\boldsymbol{n}}(\mu)^{\uP{0.4}{\!2}}}&
\text{ f"ur}\qquad\mu=1\;(1)\;\frac{M-1}{2}
\end{cases}
\label{E.5.18}
\end{equation}
ab. Damit lassen sich die Sch"atzwerte der Konfidenzintervalle nach Gleichung~(\myref{3.72}) und 
(\myref{3.73}) angeben. 

\chapter{Reellwertige Systeme}\label{E.Kap.6}

In diesem Kapitel wird darauf eingegangen, wie das in Kapitel
\myref{Resys} vorgestellte Messverfahren zur Messung reellwertiger
realer Systeme benutzt werden kann, wenn auch die deterministische St"orung 
\mbox{$u(k)$} modelliert werden soll. Au"serdem wird kurz auf drei 
weitere Varianten des RKM zur Messung reellwertiger Systeme
eingegangen. Die in diesem Kapitel gemachten Untersuchungen beziehen sich auf 
den Fall {\em eines}\/ zeitinvarianten Modellsystems (\,\mbox{$K_H\!=\!1$}\,), bei dem 
das Modellsystem ${\cal S}_{*,lin}$ in Bild \ref{E.b1h} weggelassen wird. 
Bei einem reellwertigen System macht es n"amlich keinen Sinn, ein System zu 
modellieren, das von der konjugierten Erregung gespeist wird. Von dem 
Approximationsfehlerprozess wird angenommen, dass er station"ar ist 
(\,\mbox{$K_{\Phi}\!=\!1$}\,). Die Modifikationen, die sich bei einem 
periodisch zeitvarianten Modellsystem oder einem zyklostation"aren 
Approximationsfehlerprozess ergeben w"urden, m"oge sich der Leser anhand 
der in Kapitel \ref{E.Kap.2} und \ref{E.Kap.3} und der in diesem Kapitel 
angegeben "Uberlegungen selbst herleiten.

\section{Erste Variante des RKM zur Messung reellwertiger Systeme}\label{E.Kap.6.1}

Die Art der Erregung des Systems und der Fensterung und Fouriertransformation 
des Ausgangsprozesses bleibt gegen"uber Kapitel \myref{Resys} unver"andert. Auch die in den 
Gleichungen~(\myref{4.1}) und (\myref{4.2}) genannten Symmetrien der Spektren 
bleiben erhalten. Der Approximationsfehler, dessen zweites Moment zur Systemapproximation 
minimiert wird, ist nun jedoch nach Gleichung~(\ref{E.2.17}) definiert, f"ur die sich 
im Fall eines zeitinvarianten Modellsystems 
\begin{equation}
\boldsymbol{n}(k)\,=\,
\boldsymbol{y}(k)-
\frac{1}{M}\cdoT\Sum{\mu=0}{M-1}
H\big({\T\mu\CdoT\frac{2\pi}{M}}\big)\CdoT
\boldsymbol{V}(\mu)\cdot
e^{j\cdot\frac{2\pi}{M}\cdot\mu\cdot k}-u(k),
\label{E.6.1}
\end{equation}
ergibt. Der  zu minimierende Term (\myref{2.10}) ist ebenfalls um das Modell der 
deterministischen St"orung zu erweitern:
\begin{equation}
\text{E}\big\{\,|\boldsymbol{n}(k)|^2\big\}\,=\,
\text{E}\Bigg\{\,\bigg|\,\boldsymbol{y}(k)-
\frac{1}{M}\cdoT\Sum{\mu=0}{M-1}H\big({\T\mu\CdoT\frac{2\pi}{M}}\big)\CdoT 
\boldsymbol{V}(\mu)\CdoT e^{j\cdot\frac{2\pi}{M}\cdot\mu\cdot k}\,-\,u(k)\,
\bigg|^2\Bigg\}\,\stackrel{!}{=}\,\text{minimal}\,.
\label{E.6.2}
\end{equation}
Wenn man voraussetzt, dass die Varianzen aller $M$ Zufallsgr"o"sen 
\mbox{$\boldsymbol{V}\!(\mu)$} von null verschieden sind, erh"alt man mit 
\begin{equation}
H\big({\T\mu\CdoT\frac{2\pi}{M}}\big)\;=\;
\frac{\D \text{E}\Big\{
\big(\boldsymbol{V}(\mu)\!-\!
\text{E}\{\boldsymbol{V}(\mu)\}\big)^{\!*}\!\CdoT
\boldsymbol{Y}_{\!\!\!f}(\mu)\Big\}}
{\D \text{E}\Big\{\big|\boldsymbol{V}(\mu)\!-\!
\text{E}\{\boldsymbol{V}(\mu)\}\big|^2\Big\}}
\qquad\qquad\forall\qquad\mu=0\;(1)\;M\!-\!1\quad{}
\label{E.6.3}
\end{equation}
die L"osung f"ur die theoretischen Werte der "Ubertragungsfunktion, die 
hier ebenfalls die Symmetrie~(\myref{4.3}) aufweist. Damit zeigen auch die 
Optimalwerte des Spektrums der gefensterten deterministischen St"orung nach 
Gleichung~(\ref{E.2.29}), f"ur die sich hier 
\begin{equation}
U_{\!f}(\mu)\;=\;
\text{E}\big\{\boldsymbol{Y}_{\!\!\!f}(\mu)\big\}-
\text{E}\big\{\boldsymbol{V}(\mu)\big\}\CdoT
H\big({\T\mu\CdoT\frac{2\pi}{M}}\big)
\qquad\qquad\forall\qquad \mu=0\;(1)\;M\!-\!1\quad{}
\label{E.6.4}
\end{equation}
ergibt,  dieselbe Symmetrie und die deterministische St"orung nach 
Gleichung (\ref{E.2.19}), f"ur die sich hier\vspace{-6pt}
\begin{equation}
u(k)\;=\;\text{E}\{\boldsymbol{y}(k)\}-
\frac{1}{M}\cdoT\Sum{\mu=0}{M-1}
H\big({\T\mu\CdoT\frac{2\pi}{M}}\big)\CdoT
\text{E}\big\{\boldsymbol{V}(\mu)\big\}\CdoT
e^{j\cdot\frac{2\pi}{M}\cdot\mu\cdot k}
\quad\forall\quad k=0\;(1)\;F\!-\!1.
\label{E.6.5}
\end{equation}
ergibt, ist reell. Somit ist auch der Approximationsfehlerprozess 
\begin{equation}
\boldsymbol{n}(k)\,=\,\boldsymbol{y}(k)-\boldsymbol{x}(k)-u(k)
\label{E.6.6}
\end{equation}
reell. Bei einem reellwertigen System liefert also die
optimale Approximierung immer ein reellwertiges Modellsystem, dem sich
ausgangsseitig eine reelle St"orung "uberlagert, deren erstes Moment
\mbox{$u(k)$} ebenfalls reell ist. Zur vollst"andigen Beschreibung der
zweiten zentralen Momente des Approximationsfehlerprozesses gen"ugt
daher die Angabe der reellen Autokorrelationsfolge, aus der man im
Fall eines station"aren Approximationsfehlerprozesses durch diskrete
Fouriertransformation das geradesymmetrische reelle LDS
\mbox{$\Phi_{\boldsymbol{n}}(\Omega)$} gewinnt. Das bei einem
komplexwertigen System noch anzugebende KLDS
\mbox{$\Psi_{\boldsymbol{n}}(\Omega)$} ist hier identisch mit dem LDS.
Auf die Messung der Stufenapproximation
\mbox{$\Bar{\Psi}_{\boldsymbol{n}}(\mu)$} oder deren N"aherung
\mbox{$\Tilde{\Psi}_{\boldsymbol{n}}(\mu)$}
kann daher bei einem reellwertigen System verzichtet werden. Auch bei
einem reellwertigen realen System liefert die Minimierung des
zweiten Moments des Approximationsfehlers eine L"osung, bei der die
Orthogonalit"at des Spektrums der Erregung und des Spektrums des
gefensterten Approximationsfehlers gegeben ist, so dass auch hier Gleichung
(\myref{2.30}) erf"ullt ist.

Um Sch"atzwerte f"ur die optimalen Regressionskoeffizienten durch eine Messung
zu erhalten, erregt man das System mit dem in Kapitel~\myref{RKM} beschriebenen
Verfahren, also  mit $L$ reellen Testsignalsequenzen, die bereichsweise
mit $M$ periodisch fortgesetzt sind. Diese d"urfen hier jedoch mit einem 
Zufallsvektor \mbox{$\Vec{\boldsymbol{v}}$} erzeugt werden, der ein 
zeitabh"angiges erstes Moment aufweist. Was man misst, ist nur der Realteil
der $L$ Stichprobenelemente des Ausgangssignals des realen Systems.
\vadjust{\penalty-100}Der Imagin"arteil der $L$ Systemausgangssignale wird bei der Berechnung
der Messwerte und ihrer Varianzen und Kovarianzen zu null gesetzt.
Da jedes Stichprobenelement (\,=~Signalabschnitt einer Einzelmessung\,)
sowohl am Systemein- als auch am -ausgang reell ist, weisen beide
Stichprobenvektoren \mbox{$\Hat{\Vec{V}}(\mu)$} und
\mbox{$\Hat{\Vec{Y}}_{\!\!f}(\mu)$} die in Gleichung~(\myref{4.4}) 
genannte Symmetrie auf. Die $M$ Gleichungssysteme (\ref{E.3.2}) vereinfachen sich hier zu 
\begin{equation}
\Hat{H}(\mu)\CdoT\Vec{V}(\mu)\;+\;
\Hat{U}_{\!f}(\mu)\cdot\Vec{1} \;=\; \Vec{Y}_{\!f}(\mu)
\qquad\qquad\forall\qquad \mu=0\;(1)\;M\!-\!1.
\label{E.6.7}
\end{equation}
Wir erhalten die $M$ Ausgleichsl"osungen 
\begin{equation}
\Hat{H}(\mu)\;=\;\Hat{C}_{\boldsymbol{Y}_{\!\!\!f}(\mu),\boldsymbol{V}(\mu)}\Cdot
\Hat{C}_{\boldsymbol{V}(\mu),\boldsymbol{V}(\mu)}^{\,-1}
\qquad\qquad\forall\qquad\mu=0\;(1)\;M\!-\!1
\label{E.6.8}
\end{equation}
und
\begin{equation}
\Hat{U}_{\!f}(\mu)\;=\;
\frac{1}{L}\cdot
\Big(\,\Vec{Y}_{\!f}(\mu)-\Hat{H}(\mu)\CdoT\Vec{V}(\mu)\,\Big)\cdot
\Vec{1}^{\,\Hh}
\qquad\qquad\forall\qquad\mu=0\;(1)\;M\!-\!1
\label{E.6.9}
\end{equation}
mit den empirischen Varianzen
\begin{gather}
\Hat{\underline{C}}_{\boldsymbol{V}(\mu_1),\boldsymbol{V}(\mu_2)}\;=\;
\frac{1}{L\!-\!1}\cdot
\Vec{V}(\mu_1)\cdot\underline{1}_{\bot}\Cdot\Vec{V}(\mu_2)^{\Hh}\;=
\label{E.6.10}\\*[4pt]
=\;\frac{1}{L\!-\!1}\cdot\bigg(\,\Sum{\lambda=1}{L}\,
V_{\lambda}(\mu_1)\CdoT V_{\lambda}(\mu_2)^{\Kk}-\,
\frac{1}{L}\cdot
\Sum{\lambda=1}{L}V_{\lambda}(\mu_1)\cdot
\Sum{\lambda=1}{L}V_{\lambda}(\mu_2)^{\Kk}\bigg)
\notag\\*[4pt]
\forall\qquad\mu_1=0\;(1)\;M\!-\!1\quad\text{und}
\quad\mu_2=0\;(1)\;M\!-\!1,\notag
\end{gather}
und den empirischen Kovarianzen 
\begin{gather}
\Hat{\underline{C}}_{\boldsymbol{Y}_{\!\!\!f}(\mu_1),\boldsymbol{V}(\mu_2)}\;=\;
\frac{1}{L\!-\!1}\cdot
\Vec{Y}_{\!f}(\mu_1)\cdot\underline{1}_{\bot}\Cdot\Vec{V}(\mu_2)^{\Hh}\;=
\label{E.6.11}\\*[4pt]
=\;\frac{1}{L\!-\!1}\cdot\bigg(\,\Sum{\lambda=1}{L}\,
Y_{\!f,\lambda}(\mu_1)\CdoT V_{\lambda}(\mu_2)^{\Kk}-\,
\frac{1}{L}\cdot
\Sum{\lambda=1}{L}Y_{\!f,\lambda}(\mu_1)\cdot
\Sum{\lambda=1}{L}V_{\lambda}(\mu_2)^{\Kk}\bigg)
\notag\\*[4pt]
\forall\qquad\mu_1=0\;(1)\;M\!-\!1\quad\text{und}
\quad\mu_2=0\;(1)\;M\!-\!1\notag
\end{gather}
Diese Ausgleichsl"osungen besitzen dann ebenfalls die Symmetrieeigenschaften
\begin{equation}
\Hat{H}(\mu)=\Hat{H}(\!-\mu)^{\Kk}
\quad\text{ bzw. }\quad
\Hat{U}_{\!f}(\mu)=\Hat{U}_{\!f}(\!-\mu)^{\Kk}.
\label{E.6.12}
\end{equation}
Daher gen"ugt es bei geradem $M$ --- wovon wir im weiteren ausgehen ---
die jeweils \mbox{$M/2+1$} Werte der L"osung f"ur \mbox{$\mu=0\;(1)\;M/2$}
zu berechnen. Mit den Ausgleichsl"osungen der "Ubertragungsfunktion 
erh"alt man die reellen Messwerte der deterministischen St"orung:
\begin{equation}
\Hat{u}(k)\;=\;
\frac{1}{L}\CdoT\Vec{y}(k)\cdoT\Vec{1}^{\,\Hh}\!-
\frac{1}{M\CdoT L}\cdoT\!\Sum{\mu=0}{M-1}
\Hat{H}(\mu)\CdoT\Vec{V}(\mu)\CdoT\Vec{1}^{\,\Hh}\!\Cdot
e^{j\cdot\frac{2\pi}{M}\cdot\mu\cdot k}
\quad\forall\quad k=0\;(1)\;F\!-\!1.
\label{E.6.13}
\end{equation}
Da wir bisher
keine Modifikationen im Messverfahren vorgenommen haben, braucht
die Erwartungstreue der Messwerte \mbox{$\Hat{\boldsymbol{H}}(\mu)$},
\mbox{$\Hat{\boldsymbol{U}}_{\!f}(\mu)$} und \mbox{$\Hat{\boldsymbol{u}}(k)$}
\vadjust{\penalty-100}nicht gesondert gezeigt zu werden. Die beim Einsetzen der Messwerte
in die Gleichungssysteme (\ref{E.6.7}) verbleibenden Fehlervektoren
\mbox{$\Hat{\Vec{N}}_{\!f}(\mu)$} der Ausgleichsl"osung, die sich
wieder nach Gleichung (\ref{E.3.25}) berechnen, sind ebenfalls in
derselben Art symmetrisch wie alle anderen Stichprobenvektoren
der Spektren der Signale am reellwertigen System. 

Auch bei einem reellwertigen System verwenden wir f"ur die Absch"atzung der 
Stufenapproximation des LDS die immer reellen und erwartungstreuen Messwerte
\mbox{$\Hat{\Phi}_{\boldsymbol{n}}(\mu)$} nach Gleichung (\ref{E.3.29.a}).
Dabei verwenden wir weiterhin die Matrix \mbox{$\underline{V}_{\bot}\!(\mu)$}, 
die hier nun die beiden Vektoren $\Vec{1}$ und \mbox{$\Vec{V}(\mu)$} 
als Eigenvektoren zum Eigenwert Null haben muss:
\begin{equation}
\Vec{V}(\mu)\cdot\underline{V}_{\bot}\!(\mu)\;=\;\Vec{0}
\qquad\wedge\qquad
\Vec{1}\cdot\underline{V}_{\bot}\!(\mu)\;=\;\Vec{0}.
\label{E.6.14}
\end{equation}
Bei der Konstruktion dieser 
Matrix nach Gleichung (\ref{E.3.28}) ergibt sich nun jedoch ein wesentlicher
Unterschied. Da bei einem reellwertigen System die beiden Zeilenvektoren
\mbox{$\Vec{V}(\mu)$} und \mbox{$\Vec{V}(\!-\mu)^{\Kk}$} der immer gleich sind, 
enth"alt die Matrix \mbox{$\Breve{\underline{V}}(\mu)$} nur einen Zeilenvektor:
\begin{equation}
\Breve{\underline{V}}(\mu)\;=\;\Vec{V}(\mu)
\qquad\qquad\forall\qquad\mu=0\;(1)\;M\!-\!1
\label{E.6.15}
\end{equation}
Die weitere Berechnung der Matrix \mbox{$\underline{V}_{\bot}\!(\mu)$} kann
dann wieder mit Hilfe der Gleichungen~(\ref{E.3.27}) und (\ref{E.3.28}) erfolgen und ergibt:
\begin{gather}
\underline{V}_{\bot}\!(\mu)\;=\;
\underline{V}_{\bot}\!(\mu)^{n}\;=\;
\underline{V}_{\bot}\!(\mu)^{\Hh}\;=\;
\underline{1}_{\bot}-\frac{1}{L\!-\!1}\cdot
\underline{1}_{\bot}\!\CdoT\Vec{V}(\mu)^{\Hh}\!\Cdot
\Hat{C}_{\boldsymbol{V}(\mu),\boldsymbol{V}(\mu)}^{\,-1}\CdoT
\Vec{V}(\mu)\CdoT\underline{1}_{\bot}
\notag\\*[2pt]
{\T\forall\qquad \mu=0\;(1)\;\frac{M}{2}\qquad\wedge\qquad n\in\mathbb{N}}.
\label{E.6.16}
\end{gather}
Diese Matrix erf"ullt die Gleichungen~(\ref{E.3.33}) und hat die Spur \mbox{$L\!-\!2$}.
Indem wir die modifiziert konstruierte Matrix \mbox{$\underline{V}_{\bot}\!(\mu)$} 
in die Gleichung~(\ref{E.3.29.a}) einsetzen, erhalten wir mit \mbox{$K(\mu)\!=\!1$} 
die in Gleichung~(\ref{E.3.34.a}) angegebenen Messwerte:
\begin{gather}
\Hat{\Phi}_{\boldsymbol{n}}(\mu)\;=\;
\frac{\;\Hat{\Vec{N}}_{\!f}(\mu)\CdoT
\Hat{\Vec{N}}_{\!f}(\mu)^{\HH}\;}
{M\CdoT(L\!-\!2)}\;=
\label{E.6.17}\\[5pt]
=\;\frac{\;\Vec{Y}_{\!f}(\mu)\cdot
\underline{V}_{\bot}\!(\mu)\cdot
\Vec{Y}_{\!f}(\mu)^{\HH}\;}
{M\CdoT(L\!-\!2)}\;=\;
\frac{\;\Vec{N}_{\!f}(\mu)\cdot
\underline{V}_{\bot}\!(\mu)\cdot
\Vec{N}_{\!f}(\mu)^{\HH}\;}
{M\CdoT(L\!-\!2)}\;=
\notag\\[5pt]
=\;\frac{1}{M}\cdot\frac{L\!-\!1}{L\!-\!2}\cdot\Big(\!
\Hat{C}_{\boldsymbol{Y}_{\!\!\!f}(\mu),\boldsymbol{Y}_{\!\!\!f}(\mu)}-
\Hat{C}_{\boldsymbol{Y}_{\!\!\!f}(\mu),\boldsymbol{V}(\mu)}\Cdot
\Hat{C}_{\boldsymbol{V}(\mu),\boldsymbol{V}(\mu)}^{\uP{0.4}{-1}}\CdoT
\Hat{C}_{\boldsymbol{Y}_{\!\!\!f}(\mu),\boldsymbol{V}(\mu)}^{\,\Kk}\Big)\;=
\notag\\[5pt]
=\;\frac{1}{M}\cdot\frac{L\!-\!1}{L\!-\!2}\cdot\Big(\!
\Hat{C}_{\boldsymbol{Y}_{\!\!\!f}(\mu),\boldsymbol{Y}_{\!\!\!f}(\mu)}-
\big|\Hat{H}(\mu)\big|^2\Cdot
\Hat{C}_{\boldsymbol{V}(\mu),\boldsymbol{V}(\mu)}\Big)
\notag\\[3pt]
\forall\quad\mu=0\;(1)\;M\!-\!1.\notag
\end{gather}
Dabei wurden zuletzt die empirischen Varianzen nach Gleichung (\ref{E.3.35.a}) 
mit \mbox{$\mu_1\!=\!\mu_2$} verwendet.

Die Varianz der Messwerte der "Ubertragungsfunktion berechnet sich mit 
Gleichung~(\ref{E.3.40}). Der Vektor \mbox{$\Hat{\Vec{H}}(\mu)$} enth"alt 
\vadjust{\penalty-100}in unserem Fall nur ein Element, n"amlich den Messwert \mbox{$\Hat{H}(\mu)$}. Auch die Kovarianzmatrix 
\mbox{$\Hat{\underline{\boldsymbol{C}}}_{\Tilde{\Vec{\boldsymbol{V}}}(\mu),\Tilde{\Vec{\boldsymbol{V}}}(\mu)}$}
ist hier nun eine skalare Gr"o"se, n"amlich die nach Gleichung~(\ref{E.6.10}) berechnete 
Varianz \mbox{$\Hat{\boldsymbol{C}}_{\boldsymbol{V}(\mu),\boldsymbol{V}(\mu)}$} 
des Spektralwertes der Erregung. Der Einheitsvektor \mbox{$\Vec{E}_n$}, dessen n-tes Element 
eins ist, entartet hier zu dem skalaren Wert $1$. Damit ergibt sich die 
Messwertvarianz
\begin{equation}
C_{\Hat{\boldsymbol{H}}(\mu),\Hat{\boldsymbol{H}}(\mu)}\;=\;
\frac{M}{L\!-\!1}\cdot\text{E}\big\{
\Hat{\boldsymbol{C}}_{\boldsymbol{V}(\mu),\boldsymbol{V}(\mu)}^{\uP{0.4}{\!-1}}\,\big\}\cdot
\Tilde{\Phi}_{\boldsymbol{n}}(\mu)
\qquad\forall\quad\mu=0\;(1)\;M\!-\!1.
\label{E.6.18}
\end{equation}
Die Messwerte der "Ubertragungsfunktion sind f"ur die beiden diskreten
Frequenzen \mbox{$\mu\!=\!0$} und \mbox{$\mu\!=\!M/2$} immer reell. F"ur diese
beiden Messwerte ist die Messwertkovarianz nach Gleichung~(\ref{E.3.43}) gleich der 
eben berechneten Messwertvarianz. F"ur alle anderen Frequenzen ist die Messwertkovarianz 
in guter N"aherung null. Als erwartungstreue Sch"atzwerte f"ur die Messwertvarianz erhalten 
wir mit Gleichung~(\ref{E.3.44.a}): 
\begin{equation}
\Hat{C}_{\Hat{\boldsymbol{H}}(\mu),\Hat{\boldsymbol{H}}(\mu)}\;=\;
\frac{M}{L\!-\!1}\cdot
\Hat{C}_{\boldsymbol{V}(\mu),\boldsymbol{V}(\mu)}^{\uP{0.4}{\!-1}}\cdot
\Hat{\Phi}_{\boldsymbol{n}}(\mu)
\qquad\forall\quad\mu=0\;(1)\;M\!-\!1.
\label{E.6.19}
\end{equation}
Gleichung~(\ref{E.3.44.b}) liefert f"ur die beiden 
Frequenzen \mbox{$\mu\!=\!0$} und \mbox{$\mu\!=\!M/2$} exakt dieselben Sch"atzwerte 
f"ur die Messwertkovarianz. F"ur alle anderen Frequenzen wird die Messwertkovarianz 
mit null abgesch"atzt. Damit wird f"ur die beiden Frequenzen \mbox{$\mu\!=\!0$} 
und \mbox{$\mu\!=\!M/2$} die L"ange der k"urzeren Halbachse der Konfidenzellipse zu null. 
Es ist daher f"ur diese beiden diskreten Frequenzen sinnvoll, statt der 
Konfidenzellipsen, Konfidenzintervalle analog zu den Konfidenzintervallen 
der LDS-Messwerte nach Gleichung (\myref{3.73}) anzugeben. 
Die halbe Intervallbreite wird mit Gleichung (\myref{3.72}) abgesch"atzt, 
wobei hier die Sch"atzwerte der Varianzen von 
\mbox{$\Hat{\boldsymbol{\Phi}}_{\boldsymbol{n}}(\mu)$} durch die Sch"atzwerte
der Varianzen von \mbox{$\Hat{\boldsymbol{H}}(\mu)$} zu ersetzen sind. 
Bei den komplexen Messwerten der "Ubertragungsfunktion aller anderen 
Frequenzen wird die Messwertkovarianz bei Verwendung einer hoch frequenzselektiven
Fensterfolge gegen"uber der Messwertvarianz wieder vernachl"assigbar klein,
so dass man auch beim reellwertigen System Konfidenzkreise erh"alt, deren
Radius man mit Gleichung (\myref{3.80.c}) aus der Messwertvarianz und dem 
gew"unschten Konfidenzniveau absch"atzt. 

F"ur die Messwerte \mbox{$\Hat{\boldsymbol{U}}_{\!\!f}(\mu)$} gilt dasselbe.
Auch hier sind die Messwerte f"ur \mbox{$\mu\!=\!0$} und \mbox{$\mu\!=\!M/2$}
reell, so dass man Konfidenzintervalle angibt, deren halbe Intervallbreiten
sich nach Gleichung~(\myref{3.72}) mit dem Sch"atzwerten der Messwertvarianzen
nach Gleichung~(\ref{E.3.48.a}) absch"atzen lassen. In unseren Fall ergibt sich 
die Messwertvarianz mit Gleichung~(\ref{E.3.47.a}) zu:\vspace{-16pt}
\begin{equation}
C_{\Hat{\boldsymbol{U}}_{\!\!f}(\mu),\Hat{\boldsymbol{U}}_{\!\!f}(\mu)}\;=\;
\frac{M}{L\!-\!1}\cdot\text{E}\bigg\{
\frac{\Vec{\boldsymbol{V}}(\mu)\CdoT\Vec{\boldsymbol{V}}(\mu)^{\Hh}}
{L\CdoT\Hat{\boldsymbol{C}}_{\boldsymbol{V}(\mu),\boldsymbol{V}(\mu)}}\bigg\}
\cdot\Tilde{\Phi}_{\boldsymbol{n}}(\mu)
\qquad\forall\quad\mu=0\;(1)\;M\!-\!1.
\label{E.6.20}
\end{equation}
Die Messwertvarianz wird minimal, wenn der Erwartungswert des 
Quotienten aus dem empirischen, nichtzentralen zweiten Moment 
\mbox{$\Vec{\boldsymbol{V}}(\mu)\CdoT\Vec{\boldsymbol{V}}(\mu)^{\Hh}/L$} 
und dem empirischen, zentralen zweiten Moment 
\mbox{$\Hat{\boldsymbol{C}}_{\boldsymbol{V}(\mu),\boldsymbol{V}(\mu)}$} 
minimal wird. Daher sollte man, wenn m"oglich, als Erregung eine Zufallsprozess 
w"ahlen, dessen Spektralwerte mittelwertfrei sind. 
Gleichung~(\ref{E.3.48.a}) liefert die erwartungstreuen \vadjust{\penalty-100}Sch"atzwerte\vspace{0pt}
\begin{equation}
\Hat{C}_{\Hat{\boldsymbol{U}}_{\!\!f}(\mu),\Hat{\boldsymbol{U}}_{\!\!f}(\mu)}\;=\;
\frac{M}{L\!-\!1}\cdot
\frac{\Vec{V}(\mu)\CdoT\Vec{V}(\mu)^{\Hh}}{L\CdoT
\Hat{C}_{\boldsymbol{V}(\mu),\boldsymbol{V}(\mu)}}\CdoT
\Hat{\Phi}_{\boldsymbol{n}}(\mu)
\qquad\forall\quad\mu=0\;(1)\;M\!-\!1.
\label{E.6.21}
\end{equation}
Die Radien der Konfidenzkreise der komplexen Messwerte \mbox{$\Hat{\boldsymbol{U}}_{\!\!f}(\mu)$}
aller anderen Frequenzen sch"atzt man mit dem gew"unschten Konfidenzniveau und den
Sch"atzwerten der Messwertvarianzen ab, indem man in Gleichung (\myref{3.80.c})
die Messwertvarianzen
\mbox{$\Hat{C}_{\Hat{\boldsymbol{H}}(\mu),\Hat{\boldsymbol{H}}(\mu)}$}
durch die in Gleichung~(\ref{E.6.21}) berechneten Werte ersetzt.

Die Messwerte \mbox{$\Hat{\boldsymbol{u}}(k)$} sind f"ur alle Zeitpunkte
\mbox{$k=0\;(1)\;F\!-\!1$} reell. Mit Gleichung~(\ref{E.3.52.a}) ergibt 
sich die zeitunabh"angige N"aherung
\begin{equation}
C_{\Hat{\boldsymbol{u}}(k),\Hat{\boldsymbol{u}}(k)}\;\approx\;
\frac{1}{M^2}\cdoT\Sum{\mu=0}{M-1}
C_{\Hat{\boldsymbol{U}}_{\!\!f}(\mu),\Hat{\boldsymbol{U}}_{\!\!f}(\mu)}
\qquad\forall\qquad k=0\;(1)\;F\!-\!1 
\label{E.6.22}
\end{equation}
f"ur die Varianz der Messwerte \mbox{$\Hat{\boldsymbol{u}}(k)$}, die mit Gleichung~(\ref{E.3.53.a}) aus den 
Sch"atzwerten der Varianzen der Messwerte \mbox{$\Hat{\boldsymbol{U}}_{\!\!f}(\mu)$} zu
\begin{equation}
\Hat{C}_{\Hat{\boldsymbol{u}}(k),\Hat{\boldsymbol{u}}(k)}\;=\;
\frac{1}{M^2}\cdot\Sum{\mu=0}{M-1}
\Hat{C}_{\Hat{\boldsymbol{U}}_{\!\!f}(\mu),\Hat{\boldsymbol{U}}_{\!\!f}(\mu)}
\qquad\forall\qquad k=0\;(1)\;F\!-\!1, 
\label{E.6.23}
\end{equation}
abgesch"atzt werden kann. Somit kann man hier ein zeitunabh"angiges 
Konfidenzintervall nach Gleichung (\myref{3.73}) absch"atzen, wobei 
dort die halbe Intervallbreite nach Gleichung (\myref{3.72}) mit der 
Messwertvarianz nach Gleichung (\ref{E.6.23}) eingesetzt wird.

Die Varianzen der Messwerte
\mbox{$\Hat{\boldsymbol{\Phi}}_{\boldsymbol{n}}(\mu)$} lassen sich mit
den Gleichungen (\ref{E.3.56.a}) und (\ref{E.3.60.a}) berechnen. Da wir 
Matrizen \mbox{$\underline{\boldsymbol{V}}_{\bot}\!(\mu)$} verwenden,
die die Gleichungen~(\ref{E.3.33}) erf"ullen, sind die in den 
Gleichungen~(\ref{E.3.56.a}) und  (\ref{E.3.60.a}) vor den Produkten und 
Betragsquadraten der theoretischen Werte des LDS bzw. KLDS als Vorfaktoren 
auftretenden Erwartungswerte alle gleich 
\mbox{$\big(L\!-\!1\!-\!K(\mu)\big)^{\!-1}=(L\!-\!2)^{\uP{0.4}{\!-1}}$}. Unter Ber"ucksichtigung 
der f"ur reelle Prozesse immer g"ultigen  Beziehung 
\mbox{$\Tilde{\Phi}_{\boldsymbol{n}}(\mu)\!=\!\Tilde{\Phi}_{\boldsymbol{n}}(\!-\mu)\!=\!\Tilde{\Psi}_{\boldsymbol{n}}(\mu)$}
erhalten wir
\begin{equation}
C_{\Hat{\boldsymbol{\Phi}}_{\!\boldsymbol{n}}(\mu),\Hat{\boldsymbol{\Phi}}_{\!\boldsymbol{n}}(\mu)}\;=\;
\begin{cases}
{\D\;\frac{2}{L\!-\!2}\cdot\Tilde{\Phi}_{\boldsymbol{n}}(\mu)^{\uP{0.4}{\!2}}}&
\text{ f"ur }\qquad\mu\in\big\{0\,;\frac{M}{2}\big\}\\[8pt]
{\D\;\frac{1}{L\!-\!2}\cdot\Tilde{\Phi}_{\boldsymbol{n}}(\mu)^{\uP{0.4}{\!2}}}&
\text{ f"ur}\qquad\mu=1\;(1)\;\frac{M-1}{2}.
\end{cases}
\label{E.6.24}
\end{equation}
Wir verwenden die in den Gleichungen~(\ref{E.3.58.a}) und (\ref{E.3.61.a}) angegebenen erwartungstreuen Sch"atzwerte, f"ur die sich in unserem Fall
\begin{equation}
\Hat{C}_{\Hat{\boldsymbol{\Phi}}_{\!\boldsymbol{n}}(\mu),\Hat{\boldsymbol{\Phi}}_{\!\boldsymbol{n}}(\mu)}\;=\;
\begin{cases}
{\D\;\frac{2}{L}\cdot\Hat{\Phi}_{\boldsymbol{n}}(\mu)^{\uP{0.4}{\!2}}}&
\text{ f"ur}\qquad\mu\in\big\{0\,;\frac{M}{2}\big\}\\[8pt]
{\D\;\frac{1}{L\!-\!1}\cdot\Hat{\Phi}_{\boldsymbol{n}}(\mu)^{\uP{0.4}{\!2}}}&
\text{ f"ur}\qquad\mu=1\;(1)\;\frac{M-1}{2}
\end{cases}
\label{E.6.25}
\end{equation}
ergibt. Mit deren Hilfe sch"atzt man die halbe Breite der Konfidenzintervalle nach 
Gleichung (\myref{3.73}) mit Gleichung (\myref{3.72}) ab.

\section[Weitere Varianten zur Messung reellwertiger Systeme]{Weitere Varianten zur Messung reellwertiger\\Systeme}\label{E.Kap.6.2}

Es wurden drei weitere Varianten des RKM zur Messung reellwertiger Systeme
untersucht. Diese sollen nun kurz vorgestellt werden, ohne dabei
alle Messwerte, deren Erwartungswerte, Varianzen und Kovarianzen
explizit anzugeben.

Bei allen drei Varianten werden die am realen reellwertigen System 
bei aufeinanderfolgenden Einzelmessungen auftretenden Ein- und 
Ausgangssignale jeweils als Real- und Imagin"arteil zweier 
komplexer Signalsequenzen zusammengefasst, und jeweils 
als eine konkrete Realisierung zweier entsprechender komplexer 
Zufallsvektoren am Ein- und Ausgang des Systems angesehen. 
Mit diesen komplexen Zufallsvektoren kann nun die Approximation 
mit einem komplexwertigen Modellsystem, mit einer komplexen 
deterministischen St"orung und mit einem komplexen Approximationsfehler 
vorgenommen werden, wie dies in Kapitel \ref{E.Kap.2.2} geschehen ist. 
Dabei muss die Erregung des Systems und die gegebenfalls zuf"allige 
Auswahl der Zeitintervalle der Einzelmessungen so gew"ahlt werden, 
dass sich Optimall"osungen der theoretischen komplexen Regression ergeben, 
die sich in die Optimall"osungen der theoretischen reellen Regression 
der in Unterkapitel \ref{E.Kap.6.1} geschilderten Systemapproximation mit reellen 
Zufallsvektoren "uberf"uhren lassen. Diese Optimall"osungen 
sind schlie"slich die Gr"o"sen, die man zu messen w"unscht. Wenn 
schon die theoretischen Optimall"osungen der komplexen Regression 
abweichen, wird man nicht erwarten k"onnen, dass die Erwartungswerte 
der Messergebnisse --- also der empirisch bestimmten optimalen 
Regressionskoeffizienten --- die richtigen Optimall"osungen 
der reellen Regression sein werden. Desweiteren muss sich ein 
Zufallsvektor f"ur den komplexen Approximationsfehlerprozess ergeben, 
dessen Real- und Imagin"arteil unabh"angig sind, wobei diese beiden Anteile 
dieselbe Verbundverteilung besitzen m"ussen, wie der Zufallsvektor 
des reellen Approximationsfehlerprozesses bei der Systemapproximation 
in Unterkapitel \ref{E.Kap.6.1}. Bei der reellen Regression waren 
die Stichprobenelemente des Zufallsvektors des reellen 
Approximationsfehlerprozesses zweier aufeinanderfolgender Einzelmessungen 
unabh"angig. Da die beiden Stichprobenelemente zweier aufeinanderfolgender 
Einzelmessungen nun als Real- und Imagin"arteil eines komplexen 
Stichprobenelements betrachtet werden, m"ussen die Real- und Imagin"arteile 
des komplexen Zufallsvektors unabh"angig sein und die gleiche 
Verbundverteilung besitzen wie der reelle Zufallsvektor, wenn man 
erwarten will, dass die aus den Stichproben berechneten Messwerte 
der stochastischen Eigenschaften des Approximationsfehlerprozesses 
die gew"unschten Erwartungswerte besitzen. Nur wenn sich eine Erregung 
finden l"asst, die diese Forderungen erf"ullt, kann man bei den drei 
folgenden Varianten des RKM erwarten, dass die Messergebnisse 
zu einer ad"aquaten Beschreibung des realen Systems im typischen 
Betriebszustand geeignet sind.  Bei den ersten beiden Varianten 
kann dies gegebenfalls unter Beachtung der zuf"alligen Auswahl 
der Zeitintervalle der Messung erreicht werden, indem man den Real- 
und Imagin"arteil der Erregung als eine konkrete Stichprobe aus einem 
Zufallsvektor mit unabh"angigen Real- und Imagin"arteilzufallsvektoren 
gleicher Verbundverteilung gewinnt. Man gewinnt also die 
Testsignalsequenzen auf dieselbe Weise, wie bei dem in Unterkapitel~\ref{E.Kap.6.1} 
dargestellten Messverfahren, interpretiert aber die Signalsequenzen 
aufeinanderfolgender Einzelmessungen nicht mehr als zwei vollst"andige 
Signale sondern als Real- und Imagin"arteil eines einzigen Signals. 
Bei der dritten Variante ist dies nicht m"oglich, da man hier
--- wie wir noch sehen werden --- eine spezielle komplexe Erregung,
die aus abh"angigen Real- und Imagin"arteilzufallsvektoren gewonnen
wird, verwendet.

Im weiteren werden die theoretisch optimalen Regressionskoeffizienten
der Systemapproximation mit dem reellwertigen Modellsystem nach Unterkapitel
\ref{E.Kap.6.1} mit dem Index \mbox{$\rule{0pt}{1ex}_{\text{reell}}$} gekennzeichnet,
w"ahrend die bisher verwendeten Formelzeichen die entsprechenden Gr"o"sen der
Systemapproximation mit den komplexen Zufallsvektoren bezeichnen.
Bei allen Varianten ergibt sich unter den ebengenannten Voraussetzungen
ein theoretisches komplexes Modellsystem, bei dem die
Anteile der "Ubertragungsfunktion, die die Real- und Imagin"arteile der
komplexen Zufallsvektoren des Ein- und Ausgangs kreuzweise
verkn"upfen, null sind, weil derartige Verkn"upfungen bei einer
Systemapproximation mit reellen Zufallsvektoren nicht vorkommen
k"onnen, weil die aufeinanderfolgenden Einzelmessungen unabh"angig sind.
Die Verkn"upfung der Realteile der komplexen Zufallsvektoren
des Ein- und Ausgangs ist identisch mit der Verkn"upfung
der Imagin"arteile der komplexen Zufallsvektoren
des Ein- und Ausgangs. Die wiederum ist identisch mit
der Verkn"upfung \mbox{$H_{\text{reell}}(\Omega)$} der reellen
Zufallsvektoren des Ein- und Ausgangs bei der Systemapproximation
mit reellen Signalen. Daher erh"alt man bei allen drei weiteren
RKM-Varianten zur Messung reellwertiger Systeme ein komplexwertiges
Modellsystem mit\vspace{0pt}
\begin{equation}
\qquad\qquad H(\Omega)\;=\;H_{\text{reell}}(\Omega)\;=\;
H_{\text{reell}}(\!-\Omega)^{\Kk}\;=\;H(\!-\Omega)^{\Kk}\!,
\label{E.6.26}
\end{equation}
wenn die oben genannten Voraussetzungen eingehalten werden.
Da die Regressionskoeffizienten \mbox{$u_{\text{reell}}(k)$} der 
deterministischen St"orung bei der Regression mit den
reellen Zufallsvektoren bei dem komplexen Systemmodell im
Real- und Imagin"arteil des komplexen Zufallsvektors des 
Systemausgangs identisch vorhanden sind, gilt f"ur die 
Optimall"osungen der Regressionskoeffizienten
\mbox{$u(k)$} der Regression mit den komplexen
Modellzufallsvektoren\vspace{-16pt}
\begin{gather}
\qquad\qquad\qquad\qquad
u(k)\;=\;u_{\text{reell}}(k)+j\CdoT u_{\text{reell}}(k)\;=\;
(1\!+\!j)\CdoT u_{\text{reell}}(k)\;=\notag\\*[4pt]
=\;(1\!+\!j)\CdoT u_{\text{reell}}(k)^{\Kk}\,=\;
j\CdoT(1\!-\!j)\CdoT u_{\text{reell}}(k)^{\Kk}\,=\;
j\CdoT\big((1\!+\!j)\CdoT u_{\text{reell}}(k)\big)^{\!\Kk}\,=\;
j\CdoT u(k)^{\Kk}\!.
\label{E.6.27}
\end{gather}
F"ur die Regressionskoeffizienten \mbox{$U_{\!f}(\mu)$} des Spektrums der
gefensterten komplexen deterministischen St"orung gilt entsprechend\vspace{0pt}
\begin{gather}
U_{\!f}(\mu)\;=\;U_{\!f,\text{reell}}(\mu)+
j\CdoT U_{\!f,\text{reell}}(\mu)\;=\;
(1\!+\!j)\CdoT U_{\!f,\text{reell}}(\mu)\;=\notag\\*[4pt]
=\;(1\!+\!j)\CdoT U_{\!f,\text{reell}}(\!-\mu)^{\Kk}\,=\;
j\CdoT\big((1\!+\!j)\CdoT U_{\!f,\text{reell}}(\!-\mu)\big)^{\!\Kk}\,=\;
j\CdoT U_{\!f}(\!-\mu)^{\Kk}\!.
\label{E.6.28}
\end{gather}
Wenn die Voraussetzungen f"ur die Anwendbarkeit der drei RKM-Varianten
gegeben sind, gilt f"ur die zweiten Momente des Spektrums des gefensterten
komplexen Approximationsfehlers\vspace{-12pt}
\begin{subequations}\label{E.6.29}
\begin{flalign}
&&\Tilde{\Phi}_{\boldsymbol{n}}(\mu)&\;=\;
\Tilde{\Phi}_{\boldsymbol{n}}(\!-\mu)\;=\;
2\cdot\Tilde{\Phi}_{\boldsymbol{n},\text{reell}}(\mu)\;=\;
2\cdot\Tilde{\Psi}_{\boldsymbol{n},\text{reell}}(\mu)&&
\label{E.6.29.a}\\*[4pt]
\text{und}&&\Tilde{\Psi}_{\boldsymbol{n}}(\mu)&\;=\;0,&&
\label{E.6.29.b}
\end{flalign}
\end{subequations}
wobei \mbox{$\Tilde{\Phi}_{\boldsymbol{n},\text{reell}}(\mu)=
\Tilde{\Psi}_{\boldsymbol{n},\text{reell}}(\mu)$} die entsprechenden zweiten
Momente des Spektrums des gefensterten reellen Approximationsfehlers
sind, der sich bei der Approximation des realen reellwertigen Systems mit
den reellen Zufallsvektoren ergibt.

Bei der ersten der drei weiteren RKM-Varianten wird die Messung nun genauso
durchgef"uhrt, wie wenn ein komplexwertiges System vorliegen w"urde.
Man berechnet also zun"achst die Messwerte, die im Kapitel \ref{E.Kap.3}
angegeben worden sind. Da die konkreten Realisierungen der Signale der
Einzelmessungen nun zu komplexen Signalsequenzen zusammengefasst werden,
weisen deren Spektren nicht mehr die Symmetrien auf, die den Spektren
reeller Signale eigen sind. Somit weisen auch die Messwerte nicht die
Symmetrien der entsprechenden theoretischen Gr"o"sen auf. Lediglich
die Erwartungswerte der Messwerte entsprechen diesen Symmetrien, da die
Erwartungstreue der Messwerte weiterhin gegeben ist. Man berechnet sich
daher aus den bisher gewonnenen Messwerten des komplexen Systemmodells
Messwerte f"ur die eigentlich gesuchten Regressionskoeffizienten
des reellen Systemmodells, die die gew"unschten Symmetrien aufweisen.
Man berechnet sich daher die Messwerte
\begin{subequations}\label{E.6.30}
\begin{align}
\Hat{H}_{\text{reell}}(\mu)&\,=\,
\frac{\,\Hat{H}(\mu)\!+\!\Hat{H}(\!-\mu)^{\!\Kk}}{2},
\label{E.6.30.a}\\*[4pt]
\Hat{U}_{\!f,\text{reell}}(\mu)&\,=\,\frac{1}{2}\CdoT\bigg(
\frac{\,\Hat{U}_{\!f}(\mu)\,}{1\!+\!j}\!+\!
\frac{\,j\Cdot\Hat{U}_{\!f}(\!-\mu)^{\!\Kk}\,}{1\!+\!j}\bigg)=
\frac{1}{4}\CdoT\Big((1\!-\!j)\CdoT\Hat{U}_{\!f}(\mu)\!+\!
(1\!+\!j)\CdoT\Hat{U}_{\!f}(\!-\mu)^{\!\Kk}\Big),
\label{E.6.30.b}\\[6pt]
\Hat{u}_{\text{reell}}(k)&\,=\,\frac{1}{2}\cdot\bigg(
\frac{\,\Hat{u}(k)\,}{1\!+\!j}\!+\!
\frac{\,j\CdoT\Hat{u}(k)^{\!\Kk}\,}{1\!+\!j}\bigg)\,=\,
\frac{1}{4}\CdoT\Big((1\!-\!j)\CdoT\Hat{u}(k)\!+\!
(1\!+\!j)\CdoT\Hat{u}(k)^{\!\Kk}\Big)
\label{E.6.30.c}\\*[-12pt]\intertext{und\vspace{-6pt}}
\Hat{\Phi}_{\boldsymbol{n},\text{reell}}(\mu)&\,=\,
\frac{\,\Hat{\Phi}_{\boldsymbol{n}}(\mu)\!+\!
\Hat{\Phi}_{\boldsymbol{n}}(\!-\mu)\,}{4}.
\label{E.6.30.d}
\end{align}
\end{subequations}
Wenn man von diesen Messwerten nun die Varianzen berechnet, wird man
feststellen, dass man bestenfalls dieselben Messwertvarianzen erh"alt,
die man bei der in Unterkapitel \ref{E.Kap.6.1} dargestellten Version
des RKM zur Messung reellwertiger Systeme erh"alt, wenn man dort 
insgesamt eine Einzelmessung am realen System weniger durchf"uhrt.
Diese minimale Messwertvarianz erh"alt man, wenn die Spektralwerte
der komplexen Erregung bei positiven und negativen Frequenzen gleiche
Varianz aufweisen, \vadjust{\penalty-100}und unabh"angig sind. Bei dieser Variante des RKM
mit der in den letzten Gleichungen dargestellten abschlie"senden
Messwertmittelung ist f"ur je zwei Einzelmessungen am realen System
sowohl eine DFT des komplexen Eingangssignals als auch eine DFT des
komplexen Ausgangssignals zu berechnen. In Kapitel \ref{E.Kap.8} wird gezeigt,
dass man auch bei der in Unterkapitel \ref{E.Kap.6.1} vorgestellten
RKM-Variante mit derselben Anzahl an diskreten Fouriertransformationen
auskommt. Da die letztgenannte Variante jedoch bei minimal besserer
Messwertvarianz einen geringeren Speicherbedarf aufweist, und deren
Anwendbarkeit auch nicht solch starken Restriktionen hinsichtlich
der nicht immer gegebenen Unabh"angigkeit der Einzelmessungen
unterliegt, ist diese vorzuziehen.

Bei einer weiteren RKM Variante n"utzt man die Symmetrien, die in den
Regressionskoeffizienten des komplexen Systemmodells vorhanden sein
m"ussen dadurch aus, dass man diese schon bei der Aufstellung
der Gleichungen, f"ur die die Ausgleichsl"osungen berechnet werden
sollen, ber"ucksichtigt. Mit den komplexen Stichprobenvektoren
\begin{subequations}\label{E.6.31}
\begin{align}
\Vec{V}(\mu)&=
\Big[V_1(\mu)\!+\!j\CdoT V_2(\mu),\,\ldots,
V_{2\cdot\Tilde{\lambda}-1\!}(\mu)\!+\!
j\CdoT V_{2\cdot\Tilde{\lambda}}(\mu),\,\ldots,
V_{L/2-1\!}(\mu)\!+\!j\CdoT V_{L/2}(\mu)\Big]
\label{E.6.31.a}\\*[-6pt]\intertext{und\vspace{-3pt}}
\Vec{Y}_{\!\!f}(\mu)&=
\Big[Y_{\!\!f,1\!}(\mu)\!+\!j\CdoT\!Y_{\!\!f,2}(\mu),\ldots\!,
Y_{\!\!f,2\cdot\Tilde{\lambda}-1\!}(\mu)\!+\!
j\CdoT\!Y_{\!\!f,2\cdot\Tilde{\lambda}\!}(\mu),\ldots\!,
Y_{\!\!f,L/2-1\!}(\mu)\!+\!j\CdoT\!Y_{\!\!f,L/2}(\mu)\!\Big]\!\!
\label{E.6.31.b}\\*[4pt]
&\qquad\qquad\qquad\qquad\forall\qquad\qquad\mu=0\;(1)\;M\!-\!1
\notag
\end{align}
\end{subequations}
die jeweils aus $L/2$ Elementen bestehen, erh"alt man zun"achst f"ur alle
$M$ Frequenzen die Gleichungen (\ref{E.6.7}), die jetzt aber alle die Dimension
\mbox{$1\!\times\!(L/2)$} haben, und die $M$ Unbekannten \mbox{$\Hat{H}(\mu)$}
sowie die $M$ Unbekannten \mbox{$\Hat{U}_{\!f}(\mu)$} des komplexen
Modellsystems enthalten. Indem man hier $\mu$ durch $-\mu$ ersetzt und
die Gleichungen konjugiert, erh"alt man mit
\begin{equation}
\Hat{H}(\!-\mu)^{\!\Kk}\!\CdoT\Vec{V}(\!-\mu)^{\!\Kk}\!+
\Hat{U}_{\!f}(\!-\mu)^{\!\Kk}\!\CdoT\Vec{1}\;=\;
\Vec{Y}_{\!f}(\!-\mu)^{\!\Kk}
\label{E.6.32}
\end{equation}
ebenfalls wieder $M$ redundante Gleichungen der Dimension
\mbox{$1\!\times\!(L/2)$} mit denselben Unbekannten. Nun
fordert man, dass die empirischen Regressionskoeffizienten dieselben
Symmetrien aufweisen m"ussen,
wie die theoretische Regressionskoeffizienten nach Gleichung
(\ref{E.6.27}) und (\ref{E.6.29}). Man setzt daher in die vorigen Gleichungen
und die Gleichungen (\ref{E.6.7}) die Terme\vspace{-16pt}
\begin{subequations}\label{E.6.33}\begin{flalign}
&&\Hat{H}(\mu)&\;=\;\Hat{H}(\!-\mu)^{\!\Kk}\,=\;\Hat{H}_{\text{reell}}(\mu)&&
\label{E.6.33.a}\\*[4pt]
\text{und}&&
\Hat{U}_{\!f}(\mu)&\;=\;j\CdoT\Hat{U}_{\!f}(\!-\mu)^{\!\Kk}\,=\;
(1\!+\!j)\CdoT\Hat{U}_{\!f,\text{reell}}(\mu)&&
\label{E.6.33.b}
\end{flalign}
\end{subequations}
ein, und fasst dann alle Gleichungen mit denselben Unbekannten zusammen.
Man erh"alt so die \mbox{$M/2\!+\!1$} Gleichungen:
\begin{gather}
\Hat{H}_{\text{reell}}(\mu)\CdoT
\Big[\,\Vec{V}(\mu)\,,\,\,\Vec{V}(\!-\mu)^{\!\Kk}\,\Big] +
\Hat{U}_{\!f,\text{reell}}(\mu)\CdoT
\Big[\,(1\!+\!j)\CdoT\Vec{1}\,,\,\,(1\!-\!j)\CdoT\Vec{1}\,\Big]\;=\;
\Big[\,\Vec{Y}_{\!f}(\mu)\,,\,\,
\Vec{Y}_{\!f}(\!-\mu)^{\!\Kk}\,\Big]\notag\\*[4pt]
\label{E.6.34}
{\T\forall\qquad\qquad\mu=0\;(1)\;\frac{M}{2}.}
\end{gather}
Die \mbox{$M/2\!-\!1$} Gleichungen f"ur
\mbox{$\mu=\frac{M}{2}\!+\!1\;(1)\;M\!-\!1$} lassen sich durch
Konjugieren und Vertauschen der Reihenfolge der Elemente
Zeilenvektoren in die angegebenen Gleichungen "uberf"uhren und
sind daher redundant. Mit diesen Gleichungen kann man nun die 
gesuchten Messwerte als Ausgleichsl"osungen bestimmen. In den 
$F$ Gleichungen (\ref{E.6.13}) f"ur die $F$ Unbekannten
\mbox{$\Hat{u}(k)$} der deterministischen St"orung des komplexen
Modellsystems ist die Mittelungsanzahl $L/2$ statt $L$ einzusetzen. 
Au"serdem ist dort f"ur die komplexen empirischen Mittelwerte 
\begin{subequations}\label{E.6.35}
\begin{align}
\frac{2\CdoT\Vec{V}(\mu)\CdoT\Vec{1}^{\,\Hh}}{L}&\;=\;
\frac{2}{L}\cdoT\Sum{\Tilde{\lambda}=1}{L/2}
\Big(V_{2\cdot\Tilde{\lambda}-1}(\mu)+
j\CdoT V_{2\cdot\Tilde{\lambda}}(\mu)\Big)&
\forall\qquad \mu&=0\;(1)\;M\!-\!1
\label{E.6.35.a}\\*[-10pt]\intertext{und\vspace{-10pt}}
\frac{2}{L}\CdoT\Vec{y}(k)\cdoT\Vec{1}^{\,\Hh}&\;=\;
\frac{2}{L}\cdoT\Sum{\Tilde{\lambda}=1}{L/2}
\Big(y_{2\cdot\Tilde{\lambda}-1}(k)+
j\CdoT y_{2\cdot\Tilde{\lambda}}(k)\Big)&
\forall\qquad k&=0\;(1)\;F\!-\!1
\label{E.6.35.b}
\end{align}
\end{subequations}
einzusetzten. Anschlie"send werden die Gleichungen (\ref{E.6.13}) konjugiert
und mit $j$ multipliziert. Der Lauf"|index $\mu$ der Summe wird noch durch
den negativen Lauf"|index $-\mu$ substituiert, und man erh"alt unter Ausnutzung
der $M$ Periodizit"at der Summanden die $F$ Gleichungen\vspace{-8pt}
\begin{gather}
j\cdot\Hat{u}(k)^{\Kk}\;=\;j\CdoT
\frac{2}{L}\CdoT\Vec{y}(k)^{\Kk}\cdoT\Vec{1}^{\,\Tt}-
\frac{2}{M\CdoT L}\cdoT\Sum{\mu=0}{M-1}\Hat{H}(\mu)^{\Kk}\!\CdoT
j\CdoT\Vec{V}(\mu)^{\Kk}\CdoT\Vec{1}^{\,\Tt}\Cdot
e^{\!-j\cdot\frac{2\pi}{M}\cdot\mu\cdot k}\;=\notag\\[8pt]
=\;j\CdoT\frac{2}{L}\CdoT\Vec{y}(k)^{\Kk}\cdoT\Vec{1}^{\,\Tt}-
\frac{2}{M\CdoT L}\cdoT\!\!\Sum{\mu=1-M}{0}\Hat{H}(\!-\mu)^{\Kk}\!\CdoT
j\CdoT\Vec{V}(\!-\mu)^{\Kk}\CdoT\Vec{1}^{\,\Tt}\Cdot
e^{j\cdot\frac{2\pi}{M}\cdot\mu\cdot k}\;=\notag\\[8pt]
=\;j\CdoT\frac{2}{L}\CdoT\Vec{y}(k)^{\Kk}\cdoT\Vec{1}^{\,\Tt}-
\frac{2}{M\CdoT L}\cdoT\Sum{\mu=0}{M-1}\Hat{H}(\!-\mu)^{\Kk}\!\CdoT
j\CdoT\Vec{V}(\!-\mu)^{\Kk}\CdoT\Vec{1}^{\,\Tt}\Cdot
e^{j\cdot\frac{2\pi}{M}\cdot\mu\cdot k}\notag\\*[6pt]
\forall\qquad\qquad k=0\;(1)\;F\!-\!1.
\label{E.6.36}
\end{gather}
Wieder fordert man, dass auch die empirischen Regressionskoeffizienten
der deterministischen St"orung des komplexen Systemmodells dieselben
Symmetrien aufweisen, wie die theoretischen Regressionskoeffizienten
nach Gleichung (\ref{E.6.28}). Man setzt daher in die letzten 
Gleichungen und die Gleichungen (\ref{E.6.13}) die Terme
\begin{equation}
\Hat{u}(k)\;=\;j\CdoT\Hat{u}(k)^{\Kk}\;=\;
(1\!+\!j)\CdoT\Hat{u}_{\text{reell}}(k)
\label{E.6.37}
\end{equation}
f"ur die zu bestimmenden Werte der deterministischen St"orung und Gleichung
(\ref{E.6.33.a}) f"ur die empirischen Werte der "Ubertragungsfunktion ein,
und fasst bei den Gleichungen (\ref{E.6.13}) und (\ref{E.6.36}) alle Gleichungen
mit denselben Unbekannten \mbox{$\Hat{u}_{\text{reell}}(k)$}
zusammen. Man erh"alt die $F$ Gleichungen der Dimension \mbox{$1\!\times\!2$}
\begin{gather}
(1\!+\!j)\CdoT\Hat{u}_{\text{reell}}(k)\CdoT\Vec{1}\;=
\label{E.6.38}\\[0pt]
=\frac{2}{L}\CdoT\Big[\,\Vec{y}(k)\cdoT\Vec{1}^{\,\Hh}\,,\,\,j\CdoT\Vec{y}(k)^{\Kk}\cdoT\Vec{1}^{\,\Tt}\,\Big]-
\frac{2}{M\CdoT L}\cdoT\Sum{\mu=0}{M-1}\Hat{H}_{\text{reell}}(\mu)\CdoT
\Big[\Vec{V}(\mu)\CdoT\Vec{1}^{\,\Hh}\,,\,j\CdoT\Vec{V}(\!-\mu)^{\Kk}\CdoT\Vec{1}^{\,\Tt}\,\Big]
\CdoT e^{j\cdot\frac{2\pi}{M}\cdot\mu\cdot k}\notag
\end{gather}
deren Ausgleichsl"osung sich zu
\begin{gather}
\Hat{u}_{\text{reell}}(k)\;=
\label{E.6.39}\\[0pt]
=\frac{
\Big[\Vec{y}(k)\CdoT\!\Vec{1}^{\,\Hh},\,
j\CdoT\Vec{y}(k)^{\Kk}\!\CdoT\Vec{1}^{\uP{0.6}{T}}\Big]\!\CdoT
\Vec{1}^{\,\Hh}\!}{L\CdoT(1\!+\!j)}-\frac{1}{M}\cdoT\Sum{\mu=0}{M-1}\!
\Hat{H}_{\text{reell}}(\mu)\CdoT\frac{
\Big[\Vec{V}(\mu)\CdoT\Vec{1}^{\,\Hh},\,
j\CdoT\Vec{V}(\!-\mu)^{\Kk}\!\CdoT\Vec{1}^{\uP{0.6}{T}}\Big]\!
\CdoT\Vec{1}^{\,\Hh}\!}{L\CdoT(1\!+\!j)}\CdoT
e^{j\cdot\frac{2\pi}{M}\cdot\mu\cdot k}=
\notag\\[8pt]
=\frac{1}{L}\cdoT\Sum{\Tilde{\lambda}=1}{L/2}\big(
y_{2\cdot\Tilde{\lambda}-1\!}(k)\!+\!
y_{2\cdot\Tilde{\lambda}\!}(k)\big)-
\frac{1}{M}\cdoT\Sum{\mu=0}{M-1}\Hat{H}_{\text{reell}}(\mu)\CdoT
\frac{1}{L}\CdoT\!\Sum{\Tilde{\lambda}=1}{L/2}
\big(V_{2\cdot\Tilde{\lambda}-1\!}(\mu)\!+\!
V_{2\cdot\Tilde{\lambda}\!}(\mu)\big)\CdoT
e^{j\cdot\frac{2\pi}{M}\cdot\mu\cdot k}=\notag\\[8pt]
=\;\frac{1}{L}\cdoT\Vec{y}_{\text{reell}}(k)\cdoT\Vec{1}^{\,\Hh}-
\frac{1}{M\CdoT L}\cdoT\Sum{\mu=0}{M-1}\Hat{H}_{\text{reell}}(\mu)\CdoT
\Vec{V}_{\text{reell}}(\mu)\CdoT\Vec{1}^{\,\Hh}\cdot
e^{j\cdot\frac{2\pi}{M}\cdot\mu\cdot k}\notag
\end{gather}
berechnet. Bis auf den Unterschied, dass die $M$ Messwerte der
"Ubertragungsfunktion nun die Ausgleichsl"osungen anderer
Gleichungssysteme sind, berechnen sich die Messwerte der
deterministische St"orung also identisch wie bei der Version des RKM,
die in Unterkapitel \ref{E.Kap.6.1} vorgestellt wurde. Die Messwerte
\mbox{$\Hat{\Phi}_{\boldsymbol{n}}(\mu)$} des Betragsquadrats des Spektrums
des gefensterten komplexen Approximationsfehlers erhalten wir, indem wir
das Betragsquadrat der euklidischen L"ange des Differenzvektors, der
sich beim Einsetzen der Ausgleichsl"osung jeweils als Differenz der
rechten und der linken Seite in den Gleichungen (\ref{E.6.34}) ergibt,
durch die um $2$ reduzierte Spaltendimension des Gleichungssystems
dividieren. Die Spaltendimension des Gleichungssystems ist die Anzahl
$L$ der Einzelmessungen am realen System. Diese ist um $2$ zu reduzieren,
da die linke Seite jeder Gleichung jeweils einen zweidimensionalen
Unterraum des $L$-dimensionalen Raums aufspannt.
\begin{gather*}\label{E.6.40}
\begin{flalign}
&\Hat{\Phi}_{\boldsymbol{n}}(\mu)\;=\;\frac{1}{L\!-\!2}\cdoT{}&&
\end{flalign}\\*\begin{flalign*}
&&&{}\Cdot\bigg(\Big[\Vec{Y}_{\!f}(\mu)\,,\,\,
\Vec{Y}_{\!f}(\!-\mu)^{\!\Kk}\Big]\!-\!
\Hat{H}_{\text{reell}}(\mu)\CdoT
\Big[\Vec{V}(\mu)\,,\,\,\Vec{V}(\!-\mu)^{\!\Kk}\Big]\!-\!
\Hat{U}_{\!f,\text{reell}}(\mu)\CdoT
\Big[(1\!+\!j)\CdoT\Vec{1}\,,\,\,(1\!-\!j)\CdoT\Vec{1}\,\Big]\bigg)\CdoT{}
\\*[2pt]
&&&{}\Cdot\bigg(\Big[\Vec{Y}_{\!f}(\mu)\,,\,\,
\Vec{Y}_{\!f}(\!-\mu)^{\!\Kk}\Big]\!-\!
\Hat{H}_{\text{reell}}(\mu)\CdoT
\Big[\Vec{V}(\mu)\,,\,\,\Vec{V}(\!-\mu)^{\!\Kk}\Big]\!-\!
\Hat{U}_{\!f,\text{reell}}(\mu)\CdoT
\Big[(1\!+\!j)\CdoT\Vec{1}\,,\,\,(1\!-\!j)\CdoT\Vec{1}\,\Big]\bigg)^{\!\!\HH}
\end{flalign*}
\end{gather*}
Da die Differenzvektoren, die sich mit $-\mu$ statt mit $\mu$ ergeben,
lediglich konjugiert und permutiert sind, sind diese gleich lang, und
die Symmetrie, die die theoretischen Werte nach Gleichung (\ref{E.6.29}) 
aufweisen, ist auch bei diesen Messwerten vorhanden. Desweiteren besagt
die Gleichung (\ref{E.6.29}), dass die zweiten Momente des Spektrums des
gefensterten reellen Approximationsfehlers halb so gro"s sind wie die
zweiten Momente des Spektrums des gefensterten komplexen
Approximationsfehlers. Daher sind die Messwerte
\begin{equation}
\Hat{\Phi}_{\boldsymbol{n},\text{reell}}(\mu)\;=\;
\frac{1}{2}\cdot\Hat{\Phi}_{\boldsymbol{n}}(\mu)
\label{E.6.41}
\end{equation}
erwartungstreue Sch"atzwerte f"ur die gesuchten Gr"o"sen
\mbox{$\Tilde{\Phi}_{\boldsymbol{n},\text{reell}}(\mu)$}, wenn
die Messwerte \mbox{$\Hat{\Phi}_{\boldsymbol{n}}(\mu)$} die Gr"o"sen
\mbox{$\Tilde{\Phi}_{\boldsymbol{n}}(\mu)$} erwartungstreu absch"atzen.

Diese Variante des RKM zur Messung reellwertiger Systeme besteht nun
also darin, mit Hilfe der komplexen und nicht symmetrischen
Stichprobenvektoren gem"a"s der Gleichungen (\ref{E.6.31})
die Messwerte \mbox{$\Hat{H}_{\text{reell}}(\mu)$} und
\mbox{$\Hat{U}_{f,\text{reell}}(\mu)$} als Ausgleichsl"osungen der
Gleichungssysteme (\ref{E.6.34}) zu bestimmen.
Mit diesen Ausgleichsl"osungen werden einerseits mit Gleichung 
(\ref{E.6.39}) die Messwerte \mbox{$\Hat{u}_{\text{reell}}(k)$} als
Ausgleichsl"osungen der Gleichungssysteme (\ref{E.6.38}), mit den empirischen
Mittelwerten gem"a"s der Gleichungen (\ref{E.6.35}), und andererseits die
Messwerte \mbox{$\Hat{\Phi}_{\boldsymbol{n},\text{reell}}(\mu)$}
nach Gleichung (\ref{E.6.40}) und (\ref{E.6.41}) berechnet. Man kann nun
zeigen, dass sich die Gleichungssysteme (\ref{E.6.34}) durch Multiplikation
mit einer unit"aren Matrix von rechts und durch Division durch $\sqrt{2\;}$
auf die Gleichungssysteme (\ref{E.6.7}) "uberf"uhren lassen, die bei der im
Unterkapitel \ref{E.Kap.6.1} vorgestellt RKM-Variante auftreten. Daher liefern
diese beiden RKM-Varianten wenigstens theoretisch --- also abgesehen
von Rechenungenauigkeiten --- dieselben Messwerte. Daher sind auch die
Erwartungswerte, Varianzen und Kovarianzen der Messwerte beider
RKM-Varianten gleich. Da das eben beschriebene Verfahren aber sowohl
hinsichtlich des Speicherbedarfs als auch hinsichtlich der Anzahl der
Gleitkommaoperationen (\,flops\,) aufwendiger ist, als das Verfahren
nach Unterkapitel \ref{E.Kap.6.1}, ist letztgenanntes vorzuziehen.

Bei den bisher vorgestellten drei RKM-Varianten ein reellwertiges System
zu messen, kann man immer zur Erregung in zwei aufeinanderfolgenden
Einzelmessungen Signalsequenzen verwenden, die konkrete Realisierungen
unabh"angiger reeller Zufallsvektoren sind. Andererseits kann man bei
fast allen realen Systemen in zwei aufeinanderfolgenden Einzelmessungen
auch den Real- und Imagin"arteil einer komplexen Signalsequenz zur
Erregung verwenden, die eine konkrete Realisierung eines komplexen
Zufallsvektors ist, der abh"angige Real- und Imagin"arteilvektoren
aufweist, ohne dass dabei die Gefahr besteht, dass dadurch die
Messergebnisse verf"alscht werden. Nur bei solchen Systemen l"asst
sich die nun folgende und letzte dem Autor bekannte RKM-Variante
einsetzen. Bei dieser Variante werden zur Erregung sogenannte
Halbbandsignale verwendet. Bei diesen Signalen sind die Zufallsgr"o"sen
der Spektralwerte der Erregung f"ur negative Frequenzen null.
\begin{equation}
\boldsymbol{V}\!(\mu)\;=\;0
{\T\qquad\qquad\forall\qquad\mu=\frac{M}{2}\!+\!1\;(1)\;M\!-\!1.}
\label{E.6.42}
\end{equation}
Die Messwerte der "Ubertragungsfunktion werden mit Gleichung (\ref{E.6.7}) nur
f"ur die positiven Frequenzen mit \mbox{$\mu=1\;(1)\;M/2\!-\!1$} berechnet.
Die Messwerte der restlichen Frequenzen \mbox{$\mu=M/2+1\;(1)\;M\!-\!1$}
ergeben sich durch konjugierte Spiegelung. Bei diesen Frequenzen
kann man die Messwerte \mbox{$\Hat{U}_{\!f,\text{reell}}(\mu)$} und
\mbox{$\Hat{\Phi}_{\boldsymbol{n},\text{reell}}(\mu)$} "uber eine reine
Spektralsch"atzung bestimmen, da dort kein erregungsabh"angiger
Signalanteil am Ausgang der linearen Modellsystems vorhanden ist.
Bei den beiden Frequenzen \mbox{$\mu\!=\!0$} und \mbox{$\mu\!=\!M/2$}
muss man die vollst"andige Messung der Variante nach Unterkapitel
\ref{E.Kap.6.1} verwenden, wenn man an allen Messwerten interessiert ist.
Mit diesen Messwerten kann man dann auch die Messwerte der deterministischen
St"orung nach Gleichung (\ref{E.6.13}) berechnen. Wenn man sich die
Messwertvarianzen dieses Verfahrens berechnet, so stellt man fest,
dass man bei gleicher Anzahl von Einzelmessungen fast die
doppelte Messwertvarianz erh"alt. Bei gleicher Messgenauigkeit hat
man daher also fast doppelt so viele Einzelmessung durchzuf"uhren.
Daher ist diese Variante f"ur die Messung der "Ubertragungsfunktion 
eines reellen Systems im gesamten Frequenzbereich \mbox{$\mu=1\;(1)\;M\!-\!1$}
nicht empfehlenswert.

Das zuletzt angef"uhrte Verfahren ist das in \cite{Sch/H} vorgestellte
Halbbandmessverfahren, wobei hier lediglich die deterministische
St"orung erg"anzt ist. Dieses Messverfahren stellt dort den 
Spezialfall eines Messverfahren dar, mit dessen Hilfe man mit
relativ geringem Aufwand die "Ubertragungsfunktion des Modellsystems
innerhalb eines schmalen Frequenzbandes mit hoher Frequenzauf"|l"osung messen
kann. Es wird dann allgemein mit schmalbandigen Teilbandsignalen
erregt, und bei der Messung der Ausgangssignale eine geeignete
Unterabtastung vorgenommen, wodurch bei gleichem Frequenzabstand
der Messwerte der "Ubertragungsfunktion eine immense Reduktion des
Rechenaufwands erzielt wird. Leider k"onnen bei dieser
Schmalband-RKM-Variante die Messwerte
\mbox{$\Hat{\Phi}_{\boldsymbol{n}}(\mu)$} des Spektrums des
Approximationsfehlers wegen der Unterabtastung keine Aussage
"uber das LDS einer im realen System vorhandenen St"orung liefern,
sondern lediglich zur Absch"atzung der Messwertvarianz benutzt werden.
Da aber die Hauptvorteile der Fensterung gerade bei der Messung des LDS
liegen, und daher der Einsatz einer Fensterfolge bei der schmalbandigen
RKM-Variante nur in den seltensten F"allen sinnvoll erscheint, und da
die Halbband-Variante bei gleicher Messwertvarianz einen h"oheren
Rechenaufwand aufweist, als die vorher beschriebenen Varianten,
"uberlasse ich es dem Leser sich herzuleiten, wie sich dieses
vor allem in der schmalbandigen Variante empfehlenswerte Messverfahren um
die Fensterung und die Modellierung der deterministischen St"orung
erweitern l"asst.

\chapter{Ablaufs"ubersicht f"ur eine Variante des RKM}\label{E.Kap.7}

Es zeigt sich, dass alle am Anfang des Kapitels~\ref{E.Kap.3} 
aufgelisteten Messwerte und die dazugeh"origen Konfidenzgebiete aus Werten 
berechnet werden k"onnen, die man auf einfache Weise durch die Akkumulation 
von einfachen Produkten "uber alle $L$ Einzelmessungen erh"alt. Wenn man
eine Fensterfolge verwendet, deren Spektrum die nach Gleichung~(\myref{2.27})
geforderten "aquidistanten Nullstellen am Einheitskreis aufweist,
ist es auch beim RKM mit Fensterung nicht notwendig, alle Erregungen
bzw. Systemantworten aller Einzelmessungen abzuspeichern, um
am Ende der Messung mit rechenintensiven Matrixoperationen die Messwerte
und Konfidenzgebiete zu erhalten. Hier m"ochte ich nun auf"|listen, in
welchen Schritten die Messung mit dem RKM mit Fensterung in der Praxis
durchgef"uhrt werden kann, und wieviel Speicher dazu ben"otigt wird.
Dabei m"ochte ich mich auf den Fall eines komplexwertigen realen Systems 
beschr"anken, das sich, wie in Bild \ref{E.b1h} gezeigt, durch ein zeitinvariantes 
(\,\mbox{$K_H\!=\!1$}\,) Modellsystem ${\cal S}_{lin}$, beschrieben durch die 
"Ubertragungsfunktion \mbox{$H(\mu)$}, und ein zweites zeitinvariantes Modellsystem 
${\cal S}_{*,lin}$, das von dem konjugierten Eingangssignal erregt wird und 
durch die "Ubertragungsfunktion \mbox{$H_*(\mu)$} beschrieben wird, modellieren l"asst. 
Da es sich hier um eindimensionale "Ubertragungsfunktionen handelt, wird auf die in 
den Kapiteln \ref{E.Kap.2} und \ref{E.Kap.3} verwendete Doppelindizierung verzichtet.
Hier enth"alt der nach Gleichung (\ref{E.2.15}) definierte Zufallsvektor 
\mbox{$\Tilde{\Vec{\boldsymbol{V}}}(\mu)$} lediglich die zwei Zufallsgr"o"sen 
\mbox{$\boldsymbol{V}\!(\mu)$} und \mbox{$\boldsymbol{V}\!(\!-\mu)^{\Kk}$}. 
Die komplexe, deterministische und zeitabh"angige St"orung \mbox{$u(k)$} wird 
modelliert. Von dem Approximationsfehlerprozess wird angenommen, dass er station"ar 
(\,\mbox{$K_{\Phi}\!=\!1$}\,) ist, so dass die Messung eines eindimensionalen LDS 
\mbox{$\Tilde{\Phi}_{\boldsymbol{n}}(\mu)$} und eines ebenfalls eindimensionalen 
KLDS \mbox{$\Tilde{\Psi}_{\boldsymbol{n}}(\mu)$} ausreicht. Auch hier wird 
auf die Doppelindizierung verzichtet. Der Zufallsspaltenvektor 
\mbox{$\Breve{\Vec{\boldsymbol{V}}}(\mu)$}, der bei der Berechnung der Messwerte 
f"ur das LDS und das KLDS verwendet wird ist gleich dem Zufallsspaltenvektor 
\mbox{$\Tilde{\Vec{\boldsymbol{V}}}(\mu)$}. Desweiteren sei $M$ gerade, und somit 
die N"aherungen, bei denen die Ausblendeigenschaft des Spektrums der Fensterfolge 
eingegangen ist, bei Verwendung einer hoch frequenzselektiven Fensterfolge sehr gut 
erf"ullt sind. 

In der folgenden Auf"|listung sind die Punkte, die mit $\emptyset$ gekennzeichnet sind, 
Kommentare und beschreiben meist die Berechnung von Werten,
die sich aus den davor berechneten Werten unmittelbar (\,z.~B. aufgrund einer
Symmetrieeigenschaft\,) ergeben. Sie brauchen in einem Programm {\em nicht}\/
explizit berechnet zu werden, da diese Werte bei den weiteren Berechnungen
ggf. durch die bereits berechneten Werte ersetzt werden. Die Gleichheitszeichen 
"`$=$"' sind meist als Wertzuweisungen, und nicht als Gleichungen im mathematischen 
Sinne zu verstehen, wie dies in den meisten Programmiersprachen "ublich ist.
Wie bisher ist bei Multiplikationen ggf. die Matrixmultiplikation gemeint,
und der Exponent \mbox{$(...)^{-1}$} bedeutet bei Matrizen eine Invertierung.
Unter einem Speicherplatz, wird im weiteren der Speicherbedarf verstanden,
der bei dem verwendeten Rechner f"ur die Speicherung einer reellen 
Gleitkommazahl ben"otigt wird.
\begin{enumerate}
\item Festlegen der Anzahl der zu messenden Frequenzen $M$.
\item Festlegen der Einschwingtakte $E$ des zu messenden Systems.
      Wenn man die Einschwingzeit nicht auf Grund heuristischer "Uberlegungen
      exakt vorhersehen kann (\,z.~B. Messung eines FIR-Systems\,), sollte
      man diese Zeit so gro"s w"ahlen, dass auch im schlimmsten Fall
      damit zu rechnen ist, dass die transienten Vorg"ange innerhalb
      dieser Zeit soweit abgeklungen sind, dass diese bei der Messung
      bedeutungslos werden. Im Zweifelsfall sollte man die Messung mit einer
      ver"anderten Einschwingzeit wiederholen, um festzustellen, ob sich die
      Messergebnisse dadurch ver"andern.
\item Wahl und Berechnung einer Fensterfolge der L"ange $F$.
      In den Kapiteln~\myref{Algo} und \ref{E.Kap.10.1} werden Algorithmen 
      vorgestellt, mit denen man Fensterfolgen berechnen kann, die alle 
      Bedingungen erf"ullen, die an die Fensterfolge gestellt wurden. 
      Bei diesen Fensterfolgen ist die Fensterl"ange \mbox{$F=N\CdoT M$} 
      ein ganzzahliges Vielfaches $N$ der Anzahl $M$ der Frequenzpunkte, 
      f"ur die die Messwerte gewonnen werden sollen. Die Berechnung des 
      Fensters erfolgt am besten mit dem in Kapitel~\myref{Algo} 
      vorgestellen Algorithmus, der die Variante des Fensters berechnet, 
      das mir am geeignetesten f"ur die Verwendung beim RKM erscheint. 
      Bei diesem Fenster ist es so, dass mit steigendem $N$ die 
      Frequenzselektivit"at und die Potenz des Abfalls des Betrags 
      des Spektrums der Fensterfolge f"ur zunehmende Frequenzen ansteigt.
\[
\text{\tt f}(k) \;=\; \text{\tt fenster}( N, M )
\quad\text{ f"ur }\quad k=0\;(1)\;F\!-\!1
\]
      Da die Fensterfolge reell ist, kann sie auf einem Speicher {\tt f}
      mit $F$ Speicherpl"atzen abgelegt werden.
\item[$\emptyset$) ] Es kann auch jede andere Fensterfolge beliebiger 
      L"ange verwendet werden. Wird jedoch die Bedingungen~(\myref{2.27})
      nicht erf"ullt, so ist die Messung mit einem anderen 
      Algorithmus durchzuf"uhren, der die Pseudoinverse 
      einer wesentlich gr"o"seren Matrix eines Gleichungssystems, 
      das dem Gleichungssystem~(\myref{3.4}) "ahnelt, berechnet. Wird die
      Bedingung~(\myref{2.20}) nicht erf"ullt, oder verwendet 
      man eine Fensterfolge mit geringer Frequenzselektivit"at, 
      \vadjust{\penalty-100}so werden die Messwerte unn"otig stark verrauscht sein, die 
      Messwerte \mbox{$\Hat{\Phi}_{\boldsymbol{n}}(\mu)$} und 
      \mbox{$\Hat{\Psi}_{\boldsymbol{n}}(\mu)$} werden die gew"unschten 
      Gr"o"sen \mbox{$\Bar{\Phi}_{\boldsymbol{n}}(\mu)$} und 
      \mbox{$\Bar{\Psi}_{\boldsymbol{n}}(\mu)$} 
      nur schlecht ann"ahern, und die Sch"atzwerte der Varianzen und 
      Kovarianzen, die eine N"aherung enthalten, die nur f"ur hoch 
      frequenzselektive Fensterfolgen g"ultig ist, werden nicht mehr 
      erwartungstreu sein. 
\item \label{E.I.7.4}Bereitstellen von $8$ Akkumulatorfeldern. Alle Akkumulatorfelder
      werden zu null initialisiert:\vspace{-12pt}
\[\begin{array}{rcll}
\text{\tt Akku\_y}(k)      & = & 0 &
\quad\text{ f"ur }\quad   k=0\;(1)\;F\!-\!1\\
\text{\tt Akku\_V}(\mu)    & = & 0 &
\quad\text{ f"ur }\quad \mu=0\;(1)\;M\!-\!1\\
\text{\tt Akku\_VQ}(\mu)   & = & 0 &
\quad\text{ f"ur }\quad \mu=0\;(1)\;M\!-\!1\\
\text{\tt Akku\_VV}(\mu)   & = & 0 &
\quad\text{ f"ur }\quad \mu=0\;(1)\;M/2\\
\text{\tt Akku\_YQ}(\mu)   & = & 0 &
\quad\text{ f"ur }\quad \mu=0\;(1)\;M\!-\!1\\
\text{\tt Akku\_YY}(\mu)   & = & 0 &
\quad\text{ f"ur }\quad \mu=0\;(1)\;M/2\\
\text{\tt Akku\_YV}(\mu)   & = & 0 &
\quad\text{ f"ur }\quad \mu=0\;(1)\;M\!-\!1\\
\text{\tt Akku\_VY}(\mu)   & = & 0 &
\quad\text{ f"ur }\quad \mu=0\;(1)\;M\!-\!1
\end{array}\]
     Das Akkumulatorfeld\vspace{-2pt}
\[\begin{array}{rccl}
\text{\tt Akku\_y}    & \text{ ben"otigt } &
2\CdoT F & \text{ Speicherpl"atze, }\\
\text{\tt Akku\_V}    & \text{ ben"otigt } &
2\CdoT M & \text{ Speicherpl"atze, }\\
\text{\tt Akku\_VQ}   & \text{ ben"otigt } &
M & \text{ Speicherpl"atze, }\\
\text{\tt Akku\_VV}   & \text{ ben"otigt } &
M\!+\!2 & \text{ Speicherpl"atze, }\\
\text{\tt Akku\_YQ}   & \text{ ben"otigt } &
M & \text{ Speicherpl"atze, }\\
\text{\tt Akku\_YY}   & \text{ ben"otigt } &
M\!+\!2 & \text{ Speicherpl"atze, }\\
\text{\tt Akku\_YV}   & \text{ ben"otigt } &
2\CdoT M & \text{ Speicherpl"atze und }\\
\text{\tt Akku\_VY}   & \text{ ben"otigt } &
2\CdoT M & \text{ Speicherpl"atze. }
\end{array}\]
\item Zuf"allige Auswahl der $L$ Zeitintervalle (\,Index $\lambda$\,) in denen
      das reale System erregt werden soll. Die zuf"allige Auswahl kann dabei
      in der Art erfolgen, wie dies im Kapitel~\ref{E.Kap.2.1} beschrieben worden
      ist. In vielen F"allen kann man die zuf"allige Auswahl auch in der Art
      realisieren, dass man zwischen zwei Einzelmessungen eine Pause einer
      zuf"alligen L"ange einschiebt. In diesem Fall braucht $L$ nicht zu Beginn
      der Messung festgelegt werden. Oft (\,z.~B. bei Rechnersimulationen\,)
      kann auf die zuf"allige Auswahl der Zeitintervalle ganz verzichtet
      werden. Bei Rechnersimulationen kann man die zuf"allige Lage der
      Zeitintervalle meist durch eine zuf"allige Zeitverschiebung der
      simulierten St"orung realisieren.
\item Initialisierung des Z"ahlers der Einzelmessungen.
\[
\lambda = 0
\]
\item \label{E.I.7.7}Inkrementieren des Z"ahlers der Einzelmessungen.
\begin{gather*}
\lambda = \lambda + 1
\displaybreak[3]\\
\end{gather*}\vspace{-49pt}
\item Erzeugung des Testsignals f"ur die Einzelmessung. 
      Mit Hilfe eines geeigneten Zufallssignalgenerators wird eine 
      Periode der L"ange $M$ des periodischen Zufallssignals f"ur
      \mbox{$k=0\;(1)\;M\!-\!1$} erzeugt. Dieses wird auf einem Speicher
      {\tt v} mit $2\CdoT M$ Speicherpl"atzen geschrieben.
      Die DFT des Testsignals {\tt v} liefert die Spektralwerte
      \mbox{$V_{\lambda}(\mu)$}, die auf dem Speicher {\tt V} mit
      $2\CdoT M$ Speicherpl"atzen abgespeichert werden, wobei die
      Spektralwerte, die sich evtl. noch von der vorigen Einzelmessung
      auf diesem Speicher befinden, "uberschrieben werden.\vspace{-5pt}
\[
\text{\tt V}(\mu) \;=\; \text{f"|ft}_M\big(\text{\tt v}(k)\big)
\qquad\text{ f"ur }\quad\mu=0\;(1)\;M\!-\!1
\]
      Wahlweise kann auch das Spektrum mit Hilfe eines Zufallsgenerators
      gewonnen werden, und das Testsignal durch eine inverse DFT erzeugt
      werden.
\[
\text{\tt v}(k) \;=\; \text{if"|ft}_M\big(\text{\tt V}(\mu)\big)
\qquad\text{ f"ur }\quad k=0\;(1)\;M\!-\!1
\]
      Das Testsignal {\tt v} wird in dem f"ur diese Einzelmessung
      zu Beginn ausgew"ahlten Zeitintervall auf den Eingang des zu
      messenden Systems gegeben indem es in beide Richtungen
      periodisch fortgesetzt wird. Dieses Zeitintervall wurde mit
      \mbox{$k\in[-E;F\!-\!1]$} bezeichnet, wobei $k$ die auf dieses
      Intervall bezogene Zeit ist. Es ist darauf zu achten, dass die
      gew"unschten stochastischen Eigenschaften des Testsignals auch
      in den Umgebungen der Zeitpunkte erf"ullt sind, an denen die
      einzelnen Perioden {\tt v} des Signals aneinandergef"ugt wurden.
      Dies kann vor allen bei der Generierung des Signals im Zeitbereich
      Probleme bereiten.
\item Messung der $F$ Abtastwerte \mbox{$y_{\lambda}(k)$} des
      Systemausgangssignals der Einzelmessung $\lambda$ im Zeitintervall
     \mbox{$k\in[0;F\!-\!1]$}.
      Dieser Signalausschnitt wird auf dem Speicher {\tt y} mit $2\CdoT F$
      Speicherpl"atzen abgelegt, wobei die Signalwerte,
      die sich evtl. noch von der vorigen Einzelmessung auf diesem Speicher
      befinden, "uberschrieben werden.
\[
\text{\tt y}(k) \;=\; y_{\lambda}(k)
\qquad\text{ f"ur }\quad k=0\;(1)\;F\!-\!1
\]
\item Fensterung des Systemausgangssignals und anschlie"sende DFT.
      Dies liefert nach Gleichung~(\myref{2.25}) das Spektrum \mbox{$Y_{\!f,\lambda}(\mu)$}.
      Dieses Spektrum wird auf dem Speicher {\tt Y} mit $2\CdoT M$
      Speicherpl"atzen abgelegt, wobei die Spektralwerte, die sich
      evtl. noch von der vorigen Einzelmessung auf diesem Speicher
      befinden, "uberschrieben werden.
\begin{gather*}
y_{f,\lambda}(k) \;=\;
\Sum{\kappa=0}{N}\text{\tt f}(k\!+\!\kappa\CdoT M)\cdot
\text{\tt y}(k\!+\!\kappa\CdoT M)
\qquad\text{ f"ur }\quad k=0\;(1)\;M\!-\!1\\*[6pt]
\text{\tt Y}(\mu) \;=\; \text{f"|ft}_M\big(y_{f,\lambda}(k)\big)
\qquad\text{ f"ur }\quad\mu=0\;(1)\;M\!-\!1
\end{gather*}
\item \label{E.I.7.11}Akkumulieren des Spektrums der Erregung.
      Die Werte \mbox{$V_{\lambda}(\mu)$} werden
      zum Inhalt des Akkumulatorfeldes {\tt Akku\_V} addiert.
\begin{gather*}
\text{\tt Akku\_V}(\mu) \;=\; \text{\tt Akku\_V}(\mu) + \text{\tt V}(\mu)
\qquad\text{ f"ur }\quad\mu=0\;(1)\;M\!-\!1
\displaybreak[3]\\
\end{gather*}\vspace{-54pt}
\item Die Betragsquadrate \mbox{$|V_{\lambda}(\mu)|^2$} werden
      zum Inhalt des Akkumulatorfeldes {\tt Akku\_VQ} addiert.\vspace{-3pt}
\[
\text{\tt Akku\_VQ}(\mu) \;=\; \text{\tt Akku\_VQ}(\mu) + |\text{\tt V}(\mu)|^2
\qquad\text{ f"ur }\quad\mu=0\;(1)\;M\!-\!1
\]
\item \label{E.I.7.13}Die Produkte \mbox{$V_{\lambda}(\mu)\CdoT V_{\lambda}(\!-\mu)$}
      werden zum Inhalt des Akkumulatorfeldes {\tt Akku\_VV} addiert.\vspace{-3pt}
\begin{gather*}
\text{\tt Akku\_VV}(\mu) \;=\;
\text{\tt Akku\_VV}(\mu) + \text{\tt V}(\mu)\cdot\text{\tt V}(M\!-\!\mu)
\qquad\text{ f"ur }{\T\quad\mu=1\;(1)\;\frac{M}{2}}\\*[6pt]
\text{\tt Akku\_VV}(0) \;=\; \text{\tt Akku\_VV}(0) + \text{\tt V}(0)^2
\end{gather*}
\item[$\emptyset$) ]\ \vspace{-18pt}
\[
\text{\tt Akku\_VV}(M\!-\!\mu)\;=\;\text{\tt Akku\_VV}(\mu)
\qquad\text{ sonst}
\]
\item \label{E.I.7.14}Akkumulieren des Systemausgangssignals.
      Die Werte \mbox{$y_{\lambda}(k)$} werden
      zum Inhalt des Akkumulatorfeldes {\tt Akku\_y} addiert.
\[
\text{\tt Akku\_y}(k) \;=\; \text{\tt Akku\_y}(k) + \text{\tt y}(k)
\qquad\text{ f"ur }\quad k=0\;(1)\;F\!-\!1
\]
\item \label{E.I.7.15}Die Betragsquadrate \mbox{$|Y_{\!f,\lambda}(\mu)|^2$} werden
      zum Inhalt des Akkumulatorfeldes {\tt Akku\_YQ} addiert.\vspace{-3pt}
\[
\text{\tt Akku\_YQ}(\mu) \;=\; \text{\tt Akku\_YQ}(\mu) + |\text{\tt Y}(\mu)|^2
\qquad\text{ f"ur }\quad\mu=0\;(1)\;M\!-\!1
\]
\item \label{E.I.7.16}Die Produkte \mbox{$Y_{\!f,\lambda}(\mu)\CdoT Y_{\!f,\lambda}(\!-\mu)$} 
      werden zum Inhalt des Akkumulatorfeldes {\tt Akku\_YY} addiert.\vspace{-3pt}
\begin{gather*}
\text{\tt Akku\_YY}(\mu) \;=\; \text{\tt Akku\_YY}(\mu) +
\text{\tt Y}(\mu)\cdot\text{\tt Y}(M\!-\!\mu)
\qquad\text{ f"ur }{\T\quad\mu=1\;(1)\;\frac{M}{2}}\\*[6pt]
\text{\tt Akku\_YY}(0) \;=\; \text{\tt Akku\_YY}(0) + \text{\tt Y}(0)^2
\end{gather*}
\item[$\emptyset$) ]\ \vspace{-18pt}
\[
\text{\tt Akku\_YY}(M\!-\!\mu)\;=\;\text{\tt Akku\_YY}(\mu)
\qquad\text{ sonst}
\]
\item Die Produkte \mbox{$Y_{\!f,\lambda}(\mu)\CdoT V_{\lambda}(\mu)^{\Kk}$} 
      werden zum Inhalt des Akkumulatorfeldes {\tt Akku\_YV} addiert.\vspace{-3pt}
\[
\text{\tt Akku\_YV}(\mu) \;=\; \text{\tt Akku\_YV}(\mu) +
\text{\tt Y}(\mu)\cdot\text{\tt V}(\mu)^{\Kk}
\qquad\text{ f"ur }\quad\mu=0\;(1)\;M\!-\!1
\]
\item \label{E.I.7.18}Die Produkte \mbox{$V_{\lambda}(\!-\mu)\CdoT Y_{\!f,\lambda}(\mu)$}
      werden zum Inhalt des Akkumulatorfeldes {\tt Akku\_VY} addiert.\vspace{-3pt}
\begin{gather*}
\text{\tt Akku\_VY}(\mu) \;=\; \text{\tt Akku\_VY}(\mu) +
\text{\tt V}(M\!-\!\mu)\cdot\text{\tt Y}(\mu)
\qquad\text{ f"ur }\quad\mu=1\;(1)\;M\!-\!1\\*[6pt]
\text{\tt Akku\_VY}(0) \;=\; \text{\tt Akku\_VY}(0) +
\text{\tt V}(0)\cdot\text{\tt Y}(0)
\end{gather*}\pagebreak[3]
\item Weitere Einzelmessungen durchf"uhren, indem man die
      Punkte~\ref{E.I.7.7} bis \ref{E.I.7.18} wiederholt.
      Entweder man f"uhrt eine konstante Anzahl von Einzelmessungen
      durch, oder man entscheidet anhand eines Kriteriums, das man aus den
      Messwerten gewinnt, und dessen Berechnung in den folgenden Punkten
      beschrieben wird, ob weitere Einzelmessungen durchzuf"uhren sind.
      Besonders geeignet erscheint es mir, nach einer Mindestanzahl
      von Einzelmessungen, die Singul"arwerte der im Punkt~\ref{E.I.7.23} 
      berechneten empirischen \mbox{$2\!\times\!2$} Kovarianzmatrizen 
      zu berechnen, und so lange weitere Einzelmessungen durchzuf"uhren,
      bis die Singul"arwerte bei allen diskreten Frequenzen $\mu$ innerhalb 
      eines Toleranzbereiches um die theoretischen Werte der Singul"arwerte 
      des Spektrums der Erregung liegen. Man kann jedoch an jeder Stelle
      der folgenden Messwertberechnung zu Punkt~\ref{E.I.7.7}
      zur"uckspringen und weitere Einzelmessungen durchf"uhren, wenn man
      z.~B. mit der Qualit"at der bisher erzielten Messergebnisse nicht
      zufrieden ist.
\item Wenn die Mittelungsanzahl $L$ nicht zu Beginn der Messung festgelegt
      worden ist, wird $L$ auf die tats"achlich durchgef"uhrte Anzahl
      der Einzelmessungen gesetzt.
\[
L = \lambda
\]
      Auch f"ur die Berechnung eines Abbruchkriteriums muss immer die
      Anzahl der bis dahin tats"achlich durchgef"uhrten Einzelmessungen
      verwendet werden.
\item \label{E.I.7.21}Berechnung der empirischen Varianzen
      \mbox{$\Hat{C}_{\boldsymbol{V}(\mu),\boldsymbol{V}(\mu)}$}
      des Spektrums der Erregung. Es handelt sich dabei um die Hauptdiagonalelemente 
      der Kovarianzmatrizen nach Gleichung~(\ref{E.3.10}) bzw. (\ref{E.3.27}).
      Das \mbox{$L\CdoT(L\!-\!1)$}-fache der empirischen Varianzen wird auf
      einem Speicher {\tt C\_VQ} mit $M$ Speicherpl"atzen abgelegt.
\[
\text{\tt C\_VQ}(\mu) \;=\; L\cdot\text{\tt Akku\_VQ}(\mu) -
|\text{\tt Akku\_V}(\mu)|^2
\qquad\text{ f"ur }\quad\mu=0\;(1)\;M\!-\!1
\]
\item \label{E.I.7.22}Berechnung der empirischen Kovarianzen
      \mbox{$\Hat{C}_{\boldsymbol{V}(\mu),\boldsymbol{V}(-\mu)^{\Kk}}$}
      des Spektrums der Erregung bei positiver Frequenz und des konjugierten
      Spektrums bei negativer Frequenz. Es handelt sich dabei um die Nebendiagonalelemente 
      der Kovarianzmatrizen nach Gleichung~(\ref{E.3.10}) bzw. (\ref{E.3.27}). Das 
      \mbox{$L\CdoT(L\!-\!1)$}-fache der empirischen Kovarianzen wird auf einem 
      Speicher {\tt C\_VV} mit $M\!+\!2$ Speicherpl"atzen abgelegt.
\begin{gather*}
\text{\tt C\_VV}(\mu) \;=\; L\cdot\text{\tt Akku\_VV}(\mu) -
\text{\tt Akku\_V}(\mu) \cdot \text{\tt Akku\_V}(M\!-\!\mu)
\qquad\text{ f"ur }{\T\quad\mu=1\;(1)\;\frac{M}{2}}\\*[14pt]
\text{\tt C\_VV}(0) \;=\; L\cdot\text{\tt Akku\_VV}(0)-
\text{\tt Akku\_V}(0)^2
\end{gather*}
\item[$\emptyset$) ]\ \vspace{-18pt}
\[
\text{\tt C\_VV}(M\!-\!\mu)\;=\;\text{\tt C\_VV}(\mu)
\qquad\text{ sonst}
\]
\item \label{E.I.7.23}Berechnung der Singul"arwerte \mbox{$s(\mu)$} 
      der empirischen Kovarianzmatrizen
      \mbox{$\Hat{\underline{C}}_{\Tilde{\Vec{\boldsymbol{V}}}(\mu),\Tilde{\Vec{\boldsymbol{V}}}(\mu)}$} 
      nach Gleichung~(\ref{E.3.10}) bzw. (\ref{E.3.27}). Die Singul"arwerte sind die immer
      nichtnegativen, reellen L"osungen der folgenden quadratischen Gleichungen
\begin{gather*}
\big(\,\text{\tt C\_VQ}(\mu)-L\CdoT(L\!-\!1)\cdot s(\mu)\,\big)\cdot
\big(\,\text{\tt C\_VQ}(M\!-\!\mu)-L\CdoT(L\!-\!1)\cdot s(\mu)\,\big)\;=\;
|\text{\tt C\_VV}(\mu)|^2\\*[12pt]
\text{ f"ur }{\T\quad\mu=1\;(1)\;\frac{M}{2}}\\*[16pt]
s(0)\;=\;\frac{\text{\tt C\_VQ}(0)\pm|\text{\tt C\_VV}(0)|}{L\CdoT(L\!-\!1)}
\end{gather*}
\item[$\emptyset$) ]\ \vspace{-18pt}
\[
s(M\!-\!\mu)\;=\;s(\mu)
\qquad\text{ sonst}
\]
      Anhand der Singul"arwerte kann entschieden werden, ob weitere
      Einzelmessungen durchgef"uhrt werden sollen, indem man zum 
      Punkt~\ref{E.I.7.7} zur"uckspringt.
\item \label{E.I.7.24}Berechnung der inversen empirischen Kovarianzmatrizen
      \mbox{$\Hat{\underline{C}}_{\Breve{\Vec{\boldsymbol{V}}}(\mu),\Breve{\Vec{\boldsymbol{V}}}(\mu)}^{\uP{0.4}{\!-1}}=
      \Hat{\underline{C}}_{\Tilde{\Vec{\boldsymbol{V}}}(\mu),\Tilde{\Vec{\boldsymbol{V}}}(\mu)}^{\uP{0.4}{\!-1}}$}.
      F"ur die inversen Kovarianzmatrizen mit \mbox{$\mu=1\;(1)\;M/2\!-\!1$} 
      werden je vier Speicherpl"atze ben"otigt, da diese \mbox{$2\!\times\!2$}
      Matrizen hermitesch sind und reelle Hauptdiagonalelemente aufweisen. Bei 
      den beiden Frequenzen \mbox{$\mu\!=\!0$} und \mbox{$\mu=M/2$} sind die 
      Diagonalelemente der inversen Kovarianzmatrizen gleich, so dass man insgesamt 
      also mindestens \mbox{$2\CdoT M\!+\!2$} Speicherpl"atze ben"otigt.
      Hier werden die Bindungen der Elemente der inversen Kovarianzmatrizen
      jedoch nicht ausgenutzt. Es werden die durch \mbox{$L\CdoT(L\!-\!1)$}
      dividierten, inversen, empirischen Kovarianzmatrizen auf den Feldern
      {\tt K\_VV} mit \mbox{$4\CdoT M\!+\!8$} Speicherpl"atzen abgelegt.
\begin{gather*}
\text{\tt K\_VV}(\mu)\;=\;
\begin{bmatrix}
\;\text{\tt K\_VV}(\mu,1,1)\;&\;\text{\tt K\_VV}(\mu,1,2)\;\\
\;\text{\tt K\_VV}(\mu,2,1)\;&\;\text{\tt K\_VV}(\mu,2,2)\;
\end{bmatrix}\;=\;
\begin{bmatrix}
\;\text{\tt C\_VQ}(\mu)\;&\;\text{\tt C\_VV}(\mu)\;\\
\;\text{\tt C\_VV}(\mu)^{\Kk}\;&\;\text{\tt C\_VQ}(M\!-\!\mu)\;
\end{bmatrix}^{-1}\\*[8pt]
\text{ f"ur }{\T\quad\mu=1\;(1)\;\frac{M}{2}}\\[16pt]
\text{\tt K\_VV}(0)\;=\;
\begin{bmatrix}
\;\text{\tt K\_VV}(0,1,1)\;&\;\text{\tt K\_VV}(0,1,2)\;\\
\;\text{\tt K\_VV}(0,2,1)\;&\;\text{\tt K\_VV}(0,2,2)\;
\end{bmatrix}\;=\;
\begin{bmatrix}
\;\text{\tt C\_VQ}(0)\;&\;\text{\tt C\_VV}(0)\;\\
\;\text{\tt C\_VV}(0)^{\Kk}\;&\;\text{\tt C\_VQ}(0)\;
\end{bmatrix}^{-1}
\end{gather*}
\item[$\emptyset$) ]\ \vspace{-28pt}
\[
\text{\tt K\_VV}(M\!-\!\mu)\;=\;
\begin{bmatrix}\;0\;&\;1\;\\\;1\;&\;0\;\end{bmatrix}\cdot
\text{\tt K\_VV}(\mu)^{\Tt}\cdot
\begin{bmatrix}\;0\;&\;1\;\\\;1\;&\;0\;\end{bmatrix}
\qquad\text{ sonst}
\]\vadjust{\penalty-500}
\item \label{E.I.7.25}Berechnung des empirischen Mittelwertes des Spektrums des gefensterten
      Signals am Ausgang. Das $L$-fache dieser Werte wird auf einem
      Speicher {\tt Y\_mittel} mit\linebreak \mbox{$2\CdoT M$}
      Speicherpl"atzen abgelegt.
\begin{gather*}
\overline{y}_f(k) \;=\;
\Sum{\kappa=0}{N}\,\text{\tt f}(k\!+\!\kappa\CdoT M)
\cdot\text{\tt Akku\_y}(k\!+\!\kappa\CdoT M)
\qquad\text{ f"ur }\quad k=0\;(1)\;M\!-\!1\\*[8pt]
\text{\tt Y\_mittel}(\mu)\;=\;
\text{f"|ft}_M\big(\overline{y}_f(k)\big)
\qquad\text{ f"ur }\quad\mu=0\;(1)\;M\!-\!1
\end{gather*}
\item \label{E.I.7.26}Berechnung der empirischen Varianzen
      \mbox{$\Hat{C}_{\boldsymbol{Y}_{\!\!\!f}(\mu),\boldsymbol{Y}_{\!\!\!f}(\mu)}$}
      des Spektrums des gefensterten Systemausgangssignals mit Gleichung~(\ref{E.3.35.a}).
      Das \mbox{$L\CdoT(L\!-\!1)$}-fache der empirischen Varianzen
      wird auf einem Speicher {\tt C\_YQ} mit $M$ Speicherpl"atzen abgelegt.
\[
\text{\tt C\_YQ}(\mu) \;=\; L\cdot\text{\tt Akku\_YQ}(\mu) - |\text{\tt Y\_mittel}(\mu)|^2
\qquad\text{ f"ur }\quad\mu=0\;(1)\;M\!-\!1
\]
\item \label{E.I.7.27}Berechnung der empirischen Kovarianzen
      \mbox{$\Hat{C}_{\boldsymbol{Y}_{\!\!\!f}(\mu),\boldsymbol{Y}_{\!\!\!f}(-\mu)^{\Kk}}$} des
      Spektrums des gefensterten Systemausgangssignals bei positiver
      und des konjugierten Spektrums bei negativer Frequenz mit Gleichung~(\ref{E.3.35.b}). Das
      \mbox{$L\CdoT(L\!-\!1)$}-fache der empirischen Kovarianzen wird auf
      einem Speicher {\tt C\_YY} mit $M\!+\!2$ Speicherpl"atzen abgelegt.
\begin{gather*}
\text{\tt C\_YY}(\mu) = L\cdot\text{\tt Akku\_YY}(\mu)-
\text{\tt Y\_mittel}(\mu)\cdot\text{\tt Y\_mittel}(M\!-\!\mu)
\quad\;\text{f"ur}{\T\quad\mu=1\;(1)\;\frac{M}{2}}\\*[10pt]
\text{\tt C\_YY}(0) \;=\; L\cdot\text{\tt Akku\_YY}(0)-
\text{\tt Y\_mittel}(0)^2
\end{gather*}
\item[$\emptyset$) ]\ \vspace{-18pt}
\[
\text{\tt C\_YY}(M\!-\!\mu)\;=\;\text{\tt C\_YY}(\mu)
\qquad\text{ sonst}
\]
\item \label{E.I.7.28}Berechnung der empirischen Kreuzkovarianzen
      \mbox{$\Hat{C}_{\boldsymbol{Y}_{\!\!\!f}(\mu),\boldsymbol{V}(\mu)}$}
      des Spektrums des gefensterten Systemausgangssignals und
      der Erregung. Es handelt sich dabei um die ersten Elemente der Kovarianzvektoren 
      nach Gleichung~(\ref{E.3.11}). Das \mbox{$L\CdoT(L\!-\!1)$}-fache
      der empirischen Kreuzkovarianzen wird auf einem Speicher {\tt C\_YV}
      mit \mbox{$2\CdoT M$} Speicherpl"atzen abgelegt.
\[
\text{\tt C\_YV}(\mu) \;=\; L\cdot\text{\tt Akku\_YV}(\mu) -
\text{\tt Y\_mittel}(\mu) \cdot \text{\tt Akku\_V}(\mu)^{\Kk}\!\!
\qquad\text{f"ur}\quad\mu=0\;(1)\;M\!-\!1
\]
\item \label{E.I.7.29}Berechnung der empirischen Kreuzkovarianzen
      \mbox{$\Hat{C}_{\boldsymbol{V}(-\mu),\boldsymbol{Y}_{\!\!\!f}(\mu)^{\Kk}}$}
      des konjugierten Spektrums des gefensterten Systemausgangssignals
      bei positiver und des Spektrums der Erregung bei negativer Frequenz.
      Es handelt sich dabei um die zweiten Elemente der Kovarianzvektoren 
      nach Gleichung~(\ref{E.3.11}). Das \mbox{$L\CdoT(L\!-\!1)$}-fache der 
      empirischen Kreuzkovarianzen wird auf einem Speicher {\tt C\_VY} mit 
      \mbox{$2\CdoT M$} Speicherpl"atzen abgelegt.
\begin{gather*}
\text{\tt C\_VY}(\mu)=L\cdot\text{\tt Akku\_VY}(\mu) -
\text{\tt Akku\_V}(M\!-\!\mu)\cdot\text{\tt Y\_mittel}(\mu) 
\quad\;\text{f"ur}\quad\mu=1\;(1)\;M\!-\!1\\*[10pt]
\text{\tt C\_VY}(0) \;=\; L\cdot\text{\tt Akku\_VY}(0) -
\text{\tt Akku\_V}(0) \cdot \text{\tt Y\_mittel}(0)
\end{gather*}
\item \label{E.I.7.30}Berechnung der Messwerte \mbox{$\Hat{H}(\mu)$} und \mbox{$\Hat{H}_*(\mu)$} der
      beiden "Ubertragungsfunktionen mit Gleichung~(\ref{E.3.8}). Die Werte der
      "Ubertragungsfunktionen werden auf zwei Speichern {\tt H} und {\tt HS} mit
      je \mbox{$2\CdoT M$} Speicherpl"atzen abgelegt.
\[
\begin{bmatrix}\;\text{\tt H}(\mu)\;&\;\text{\tt HS}(\mu)\;\end{bmatrix} \;=\;
\begin{bmatrix}\;\text{\tt C\_YV}(\mu)\;&\;\text{\tt C\_VY}(\mu)\;\end{bmatrix}\cdot
\text{\tt K\_VV}(\mu)
\qquad\text{ f"ur }\quad\mu=0\;(1)\;M\!-\!1
\]
\item \label{E.I.7.31}Berechnung des empirischen Mittelwertes des Spektrums der Summe der Signale am
      Ausgang der beiden linearen Modellsysteme. Das $L$-fache dieser Werte wird auf
      einem Speicher {\tt X\_mittel} mit \mbox{$2\CdoT M$} Speicherpl"atzen
      abgelegt.
\begin{gather*}
\text{\tt X\_mittel}(\mu) \;=\;
\begin{bmatrix}\;\text{\tt H}(\mu)\;&\;\text{\tt HS}(\mu)\;\end{bmatrix}\cdot 
\begin{bmatrix}\;\text{\tt Akku\_V}(\mu)\;\\\;\text{\tt Akku\_V}(M\!-\!\mu)^{\Kk}\;\end{bmatrix}
\qquad\text{ f"ur }\quad\mu=1\;(1)\;M\!-\!1\\*[10pt]
\text{\tt X\_mittel}(0) \;=\;
\begin{bmatrix}\;\text{\tt H}(0)\;&\;\text{\tt HS}(0)\;\end{bmatrix}\cdot 
\begin{bmatrix}\;\text{\tt Akku\_V}(0)\;\\\;\text{\tt Akku\_V}(0)^{\Kk}\;\end{bmatrix}
\end{gather*}
\item \label{E.I.7.32}Berechnung der Messwerte \mbox{$\Hat{U}_{\!f}(\mu)$} des Spektrums der
      gefensterten St"orung gem"a"s Gleichung~(\ref{E.3.5}).
      Diese Werte werden f"ur die weiteren Berechnungen nicht mehr ben"otigt.
\[
\text{\tt U}(\mu) \;=\;
\frac{1}{L}\cdot\big(\,\text{\tt Y\_mittel}(\mu) -
 \text{\tt X\_mittel}(\mu)\,\big)
\qquad\text{ f"ur }\quad\mu=0\;(1)\;M\!-\!1
\]
\item Berechnung des empirischen Mittelwertes des Signals am Ausgang des
      linearen Modellsystems. Das $L$-fache dieser Werte wird auf einem
      Speicher {\tt x\_mittel} mit\linebreak \mbox{$2\CdoT M$}
      Speicherpl"atzen abgelegt.
\[
\text{\tt x\_mittel}(k)=\text{if"|ft}_M\big( \text{\tt X\_mittel}(\mu)\big)
\quad\text{ f"ur}\quad\mu=0\;(1)\;M\!-\!1\quad\wedge\quad k=0\;(1)\;M\!-\!1
\]
\item \label{E.I.7.34}Berechnung der Messwerte \mbox{$\Hat{u}(k)$} der
      deterministischen St"orung mit Hilfe der Gleichung~(\ref{E.3.14}).
      Diese Werte werden f"ur die weiteren Berechnungen nicht mehr ben"otigt.
\begin{gather*}
\text{\tt u}(k) \;=\;  \frac{1}{L}\cdot
\big(\,\text{\tt Akku\_y}(k) - \text{\tt x\_mittel}(k\!-\!\kappa\CdoT M)\,\big)\\*[14pt]
\text{ f"ur }\quad k=0\;(1)\;F\!-\!1\quad\wedge\quad 0\le k\!-\!\kappa\CdoT M<M
\end{gather*}
\item \label{E.I.7.35}Berechnung der Messwerte \mbox{$\Hat{\Phi}_{\boldsymbol{n}}(\mu)$} 
      nach Gleichung~(\ref{E.3.34.a}) zur Beschreibung des LDS. Diese Werte 
      werden auf einem Speicher {\tt Phi} mit $M$ Speicherpl"atzen abgelegt.
\begin{gather*}
\text{\tt Phi}(\mu) \;=\;
\frac{\;\text{\tt C\_YQ}(\mu)-
\left[\text{\tt C\_YV}(\mu),\,\text{\tt C\_VY}(\mu)\right]
\cdot\text{\tt K\_VV}(\mu)\cdot
\left[\text{\tt C\_YV}(\mu),\,\text{\tt C\_VY}(\mu)\right]^{\Hh}}
{M\cdot L\CdoT(L\!-\!3)}\\*[14pt]
\text{ f"ur }{\T\quad\mu=0\;(1)\;M\!-\!1}
\end{gather*}
\item \label{E.I.7.36}Berechnung der Messwerte \mbox{$\Hat{\Psi}_{\boldsymbol{n}}(\mu)$}
      nach Gleichung~(\ref{E.3.34.b}) zur Beschreibung des KLDS. Diese Werte 
      werden auf einem Speicher {\tt Psi} mit $M\!+\!2$ Speicherpl"atzen abgelegt.
\begin{gather*}
\text{\tt Psi}(\mu) \;=\;
\frac{\;\text{\tt C\_YY}(\mu)-
\left[\text{\tt C\_YV}(\mu),\text{\tt C\_VY}(\mu)\right]\cdot
\text{\tt K\_VV}(\mu)\cdot
\left[\text{\tt C\_VY}(M\!-\!\mu),\,\text{\tt C\_YV}(M\!-\!\mu)\right]^{\Tt}}
{M\CdoT L\CdoT(L\!-\!3)}\\*[8pt]
\text{f"ur }{\T\quad\mu=1\;(1)\;\frac{M}{2}}\\[10pt]
\text{\tt Psi}(0) \;=\;
\frac{\;\text{\tt C\_YY}(0)-
\left[\text{\tt C\_YV}(0),\,\text{\tt C\_VY}(0)\right]\cdot
\text{\tt K\_VV}(0)\cdot
\left[\text{\tt C\_VY}(0),\,\text{\tt C\_YV}(0)\right]^{\Tt}}
{M\CdoT L\CdoT(L\!-\!3)}
\end{gather*}
\item[$\emptyset$) ]\ \vspace{-18pt}
\[
\text{\tt Psi}(M\!-\!\mu)\;=\;\text{\tt Psi}(\mu)
\qquad\text{ sonst}
\]
\item \label{E.I.7.37}Berechnung der Varianzen 
      \mbox{$\Hat{C}_{\Hat{\boldsymbol{H}}(\mu),\Hat{\boldsymbol{H}}(\mu)}$}
      der Messwerte der "Ubertragungsfunktion \mbox{$H\!(\mu)$} gem"a"s Gleichung~(\ref{E.3.44.a}). 
      Diese Werte werden auf einem Speicher
      {\tt C\_HQ} mit \mbox{$M$} Speicherpl"atzen abgelegt.
\[
\text{\tt C\_HQ}(\mu) \;=\;M\CdoT L\cdot\text{\tt K\_VV}(\mu,1,1)\cdot\text{\tt Phi}(\mu)
\qquad\text{ f"ur }\quad\mu=0\;(1)\;M\!-\!1
\]
\item \label{E.I.7.38}Berechnung der Varianzen 
      \mbox{$\Hat{C}_{\Hat{\boldsymbol{H}}_*(\mu),\Hat{\boldsymbol{H}}_*(\mu)}$}
      der Messwerte der "Ubertragungsfunktion \mbox{$H_*(\mu)$} gem"a"s Gleichung~(\ref{E.3.44.a}). 
      Diese Werte werden auf einem Speicher
      {\tt C\_HSQ} mit \mbox{$M$} Speicherpl"atzen abgelegt.
\[
\text{\tt C\_HSQ}(\mu) \;=\;M\CdoT L\cdot\text{\tt K\_VV}(\mu,2,2)\cdot\text{\tt Phi}(\mu)
\qquad\text{ f"ur }\quad\mu=0\;(1)\;M\!-\!1
\]
\item \label{E.I.7.39}Berechnung der Kovarianzen
      \mbox{$\Hat{C}_{\Hat{\boldsymbol{H}}(\mu),\Hat{\boldsymbol{H}}(\mu)^{\Kk}}$}
      der Messwerte der "Ubertragungsfunk\-tion \mbox{$H(\mu)$} gem"a"s Gleichung~(\ref{E.3.44.b}).
      Bei Verwendung einer hoch frequenzselektiven Fensterfolge sind nur
      die Werte f"ur \mbox{$\mu\!=\!0$} und \mbox{$\mu=M/2$} nennenswert von null
      verschieden. In Gleichung~(\ref{E.3.44.b}) ergibt das Produkt aus der Matrix 
      in der Mitte und der rechten, inversen Matrix die Permutationsmatrix 
      \mbox{$\Big[\begin{smallmatrix}0&1\\1&0\end{smallmatrix}\Big]$}. Die Einheitsvektoren 
      links und rechts greifen ein Element der verbleibenden inversen, konjugierten Matrix heraus.
      Die Kovarianzen werden auf einem Speicher {\tt C\_HH} mit
      \mbox{$4$} Speicherpl"atzen abgelegt.
\[
\text{\tt C\_HH}(\mu) \;=\; 
M\CdoT L \cdot \text{\tt K\_VV}(\mu,2,1)^{\Kk}\Cdot\text{\tt Psi}(\mu)
\qquad\text{ f"ur }\quad{\T\mu\in\big\{0;\frac{M}{2}\big\}}
\]
\item \label{E.I.7.40}Berechnung der Kovarianzen
      \mbox{$\Hat{C}_{\Hat{\boldsymbol{H}}_*(\mu),\Hat{\boldsymbol{H}}_*(\mu)^{\Kk}}$}
      der Messwerte der "Ubertragungsfunk\-tion \mbox{$H_*(\mu)$} gem"a"s Gleichung~(\ref{E.3.44.b}).
      Bei Verwendung einer hoch frequenzselektiven Fensterfolge sind nur
      die Werte f"ur \mbox{$\mu\!=\!0$} und \mbox{$\mu=M/2$} nennenswert von null
      verschieden. In Gleichung~(\ref{E.3.44.b}) ergibt das Produkt aus der Matrix 
      in der Mitte und der rechten, inversen Matrix die Permutationsmatrix 
      \mbox{$\Big[\begin{smallmatrix}0&1\\1&0\end{smallmatrix}\Big]$}. \pagebreak[2]Die Einheitsvektoren 
      links und rechts greifen ein Element der verbleibenden inversen, konjugierten Matrix heraus.
      Die Kovarianzen werden auf einem Speicher {\tt C\_HSHS} mit
      \mbox{$4$} Speicherpl"atzen abgelegt.
\[
\text{\tt C\_HSHS}(\mu) \;=\; 
M\CdoT L \cdot \text{\tt K\_VV}(\mu,1,2)^{\Kk}\Cdot\text{\tt Psi}(\mu)
\qquad\text{ f"ur }\quad{\T\mu\in\big\{0;\frac{M}{2}\big\}}
\]
\item \label{E.I.7.41}Berechnung der Varianzen
      \mbox{$\Hat{C}_{\Hat{\boldsymbol{U}}_{\!\!f}(\mu),\Hat{\boldsymbol{U}}_{\!\!f}(\mu)}$}
      der Messwerte des Spektrums der gefensterten St"orung gem"a"s
      Gleichung~(\ref{E.3.48.a}). In dieser Gleichung ergibt das Produkt aus der Matrix 
      in der Mitte und der rechten, inversen Matrix die Einheitsmatrix. 
      Die Messwertvarianzen werden auf einem Speicher {\tt C\_UQ} mit 
      \mbox{$M$} Speicherpl"atzen abgelegt.
\begin{gather*}
\text{\tt C\_UQ}(\mu) \;=\;
\frac{M}{L}\cdot\Bigg(1+
\begin{bmatrix}\text{\tt Akku\_V}(\mu)\\\text{\tt Akku\_V}(M\!-\!\mu)^{\Kk}\end{bmatrix}^{\Hh}
\!\Cdot\text{\tt K\_VV}(\mu)\cdot
\begin{bmatrix}\text{\tt Akku\_V}(\mu)\\\text{\tt Akku\_V}(M\!-\!\mu)^{\Kk}\end{bmatrix}\Bigg)
\cdot\text{\tt Phi}(\mu)\\*[6pt]
\qquad\text{ f"ur }\quad\mu=1\;(1)\;M\!-\!1\\*[12pt]
\text{\tt C\_UQ}(0) \;=\;
\frac{M}{L}\cdot\Bigg(\;1+
\begin{bmatrix}\text{\tt Akku\_V}(0)\;\\\;\text{\tt Akku\_V}(0)^{\Kk}\end{bmatrix}^{\Hh}
\Cdot\text{\tt K\_VV}(0)\cdot
\begin{bmatrix}\text{\tt Akku\_V}(0)\;\\\;\text{\tt Akku\_V}(0)^{\Kk}\end{bmatrix}\Bigg)
\cdot\text{\tt Phi}(0)
\end{gather*}
\item \label{E.I.7.42}Berechnung der Kovarianzen
     \mbox{$\Hat{C}_{\Hat{\boldsymbol{U}}_{\!\!f}(\mu),\Hat{\boldsymbol{U}}_{\!\!f}(\mu)^{\Kk}}$}
      der Messwerte des Spektrums der gefensterten St"orung gem"a"s
      Gleichung~(\ref{E.3.48.b}). Bei Verwendung einer hoch
      frequenzselektiven Fensterfolge sind nur die Kovarianzen f"ur
      \mbox{$\mu\!=\!0$} und \mbox{$\mu=M/2$} nennenswert von null
      verschieden. Da jedoch f"ur die Berechnung der Kovarianzen
      der deterministischen St"orung im Zeitbereich alle Kovarianzen
     \mbox{$\Hat{C}_{\Hat{\boldsymbol{U}}_{\!\!f}(\mu),\Hat{\boldsymbol{U}}_{\!\!f}(-\mu)^{\Kk}}$}
      ben"otigt werden, und diese f"ur die beiden angegebenen Frequenzen mit
      den obengenannten "ubereinstimmen, werden gleich alle berechnet.
      In Gleichung~(\ref{E.3.48.b}) ergibt das Produkt aus der Matrix 
      in der Mitte und der rechten, inversen Matrix die Permutationsmatrix 
      \mbox{$\Big[\begin{smallmatrix}0&1\\1&0\end{smallmatrix}\Big]$}. 
      Die Kovarianzen werden auf einem Speicher {\tt C\_UU} mit
      \mbox{$M\!+\!2$} Speicherpl"atzen abgelegt.
\begin{gather*}
\text{\tt C\_UU}(\mu) \;=\; 
\frac{M}{L}\cdot\Bigg(\;1+
\begin{bmatrix}\text{\tt Akku\_V}(\mu)\\\text{\tt Akku\_V}(M\!-\!\mu)^{\Kk}\end{bmatrix}^{\Hh}
\!\!\Cdot\text{\tt K\_VV}(\mu)\cdot
\begin{bmatrix}\text{\tt Akku\_V}(\mu)\\\text{\tt Akku\_V}(M\!-\!\mu)^{\Kk}\end{bmatrix}\Bigg)
\cdot\text{\tt Psi}(\mu)\\*[6pt]
\qquad\text{ f"ur }{\T\quad\mu=1\;(1)\;\frac{M}{2}}\\*[12pt]
\text{\tt C\_UU}(0) \;=\; \frac{M}{L}\cdot\Bigg(\;1+
\begin{bmatrix}\text{\tt Akku\_V}(0)\;\\\;\text{\tt Akku\_V}(0)^{\Kk}\end{bmatrix}^{\Hh}
\Cdot\text{\tt K\_VV}(0)\cdot
\begin{bmatrix}\text{\tt Akku\_V}(0)\;\\\;\text{\tt Akku\_V}(0)^{\Kk}\end{bmatrix}\Bigg)
\cdot\text{\tt Psi}(0)
\end{gather*}
\item[$\emptyset$) ]\ \vspace{-18pt}
\[
\text{\tt C\_UU}(M\!-\!\mu)=\text{\tt C\_UU}(\mu)
\qquad\text{ sonst}
\]
\item \label{E.I.7.43}Berechnung der Varianz
      \mbox{$\Hat{C}_{\Hat{\boldsymbol{u}}(k),\Hat{\boldsymbol{u}}(k)}$}
      der Messwerte der St"orung mit Hilfe der Gleichung~(\ref{E.3.53.a}).
      Hier handelt es sich um den reellen Wert der zeitunabh"angigen
      N"aherung. Er wird auf einem Speicherplatz {\tt C\_uQ} abgelegt.
\[
\text{\tt C\_uQ}\;=\;\frac{1}{M^2}\cdot\Sum{\mu=0}{M-1}\,\text{\tt C\_UQ}(\mu)
\]
\item \label{E.I.7.44}Berechnung der Kovarianz
      \mbox{$\Hat{C}_{\Hat{\boldsymbol{u}}(k),\Hat{\boldsymbol{u}}(k)^{\Kk}}$}
      der Messwerte der St"orung mit Hilfe der Gleichung~(\ref{E.3.53.b}). Hier
      handelt es sich um den komplexen Wert der zeitunabh"angigen N"aherung.
      Er wird auf zwei Speicherpl"atzen {\tt C\_uu} abgelegt.
\[
\text{\tt C\_uu}\;=\;\frac{1}{M^2}\cdot\Sum{\mu=0}{M-1}\,\text{\tt C\_UU}(\mu)
\]
\item \label{E.I.7.45}Berechnung der Varianzen 
      \mbox{$\Hat{C}_{\Hat{\boldsymbol{\Phi}}_{\!\boldsymbol{n}}(\mu),\Hat{\boldsymbol{\Phi}}_{\!\boldsymbol{n}}(\mu)}$}
      der LDS-Messwerte nach Gleichung~(\ref{E.3.58.a}) und (\ref{E.3.61.a}). Diese Werte werden auf einem Speicher
      {\tt C\_PhiQ} mit \mbox{$M$} Speicherpl"atzen abgelegt.
\begin{gather*}
\text{\tt C\_PhiQ}(\mu) \;=\;
\frac{L-5}{(L\!-\!1)\CdoT(L\!-\!4)}\cdot\text{\tt Phi}(\mu)^2\;+\;
\frac{L\!-\!3}{(L\!-\!1)\CdoT(L\!-\!4)}\cdot|\text{\tt Psi}(\mu)|^2\\*[9pt]
\text{ f"ur }{\T\quad\mu\in\big\{0;\frac{M}{2}\big\}}\\*[20pt]
\text{\tt C\_PhiQ}(\mu) \;=\; \frac{1}{L\!-\!2}\cdot\text{\tt Phi}(\mu)^2
\qquad\text{ f"ur }
{\T\quad\mu=1\;(1)\;M\!-\!1\quad\wedge\quad\mu\neq\frac{M}{2}}
\end{gather*}

\item \label{E.I.7.46}Berechnung der Varianzen
      \mbox{$\Hat{C}_{\Hat{\boldsymbol{\Psi}}_{\!\boldsymbol{n}}(\mu),\Hat{\boldsymbol{\Psi}}_{\!\boldsymbol{n}}(\mu)}$}
      der KLDS-Messwerte nach Gleichung~(\ref{E.3.58.c}) und (\ref{E.3.61.c}). Diese Werte werden auf einem
      Speicher {\tt C\_PsiQ} mit \mbox{$M/2\!+\!1$} Speicherpl"atzen abgelegt.
\begin{gather*}
\text{\tt C\_PsiQ}(\mu) \;=\; \frac{2\cdot(L\!-\!3)}
{(L\!-\!1)\CdoT(L\!-\!4)}\cdot\text{\tt Phi}(\mu)^2\;-\;
\frac{2}{(L\!-\!1)\CdoT(L\!-\!4)}\cdot|\text{\tt Psi}(\mu)|^2\\*[8pt]
\text{ f"ur }{\T\quad\mu\in\big\{0;\frac{M}{2}\big\}}\\[14pt]
\text{\tt C\_PsiQ}(\mu) \;=\; \frac{(L\!-\!3)}{(L\!-\!2)\CdoT(L\!-\!4)}
\cdot\text{\tt Phi}(\mu)\cdot\text{\tt Phi}(M\!-\!\mu)\;-\;
\frac{1}{(L\!-\!2)\CdoT(L\!-\!4)}\cdot|\text{\tt Psi}(\mu)|^2\\*[8pt]
\text{ f"ur }{\T\quad\mu=1\;(1)\;\frac{M}{2}\!-\!1}
\end{gather*}
\item[$\emptyset$) ]\ \vspace{-18pt}
\[
\text{\tt C\_PsiQ}(M\!-\!\mu)\;=\;\text{\tt C\_PsiQ}(\mu)
\qquad\text{ sonst}
\]
\item \label{E.I.7.47}Berechnung der Kovarianzen 
      \mbox{$\Hat{C}_{\Hat{\boldsymbol{\Psi}}_{\!\boldsymbol{n}}(\mu),\Hat{\boldsymbol{\Psi}}_{\!\boldsymbol{n}}(\mu)^{\Kk}}$}
      der KLDS-Messwerte nach Gleichung (\ref{E.3.58.d}) und (\ref{E.3.61.d}). Diese Werte werden auf einem
      Speicher {\tt C\_PsiPsi} mit \mbox{$M\!+\!2$} Speicherpl"atzen abgelegt.
\begin{align*}
\text{\tt C\_PsiPsi}(\mu)&\;=\; \frac{2}{L\!-\!1}\cdot\text{\tt Psi}(\mu)^2
\qquad\text{ f"ur }{\T\quad\mu\in\big\{0;\frac{M}{2}\big\}}\\[14pt]
\text{\tt C\_PsiPsi}(\mu)&\;=\; \frac{1}{L\!-\!2}\cdot\text{\tt Psi}(\mu)^2
\qquad\text{ f"ur }{\T\quad\mu=1\;(1)\;\frac{M}{2}\!-\!1}
\end{align*}
\item[$\emptyset$) ]\ \vspace{-22pt}
\[
\text{\tt C\_PsiPsi}(M\!-\!\mu)\;=\;\text{\tt C\_PsiPsi}(\mu)
\qquad\text{ sonst}
\]
\item Festlegen des Parameters $\alpha$ des Konfidenzniveaus
      \mbox{$1\!-\!\alpha$}.
\item \label{E.I.7.49}Konfidenzgebiete der Messwerte der "Ubertragungsfunktion 
      des linearen Teilsystems ${\cal S}_{lin}$. Bei allen Frequenzen 
      au"ser bei \mbox{$\mu\!=\!0$} und \mbox{$\mu=M/2$}
      ergeben sich Kreise um die Messwerte \mbox{$\Hat{H}(\mu)$}.
      Der Radius dieser Kreise ist gleich dem Betrag der Halbachsen
      der Konfidenzellipsen gem"a"s der Gleichungen~(\myref{3.80}):
\[
\begin{aligned}
\text{\tt A\_1\_H}(\mu)\;=\;
&\sqrt{-\ln(\alpha)\cdot\text{\tt C\_HQ}(\mu)\;}\\[10pt]
\text{\tt A\_2\_H}(\mu) \;=\; &j\cdot\text{\tt A\_1\_H}(\mu)
\end{aligned}
\qquad\text{ f"ur }
{\T\quad\mu=1\;(1)\;M\!-\!1\quad\wedge\quad\mu\neq\frac{M}{2}}
\]
      Bei den Frequenzen \mbox{$\mu\!=\!0$} und \mbox{$\mu=M/2$}
      ergeben sich Konfidenzellipsen um die Messwerte \mbox{$\Hat{H}(\mu)$}.
      Die Halbachsen der Ellipsen sind:
\begin{align*}
\text{\tt A\_1\_H}(\mu)\;=\;\phantom{j\cdot}&
\sqrt{\:-\ln(\alpha)\cdot\big(\:\text{\tt C\_HQ}(\mu)+
|\text{\tt C\_HH}(\mu)|\:\big)\:}
\cdot e^{\frac{j}{2}\cdot\winkel\{\text{\tt C\_HH}(\mu)\}}\\*[10pt]
\text{\tt A\_2\_H}(\mu)\;=\;j\cdot&
\sqrt{\:-\ln(\alpha)\cdot\big(\:\text{\tt C\_HQ}(\mu)-
|\text{\tt C\_HH}(\mu)|\:\big)\:}
\cdot e^{\frac{j}{2}\cdot\winkel\{\text{\tt C\_HH}(\mu)\}}\\*[8pt]
&\text{ f"ur }{\T\quad\mu\in\big\{0;\frac{M}{2}\big\}}
\end{align*}
\item \label{E.I.7.50}Konfidenzgebiete der Messwerte der "Ubertragungsfunktion 
      des linearen Teilsystems ${\cal S}_{*,lin}$. Bei allen Frequenzen 
      au"ser bei \mbox{$\mu\!=\!0$} und \mbox{$\mu=M/2$}
      ergeben sich Kreise um die Messwerte \mbox{$\Hat{H}_*(\mu)$}.
      Der Radius dieser Kreise ist gleich dem Betrag der Halbachsen
      der Konfidenzellipsen gem"a"s der Gleichungen~(\myref{3.80}):
\[
\begin{aligned}
\text{\tt A\_1\_HS}(\mu)\;=\;
&\sqrt{-\ln(\alpha)\cdot\text{\tt C\_HSQ}(\mu)\;}\\[10pt]
\text{\tt A\_2\_HS}(\mu) \;=\; &j\cdot\text{\tt A\_1\_HS}(\mu)
\end{aligned}
\qquad\text{ f"ur }
{\T\quad\mu=1\;(1)\;M\!-\!1\quad\wedge\quad\mu\neq\frac{M}{2}}
\]
      Bei den Frequenzen \mbox{$\mu\!=\!0$} und \mbox{$\mu=M/2$}
      ergeben sich Konfidenzellipsen um die Messwerte \mbox{$\Hat{H}_*(\mu)$}.
      Die Halbachsen der Ellipsen sind:
\begin{align*}
\text{\tt A\_1\_HS}(\mu)\;=\;\phantom{j\cdot}&
\sqrt{\:-\ln(\alpha)\cdot\big(\:\text{\tt C\_HSQ}(\mu)+
|\text{\tt C\_HSHS}(\mu)|\:\big)\:}
\cdot e^{\frac{j}{2}\cdot\winkel\{\text{\tt C\_HSHS}(\mu)\}}\\*[10pt]
\text{\tt A\_2\_HS}(\mu)\;=\;j\cdot&
\sqrt{\:-\ln(\alpha)\cdot\big(\:\text{\tt C\_HSQ}(\mu)-
|\text{\tt C\_HSHS}(\mu)|\:\big)\:}
\cdot e^{\frac{j}{2}\cdot\winkel\{\text{\tt C\_HSHS}(\mu)\}}\\*[8pt]
&\text{ f"ur }{\T\quad\mu\in\big\{0;\frac{M}{2}\big\}}
\end{align*}
\item \label{E.I.7.51}Konfidenzgebiete der Messwerte des Spektrums der gefensterten St"orung.
      Bei allen Frequenzen au"ser bei \mbox{$\mu\!=\!0$} und \mbox{$\mu=M/2$}
      ergeben sich Kreise um die Messwerte \mbox{$\Hat{U}_{\!f}(\mu)$}.
      Der Radius dieser Kreise ist gleich dem Betrag der Halbachsen
      der Konfidenzellipsen, die sich analog zu den Gleichungen~(\myref{3.80}) berechnen:
\[
\begin{aligned}
\text{\tt A\_1\_U}(\mu)\;=\;
&\sqrt{-\ln(\alpha)\cdot\text{\tt C\_UQ}(\mu)\;}\\[10pt]
\text{\tt A\_2\_U}(\mu) \;=\; &j\cdot\text{\tt A\_1\_U}(\mu)
\end{aligned}
\qquad\text{ f"ur }
{\T\quad\mu=1\;(1)\;M\!-\!1\quad\wedge\quad\mu\neq\frac{M}{2}}
\]
      Bei den Frequenzen \mbox{$\mu\!=\!0$} und \mbox{$\mu=M/2$}
      ergeben sich Konfidenzellipsen um die Messwerte \mbox{$\Hat{U}_{\!f}(\mu)$}.
      Die Halbachsen der Ellipsen sind:
\begin{align*}
\text{\tt A\_1\_U}(\mu)\;=\;\phantom{j\cdot}&
\sqrt{\:-\ln(\alpha)\cdot\big(\:\text{\tt C\_UQ}(\mu)+
|\text{\tt C\_UU}(\mu)|\:\big)\:}
\cdot e^{\frac{j}{2}\cdot\winkel\{\text{\tt C\_UU}(\mu)\}}\\*[10pt]
\text{\tt A\_2\_U}(\mu)\;=\;j\cdot&
\sqrt{\:-\ln(\alpha)\cdot\big(\:\text{\tt C\_UQ}(\mu)-
|\text{\tt C\_UU}(\mu)|\:\big)\:}
\cdot e^{\frac{j}{2}\cdot\winkel\{\text{\tt C\_UU}(\mu)\}}\\*[8pt]
&\text{ f"ur }{\T\quad\mu\in\big\{0;\frac{M}{2}\big\}}
\end{align*}
\item \label{E.I.7.52}Konfidenzgebiete der Messwerte der deterministischen St"orung.
      Die Konfidenzgebiete werden f"ur alle Zeitpunkte gleich abgesch"atzt.
      Die zeitunabh"angigen Halbachsen der Ellipsen berechnen sich analog zu den 
      Gleichungen~(\myref{3.80}) und sind:
\begin{align*}
\text{\tt A\_1\_u}\;=\;\phantom{j\cdot}&
\sqrt{\:-\ln(\alpha)\cdot\big(\:\text{\tt C\_uQ}+
|\text{\tt C\_uu}|\:\big)\:}
\cdot e^{\frac{j}{2}\cdot\winkel\{\text{\tt C\_uu}\}}\\*[10pt]
\text{\tt A\_2\_u}\;=\;j\cdot&
\sqrt{\:-\ln(\alpha)\cdot\big(\:\text{\tt C\_uQ}-
|\text{\tt C\_uu}|\:\big)\:}
\cdot e^{\frac{j}{2}\cdot\winkel\{\text{\tt C\_uu}\}}
\end{align*}

\item \label{E.I.7.53}Konfidenzintervalle der LDS-Messwerte.
      Da diese Messwerte immer reell sind, \linebreak ergeben sich Konfidenzintervalle
      \mbox{$\Big[\,\Hat{\Phi}_{\boldsymbol{n}}(\mu)-\Hat{A}_{\Phi}(\mu)\,;\;
      \Hat{\Phi}_{\boldsymbol{n}}(\mu)+\Hat{A}_{\Phi}(\mu)\;\Big]$},
      die \linebreak sym\-me\-trisch zum Messwert liegen. Die halbe Intervallbreite
      berechnet sich mit \linebreak Gleichung~(\myref{3.72}) zu:
\[
\text{\tt A\_Phi}(\mu) \;=\;
\sqrt{2\;}\cdot\sqrt{\:\text{\tt C\_PhiQ}(\mu)\:}\cdot\text{erfc}^{-1}(\alpha)
\qquad\text{ f"ur }\quad\mu=0\;(1)\;M\!-\!1
\]
\item \label{E.I.7.54}Konfidenzgebiete der KLDS-Messwerte.
      Bei allen Frequenzen ergeben sich Konfidenzellipsen
      um die Messwerte \mbox{$\Hat{\Psi}_{\boldsymbol{n}}(\mu)$}.
      Die Halbachsen der Ellipsen, die sich analog zu den 
      Gleichungen~(\myref{3.80}) berechnen, sind:
\begin{align*}
\text{\tt A\_1\_Psi}(\mu)\;=\;\phantom{j\cdot}&
\sqrt{\:-\ln(\alpha)\cdot\big(\:\text{\tt C\_PsiQ}(\mu)+
|\text{\tt C\_PsiPsi}(\mu)|\:\big)\:}
\cdot e^{\frac{j}{2}\cdot\winkel\{\text{\tt C\_PsiPsi}(\mu)\}}\\*[10pt]
\text{\tt A\_2\_Psi}(\mu)\;=\;j\cdot&
\sqrt{\:-\ln(\alpha)\cdot\big(\:\text{\tt C\_PsiQ}(\mu)-
|\text{\tt C\_PsiPsi}(\mu)|\:\big)\:}
\cdot e^{\frac{j}{2}\cdot\winkel\{\text{\tt C\_PsiPsi}(\mu)\}}\\*[8pt]
&\text{ f"ur }{\T\quad\mu=0\;(1)\;\frac{M}{2}}
\end{align*}
\item[$\emptyset$) ]\ \vspace{-28pt}
\[
\begin{aligned}
\text{\tt A\_1\_Psi}(\mu)\;=\;&\text{\tt A\_1\_Psi}(M\!-\!\mu)\\
\text{\tt A\_2\_Psi}(\mu)\;=\;&\text{\tt A\_2\_Psi}(M\!-\!\mu)
\end{aligned}
\qquad\text{ sonst}
\]
\end{enumerate}
Unter den Voraussetzungen, die im Laufe der Herleitung des RKM mit Fensterung
f"ur komplexwertige Systeme gemacht worden sind,
sind die so gewonnenen Messwerte \mbox{$\Hat{\boldsymbol{H}}(\mu)$},
\mbox{$\Hat{\boldsymbol{U}}_{\!f}(\mu)$},
\mbox{$\Hat{\boldsymbol{u}}(k)$},
\mbox{$\Hat{\boldsymbol{\Phi}}_{\!\boldsymbol{n}}(\mu)$} und
\mbox{$\Hat{\boldsymbol{\Psi}}_{\!\boldsymbol{n}}(\mu)$} erwartungstreu
und konsistent. Die Qualit"at der Messung kann mit Hilfe der angegebenen
Konfidenzgebiete und Intervalle abgesch"atzt werden.

\chapter{Aufwandsabsch"atzung f"ur andere Varianten des RKM}\label{E.Kap.8}

In diesem Kapitel soll nun untersucht werden, wie der im letzten Kapitel 
geschilderte Ablauf zur Berechnung der Messwerte zu modifizieren 
ist, wenn man das periodisch zeitvariante Modellsystem mit \mbox{$K_H\!>\!1$} 
bei vorliegen einer zyklostation"aren St"orung mit \mbox{$K_{\Phi}\!>\!1$} verwendet. 
Im Anschluss daran werden die Varianten mit den reduzierten Systemmodellen nach \cite{Diss} 
ebenso untersucht, wie auch die Varianten zur Messung an reellwertigen Systemen nach Kapitel 
\ref{E.Kap.6} und zur Spektralsch"atzung nach Kapitel \ref{E.Kap.5}.

Wird ein periodisch zeitvariantes Modellsystem angesetzt, und ist der Approximationsfehlerprozess
zyklostation"ar, so ben"otigt man zus"atzliche Akkumulatorfelder um die Kovarianzen
der Spektralwerte bei der Frequenz $\mu$ mit den Spektralwerten
bei den um Vielfache von $M/K_S$ verschobenen Frequenzen empirisch ermitteln zu k"onnen. 
Man muss nun in Punkt \ref{E.I.7.4} die Akkumulatorfelder {\tt Akku\_VQ}, {\tt Akku\_VV}, 
{\tt Akku\_YQ}, {\tt Akku\_YY}, {\tt Akku\_YV} und {\tt Akku\_VY} mit $K_S$-facher 
Gr"o"se bereitstellen, um alle Produkte\newline 
\mbox{$V_{\lambda}(\mu)\CdoT V_{\lambda}\big({\T\mu\!+\!\Hat{\mu}\CdoT\frac{M}{K_S}}\big)^{\Kk}$},\newline 
\mbox{$V_{\lambda}(\mu)\CdoT V_{\lambda}\big({\T-\mu\!-\!\Hat{\mu}\CdoT\frac{M}{K_S}}\big)$},\newline 
\mbox{$Y_{\!f,\lambda}(\mu)\CdoT Y_{\!f,\lambda}\big({\T\mu\!+\!\Hat{\mu}\CdoT\frac{M}{K_S}}\big)^{\Kk}$},\newline 
\mbox{$Y_{\!f,\lambda}(\mu)\CdoT Y_{\!f,\lambda}\big({\T-\mu\!-\!\Hat{\mu}\CdoT\frac{M}{K_S}}\big)$},\newline 
\mbox{$Y_{\!f,\lambda}(\mu)\CdoT V_{\lambda}\big({\T\mu\!+\!\Hat{\mu}\CdoT\frac{M}{K_S}}\big)^{\Kk}$}
und \newline 
\mbox{$Y_{\!f,\lambda}(\mu)\CdoT V_{\lambda}\big(\!{\T-\mu\!-\!\Hat{\mu}\CdoT\frac{M}{K_S}}\big)$}\newline 
aller Einzelmessungen in der Messschleife der Punkte \ref{E.I.7.7} bis \ref{E.I.7.18} akkumulieren zu k"onnen. 
In den Punkten \ref{E.I.7.21}, \ref{E.I.7.22} und \ref{E.I.7.26} bis \ref{E.I.7.29} wird man dann $K_S$ mal 
soviele entsprechende \mbox{$L\CdoT(L\!-\!1)$}-fache Kovarianzen {\tt C\_VQ}, {\tt C\_VV}, {\tt C\_YQ}, {\tt C\_YY}, 
{\tt C\_YV} und {\tt C\_VY} erhalten, indem man von den $L$-fachen Akkumulatoren {\tt Akku\_VQ}, 
{\tt Akku\_VV}, {\tt Akku\_YQ}, {\tt Akku\_YY}, {\tt Akku\_YV} und {\tt Akku\_VY} die Produkte \newline 
\mbox{$\text{\tt Akku\_V}(\mu)\CdoT\text{\tt Akku\_V}\big({\T\mu\!+\!\Hat{\mu}\CdoT\frac{M}{K_S}}\big)^{\Kk}$}, \newline 
\mbox{$\text{\tt Akku\_V}(\mu)\CdoT\text{\tt Akku\_V}\big(\!{\T-\mu\!-\!\Hat{\mu}\CdoT\frac{M}{K_S}}\big)$}, \newline 
\mbox{$\text{\tt Y\_mittel}(\mu)\CdoT\text{\tt Y\_mittel}\big({\T\mu\!+\!\Hat{\mu}\CdoT\frac{M}{K_S}}\big)^{\Kk}$}, \newline 
\mbox{$\text{\tt Y\_mittel}(\mu)\CdoT\text{\tt Y\_mittel}\big(\!{\T-\mu\!-\!\Hat{\mu}\CdoT\frac{M}{K_S}}\big)$}, \newline 
\mbox{$\text{\tt Y\_mittel}(\mu)\CdoT\text{\tt Akku\_V}\big({\T\mu\!+\!\Hat{\mu}\CdoT\frac{M}{K_S}}\big)^{\Kk}$} und \newline 
\mbox{$\text{\tt Y\_mittel}(\mu)\CdoT\text{\tt Akku\_V}\big(\!{\T-\mu\!-\!\Hat{\mu}\CdoT\frac{M}{K_S}}\big)$} 
subtrahiert. 

Da $K_S$ ein ganzzahliges Vielfaches von $K_H$ ist, ben"otigt man 
zur Berechnung der Messwerte der beiden bifrequenten "Ubertragungsfunktionen mit Hilfe der 
Gleichung (\ref{E.3.8}) keine weiteren \mbox{$L\CdoT(L\!-\!1)$}-fachen Kovarianzen. 
Alle dort in dem Kovarianzvektor \mbox{$\Hat{\Vec{C}}_{\boldsymbol{Y}_{\!\!\!f}(\mu),\Tilde{\Vec{\boldsymbol{V}}}(\mu)}$} 
und der Kovarianzmatrix \mbox{$\Hat{\underline{C}}_{\Tilde{\Vec{\boldsymbol{V}}}(\mu),\Tilde{\Vec{\boldsymbol{V}}}(\mu)}$}
vorkommenden Kovarianzen sind, abgesehen von dem Faktor \mbox{$L\CdoT(L\!-\!1)$}, der sich sowieso herausk"urzt, 
in den Feldern {\tt C\_VQ}, {\tt C\_VV}, {\tt C\_YV} und {\tt C\_VY} enthalten. 

Es sei darauf hingewiesen, dass die Zufallsvektoren
\mbox{$\Tilde{\Vec{\boldsymbol{V}}}(\mu)$} nach Gleichung~(\ref{E.2.15}) bei einer
Erh"ohung von $\mu$ um ein ganzzahliges Vielfaches von \mbox{$M/K_H$}
lediglich permutiert werden. Die dabei auftretende Permutationsmatrix l"asst sich 
mit Hilfe der Permutationsmatrix
\begin{equation}
\underline{Q}\;=\;
\begin{bmatrix}
0&1&0&\cdots&\cdots&0\\
\vdots&0&1&0&\cdots&0\\
\vdots&&\myddots&\myddots&\myddots&\vdots\\
\vdots&&&0&1&0\\
0&\cdots&\cdots&\cdots&0&1\\
1&0&\cdots&\cdots&\cdots&0
\end{bmatrix}
\label{E.8.1}
\end{equation}
gem"a"s\vspace{-8pt}
\begin{equation}
\underline{P}\;=\;
\begin{bmatrix}
\underline{Q}&\underline{0}\\
\underline{0}&\underline{Q}
\end{bmatrix}
\label{E.8.2}
\end{equation}
konstruieren. F"ur die Zufallsvektoren gilt dann:
\begin{equation}
\Tilde{\Vec{\boldsymbol{V}}}(\mu)\;=\;
\Tilde{\Vec{\boldsymbol{V}}}\big({\T\Tilde{\mu}\!+\!\Hat{\mu}\CdoT\frac{M}{K_H}}\big)\;=\;
\underline{P}^{\uP{0.4}{\,\Hat{\mu}}}\cdot\Tilde{\Vec{\boldsymbol{V}}}(\Tilde{\mu}).
\label{E.8.3}
\end{equation}
Daher sind die Kovarianzmatrizen aller Gleichungssysteme (\ref{E.2.25})
mit Werten von $\mu$, die sich um ein ganzzahliges Vielfaches von
\mbox{$M/K_H$} unterscheiden, ebenfalls permutierte Versionen der
Kovarianzmatrix die man mit \mbox{$0\le\mu<M/K_H$} erh"alt.
\begin{gather}
\underline{C}_{\Tilde{\Vec{\boldsymbol{V}}}(\mu),\Tilde{\Vec{\boldsymbol{V}}}(\mu)}\;=\;
\text{E}\bigg\{\Big(\Tilde{\Vec{\boldsymbol{V}}}(\mu)\!-\!
\text{E}\big\{\Tilde{\Vec{\boldsymbol{V}}}(\mu)\big\}\Big)\CdoT
\Big(\Tilde{\Vec{\boldsymbol{V}}}(\mu)\!-\!
\text{E}\big\{\Tilde{\Vec{\boldsymbol{V}}}(\mu)\big\}\Big)^{\!\HH}\bigg\}\;=
\notag\\[2pt]
=\;\text{E}\bigg\{\Big(\Tilde{\Vec{\boldsymbol{V}}}\big({\T\Tilde{\mu}\!+\!\Hat{\mu}\CdoT\frac{M}{K_H}}\!\big)\!-\!
\text{E}\big\{\Tilde{\Vec{\boldsymbol{V}}}\big({\T\Tilde{\mu}\!+\!\Hat{\mu}\CdoT\frac{M}{K_H}}\!\big)\big\}\Big)\CdoT
\Big(\Tilde{\Vec{\boldsymbol{V}}}\big({\T\Tilde{\mu}\!+\!\Hat{\mu}\CdoT\frac{M}{K_H}}\!\big)\!-\!
\text{E}\big\{\Tilde{\Vec{\boldsymbol{V}}}\big({\T\Tilde{\mu}\!+\!\Hat{\mu}\CdoT\frac{M}{K_H}}\!\big)\big\}\Big)^{\!\HH}\bigg\}\;=
\notag\\[2pt]
=\;\text{E}\bigg\{\Big(\underline{P}^{\uP{0.4}{\,\Hat{\mu}}}\CdoT
\Tilde{\Vec{\boldsymbol{V}}}(\Tilde{\mu})\!-\!
\text{E}\big\{\underline{P}^{\uP{0.4}{\,\Hat{\mu}}}\CdoT
\Tilde{\Vec{\boldsymbol{V}}}(\Tilde{\mu})\big\}\Big)\CdoT
\Big(\underline{P}^{\uP{0.4}{\,\Hat{\mu}}}\CdoT\Tilde{\Vec{\boldsymbol{V}}}(\Tilde{\mu})-
\text{E}\big\{\underline{P}^{\uP{0.4}{\,\Hat{\mu}}}\CdoT
\Tilde{\Vec{\boldsymbol{V}}}(\Tilde{\mu})\big\}\Big)^{\!\HH}\bigg\}\;=
\notag\\[2pt]
=\;\text{E}\bigg\{\,\underline{P}^{\uP{0.4}{\,\Hat{\mu}}}\Cdot
\Big(\Tilde{\Vec{\boldsymbol{V}}}(\Tilde{\mu})\!-\!
\text{E}\big\{\Tilde{\Vec{\boldsymbol{V}}}(\Tilde{\mu})\big\}\Big)\cdot
\Big(\Tilde{\Vec{\boldsymbol{V}}}(\Tilde{\mu})\!-\!
\text{E}\big\{\Tilde{\Vec{\boldsymbol{V}}}(\Tilde{\mu})\big\}\Big)^{\!\HH}
\Cdot\big(\underline{P}^{\uP{0.4}{\,\Hat{\mu}}}\big)^{\HH}\bigg\}\;=
\notag\\[2pt]
=\;\underline{P}^{\uP{0.4}{\,\Hat{\mu}}}\cdot
\text{E}\bigg\{\Big(\Tilde{\Vec{\boldsymbol{V}}}(\Tilde{\mu})\!-\!
\text{E}\big\{\Tilde{\Vec{\boldsymbol{V}}}(\Tilde{\mu})\big\}\Big)\CdoT
\Big(\Tilde{\Vec{\boldsymbol{V}}}(\Tilde{\mu})\!-\!
\text{E}\big\{\Tilde{\Vec{\boldsymbol{V}}}(\Tilde{\mu})\big\}\Big)^{\!\HH}\bigg\}
\cdot\big(\underline{P}^{\uP{0.4}{\,\Hat{\mu}}}\big)^{\HH}\;=
\notag\\[2pt]
=\;\underline{P}^{\uP{0.4}{\,\Hat{\mu}}}\cdot
\underline{C}_{\Tilde{\Vec{\boldsymbol{V}}}(\Tilde{\mu}),\Tilde{\Vec{\boldsymbol{V}}}(\Tilde{\mu})}
\cdot\big(\underline{P}^{\uP{0.4}{\,\Hat{\mu}}}\big)^{\HH}
\label{E.8.4}
\end{gather}
Dieselbe Permutationssymmetrie gilt auch f"ur die Stichprobenmatrizen
\begin{equation}
\Tilde{\underline{V}}(\mu)\;=\;
\Tilde{\underline{V}}\big({\T\Tilde{\mu}\!+\!\Hat{\mu}\CdoT\frac{M}{K_H}}\big)
\;=\;\underline{P}^{\uP{0.4}{\,\Hat{\mu}}}\cdot\Tilde{\underline{V}}(\Tilde{\mu})
\label{E.8.5}
\end{equation}
und die empirischen Kovarianzmatrizen
\begin{equation}
\Hat{\underline{C}}_{\Tilde{\Vec{\boldsymbol{V}}}(\mu),\Tilde{\Vec{\boldsymbol{V}}}(\mu)}\;=\;
\Hat{\underline{C}}_{\Tilde{\Vec{\boldsymbol{V}}}(\Tilde{\mu}+\Hat{\mu}\cdot\frac{M}{K_H}),\Tilde{\Vec{\boldsymbol{V}}}(\Tilde{\mu}+\Hat{\mu}\cdot\frac{M}{K_H})}\;=\;
\underline{P}^{\uP{0.4}{\,\Hat{\mu}}}\Cdot
\Hat{\underline{C}}_{\Tilde{\Vec{\boldsymbol{V}}}(\Tilde{\mu}),\Tilde{\Vec{\boldsymbol{V}}}(\Tilde{\mu})}\cdot
\big(\underline{P}^{\uP{0.4}{\,\Hat{\mu}}}\big)^{\HH}
\label{E.8.6}
\end{equation}
in den L"osungen (\ref{E.3.8}) f"ur die beiden bifrequenten "Ubertragungsfunktionen 
mit Werten von $\mu$, die sich um ein ganzzahliges Vielfaches von \mbox{$M/K_H$} 
unterscheiden. Eine weitere Symmetrie ergibt sich f"ur negative Frequenzen.
Mit der Antidiagonalmatrix
\begin{equation}
\underline{\overline{D}}\;=\;
\begin{bmatrix}
0&0&\cdots&\cdots&0&0&1\\
0&0&\cdots&\cdots&0&1&0\\
\vdots&\vdots&&\adots&1&0&0\\
\vdots&\vdots&\adots&\adots&\adots&\vdots&\vdots\\
0&0&1&\adots&&\vdots&\vdots\\
0&1&0&\cdots&\cdots&0&0\\
1&0&0&\cdots&\cdots&0&0
\end{bmatrix}
\label{E.8.7}
\end{equation}
gilt f"ur die Zufallsvektoren der Erregung und deren Stichprobenmatrizen
\begin{equation}
\Tilde{\Vec{\boldsymbol{V}}}(-\mu)^{\Kk}\,=\;\underline{\overline{D}}\cdot\underline{P}\cdot\Tilde{\Vec{\boldsymbol{V}}}(\mu)
\quad\text{und}\quad
\Tilde{\underline{V}}(-\mu)^{\Kk}\,=\;\underline{\overline{D}}\cdot\underline{P}\cdot\Tilde{\underline{V}}(\mu),
\label{E.8.8}
\end{equation}
und somit ergeben sich die Permutationssymmetrien
\begin{gather}
\underline{C}_{\Tilde{\Vec{\boldsymbol{V}}}(-\mu),\Tilde{\Vec{\boldsymbol{V}}}(-\mu)}^{\;\Kk}\;=\;
\underline{\overline{D}}\cdot\underline{P}\cdot
\underline{C}_{\Tilde{\Vec{\boldsymbol{V}}}(\mu),\Tilde{\Vec{\boldsymbol{V}}}(\mu)}\cdot
\underline{P}^{\Hh}\cdot\underline{\overline{D}}^{\Hh}
\qquad\text{ und}\notag\\[6pt]
\Hat{\underline{C}}_{\Tilde{\Vec{\boldsymbol{V}}}(-\mu),\Tilde{\Vec{\boldsymbol{V}}}(-\mu)}^{\;\Kk}\;=\;
\underline{\overline{D}}\cdot\underline{P}\cdot
\Hat{\underline{C}}_{\Tilde{\Vec{\boldsymbol{V}}}(\mu),\Tilde{\Vec{\boldsymbol{V}}}(\mu)}\cdot
\underline{P}^{\Hh}\cdot\underline{\overline{D}}^{\Hh}
\label{E.8.9}
\end{gather}
f"ur die theoretischen und die empirischen Kovarianzmatrizen. Man braucht
zur Berechnung der beiden bifrequenten "Ubertragungsfunktionen
letztendlich nur etwa \mbox{$M/(2\CdoT K_H)$} empirische Kovarianzmatrizen,
deren Dimension hier jeweils \mbox{$(2\CdoT K_H)\times(2\CdoT K_H)$} ist,
zu invertieren.

Zur Berechnung der Messwerte des LDS und des KLDS gem"a"s der Gleichungen (\ref{E.3.34}) 
ben"otigt man  u.~a. auch die Matrizen \mbox{$\underline{V}_{\bot}\!(\mu)$} nach 
Gleichung (\ref{E.3.28}). Wenn man diese Matrizen nach dem in Kapitel~\ref{E.Kap.3} 
beschriebenen Verfahren konstruiert, so dass die Gleichungen~(\ref{E.3.33}) 
erf"ullt sind, ist es bei der Berechnung der Messwerte erforderlich, 
\vadjust{\penalty-200}die in den Matrizen \mbox{$\underline{V}_{\bot}\!(\mu)$} nach Gleichung 
(\ref{E.3.28}) auftretenden empirischen Kovarianzmatrizen
\mbox{$\Hat{\underline{C}}_{\Breve{\Vec{\boldsymbol{V}}}(\mu),\Breve{\Vec{\boldsymbol{V}}}(\mu)}$}
nach Gleichung (\ref{E.3.27}) zu bestimmen, die alle empirischen 
Kovarianzen eines Satzes von linear unabh"angigen Zufallgr"o"sen 
untereinander enthalten, die aus den Zufallgr"o"sen
\mbox{$\boldsymbol{V}\!\big({\T\mu\!+\!\Breve{\mu}\CdoT\frac{ M}{K_S}}\big)$} und 
\mbox{$\boldsymbol{V}\!\big(\!{\T-\mu\!-\!\Breve{\mu}\CdoT\frac{ M}{K_S}}\big)^{\Kk}$} 
f"ur alle \mbox{$\Breve{\mu}=0\;(1)\;K_S\!-\!1$} entnommen wurden.
Nimmt man an, dass all diese Zufallgr"o"sen linear unabh"angig sind,
so ist $K(\mu)=2\CdoT K_S$ und es sind f"ur jede der Frequenzen
\mbox{$\mu=0\;(1)\;(M\!-\!K_S)/(2\cdoT K_S)$} genau
\mbox{$4\cdoT K_S^2$} empirische Kovarianzen des Spektrums der Erregung zu 
berechnen. Die Kovarianzmatrizen f"ur die restlichen Frequenzen erh"alt man 
dann durch geeignete Permutation aus den Kovarianzmatrizen 
\mbox{$\Hat{\underline{C}}_{\Breve{\Vec{\boldsymbol{V}}}(\mu),\Breve{\Vec{\boldsymbol{V}}}(\mu)}$}
bei den angegebenen Frequenzen. Wie man das \mbox{$L\CdoT(L\!-\!1)$}-fache der
in den Kovarianzmatrizen auftretenden empirischen Kovarianzen mit Hilfe 
der Akkumulatorfelder {\tt Akku\_VQ}, {\tt Akku\_VV} und {\tt Akku\_V} berechnet, 
wurde oben bereits angegeben. Aus den \mbox{$L\CdoT(L\!-\!1)$}-fachen Kovarianzen 
{\tt C\_VQ} und {\tt C\_VV} sucht man sich jeweils die f"ur die Frequenz $\mu$
in der Kovarianzmatrix enthaltenen empirischen Kovarianzen heraus. 

$K_S$ ist als das kleinste gemeinsame Vielfache der beiden Perioden $K_H$ und 
$K_{\Phi}$ definiert. Somit kann die Zahl der Elemente der empirischen Kovarianzmatrix 
gro"s werden, wenn beide Perioden teilerfremd sind. Da die empirische Kovarianzmatrix
invertiert werden muss, wird dann nicht nur der Speicherbedarf, sondern
auch der Rechenaufwand relativ gro"s. Bei vielen periodisch zeitvarianten
Systemen ist die Periode der "uberlagerten zuf"alligen St"orung mit der
Periode des zeitvarianten Systems zyklostation"ar. In diesem Fall h"alt
sich der Speicher- und Rechenaufwand noch in vertretbaren Grenzen.

Zur Berechnung der Messwerte
\mbox{$\Hat{\Phi}_{\boldsymbol{n}}\big({\T\mu,\mu\!+\!\Tilde{\mu}\CdoT\frac{M}{K_{\Phi}}}\big)$} und
\mbox{$\Hat{\Psi}_{\boldsymbol{n}}\big({\T\mu,\mu\!+\!\Tilde{\mu}\CdoT\frac{M}{K_{\Phi}}}\big)$}
ben"otigt man f"ur jede Frequenz $\mu$ alle empirischen
Kreuzkovarianzen der Spektralwerte des gefensterten Ausgangssignals
\mbox{$\Vec{Y}_{\!f}\big({\T\mu\!+\!\Tilde{\mu}\CdoT\frac{M}{K_{\Phi}}}\big)$} 
und der Zufallgr"o"sen
\mbox{$\boldsymbol{V}\!\big({\T\mu\!+\!\Breve{\mu}\CdoT\frac{ M}{K_S}}\big)$} und 
\mbox{$\boldsymbol{V}\!\big(\!{\T-\mu\!-\!\Breve{\mu}\CdoT\frac{ M}{K_S}}\big)^{\Kk}$} 
des Spektrums der Erregung. Wie man das \mbox{$L\CdoT(L\!-\!1)$}-fache dieser empirische
Kovarianzen {\tt C\_YV} und {\tt C\_VY}  mit Hilfe der Akkumulatorfelder {\tt Akku\_YV}, 
{\tt Akku\_VY}, {\tt Akku\_V} und {\tt Y\_mittel} berechnet, ist oben angegeben. Auch 
hier sucht man sich die f"ur die Messwerte der Frequenz $\mu$ jeweils ben"otigten
Kreuzkovarianzen heraus. Desweiteren werden f"ur jede Frequenz $\mu$ noch
Sch"atzwerte f"ur alle Kovarianzen der Zufallsgr"o"sen 
\mbox{$\Vec{Y}_{\!f}\big({\T\mu\!+\!\Tilde{\mu}\CdoT\frac{M}{K_{\Phi}}}\big)$} und 
\mbox{$\Vec{Y}_{\!f}\big(\!{\T-\mu\!-\!\Tilde{\mu}\CdoT\frac{M}{K_{\Phi}}}\big)$} 
untereinander ben"otigt, die man analog f"ur jede Frequenz $\mu$ mit Hilfe der 
Akkumulatorfelder {\tt Akku\_YQ}, {\tt Akku\_YY} und {\tt Y\_mittel} berechnet, und 
deren \mbox{$L\CdoT(L\!-\!1)$}-faches man mit etwa \mbox{$K_{\Phi}$}-fachen Speichbedarf 
auf den Feldern {\tt C\_YQ} und {\tt C\_YY} abspeichert.

Unabh"angig davon, welches Modellsystem man verwendet
(\,zeitinvariant oder periodisch zeitvariant und evtl. das
von dem konjugierten Eingangssignal erregte Modellsystem\,),
sind die f"ur die Berechnung der "Ubertragungsfunktionen und
der deterministischen St"orung sowie ihres Spektrums ben"otigten
Kovarianzen bereits in den Kovarianzen, die man zur
Berechnung des LDS und KLDS braucht, enthalten.

\pagebreak
Es sei noch darauf hingewiesen, dass die Matrixprodukte der
bilinearen Formen, die als Vorfaktoren vor den Messwerten 
\mbox{$\Hat{\Phi}_{\boldsymbol{n}}\big({\T\mu,\mu\!+\!\Tilde{\mu}\CdoT\frac{M}{K_{\Phi}}}\big)$} und
\mbox{$\Hat{\Psi}_{\boldsymbol{n}}\big({\T\mu,\mu\!+\!\Tilde{\mu}\CdoT\frac{M}{K_{\Phi}}}\big)$} 
bei den Messwertkovarianzen der "Ubertragungsfunktionen gem"a"s der Gleichungen~(\ref{3.44})
stehen, sich wegen der obengenannten Permutationseigenschaften ebenfalls durch eine Permutation 
ineinander "uberf"uhren lassen, wenn man $\mu$ um
ein Vielfaches von \mbox{$M/K_H$} erh"oht.
Bei den Messwertkovarianzen des Spektrums der deterministischen St"orung
gem"a"s der Gleichungen~(\ref{E.3.48}) bewirkt einen Verschiebung
von $\mu$ um ein Vielfaches von \mbox{$M/K_H$} gar keine
Ver"anderung der bilinearen Formen vor den Messwerten des LDS und KLDS.
Im Fall, dass $K_H$ ein ganzzahliges Vielfaches von $K_{\Phi}$ ist,
ergeben sich somit Vorfaktoren, die von~$\Tilde{\mu}$ unabh"angig sind.
Bei den Messwertvarianzen ist nach der Permutation auch noch die empirischen 
Kovarianzmatrix in der Mitte des Matrixprodukts in der Gleichung~(\ref{E.3.48.a})
gleich der empirischen Kovarianzmatrix
\mbox{$\Hat{\underline{C}}_{\Tilde{\Vec{\boldsymbol{V}}}(\mu),\Tilde{\Vec{\boldsymbol{V}}}(\mu)}$}, 
so dass diese sich mit einer der beiden inversen Kovarianzmatrizen 
kompensiert. Wenn man das von der konjugierten Erregung gespeiste 
Modellsystem mit der bifrequenten "Ubertragungsfunktion
\mbox{$H_*\!\big({\T \mu,\mu\!+\!\Tilde{\mu}\CdoT\frac{M}{K_H}}\big)$} 
verwendet, erh"alt man nach einer weiteren Permutation bei den Messwertkovarianzen
nach Gleichung~(\ref{E.3.48.b}) dieselben bilinearen Formen wie bei den 
Messwertvarianzen nach Gleichung~(\ref{E.3.48.a}), und sowohl bei den Messwertvarianzen, 
als auch bei den Messwertkovarianzen, kann die eine Inverse wieder gek"urzt werden, so dass
\mbox{$\Hat{\Phi}_{\boldsymbol{n}}\big({\T\mu,\mu\!+\!\Tilde{\mu}\CdoT\frac{M}{K_{\Phi}}}\big)$}
mit demselben Vorfaktor in die Messwertvarianzen eingeht, wie
\mbox{$\Hat{\Psi}_{\boldsymbol{n}}\big({\T\mu,\mu\!+\!\Tilde{\mu}\CdoT\frac{M}{K_{\Phi}}}\big)$}
in die Messwertkovarianzen.

Nun soll angedeutet werden, welche Vereinfachungen an dem im
Kapitel~\ref{E.Kap.7} geschilderten Ablauf zur Berechnung der Messwerte
vorgenommen werden k"onnen, wenn man die reduzierten Systemmodelle
nach Kapitel~\myref{RKM} bis \myref{Resys} und Kapitel \ref{E.Kap.5} 
und \ref{E.Kap.6} verwendet. Wie in Kapitel~\ref{E.Kap.7} beschr"anken wir 
uns auch hier auf den Fall zeitinvarianter Modellsysteme ${\cal S}_{lin}$ 
und ${\cal S}_{*,lin}$ mit \mbox{$K_H\!=\!1$} und nehmen an, dass ein station"arer 
Approximationsfehlerprozess mit \mbox{$K_{\Phi}\!=\!1$} vorliegt.
 
Wenn man davon ausgehen kann, dass ein mittelwertfreier Approximationsfehler 
vorliegt, kann man auf die Modellierung der deterministischen St"orung 
verzichten, wie dies in \cite{Diss} geschehen ist. Ein erstes Moment 
im Ausgangssignal kann unter dieser Annahme nur durch ein durch das lineare 
Modellsystem verzerrtes erstes Moment der Erregung verursacht sein. 
Da das erste Moment der Erregung in derselben Art verzerrt wird, wie 
die in den Stichprobenelementen enthaltenen zuf"alligen Abweichungen 
von diesem ersten Moment, braucht man den Mittelwert des Ausgangssignals 
nun nicht getrennt behandeln.  Die Akkumulatorfelder {\tt Akku\_y} und
{\tt Akku\_V}, die bisher zur Bestimmung der empirischen Mittelwerte
am Systemein- und -ausgang verwendet wurden, werden daher nicht mehr
ben"otigt, so dass sich ein gegen"uber dem vollst"andigen Systemmodell
reduzierter Rechenaufwand ergibt. Zur Berechnung der Messwert und der
Sch"atzwerte der Messwert"-(ko)"-varianzen werden bei dem reduzierten
Systemmodell ohne deterministische St"orung die Akkumulatorfelder

\[\begin{array}{rccl}
\text{\tt Akku\_VQ}&\text{ mit }&M       &\text{ Speicherpl"atzen,}\\
\text{\tt Akku\_VV}&\text{ mit }&M\!+\!2 &\text{ Speicherpl"atzen,}\\
\text{\tt Akku\_YQ}&\text{ mit }&M       &\text{ Speicherpl"atzen,}\\
\text{\tt Akku\_YY}&\text{ mit }&M\!+\!2 &\text{ Speicherpl"atzen,}\\
\text{\tt Akku\_YV}&\text{ mit }&2\CdoT M&\text{ Speicherpl"atzen und}\\
\text{\tt Akku\_VY}&\text{ mit }&2\CdoT M&\text{ Speicherpl"atzen}
\end{array}\]
ben"otigt. Das $L$-fache dieser Akkumulatoren tritt in allen Berechnungen
nun an die Stelle der $L\CdoT(L\!-\!1)$-fachen Kovarianzen und
Kovarianzmatrizen. Die Punkte \ref{E.I.7.11}, \ref{E.I.7.14}, \ref{E.I.7.21}, \ref{E.I.7.22}, \ref{E.I.7.25} 
-- \ref{E.I.7.29}, \ref{E.I.7.31} -- \ref{E.I.7.34}, \ref{E.I.7.41} -- \ref{E.I.7.44}, 
\ref{E.I.7.51} und \ref{E.I.7.52} entfallen ersatzlos. Bei der Berechnung der Vorfaktoren der 
Messwerte \mbox{$\Hat{\Phi}_{\boldsymbol{n}}(\mu)$} und \mbox{$\Hat{\Psi}_{\boldsymbol{n}}(\mu)$} 
in den Punkten \ref{E.I.7.35} und \ref{E.I.7.36} ist jeweils ein um $1$ 
erh"ohter Matrixrang zu ber"ucksichtigen, indem man dort im Nenner den Faktor 
$(L\!-\!3)$ durch $(L\!-\!2)$ ersetzt. Auch bei der Berechnung der Varianzen und Kovarianzen 
dieser Messwerte in den Punkten \ref{E.I.7.45} bis \ref{E.I.7.47} ist der erh"ohte 
Matrixrang zu ber"ucksichtigen, indem man dort jeweils $L$ durch $L\!+\!1$ ersetzt.

In \cite{Diss} wurde nur das eine zeitinvariante Modellsystem ${\cal S}_{lin}$ 
angesetzt. Auch bei dieser Variante ben"otigt man alle eben genannten Akkumulatorfelder.
Die Berechnung der "Ubertragungsfunktion in Punkt \ref{E.I.7.30} erfolgt hier jedoch 
nach Gleichung (\myref{3.14}):
\[\text{\tt H}(\mu)\;=\;\text{\tt Akku\_YV}(\mu) / \text{\tt Akku\_VQ}(\mu).\]
In Punkt \ref{E.I.7.37} ergibt sich mit Gleichung (\myref{3.59}) die Messwertvarianz
\[\text{\tt C\_HQ}(\mu) \;=\; M/\text{\tt Akku\_VQ}(\mu)\cdot\text{\tt Phi}(\mu),\]
und mit Gleichung (\myref{3.64}) in Punkt \ref{E.I.7.39} die Messwertkovarianz
\[\text{\tt C\_HH}(\mu) \;=\; M\cdot\text{\tt Akku\_VQ}(\mu)^{\Kk}/\text{\tt Akku\_VQ}(\mu)^2\Cdot\text{\tt Psi}(\mu).\]
Abgesehen davon, dass die Punkte \ref{E.I.7.38}, \ref{E.I.7.40} und \ref{E.I.7.50} nun entfallen, bleibt die 
Berechnung der weiteren Messwerte und ihrer Varianzen und Kovarianzen unver"andert, wobei 
die Erh"ohung des Matrixranges, wie eben beschrieben, zu ber"ucksichtigen ist.

Bei dem Spektralsch"atzverfahren komplexer Prozesse in Kapitel \ref{E.Kap.5}, bei dem 
das erste und das zweite zentrale Moment getrennt berechnet werden, wird kein lineares 
Modellsystem angesetzt. Zur Berechnung der Messwerte und der Sch"atzwerte
der Messwert"-(ko)"-varianzen werden dann nur die Akkumulatorfelder
\[\begin{array}{rccl}
\text{\tt Akku\_y} &\text{ mit }&2\CdoT F&\text{ Speicherpl"atzen,}\\
\text{\tt Akku\_YQ}&\text{ mit }&M       &\text{ Speicherpl"atzen und}\\
\text{\tt Akku\_YY}&\text{ mit }&M\!+\!2 &\text{ Speicherpl"atzen}
\end{array}\]
ben"otigt. Die anderen Akkumulatorfelder werden nicht ben"otigt,
da diese zur Akkumulation von Produkten ben"otigt wurden, die Faktoren
enthielten, die von der Erregung abh"angig waren. Im Fall der
reinen Spektralsch"atzung gibt es entweder keine Erregung, oder sie
wird gefiltert dem zu vermessenden Zufallsprozess zugeschlagen.
Das erste Moment ist die deterministische St"orung, die sich 
nach Gleichung (\ref{E.5.4}) in Punkt \ref{E.I.7.34} nun ohne 
$\text{\tt x\_mittel}(k\!-\!\kappa\CdoT M)\,\big)$ zu 
\[\text{\tt u}(k)\;=\;\text{\tt Akku\_y}(k)/L\]
berechnet. In Punkt \ref{E.I.7.32} wird deren Spektrum analog mit Gleichung (\ref{E.5.5}) als
\[\text{\tt U}(\mu)\;=\;\text{\tt Y\_mittel}(\mu)/L\]
berechnet. Die Varianzen und Kovarianzen des Spektrums der deterministischen St"orung
werden in den Punkten \ref{E.I.7.41} und \ref{E.I.7.42} nun mit den Gleichungen (\ref{E.5.8}) 
wesentlich einfacher als 
\[\text{\tt C\_UQ}(\mu)\;=\;\frac{M}{L}\cdot\text{\tt Phi}(\mu)
\qquad\text{ und }\qquad
\text{\tt C\_UU}(\mu)\;=\;\frac{M}{L}\cdot\text{\tt Psi}(\mu)\]
berechnet. Die Varianzen und Kovarianzen der deterministischen St"orung selbst werden unver"andert 
in den Punkten \ref{E.I.7.43} und \ref{E.I.7.44} bestimmt.
Die Berechnung der Messwerte \mbox{$\Hat{\Phi}_{\boldsymbol{n}}(\mu)$} und 
\mbox{$\Hat{\Psi}_{\boldsymbol{n}}(\mu)$} in den Punkten \ref{E.I.7.35} und \ref{E.I.7.36}
erfolgt hier gem"a"s der Gleichungen~(\ref{E.5.6}):
\[\text{\tt Phi}(\mu)\;=\;\frac{\;\text{\tt C\_YQ}(\mu)}{M\cdot L\CdoT(L\!-\!1)}
\qquad\text{ und }\qquad
\text{\tt Psi}(\mu)\;=\;\frac{\;\text{\tt C\_YY}(\mu)}{M\cdot L\CdoT(L\!-\!1)}.\]
Bei der Berechnung der Varianzen und Kovarianzen dieser Messwerte in den Punkten 
\ref{E.I.7.45} bis \ref{E.I.7.47} ist nun ein um $2$ erh"ohter Matrixrang zu 
ber"ucksichtigen. Indem man dort jeweils $L$ durch $L\!+\!2$ ersetzt erh"alt man die 
in den Gleichungen~(\ref{E.5.13}) angegebenen Werte.

Wird bei der Spektralsch"atzung eines komplexen Prozesses \mbox{$\boldsymbol{y}(k)$} 
das erste Moment nicht separat gemessen, so wird auch der Akkumulator {\tt Akku\_y}
nicht ben"otigt, da dieser ja f"ur die Bestimmung des empirischen
Mittelwertes vorgesehen ist. Man ersetzt die Punkte \ref{E.I.7.35} und \ref{E.I.7.36}
durch die in den Gleichungen (\myref{5.1}) und (\myref{5.2}) angegebenen Terme 
und erh"alt:
\[\text{\tt Phi}(\mu)\;=\;\frac{\;\text{\tt Akku\_YQ}(\mu)}{M\cdot L}
\qquad\text{ und }\qquad
\text{\tt Psi}(\mu)\;=\;\frac{\;\text{\tt Akku\_YY}(\mu)}{M\cdot L}.\]
Da hier nun ein um $3$ erh"ohter Matrixrang zu ber"ucksichtigen ist, ist bei der Berechnung 
der Varianzen und Kovarianzen dieser Messwerte in den Punkten \ref{E.I.7.45} bis \ref{E.I.7.47} 
jeweils $L$ durch $L\!+\!3$ zu ersetzen, um die in den Gleichungen~(\myref{5.4}) angegebenen Werte 
zu erhalten.

Bei Verwendung eines Mehrtonsignals ergibt sich
der Vorteil, dass die Summe "uber alle Betragsquadratspektren
des Eingangssignals aller Einzelmessungen, die sonst mit
Hilfe des Akkumulators {\tt Akku\_VQ} berechnet werden muss,
entfallen kann, da diese dann das $L$-fache des bei allen 
Einzelmessungen konstanten Betragsquadrats des erregenden Spektrums ist.
Bei Verwendung des Chirpsignals ergibt sich hinsichtlich des Rechenaufwands
nur dann ein weiterer wesentlicher Vorteil, wenn man lediglich
\mbox{$\Hat{H}(\mu)$}, \mbox{$\Hat{H}_*(\mu)$}, \mbox{$\Hat{U}_{\!f}(\mu)$} 
und \mbox{$\Hat{u}(k)$} messen will. Dann werden zum Einen statt des 
Akkumulators {\tt Akku\_V} mit \mbox{$2\CdoT M$} Speicherpl"atzen nur 
zwei Speicherpl"atze f"ur die Akkumulation des Real- und Imagin"arteils 
des zuf"alligen Drehfaktors ben"otigt. Zum Anderen, wird man auf den Akkumulatoren
{\tt Akku\_YV} und {\tt Akku\_VY} nicht die in Kapitel~\ref{E.Kap.7} angegebenen
Spektralwertprodukte aufsummieren, sondern stattdessen jeweils die ebenfalls 
$M$ Werte \mbox{$y_{f,\lambda}(k)$} des gefensterten und blockweise
"uberlagerten Zeitsignals der Einzelmessungen, die man mit dem zuf"alligen
Drehfaktor \mbox{$e^{-j\cdot\Hat{\varphi}_{\lambda}}$} 
bzw. \mbox{$e^{j\cdot\Hat{\varphi}_{\lambda}}$} dieser Einzelmessung
multipliziert. Die beiden f"ur die weitere Berechnung der Messwerte notwendigen
DFTs werden dann nur einmal am Ende der Messschleife (\,Punkt~\ref{E.I.7.7}
bis \ref{E.I.7.18} in Kapitel~\ref{E.Kap.7}\,) durchgef"uhrt. Das Ergebnis
der einen DFT wird man anschlie"send mit dem Spektrum des konstanten,
nicht zuf"alligen Spektralanteils des Chirpsignals multiplizieren, und das Ergebnis
der anderen DFT mit dem konjugierten Spektrum. Wenn man jedoch das LDS, das KLDS 
oder die Messwertvarianzen und -kovarianzen bestimmen will, ben"otigt man auch 
die Akkumulatorfelder {\tt Akku\_YQ} und {\tt Akku\_YY}. Da zu deren Berechnung 
bei jeder Einzelmessung die DFT notwendig ist, die eben vermieden werden sollte,
bringt in diesem Fall die Verwendung des Chirpsignals --- abgesehen von dem guten 
Crest-Faktor --- keine Vorteile.

Die im weiteren dargestellten Modifikationen des Messablaufs nach
Kapitel~\ref{E.Kap.7} ergeben sich bei der Messung eines reellwertigen
Systems, bzw. bei der Spektralsch"atzung reeller Prozesse, wobei auch f"ur
die reduzierten Modellsysteme kurz auf die Besonderheiten, die sich
bei der Berechnung der Messwerte reeller Systeme und Signale ergeben,
eingegangen werden soll.

Zur Berechnung der Messwert werden bei einem vollst"andigen
reellwertigen Modellsystem die Akkumulatorfelder
\[\begin{array}{rccl}
\text{\tt Akku\_y} &\text{ mit }&F    &\text{ Speicherpl"atzen,}\\
\text{\tt Akku\_V} &\text{ mit }&M    &\text{ Speicherpl"atzen,}\\
\text{\tt Akku\_VQ}&\text{ mit }&M/2+1&\text{ Speicherpl"atzen,}\\
\text{\tt Akku\_YQ}&\text{ mit }&M/2+1&\text{ Speicherpl"atzen und}\\
\text{\tt Akku\_YV}&\text{ mit }&M    &\text{ Speicherpl"atzen}
\end{array}\]
ben"otigt. Dabei wurde ber"ucksichtigt, dass die Akkumulatorfelder
{\tt Akku\_VV}, {\tt Akku\_YY} und {\tt Akku\_VY} nicht ben"otigt werden.
Diese w"urden n"amlich denselben Inhalt wie die  Akkumulatorfelder
{\tt Akku\_VQ}, {\tt Akku\_YQ} und {\tt Akku\_YV} enthalten.
Bei der Berechnung des Speicherbedarfs der Akkumulatorfelder wurde
die Symmetrie und die evtl. vorhandene Reellwertigkeit der Spektralwerte
und Zeitsignale mit einkalkuliert. Im Messablaufs nach Kapitel~\ref{E.Kap.7}
brauchen die Punkte \ref{E.I.7.13}, \ref{E.I.7.16}, \ref{E.I.7.18}, 
\ref{E.I.7.22} -- \ref{E.I.7.24}, \ref{E.I.7.27}, \ref{E.I.7.29}, \ref{E.I.7.36}, 
\ref{E.I.7.38} -- \ref{E.I.7.40}, \ref{E.I.7.42}, \ref{E.I.7.44}, \ref{E.I.7.46}, 
\ref{E.I.7.47}, \ref{E.I.7.50} und \ref{E.I.7.54} nicht berechnet zu werden. Die Werte der 
"Ubertragungsfunktion berechnen sich in Punkt \ref{E.I.7.30} nun mit Gleichung 
(\ref{E.6.8}) zu:
\[\text{\tt H}(\mu)\;=\;\text{\tt C\_YV}(\mu) / \text{\tt C\_VQ}(\mu).\]
F"ur den $L$-fachen Mittelwert des Spektrums am Ausgang des linearen 
Modellsystems ist in Punkt \ref{E.I.7.31} 
\[\text{\tt X\_mittel}(\mu)\;=\;\text{\tt H}(\mu)\cdot\text{\tt Akku\_V}(\mu)\]
einzusetzen. Bei der Berechnung der Messwerte \mbox{$\Hat{\Phi}_{\boldsymbol{n}}(\mu)$}
ist nun der Matrixrang $L\!-\!2$ zu zu verwenden, und somit erh"alt man mit Gleichung 
(\ref{E.6.17}) in Punkt \ref{E.I.7.35} die Messwerte:
\[\text{\tt Phi}(\mu)\;=\;\frac{\;\text{\tt C\_YQ}(\mu)-|\text{\tt C\_YV}(\mu)|^2/\text{\tt C\_VQ}(\mu)}{M\cdot L\CdoT(L\!-\!2)}.\]
In Punkt \ref{E.I.7.37} ergibt sich nach Gleichung (\ref{E.6.19}) f"ur die Varianz der Messwerte 
der "Ubertragungsfunktion:
\[\text{\tt C\_HQ}(\mu)\;=\;M\CdoT L/\text{\tt C\_VQ}(\mu)\cdot\text{\tt Phi}(\mu),\]
und mit Gleichung (\ref{E.6.21}) wird in Punkt \ref{E.I.7.41} die Varianz der Messwerte
des Spektrums der deterministischen St"orung zu:
\[\text{\tt C\_UQ}(\mu)\;=\;M\cdot\text{\tt Akku\_VQ}(\mu)/\text{\tt C\_VQ}(\mu)\cdot\text{\tt Phi}(\mu).\]
Die Varianz der Messwerte des LDS wird mit Gleichung (\ref{E.6.25}) in Punkt \ref{E.I.7.45} zu 
\begin{gather*}
\text{\tt C\_PhiQ}(\mu)\;=\;\frac{2}{L}\cdot\text{\tt Phi}(\mu)^2
\qquad\text{ f"ur }{\T\quad\mu\in\big\{0;\frac{M}{2}\big\}}\\
\text{ und }\qquad
\text{\tt C\_PhiQ}(\mu)\;=\;\frac{1}{L\!-\!1}\cdot\text{\tt Phi}(\mu)^2\qquad\text{ sonst}
\end{gather*}
abgesch"atzt. Da die Werte der "Ubertragungsfunktion f"ur die beiden Frequenzen 
\mbox{$\mu=0$} und \mbox{$\mu=\frac{M}{2}$} immer reell sind, ergeben sich hier keine Konfidenzellipsen 
sondern Konfidenzintervalle deren Breite man analog zu Gleichung~(\myref{3.72}) in Punkt \ref{E.I.7.49} als
\[\text{\tt A\_H}(\mu)\;=\;\sqrt{2\;}\cdot\sqrt{\:\text{\tt C\_HQ}(\mu)\:}\cdot\text{erfc}^{-1}(\alpha)
\qquad\text{ f"ur }{\T\quad\mu\in\big\{0;\frac{M}{2}\big\}}\]
angibt. F"ur dieselben Frequenzen sind auch die Spektralwerte der deterministischen St"orung immer reell. 
Daher werden in Punkt \ref{E.I.7.51} die Breiten der Konfidenzintervalle mit
\[\text{\tt A\_U}(\mu)\;=\;\sqrt{2\;}\cdot\sqrt{\:\text{\tt C\_UQ}(\mu)\:}\cdot\text{erfc}^{-1}(\alpha)
\qquad\text{ f"ur }{\T\quad\mu\in\big\{0;\frac{M}{2}\big\}}\]
abgesch"atzt. Auch f"ur die reelle deterministischen St"orung werden in Punkt \ref{E.I.7.52} 
die zeit"-un"-ab"-h"an"-gigen Konfidenzintervalle
\[\text{\tt A\_u}(\mu)\;=\;\sqrt{2\;}\cdot\sqrt{\:\text{\tt C\_uQ}(\mu)\:}\cdot\text{erfc}^{-1}(\alpha)\]
angegeben. Die Konfidenzintervalle der LDS-Messwerte in Punkt \ref{E.I.7.53} bleiben unver"andert.
Alle Punkte der Messablaufsliste, die zur Berechnung solcher Werte
dienen, die durch eine einfache Symmetrie auf bereits berechnete Werte
zur"uckf"uhrbar sind, kann man "uberspringen. Sollten diese dann nicht
berechneten Werte in weiteren Kalkulationen ben"otigt werden, so sind
sie durch die entsprechenden symmetrischen Werte zu ersetzen.

Bei komplexwertigen Systemen ist bei jeder Einzelmessung
jeweils eine DFT f"ur das gefensterte Systemein- und -ausgangssignal
durchzuf"uhren. Da diese Signale nun reell sind, kann man die beiden
DFTs zweier aufeinanderfolgender Einzelmessungen jeweils zu einer
DFT zusammenfassen. F"ur das Systemausgangssignal sei dies kurz
erl"autert. Aus den beiden reellen gefensterten Signalen zweier
aufeinanderfolgender Einzelmessungen wird das komplexe Signal
\begin{equation}
\Tilde{y}_{f,\lambda}(k)\;=\;
y_{f,\lambda-1}(k)+j\cdot y_{f,\lambda}(k)
\qquad\qquad\forall\qquad\lambda=2\;(2)\;L
\label{E.8.10}
\end{equation}
gebildet und einer DFT unterworfen, so dass man
\mbox{$\Tilde{Y}_{f,\lambda}(\mu)$} erh"alt. Mit Hilfe der
Symmetrieeigenschaften der Spektren reeller Signale kann man die
Spektren der einzelnen Anteile nach der DFT des komplexen Signals wieder
trennen.
\begin{align}
Y_{\!f,\lambda-1}(\mu)&=\;
\frac{\Tilde{Y}_{f,\lambda}(\mu)+\Tilde{Y}_{f,\lambda}(\!-\mu)^{\Kk}}{2}
\notag\\[4pt]
Y_{\!f,\lambda}(\mu)\;&=\;
\frac{\Tilde{Y}_{f,\lambda}(\mu)-\Tilde{Y}_{f,\lambda}(\!-\mu)^{\Kk}}{2\cdot j}
\notag\\[4pt]
\label{E.8.11}
\forall\qquad&\qquad\lambda=2\,(2)\,L
\end{align}
Will man sich die Division durch $2$ ersparen, so verwendet man einfach die
halbierte Fensterfolge. Die Division durch $j$ ist keine echte komplexe
Division. Sie macht lediglich aus dem Imagin"arteil den Realteil und
aus dem Realteil den negativen Imagin"arteil. Insgesamt kann man sich so die
H"alfte aller bei der DFT auftretenden komplexen Multiplikationen sparen.

Wenn man bei einem reellwertigen System die deterministische St"orung
nicht modelliert, werden zur Berechnung der Messwert die Akkumulatorfelder
{\tt Akku\_y} und {\tt Akku\_V} nicht ben"otigt, die f"ur die Bestimmung
der empirischen Mittelwerte der Spektren der Signale am Systemein- und
-ausgang vorgesehen waren. Da wir jetzt annehmen, dass das erste
Moment des Approximationsfehlers null ist, gehen wir davon aus, dass
ein erstes Moment im Ausgangssignal durch ein erstes Moment in der
Erregung des linearen Modellsystems verursacht wird. Da das erste Moment
der Erregung durch das lineare Modellsystem genauso verzerrt wird wie
die zuf"alligen Anteile der Erregung, braucht man den Mittelwert des
Ausgangssignals nicht separat behandeln. Zur Berechnung der Messwert und der
Sch"atzwerte der Messwert"-(ko)"-varianzen werden bei dem reduzierten
Systemmodell ohne deterministische St"orung nur  die Akkumulatorfelder
\[\begin{array}{rccl}
\text{\tt Akku\_VQ}&\text{ mit }&M/2+1&\text{ Speicherpl"atzen,}\\
\text{\tt Akku\_YQ}&\text{ mit }&M/2+1&\text{ Speicherpl"atzen und}\\
\text{\tt Akku\_YV}&\text{ mit }&M    &\text{ Speicherpl"atzen}
\end{array}\]
ben"otigt. Das $L$-fache dieser Akkumulatoren tritt in allen Berechnungen
nun an die Stelle der $L\CdoT(L\!-\!1)$-fachen Kovarianzen und
Kovarianzmatrizen. Zus"atzlich entfallen hier noch die Punkte \ref{E.I.7.11}, \ref{E.I.7.14}, 
\ref{E.I.7.21}, \ref{E.I.7.25}, \ref{E.I.7.26}, \ref{E.I.7.28}, \ref{E.I.7.31} -- \ref{E.I.7.34}, \ref{E.I.7.41}, \ref{E.I.7.43}, 
\ref{E.I.7.51} und \ref{E.I.7.52} ersatzlos. Bei der Berechnung der Vorfaktoren der 
Messwerte \mbox{$\Hat{\Phi}_{\boldsymbol{n}}(\mu)$} in Punkt \ref{E.I.7.35} ist 
gegen"uber dem Fall, dass bei einem reellen System die deterministische St"orung modelliert wird, jeweils ein um $1$ 
erh"ohter Matrixrang zu ber"ucksichtigen, indem man dort im Nenner den Faktor 
$(L\!-\!2)$ durch $(L\!-\!1)$ ersetzt. So erh"alt man die in Gleichung~(\myref{4.8})
angegebene Messwerte
\[\text{\tt Phi}(\mu)\;=\;\frac{\;\text{\tt Akku\_YQ}(\mu)-|\text{\tt Akku\_YV}(\mu)|^2/\text{\tt Akku\_VQ}(\mu)}{M\cdot (L\!-\!1)}.\]
Auch bei der Berechnung der Varianzen dieser Messwerte in Punkt \ref{E.I.7.45} ist der erh"ohte 
Matrixrang zu ber"ucksichtigen, indem man dort jeweils $L$ durch $L\!+\!1$ ersetzt.
Die f"uhrt zu den in den Gleichungen~(\myref{4.10}) genannten Werten
\begin{gather*}
\text{\tt C\_PhiQ}(\mu)\;=\;\frac{2}{L\!+\!1}\cdot\text{\tt Phi}(\mu)^2
\qquad\text{ f"ur }{\T\quad\mu\in\big\{0;\frac{M}{2}\big\}}\\
\text{ und }\qquad
\text{\tt C\_PhiQ}(\mu)\;=\;\frac{1}{L}\cdot\text{\tt Phi}(\mu)^2\qquad\text{ sonst.}
\end{gather*}
Da die Werte der "Ubertragungsfunktion f"ur die beiden Frequenzen 
\mbox{$\mu=0$} und \mbox{$\mu=\frac{M}{2}$} immer reell sind, ergeben sich hier keine Konfidenzellipsen 
sondern Konfidenzintervalle deren Breite man analog zu Gleichung~(\myref{3.72}) in Punkt \ref{E.I.7.49} als
\[\text{\tt A\_H}(\mu)\;=\;\sqrt{2\;}\cdot\sqrt{\:\text{\tt C\_HQ}(\mu)\:}\cdot\text{erfc}^{-1}(\alpha)
\qquad\text{ f"ur }{\T\quad\mu\in\big\{0;\frac{M}{2}\big\}}\]
angibt. Die Konfidenzintervalle der LDS-Messwerte in Punkt \ref{E.I.7.53} bleiben unver"andert.
Auch hier kann man jeweils die DFT der Signale zweier Einzelmessungen
gem"a"s Gleichung~(\ref{E.8.10}) zu einer DFT zusammenfassen, und mit
Gleichung~(\ref{E.8.11}) die Spektren der Signale der Einzelmessungen
wieder trennen.

Zur Berechnung der Messwerte werden bei der Spektralsch"atzung
reeller Prozesse mit getrennter Messung des ersten und des
zweiten zentralen Moments nur die zwei Akkumulatorfelder
\[\begin{array}{rccl}
\text{\tt Akku\_y} &\text{ mit }&F    &\text{ Speicherpl"atzen und}\\
\text{\tt Akku\_YQ}&\text{ mit }&M/2+1&\text{ Speicherpl"atzen}
\end{array}\]
ben"otigt, die man wie in den Punkten \ref{E.I.7.14} und \ref{E.I.7.15}
angegeben akkumuliert. Weiterhin unver"andert sind noch die Punkte 
\ref{E.I.7.25} und \ref{E.I.7.26} zu berechnenden. Punkt \ref{E.I.7.32}
liefert mit Gleichung (\ref{E.5.5}) das Spektrum der deterministischen St"orung als
\[\text{\tt U}(\mu)\;=\;\text{\tt Y\_mittel}(\mu)/L.\]
Die Varianzen des Spektrums der deterministischen St"orung
werden in Punkt \ref{E.I.7.41} nun mit Gleichung (\ref{E.5.8.a}) 
wesentlich einfacher als 
\[\text{\tt C\_UQ}(\mu)\;=\;\frac{M}{L}\cdot\text{\tt Phi}(\mu)\]
berechnet. Die zeitunabh"angige Varianz der deterministischen St"orung selbst wird unver"andert 
in Punkt \ref{E.I.7.43} bestimmt.
Die Berechnung der Messwerte \mbox{$\Hat{\Phi}_{\boldsymbol{n}}(\mu)$} in Punkt \ref{E.I.7.35} 
erfolgt hier gem"a"s der Gleichung~(\ref{E.5.6.a}):
\[\text{\tt Phi}(\mu)\;=\;\frac{\;\text{\tt C\_YQ}(\mu)}{M\cdot L\CdoT(L\!-\!1)}.\]
Bei der Berechnung der Varianzen dieser Messwerte in Punkt \ref{E.I.7.45} f"uhrt der 
ge"anderte Matrixrang zu den in Gleichung (\ref{E.5.18}) angegebenen Werten 
\begin{gather*}
\text{\tt C\_PhiQ}(\mu)\;=\;\frac{2}{L\!+\!1}\cdot\text{\tt Phi}(\mu)^2
\qquad\text{ f"ur }{\T\quad\mu\in\big\{0;\frac{M}{2}\big\}}\\
\text{ und }\qquad
\text{\tt C\_PhiQ}(\mu)\;=\;\frac{1}{L}\cdot\text{\tt Phi}(\mu)^2\qquad\text{ sonst.}
\end{gather*}
F"ur die beiden Frequenzen \mbox{$\mu=0$} und \mbox{$\mu=\frac{M}{2}$} sind die Spektralwerte der 
deterministischen St"orung immer reell. Daher werden in Punkt \ref{E.I.7.51} die Breiten der Konfidenzintervalle mit
\[\text{\tt A\_U}(\mu)\;=\;\sqrt{2\;}\cdot\sqrt{\:\text{\tt C\_UQ}(\mu)\:}\cdot\text{erfc}^{-1}(\alpha)
\qquad\text{ f"ur }{\T\quad\mu\in\big\{0;\frac{M}{2}\big\}}\]
abgesch"atzt. Auch f"ur die reelle deterministischen St"orung werden in Punkt \ref{E.I.7.52} 
die zeitunabh"angigen Konfidenzintervalle
\[\text{\tt A\_u}(\mu)\;=\;\sqrt{2\;}\cdot\sqrt{\:\text{\tt C\_uQ}(\mu)\:}\cdot\text{erfc}^{-1}(\alpha)\]
angegeben. Die Konfidenzintervalle der LDS-Messwerte in Punkt \ref{E.I.7.53} bleiben unver"andert.
Auch hier kann man jeweils die DFT der Signale zweier Einzelmessungen
gem"a"s Gleichung~(\ref{E.8.10}) zu einer DFT zusammenfassen, und mit
Gleichung~(\ref{E.8.11}) die Spektren der Signale der Einzelmessungen
wieder trennen.

Verzichtet man bei der Spektralsch"atzung reeller Prozesse auf die Messung des 
ersten Momentes des Prozesses \mbox{$\boldsymbol{y}(k)$}, weil man von diesem
wei"s, dass er mittelwertfrei ist, so braucht man nur das eine
Akkumulatorfeld {\tt Akku\_YQ}, da das Akkumulatorfeld {\tt Akku\_y}
f"ur die Bestimmung des empirischen Mittelwertes vorgesehen ist,
der nun mit null abgesch"atzt wird.
Die Berechnung der Messwerte \mbox{$\Hat{\Phi}_{\boldsymbol{n}}(\mu)$} in Punkt \ref{E.I.7.35} 
erfolgt hier gem"a"s der Gleichung~(\myref{5.1}):
\[\text{\tt Phi}(\mu)\;=\;\frac{\;\text{\tt Akku\_YQ}(\mu)}{M\cdot L}.\]
Bei der Berechnung der Varianzen dieser Messwerte in Punkt \ref{E.I.7.45} f"uhrt der 
ge"anderte Matrixrang zu den in Gleichung (\myref{5.7}) angegebenen Werten 
\begin{gather*}
\text{\tt C\_PhiQ}(\mu)\;=\;\frac{2}{L\!+\!2}\cdot\text{\tt Phi}(\mu)^2
\qquad\text{ f"ur }{\T\quad\mu\in\big\{0;\frac{M}{2}\big\}}\\
\text{ und }\qquad
\text{\tt C\_PhiQ}(\mu)\;=\;\frac{1}{L\!+\!1}\cdot\text{\tt Phi}(\mu)^2\qquad\text{ sonst.}
\end{gather*}
Die Konfidenzintervalle der LDS-Messwerte in Punkt \ref{E.I.7.53} bleiben unver"andert.
Auch hier kann man jeweils die DFT der Signale zweier Einzelmessungen
gem"a"s Gleichung~(\ref{E.8.10}) zu einer DFT zusammenfassen, und mit
Gleichung~(\ref{E.8.11}) die Spektren der Signale der Einzelmessungen
wieder trennen.

\chapter{Beispiele f"ur RKM Messergebnisse}\label{E.Kap.9}

\section[Varianz und Kovarianz der Messwerte der deterministischen St"orung]{Varianz und Kovarianz der Messwerte \\der deterministischen St"orung}\label{E.Kap.9.1}

\begin{figure}[btp]
\begin{center}
{ 
\begin{picture}(454,599)

\input{mbild5c1}
\put(25,540){\makebox(0,0)[r]{\small$0,001$}}
\put(25,502){\makebox(0,0)[rb]{\small$0$}}
\put(25,460){\makebox(0,0)[r]{\small$-0,001$}}
\put(32,487){\makebox(0,0)[bl]{\small$0$}}
\put(110,487){\makebox(0,0)[b]{\small$50$}}
\put(190,487){\makebox(0,0)[b]{\small$100$}}
\put(270,487){\makebox(0,0)[b]{\small$150$}}
\put(350,487){\makebox(0,0)[b]{\small$200$}}
\put(430,487){\makebox(0,0)[b]{\small$250$}}
\put(455,495){\makebox(0,0)[tr]{$k$}}
\put(0,590){\makebox(0,0)[l]{
$\Hat{C}_{\Hat{\boldsymbol{u}}(k),\Hat{\boldsymbol{u}}(k),\text{gemessen}}\,;\;
\Hat{C}_{\Hat{\boldsymbol{u}}(k),\Hat{\boldsymbol{u}}(k),\text{N"aherung, gemittelt}}$}}

\input{mbild5c2}
\put(25,340){\makebox(0,0)[r]{\small$0,001$}}
\put(25,302){\makebox(0,0)[rb]{\small$0$}}
\put(25,260){\makebox(0,0)[r]{\small$-0,001$}}
\put(32,289){\makebox(0,0)[bl]{\small$0$}}
\put(110,289){\makebox(0,0)[b]{\small$50$}}
\put(190,289){\makebox(0,0)[b]{\small$100$}}
\put(270,289){\makebox(0,0)[b]{\small$150$}}
\put(350,289){\makebox(0,0)[b]{\small$200$}}
\put(430,289){\makebox(0,0)[b]{\small$250$}}
\put(455,295){\makebox(0,0)[tr]{$k$}}
\put(0,390){\makebox(0,0)[l]{
$\Re\big\{\Hat{C}_{\Hat{\boldsymbol{u}}(k),\Hat{\boldsymbol{u}}(k)^*,\text{gemessen}}\big\};\;
\Re\big\{\Hat{C}_{\Hat{\boldsymbol{u}}(k),\Hat{\boldsymbol{u}}(k)^*,\text{N"aherung, gemittelt}}\big\}$}}

\input{mbild5c3}
\put(25,140){\makebox(0,0)[r]{\small$0,001$}}
\put(25,102){\makebox(0,0)[rb]{\small$0$}}
\put(25,60){\makebox(0,0)[r]{\small$-0,001$}}
\put(32,87){\makebox(0,0)[bl]{\small$0$}}
\put(110,87){\makebox(0,0)[b]{\small$50$}}
\put(190,87){\makebox(0,0)[b]{\small$100$}}
\put(270,87){\makebox(0,0)[b]{\small$150$}}
\put(350,87){\makebox(0,0)[b]{\small$200$}}
\put(430,87){\makebox(0,0)[b]{\small$250$}}
\put(455,95){\makebox(0,0)[tr]{$k$}}
\put(0,190){\makebox(0,0)[l]{
$\Im\big\{\Hat{C}_{\Hat{\boldsymbol{u}}(k),\Hat{\boldsymbol{u}}(k)^*,\text{gemessen}}\big\};\;
\Im\big\{\Hat{C}_{\Hat{\boldsymbol{u}}(k),\Hat{\boldsymbol{u}}(k)^*,\text{N"aherung, gemittelt}}\big\}$}}

\end{picture}}
\end{center}\vspace{-30pt}
\caption[Messwert"-(ko)"-varianzen von \mbox{$\protect\Hat{\boldsymbol{u}}(k)$}]{
Messwert"-(ko)"-varianzen von \mbox{$\protect\Hat{\boldsymbol{u}}(k)$}.\protect\\
Messung mit: $M=64$, $E=0$, $L=20$ und Fenster
nach Kapitel \myref{Algo} mit $N=4$.\protect\\
Mittelung "uber $1000$ komplette RKM-Messungen.}
\label{E.b5c}
\end{figure}
Bei der Berechnung der Messwert"-(ko)"-varianzen der deterministischen St"orung
trat in Gleichung~(\ref{E.3.52}) eine Doppelsumme der zwei-dimensionalen DFT 
"uber ein Produkt von Erwartungswerten auf. Von dieser Doppelsumme wurden bei der 
Berechnung der Messwert"-(ko)"-varianzen nur die Summanden ber"ucksichtigt, bei denen 
$\mu_2$ um ein ganzzahliges Vielfaches von \mbox{$M/K_{\Phi}$} gegen"uber 
$\mu_1$ verschoben ist. Es wurde festgestellt, dass die anderen Summanden
aufgrund der Ausblendeigenschaft des Betragsquadrats des Spektrums des Fensters
vernachl"assigt werden k"onnen. Da das Betragsquadrat des Spektrums
eines endlich langen Fensters keinen ideal rechteckf"ormigen Verlauf
haben kann, ist zwischen dem Durchlassbereich und dem Sperrbereich
des Fensterspektrums immer ein "Ubergangs\-bereich vorhanden, so dass
man im Integral nach Gleichung~(\ref{E.2.41}) auch f"ur \linebreak
\mbox{$\Hat{\mu}\!=\!\mu\!+\!\Tilde{\mu}\CdoT M/K_{\Phi}\pm1$}
noch nennenswert von null verschiedene Terme erhalten wird. 
Wenn man f"ur das Spektrum der Erregung einen nichtzentralen
(\,\mbox{E$\{\boldsymbol{V}(\mu)\}\neq 0$}\,) Zufallsvektor
\mbox{$\Vec{\boldsymbol{V}}$} verwendet, dessen Kovarianzen
\mbox{$C_{\boldsymbol{V}(\mu_2),\boldsymbol{V}(\mu_1)}$} bzw.
\mbox{$C_{\boldsymbol{V}(\mu_2)^{\Kk},\boldsymbol{V}(\mu_1)}$}
f"ur \mbox{$\mu_2\!=\!\mu_1\!+\!\Tilde{\mu}\CdoT M/K_{\Phi}\pm1$} von null 
verschieden sind, werden die in erster N"aherung vernachl"assigten Summanden 
bewirken, dass die Messwert"-(ko)"-varianzen von \mbox{$\Hat{\boldsymbol{u}}(k)$} 
nicht mehr mit $K_{\Phi}$ periodisch sind, sondern Anteile mit den 
benachbarten Kreisfrequenzen \mbox{$\Omega=\Tilde{\mu}\CdoT\frac{2\pi}{K_{\Phi}}\pm\frac{2\pi}{M}$} 
aufweisen. Dies soll nun am Beispiel des zeitinvarianten Systems \mbox{($K_H\!=\!1$)} 
mit der "Ubertragungsfunktion \mbox{$H(\Omega)\!=\!1$} demonstriert werden.

Zur Erregung verwenden wir die mit \mbox{$M\!=\!64$} periodisch fortgesetzten
Impulssequenzen \mbox{$\boldsymbol{v}(k)$}, die f"ur
\mbox{$k=1\;(1)\;M\!-\!1$} immer null sind, und nur f"ur \mbox{$k=0$}
einen normalverteilten komplexen Zufallswert mit dem Mittelwert Eins,
der Varianz Eins und dem komplexen Korrelationskoeffizienten
\mbox{$\text{E}\{\boldsymbol{v}(0)^2\}/\text{E}\{|\boldsymbol{v}(0)|^2\}
=0,\!7\!+\!0,\!3\CdoT j$} enthalten. Bei dieser Art der Erregung wird das
Spektrum \mbox{$V_{\lambda}(\mu)$} bei jeder Einzelmessung von $\mu$ unabh"angig. 
Daher sind auch die Terme, die in den Gleichungen (\ref{E.3.47}) 
vor dem LDS bzw. KLDS der gefensterten St"orung stehen, und somit in die 
Gleichungen (\ref{E.3.52}) eingehen, von den beiden beteiligten 
Frequenzen unabh"angig und von null verschieden. Eine deterministische
St"orung wird bei der Messung nicht eingespeist, so dass $u(k)\!=\!0$ gilt.
Dem periodisch fortgesetzten erregenden Zufallssignal wird eine station"are 
mittelwertfreie gau"sverteilte komplexe St"orung \mbox{($K_{\Phi}\!=\!1$)} 
mit der Varianz $0,\!01$ und dem komplexen Korrelationskoeffizienten
\mbox{$\text{E}\{\boldsymbol{n}(k)^2\}/\text{E}\{|\boldsymbol{n}(k)|^2\}
= -0,\!5\!+\!0,\!5\CdoT j$} "uberlagert. Die Rauschwerte f"ur unterschiedliche
Werte von $k$ sind unkorreliert. Das LDS dieser St"orung ist konstant
$0,\!01$ und das KLDS ist konstant \mbox{$-0,\!005\!+\!0,\!005\CdoT j$}.
Nun wurde eine Messung mit \mbox{$L\!=\!20$} Einzelmessungen mit einem Fenster 
nach Kapitel \myref{Algo} mit \mbox{$N\!=\!4$} durchgef"uhrt,
und diese $999$ mal wiederholt. Die Messwert"-(ko)"-varianzen wurden
nun aus den $1000$ Sch"atzwerten \mbox{$\Hat{u}(k)$} f"ur jeden Zeitpunkt
\mbox{$k=0\;(1)\;F\!-\!1$} empirisch bestimmt, und sind in Bild
\ref{E.b5c} "uber der diskreten Zeit $k$ aufgetragen.
Desweiteren sind die nach den Gleichungen (\ref{E.3.53}) berechneten
und "uber alle $1000$ kompletten RKM-Messungen gemittelten
zeitunabh"angigen N"aherungen der Messwert"-(ko)"-varianzen eingetragen.
Deutlich erkennt man, dass neben den zeitlich konstanten Gleichanteilen,
die durch die zeitunabh"angigen N"aherungen offensichtlich recht gut
abgesch"atzt werden, auch noch Anteile mit der Kreisfrequenz
\mbox{$\Omega=\pm2\pi/64$} vorhanden sind, w"ahrend die
h"oherfrequenten Anteile offensichtlich rauschartiger Natur sind.
Es sei noch darauf hingewiesen, dass man die systematischen Fehler,
die man durch die zeitunabh"angige N"aherung der Messwert"-(ko)"-varianzen
macht, bei Verwendung eines sinnvolleren erregenden Zufallsvektors
\mbox{$\Vec{\boldsymbol{V}}$} (\,mittelwertfreie und
unkorrelierte Spektralwerte\,) vermeiden kann.

\section{Messung eines komplexen, periodisch zeitvarianten Systems mit zyklostation"arer St"orung}\label{E.Kap.9.2}

Bei diesem Beispiel handelt es sich um die Simulation eines
Stereodecoders, bei dem ein Stereo-Multiplex-Signal, wie es im UKW-Rundfunk
"ublich ist, mit $76$\,kHz abgetastet und A/D gewandelt wird, und
anschlie"send mit einem digitalen FIR-Filter in der Art interpoliert wird,
dass einerseits die beiden Signale des linken und rechten Kanals
mit einer Abtastrate von jeweils $38$\,kHz getrennt vorliegen, und
andererseits der Pilotton bei $19$\,kHz unterdr"uckt wird. Das Signal
des linken Kanals k"onnen wir als den Realteil eines komplexen Signals
auf"|fassen, dessen Imagin"arteil das Signal des rechten Kanals ist.
Als Erregung \mbox{$v_{\lambda}(k)$} wird bei jeder Einzelmessung
ein Ausschnitt der L"ange \mbox{$M\!=\!512$} eines mittelwertfreien,
normalverteilten Zufallssignals verwendet. Die Varianz
der komplexen Erregung wird zeitunabh"angig auf
\mbox{$C_{\boldsymbol{v}(k),\boldsymbol{v}(k)}=0,\!072$}
festgelegt. Die Abtastwerte der Erregung wurden f"ur unterschiedliche
Zeitpunkte unkorreliert gew"ahlt, w"ahrend f"ur den Real- und Imagin"arteil
f"ur gleiche Zeitpunkte ein komplexer Korrelationskoeffizient von
\mbox{$C_{\boldsymbol{v}(k),\boldsymbol{v}(k)^{\Kk}}/
C_{\boldsymbol{v}(k),\boldsymbol{v}(k)}= j\CdoT0,\!7$} eingestellt wurde.
Dies entspricht bei den reellen Signalen des rechten und des linken
Stereokanals einem reellen Korrelationskoeffizienten von $0,\!7$,
wobei die Signale beider Kan"alen dieselbe Varianz aufweisen.

Aus diesen $M$ Abtastwerten wird nun das analoge Stereo-Multiplex-Signal
f"ur die Abtastzeitpunkte des AD-Wandlers im Stereodecoder berechnet.
Dabei wird angenommen, dass die Werte \mbox{$v_{\lambda}(k)$}
die Abtastwerte im Abstand $1/38$\,kHz einer Periode eines periodischen
bandbegrenzten analogen komplexen Signals seien. 
In beiden Stereokan"alen wird zun"achst ein analoges Tiefpassfilter elften
Grades mit Tschebyscheff-Verhalten im Durchlassbereich bei der Berechnung
des Stereo-Multiplex-Signals simuliert. Wenn mit \mbox{$v_{TP,L}(t)$}
und \mbox{$v_{TP,R}(t)$} die beiden reellen, analogen Signale an den
Ausg"angen der beiden Tiefpassfilter bezeichnet sind, erh"alt man
das Stereo-Multiplex-Signal gem"a"s
\begin{align}
v_{Stereo}(t)\;&=\;\big(v_{TP,L}(t)+v_{TP,R}(t)\big)\;+
\label{E.9.1}\\[4pt]
&+\;\big(v_{TP,L}(t)-v_{TP,R}(t)\big)\cdot\sin(2\pi\CdoT38\,\text{kHz})\;+
\notag\\[4pt]
&+\;0,\!1\cdot\sin(2\pi\CdoT19\,\text{kHz}).\notag
\end{align}
Die Pilottonamplitude \mbox{$0,\!1$} und die Streuung der Erregung
wurden dabei in einem Verh"altnis gew"ahlt, wie es im UKW-Rundfunk
"ublich ist. Aufgrund der Periodizit"at der\linebreak Erregung l"asst sich
das Stereo-Multiplex-Signal f"ur jeden beliebigen Zeitpunkt
im Rahmen der Rechengenauigkeit als station"are L"osung exakt
berechnen, obwohl die analogen Tiefp"asse keine zeitlich begrenzte 
Impulsantwort besitzen. Der mit $76$\,kHz getaktete AD-Wandler im 
Stereodecoder wird dadurch simuliert, dass man sowohl f"ur die Zeitpunkte
\mbox{$t=(k\!+\!1/4\!+\!\Delta k_{sync})/38\,\text{kHz}$} als auch f"ur die 
Zeitpunkte \mbox{$t=(k\!+\!3/4\!+\!\Delta k_{sync})/38\,\text{kHz}$} mit 
\mbox{$k\in[-E;F\!\!-\!1]$} die Abtastwerte des Stereo-Multiplex-Signals
berechnet. F"ur jeden Zeitpunkt $k$ erh"alt man also zwei Abtastwerte.
Dabei ber"ucksichtigt der Summand \mbox{$\Delta k_{sync}$} eine evtl.
vorhandene Fehlsynchronisation des AD-Wandlers im Stereodecoder. In der 
Simulation wurde der Wert \mbox{$\Delta k_{sync}=0,\!01$} eingestellt. 
Um einen Linearit"atsfehler des AD-Wandlers zu simulieren wurde vor der 
gleichm"a"sigen und symmetrischen 16-Bit-Quantisierung, die auch im 
Beispiel des Kapitels \myref{SD-DAC} ---\,dort aber mit 12-Bit\,--- verwendet wird, 
die nichtlineare Verzerrung
\begin{equation}
0,\!01\CdoT v_{Stereo}(t)^3-0,\!002\CdoT v_{Stereo}(t)^2+v_{Stereo}(t)+0,\!002
\label{E.9.2}
\end{equation}
des Stereo-Multiplex-Signals vorgenommen. Jeder zweite Abtastwert
des Ausgangssignals des AD-Wandlers wird mit einem linearphasigen
FIR-Filter interpoliert, und liefert uns den Realteil (\,= linker Stereokanal\,) des
Ausgangssignals \mbox{$y_{\lambda}(k)$} des simulierten Systems.
Die dazwischenliegenden Abtastwerte des Ausgangssignals des AD-Wandlers
werden ebenfalls mit mit einem linearphasigen FIR-Filter interpoliert
und bilden den Imagin"arteil (\,=~rechter Stereokanal\,) von \mbox{$y_{\lambda}(k)$}.
Dabei wird das FIR-Filter verwendet, dessen Koeffizienten gerade die
zeitlich gespiegelten Koeffizienten des Filters des Realteilpfades
sind. Als Koeffizienten der beiden FIR-Filter wurden die Abtastwerte der 
kontinuierlichen Fensterautokorrelationsfunktion \mbox{$d_{\infty}(t)$} verwendet, 
die in Kapitel~\ref{E.Kap.10.7} eingef"uhrt wird. Die Fourierreihenkoeffizienten 
des Filters kann man mit dem Programm im Unterkapitel~\ref{E.Kap.11.4} mit den Parametern
\mbox{$N\!=\!34$}, \mbox{$A_0\!=\!3$} und \mbox{$A_1\!=\!15$}
ohne eine weitere frei w"ahlbare Nullstelle \mbox{$s_0$} berechnen.
Die Abtastwerte der Fensterautokorrelationsfunktion wurden mit dem Programm
in Unterkapitel~\ref{E.Kap.11.7} f"ur die normierten Zeitpunkte
\mbox{$t=(k\!+\!1/4)/38$} im Intervall \mbox{$[-1;1]$} berechnet. 
Die Fouriertransformierte der beiden FIR-Filter hat bei
der halben Abtastfrequenz $19$\,kHz eine Nullstelle, die
f"ur eine hohe D"ampfung des Pilottons sorgt. Desweiteren ist der
Betrag der Fouriertransformierten der FIR-Filter f"ur die Frequenzen
\mbox{$0\;(1)\;15\,kHz$} ---\,diese liegen innerhalb der UKW-Bandbreite\,---
bei geeigneter Normierung eins.

Damit ist das simulierte System beschrieben, und es wird nun kurz
auf die Parametereinstellung der RKM-Messung eingegangen. Da das
Stereo-Multiplex-Signal im AD-Wandler des Stereodecoders mit der vierfachen
Frequenz des Pilottons abgetastet wird, und jeweils zwei reelle Abtastwerte
einen komplexen Wert des Ausgangssignals \mbox{$y_{\lambda}(k)$}
liefern, f"uhrt eine Erh"ohung von $k$ um zwei zu einer unver"anderten
Aussteuerung der leicht nichtlinearen Kennlinie des AD-Wandler.
Daher kann angenommen werden, dass sich der Stereodecoder
durch ein periodisch zeitvariantes System mit der Periode \mbox{$K_H\!=\!2$}
gut modellieren l"asst. Vom Approximationsfehler kann erwartet werden,
dass er mit der Periode \mbox{$K_{\Phi}\!=\!2$} zyklostation"ar ist. 
Da zu vermuten ist, dass der simulierte Synchronisationsfehler bei der 
Abtastung des Stereo-Multiplex-Signals zu einem u.\,U. unsymmetrischen 
"Ubersprechen der beiden Stereo-Kan"ale f"uhrt, wurde auch das von der konjugierten
Erregung \mbox{$\boldsymbol{v}(k)^{\Kk}$} gespeiste Modellsystem  ${\cal S}_{*,lin}$
angesetzt. Bei der Berechnung der Messwerte f"ur das LDS und KLDS
wurde die Variante gem"a"s der Gleichungen~(\ref{E.3.34}) gew"ahlt.
Da bei der Simulation des Systems bereits die Abtastwerte des
periodischen Stereo-Multiplex-Signals berechnet werden, braucht die
Einschwingzeit der Tschebyscheff-Tiefp"asse bei der Messung nicht
ber"ucksichtigt zu werden. Es gen"ugt daher wenn man die Einschwingzeit $E$
gr"o"ser gleich der L"ange $76$ der FIR-Interpolationsfilter w"ahlt.
Bei der Messung wurde \mbox{$E\!=\!2048$} eingestellt. Als Fensterfolge
beim RKM wurde die in~\myref{Algo} vorgestellte Fensterfolge mit
\mbox{$N\!=\!4$} verwendet. Es wurde "uber \mbox{$L\!=\!50000$}
Einzelmessungen gemittelt. Die L"ange der DFT betrug \mbox{$M\!=\!512$},
so dass alle Spektren im Raster \mbox{$2\pi/M$} gemessen wurden.
Das Konfidenzniveau wurde auf \mbox{$1\!-\!\alpha=90\%$} festgelegt.

Bild~\ref{E.b6aa} zeigt die gemessene deterministische St"orung. 
Im unteren Teilbild ist einerseits der Betrag des gemessenen
Spektrums ---\,verrauschte Kurve\,--- und andererseits die gemessene
Messwertstreuung ---\,glatte Kurve\,--- halblogarithmisch dargestellt.
Au"ser bei der Frequenz \mbox{$\Omega\!=\!0$} liegen alle Messwerte
in der Gr"o"senordnung der Messwertstreuung, und sind daher als null
anzusehen, oder zumindest wohl kleiner als \mbox{$-90\,$dB}.
Da sowohl die deterministische St"orung als auch der zuf"allige
Approximationsfehler die beiden FIR-Interpolationsfilter durchlaufen,
ist eine Absenkung des Spektrums und der Messwertstreuung im Bereich
um die Frequenz \mbox{$\Omega\!=\!\pi\,\Hat{=}\,19$\,kHz} zu beobachten.
Bei \mbox{$\Omega\!=\!0$} ergibt sich der Wert \mbox{$2,\!9121\,$dB}.
Somit ist die deterministische St"orung praktisch zeitlich
konstant. Dies erkennt man auch in den beiden oberen Teilbildern.
Dort sind einmal f"ur die geraden Zeitpunkte $k$ und zum zweiten
f"ur ungerades $k$ die Werte \mbox{$\Hat{u}(k)$} als Punkte in der komplexen
Ebene dargestellt. An der Skalierung der Achsen erkennt man, wie
nahe beieinander diese Punkte f"ur die unterschiedlichen Werte von $k$
liegen. Da man bei einem zyklostation"aren Approxima\-tions\-fehler
mit \mbox{$K_{\Phi}\!=\!2$} Konfidenzellipsen f"ur die Messwerte
\mbox{$\Hat{u}(k)$} erh"alt, die sich periodisch wiederholen,
wurde f"ur die geraden und die ungeraden Zeitpunkte $k$ je eine
Konfidenzellipse bei einem Messwert eingetragen. Die beiden
Konfidenzellipsen unterscheiden sich doch erheblich hinsichtlich ihrer
Neigung, was ein erster Hinweis auf die Zyklostationarit"at der
Momente des Approximationsfehlers ist.

Die beiden oberen Teilbilder des Bildes~\ref{E.b6ab} zeigen die beiden
Impulsantworten \mbox{$\Hat{\Tilde{h}}_{\kappa}(k)$}
des periodisch zeitvarianten Teilmodellsystems ${\cal S}_{lin}$, das von der
{\em nicht}\/ konjugierten Erregung \mbox{$\boldsymbol{v}(k)$}
gespeist wird.  Hier ist die Antwort \mbox{$\Hat{\Tilde{h}}_0(k)$}
des Systems auf den Impuls \mbox{$\gamma_0(k)$} mit kleinen Kreisen,
und die Antwort \mbox{$\Hat{\Tilde{h}}_1(k)$} des Systems auf den
verschobenen Impuls \mbox{$\gamma_0(k\!-\!1)$} mit kleinen Kreuzchen 
dargestellt. Das oberste Teilbild zeigt die ersten $100$ Werte
des Realteils, w"ahrend darunter die entsprechenden Imagin"arteilwerte
zu sehen sind. F"ur alle Messwerte der Impulsantworten,
die nach Kapitel~\ref{E.Kap.4.1} aus den Messwerten der bifrequenten
"Ubertragungsfunktion berechnet wurden, ergab sich eine empirische
Messwertstreuung zwischen \mbox{$3,\!3\CdoT10^{-7}$} und
\mbox{$3,\!4\CdoT10^{-7}$}. Die Imagin"arteile der Messwerte der
Impulsantworten liegen somit in der Gr"o"senordnung der Messwertstreuung.
Daher kann das von \mbox{$\boldsymbol{v}(k)$} erregte Teilsystem
im Rahmen der Messgenauigkeit als reellwertig angesehen werden.
Die in Bild~\ref{E.b3a} eingezeichneten Teilsysteme $h_{\beta,\kappa}(k)$ und
$h_{\gamma,\kappa}(k)$ liefern also einen vernachl"assigbar kleinen Beitrag
zu der periodisch zeitvarianten Impulsantwort des von
\mbox{$\boldsymbol{v}(k)$} erregten Teilsystems, so dass bei
diesem Teilsystem kein nennenswertes "Ubersprechen zwischen den
beiden Stereokan"alen zu beobachten ist. Bei den Realteilen der
beiden Impulsantworten ist im obersten Teilbild lediglich eine
Verschiebung um einen Takt zu beobachten, wie man dies auch bei
einem zeitinvarianten System erwarten w"urde. Die Unterschiede
der beiden Impulsantworten \mbox{$\Hat{\Tilde{h}}_0(k)$} und
\mbox{$\Hat{\Tilde{h}}_1(k\!+\!1)$} sind so gering, dass sie bei
dieser Art der Darstellung nicht zu erkennen sind. 

Die Betr"age der Abtastwerte der bifrequenten "Ubertragungsfunktion sind in den unteren
beiden Teilbildern des Bildes~\ref{E.b6ab} in halblogarithmischer Darstellung
zu sehen. Die dicke Kurve im vorletzten Teilbild zeigt die Hauptdiagonale
der bifrequenten "Ubertragungsfunktion mit dem gleichen Argument $\mu$ in
beiden Variablen. Im Bereich niedriger normierter Kreisfrequenzen mit
\mbox{$|\Omega|\le15/19\CdoT\pi$} liegt diese Kurve so nahe bei 
\mbox{$0\,$dB}, dass sie in diesem Bereich konstant erscheint. 
Deshalb wurde auch eine um den Faktor $1000$ gespreizte Darstellung als
d"unne Kurve in dieses Teilbild eingezeichnet. Man erkennt nun
das typische Verhalten der Tschebyscheff-Tiefp"asse, die bei der
Erzeugung des Stereo-Multiplex-Signals verwendet wurden. Die Betr"age der
"Ubertragungsfunktionen der FIR-Filter, die zur Interpolation der
Abtastwerte des Stereo-Multiplex-Signals eingesetzt werden, liegen
innerhalb der Nutzbandbreite bis $15$\,kHz offensichtlich so nahe bei
eins, dass nur die "Ubertragungsfunktionen der identischen
Tschebyscheff-Tiefp"asse im linken und rechten Stereokanal
im Betrag der Hauptdiagonale der bifrequenten "Uber\-tra\-gungs\-funk\-tion
des komplexen Signals ihre Wirkung zeigen. Als letzte Kurve ist in
diesem Teilbild auch noch die Varianz dieser Messwerte eingezeichnet.
Sie liegt im gesamten Frequenzbereich unterhalb von
\mbox{$4\CdoT10^{-11}\,\Hat{\approx}\,104\,$dB} und nimmt ab, je n"aher
man zur halben Abtastfrequenz \mbox{$19$\,kHz$\,\Hat{=}\,\Omega\!=\!\pi$}
kommt. Hier wirkt sich aus, dass auch der Approximationsfehlerprozess
die beiden ausgangsseitigen FIR-Interpolationsfilter durchl"auft,
die zugleich den Pilotton bei \mbox{$19$\,kHz} unterdr"ucken.

Das unterste Teilbild in Bild \ref{E.b6ab} zeigt den Betrag des Anteils der 
bifrequenten "Ubertragungsfunktion, bei dem das zweite Argument um $\pi$ 
gr"o"ser ist als das erste Argument, was bei der gew"ahlten Abtastung der
bifrequenten "Ubertragungsfunktion zu einer Argumentverschiebung
von $M/2$ f"uhrt. Dieser Anteil spiegelt das periodisch zeitvariante
Verhalten des Systems wider, und w"are bei einem zeitinvarianten System
konstant null. In der gew"ahlten halblogarithmischen Darstellung liegt
dieser Anteil in einem weiten Frequenzbereich etwa bei \mbox{$-74$\,dB}.
Im Bereich niedriger Frequenzen f"allt dieser Anteil bis in die
Gr"o"senordnung der ebenfalls dargestellten Messwertstreuung ab.
Man erkennt, dass in diesem Frequenzbereich die Messkurve des
Betrags der bifrequenten "Ubertragungsfunktion stark verrauscht erscheint,
w"ahrend sie im restlichen Frequenzbereich doch relativ glatt verl"auft.

Die beiden Impulsantworten \mbox{$\Hat{\Tilde{h}}_{*,\kappa}(k)$}
des periodisch zeitvarianten Teilmodellsystems ${\cal S}_{*,lin}$, 
das von der konjugierten Erregung \mbox{$\boldsymbol{v}(k)^{\Kk}$}
gespeist wird, sowie die Betr"age der beiden Anteile der bifrequenten
"Ubertragungsfunktion dieses Teilmodellsystems sind in Bild~\ref{E.b6ac}
dargestellt. Die Streuung der Messwerte der Impulsantworten liegt auch
hier bei allen Messwerten zwischen \mbox{$3,\!3\CdoT10^{-7}$} und
\mbox{$3,\!4\CdoT10^{-7}$}. Da nun der Imagin"arteil der beiden
Impulsantworten um Gr"o"senordnungen gr"o"ser ist als die
Messwertstreuung, bewirkt dieses Teilsystem ein "Ubersprechen zwischen
den beiden Stereokan"alen. Bei den beiden Graphiken der Betr"age der
beiden Anteile der bifrequenten "Ubertragungsfunktion wurde jeweils als
untere Kurve wieder die Messwertstreuung eingezeichnet.

Mit Hilfe der Gleichungen~\ref{E.2.1} lassen sich aus den Messwerten
\mbox{$\Hat{\Tilde{h}}_{\kappa}(k)$} und
\mbox{$\Hat{\Tilde{h}}_{*,\kappa}(k)$} der der zeitvarianten 
Impulsantworten  der beiden Teilsysteme auch Messwerte f"ur die 
vier Impulsantworten $h_{\alpha,\kappa}(k)$ bis $h_{\delta,\kappa}(k)$
der reellwertigen Systeme in Bild~\ref{E.b3a} berechnen.
Diese sind nun ebenfalls periodisch zeitvariant. Auf eine graphische
Darstellung der Impulsantworten und der "Ubertragungsfunktionen
dieser reellwertigen Systeme wird verzichtet.

Die beiden oberen Teilbilder der Bilder~\ref{E.b6ad} und \ref{E.b6ae}
zeigen jeweils den Real- und Imagin"arteil der Messwertfolgen
\mbox{$\Hat{\phi}_{\boldsymbol{n}}(k\!+\!\kappa,k)$} und
\mbox{$\Hat{\psi}_{\boldsymbol{n}}(k\!+\!\kappa,k)$}, mit deren Hilfe man die beiden
Kovarianzfolgen \mbox{$\text{E}\big\{\boldsymbol{n}(k\!+\!\kappa)\CdoT
\boldsymbol{n}(k)^{\Kk}\big\}$} und \mbox{$\text{E}\big\{
\boldsymbol{n}(k\!+\!\kappa)\CdoT\boldsymbol{n}(k)\big\}$}
absch"atzen kann. Die Messwertfolgen wurden aus den Messwerten
\mbox{$\Hat{\Phi}_{\boldsymbol{n}}\big({\T\mu,\mu\!+\!\Tilde{\mu}\CdoT\frac{M}{K_{\Phi}}}\big)$}
und \mbox{$\Hat{\Psi}_{\boldsymbol{n}}\big({\T\mu,\mu\!+\!\Tilde{\mu}\CdoT\frac{M}{K_{\Phi}}}\big)$}
f"ur die N"aherungen des LDS bzw. KLDS mit Hilfe einer zweidimensionalen
DFT berechnet, wie dies in Kapitel~\ref{E.Kap.4.2} beschrieben ist.
Da bei einem zyklostation"aren Approximationsfehlerprozess
die Kovarianzfolgen nicht nur von dem freien Parameter $\kappa$
der Differenz der beiden betrachteten Zeitpunkte des Prozesses
abh"angen, sondern auch vom absoluten Zeitpunkt $k$, wobei sich
die Kovarianzfolgen mit $K_{\Phi}$ periodisch in $k$ wiederholen,
sind in jedem Teilbild die \mbox{$K_{\Phi}\!=\!2$} Folgen dargestellt,
die sich mit \mbox{$k\!=\!0$} und \mbox{$k\!=\!1$} ergeben. Erstere 
sind jeweils mit kleinen Kreisen und letztere mit kleinen Kreuzchen
markiert. Von den Kovarianzfolgen wurden nur die Werte mit
\mbox{$|\kappa|\le50$} dargestellt. Im obersten Teilbild in
Bild~\ref{E.b6ad} kann man an der Stelle \mbox{$\kappa\!=\!0$}
die beiden Werte der Varianz des Approximationsfehlerprozesses
ablesen, die sich in $k$ periodisch abwechseln. Diese liegen deutlich
"uber dem Wert \mbox{$7,\!76\CdoT10^{-11}$}, der sich bei einer
Abtastung des Stereo-Multiplex-Signals mit einem AD-Wandler mit einer
perfekt gleichm"a"sigen Kennlinie ergeben w"urde, wenn man die Varianz
als \mbox{$1/12$} des Quadrats der Quantisierungsstufenh"ohe berechnet.
Somit "uberwiegt also der Anteil des Approximationsfehlerprozesses,
der durch die simulierte Nichtlinearit"at der AD-Wandlerkennlinie nach
Gleichung~(\ref{E.9.2}) verursacht wird. In den unteren beiden
Teilbildern der Bilder~\ref{E.b6ad} und \ref{E.b6ae}
besteht die obere Kurve jeweils aus den Messwerten f"ur die
St"arken der Impulslinien der N"aherung des bifrequenten LDS bzw. KLDS,
die in Kapitel~\ref{E.2.3} beschrieben sind. Die Kurven in den
vorletzten Teilbildern sch"atzen jeweils die St"arken der Impulslinien
mit \mbox{$\Omega_2\!=\!\Omega_1$} ab, w"ahrend die Kurven in den
untersten Teilbildern Messwerte f"ur die St"arken der Impulslinien mit
\mbox{$\Omega_2\!=\!\Omega_1\!+\!\pi$} sind. Damit die Zuverl"assigkeit
dieser Messkurven beurteilt werden kann, enthalten diese Teilbilder
zus"atzlich die Kurven der Messwertvarianzsch"atzwerte, die immer unterhalb der
Kurven der Messwerte liegen. Durch die Mittelung "uber \mbox{$L\!=\!50000$}
Einzelmessung konnte erreicht werden, dass die Betr"age der Messwerte
des LDS und KLDS immer deutlich gr"o"ser als die Messwertstreuungen
waren, und man somit relativ zuverl"assige Messwerte erhalten konnte.
\input{bild6a.tex}

\chapter{Erg"anzungen zum Fenster}\label{E.Kap.10}

\section[Algorithmus f"ur Fenster mit frei w"ahlbaren Nullstellen]{Algorithmus f"ur Fenster mit frei w"ahlbaren \\Nullstellen}\label{E.Kap.10.1}

In diesem Kapitel wird der in Kapitel \myref{Algo} vorgestellte Algorithmus zur 
Konstruktion der Fensterfolge so erweitert, dass es m"oglich wird, die dort nicht 
genutzten Freiheitsgrade dazu zu verwenden, weitere Fensterfolgen zu konstruieren, 
die ebenfalls die Eigenschaften (\myref{2.20}) und (\myref{2.27}) erf"ullen, 
die den Einsatz der Fensterfolge beim RKM erm"oglichen. Auch die hier berechneten
Fensterfolgen sind {\em reell}, so dass an einigen Stellen auf das Konjugieren 
verzichtet werden kann, und die im Spektrum vorhandene Symmetrie ausgenutzt wird, 
ohne dass darauf extra hingewiesen wird. Desweiteren sind auch hier die Fensterfolgen 
in der Regel {\em nicht}\/ linearphasig, und enthalten meist auch Werte kleiner null. 

Die in Kapitel \myref{Fen} beschriebene prinzipielle Konstruktion der Fensterfolge wird 
im wesentlichen beibehalten und lediglich etwas erweitert. So kann sowohl die zeitdiskrete
Fensterfolge wie auch die zugrundeliegende zeitdiskrete, reelle Basisfensterfolge 
\mbox{$g(k)$} prinzipiell im gesamten Intervall \mbox{$k\in[0;F\!-\!1]$} von null 
verschiedene Werte annehmen. Mit dem Parameter $A_0$, f"ur den \mbox{$0\!\le\!A_0\!<\!N$} 
gelten muss, kann man festlegen, dass die letzten $A_0$ Werte innerhalb des Intervalls 
null sein sollen:
\begin{gather}
g(k)=0\qquad\qquad\qquad\forall\qquad k<0\quad\vee\quad k\ge F\!-\!A_0
\notag\\
g(0) \neq 0 \qquad\wedge\qquad g(F\!-\!A_0\!-\!1) \neq 0.
\label{E.10.1}
\end{gather}
Die L"ange $F$ soll auch hier ein ganzzahliges Vielfaches $N$ der DFT-L"ange $M$ 
beim RKM sein. Bis auf einen konstanten Faktor --- den man nicht fest vorgibt, da
er nie explizit bestimmt werden muss --- wird die Basisfensterfolge durch die nun
\mbox{$F\!-\!A_0\!-\!1$} Nullstellen des Polynoms der Z-Transformierten der 
Basisfensterfolge festgelegt:
\begin{equation}
z^{F-A_0-1}\Cdot G(z)\;=\;z^{F-A_0-1}\cdoT\Sum{k=0}{F-A_0-1}\,g(k)\CdoT z^{\!-k}.
\label{E.10.2}
\end{equation}\pagebreak[3]

Dieses Polynom l"asst sich als Produkt zweier Polynome darstellen:
\begin{equation}
z^{F-A_0-1}\Cdot G(z)\;=\;z^{F-N+2\cdoT A_1}\Cdot G_1(z)\cdot z^{N-A_0-2\cdoT A_1-1}\Cdot G_2(z).
\label{E.10.3}
\end{equation}
Die \mbox{$F\!-\!N\!+\!2\CdoT A_1$} Nullstellen des ersten Polynoms 
\mbox{$z^{F-N+2\cdoT A_1}\Cdot G_1(z)$} enthalten die in Kapitel \myref{Fen} 
genannten, fest vorgegeben Nullstellen am Einheitskreis, sowie weitere $2\CdoT A_1$ 
Nullstellen, die sich am Einheitskreis im gleichen Frequenzraster daran anschlie"sen: 
\begin{equation}
G\big(e^{j\cdot\frac{\pi}{F}\cdot\nu}\big)\;=\;0
\qquad\qquad\text{ f"ur}
\qquad\nu\,=\,N\!+\!1\!-\!2\CdoT A_1\;(2)\;2\CdoT F\!-\!N\!-\!1\!+\!2\CdoT A_1.
\label{E.10.4}
\end{equation}
Die \mbox{$N\!-\!A_0\!-2\CdoT A_1\!-\!1$} Nullstellen des zweiten Polynoms 
\mbox{$z^{N-A_0-2\cdoT A_1-1}\Cdot G_2(z)$} sind weitestgehend frei w"ahlbar.
Sie d"urfen jedoch nicht bei \mbox{$z\!=\!0$} oder bei 
\mbox{$z\!=\!e^{\pm j\cdot\frac{\pi}{F}\cdot(N-1-2\cdoT A_1)}$} liegen, 
da sie in diesen F"allen gegebenfalls durch eine Erh"ohung von $A_0$ bzw. $A_1$ 
zu ber"ucksichtigen w"aren. Wird gew"unscht, dass irgendwelche der in der 
letzten Gleichung genannten Nullstellen des ersten Polynoms mit gr"o"serer als 
nur einfacher Vielfachheit auftreten, so sind die dazu ben"otigten zus"atzlichen
Nullstellen ebenfalls dem zweiten Polynom zuzuordnen. Desweiteren m"ussen Nullstellen, 
die einen von null verschiedenen Imagin"arteil besitzen, auf Grund der 
Reellwertigkeit der Basisfensterfolge \mbox{$g(k)$} als zueinander konjugiert 
komplexe Paare auftreten. Die Nullstellen des zweiten Polynoms werden im weiteren 
mit \mbox{$z_{0,\rho}=|z_{0,\rho}|\CdoT e^{j\cdot\psi_{0,\rho}} $} mit 
\mbox{$\rho=1\;(1)\;N\!-\!1\!-\!A_0\!-\!2\CdoT\!A_1$} bezeichnet. Die in 
Kapitel \myref{Algo} vorgestellte Fensterfolge stellt somit den Spezialfall mit 
\mbox{$A_0=N\!-\!1$} und \mbox{$A_1\!=\!0$} dar, bei dem 
keine zus"atzlichen Nullstellen frei gew"ahlt werden.

F"ur das erste Polynom kann man 
\begin{equation}
z^{F-N+2\cdoT A_1}\Cdot G_1(z)\;=\;\frac{\D z^F\!+(-1)^N}
{\T\;\;\Prod{\nu_2=\frac{1-N}{2}+A_1}{\frac{N-1}{2}-A_1}\!\!\D
\big(z\!-\!e^{j\cdot\frac{2\pi}{F}\cdot\nu_2}\big)\;}
\label{E.10.5}
\end{equation}
schreiben, wobei f"ur das Produkt im Nenner wieder die in \cite{Diss} in 
der Liste der Formelzeichen angegebene Definition verwendet wird, die es 
zul"asst, dass der Lauf"|index $\nu_2$ auch Werte annehmen kann, die nicht 
ganzzahlig sind. Das zweite Polynom kann in Form seiner Produktdarstellung als
\begin{equation}
z^{N-1-A_0-2\cdoT A_1}\Cdot G_2(z)\;=\;
\Prod{\rho=1}{N-1-A_0-2\cdoT A_1}\!\!(z\!-\!z_{0,\rho})
\label{E.10.6}
\end{equation}
geschrieben werden. 

Wie in \cite{Diss} wird durch die "Uberlagerung der verschobenen
Betragsquadrate des Spektrums \mbox{$G\big(e^{j\Omega}\big)$}
das Betragsquadrat des Spektrums \mbox{$F(\Omega)$} gewonnen.\pagebreak[3] 
Somit "uberlagern sich bei den Frequenzen \mbox{$\Omega=\nu\CdoT2\pi/F$} 
mit \mbox{$\nu=N\!-\!A_1\;(1)\;F\!-\!N\!+\!A_1$} 
jeweils die doppelten Nullstellen, die im Polynom
\mbox{$z^{F-N+2\cdoT A_1}\Cdot G_1(z)\CdoT G_1(z^{\!-1})$} und damit 
auch im Polynom \mbox{$z^{F-A_0-1}\Cdot G(z)\CdoT G(z^{\!-1})$}
vorhanden sind. \mbox{$z^{F-A_0-1}\Cdot\big|F\big(\!-j\CdoT\ln(z)\big)\big|^2$}
hat daher bei diesen Frequenzen doppelte Nullstellen. Diese doppelten
Nullstellen am Einheitskreis werden bei der Aufspaltung in einen
minimalphasigen Anteil \mbox{$z^{F-A_0-1}\Cdot F\big(\!-j\Cdot\ln(z)\big)$}
und den Rest \mbox{$F\big(j\Cdot\ln(z)\big)$} als einfache Nullstellen
jedem der beiden Polynome zugeordnet. Daher gilt
die gegen"uber Gleichung (\myref{6.15}) modifizierte Gleichung:
\begin{equation}
F\Left({\T\nu\CdoT\frac{2\pi}{F}}\right)=0
\qquad\qquad\forall\qquad\nu=N\!-\!A_1\;(1)\;F\!-\!N\!+\!A_1.
\label{E.10.7}
\end{equation}
Die in Gleichung (\myref{6.15}) aufgef"uhrten Nullstellen sind hier ebenfalls
enthalten.

Auch hier ist die Fensterfolge durch die Werte des Spektrums
\mbox{$F(\Omega)$} bei den Frequenzen \mbox{$\Omega=\nu\CdoT2\pi/F$} mit
\mbox{$\nu=0\;(1)\;F\!-\!1$} vollst"andig festgelegt. Nach Gleichung (\ref{E.10.7}) 
sind die meisten Werte des Spektrums bei diesen Frequenzen null.
Daher und wegen der Reellwertigkeit von \mbox{$f(k)$} gen"ugt es das
Spektrum der Fensterfolge bei den \mbox{$N\!-\!A_1$} Frequenzen
\mbox{$\Omega=\nu\CdoT2\pi/F$} mit \mbox{$\nu=0\;(1)\;N\!-\!1\!-\!A_1$}
zu berechnen. Dies sind $A_1$ Werte weniger als bei der Fensterfolge nach 
Kapitel \myref{Algo}. In Gleichung (\myref{6.16}) ist daher in den Grenzen des 
Summationsindex \mbox{$N\!-\!1$} durch \mbox{$N\!-\!1\!-\!A_1$} zu ersetzen.

Die L"ange der zu berechnenden Fensterfolge \mbox{$f(k)$} kann auch hier
angegeben werden. Die L"ange der Basisfensterfolge ist \mbox{$F\!-\!A_0$}.
Das Polynom \mbox{$z^{F-A_0-1}\Cdot G(z)\CdoT G(z^{\!-1})$} ist dann vom Grad
\mbox{$2\CdoT(F\!-\!A_0\!-\!1)$}. Da die Koeffizienten \mbox{$d(k)$} des
Polynoms \mbox{$z^{F-A_0-1}\Cdot D(z)$} nach Gleichung (\myref{6.10}) durch
Multiplikation der Fensterautokorrelationsfolge \mbox{$g_Q(k)$}
mit der periodisch fortgesetzten si-Funktion entstanden sind, ist
der Grad von \mbox{$z^{F-A_0-1}\Cdot D(z)$} ebenfalls gleich
\mbox{$2\CdoT(F\!-\!A_0\!-\!1)$}. Durch Aufspaltung dieses Polynoms
in den minimalphasigen Anteil und den Rest entsteht das Polynom
\mbox{$z^{F-A_0-1}\Cdot F\big(\!-j\Cdot\ln(z)\big)$} von halbem Grad.
Da die Koeffizienten dieses Polynoms die Werte der Fensterfolge
\mbox{$f(k)$} sind, ist deren L"ange gleich der L"ange der
Basisfensterfolge also \mbox{$F\!-\!A_0$}.

Nach Gleichung (\myref{6.7}) und Gleichung (\myref{6.13}) erh"alt man aus
\mbox{$|G(z)|^2$} durch Verschiebung und "Uberlagerung das Betragsquadrat des
Spektrums der Fensterfolge. Wenn man f"ur \mbox{$G_1(z)$} Gleichung (\ref{E.10.5})
und f"ur \mbox{$G_2(z)$} Gleichung (\ref{E.10.6}) eingesetzt, erh"alt man analog zu 
den Gleichungen (\myref{6.17}) und (\myref{6.19}) daf"ur: 
\begin{gather}
F\big(\!-j\Cdot\ln(z)\big)\CdoT F\big(j\Cdot\ln(z)\big)\;\sim\;
D(z)\;=
\label{E.10.8}\\[6pt]
=\Sum{\nu_1=\frac{1-N}{2}}{\frac{N-1}{2}}\!
G_1\big(z\CdoT e^{\!-j\cdot\frac{2\pi}{F}\cdot\nu_1}\big)\CdoT
G_1\big(z^{\!-1}\!\CdoT e^{j\cdot\frac{2\pi}{F}\cdot\nu_1}\big)\CdoT
G_2\big(z\CdoT e^{\!-j\cdot\frac{2\pi}{F}\cdot\nu_1}\big)\CdoT
G_2\big(z^{\!-1}\!\CdoT e^{j\cdot\frac{2\pi}{F}\cdot\nu_1}\big)\;=
\notag\\[10pt]\begin{align}
=\!\Sum{\nu_1=\frac{1-N}{2}}{\frac{N-1}{2}}&
\frac{\D z^F\!\CdoT e^{\!-j\cdot2\pi\cdot\nu_1}+(-1)^N}
{\T\Prod{\nu_2=\frac{1-N}{2}+A_1}{\frac{N-1}{2}-A_1}\!
\Big(z\CdoT e^{\!-j\cdot\frac{2\pi}{F}\cdot\nu_1}-
e^{j\cdot\frac{2\pi}{F}\cdot\nu_2}\Big)}\cdoT
\!\!\Prod{\rho=1}{N-1-A_0-2\cdoT A_1}\!
\Big(z\CdoT e^{\!-j\cdot\frac{2\pi}{F}\cdot\nu_1}-z_{0,\rho}\Big)\cdot{}
\tag{\mbox{$\ast$}}\\*[6pt]
{}\cdot\;{}&
\frac{z^{\!-F}\!\CdoT e^{j\cdot2\pi\cdot\nu_1}+(-1)^N}
{\T\Prod{\nu_2=\frac{1-N}{2}+A_1}{\frac{N-1}{2}-A_1}\!
\Big(z^{\!-1}\!\CdoT e^{j\cdot\frac{2\pi}{F}\cdot\nu_1}-
e^{j\cdot\frac{2\pi}{F}\cdot\nu_2}\Big)}\cdoT
\!\!\Prod{\rho=1}{N-1-A_0-2\cdoT A_1}\!
\Big(z^{\!-1}\!\CdoT e^{j\cdot\frac{2\pi}{F}\cdot\nu_1}-z_{0,\rho}\Big)\;=
\rule{14pt}{0pt}\notag\end{align}\notag\\[18pt]\begin{flalign*}
&=\;\underbrace{
\frac{-z^F\!+2-z^{\!-F}}{\T\;\Prod{\nu_3=1-N+A_1}{N-1-A_1}\!\!
\Big(\!-z\CdoT e^{\!-j\cdot\frac{2\pi}{F}\cdot\nu_3}+2-
z^{\!-1}\!\CdoT e^{j\cdot\frac{2\pi}{F}\cdot\nu_3}\Big)\;}}_{
\D=D_E(z)}\;\cdot{}&&
\end{flalign*}\notag\\*[6pt]
{}\cdot\Sum{\nu_1=\frac{1-N}{2}}{\frac{N-1}{2}}\;\Prod{\nu_2}{}
\Big(\!-z\CdoT e^{\!-j\cdot\frac{2\pi}{F}\cdot\nu_2}+2-
z^{\!-1}\!\CdoT e^{j\cdot\frac{2\pi}{F}\cdot\nu_2}\Big)\cdot{}\rule{43pt}{0pt}
\notag\\*[8pt]\begin{flalign*}
&&\underbrace{\rule{70pt}{0pt}
{}\cdot\!\!\Prod{\rho=1}{N-1-A_0-2\cdoT A_1}\bigg(
\Big(z\CdoT e^{\!-j\cdot\frac{2\pi}{F}\cdot\nu_1}-z_{0,\rho}\Big)\CdoT
\Big(z^{\!-1}\!\CdoT e^{j\cdot\frac{2\pi}{F}\cdot\nu_1}-z_{0,\rho}^*\Big)\bigg)}
_{\D=D_{\overline{E}}(z)}&
\end{flalign*}\notag
\end{gather}
Setzt man wieder \mbox{$z=e^{j\cdot\frac{2\pi}{F}\cdot\nu}$} mit
\mbox{$\nu=0\;(1)\;N\!-\!1\!-\!A_1$} in die mit
$(\ast)$ gekennzeichnete Form ein, so erh"alt man f"ur das Betragsquadrat
des Spektrums der Fensterfolge bis auf einen konstanten
Faktor den gegen"uber Gleichung (\myref{6.18}) umfangreicheren
Ausdruck
\begin{gather}
\big|F\big({\T\nu\CdoT\frac{2\pi}{F}}\big)\big|^2\;=\;
F\big({\T\nu\CdoT\frac{2\pi}{F}}\big)\CdoT
F\big({\T-\nu\CdoT\frac{2\pi}{F}}\big)\;\sim\;
D\big(e^{j\cdot\frac{2\pi}{F}\cdot\nu}\big)\;=
\label{E.10.9}\\*[10pt]\begin{flalign*}
& F^2\cdot 4^{1-N+2\cdoT A_1}\cdoT\!\!\!
\Sum{\nu_1=\max(\nu+A_1,0)-\frac{N-1}{2}}{\min(\nu-A_1,0)+\frac{N-1}{2}}\;\;
\Prod{\substack{\nu_2=-\frac{N-1}{2}+A_1\\\nu_2\neq\nu-\nu_1}}
{\frac{N-1}{2}-A_1}\!\!
\sin\!\big({\T(\nu\!-\!\nu_2\!-\!\nu_1)\CdoT\frac{\pi}{F}}\big)^{\!-2}\cdot{}&&
\end{flalign*}\notag\\*[8pt]\begin{flalign*}
&&{}\cdot\!\!
\Prod{\rho=1}{N-1-A_0-2\cdoT A_1}\!
\Big(\big(1\!-\!|z_{0,\rho}|\big)^{\uP{0.3}{\!2}}\!+4\CdoT|z_{0,\rho}|\CdoT
\sin\!\big({\T(\nu\!-\!\nu_1)\CdoT
\frac{\pi}{F}-\frac{\psi_{0,\rho}}{2}}\big)^{\uP{0.3}{\!2}}\Big).&
\end{flalign*}\notag
\end{gather}
Zum einen sind hier die Abstandsquadrate zu den
frei w"ahlbaren Nullstellen $z_{0,\rho}$, deren Nullstellenwinkel
mit $\psi_{0,\rho}$ bezeichnet seien, neu hinzugekommen. Zum anderen wurden 
die Grenzen der Summen- und Produktindizes modifiziert, weil durch die\pagebreak[3] 
\mbox{$2\CdoT\!A_1$} zus"atzlichen Nullstellen am Einheitskreis nun
mehr Summanden null sind, und weniger Faktoren in den einzelnen
Summanden vorhanden sind. Auch hier braucht der nun modifizierte
Vorfaktor \mbox{$F^2\cdot 4^{1-N+2\cdoT A_1}$} wegen der abschlie"senden
Normierung auf \mbox{$F(0)\!=\!M$} nicht explizit berechnet zu werden.
Die positiven Wurzeln der mit der letzten Gleichung berechneten Werte
sind bis auf einen konstanten Faktor die Betr"age der Spektralwerte
und somit im wesentlichen die Betr"age Fourierreihenkoeffizienten der
Fensterfolge.

Um in Gleichung (\ref{E.10.8}) von der Form $(\ast)$ auf die endg"ultige
Form zu kommen, wurde wieder auf den gemeinsamen Hauptnenner
aller Summanden erweitert. Dieser wurde zusammen mit dem von
$\nu_1$ unabh"angigen Z"ahler vor die Summe gezogen.
Der Lauf"|index $\nu_2$ des in der Summe stehenden Produkts nimmt bei jedem
Summanden jeweils wieder die \mbox{$N\!-\!1$} Werte an, die das Produkt
entstehen lassen, das zur Erweiterung auf den Hauptnenner ben"otigt wird.
Beim Summanden mit dem Lauf"|index $\nu_1$ nimmt $\nu_2$ also die Werte
\mbox{$1\!-\!N\!+\!A_1\;(1)\;N\!-\!1\!-\!A_1$} ohne die Werte
\mbox{$\nu_1\!+\!A_1\!-\!(N\!-\!1)/2\;\;(1)\;\;\nu_1\!-\!A_1\!+\!(N\!-\!1)/2$}
an.

Der Anteil \mbox{$D_E(z)$} ist mit
\mbox{$z^{F-2\cdot N+2\cdot A_1+1}$}
multipliziert ein Polynom aus doppelten Nullstellen am
Einheitskreis bei \mbox{$\Omega=\nu\CdoT2\pi/F$} mit
\mbox{$\nu=N\!-\!A_1\;(1)\;F\!-\!N\!+\!A_1$}. Jede dieser
doppelten Nullstellen tritt als einfache Nullstelle im
minimalphasigen Anteil von \mbox{$D_E(z)$} auf. Die Phase
\mbox{$(F\!-\!2\CdoT\!N\!+\!1\!+\!2\CdoT\!A_1)\CdoT\Omega/2$} dieses
Anteils ist linear. Die Gruppenlaufzeit ist nun um $A_1$ gr"o"ser als
bei der in \cite{Diss} vorgestellten Fensterfolge. Im Bereich
der interessierenden Frequenzen \mbox{$|\Omega|<(N\!-\!A_1)\CdoT2\pi/F$}
besitzt \mbox{$D_E(z)$} keine Nullstellen, so dass die Phase des
minimalphasigen Anteils von \mbox{$D_E(z)$} in diesem Intervall
keine $\pi$-Spr"unge hat.

Die Summe \mbox{$D_{\overline{E}}(z)$} ist auch hier positiv und weist am
Einheitskreis keine Nullstellen auf, weil dort alle Summanden nichtnegativ sind,
und nie alle $N$ Summanden gleichzeitig null werden. Dies ist dadurch
sichergestellt, dass f"ur \mbox{$z=e^{\pm j\cdot(N-1-2\cdoT A_1)\cdot\pi/F}$}
das Polynom \mbox{$z^{N-1-A_0-2\cdoT A_1}\Cdot G_2(z)$} von null verschieden
ist (\,siehe Einschr"ankung f"ur die Wahl von $z_{0,\rho}$ im Anschluss an 
Gleichung (\ref{E.10.4})\,), und nur maximal \mbox{$N-1$} Nullstellen frei 
gew"ahlt werden k"onnen, und somit bei $N$ unterschiedlichen Rotationen 
nicht bei allen Summanden eine Nullstelle auf demselben Punkt des Einheitskreises 
liegen kann. Um bei der Berechnung der Phase des minimalphasigen Anteils von
\mbox{$D_{\overline{E}}(z)$} wieder ein Cepstrum brauchbarer L"ange
zu erhalten empfiehlt es sich auch hier vorher eine Bilineartransformation
mit der Substitution von $z$ nach Gleichung (\myref{6.20}) durchzuf"uhren.
\begin{gather*}\label{E.10.10}
\begin{flalign}
&D_{\overline{E}}(z)\;=
\Sum{\nu_1=\frac{1-N}{2}}{\frac{N-1}{2}}\;\Prod{\nu_2}{}
\Big(\!-z\CdoT e^{\!-j\cdot\frac{2\pi}{F}\cdot\nu_2}+2-
z^{\!-1}\!\CdoT e^{j\cdot\frac{2\pi}{F}\cdot\nu_2}\Big)\cdot{}&&
\end{flalign}\\*[-8pt]\begin{flalign*}
&&{}\cdoT\Prod{\rho=1}{N-1-A_0-2\cdoT A_1}\bigg(
\big(z\CdoT e^{\!-j\cdot\frac{2\pi}{F}\cdot\nu_1}\!-\!z_{0,\rho}\big)\CdoT
\big(z^{\!-1}\!\CdoT e^{j\cdot\frac{2\pi}{F}\cdot\nu_1}\!-\!z_{0,\rho}^*\big)\bigg)\;=&
\end{flalign*}\\[10pt]\begin{flalign*}
&=\;\underbrace{\bigg(\frac{\Tilde{z}}{\,(1\!+\!(1\!-\!c)\CdoT \Tilde{z})\CdoT
(\Tilde{z}\!+\!(1\!-\!c))\,}\bigg)^{\!\!2\cdot(N-1-A_1)-A_0}\!\!}_{
\D=\widetilde{D}_P(\Tilde{z})}\;\cdot{}&&
\end{flalign*}\\*[-4pt]\begin{flalign*}
&&{}\!\!\!\!\!\CdoT\underbrace{
\Sum{\nu_1=\frac{1-N}{2}}{\frac{N-1}{2}}\Prod{{}\;\nu_2}{}\!
\big(K_{\nu_2}\!\CdoT(\Tilde{z}\!-\!\Tilde{z}_{\nu_2})\CdoT
(\Tilde{z}^{-1}\!\!\!-\!\Tilde{z}_{\nu_2}^*)\big)
\,\cdoT\!\!\!\!\!\!\!
\Prod{\rho=1}{N-1-A_0-2\cdoT A_1}\!\!\!\!\!\!\!\!\!\!\!\!\big(
(\Tilde{z}\CdoT \Tilde{z}_{N,\rho,\nu_1}\!-\!\Tilde{z}_{Z,\rho,\nu_1})\CdoT
(\Tilde{z}^{-1}\!\CdoT\Tilde{z}_{N,\rho,\nu_1}^*\!-\!\Tilde{z}_{Z,\rho,\nu_1}^*)\big)}
_{\D=\widetilde{D}_N(\Tilde{z})}.&
\end{flalign*}
\end{gather*}
Wieder sind die Nullstellen $\Tilde{z}_{\nu_2}$ die  Nullstellen, die 
durch die Bilineartransformation der Nullstellen am Einheitskreis bei
\mbox{$z\!=\!e^{j\cdot\frac{2\pi}{F}\cdot\nu_2}$} entstanden sind.
Sie berechnen sich weiterhin nach Gleichung (\myref{6.24}) mit dem
Nullstellenwinkelanteil $\Tilde{\psi}_{\nu_2}$ nach Gleichung (\myref{6.26}).
Auch die Konstante $K_{\nu_2}$ berechnet sich unver"andert nach
Gleichung (\myref{6.25}). Neu hinzugekommen ist das Produkt der Faktoren mit den
Nullstellen bei \mbox{$\Tilde{z}_{Z,\rho,\nu_1}/\Tilde{z}_{N,\rho,\nu_1}$}
und \mbox{$\Tilde{z}_{N,\rho,\nu_1}^*/\Tilde{z}_{Z,\rho,\nu_1}^*$}.
Diese Nullstellen berechnen sich als die mit Gleichung (\myref{6.21})
Bilineartransformierten der mit dem Drehfaktor
\mbox{$e^{j\cdot\frac{2\pi}{F}\cdot\nu_1}$} multiplizierten
Nullstellen $z_{0,\rho}$.
\begin{gather}
\Tilde{z}_{Z,\rho,\nu_1}=
z_{0,\rho}\cdot e^{j\cdot\frac{2\pi}{F}\cdot\nu_1}-(1\!-\!c)
\quad\text{und}\quad
\Tilde{z}_{N,\rho,\nu_1}=
1-(1\!-\!c)\cdot z_{0,\rho}\cdot e^{j\cdot\frac{2\pi}{F}\cdot\nu_1}
\label{E.10.11}
\end{gather}
Im weiteren sei $\Tilde{\psi}_{0,\rho,\nu_1}$ der Winkel der Nullstelle
\mbox{$\Tilde{z}_{Z,\rho,\nu_1}/\Tilde{z}_{N,\rho,\nu_1}$}, der sich
als Differenz der Winkel des Z"ahlers \mbox{$\Tilde{z}_{Z,\rho,\nu_1}$}
und des Nenners \mbox{$\Tilde{z}_{N,\rho,\nu_1}$} der Nullstelle
mit Hilfe der $4$-Quadran\-ten Arkustangensfunktion berechnen l"asst.

In Gleichung (\ref{E.10.10}) steht wieder ein Term vor der
Summe, dessen Nullstellen alle bei \mbox{$\Tilde{z}\!=\!0$}, und 
dessen Polstellen bei \mbox{$\Tilde{z}=c\!-\!1$} und
\mbox{$\Tilde{z}=1/(c\!-\!1)$} liegen. Dieser Anteil liefert f"ur
das minimalphasige Polynom wieder einen Phasenbeitrag, der sich nach
Gleichung (\myref{6.28}) berechnet, und der hier mit der 
ge"anderten Vielfachheit \mbox{$2\CdoT(N\!-\!1\!-\!A_1)\!-\!A_0$}
statt \mbox{$N\!-\!1$} in \cite{Diss} multipliziert zu addieren ist.

Nun m"ussen wir wieder die Phase des minimalphasigen Anteils des
Summenanteils \mbox{$\widetilde{D}_N(\Tilde{z})$} in Gleichung (\ref{E.10.10})
berechnen. Diese Summe ist f"ur \mbox{$\Tilde{z}=e^{j\widetilde{\Omega}}$}
wieder positiv reell und l"asst sich mit
\mbox{$\widetilde{\Omega}=\eta\CdoT2\pi/\widetilde{M}$} f"ur alle
ganzzahligen Werte $\eta$ bis auf einen konstanten Faktor in der Form
\begin{gather*}\label{E.10.12}
\begin{flalign}
&\widetilde{D}_N(e^{j\widetilde{\Omega}})\;\sim\!
\Sum{\nu_1=\frac{1-N}{2}}{\frac{N-1}{2}}\;
\Prod{\nu_2}{}\Big(K_{\nu_2}\Cdot\sin\!\big({\T
\frac{\widetilde{\Omega}}{2}-\nu_2\CdoT\frac{\pi}{F}-
\Tilde{\psi}_{\nu_2}}\big)^{\!2}\Big)\cdot{}&&
\end{flalign}\\*[6pt]\begin{flalign*}
&&{}\cdoT\!\Prod{\rho=1}{N-1-A_0-2\cdoT A_1}\!\!
\Big(\big(\,|\Tilde{z}_{N,\rho,\nu_1}|\!-\!
|\Tilde{z}_{Z,\rho,\nu_1}|\,\big)^{\!2}\!+
4\CdoT|\Tilde{z}_{N,\rho,\nu_1}|\CdoT
|\Tilde{z}_{Z,\rho,\nu_1}|\CdoT\sin\!\big({\T
\frac{\widetilde{\Omega}-\Tilde{\psi}_{0,\rho,\nu_1}}{2}}\big)^{\!2}\Big)\;=&
\end{flalign*}\\[10pt]\begin{flalign*}
&=\!\Sum{\nu_1=\frac{1-N}{2}}{\frac{N-1}{2}}\;
\Prod{\nu_2}{}\bigg(K_{\nu_2}\!\CdoT
\sin\!\bigg(\frac{\pi}{\widetilde{M}}\CdoT
\Big(\eta\!-\!\widetilde{M}\CdoT\big({\T\frac{\nu_2}{F}\!+\!
\frac{\Tilde{\psi}_{\nu_2}}{\pi}}\big)\Big)\bigg)^{\!\!2}\,\bigg)\cdot{}&&
\end{flalign*}\\*[4pt]\begin{flalign*}
&&{}\cdoT\!\Prod{\rho=1}{N-1-A_0-2\cdoT A_1}\!\!
\bigg(\big(\,|\Tilde{z}_{N,\rho,\nu_1}|\!-\!
|\Tilde{z}_{Z,\rho,\nu_1}|\big)^{\!2}\!+
4\CdoT|\Tilde{z}_{N,\rho,\nu_1}|\CdoT|\Tilde{z}_{Z,\rho,\nu_1}|\CdoT
\sin\!\Big({\T\frac{\pi}{\widetilde{M}}\CdoT\big(\eta\!-\!
\frac{\widetilde{M}}{2\pi}\CdoT
\Tilde{\psi}_{0,\rho,\nu_1}\big)}\Big)^{\!\!2}\,\bigg)&
\end{flalign*}
\end{gather*}
berechnen. Auch hier kann bei jedem Faktor wieder eine Reduktion der ganzen
Zahl $\eta$ um ein Vielfaches von $\widetilde{M}$ in der Art vorgenommen werden,
dass der Betrag des Arguments der Sinusfunktion betraglich kleiner
als \mbox{$\pi/2$} bleibt. Es zeigte sich jedoch, dass man auf
diese Art der $2\pi$-Reduktion verzichten kann, da dies zu keiner
nennenswerten Zunahme der Fehler in der Fensterfolge f"uhrt.
Der konstanter Faktor, der bisher nicht
festgelegt wurde, wird wieder zur Optimierung der Genauigkeit bei
der Berechnung des Logarithmus genutzt.

Eine Besonderheit ergibt sich bez"uglich der Genauigkeit der Berechnung
des Z"ahlers und des Nenners der bilineartransformierten Nullstelle
$z_{0,\rho}$. Nach Gleichung (\ref{E.10.11}) werden diese Gr"o"sen als Differenz
berechnet. Daher wird der relative Fehler dieser Gr"o"sen sehr gro"s, wenn
deren Wert selbst sehr klein wird. Dadurch kann auch der Nullstellenwinkel
in diesen F"allen nicht mehr exakt berechnet werden. Da jedoch der
Beitrag einer frei gew"ahlten Nullstelle $z_{0,\rho}$ zur Summe, die das
Betragsquadratspektrum bildet, immer dann klein wird, wenn
$\Tilde{z}_{Z,\rho,\nu_1}$ oder $\Tilde{z}_{N,\rho,\nu_1}$ besonders
schlecht berechnet werden kann, wirkt sich diese Besonderheit
nicht wesentlich auf die Genauigkeit der Berechnung der Phase aus.

Nach Gleichung (\myref{6.32}) liefert uns eine 
inverse FFT des nat"urlichen Logarithmus des Terms 
\mbox{$\widetilde{D}_N\big(e^{j\cdot\eta\cdot\frac{2\pi}{\widetilde{M}}}\big)$} 
das doppelte Cepstrum. Damit das Cepstrum m"oglichst rasch abklingt, muss der 
Parameter $c$ geeignet gew"ahlt werden. Da der optimale Parameter $c$ von der 
Lage der nach der "Uberlagerung vorhandenen, unbekannten Nullstellen $\zu$ 
abh"angig ist, die wiederum von den frei gew"ahlten Nullstellen $z_{0,\rho}$ 
abh"angen, kann hier keine optimale Einstellung empirisch gewonnen werden.

Ich schlage daher vor, den Bilineartransformationsparameter $c$ und
die L"ange $\widetilde{M}$ der inversen FFT nach den Gleichungen
(\myref{6.30}) und (\myref{6.31}) einzustellen, die Berechnung des Cepstrums
durchzuf"uhren, zu kontrollieren ob die L"ange der inversen FFT zu
kurz war und gegebenenfalls die L"ange der inversen FFT zu verdoppeln,
den Parameter $c$ zu ver"andern und die Berechnung der Summe
\mbox{$\widetilde{D}_N\big(e^{j\widetilde{\Omega}}\big)$} nach
Gleichung (\ref{E.10.12}) zu wiederholen. Dies wird solange durchgef"uhrt,
bis ein Abbruchskriterium feststellt, dass die L"ange der inversen
FFT gro"s genug ist. Um ein Abbruchskriterium zu erhalten, wird der
Betrag des berechneten Cepstrums mit einer abfallenden Grenze verglichen,
die der Betrag des Cepstrums nicht "uberschreiten darf. Als Grenze wird die
feinere der beiden oberen Absch"atzungen nach Ungleichung (\myref{6.38}) f"ur
den Betrag eines Cepstrums verwendet. Dabei wird als schlimmster Fall
angenommen, dass alle \mbox{$2\CdoT(N\!-\!1\!-\!A_1)\!-\!A_0$} Nullstellen
des minimalphasigen Polynoms, von dem das Cepstrum berechnet wird,
den gleichen Betrag haben. Damit ergibt sich f"ur die Grenze
\mbox{${\D(2\CdoT(N\!-\!1\!-\!A_1)\!-\!A_0)/k \cdot|\zu|^k\!}$}.
Der Betrag der unbekannten Nullstelle in dieser Grenze wird nun
so gew"ahlt, dass die Grenze f"ur \mbox{$k=\widetilde{M}/2$} auf
einen Sch"atzwert $\Bar{\sigma}$ f"ur den Mindestwert der Streuung des
Rauschsockels abgefallen ist, der bei der Berechnung des Cepstrums mit
der inversen FFT unweigerlich entsteht. F"ur die Grenze ergibt sich daher:
\begin{equation}
\big|C(k)\big|\;\stackrel{\T!}{<}\;
\frac{{}\;2\CdoT\! N\!-\!2\!-\!2\CdoT\!A_1\!-\!A_0\;{}}{k}\cdot\bigg(
\frac{\widetilde{M}\cdot\Bar{\sigma}}
{{}\;2\CdoT(2\CdoT\! N\!-\!2\!-\!2\CdoT\! A_1\!-\!A_0)\;{}}
\bigg)^{\T\!\frac{2\cdot k}{\widetilde{M}}}.
\label{E.10.13}
\end{equation}
Um einen Sch"atzwert f"ur den Mindestwert der Streuung des
Rauschsockels zu erhalten, wird angenommen, dass der Logarithmus von
\mbox{$\widetilde{D}_N\big(e^{j\widetilde{\Omega}}\big)$} f"ur jede Frequenz
mit einer Varianz, die sich mit dem "`$Q^2/12\,$"'-Modell berechnet,
verrauscht ist. Als Quantisierungsstufenh"ohe wird
\mbox{$\max\!\big(\,\big|\ln\!\big(
\widetilde{D}_N\big(e^{j\widetilde{\Omega}}\big)\,\big)\big|\,,
 1\,\big)\cdot\varepsilon$} eingesetzt (\,siehe Anhang \myref{log}\,).
Diese Varianz wird "uber alle  $\widetilde{M}$ Frequenzen gemittelt.
Die mittlere Rauschleistung, die sich nach einer
idealen FFT --- ohne Quantisierung der Drehfaktoren und ohne Arithmetikfehler
--- ergibt, ist um den Faktor \mbox{$1/\widetilde{M}$} kleiner. Die halbe
Wurzel der mittleren Rauschleistung nach der idealen FFT wird als Sch"atzwert
f"ur den Mindestwert der Streuung des Rauschsockels im Cepstrum
verwendet.
\begin{equation}
\Bar{\sigma}\;=\;\frac{\varepsilon}{{}\;\widetilde{M}\CdoT\sqrt{48\;}\;{}}\cdot
\sqrt{\Sum{\eta=0}{\widetilde{M}-1}\max\!\bigg(\ln\!\Big(
\widetilde{D}_N\big(e^{j\cdot\frac{2\pi}{\widetilde{M}}\cdot\eta}\big)
\Big)^{\!2}\!, 1\bigg)\;}
\label{E.10.14}
\end{equation}
Dabei ber"ucksichtigt der Faktor \mbox{$1/2$}, dass das Cepstrum sich
aus dem Betrag des minimalphasigen Anteils berechnet, und nicht aus
den Betragsquadrat. Haben alle unbekannten Nullstellen des Spektrums,
dessen Cepstrum berechnet wurde, einen Betrag, der kleiner ist als der
f"ur die Berechnung der Grenze verwendete Nullstellenbetrag, so liegt
der Betrag des Cepstrums sicher unterhalb der Grenze. Ist jedoch auch nur
eine der unbekannten Nullstellen betraglich gr"o"ser, so wird das Cepstrum
die Grenze fr"uher oder sp"ater "uberschreiten, da dann das berechnete
Cepstrum einen im wesentlichen exponentiellen Anteil besitzt, der langsamer
abklingt als die Grenze. Sofern dieses "Uberschreiten der Grenze nicht
stattfindet, bzw. erst stattfindet, wenn sowohl das Cepstrum als auch die
Grenze unterhalb eines Sch"atzwertes $\Hat{\sigma}$ f"ur den maximal m"oglichen
Rauschwert liegen, wird angenommen, dass die L"ange der inversen FFT bei der
Berechnung des Cepstrums gro"s genug gew"ahlt war. In diesem Fall kann mit der
Berechnung der Phase durch die Auswertung der Sinusreihe wie in \cite{Diss} 
fortgefahren werden.

Um einen Sch"atzwert f"ur dem maximal m"oglichen Rauschwert im Cepstrum
zu erhalten, nehmen wir an, dass der Logarithmus von
\mbox{$\widetilde{D}_N\big(e^{j\widetilde{\Omega}}\big)$} bei jeder Frequenz,
bei der er berechnet worden ist, einen absoluten Fehler der Gr"o"se
\mbox{$\max\!\big(\,\big|\ln\!\big(
\widetilde{D}_N\big(e^{j\widetilde{\Omega}}\big)
\,\big)\big|\,, 1\,\big) \cdot\varepsilon$} aufweist. Schlimmstenfalls
k"onnen sich die absoluten Fehler aller \mbox{$\widetilde{M}$}
Frequenzen bei der Berechnung der FFT betragsm"a"sig addieren. 
Wenn eine Radix-\mbox{$\log_2(\widetilde{M})$} FFT durchgef"uhrt
wird, entstehen in den einzelnen Stufen zus"atzliche Fehler. Als
"ubelsten Fall nehmen wir an, dass sich bei jeder Stufe der FFT ein
zus"atzlicher absoluter Fehler betraglich addiert, der maximal
genausogro"s ist, wie die ebengenannte Summe der Betr"age der absoluten
Fehler der Logarithmuswerte aller Frequenzen. Da bei der inversen
FFT der Faktor \mbox{$1/\widetilde{M}$} auftritt, wird sich zum einen maximal
noch einmal derselbe Fehler addieren, und zum anderen auch der absolute
Fehler entsprechend mit diesem Faktor verringern. Daher kann man absch"atzen,
dass der absolute Fehler einen Maximalwert von
\begin{equation}
\Hat{\sigma}\;=\;
\frac{\;\varepsilon\cdot\big(\,2\!+\!\log_2(\widetilde{M})\,\big)\;}
{2\cdot\widetilde{M}}\,\cdot
\Sum{\eta=0}{\widetilde{M}-1}\max\!\bigg(\,\Big|\ln\!\Big(
\widetilde{D}_N\big(e^{j\cdot\frac{2\pi}{\widetilde{M}}\cdot\eta}\big)
\Big)\Big|\,, 1\bigg)
\label{E.10.15}
\end{equation}
nicht "uberschreiten wird. Wieder ber"ucksichtigt der Faktor \mbox{$1/2$},
dass das Cepstrum halb so gro"s ist, wie die Werte, die man durch eine
FFT aus den Logarithmen der Betragsquadrate der Spektralwerte des
minimalphasigen Anteils erh"alt. Durch diesen relative hoch gesch"atzten
maximal m"oglichen Wert, der durch Rauschen im Cepstrum entstehen kann,
wird vermieden, dass ein "`Ausrei"ser"' im Rauschsockel zu einer unn"otigen
Verl"angerung der FFT-L"ange $\widetilde{M}$ f"uhrt. Auch ergaben sich dadurch
keine Probleme, weil der Betrag des berechneten Cepstrums die Grenze im
Falle einer falschen FFT-L"ange immer bereits deutlich vor erreichen des
Sch"atzwertes f"ur das Maximum des Rauschsockels "uberschritten hat. Selbst
bei kritischen FFT-L"angen $\widetilde{M}$ war das Cepstrum bei diesem
Abbruchkriterium immer bis zum wahren Rauschsockel abgesunken, wenn die
FFT-L"ange als hinreichend lang detektiert worden ist.

\begin{figure}[!t]
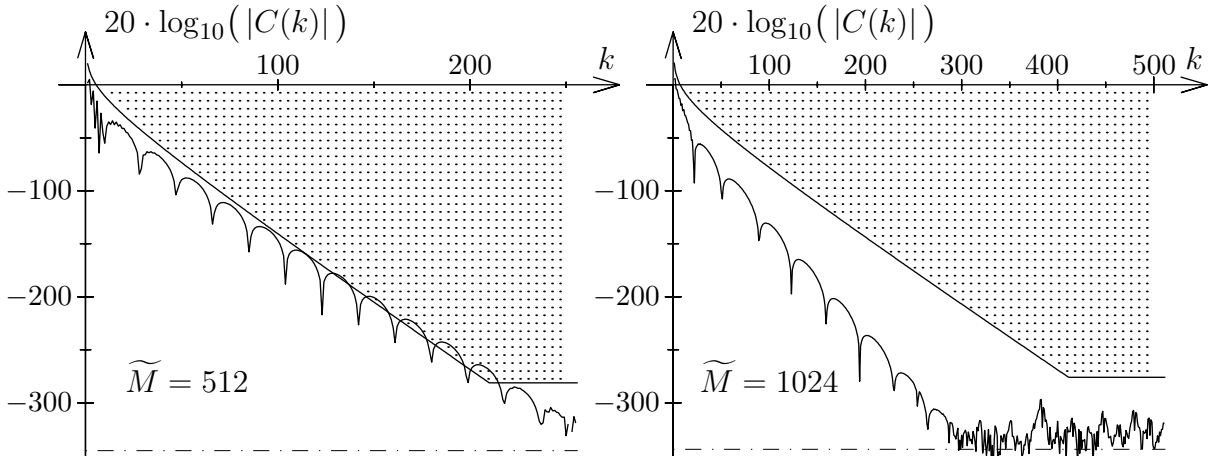

\begin{center}
{ 
\begin{picture}(450,171)

\input{mbild4f1.tex}
\put(37,163){\makebox(0,0)[l]{$
20\cdot\log_{10}\!\big(\,|C(k)|\,\big)$}}
\put(25,100){\makebox(0,0)[r]{\small$-100$}}
\put(25,60){\makebox(0,0)[r]{\small$-200$}}
\put(25,20){\makebox(0,0)[r]{\small$-300$}}
\put(102,144){\makebox(0,0)[b]{\small$100$}}
\put(174,144){\makebox(0,0)[b]{\small$200$}}
\put(225,146){\makebox(0,0)[b]{$k$}}
\put(45,30){\makebox(0,0)[l]{$\widetilde{M}=512$}}

\input{mbild4f2.tex}
\put(257,163){\makebox(0,0)[l]{$
20\cdot\log_{10}\!\big(\,|C(k)|\,\big)$}}
\put(245,100){\makebox(0,0)[r]{\small$-100$}}
\put(245,60){\makebox(0,0)[r]{\small$-200$}}
\put(245,20){\makebox(0,0)[r]{\small$-300$}}
\put(286,144){\makebox(0,0)[b]{\small$100$}}
\put(322,144){\makebox(0,0)[b]{\small$200$}}
\put(358,144){\makebox(0,0)[b]{\small$300$}}
\put(390,144){\makebox(0,0)[b]{\small$400$}}
\put(430,144){\makebox(0,0)[b]{\small$500$}}
\put(445,146){\makebox(0,0)[b]{$k$}}
\put(260,30){\makebox(0,0)[l]{$\widetilde{M}=1024$}}

\end{picture}}
\end{center}\vspace{-15pt}
\caption{Abklingen des Betrags des Cepstrums einer Fensterfolge.}
\label{E.b4f}
\rule{\textwidth}{0.5pt}\vspace{-10pt}
\end{figure}

In Bild \ref{E.b4f} ist links ein Grenzfall des Abklingens
des Betrags des Cepstrums abgebildet, bei dem die L"ange $\widetilde{M}$
gerade noch f"ur zu klein befunden wurde. Au"ser dem Betrag des Cepstrums
ist noch der Sch"atzwert f"ur die minimale Streuung des Rauschsockels,
sowie die nach unten auf dem Sch"atzwert f"ur den Maximal m"oglichen
Rauschwert begrenzte Grenze, anhand der entschieden wird ob das Cepstrum
schnell genug abklingt, dargestellt. Da in diesem Fall der Betrag des
Cepstrums auch in dem punktiert hervorgehobenen Gebiet liegt, wird die
FFT-L"ange verdoppelt. Bild \ref{E.b4f} zeigt rechts den sich danach
ergebenden Betrag des Cepstrums, der als rasch genug abklingend
erkannt wird. Bei der Berechnung dieser Cepstra waren die
Parameter \mbox{$N\!=\!5$}, \mbox{$M\!=\!256$}, \mbox{$A_0\!=\!0$},
\mbox{$A_1\!=\!1$} und die zwei frei w"ahlbaren Nullstellen
\mbox{$z_{0}=0,7\cdot e^{\pm0,\!5j}$} eingestellt worden. Zun"achst
betrug die FFT-L"ange \mbox{$\widetilde{M}\!=\!512$} und der
Parameter der Bilineartransformation \mbox{$c\!=\!0,\!06253$}.
Nach der Verdoppelung der FFT-L"ange waren die entsprechenden Werte
\mbox{$\widetilde{M}\!=\!1024$} und \mbox{$c\!=\!0,\!121$}.

Wird eine zu kleine FFT-L"ange $\widetilde{M}$ festgestellt,
so wird $\widetilde{M}$ verdoppelt, und es kann auch gleich der
Bilineartransformationsparameter $c$ ver"andert werden. Eine Ver"anderung
des Parameters $c$ entspricht einer weiteren Bilineartransformation der bereits
bilineartransformierten Funktion \mbox{$\widetilde{D}_N(\Tilde{z})$}. Der
Parameter dieser weiteren Bilineartransformation sei mit \mbox{$\Delta c$}
bezeichnet, und entspreche dem Term \mbox{$1\!-\!c$} bei der
Bilineartransformation nach Gleichung (\myref{6.20}) bzw. (\myref{6.21}).
Er liegt daher im Bereich zwischen $1$ und $-1$. Den neuen
Wert $c_{\text{neu}}$ erh"alt man aus \mbox{$\Delta c$}
und aus dem alten Wert $c_{\text{alt}}$ gem"a"s
\begin{equation}
c_{\text{neu}}\;=\;
\frac{c_{\text{alt}}\cdot(1\!-\!\Delta c)}
{{}\;1+(1\!-\!c_{\text{alt}})\CdoT\Delta c\;{}}
\label{E.10.16}
\end{equation}
Nun wird \mbox{$\Delta c$} so gew"ahlt, dass die betragsm"a"sig gr"o"ste
Nullstelle $\zu$ des Spektrums, dessen Cepstrum mit
$c_{\text{alt}}$ berechnet wurde, die also den Anteil
im Cepstrum hervorruft, der am langsamsten abklingt, weiter entfernt vom
Einheitskreis liegt. Der Abstand, den diese Nullstelle von Einheitskreis hat,
ist \mbox{$1\!-\!\max(\,|\zu|\,)$}. Nach der Verdoppelung der
FFT-L"ange und nach der "Anderung des Parameters $c$ kann u.U. eine andere
Nullstelle den gr"o"sten Betrag aufweisen. Der Abstand dieser Nullstelle vom
Einheitskreis darf dann niemals kleiner sein als der halbe Abstand vor der
"Anderung des Parameters $c$, da sonst durch die Verdoppelung der FFT-L"ange
nichts gewonnen w"are. Man kann n"aherungsweise zeigen, dass diese Bedingung
erf"ullt ist, wenn \mbox{$|\Delta c|<1/3$} ist. Desweiteren kann man zeigen,
dass sich die unbekannten Nullstellen bei einer "Anderung des
Bilineartransformationsparameters auf Kreisen bewegen, die durch
die Punkte \mbox{$z=\pm1$} gehen. Somit wird der Abstand einer
Nullstelle vom Einheitskreis am gr"o"sten, wenn die Nullstelle
durch die ver"anderte Bilineartransformation auf die imagin"are
Achse abgebildet wird. Au"serdem kann man wieder n"aherungsweise zeigen,
dass eine Nullstelle, die sehr nahe am Einheitskreis liegt und einen Winkel
$\psiu$ in der N"ahe von \mbox{$\pi/2$} hat, bei der Wahl von
\mbox{$\Delta c=1\!-\!\psiu\Cdot2/\pi$} auf die imagin"are Achse
abgebildet wird. \mbox{$\Delta c$} ist daher abh"angig vom Winkel der
betraglich gr"o"sten Nullstelle so zu w"ahlen, dass \mbox{$\Delta c$}
einerseits betraglich auf \mbox{$1/3$} begrenzt ist, und andererseits f"ur
Winkel in der N"ahe von \mbox{$\pi/2$} n"aherungsweise der lineare
Zusammenhang zwischen \mbox{$\Delta c$} und $\psiu$ gegeben ist. Die Funktion
\begin{equation}
\Delta c\;=\;\frac{1}{3}\cdot\sin\!\Big({\T
\big(\frac{2}{\pi}\big)^{\!2}\Cdot\psiu^{\uP{-0.6}{\!\!\!3}}\!-
\frac{6}{\pi}\cdot\psiu^{\uP{-0.6}{\!\!\!2}}\!+
\psiu+\frac{\pi}{2}}\Big)
\label{E.10.17}
\end{equation}
erf"ullt diese Bedingungen.

Um \mbox{$\Delta c$} berechnen zu k"onnen, muss zun"achst der Winkel dieser
unbekannten Nullstelle anhand des berechneten Cepstrums bestimmt werden. Das
Cepstrum, das mit der inversen FFT berechnet worden ist, ergibt sich als die
durch Faltung mit einem Impulskamm mit der Periode $\widetilde{M}$ periodisch
fortgesetzte Folge \mbox{$1/|k|\CdoT\sum\zu^{\uP{-0.4}{\!\!|k|}}$}.
Wenn man aus dieser periodischen Folge einen Ausschnitt mit
\mbox{$\widetilde{M}/4\le k<\widetilde{M}/2$} betrachtet, die
"Uberfaltungsfehler vernachl"assigt und mit $k$ multipliziert
erh"alt man die endliche Folge
\begin{equation}
\sum\;\zu^{\uP{-0.4}{\!\!k}}\;=\;
\sum\;\zu^{\uP{-0.1}{\!\!\T\frac{\widetilde{M}}{4}}}\cdot
\zu^{\uP{-0.1}{\!\!k-\frac{\scriptstyle\widetilde{M}}{\scriptstyle4}}}\;=\;
\sum\;\zu^{\uP{-0.1}{\!\!\frac{\widetilde{M}}{4}}}\cdot
\zu^{\uP{-0.4}{\!\!\Tilde{k}}},
\label{E.10.18}
\end{equation}
also mit der substituierten Zeitvariable \mbox{$\Tilde{k}=k-\widetilde{M}/4$},
die "Uberlagerung abklingender Exponentialfolgen, die mit
\mbox{$\zu^{\uP{-0.4}{\!\!\widetilde{M}/4}}$} gewichtet sind.
Die Z-Transformierte dieser Folge mit \mbox{$\Tilde{k}\ge0$}
entspricht der Partialbruchzerlegung
\mbox{$\sum\;\zu^{\uP{-0.6}{\!\widetilde{M}/4}}\cdot
z/(z\!-\!\zu)$}. F"ur \mbox{$z\!=\!e^{j\Omega}$} wird dieser
Ausdruck sein betragliches Maximum in der Gegend der Frequenz haben,
deren Polstelle am n"achsten am Einheitskreis liegt, da dann sowohl
der Vorfaktor \mbox{$\zu^{\uP{-0.6}{\!\widetilde{M}/4}}$} maximal, als auch
der Nenner \mbox{$e^{j\Omega}\!-\!\zu$} minimal wird. Dies trifft
vor allem dann zu, wenn die Nullstelle mit den gr"o"sten Betrag deutlich n"aher
am Einheitskreis liegt als alle anderen. F"uhrt man die Z-Transformation nur mit
der auf \mbox{$0\le\Tilde{k}<\widetilde{M}/4$} zeitlich begrenzen Folge durch,
so entspricht das im Spektrum der Faltung mit der periodischen fortgesetzten
si-Funktion mit dem Nullstellenabstand \mbox{$8\CdoT\pi/\widetilde{M}$}. Auch
das Maximum des gefalteten Spektrums liegt dann etwa bei derselben Frequenz.
F"ur die Frequenzen im Raster \mbox{$8\CdoT\pi/\widetilde{M}$} l"asst sich die
Z-Transformation durch eine FFT der L"ange \mbox{$\widetilde{M}/4$} berechnen.
Aus dem Ergebnis dieser FFT kann man die Lage des Maximums und damit einen
Sch"atzwert f"ur den Winkel der Nullstelle mit dem gr"o"sten Betrag
bestimmen. Den gesch"atzten Winkel setzt man in Gleichung (\ref{E.10.17}) ein und
das Ergebnis wiederum in Gleichung (\ref{E.10.16}) und erh"alt den neuen Wert
f"ur $c$. Mit der doppelten FFT-L"ange und dem modifizierten $c$ berechnet man
das Cepstrum erneut. Diesen Vorgang wiederholt man solange, bis das Cepstrum
die n"otige L"ange hat.

Nachdem das Cepstrum und damit die Phase analog zu der ersten Variante
berechnet worden ist, erh"alt man nach der Multiplikation mit dem Betrag
wieder die Spektralwerte \mbox{$F(\nu\CdoT2\pi/F)$}, aus welchen man
wieder die Fensterfolge \mbox{$f(k)$} gewinnt, indem man die Fourierreihe
nach Gleichung (\myref{6.16}) auswertet, wobei man die Anteile der \pagebreak[2]
Sinus- und Kosinusreihe wieder in umgekehrter Reihenfolge akkumuliert,
wie dies in \cite{Diss} beschrieben ist. Dabei sind
nun weitere \mbox{$2\CdoT\! A_1$} Fourierreihenkoeffizienten null,
so dass sich eine verk"urzte Reihe ergibt. Auch hier empfiehlt es sich
{\em nicht}, die letzten $A_0$ Werte der Fensterfolge abschlie"send zu null
zu setzen, auch wenn man wei"s, dass diese eigentlich exakt null sein
m"ussten, weil dadurch die Nullstellenbedingung (\myref{2.27}) schlechter
erf"ullt werden w"urde. Ein kommentierter, und auf das wesentliche
gek"urzter Auszug aus einem Programm, das Fensterfolgen mit dem hier 
beschriebenen Verfahren mit automatischer Bestimmung der ben"otigten 
FFT-L"ange $\widetilde{M}$ und des Bilineartransformationsparameters $c$ 
berechnet, ist in Kapitel \ref{E.Kap.11.1} abgedruckt.

\section{Berechnung des Spektrums der Fensterfolge}\label{E.Kap.10.2}

Will man das Spektrum der Fensterfolge f"ur beliebige Frequenzen
$\Omega$ berechnen, so k"onnte man entweder die Fourierreihe des
Spektrums, deren Koeffizienten die Werte der Fensterfolge sind,
f"ur die gew"unschten Frequenzen auswerten, oder die periodische
si-Funktion mit den bisher berechneten Fourierreihenkoeffizienten
\mbox{$F(\nu\CdoT2\pi/F)/F$} der Fensterfolge falten. Der absolute Fehler
ist aber im ersten Fall auf etwa \mbox{$M\CdoT\varepsilon$}, und im
zweiten Fall auf etwa \mbox{$\varepsilon/\big(N\CdoT\sin(\Omega/2)\big)$}
begrenzt, so dass man in beiden F"allen f"ur gro"se Sperrd"ampfung nur mehr
einen Rauschsockel berechnen kann, wenn $M$ gro"s ist und die Sperrd"ampfung
mit einer h"oheren Potenz in \mbox{$\sin(\Omega/2)\big)$} ansteigt. Es ist
daher sinnvoller den Betrag und die Phase des Spektrums getrennt
zu berechnen. Dies geschieht in "ahnlicher Art, wie auch die
Fourierreihenkoeffizienten \mbox{$F(\nu\CdoT2\pi/F)/F$} berechnet
worden sind. Dazu berechnet man die Phase wieder aus dem Cepstrum,
das mit dem ebengenannten Algorithmus berechnet wird, aber
diesmal nicht bei den Frequenzen \mbox{$\Omega=\nu\CdoT2\pi/F$}
mit \mbox{$\nu=1\;(1)\;N\!-\!1\!-\!A_1$}, sondern bei den Frequenzen,
f"ur die das Spektrum berechnet werden soll. Der Betrag wird durch
vorzeichenrichtiges radizieren des Betragsquadrats gewonnen.
Da \mbox{$F\big(\!-j\Cdot\ln(z)\big)$} am Einheitskreis nur einfache
Nullstellen im "aquidistanten Raster aufweist, ist das Vorzeichen
bei der Radizierung auf einfache Weise aus $\Omega$ bestimmbar.
Das Betragsquadrat von \mbox{$F(\Omega)$} wird durch "Uberlagerung
der verschobenen Betragsquadrate von \mbox{$G\big(e^{j\Omega}\big)$}
berechnet. Dabei ist jeweils die Polstelle von
\mbox{$G_1\big(z\CdoT e^{\!-j\cdot\frac{2\pi}{F}\cdot\nu_1}\big)$}
nach Gleichung (\ref{E.10.5}), die dem Punkt \mbox{$z\!=\!e^{j\Omega}$},
f"ur den das Spektrum berechnet werden soll, am n"achsten liegt,
mit der entsprechenden Nullstelle im Z"ahler zu k"urzen, so dass
sich f"ur \mbox{$z\!=\!e^{j\Omega}$} im Z"ahler die periodisch
fortgesetzte, und in der Frequenz entsprechend verschobene
si-Funktion ergibt. Deren Betragsquadrat ist mit dem Betragsquadrat
des Produktes der Abst"ande zu den restlichen Polstellen zu dividieren
und mit dem Betragsquadrat des Produktes der Abst"ande zu den verschobenen \pagebreak[3]
frei gew"ahlten Nullstellen zu multiplizieren. Da auf diese Weise eine
Summe von positiven Produkten, deren Faktoren mit h"ochstm"oglicher
relativer Genauigkeit berechnet worden sind, gebildet wird, ist so
das Spektrum auch f"ur den Sperrbereich akkurat berechenbar.
In Kapitel \ref{E.Kap.11.2} ist ein Programmkern angegeben, mit dessen 
Hilfe die Fouriertransformierte bis zu einer Sperrd"ampfung von etwa 
\mbox{$-10\cdot\log_{10}(\text{\tt realmin})$} berechnet werden kann. 
{\tt realmin} ist dabei die kleinste positive Zahl, die am Rechner 
als Gleitkommazahl mit voller Genauigkeit dargestellt werden kann. 

\begin{figure}[btp]
\begin{center}
{ 
\begin{picture}(450,600)

\input{mbild4g}
\put(246,581){\makebox(0,0)[b]{Variante 1: FFT}}
\put(15,544){\vector(0,1){32}}
\put(15,534){\makebox(0,0)[t]{\rotatebox{90}{$20\cdot\log_{10}\big(\,|F(\Omega)|\,\big)$}}}
\put(410,401){\vector(1,0){37}}
\put(405,401){\makebox(0,0)[r]{$\Omega$}}

\put(246,386){\makebox(0,0)[b]{Variante 2: Auswertung der DFT-Summe}}
\put(15,349){\vector(0,1){32}}
\put(15,339){\makebox(0,0)[t]{\rotatebox{90}{$20\cdot\log_{10}\big(\,|F(\Omega)|\,\big)$}}}
\put(410,206){\vector(1,0){37}}
\put(405,206){\makebox(0,0)[r]{$\Omega$}}

\put(246,191){\makebox(0,0)[b]{Variante 3: Berechnungsmethode laut Text}}
\put(15,154){\vector(0,1){32}}
\put(15,144){\makebox(0,0)[t]{\rotatebox{90}{$20\cdot\log_{10}\big(\,|F(\Omega)|\,\big)$}}}
\put(410,11){\vector(1,0){37}}
\put(405,11){\makebox(0,0)[r]{$\Omega$}}

\end{picture}}
\end{center}\vspace{-10pt}
\caption{Betrag des Spektrums der Fensterfolge mit \mbox{$M\!=\!32$},
\mbox{$N\!=\!13$}, \mbox{$A_0\!=\!12$} und \mbox{$A_1\!=\!0$}.}
\label{E.b4g}
\end{figure}
Bild \ref{E.b4g} zeigt an dem Beispiel mit \mbox{$M\!=\!32$},
\mbox{$N\!=\!13$}, \mbox{$A_0\!=\!12$} und \mbox{$A_1\!=\!0$} den Betrag des
Spektrums der Fensterfolge. Dabei wurde das Spektrum auf drei Arten berechnet.
Zuerst wurde eine FFT durchgef"uhrt, dann wurde die diskrete
Fouriertransformierte von \mbox{$f(k)$} durch Auswertung der Summenformel
der DFT berechnet, und zuletzt wurde der oben erw"ahnten Algorithmus
zur Berechnung des Spektrums f"ur beliebige Frequenzen eingesetzt,
der alle Spektralwerte mit etwa gleicher relativer Genauigkeit berechnet.
Bei der Berechnungsvariante mit der FFT entstehen systematische
Fehler die nach \cite{Heute} durch die Quantisierung der Drehfaktoren zum
Teil zu erkl"aren sind. Bei der Variante mit der Auswertung der
DFT-Summenformel ist der typische Rauschsockel zu erkennen, der teils
durch die begrenzte Wortl"ange der Werte von \mbox{$f(k)$} und teils
durch die Rundungen der Berechnung selbst verursacht wird.
Wird das in der Wortl"ange begrenzte Fenster bei dem Vorgang der Fensterung
mit einem Signal multipliziert, das auf dieselbe Wortl"ange quantisiert ist,
so kann man nicht erwarten, dass die hohe theoretische Sperrd"ampfung
im Bereich von \mbox{$\Omega\!=\!\pi$} zu einer Verbesserung bei der Messung
des LDS beitr"agt. Man kann absch"atzen, dass in unserem Beispiel oberhalb
einer Frequenz von etwa \mbox{$3,\!5\cdot2\pi/M$} der theoretische
Betrag des Spektrums unterhalb des Rauschsockels liegt.\vspace{20pt}

\section{Berechnung der AKF der Fensterfolge}\label{E.Kap.10.3}

Wenn man die Autokorrelationsfolge \mbox{$d(k)$} der Fensterfolge
berechnen will, hat man mehrere M"oglichkeiten. Die erste besteht
darin, die Faltungsoperation als Summe f"ur jeden Zeitpunkt
\mbox{$k=A_0\!-\!F\;(1)\;F\!-\!A_0$} zu berechnen. Diese M"oglichkeit
kann aber f"ur gro"se Fensterl"angen $F$ sehr zeitaufwendig sein.
Zum zweiten kann man auch zun"achst die \mbox{$2\CdoT F$} Werte
des Betragsquadrats des Spektrums der Fensterfolge f"ur alle Frequenzen
im Abstand $\pi/F$ mit dem im letzten Abschnitt beschrieben
Verfahren bestimmen. Von den \mbox{$2\CdoT F$} Spektralwerten
k"onnen h"ochstens \mbox{$2\CdoT(F\!-\!N\!+\!A_1)\!+\!1$} Werte
von null verschieden sein.  Aus diesen kann man mit einer DFT die
Autokorrelationsfolge berechnen. Diese Art der Berechnung hat zum
einen den Nachteil, dass bei Verwendung einer FFT die Werte ungenau
berechnet werden, und dass zum anderen zuerst zus"atzliche Spektralwerte
berechnet werden m"ussen, obwohl die \mbox{$N\!-\!A_1$} Spektralwerte,
die zur Berechnung der Fensterfolge ben"otigt wurden, diese
vollst"andig beschreiben. Bei der dritten Art der Berechnung der
Autokorrelationsfolge \mbox{$d(k)$} braucht man nur diese
\mbox{$N\!-\!A_1$} Spektralwerte. Die Fensterautokorrelationsfolge
l"asst sich mit den Sinusreihenkoeffizienten
\begin{equation}
S_{\nu}=\!\Sum{\substack{\Tilde{\nu}=1+A_1-N\\\Tilde{\nu}\neq\nu}}{N-A_1-1}
\bigg(\frac{\;\Re\big\{F\big({\T\nu\CdoT\frac{2\pi}{F}}\big)^*\!\CdoT
F\big({\T\Tilde{\nu}\CdoT\frac{2\pi}{F}}\big)\big\}\;}
{F^2\CdoT M\CdoT\tan\big(\frac{\pi}{F}\CdoT(\Tilde{\nu}\!-\!\nu)\big)}+
\frac{\;\Im\big\{F\big({\T\nu\CdoT\frac{2\pi}{F}}\big)^*\!\CdoT
F\big({\T\Tilde{\nu}\CdoT\frac{2\pi}{F}}\big)\big\}\;}{F^2\CdoT M}\bigg)
\label{E.10.19}
\end{equation}

f"ur \mbox{$0\!\le\!k\!<\!F$} n"amlich als

\begin{equation}
d(k)=\!\Sum{\nu=1+A_1-N}{N-A_1-1}\!\!
S_{\nu}\cdot\sin\big({\T\frac{2\pi}{F}\CdoT\nu\CdoT k}\big)+
\frac{F\!-\!k}{F^2\CdoT M}\cdoT\!\!\Sum{\nu=1+A_1-N}{N-A_1-1}\!
\big|F\big({\T\nu\CdoT\frac{2\pi}{F}}\big)\big|^2\Cdot
\cos\big({\T\frac{2\pi}{F}\CdoT\nu\CdoT k}\big)
\label{E.10.20}
\end{equation}
schreiben. F"ur \mbox{$k\!\ge\!F$} ist sie null, und f"ur 
negatives $k$ ist sie geradesymmetrisch. Im "ubrigen l"asst sich die
Autokorrelationsfolge jeder Fensterfolge auf diese Weise berechnen,
wobei die Summationsgrenzen entsprechend so zu modifizieren sind, 
dass alle von null verschiedenen Fourierreihenkoeffizienten
(\,ggf. sind das alle Werte der DFT der Fensterfolge\,)
ber"ucksichtigt werden. Ein kurzer Auszug aus einer Liste eines Programms,
das die Fensterautokorrelationsfolge \mbox{$d(k)$} auf diese Art berechnet,
ist in  Kapitel \ref{E.Kap.11.3} zu finden.\vspace{20pt}

\section{Fenster mit weiteren Nullstellen am Einheitskreis im $2\pi/F$ Raster}\label{E.Kap.10.4}

Nun will ich mich der Frage widmen, wohin die frei w"ahlbaren Nullstellen
der Z-Trans\-for\-mier\-ten der Basisfensterfolge \mbox{$g(k)$} gelegt werden
sollten, um die Sperrd"ampfung f"ur gro"se Werte von $N$ in der N"ahe des
Durchlassbereichs auf Kosten der Sperrd"ampfung in der Umgebung von
\mbox{$\Omega\!=\!\pi$} zu erh"ohen.

\begin{figure}[btp]
\begin{center}
{ 
\begin{picture}(450,600)

\input{mbild4h}
\put(60,578){\makebox(0,0)[lb]{Fensterfolge mit \mbox{$M\!=\!32$},
\mbox{$N\!=\!13$}, \mbox{$A_0\!=\!12$} und \mbox{$A_1\!=\!0$}}}
\put(15,550){\vector(0,1){24}}
\put(15,540){\makebox(0,0)[t]{\rotatebox{90}{$20\cdot\log_{10}\big(\,|F_0(\Omega)|\,\big)$}}}
\put(395,400){\vector(1,0){48}}
\put(390,400){\makebox(0,0)[r]{$\Omega$}}

\put(60,385){\makebox(0,0)[lb]{Fensterfolge mit \mbox{$M\!=\!32$},
\mbox{$N\!=\!13$}, \mbox{$A_0\!=\!10$} und \mbox{$A_1\!=\!1$}}}
\put(15,357){\vector(0,1){24}}
\put(15,347){\makebox(0,0)[t]{\rotatebox{90}{$20\cdot\log_{10}\big(\,|F_1(\Omega)|\,\big)$}}}
\put(395,207){\vector(1,0){48}}
\put(390,209){\makebox(0,0)[r]{$\Omega$}}

\put(60,191){\makebox(0,0)[lb]{Differenz der D"ampfungen der
beiden Fensterfolgen mit \mbox{$A_1\!=\!1$} und \mbox{$A_1\!=\!0$}}}
\put(15,164){\vector(0,1){24}}
\put(15,154){\makebox(0,0)[t]{\rotatebox{90}{$20\cdot\log_{10}\big(\,|F_0(\Omega)/F_1(\Omega)|\,\big)$}}}
\put(395,14){\vector(1,0){48}}
\put(390,14){\makebox(0,0)[r]{$\Omega$}}

\end{picture}}
\end{center}\vspace{-10pt}
\caption{"Anderung der Sperrd"ampfung durch
\mbox{$A_1\!=\!1$} statt \mbox{$A_1\!=\!0$}.}
\label{E.b4h}
\end{figure}
Der gew"unschte Durchlassbereich des Spektrums der Fensterfolge wird durch
die zu approximierende mit $2\pi$ periodische Rechteckfunktion, die im
Intervall \mbox{$(-\pi;\pi]$} f"ur \mbox{$|\Omega|<\pi/M$} den
Wert $M^2$ annimmt, und sonst null ist, bestimmt. Das Spektrum
der Fensterfolge mit \mbox{$A_0=N\!-\!1$} und \mbox{$A_1\!=\!0$},
die mit dem in Kapitel \myref{Algo} vorgestellten Algorithmus
berechnet worden ist, hat ihre erste Nullstelle am Einheitskreis erst
bei der Frequenz \mbox{$\Omega=2\pi/M$} also erst bei der doppelten
Grenzfrequenz des gew"unschten Durchlassbereichs. Daher ist die
Sperrd"ampfung unmittelbar au"serhalb des Durchlassbereichs nicht 
so hoch, wie dies u.~U. gew"unscht wird. Das gilt besonders dann, 
wenn es beim RKM auf eine gute Trennung unmittelbar benachbarter 
Frequenzpunkte \mbox{$\mu\CdoT2\pi/M$} ankommt.
Stellt man nun fest, dass die D"ampfung der Fensterfolge
mit \mbox{$A_0=N\!-\!1$} und \mbox{$A_1\!=\!0$} bei \mbox{$\Omega\!=\!\pi$} \pagebreak[2]
so hoch ist, dass das Spektrum der Fensterfolge in einem gro"sen
Frequenzbereich unterhalb des Rauschsockels liegt, so kann man 
mit \mbox{$A_0=N\!-\!3$} und \mbox{$A_1\!=\!1$} eine Fensterfolge berechnen,
deren Spektrum bei \mbox{$\Omega=(N\!-\!1)\CdoT2\pi/F$} eine weitere
Nullstelle aufweist, die bewirkt, dass das Betragsquadrat des Spektrums 
im Bereich \mbox{$\pi/M<|\Omega|<2\CdoT\pi/M$} kleiner ist
als bei der Fensterfolge nach \cite{Diss}. Man erkauft sich die h"ohere
Sperrd"ampfung in diesem Frequenzbereich jedoch auf Kosten eines Abfalls
des Betrags des Spektrums f"ur h"ohere Frequenzen mit einer Potenz, die
um zwei Grade niedriger ist, so dass die Sperrd"ampfung oberhalb der Frequenz
\mbox{$\Omega=2\pi/M$} bald unter die Sperrd"ampfung der zuerst
berechneten Fensterfolge abf"allt. Bild \ref{E.b4h} zeigt,
an dem Beispiel \mbox{$M\!=\!32$}, \mbox{$N\!=\!13$}, \mbox{$A_0\!=\!10$}
und \mbox{$A_1\!=\!1$} wie sich die Sperrd"ampfung der Festerfolge durch
die zwei zus"atzlichen Nullstellen der Z-Transformierten der Basisfensterfolge
am Einheitskreis ver"andert.\vspace{10pt}

\begin{figure}[btp]
\begin{center}
{ 
\begin{picture}(450,245)

\input{mbild5o}
\put(45,243){\makebox(0,0)[rt]{$\big|F(\Omega)\big|$}}
\put(48,216){\makebox(0,0)[rt]{$M$}}
\put(48,161){\makebox(0,0)[r]{$\frac{M}{\,\sqrt{2\,}\,}$}}
\put(47,120){\makebox(0,0)[r]{$\frac{M}{2}$}}
\put(48,18){\makebox(0,0)[rt]{$0$}}
\put(140,16){\makebox(0,0)[t]{$\frac{\vphantom{M}\pi}{2\cdot M}$}}
\put(230,16){\makebox(0,0)[t]{$\frac{\vphantom{M}\pi}{M}$}}
\put(320,16){\makebox(0,0)[t]{$\frac{\vphantom{M}3\pi}{2\cdot M}$}}
\put(410,16){\makebox(0,0)[t]{$\frac{\vphantom{M}2\pi}{M}$}}
\put(450,15){\makebox(0,0)[rt]{$\Omega$}}
\put(261,122){\makebox(0,0)[l]{$A_1$}}
\put(201,183){\makebox(0,0)[r]{$A_1$}}
\put(257,120){\vector(-2,-1){30}}
\put(205,185){\vector(2,1){30}}

\end{picture}}
\end{center}\vspace{-20pt}
\caption{Betr"age der Spektren der Fensterfolgen mit \mbox{$M\!=\!1024$} und \mbox{$N\!=\!15$}.\protect\\
\mbox{$A_1=1\;(1)\;7$} als Parameter, \mbox{$A_0=14\!-\!2\CdoT A_1$}.}
\label{E.b5o}
\rule{\textwidth}{0.5pt}\vspace{-5pt plus 20pt}
\end{figure}
\begin{figure}[btp!]
\begin{center}
{ 
\begin{picture}(450,455)

\input{mbild5p}
\put(8,400){\makebox(0,0)[t]{\rotatebox{90}{$20\cdot\log_{10}\big(\,|F(\Omega)|\,\big)$}}}
\put(8,405){\vector(0,1){40}}
\put(390,10){\makebox(0,0)[r]{$20\cdot\log_{10}\big(\,|\sin(\Omega/2)|\,\big)$}}
\put(400,10){\vector(1,0){40}}
\put(308,385){\rotatebox{-7.125}{$A_1=7$, Steigung $= -1$}}
\put(305,312){\rotatebox{-20.56}{$A_1=6$, Steigung $= -3$}}
\put(311,245){\rotatebox{-32.01}{$A_1=5$, Steigung $= -5$}}
\put(333,166){\rotatebox{-41.19}{$A_1=4$, Steigung $= -7$}}
\put(292,152){\rotatebox{-48.37}{$A_1=3$, Steigung $= -9$}}
\put(252,155){\rotatebox{-53.97}{$A_1=2$, Steigung $= -11$}}
\put(232,154){\rotatebox{-58.39}{$A_1=1$, Steigung $= -13$}}

\end{picture}}
\end{center}\vspace{-14pt}
\caption{Sperrd"ampfung der Nebenmaxima der Spektren der Fensterfolgen mit \mbox{$M\!=\!1024$} und \mbox{$N\!=\!15$}. 
\mbox{$A_1=1\;(1)\;7$} als Parameter, \mbox{$A_0\!=\!14\!-\!2\CdoT A_1$}}
\label{E.b5p}
\rule{\textwidth}{0.5pt}\vspace{-10pt}
\end{figure}
Am Beispiel des Fensters mit \mbox{$M\!=\!1024$} und \mbox{$N\!=\!15$} 
wird nun gezeigt, wie sich der Betrag des Spektrums ver"andert, wenn man 
den Parameter \mbox{$A_1=1\;(1)\;7$} variiert. Bild \ref{E.b5o} zeigt
die Betr"age der Spektren dieser Fensterfolgen im Frequenzbereich
zwischen $0$ und \mbox{$2\pi/M$} in linearer Darstellung. Mit zunehmendem
Wert des Parameters $A_1$ entstehen weitere Nullstellen, die sich
unterhalb der Frequenz \mbox{$2\pi/M$} im Abstand \mbox{$2\pi/F$}
an die bereits vorhandenen Nullstellen anschlie"sen. Sie bewirken,
dass der Betrag des Spektrums der Fensterfolge mit steigendem $A_1$
bei der Frequenz \mbox{$\pi/M$} immer steiler abf"allt. Diese
Zunahme der spektralen Flankensteilheit wird jedoch mit einer
Abnahme der Sperrd"ampfung der ersten Nebenmaxima und mit einer
Abnahme der Potenz des Anstiegs der Sperrd"ampfung mit
\mbox{$\sin(\Omega/2)$} erkauft. 

Bild \ref{E.b5p} zeigt, dass die Sperrd"ampfung der Nebenmaxima der 
Spektren der Fensterfolgen f"ur \mbox{$\Omega\gg2\pi/M$} mit der Potenz
\mbox{$N\!-\!2\CdoT\!A_1$} ansteigt. Bei dieser Darstellung wurde
die Ordinate um den Faktor $8$ enger skaliert als die Abszisse.
Um die Betragsspektren im vollen Dynamikbereich (ca. 500~dB\,) der
Graphik bestimmen zu k"onnen, wurden diese nicht mit Hilfe der
DFT aus den Fensterfolgen, sondern wie in Kapitel \ref{E.Kap.10.2} 
beschrieben, berechnet.

\section{Basisfenster mit sechs zus"atzlichen Nullstellen am Einheitskreis}\label{E.Kap.10.5}

\begin{figure}[btp]
\begin{center}
{ 
\begin{picture}(450,600)

\input{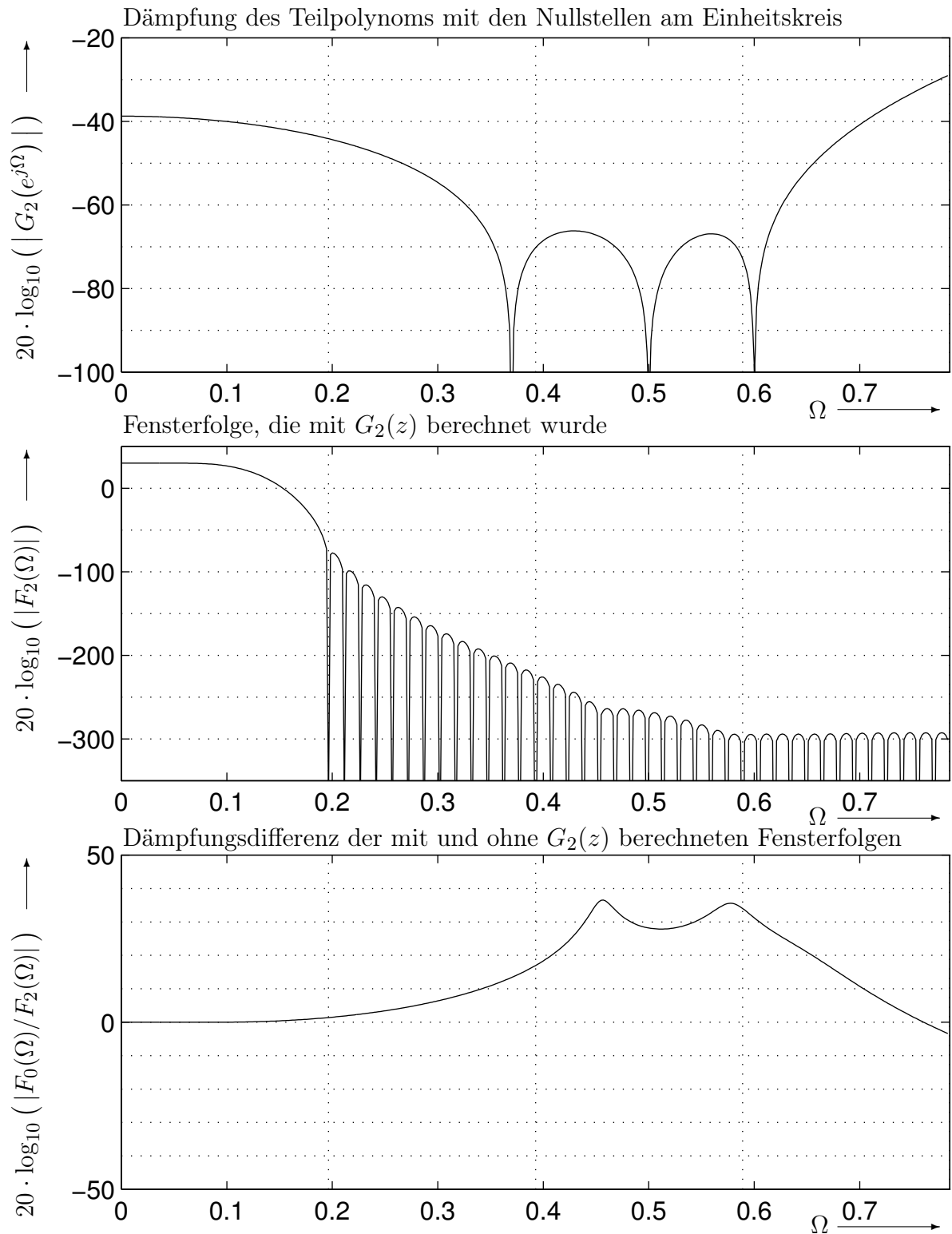}
\put(60,579){\makebox(0,0)[lb]{D"ampfung des Teilpolynoms mit den Nullstellen
am Einheitskreis}}
\put(15,550){\vector(0,1){24}}
\put(15,540){\makebox(0,0)[t]{\rotatebox{90}{$20\cdot\log_{10}\big(\,\big|\,G_2\big(e^{j\Omega}\big)\,\big|\,\big)$}}}
\put(395,400){\vector(1,0){48}}
\put(390,400){\makebox(0,0)[r]{$\Omega$}}

\put(60,386){\makebox(0,0)[lb]{Fensterfolge, die mit \mbox{$G_2(z)$}
berechnet wurde}}
\put(15,357){\vector(0,1){24}}
\put(15,347){\makebox(0,0)[t]{\rotatebox{90}{$20\cdot\log_{10}\big(\,|F_2(\Omega)|\,\big)$}}}
\put(395,207){\vector(1,0){48}}
\put(390,208){\makebox(0,0)[r]{$\Omega$}}

\put(60,191){\makebox(0,0)[lb]{D"ampfungsdifferenz der
mit und ohne \mbox{$G_2(z)$} berechneten Fensterfolgen}}
\put(15,164){\vector(0,1){24}}
\put(15,154){\makebox(0,0)[t]{\rotatebox{90}{$20\cdot\log_{10}\big(\,|F_0(\Omega)/F_2(\Omega)|\,\big)$}}}
\put(395,14){\vector(1,0){48}}
\put(391,15){\makebox(0,0)[r]{$\Omega$}}

\end{picture}}
\end{center}\vspace{-10pt}
\caption{Betrag des Spektrums der Fensterfolge mit \mbox{$A_1\!=\!0$}
und sechs weiteren Nullstellen am Einheitskreis.}
\label{E.b4i}
\end{figure}
Eine andere M"oglichkeit die Sperrd"ampfung des Fensters unmittelbar
oberhalb des gew"unschten Durchlassbereichs zu erh"ohen, besteht darin,
die frei w"ahlbaren Nullstellen der Z-Transformierten der Basisfensterfolge
geeignet zu w"ahlen. Man kann z.~B. die Nullstellen des Polynoms
\mbox{$z^{F-N+2\cdot A_1}\Cdot G_2(z)$}, das nur die frei w"ahlbaren
Nullstellen enth"alt, auf dem Einheitskreis in dem Frequenzbereich
verteilen, in dem der Betrag des Spektrums der Fensterfolge mit
\mbox{$A_0=N\!-\!1$} und \mbox{$A_1\!=\!0$} oberhalb des Rauschsockels
liegt. Dabei achtet man darauf, dass das Betragsquadrat von
\mbox{$G_2\big(e^{j\Omega}\big)$} bei niedrigen Frequenzen --- also im
Durchlassbereich --- nicht kleiner sein sollte als in dem Frequenzbereich,
in dem man die Sperrd"ampfung erh"ohen m"ochte. Durch solch eine Wahl der
Nullstellen wird das Spektrum der Basisfensterfolge in dem Frequenzbereich
zus"atzlich ged"ampft, der nach der "Uberlagerung der verschobenen
Betragsquadrate der Spektren den "Ubergangsbereich ergibt, w"ahrend
das Spektrum der Basisfensterfolge in dem Bereich angehoben wird,
aus dem der Bereich des Rauschsockels hervorgeht. Somit wird im
"Ubergangsbereich des Spektrums der Fensterfolge eine h"ohere Sperrd"ampfung 
erzielt, als bei den Fenster nach \cite{Diss}. Bild \ref{E.b4i} zeigt
an dem Beispiel mit \mbox{$M\!=\!32$}, \mbox{$N\!=\!13$}, \mbox{$A_0\!=\!6$},
\mbox{$A_1\!=\!0$} und weiteren $6$ Nullstellen am Einheitskreis,
die das Polynom
\[
G_2(z) = \big(1\!-\!2\CdoT\cos(0,\!37)\CdoT z^{\!-1}\!+\!z^{\!-2}\big)\CdoT
\big(1\!-\!2\CdoT\cos(0,\!5)\CdoT z^{\!-1}\!+\!z^{\!-2}\big)\CdoT
\big(1\!-\!2\CdoT\cos(0,\!6)\CdoT z^{\!-1}\!+\!z^{\!-2}\big)
\]
bilden, erstens den Betragsfrequenzgang dieses Polynoms, zweitens
die D"ampfung des Spektrums der damit berechneten Fensterfolge und 
drittens die Ver"anderung der D"ampfung gegen"uber der in Bild \ref{E.b4g}
dargestellten D"ampfung in dem Frequenzbereich, in dem das Spektrum
der Fensterfolgen nicht im Rauschsockel versinkt.

Bei der Berechnung der drei Fensterfolgen der letzten beiden Unterkapitel
wurden f"ur die Berechnung des Cepstrums folgende FFT-L"angen ben"otigt.
\[
{\renewcommand{\arraystretch}{1.3}
\begin{array}{|c|ccccc|}
\hline
 &\;M\;&\;N\;&\;A_0\;&\;A_1\;&\;\widetilde{M}\;\\
\hline
F_0(\Omega) & 32 & 13 & 12 & 0 & 256 \\
F_1(\Omega) & 32 & 13 & 10 & 1 & 256 \\
F_2(\Omega) &\;32\;&\;13\;& 6 & 0 &\;4096\;\\
\hline
\end{array}}
\]
Eine Kombination der beiden eben beschriebenen Methoden die frei w"ahlbaren
Nullstellen einzustellen um im "Ubergangsbereich einen besseren
D"ampfungsfrequenzgang zu erhalten, ist ebenfalls denkbar. Da es f"ur die
Wahl der Nullstellen keine allgemeine optimale L"osung geben kann, weil es
vom vorgesehenen Einsatz des Fensters abh"angt, was unter optimal
zu verstehen ist, erscheint es mir am einfachsten, die Nullstellenlage
durch Probieren zu erhalten, also durch iteratives Verschieben der
Nullstellen, Berechnen des Spektrums und Beurteilen der D"ampfung im
"Ubergangsbereich. In den meisten F"allen wird man sowieso den Parameter $N$
nicht so gro"s w"ahlen, dass das Spektrum der Fensterfolge bereits
bei niedrigen Frequenzen in den Rauschsockel absinkt, so dass man 
das in \cite{Diss} vorgestellte Fenster verwenden kann.

\section{Halbbandfilter}\label{E.Kap.10.6}

\begin{figure}[btp]
\begin{center}
{ 
\begin{picture}(450,295)

\input{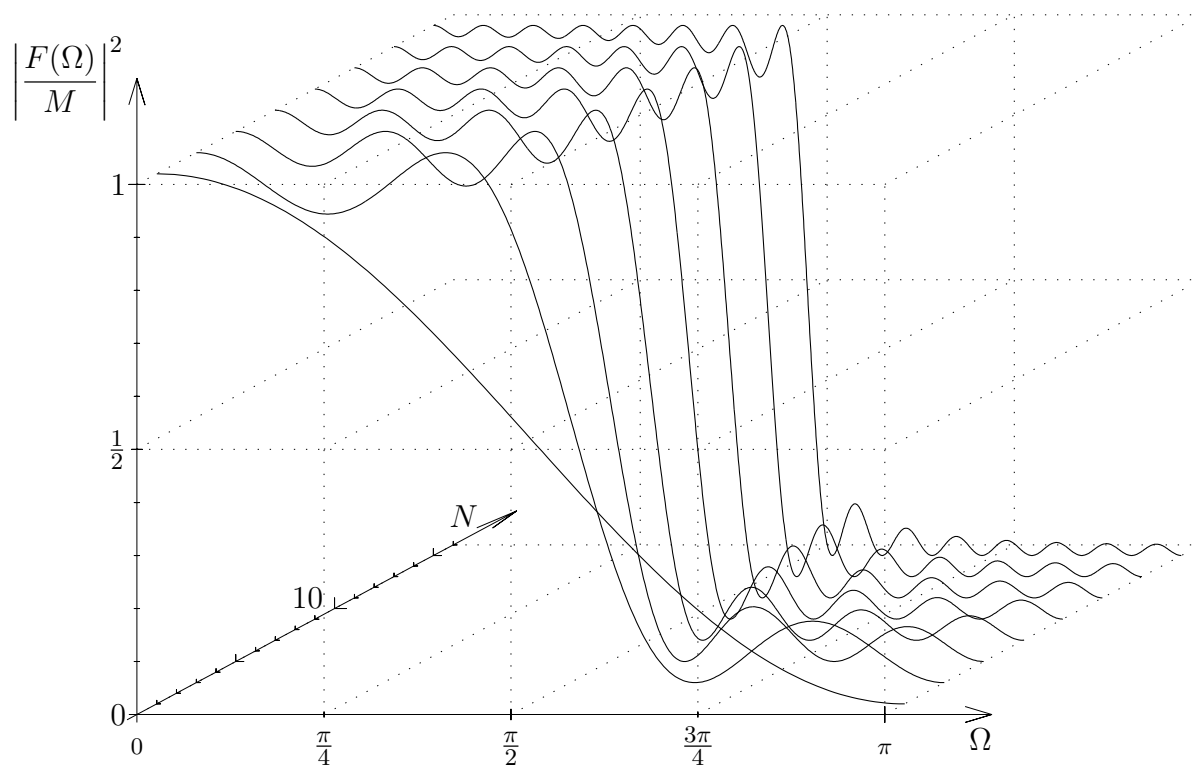}

\put(45,260){\makebox(0,0)[r]{${\D\bigg|\frac{F(\Omega)}{M}\bigg|^2}$}}
\put(46,220){\makebox(0,0)[r]{$1$}}
\put(46,120){\makebox(0,0)[r]{$\frac{1}{2}$}}
\put(46,20){\makebox(0,0)[r]{$0$}}
\put(50,16){\makebox(0,0)[t]{${\scriptstyle0}\vphantom{\frac{M}{M}}$}}
\put(120,16){\makebox(0,0)[t]{$\frac{\vphantom{M}\pi}{4\vphantom{M}}$}}
\put(190,16){\makebox(0,0)[t]{$\frac{\vphantom{M}\pi}{2\vphantom{M}}$}}
\put(260,16){\makebox(0,0)[t]{$\frac{\vphantom{M}3\pi}{4\vphantom{M}}$}}
\put(330,16){\makebox(0,0)[t]{${\scriptstyle\pi}\vphantom{\frac{M}{M}}$}}
\put(370,15){\makebox(0,0)[rt]{$\Omega$}}
\put(167,95){\makebox(0,0)[l]{$N$}}
\put(108,64){\makebox(0,0)[l]{$10$}}

\end{picture}}
\end{center}\vspace{-15pt}
\setlength{\belowcaptionskip}{5pt}
\caption{Spektren der Autokorrelationsfolgen der Fenster mit \mbox{$M\!=\!2$} (\,Halbbandfilter\,).\protect\\
\mbox{$N=1\;(2)\;15$} als Parameter; \mbox{$A_1=(N\!-\!1)/2$}.}
\label{E.b5r}
\rule{\textwidth}{0.5pt}\vspace{-10pt}
\end{figure}
Bereits in Kapitel \myref{FenBeisp} hatten wir festgestellt, dass die 
FIR-Filter, die man mit \mbox{$M\!=\!2$} erh"alt, wenn man die Werte 
der Fensterautokorrelationsfolge \mbox{$d(k)$} als Filterkoeffizienten 
verwendet, Halbbandfilter sind. Man kann nun versuchen die in Bild \myref{b5q} 
dargestellten Spektralverl"aufe g"unstig zu beeinflussen, indem man die 
bei der Konstruktion der Fensterfolge nach Kapitel \ref{E.Kap.10.1} neu 
hinzugekommenen Freiheitsgrade ausnutzt. Berechnet man die Filterkoeffizienten 
als die Werte der Autokorrelationsfolge, die man bei ungeradem $N$ erh"alt, 
wenn man $A_1$ auf den maximal m"oglichen Wert \mbox{$A_1=(N\!-\!1)/2$} setzt, 
so erh"alt man die Halbbandfilter, deren Spektren in Bild \ref{E.b5r} dargestellt 
sind. Diese zugrundeliegenden Fenster besitzen die L"ange
\mbox{$F=M\CdoT N=2\CdoT N$}. Dem entsprechend ist die L"ange der 
Fensterautokorrelationsfolge, also des Halbbandfilters, gleich  
\mbox{$2\CdoT F\!-\!1=4\CdoT N\!-\!1$}. Man erkennt, dass im Vergleich zu 
Bild \myref{b5q} der "Ubergangsbereich schm"aler ist, was jedoch
mit einer h"oheren Welligkeit in Durchlass- und Sperrbereich erkauft wird.

\begin{figure}[btp]
\begin{center}
{ 
\begin{picture}(450,226)

\input{mbild5s.tex}

\put(70,200){\makebox(0,0)[b]{
$20\CdoT\log_{10}\Big(\Big|D\big(e^{j\Omega}\big)/M\Big|\Big)$}}
\put(48,182){\makebox(0,0)[rb]{$0$}}
\put(46,150){\makebox(0,0)[r]{$-100$}}
\put(46,120){\makebox(0,0)[r]{$-200$}}
\put(46,90){\makebox(0,0)[r]{$-300$}}
\put(46,60){\makebox(0,0)[r]{$-400$}}
\put(46,30){\makebox(0,0)[r]{$-500$}}
\put(140,184){\makebox(0,0)[b]{$\frac{\vphantom{M}\pi}{4\vphantom{M}}$}}
\put(230,184){\makebox(0,0)[b]{$\frac{\vphantom{M}\pi}{2\vphantom{M}}$}}
\put(320,184){\makebox(0,0)[b]{$\frac{\vphantom{M}3\pi}{4\vphantom{M}}$}}
\put(410,184){\makebox(0,0)[b]{${\scriptstyle\pi}\vphantom{\frac{M}{M}}$}}
\put(450,185){\makebox(0,0)[rb]{$\Omega$}}
\put(315,165){\vector(-1,-1){50}}
\put(265,115){\makebox(0,0)[rt]{$A_1$}}

\end{picture}}
\end{center}\vspace{-10pt}
\setlength{\belowcaptionskip}{4pt}
\caption{Spektren der Autokorrelationsfolgen der Fenster mit \mbox{$M\!=\!2$} (\,Halbbandfilter\,).\protect\\
\mbox{$A_1=0\;(1)\;5$} als Parameter, \mbox{$N=4\CdoT\!A_1\!+\!3$}
und Nullstellen bei
\mbox{$z_{0,\rho}=e^{(-1)^{\protect\up{0.5}{\rho}}\cdot j\cdot\frac{2\pi}{F}\cdot(A_1+2)}$}\protect\\
mit \mbox{$\rho=\,1\;\,(1)\;\,2\CdoT(A_1\!+\!1)$}.}
\label{E.b5s}
\rule{\textwidth}{0.5pt}\vspace{-10pt}
\end{figure}
Es erscheint daher sinnvoll, hier den Parameter $A_1$ nicht maximal
zu w"ahlen, sondern auch einige der frei w"ahlbaren Nullstellen
$z_{0,\rho}$ dazu zu nutzen, einen Kompromiss aus einer hohen Sperrd"ampfung
und einer geringen Breite des "Ubergangsbereichs zu finden.
Es zeigte sich, dass man beispielsweise im Frequenzbereich
\mbox{$3\pi/4<\Omega\le\pi$} eine fast gleichm"a"sig hohe
minimale Sperrd"ampfung erh"alt, wenn man den Parameter
\mbox{$N=4\CdoT A_1+3$} einstellt, und als frei w"ahlbare
Nullstellen $z_{0,\rho}$ das Nullstellenpaar
\mbox{$e^{\pm j\cdot\pi\cdot\frac{A_1+2}{4\cdot A_1+3}}$}
mit der Vielfachheit \mbox{$A_1\!+\!1$} verwendet.
Das Spektrum der Basisfensterfolge besitzt dann nur Nullstellen am
Einheitskreis, wobei die niederfrequentesten Nullstellen die
Vielfachheit \mbox{$A_1\!+\!2$} aufweisen, weil die frei gew"ahlten
Nullstellen alle auf den sowieso schon vorhandenen einfachen Nullstellen
bei diesen Frequenzen zu liegen kommen. Alle weiteren Nullstellen
im Abstand \mbox{$2\pi/F$} sind einfach. F"ur die Parameterwerte
\mbox{$A_1=0\;(1)\;5$} sind die Betragsfrequenzg"ange der entsprechenden
Halbbandfilter in Bild \ref{E.b5s} halblogarithmisch dargestellt.
Die Filterl"angen der Halbbandfilter sind \mbox{$2\CdoT F\!-\!1=16\CdoT A_1+11$}. 
F"ur $A_1\!=\!5$ wird bei der Berechnung des Cepstrums eine noch tolerierbare 
FFT-L"ange von \mbox{$\widetilde{M}\!=\!2^{13}$} ben"otigt.

\section[Berechnung einer kontinuierlichen Fensterfunktion]{Berechnung einer kontinuierlichen \\Fensterfunktion}\label{E.Kap.10.7}

Dieses Unterkapitel behandelt einen Algorithmus, mit dessen Hilfe eine
kontinuierliche Fensterfunktion \mbox{$f_{\!\infty}(t)$} berechnet werden
kann, die endlich lang ist, und deren Spektrum "ahnliche Eigenschaften,
wie die in den Gleichungen (\myref{2.20}) und (\myref{2.27}) beschriebenen, besitzt.
Es wird --- wie sich im weiteren zeigen wird --- durch die Art der
Konstruktion der Fensterfunktion erreicht, dass erstens die Fensterfunktion
zeitlich gem"a"s
\begin{equation}
f_{\!\infty}(t)=0\qquad\text{ f"ur }\quad t<0\;\;\vee\;\; t\ge1
\label{E.10.21}
\end{equation}
begrenzt ist, dass zweitens das Spektrum \mbox{$F_{\!\infty}(\omega)$}
der Fensterfunktion "aquidistante Nullstellen auf der imagin"aren Achse
mit Ausnahme des Bereichs niedriger Frequenzen aufweist, so dass\vspace{-6pt}
\begin{equation}
F_{\!\infty}(2\pi\CdoT\nu)\,=\,0\qquad
\text{ f"ur }\quad\nu\in\mathbb{Z}\quad\text{ mit }
\quad\nu\neq N\!-\!1\;(1)\;N\!-\!1
\label{E.10.22}
\end{equation}
gilt, und dass drittens das Betragsquadratspektrum --- bei entsprechender
Normierung mit einem konstanten Faktor --- die Eigenschaft
\begin{equation}
\Sum{\mu=-\infty}{\infty}
\big|F_{\!\!\infty}(\omega\!-\!\mu\CdoT N\CdoT2\pi)\big|^2\,=\;1.
\label{E.10.23}
\end{equation}
besitzt. Die Fensterautokorrelationsfunktion
\mbox{$d_{\infty}(t)=f_{\!\infty}(t)\!\ast\!\!f_{\!\infty}(\!-t)$}
weist daher "aquidistante Nullstellen bei Vielfachen von \mbox{$1/N$}
au"ser bei \mbox{$t\!=\!0$} auf.

Die bisher behandelten zeitdiskreten Fensterfolgen \mbox{$f(k)$},
die von dem Parameter $M$ abh"angen, kann man als die Abtastwerte
kontinuierlicher Fensterfunktionen f"ur die Abtastzeitpunkte\vspace{-10pt}
\begin{equation}
t\,=\,\frac{k}{{}\,N\CdoT M\,{}}
\label{E.10.24}
\end{equation}
betrachten. Dabei sind die kontinuierlichen Fensterfunktionen, die ebenfalls
von $M$ abh"angen, au"serhalb des in Gleichung (\ref{E.10.21})
genannten Zeitintervalls null. Wenn man nun den Grenz"ubergang
\mbox{$M\to\infty$} durchf"uhrt, so ergibt sich die in $k$ unbegrenzte
Fensterfolge, die durch unbegrenzt feine Abtastung der kontinuierlichen
Fensterfunktion \mbox{$N\CdoT f_{\!\infty}(t)$} entsteht. Wie auch bei
den bisher behandelten Fensterfolgen, beschreiben die $N$ Spektralwerte
\mbox{$F_{\!\infty}(\nu\CdoT2\pi)$} mit \mbox{$\nu=0\;(1)\;N\!-\!1$},
die kontinuierliche Fensterfunktion vollst"andig. Die Fensterfunktion
l"asst sich analog zu Gleichung (\myref{6.16}) durch Auswertung
der Fourierreihe f"ur jeden beliebigen Zeitpunkt gem"a"s
\begin{equation}
f_{\!\infty}(t) = \begin{cases}
{\D\Sum{\nu=1+A_1-N}{N-1-A_1} F_{\!\infty}(\nu\CdoT2\pi)\cdot
e^{j\cdot2\pi\cdot\nu\cdot t}}\qquad
&\text{f"ur}\quad0\le t<1\\
\quad0&\text{sonst.}
\end{cases}
\label{E.10.25}
\end{equation}
berechnen. Die Spektralwerte \mbox{$F_{\!\infty}(\nu\CdoT2\pi)$} sind
die Grenzwerte der auf $M$ normierten Spektralwerte
\mbox{$F\big({\T\nu\CdoT\frac{2\pi}{F}}\big)$} der
zeitdiskreten Fensterfolgen f"ur \mbox{$M\to\infty$}.
Beim Grenz"ubergang ist die normierte Frequenz $\Omega$,
durch die Kreisfrequenz
\begin{equation}
\omega\;=\;M\CdoT N\CdoT\Omega
\label{E.10.26}
\end{equation}
zu ersetzten, so dass ein konstanter Wert von $\omega$ einem
konstanten Produkt aus $M$ und $\Omega$ entspricht. Wegen des
eben beschriebenen Grenzwertverhaltens, kennzeichne ich die
kontinuierliche Fensterfunktion und deren Spektrum mit dem
Index $\infty$. Die kontinuierliche Fensterfunktion l"asst
sich nat"urlich nicht beim RKM einsetzen. Dennoch halte ich es f"ur
denkbar, dass auf anderen Gebieten eine kontinuierliche Fensterfunktion
mit den eben genannten Eigenschaften von Interesse sein k"onnte.
Daher sei in diesem Kapitel ein Algorithmus aufgef"uhrt, der dem
Algorithmus zur Konstruktion der diskreten Fensterfolge mit
\mbox{$A_0=N\!-\!1$} und \mbox{$A_1\!=\!0$} sehr "ahnlich ist, und mit
dessen Hilfe die Spektralwerte \mbox{$F_{\!\infty}(\nu\CdoT2\pi)$}
berechnet werden k"onnen. Auch f"ur andere Einstellungen der Parameter 
$A_0$ und $A_1$ kann man kontinuierliche Fensterfunktion als 
Grenzwertl"osungen f"ur \mbox{$M\to\infty$} erhalten. Der Leser m"oge die 
dazu notwendigen Modifikationen selbst herleiten. In dem in Kapitel 
\ref{E.Kap.11.4} aufgelisteten Programmrumpf ist diese Erweiterung enthalten.

Um die Fensterfunktion \mbox{$f_{\!\infty}(t)$} zu erzeugen gehen wir zun"achst
wieder von einer reellen Basisfensterfunktion \mbox{$g_{\infty}(t)$} aus,
die im gleichen Zeitintervall von null verschieden sein kann, wie die
Fensterfunktion \mbox{$f_{\!\infty}(t)$}. Als Basisfensterfunktion wird ein
Ausschnitt einer periodischen Funktion verwendet. Abh"angig davon, ob $N$ eine
gerade oder ungerade ganze Zahl ist, ist die Periode zwei oder eins. In beiden
F"allen l"asst sich die Basisfensterfunktion als Fourierreihe schreiben:
\begin{equation}
g_{\infty}(t) = \begin{cases}
{\D\Sum{\nu=\frac{1-N}{2}}{\frac{N-1}{2}}\!
G_{\infty}(j\CdoT\nu\CdoT2\pi)\cdot e^{j\cdot2\pi\cdot\nu\cdot t}}\qquad
&\text{f"ur}\quad0\,\le\,t<1.\\
\quad0&\text{sonst.}
\end{cases}
\label{E.10.27}
\end{equation}
Sie ist durch die $N$ Fourierreihenkoeffizienten
\mbox{$G_{\infty}(j\CdoT\nu\CdoT2\pi)$} vollst"andig festgelegt.
Das Spektrum der Basisfensterfunktion ist dann
\begin{gather}
G_{\infty}(j\omega)\;=\; 
\Sum{\nu_2=\frac{1-N}{2}}{\frac{N-1}{2}}\!G_{\infty}(j\CdoT\nu_2\CdoT2\pi)\cdot
e^{\!-j\cdot(\frac{\omega}{2}-\nu_2\cdot\pi)}\cdot
\text{si}\big({\T\frac{\omega}{2}\!-\!\nu_2\CdoT\pi}\big)\;=
\label{E.10.28}\\*[10pt]
=\;e^{\!-j\cdot\frac{\omega-(N-1)\cdot\pi}{2}}\cdot
\sin\!\big({\T\frac{\omega-(N-1)\cdot\pi}{2}}\big)\cdot
\Sum{\nu_2=\frac{1-N}{2}}{\frac{N-1}{2}}
\frac{{}\;2\cdot G_{\infty}(j\CdoT\nu_2\CdoT2\pi)\;{}}
{\omega-\nu_2\CdoT2\pi},\notag
\end{gather}
und weist f"ur \mbox{$\omega=(2\CdoT\nu\!-\!N\!+\!1)\CdoT\pi$} mit
\mbox{$\nu\in{}\mathbb{Z}$} au"ser f"ur \mbox{$\nu=0\;(1)\;N\!-\!1$}
"aquidistante Nullstellen auf. Durch Multiplikation der auf \mbox{$-1\!<\!t\!<\!1$}
zeitlich begrenzten Basisfensterautokorrelationsfunktion\vspace{-6pt}
\begin{equation}
g_{Q,\infty}(t)\;=\;g_{\infty}(t)\ast\!g_{\infty}(\!-t)\;=\;
\frac{1}{2\pi}\CdoT\!\Int{-\infty}{\infty}\big|G_{\infty}(j\omega)\big|^2\Cdot
e^{j\cdot\omega\cdot t}\cdot d\omega
\label{E.10.29}
\end{equation}
mit der periodisch fortgesetzten si-Funktion
\begin{equation}
\frac{{}\;\sin(N\CdoT \pi\CdoT t)\;{}}{\sin(\pi\CdoT t)}
\label{E.10.30}
\end{equation}
entsteht die Funktion \mbox{$d_{\infty}(t)$}. Diese weist dieselben
Nullstellen im Abstand \mbox{$1/N$} auf, wie die periodisch fortgesetzte
si-Funktion. Im Frequenzbereich entspricht die Multiplikation der
Faltung des Betragsquadratspektrums von \mbox{$g_{\infty}(t)$} mit
einem zu \mbox{$\omega\!=\!0$} symmetrischen Impulskamm der konstanten
St"arke $2\pi$ und der L"ange $N$ mit einem Impulsabstand von $2\pi$.
Das Faltungsintegral l"asst sich daher auch als Summe mit $N$ Summanden
schreiben:
\begin{equation}
D_{\infty}(j\omega)\;=\!
\Sum{\nu_1=\frac{1-N}{2}}{\frac{N-1}{2}}
\big|G_{\infty}(j\omega\!-\!j\CdoT\nu_1\CdoT2\pi)\big|^2\!.
\label{E.10.31}
\end{equation}
Wie im zeitdiskreten Fall, kann man sich auch im kontinuierlichen Fall
anhand der Nullstellenlage der verschobenen, an der Faltung beteiligten
Spektren "uberlegen, dass das so entstandene Spektrum die Gleichung (\ref{E.10.22})
erf"ullt. Da \mbox{$\big|G_{\infty}(j\omega)\big|^2$} eine reelle,
geradesymmetrische und nichtnegative Funktion in $\omega$ ist, und da die
"Uberlagerung ebenfalls symmetrisch zu \mbox{$\omega\!=\!0$} erfolgt, ist
auch \mbox{$D_{\infty}(j\omega)$} eine reelle, geradesymmetrische und
nichtnegative Funktion in $\omega$. Daher sind alle Nullstellen von
\mbox{$D(s)$} spiegelsymmetrisch zur reellen Achse, alle Nullstellen
auf der imagin"aren Achse sind von gerader Vielfachheit und alle
Nullstellen, die nicht auf der imagin"aren Achse liegen, sind
spiegelsymmetrisch zur imagin"aren Achse. Es handelt sich somit bei
\mbox{$D_{\infty}(j\omega)$} um das Betragsquadratspektrum einer
reellen Fensterfunktion. \mbox{$d_{\infty}(t)$} l"asst sich daher
als \mbox{$f_{\!\infty}(t)\!\ast\!\!f_{\!\infty}(\!-t)$} schreiben.
Durch Wahl eines geeigneten konstanten Faktors kann f"ur das Spektrum
der Basisfensterfunktion
\begin{equation}
\Sum{\nu=\frac{1-N}{2}}{\frac{N-1}{2}}
\big|G_{\infty}(j\CdoT2\pi\CdoT\nu)\big|^2\,=\;1
\label{E.10.32}
\end{equation}
erzwungen werden. Dadurch wird erreicht, dass auch die Gleichung (\ref{E.10.23})
erf"ullt wird.

Im weiteren m"ochte ich mich auf den Fall beschr"anken, dass als
Basisfensterfunktion die \mbox{$N\!-\!1$-te} Potenz einer halben
Periode der Sinusfunktion \mbox{$\sin(\pi\CdoT t)$} mit geeigneter
Amplitude verwendet wird:
\begin{equation}
g_{\infty}(t)\;=\;\begin{cases}
\;2^{\uP{0.4}{N-1}}\Cdot
\binom{2\cdot N-2}{N-1}^{\!\!-\frac{1}{2}}\!\Cdot
\sin(\pi\CdoT t)^{\uP{0.4}{N-1}}\qquad
&\text{f"ur}\quad0\le t<1\\
\;0&\text{sonst.}
\end{cases}
\label{E.10.33}
\end{equation}
Bei der in Kapitel \myref{Algo} verwendeten diskreten Basisfensterfolge
waren die letzten \mbox{$N\!-\!1$} Werte null. Wenn wir diese Werte
als die mit steigendem $M$ immer enger beieinanderliegenden Abtastwerte
einer kontinuierlichen Basisfensterfunktion betrachten, r"ucken diese
\mbox{$N\!-\!1$} Nullstellen mit zunehmendem Parameter $M$ immer enger
zusammen, so dass man grenzwertig die in der Basisfensterfunktion
\mbox{$g_{\infty}(t)$} vorhandene \mbox{$N\!-\!1$}-fache Nullstelle erh"alt.
Die der Basisfensterfunktion zugrundeliegende periodische Funktion besitzt
die Fourierreihenkoeffizienten
\begin{equation}
G_{\infty}(j\CdoT\nu\CdoT2\pi)\;=\;
\tbinom{2\cdot N-2}{ N-1}^{\T\!\!-\frac{1}{2}}\!\Cdot
j^{\uP{0.6}{-2\cdot\nu}}\!\Cdot\tbinom{N-1}{\frac{N-1}{2}+\nu}
\qquad\text{ f"ur }{\T\quad\nu=\frac{1-N}{2}\;(1)\;\frac{N-1}{2}}.
\label{E.10.34}
\end{equation}
Wenn man dies und die Partialbruchzerlegung
\begin{equation}
\frac{\D(N\!-\!1)!\cdot\pi^{\uP{0.4}{N-1}}}
{\Prod{\nu_2=\frac{1-N}{2}}{\frac{N-1}{2}}\!
\big(\frac{\omega}{2}\!-\!\nu_2\CdoT\pi\big)}\;=\;
j^{\uP{0.8}{N-1}}\!\CdoT\Sum{\nu_2=\frac{1-N}{2}}{\frac{N-1}{2}}
\frac{\;2\CdoT\tbinom{N-1}{\frac{N-1}{2}+\nu_2}\,}
{\omega-\nu_2\CdoT2\pi}\cdot
j^{\uP{0.6}{-2\cdot\nu_2}}
\label{E.10.35}
\end{equation}
in Gleichung (\ref{E.10.28}) einsetzt, ergibt sich f"ur des Spektrum der
Basisfensterfunktion:
\begin{gather}
G_{\infty}(j\omega)\;=\;
\tbinom{2\cdot N-2}{N-1}^{\T\!\!-\frac{1}{2}}\!\Cdot
e^{\!-j\cdot\frac{\omega-(N-1)\cdot\pi}{2}}\cdot
\sin\!\big({\T\frac{\omega-(N-1)\cdot\pi}{2}}\big)\cdoT\!
\Sum{\nu_2=\frac{1-N}{2}}{\frac{N-1}{2}}
\frac{\;2\CdoT\tbinom{N-1}{\frac{N-1}{2}+\nu_2}\,}
{\omega-\nu_2\CdoT2\pi}\cdot
j^{\uP{0.6}{-2\cdot\nu_2}}\;=
\notag\\*[10pt]
=\;\frac{\D\;\pi^{\uP{0.4}{N-1}}\Cdot(N\!-\!1)!\cdot
e^{\!-j\cdot\frac{\omega}{2}}\cdot
\sin\!\big({\T\frac{\omega-\pi\cdot(N-1)}{2}}\big)\,}
{\tbinom{2\cdot N-2}{N-1}^{\T\!\frac{1}{2}}\CdoT
\Prod{\nu_2=\frac{1-N}{2}}{\frac{N-1}{2}}\!
\big(\frac{\omega}{2}\!-\!\pi\CdoT\nu_2\big)}.
\label{E.10.36}
\end{gather}Da es sich bei der Basisfensterfunktion um eine endliche,
zeitlich begrenzte und kausale Zeitfunktion handelt konvergiert deren
Laplacetransformierte f"ur alle \mbox{$s\in{}\mathbb{C}$}. Sie ergibt sich
aus der Fouriertransformierten indem man \mbox{$s\!=\!j\omega$} substituiert.
\begin{equation}
G_{\infty}(s)\;=\;
\frac{\D\,(j\CdoT2\pi)^N\Cdot(N\!-\!1)!\cdot
e^{\!-\frac{s}{2}}\Cdot
\sin\!\big({\T\frac{s-j\cdot(N-1)\cdot\pi}{2\cdot j}}\big)\,}
{\pi\CdoT\tbinom{2\cdot N-2}{N-1}^{\T\!\frac{1}{2}}\CdoT
\Prod{\nu_2=\frac{1-N}{2}}{\frac{N-1}{2}}\!
(s\!-\!j\CdoT\nu_2\CdoT2\pi)}
\label{E.10.37}
\end{equation}
Man beachte, dass sich die einfachen Polstellen auf der imagin"aren
Achse mit den einfachen Nullstellen der Sinusfunktion k"urzen lassen.

{\small Anmerkung: Bei dieser Basisfensterfunktion tritt im Z"ahler der
Laplacetransformierten kein Polynom in $s$ auf. Dadurch ergibt sich ein
asymptotischer Anstieg der Sperrd"ampfung mit der Potenz in $\omega$, die dem
Grad des Nennerpolynoms --- also $N$ --- entspricht. Im zeitdiskreten Fall
entspricht diese Graddifferenz des Z"ahler- und des Nennerpolynoms der Wahl
\mbox{$A_0=N\!-\!1$}, da dann auch dort die Z-Transformierte der Basisfensterfolge 
au"ser den "aquidistanten Nullstellen am Einheitskreis im Raster \mbox{$2\pi/F$} 
( entspricht der Sinusfunktion im Z"ahler ) keine weiteren Nullstellen enth"alt. 
H"atte man im kontinuierlichen Fall eine Basisfensterfunktion mit anderen
Fourierreihenkoeffizienten gew"ahlt, so w"urde im Z"ahler ein Polynom in $s$
auftreten dessen Grad maximal \mbox{$N\!-\!1$} w"are. Dieser Fall entspr"ache
bei der zeitdiskreten Fensterfolge der Wahl von beliebigen Nullstellen
$z_{0,\rho}$. Ebenso k"onnte man also zur Konstruktion der kontinuierlichen
Fensterfunktion maximal \mbox{$N\!-\!1$} Nullstellen in der $s$-Ebene frei
w"ahlen. Die Modifikationen, die sich daraus f"ur den im folgenden
angegebenen Algorithmus erg"aben, sind "ahnlich denen, die sich bei der
Konstruktion der zeitdiskreten Fensterfolge durch die Einf"uhrung der
frei w"ahlbaren Nullstellen ergeben haben. Auf eine Darstellung dieser
Modifikationen wird jedoch verzichtet. Das in Kapitel \ref{E.Kap.11.4} aufgelistete
Programm enth"alt die Modifikationen, die es erm"oglichen \mbox{$N\!-\!1$}
Nullstellen frei zu w"ahlen.}

Wie oben beschrieben erhalten wir das Betragsquadrat des Spektrums der
Fensterfunktion durch die "Uberlagerung der verschobenen Betragsquadrate
des Spektrums der Basisfensterfunktion. F"ur \mbox{$s\!=\!j\omega$} l"asst
sich  Betragsquadrat des Spektrums der Fensterfunktion aus
\begin{gather*}
F_{\!\infty}(-j\CdoT s)\cdot F_{\!\infty}(j\CdoT s)\;=\;D_{\infty}(s)\;=\!
\Sum{\nu_1=\frac{1-N}{2}}{\frac{N-1}{2}}
G_{\infty}(s\!-\!j\CdoT\nu_1\CdoT2\pi)\cdot
G_{\infty}(\!-s\!+\!j\CdoT\nu_1\CdoT2\pi)\;=
\notag\\[18pt]\label{E.10.38}\begin{flalign}
&=\!\Sum{\nu_1=\frac{1-N}{2}}{\frac{N-1}{2}}
\frac{\D\,(j\CdoT2\pi)^N\Cdot(N\!-\!1)!\cdot
e^{\!-\frac{s-j\cdot\nu_1\cdot2\pi}{2}}\cdot
\sin\!\big({\T\frac{s-j\cdot(N-1+2\cdot\nu_1)\cdot\pi}
{2\cdot j}}\big)\,}
{\pi\CdoT\tbinom{2\cdot N-2}{N-1}^{\T\!\frac{1}{2}}\CdoT
\Prod{\nu_2=\frac{1-N}{2}}{\frac{N-1}{2}}\!
\big(s\!-\!j\CdoT(\nu_2\!+\!\nu_1)\CdoT2\pi\big)}\;\cdot{}&&
\end{flalign}\\*[8pt]\begin{flalign*}
&&{}\cdot\;\frac{\D\,(j\CdoT2\pi)^N\Cdot(N\!-\!1)!\cdot
e^{\frac{s-j\cdot\nu_1\cdot2\pi}{2}}\cdot
\sin\!\big({\T\frac{-s-j\cdot(N-1-2\cdot\nu_1)\cdot\pi}
{2\cdot j}}\big)\,}
{\pi\CdoT\tbinom{2\cdot N-2}{N-1}^{\T\!\frac{1}{2}}\CdoT
\Prod{\nu_2=\frac{1-N}{2}}{\frac{N-1}{2}}\!
\big(-s\!-\!j\CdoT(\nu_2\!-\!\nu_1)\CdoT2\pi\big)}\;=&
\end{flalign*}\\[16pt]
=\;\frac{\D\,( j\CdoT2\pi)^{2\cdot N}\Cdot(N\!-\!1)!^{\T{}^2}\,}
{\pi^{\uP{0.4}{2}}\Cdot\tbinom{2\cdot N-2}{N-1}}\cdot
\sin\!\big({\T\frac{s}{2\cdot j}}\big)^{\!2}\cdoT\!
\Sum{\nu_1=\frac{1-N}{2}}{\frac{N-1}{2}}
\frac{1}{\;\Prod{\nu_2=\frac{1-N}{2}}{\frac{N-1}{2}}\!
\big(s\!-\!j\CdoT(\nu_2\!+\!\nu_1)\CdoT2\pi\big)^{\uP{0.4}{\!2}}\,}
\end{gather*}
berechnen. Setzt man \mbox{$s=j\CdoT\nu\CdoT2\pi$} mit
\mbox{$\nu=0\;(1)\;N\!-\!1$} ein, so erh"alt man f"ur das Betragsquadrat
der gesuchten Spektralwerte der Fensterfunktion den Ausdruck
\begin{equation}
\big|F_{\!\infty}(2\pi\CdoT\nu)\big|^2\,=\;
\frac{\Sum{\nu_1=0}{N-1-|\nu|}\tbinom{N-1}{\nu_1}^{\!\!2}{}}
{\tbinom{2\cdot N-2}{N-1}},
\label{E.10.39}
\end{equation}
der f"ur \mbox{$\nu\!=\!0$} den f"ur die G"ultigkeit von Gleichung (\ref{E.10.23})
notwendigen Wert Eins ergibt. In Gleichung (\ref{E.10.38}) erh"alt man durch
Erweiterung auf den Hauptnenner
\begin{gather}
F_{\!\infty}(\!-j\CdoT s)\CdoT F_{\!\infty}(j\CdoT s)\;=\;D_{\infty}(s)\;=
\label{E.10.40}\\*[8pt]
=\;\underbrace{\;{}-\frac{\D(2\pi)^{2\cdot N}\cdot(N\!-\!1)!^{\T{}^2}\cdot
\sin\!\big({\T\frac{s}{2\CdoT j}}\big)^{\!2}}
{\pi^{\uP{0.4}{2}}\Cdot\tbinom{2\cdot N-2}{N-1}\!
\Prod{\nu_3=1-N}{N-1}\!\!(s-\!j\CdoT\nu_3\CdoT2\pi)^2}}_{
\D=D_{E,\infty}(s)}\,\cdot\;
\underbrace{(-1)^{N-1}\CdoT\!\!\!\Sum{\nu_1=\frac{1-N}{2}}{\frac{N-1}{2}}
\Prod{{}\;\;\nu_2}{}(s\!-\!j\CdoT\nu_2\CdoT2\pi)^{\uP{0.4}{\!2}}}_{
\D=D_{\overline{E},\infty}(s)},\notag
\end{gather}
wobei der Lauf"|index $\nu_2$ des in der Summe stehenden Produkts jeweils die
\mbox{$N\!-\!1$} Werte annimmt, die das Produkt entstehen lassen, das zur
Erweiterung auf den Hauptnenner ben"otigt wird.
Beim Summanden mit dem Lauf"|index $\nu_1$ nimmt $\nu_2$ die Werte
\mbox{$1\!-\!N\;(1)\;N\!-\!1$} ohne die Werte
\mbox{$\nu_1\!+\!(1\!-\!N)/2\;(1)\;\nu_1\!+\!(N\!-\!1)/2$} an.

Bleibt noch die Phase des Spektrums der Fensterfunktion zu berechnen. Der
Anteil \mbox{$D_{E,\infty}(s)$} ist eine Funktion, deren Nullstellen
alle doppelt sind und alle auf der imagin"aren Achse bei
\mbox{$\omega=2\pi\CdoT\nu$} liegen, wobei $\nu$ eine ganze Zahl au"ser
\mbox{$\nu=1\!-\!N\;(1)\;N\!-\!1$} ist. Jede dieser doppelten Nullstellen
tritt als einfache Nullstelle im minimalphasigen Anteil auf. Die Phase des
Anteils der einfachen Nullstellen ist f"ur \mbox{$|\omega|\!<\!N\CdoT2\pi$}
null. Um zu erreichen, dass der minimalphasige Anteil von
\mbox{$D_{E,\infty}(s)$} zu einer kausalen Zeitfunktion geh"ort, ist
der Anteil der einfachen Nullstellen noch mit \mbox{$e^{\!-\frac{s}{2}}$}
zu multiplizieren (\,vgl. Gleichung (\ref{E.10.37})\,). F"ur die Frequenzen
\mbox{$\omega=2\pi\CdoT\nu$} mit \mbox{$\nu=1\!-\!N\;(1)\;N\!-\!1$}
erhalten wir daher die Phase \mbox{$\nu\CdoT\pi$}.

\mbox{$D_{\overline{E},\infty}(s)$} besitzt keine Nullstellen auf der
imagin"aren Achse, weil niemals alle f"ur \mbox{$s\!=\!j\omega$} reellen
Summanden zugleich null werden k"onnen, und alle Summanden das gleiche
Vorzeichen aufweisen. Die Phase des Anteils von \mbox{$D_{\overline{E},\infty}(s)$}, 
dessen Nullstellen links der imagin"aren Achse liegen, wird "uber das Cepstrum 
berechnet. Es empfiehlt sich wieder vor der Berechnung des Cepstrums eine 
Bilineartransformation
\begin{equation}
s\;=\;c_{\infty}\CdoT2\pi\Cdot\frac{\;\Tilde{z}\!-\!1\;}{\Tilde{z}\!+\!1}.
\label{E.10.41}
\end{equation}
mit dem Transformationsparameter \mbox{$c_{\infty}\!>\!0$} durchzuf"uhren.
Nach $\Tilde{z}$ aufgel"ost erh"alt man:
\begin{equation}
\Tilde{z}\;=\;\frac{\;c_{\infty}\CdoT2\pi\!+\!s\;}{c_{\infty}\CdoT2\pi\!-\!s}
\label{E.10.42}
\end{equation}
F"ur die Frequenzen $\omega$ und $\widetilde{\Omega}$ ergibt sich folgender Zusammenhang:
\begin{equation}
\widetilde{\Omega}=
2\CdoT\arctan\Big(c_{\infty}\Cdot\frac{\omega}{2\pi}\Big).
\label{E.10.43}
\end{equation}
F"ur die aufzuspaltende Summe der Produkte in Gleichung (\ref{E.10.40})
erh"alt man dann durch Substitution von~$s$
\begin{gather}
D_{\overline{E},\infty}(s)\;=
(-1)^{N-1}\CdoT\!\!\!\Sum{\nu_1=\frac{1-N}{2}}{\frac{N-1}{2}}
\Prod{{}\;\;\nu_2}{}(s\!-\!j\CdoT2\pi\CdoT\nu_2)^{\uP{0.4}{\!2}}\,=\;
\widetilde{D}_{\overline{E},\infty}(\Tilde{z})\;=
\label{E.10.44}\\*
=\;\underbrace{\bigg(\frac{\Tilde{z}}{(1\!+\!\Tilde{z})^2}\bigg)^{\!\!N-1}
\!\!}_{\D=\widetilde{D}_{P,\infty}(\Tilde{z})}\,\;\cdot
\underbrace{\Sum{\nu_1=\frac{1-N}{2}}{\frac{N-1}{2}}\;\Prod{\;\nu_2}{}
\big(K_{\infty,\nu_2}\Cdot
(\Tilde{z}\!-\!\Tilde{z}_{\infty,\nu_2})\CdoT
(\Tilde{z}^{-1}\!\!-\!\Tilde{z}_{\infty,\nu_2}^*)\big)}_{\D=
\widetilde{D}_{N,\infty}(\Tilde{z})},\notag
\end{gather}
mit den Nullstellen $\Tilde{z}_{\infty,\nu_2}$ die durch die
Bilineartransformation aus den Nullstellen auf der
imagin"aren Achse bei \mbox{$j\CdoT2\pi\CdoT\nu_2$} entstanden sind.
\begin{equation}
\Tilde{z}_{\infty,\nu_2}\;=\;
\frac{{}\;c_{\infty}+j\CdoT\nu_2\;{}}{c_{\infty}-j\CdoT\nu_2}\;=\;
e^{j\cdot2\cdot\arctan\!\big({\T\frac{\nu_2}{c_{\infty}}}\big)}
\label{E.10.45}
\end{equation}
F"ur die Konstanten $K_{\infty,\nu_2}$ ergeben sich die Werte
\begin{equation}
K_{\infty,\nu_2}\;=\;(2\pi)^2\Cdot(c_{\infty}^2\!+\!\nu_2^2).
\label{E.10.46}
\end{equation}
Sie lassen sich mit einem relativen Fehler von etwa $\varepsilon$ berechnen. 
Der Nullstellenwinkel \mbox{$2\CdoT\arctan(\nu_2/c_{\infty})$} liegt im
Intervall \mbox{$[-\pi;\pi]$}. Im gesamten Intervall liegt der absolute Fehler
des Nullstellenwinkels maximal in der Gr"o"senordnung von $\varepsilon$.

Die Z-Transformierte \mbox{$\widetilde{D}_{P,\infty}(\Tilde{z})$} hat eine
\mbox{$2\cdoT\!N\!-\!2$-fache} Polstelle bei \mbox{$\Tilde{z}\!=\!-1$}.
Der Anteil \mbox{$\widetilde{D}_{P,\infty}(\Tilde{z})$} wird nun
zur"ucktransformiert. Man erh"alt bis auf einen konstanten Faktor
den Term\vspace{-12pt}
\begin{equation}
D_{P,\infty}(s)\;=\;
(s\!+\!c_{\infty}\CdoT2\pi)^{N-1}\CdoT(s\!-\!c_{\infty}\CdoT2\pi)^{N-1}
\label{E.10.47}
\end{equation}
Dieser Term hat ein zur imagin"aren Achse symmetrisches Nullstellenpaar der
Vielfachheit \mbox{$N\!-\!1$} bei \mbox{$s=\pm c_{\infty}\CdoT2\pi$}.
Jeder Polynomfaktor \mbox{$s\!+\!c_{\infty}\CdoT2\pi$} einer einfachen
Nullstelle links der imagin"aren Achse liefert den Phasenbeitrag
\begin{equation}
-\arctan\!\Big(c_{\infty}\Cdot\frac{\omega}{2\pi}\Big),
\label{E.10.48}
\end{equation}
der bei der Berechnung der Phase von \mbox{$F_{\!\infty}(\nu\CdoT2\pi)$}
entsprechend mit der Vielfachheit \mbox{$N\!-\!1$} zu addieren ist.

Nun m"ussen wir noch die Phase des minimalphasigen Anteils des Summenanteils 
\mbox{$\widetilde{D}_{N,\infty}(\Tilde{z})$} in Gleichung (\ref{E.10.44}) berechnen. 
Diese Summe ist f"ur \mbox{$\Tilde{z}=e^{j\widetilde{\Omega}}$} positiv reell 
und l"asst sich mit \mbox{$\widetilde{\Omega}=\eta\CdoT2\pi/\widetilde{M}$}
f"ur alle ganzzahligen Werte $\eta$ in der Form
\begin{gather}
\widetilde{D}_{N,\infty}\big(e^{j\widetilde{\Omega}}\big)\;=
\Sum{\nu_1=\frac{1-N}{2}}{\frac{N-1}{2}}\Prod{\;\nu_2}{}
\bigg(4\CdoT K_{\infty,\nu_2}
\Cdot\sin\!\Big({\T\frac{\widetilde{\Omega}}{2}-
\arctan\!\big(\frac{\nu_2}{c_{\infty}}\big)}\Big)^{\!\!2}\bigg)\;=
\label{E.10.49}\\*[8pt]
=\Sum{\nu_1=\frac{1-N}{2}}{\frac{N-1}{2}}\Prod{\;\nu_2}{}
\bigg(4\CdoT K_{\infty,\nu_2}
\Cdot\sin\!\bigg({\T\frac{\pi}{\widetilde{M}}\CdoT
\Big(\eta\!-\!\frac{\widetilde{M}}{\pi}\CdoT
\arctan\!\big(\frac{\nu_2}{c_{\infty}}\big)\Big)}\bigg)^{\!\!2}\,\bigg)
\notag
\end{gather}
berechnen. Verwendet man die $\widetilde{M}$ ganzzahligen
$\eta$-Werte
\begin{equation}
{\T\widetilde{M}\CdoT\Big(\frac{1}{\pi}\Cdot
\arctan\!\big(\frac{\nu_2}{c_{\infty}}\big)-\frac{1}{2}\Big)\;<\;
\eta\;\le\;
\widetilde{M}\CdoT\Big(\frac{1}{\pi}\Cdot
\arctan\!\big(\frac{\nu_2}{c_{\infty}}\big)+\frac{1}{2}\Big)
\quad\text{ mit}\quad\eta\in\mathbb{Z}},
\label{E.10.50}
\end{equation}
so ist das Argument der Sinusfunktion betraglich kleiner als \mbox{$\pi/2$},
und die Fehler bei der Berechnung des Quadrats der Sinusfunktion werden
klein gehalten. Da die $\eta$-Werte ganze Zahlen sind,
die am Rechner exakt darstellbar sind, l"asst sich die Differenz
aus $\eta$ und dem durch die endliche Wortl"ange quantisierten und
normierten Winkel \mbox{$\widetilde{M}/\pi\CdoT\arctan(\nu_2/c_{\infty})$}
mit einem relativen Fehler von $\varepsilon$ berechnen.
F"ur diese Werte von $\eta$ kann daher auch das Quadrat der
Sinusfunktion mit derselben relativen Genauigkeit berechnet werden.
In der Praxis hat sich gezeigt, dass es auch hier nicht notwendig
ist, die Werte $\eta$ in der eben dargelegten Art um ganzzahlige
Vielfache von $\widetilde{M}$ zu reduzieren, da die ohne
die Reduktion berechneten Fourierreihenkoeffizienten der Fensterfunktion
sich nur unwesentlich von den Werten unterscheiden, die man bei
Ber"ucksichtigung von Gleichung (\ref{E.10.50}) erh"alt.
Die Faktoren $K_{\infty,\nu_2}$ sind positiv und k"onnen ebenfalls
mit der gew"unschten relativen Genauigkeit berechnet werden. Alle Summanden
in Gleichung (\ref{E.10.49}) sind nichtnegativ und weisen relative Fehler
in der Gr"o"senordnung von $\varepsilon$ auf. Da bei keiner Frequenz
$\widetilde{\Omega}$ alle Summanden gleichzeitig null werden ist die
"Uberlagerung aller $\nu_1$ Anteile --- also das Betragsquadrat des
Spektrums, dessen Phase berechnet werden soll --- stets echt positiv
und besitzt die geforderte relative Genauigkeit, die es erm"oglicht den
nat"urlichen Logarithmus des Betragsquadratspektrums mit
der notwendigen absoluten Genauigkeit zu berechnen. Bei der Akkumulation
der Summanden in Gleichung (\ref{E.10.49}) wird auf den konstanten Faktor
\mbox{${\D4\CdoT\max(K_{\infty,\nu_2})=16\CdoT\pi^2\Cdot\big(c_{\infty}^2\!+
(N\!-\!1)^2\big)}$} normiert, um sicherzustellen, dass der am Rechner
darstellbare Zahlenbereich nicht "uberschritten wird. Ein weiterer konstanter
Faktor, wird so gew"ahlt, dass das Maximum des Betragsquadratspektrums
gleich dem Reziprokwert des Minimums ist, so dass der nat"urliche Logarithmus
mit optimaler Genauigkeit berechnet wird. Eine inverse FFT des Logarithmus
von \mbox{$\widetilde{D}_{N,\infty}\big(e^{j\widetilde{\Omega}}\big)$}
liefert uns das doppelte, reelle und geradesymmetrische Cepstrum. Der
durch die Quantisierungsfehler der FFT entstehende Imagin"arteil wird durch
eine Realteilbildung unterdr"uckt. Die gerade Symmetrie n"utzt man aus, um die 
Quantisierungsfehler der FFT im Realteil zu verringern, indem man den 
Mittelwert des Cepstrums und des gespiegelten Cepstrums bildet. Damit das
Cepstrum m"oglichst rasch abklingt muss der Bilineartransformationsparameter
$c_{\infty}$ geeignet gew"ahlt werden. $c_{\infty}$ erh"alt man aus dem
empirisch gewonnenen $c$ nach Gleichung (\myref{6.30}) indem man die 
Bilineartransformation nach Gleichung (\myref{6.21}) mit der N"aherung
\mbox{$z=e^{j\Omega}\approx1\!+\!j\Omega=1\!+\!j\omega/M/N=1\!+\!s/M/N$} f"ur
hinreichend gro"ses $M$ und kleines $\omega$ im Durchlassbereich des Spektrums
der Fensterfunktion in die Bilineartransformation nach Gleichung (\ref{E.10.42})
"uberf"uhrt. Damit ergibt sich:
\begin{equation}
c_{\infty}\;=\;\frac{1}{2\pi}\cdot\lim_{M\to\infty}c\CdoT M\CdoT N\;=\;
\Big(\frac{N}{2}\Big)^{\T\frac{4}{3}}.
\label{E.10.51}
\end{equation}
Mit \mbox{$M\to\infty$} in Gleichung (\myref{6.31}) erh"alt man f"ur die
n"otige L"ange $\widetilde{M}$ der inversen FFT den Mindestwert
\begin{equation}
\widetilde{M}\;>\;\big(\text{ Mantissenwortl"ange }-1\,\big)
\cdot 2^{{}^{\T\ln\big(\frac{N}{3}\big)}},
\label{E.10.52}
\end{equation}
der wieder von der Mantissenwortl"ange des zur Berechnung der Fensterfunktion
verwendeten Rechners abh"angt. Wieder wird als L"ange der inversen FFT die
kleinste Zweierpotenz gew"ahlt, die die Ungleichung (\ref{E.10.52}) erf"ullt.

Die Werte des Cepstrums f"ur \mbox{$k\!>\!0$} sind die Sinusreihenkoeffizienten
der in $\widetilde{\Omega}$ periodischen schiefsymmetrischen Phasenfunktion.
Gesucht wird die Phase der Fourierreihenkoeffizienten der Fensterfunktion, also
die Phase von \mbox{$F_{\!\infty}(\omega)$} bei den Frequenzen
\mbox{$\omega=\nu\CdoT2\pi$} mit \mbox{$\nu=1\;(1)\;N\!-\!1$}. 
Mit Gleichung (\ref{E.10.43}) berechnet man die entsprechenden Frequenzen 
$\widetilde{\Omega}$. Sie sind kleiner als $\pi$ und lassen sich mit einem 
relativen Fehler von circa $\varepsilon$ berechnen. Nun werden die Werte der in
$\widetilde{\Omega}$ periodischen, schiefsymmetrischen Phasenfunktion des
minimalphasigen Anteils von \mbox{$\widetilde{D}_{N,\infty}(\Tilde{z})$}
durch Auswertung der Sinusreihe bei diesen Frequenzen $\widetilde{\Omega}$
berechnet. Bei der Auswertung der Fouriersinusreihe wird die Reihenfolge der
Summanden so gew"ahlt, dass zun"achst mit den kleinsten Summanden begonnen
wird. Dieses Vorgehen ergibt eine h"ohere Genauigkeit des Ergebnisses.

Die anschlie"sende Addition dieses Phasenanteils zu dem Phasenanteil
des in Gleichung (\ref{E.10.44}) vor die Summe gezogenen Terms
\mbox{$\widetilde{D}_{P,\infty}(\Tilde{z})$} nach Gleichung (\ref{E.10.48})
ergibt die gesuchte Phase von \mbox{$F_{\!\infty}(\nu\CdoT2\pi)$}.
Der Betrag wurde als die positive Wurzel des Ausdrucks in Gleichung
(\ref{E.10.39}) berechnet. Damit sind die Fourierreihenkoeffizienten der
Fensterfunktion berechnet.

Es sei wieder angemerkt, dass es au"ser f"ur den trivialen Fall mit 
\mbox{$N\!=\!1$} (\,Rechteckfenster\,) auch f"ur \mbox{$N=2\;(1)\;5$} 
geschlossene L"osungen f"ur \mbox{$F_{\!\infty}(\nu\CdoT2\pi)$} gibt, 
deren Berechnung sich nur f"ur \mbox{$N\!=\!2$} lohnt. Man erh"alt:

\begin{align}
F_{\!\infty}(-2\pi)&\;=\;-(1\!-\!j)/2
\notag\\
F_{\!\infty}(0)&\;=\;1
\label{E.10.53}\\
F_{\!\infty}(2\pi)&\;=\;-(1\!+\!j)/2.
\notag
\end{align}\vspace{0pt}

In Kapitel \ref{E.Kap.11.5} ist ein Hilfsprogramm abgedruckt, 
das die Werte der Fensterfunktion \mbox{$f_{\!\infty}(t)$} f"ur 
beliebiges $t$ nach Gleichung (\ref{E.10.25}) berechnet. Bei der
Berechnung von \mbox{$f_{\!\infty}(t)$} bilden die Realteile
\mbox{$2\cdot\Re\big\{\,F_{\!\infty}(\nu\CdoT2\pi)\,\big\}$}
der berechneten $N$ Werte die Koeffizienten einer Kosinusreihe, die
Imagin"arteile \mbox{$2\cdot\Im\big\{\,F_{\!\infty}(\nu\CdoT2\pi)\,\big\}$}
bilden die einer Sinusreihe. Beide Reihen bilden zusammen eine
zeitkontinuierliche, periodische und reelle Funktion in $t$.
Setzt man die so berechnete Funktion f"ur \mbox{$t\!<\!0$} und
f"ur \mbox{$t\!\ge\!1$} zu null, so erh"alt man den gesuchten
Wert der kontinuierlichen Fensterfunktion zum Zeitpunkt $t$.
Um ein genaueres Ergebnis zu erhalten wird die Reihenfolge der
Summation der einzelnen Reihenglieder wieder vertauscht, da auch
hier die ersten Koeffizienten der Reihen die gr"o"sten sind.

Das Spektrum \mbox{$F_{\!\infty}(\omega)$} kann mit dem Hilfsprogramm 
in Kapitel \ref{E.Kap.11.6} f"ur beliebiges $\omega$ hochgenau berechnet werden.
Bei der Berechnung des Spektrums der Fensterfunktion f"ur beliebige
Frequenzen geht man analog zu dem Fall der diskreten Fensterfolge vor.
Wieder berechnet man den Betrag und die Phase des Spektrums getrennt.
Die Phase wird wieder aus dem Cepstrum, das mit dem ebengenannten
Algorithmus berechnet wird, durch Auswertung der Sinusreihe aber
diesmal nicht bei den Frequenzen \mbox{$\omega=\nu\CdoT2\pi$} mit
\mbox{$\nu=1\;(1)\;N\!-\!1$}, sondern bei den Frequenzen, f"ur die
das Spektrum berechnet werden soll, gewonnen. Der Betrag wird durch
vorzeichenrichtiges Radizieren des Betragsquadrats gewonnen. Da
\mbox{$F_{\!\infty}(-j\CdoT s)$} auf der imagin"aren Achse nur einfache
Nullstellen im "aquidistanten Raster $2\pi$ aufweist, ist das Vorzeichen
bei der Radizierung auf einfache Weise aus $\omega$ bestimmbar. 

Das Betragsquadrat von \mbox{$F_{\!\infty}(\omega)$} wird durch "Uberlagerung der
verschobenen Betragsquadrate von \mbox{$G_{\infty}(j\omega)$} berechnet. Dabei
ist jeweils die Polstelle von \mbox{$G_{\infty}(s\!-\!j\CdoT\nu_1\CdoT2\pi)$},
die dem Punkt \mbox{$s\!=\!j\omega$}, f"ur den das Spektrum berechnet werden
soll, am n"achsten liegt, mit der Sinusfunktion im Z"ahler von
\mbox{$G_{\infty}(j\omega\!-j\CdoT\!\nu_1\CdoT2\pi)$} zu einer
in der Frequenz verschobenen si-Funktion in $\omega$ zusammenzufassen,
auszuwerten, zu quadrieren und mit dem Betragsquadrat des
Produktes der Abst"ande zu den restlichen Polstellen zu dividieren.
Da auf diese Weise eine Summe von positiven Produkten, deren Faktoren
mit h"ochstm"oglicher relativer Genauigkeit berechnet worden sind, gebildet
wird, ist so das Spektrum auch f"ur den Sperrbereich bis zu sehr hohen
Frequenzen akkurat berechenbar. Mit dem in Kapitel \ref{E.Kap.11.6} auszugsweise
angegebenen Programm kann die Fouriertransformierte bis zu einer
Sperrd"ampfung von etwa \mbox{$-10\CdoT\log_{10}(\text{\tt realmin})$}
berechnet werden.

\begin{figure}[btp]
\begin{center}
{ 
\begin{picture}(450,390)

\input{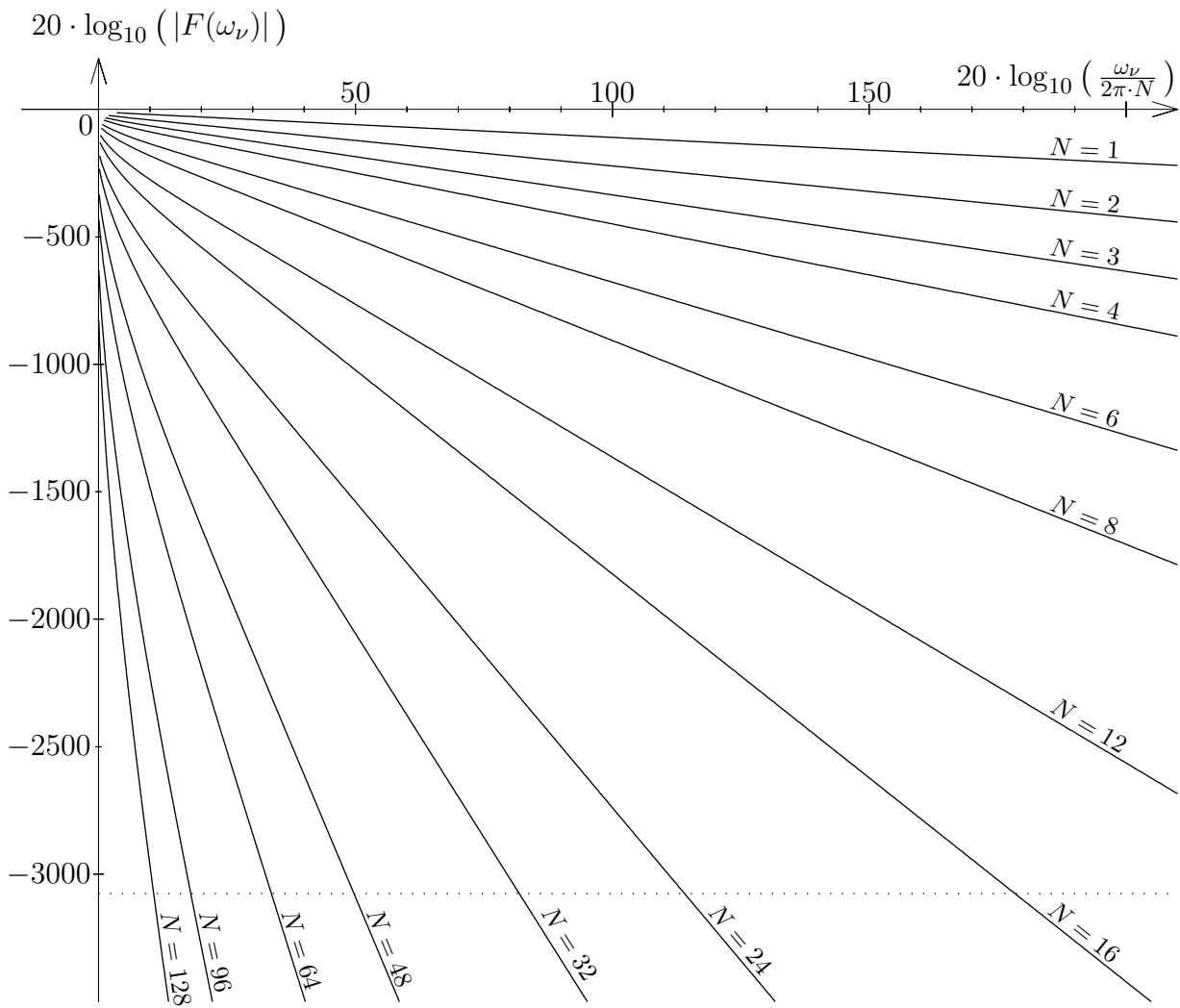}
\put(0,375){\makebox(0,0)[lb]{
$20\cdot\log_{10}\big(\,|F(\omega_{\nu})|\,\big)$}}
\put(450,355){\makebox(0,0)[rb]
{$20\cdot\log_{10}\big(\,\frac{\omega_{\nu}}{2\pi\cdot N}\,\big)$}}
\put(28,348){\makebox(0,0)[rt]{$0$}}
\put(27,300){\makebox(0,0)[r]{$-500$}}
\put(27,250){\makebox(0,0)[r]{$-1000$}}
\put(27,200){\makebox(0,0)[r]{$-1500$}}
\put(27,150){\makebox(0,0)[r]{$-2000$}}
\put(27,100){\makebox(0,0)[r]{$-2500$}}
\put(27,50){\makebox(0,0)[r]{$-3000$}}
\put(130,353){\makebox(0,0)[b]{$50$}}
\put(230,353){\makebox(0,0)[b]{$100$}}
\put(330,353){\makebox(0,0)[b]{$150$}}
\put(400,332){\rotatebox{-2.86}{\footnotesize$N=1$}}
\put(400,312){\rotatebox{-5.71}{\footnotesize$N=2$}}
\put(400,293){\rotatebox{-8.53}{\footnotesize$N=3$}}
\put(400,273){\rotatebox{-11.31}{\footnotesize$N=4$}}
\put(400,233){\rotatebox{-16.70}{\footnotesize$N=6$}}
\put(400,193){\rotatebox{-21.80}{\footnotesize$N=8$}}
\put(400,114){\rotatebox{-30.96}{\footnotesize$N=12$}}
\put(399,35){\rotatebox{-38.66}{\footnotesize$N=16$}}
\put(267,35){\rotatebox{-50.19}{\footnotesize$N=24$}}
\put(202,35){\rotatebox{-58.00}{\footnotesize$N=32$}}
\put(134,35){\rotatebox{-67.38}{\footnotesize$N=48$}}
\put(101,35){\rotatebox{-72.65}{\footnotesize$N=64$}}
\put(69,35){\rotatebox{-78.23}{\footnotesize$N=96$}}
\put(54,35){\rotatebox{-81.12}{\footnotesize$N=128$}}

\end{picture}}
\end{center}\vspace{-8pt}
\setlength{\belowcaptionskip}{2pt}
\caption{Sperrd"ampfung der Nebenmaxima der Spektren der kontinuierlichen Fensterfunktionen bei den Frequenzen
\mbox{$\omega_{\nu}=(2\CdoT\nu\!+\!1)\CdoT\pi$} mit $N$ als Parameter.}
\label{E.b5u}
\rule{\textwidth}{0.5pt}\vspace{-10pt}
\end{figure}
In Bild \ref{E.b5u} sind die Nebenmaxima der
Betr"age der Spektren der kontinuierlichen Fensterfunktionen
bei den Frequenzen \mbox{$\omega_{\nu}=(2\CdoT\nu\!+\!1)\CdoT\pi$}
mit ganzzahligem \mbox{$\nu\!>\!N$} doppelt logarithmisch "uber der
auf \mbox{$2\pi\CdoT\!N$} normierten Frequenz $\omega$ aufgetragen.
F"ur jeden der Werte $1$, $2$, $3$, $4$, $6$, $8$, $12$, $16$, $24$,
$32$, $48$, $64$, $96$ und $128$ des Fensterl"angenfaktors $N$ wurde
eine Kurve eingetragen, wobei einerseits f"ur gro"se Frequenzen nicht
alle Nebenmaxima berechnet wurden (\,es w"aren einfach zu viele\,),
und andererseits zur besseren Darstellbarkeit die bei den Nebenmaxima
berechneten Betr"age der Spektralwerte durch gerade Linien miteinander
verbunden wurden. Bei dieser Graphik mit der etwas ungew"ohnlichen Skalierung
der Ordinate sollen nun drei Dinge gezeigt werden. Zum ersten ist zu sehen, dass
der asymptotische Anstieg der Sperrd"ampfung mit der Potenz $N$
erfolgt. Im Bild \ref{E.b5u} wurde die Ordinate um den Faktor $20$
enger skaliert als die Abszisse und die Beschriftung der Kurven
wurde mit der Neigung eingetragen, die der Potenz \mbox{$\omega^{-N}$}
entspricht. Zum zweiten sieht man, dass sich das Quadrat des
Betragsfrequenzgangs wirklich bis in die Gr"o"senordnung der kleinsten
positiven Zahl {\tt realmin}, die am Rechner mit voller Genauigkeit darstellbar
ist, berechnen l"asst, wenn man diesen "uber die Summe der verschobenen
Betragsquadratspektren berechnet, wobei man deren Summanden jeweils
als Produkte der Betragsquadrate der Abst"ande zu den Nullstellen berechnet.
Zum Vergleich wurde der Wert $\sqrt{\text{\tt realmin}\,}$ als punktierte
Linie eingetragen. Drittens wird demonstriert, dass auch extrem gro"se
Werte von $N$ kein Problem f"ur den verwendeten Berechnungsalgorithmus
darstellen. Lediglich die Dauer der Berechnung nimmt dann
--- grob gesch"atzt etwa quadratisch mit $N$ --- zu. Bei \mbox{$N\!=\!128$}
wurde eine FFT-L"ange von \mbox{$\widetilde{M}\!=\!1024$} ben"otigt.

Mit dem Hilfsprogramm in Kapitel \ref{E.Kap.11.7} kann man die Autokorrelationsfunktion
\mbox{$d_{\infty}(t)$} des kontinuierlichen Fensters f"ur beliebiges
$t$ zumindest theoretisch exakt berechnen. Bei einem kontinuierlichen
Fenster kann man das Faltungsintegral, das die Fensterautokorrelationsfunktion
definiert, nur n"aherungsweise und mit einem erheblichen
Rechenaufwand berechnen, wenn man hohe Anspr"uche an die Genauigkeit
der Berechnung stellt. Auch die bei einer diskreten Fensterfolge immer
bestehende M"oglichkeit, die Fensterautokorrelationsfolge aus endlich
vielen Abtastwerten des Betragsquadrats des Fensterspektrums mit einer
DFT zu berechnen, ist hier in der Regel nicht gegeben, da wegen der zeitlichen
Begrenzung der Autokorrelationsfunktion des Fensters eine unbegrenzte Anzahl
spektraler Abtastwerte ben"otigt werden w"urde. Mit den
Sinusreihenkoeffizienten
\begin{equation}
S_{\nu}=\!\Sum{\substack{\Tilde{\nu}=1-N\\\Tilde{\nu}\neq\nu}}{N-1}
\frac{{\D\;\Re\big\{F_{\!\infty}(\nu\CdoT2\pi)^*\!\CdoT
F_{\!\infty}(\Tilde{\nu}\CdoT2\pi)\big\}}\;}
{\pi\cdot(\Tilde{\nu}\!-\!\nu)}
\label{E.10.54}
\end{equation}
kann man jedoch die Fensterautokorrelationsfunktion f"ur \mbox{$0\!\le\!t\!<\!1$} als
\begin{equation}
d_{\infty}(t)=\Sum{\nu=1-N}{N-1}S_{\nu}\cdot\sin(2\pi\CdoT\nu\CdoT t)+
(1\!-\!t)\cdoT\Sum{\nu=1-N}{N-1}
\big|F_{\!\infty}(\nu\CdoT2\pi)\big|^2\Cdot\cos(2\pi\CdoT\nu\CdoT t)
\label{E.10.55}
\end{equation}
"uber die bereits bestimmten Abtastwerte des Spektrums der
Fensterfunktion berechnen. Die gerade Symmetrie der Fensterautokorrelationsfunktion 
liefert die Funktionswerte f"ur negatives $t$. Diese Art der Berechnung der
Fensterautokorrelationsfunktion ist "ubrigens bei allen
Fensterfunktionen m"oglich, die sich als eine Periode einer
endlichen Fourierreihe darstellen lassen, wobei die Grenzen
der Summationsindizes so zu modifizieren sind, dass alle
Fourierreihenkoeffizienten ber"ucksichtigt werden. Bei dem
in Kapitel \ref{E.Kap.11.7} enthaltenen Programmauszug sind die Summationsgrenzen
so ver"andert, dass auch f"ur die hier nicht beschriebene freie
Wahl weiterer Nullstellen der Laplacetransformierten der
Basisfensterfunktion die Berechnung der Fensterautokorrelationsfunktion
m"oglich ist.

\section[Die kontinuierlichen Fensterfunktion als Grenzwertl"osung]{Die kontinuierlichen Fensterfunktion als \\Grenzwertl"osung}\label{E.Kap.10.8}

\begin{figure}[btp]
\begin{center}
{ 
\begin{picture}(450,338)

\input{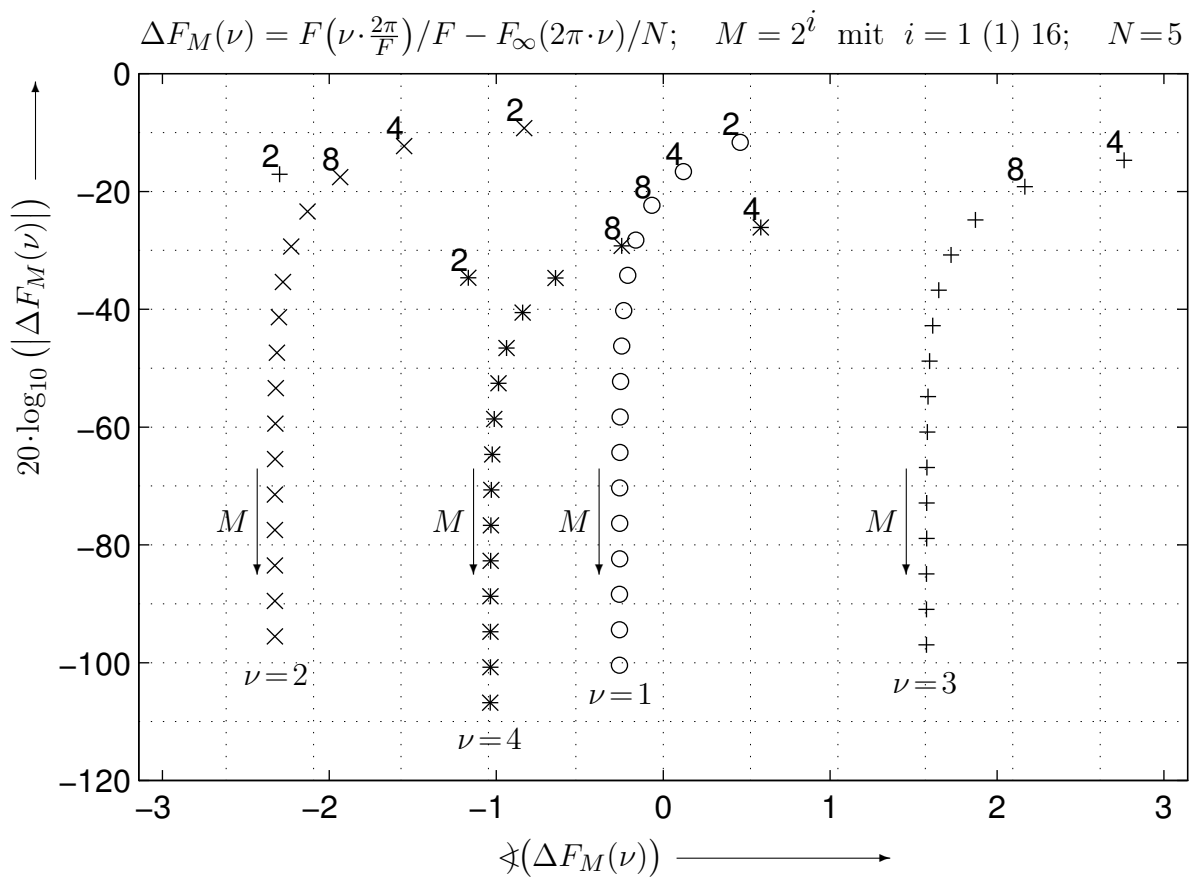}
\put(54,320){$\Delta F_M(\nu)=F\big({\T\nu\CdoT\frac{2\pi}{F}}\big)/F-
F_{\!\infty}(2\pi\CdoT\nu)/N;\quad{\D M=2^{\T i}}\;\text{ mit }\;i=1\;(1)\;16;\quad N\!=\!5$}
\put(234,79){\makebox(0,0)[t]{$\nu\!=\!1$}}
\put(105,86){\makebox(0,0)[t]{$\nu\!=\!2$}}
\put(348,83){\makebox(0,0)[t]{$\nu\!=\!3$}}
\put(185,62){\makebox(0,0)[t]{$\nu\!=\!4$}}
\put(255,13){\vector(1,0){80}}
\put(15,269){\vector(0,1){37}}
\put(6,260){\rotatebox[origin = cr]{90}{
$20\CdoT\log_{10}\big(\big|\Delta F_M(\nu)\big|\big)$}}
\put(250,13){\makebox(0,0)[r]{$\winkel\big(\Delta F_M(\nu)\big)$}}
\put(98,160){\vector(0,-1){40}}
\put(179,160){\vector(0,-1){40}}
\put(226,160){\vector(0,-1){40}}
\put(341,160){\vector(0,-1){40}}
\put(95,140){\makebox(0,0)[r]{$M$}}
\put(176,140){\makebox(0,0)[r]{$M$}}
\put(223,140){\makebox(0,0)[r]{$M$}}
\put(338,140){\makebox(0,0)[r]{$M$}}

\end{picture}}
\end{center}\vspace{-17pt}
\setlength{\belowcaptionskip}{-1pt}
\caption{Grenzwerte der Fourierreihenkoeffizienten der diskreten Fensterfolgen f"ur \mbox{$M\!\to\!\infty$}
im Vergleich zu den Fourierreihenkoeffizienten der kontinuierlichen Fensterfunktion.}
\label{E.b5t}
\rule{\textwidth}{0.5pt}\vspace{-5pt}
\end{figure}
Am Anfang des letzten Unterkapitels wurde gesagt, dass 
sich die auf $N$ normierten Fourierreihenkoeffizienten der kontinuierlichen 
Fensterfunktion als die Grenzwerte der Fourierreihenkoeffizienten der diskreten
Fensterfolgen nach Kapitel \myref{Algo} f"ur \mbox{$M\!\to\!\infty$}
ergeben. Um dies zu demonstrieren, sind in Bild \ref{E.b5t} am
Beispiel der Fensterfolgen, die man mit \mbox{$N\!=\!5$} bei dem
Algorithmus nach Kapitel \myref{Algo} erh"alt, die Abweichungen der
Fourierreihenkoeffizienten \mbox{$F\big({\T\nu\CdoT\frac{2\pi}{F}}\big)/F$}
von den auf $N$ normierten Fourierreihenkoeffizienten
\mbox{$F_{\!\infty}(2\pi\CdoT\nu)$} der kontinuierlichen Fensterfunktion 
graphisch dargestellt. Dabei wurde diese komplexe
Differenz nicht in der komplexen Ebene eingetragen, sondern es wurde
der Logarithmus des Betrages der Differenz jeweils "uber dem Winkel
der Differenz aufgetragen. Dadurch kann man auch bei gro"sen Werten
von $M$ die dann sehr kleine Differenz noch geeignet darstellen.
F"ur die Werte $M$ wurden \mbox{alle} Zweierpotenzen von $2$ bis ${\D2^{16}}$
ausgew"ahlt. Der Fourierreihenkoeffizient des Gleichanteils wurde
nicht eingetragen, da dieser immer mit dem auf $N$ normierten
Fourierreihenkoeffizienten der kontinuierlichen Fensterfunktion
"ubereinstimmt. Einerseits erkennt man, dass die Abweichungen mit
steigendem $M$ betraglich abnehmen, w"ahrend der Winkel f"ur
gro"se Werte von $M$ etwa konstant bleibt. Andererseits kann man ablesen,
dass selbst bei dem relativ gro"sen Wert von \mbox{${\D M=2^{16}}$}
noch Abweichungen in der f"unften Stelle hinter dem Komma auftreten.

\section[Augendiagramm des AKF der kontinuierlichen Fensterfunktion]{Augendiagramm des AKF der kontinuierlichen \\Fensterfunktion}\label{E.Kap.10.9}

\begin{figure}[btp]
\begin{center}
{ 
\begin{picture}(440,300)

\input{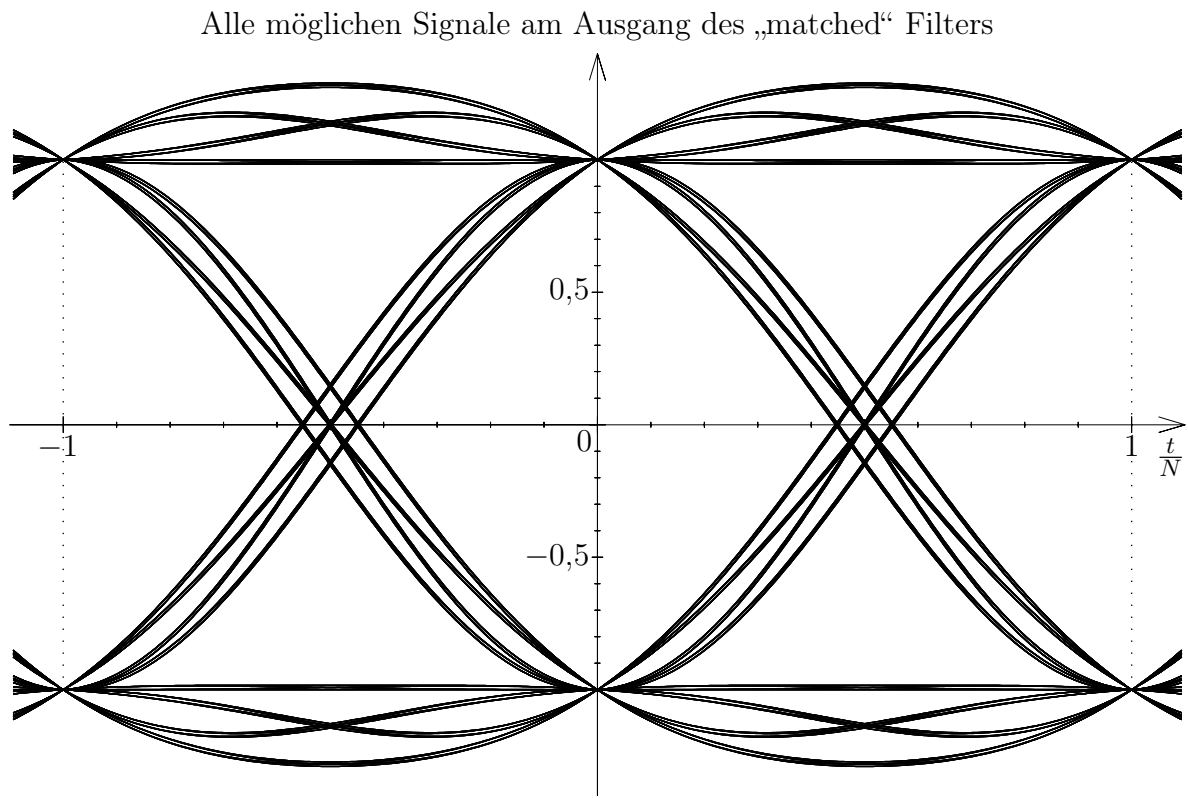}
\put(220,285){\makebox(0,0)[b]{Alle m"oglichen Signale am Ausgang des "`matched"' Filters}}
\put(440,135){\makebox(0,0)[rt]{$\frac{t}{N}$}}
\put(218,138){\makebox(0,0)[rt]{$0$}}
\put(20,136){\makebox(0,0)[t]{$\!\!-1$}}
\put(420,136){\makebox(0,0)[t]{$1$}}
\put(217,90){\makebox(0,0)[r]{$-0,\!5$}}
\put(217,190){\makebox(0,0)[r]{$0,\!5$}}

\end{picture}}
\end{center}\vspace{-12pt}
\setlength{\belowcaptionskip}{-3pt}
\caption{Augendiagramm einer ungest"orten 2ASK-PAM bei Verwendung der kontinuierlichen 
Fensterfunktion mit \mbox{$N\!=\!4$} als Sendeimpuls und als "`matched"' Filter.}
\label{E.b5v}
\rule{\textwidth}{0.5pt}\vspace{-10pt}
\end{figure}
Im Anschluss an Gleichung (\ref{E.10.23}) wurde festgestellt,
dass die Fensterautokorrelationsfunktion
\mbox{$d_{\infty}(t)=f_{\!\infty}(t)\!\ast\!\!f_{\!\infty}(\!-t)$}
"aquidistante Nullstellen bei Vielfachen von \mbox{$1/N$}
au"ser bei \mbox{$t\!=\!0$} aufweist. Sie erf"ullt also die erste
Nyquist-Bedingung. Bei der Fensterfunktion \mbox{$f_{\!\infty}(t)$} 
handelt es sich somit um einen sog. Wurzel-Nyquist-Impuls. Verwendet
man daher die Fensterfunktion als Sendeimpuls bei einer digitalen
"Ubertragung und filtert man das Empfangssignal mit einem "`matched"'-Filter,
dessen Impulsantwort \mbox{$f_{\!\infty}(-t)/N$} genau die gespiegelte
und auf $N$ normierte Fensterfunktion ist, so erh"alt man bei einem
verzerrungsfreien ungest"orten Kanal am Ausgang des Empfangsfilters
die mit der auf $N$ normierten Fensterautokorrelationsfunktion gefalteten
Sendesymbole. Bei geeigneter Abtastung kann man dann die digitalen
Sendesymbole ohne Intersymbolinterferenzen wiedergewinnen.
Um beurteilen zu k"onnen, wie gut die Fensterfunktion f"ur solch eine
"Ubertragung geeignet ist, ist in Bild~\ref{E.b5v} das
Augendiagramm f"ur eine 2ASK-PAM
(\,{\bf P}uls-{\bf A}mplituden-{\bf M}odulation mit bin"arer
Amplitudenumtastung = {\bf A}mplitude {\bf S}hift {\bf K}eying\,) aufgetragen.
Man erkennt, dass aufgrund fehlender Intersymbolinterferenzen die maximal
m"ogliche vertikale Augen"offnung erreicht werden kann, und dass sich auch
eine ganz brauchbare \mbox{horizontale} Augen"offnung ergibt. M"oglicherweise
kann man die Freiheitsgrade bei der Wahl der Basisfensterfunktion
bei der Konstruktion der Fensterfunktion nutzen, um weitere g"unstige
Eigenschaften f"ur die Anwendung als Sendeimpuls bei einer digitalen
"Ubertragung zu erzielen. Die Anwendung der Fensterfunktion als
Sendeimpuls wurde im Rahmen dieser Arbeit jedoch nicht weiter untersucht.
\newpage
\rule{0pt}{0pt}
\vfill
{\bf Beweis der Formel im Anschluss an das Inhaltsverzeichnis:}\label{proof}

Die Nullstellen des Polynoms \mbox{$z^M\!-\!1$} liegen alle auf dem Einheitskreis 
im Raster \mbox{$\frac{2\pi}{M}$}. Somit liefert eine Polynomdivision mit \mbox{$z\!-\!1$}:
\[
\Prod{\rho=1}{M-1}\Big(z\!-\!e^{j\cdot\frac{2\pi}{M}\cdot\rho}\Big)\;=\;
\frac{z^M\!-\!1}{z\!-\!1}\;=\;
\Sum{\nu=0}{M-1}z^{\nu}.
\] 
Setzt man \mbox{$z\!=\!1$} ein und bildet man auf beiden Seiten der Gleichung 
den Absolutbetrag, so erh"alt man:
\[
\bigg|\Prod{\rho=1}{M-1}\Big(1\!-\!e^{j\cdot\frac{2\pi}{M}\cdot\rho}\Big)\bigg|\;=\;
\Prod{\rho=1}{M-1}\Big|1\!-\!e^{j\cdot\frac{2\pi}{M}\cdot\rho}\Big|\;=\;
\Sum{\nu=0}{M-1}1^{\nu}\;=\;M
\] 
Mit der Eulerschen Formel k"onnen die Betr"age der Faktoren des Produkts berechnet werden:
\[
\Big|1\!-\!e^{j\cdot\frac{2\pi}{M}\cdot\rho}\Big|\;=\;
\Big|e^{j\cdot\frac{\pi}{M}\cdot\rho}\!-\!e^{\!-j\cdot\frac{\pi}{M}\cdot\rho}\Big|\;=\;
2\cdot\sin\Big(\frac{\pi}{M}\cdot\rho\Big).
\]
Setzt man dies in die vorletzte Gleichung ein und dividiert man auf beiden 
Seiten mit \mbox{$2^{M-1}$}, so erh"alt man die gesuchte Formel. 

\chapter{{\tt MATLAB}-Programmausz"uge}\label{E.Kap.11}

{\setlength{\parskip}{0ex}
Hier werden nun die entscheidenden Zeilen der Programme in der
Interpretersprache {\tt MATLAB} angegeben, die f"ur die Berechnung
\begin{itemize}
{\setlength{\itemsep}{0ex}\setlength{\parsep}{0ex}\setlength{\topsep}{0ex}
\item der Fourierreihenkoeffizienten der diskreten Fensterfolgen 
      nach Kapitel~\ref{E.Kap.10.1} sowie der Werte dieser Fensterfolgen f"ur beliebige Zeitpunkte,
\item der Spektren dieser Fensterfolgen f"ur beliebige Frequenzen,
\item der Autokorrelationsfolgen dieser Fensterfolgen f"ur beliebige Zeitpunkte,
\item der Fourierreihenkoeffizienten der kontinuierlichen\\ 
      Fensterfunktionen nach Kapitel~\ref{E.Kap.10.7},
\item der Werte dieser Fensterfunktionen f"ur beliebige Zeitpunkte,
\item der Spektren dieser Fensterfunktionen f"ur beliebige Frequenzen und
\item der Autokorrelationsfunktionen dieser Fensterfunktionen f"ur beliebige Zeitpunkte}
\end{itemize}
ben"otigt werden. Dabei wird darauf verzichtet, die Teile des 
Programms abzudrucken, die nicht f"ur die eigentliche Berechnung der 
Fenster ben"otigt werden, die aber bei einem guten Programm immer
vorhanden sein sollten, wie zum Beispiel eine "Uberpr"ufung der
Eingabeparameter oder ein ad"aquate Behandlung von Spezial- und
Ausnahmef"allen. Zun"achst werden jeweils die Programmzeilen in
\verb|Schreibmaschinenschrift| aufgelistet, wobei diese durchnummeriert
sind, um im folgenden Kommentar auf die Zeilen Bezug nehmen zu k"onnen.
Es werden dabei jeweils nur die Programmzeilen ausf"uhrlich kommentiert,
die nicht schon bei einer der vorangegangenen Programmvarianten oder in
Kapitel~\myref{MatFen1} kommentiert worden sind. Bei vielen dieser
Programme wird vorausgesetzt, dass die {\tt MATLAB}-interne globale Variable
{\tt eps} vorhanden ist, welche die relative Rechnergenauigkeit $\varepsilon$
enth"alt, und die nicht explizit an das jeweilige Programm "ubergeben werden
muss. Gleiches gilt f"ur die Variable \verb|pi|, die ---\,wie der Name schon
sagt\,--- $\pi$ enth"alt. Die Programmausz"uge sind nur stellenweise f"ur
eine besonders schnelle Berechnung der Fensterfolge optimiert. Die in
\cite{Diss} durchgef"uhrten Betrachtungen "uber die Genauigkeit der Berechnung
sind hier alle ber"ucksichtigt. Auch wurde weitgehend versucht, bei der
Art der Berechnung die in Kapitel~\ref{E.Kap.10} beschriebenen
Vorgehensweisen beizubehalten, so dass diese Programme auch dazu dienen
sollen, dem Leser den konkreten Ablauf zu zeigen, und zu demonstrieren,
dass die dort f"ur die theoretische Herleitung angegebenen Formeln
nur unwesentlich ver"andert "ubernommen werden k"onnen.}

\section{Die diskreten Fensterfolgen}\label{E.Kap.11.1}

Neben dem Fensterl"angenfaktor $N$ und dem Abstand $M$ der Nullstellen 
der Fensterautokorrelationsfolge, werden die Anzahl $A_1$ der zus"atzlichen 
Nullstellen der Z-Transformierten \mbox{$G(z)$} der Basisfensterfolge
\mbox{$g(k)$} am Einheitskreis im Raster $2\pi/F$, die Anzahl $A_0$ der
abschlie"senden Nullwerte der Fensterfolge, sowie die frei w"ahlbaren 
Nullstellen $z_{0,\rho}$ von \mbox{$G(z)$} ben"otigt. Die ganzzahligen 
Parameter $N$, $M$, $A_1$ und $A_0$ m"ussen als skalare Gr"o"sen 
(\,also als \mbox{$1\!\times\!1$} Matrizen\,) \verb|N|, \verb|M|, 
\verb|A_1| und \verb|A_0| beim Programmaufruf angegeben werden. F"ur diese 
Parameter m"ussen die Bedingungen \mbox{$N\!>\!1$}, \mbox{$M\!>\!1$}, 
\mbox{$0\!\le\!A_1\!<\!N/2$} und \mbox{$0\!\le\!A_0\!<\!N\!-\!1\!-\!2\CdoT\!A_1$}
erf"ullt sein. Die Nullstellen $z_{0,\rho}$ bilden die Elemente des
Vektors \verb|z_0|, der ebenfalls beim Programmaufruf zu "ubergeben ist.
Dieser Vektor muss genau \mbox{$N\!-\!1\!-\!A_0\!-2\CdoT\!A_1$}
Elemente enthalten. Nullstellen, die nicht reell sind, m"ussen
als zueinander konjugiert komplexe Paare in \verb|z_0| vorhanden sein.
Um eine m"oglichst gute Genauigkeit zu erzielen, sollten Nullstellen
au"serhalb des Einheitskreises durch die am Einheitskreis gespiegelten
ersetzt worden sein, und die Nullstellen sollten nach aufsteigendem Abstand
zum Punkt \mbox{$z\!=\!1$} sortiert sein. Als Ergebnis werden von diesem
Programm die Werte der Fensterfolge auf dem Vektor \verb|f_k| sowie die
Fourierreihenkoeffizienten f"ur \mbox{$\nu=0\;(1)\;N\!-\!A_1\!-\!1$}
auf dem Vektor \verb|F_nu| zur"uckgegeben.
{\renewcommand{\labelenumi}{{\footnotesize\arabic{enumi}:}}
\begin{enumerate}
\setlength{\itemsep}{-7pt plus1pt minus0pt}
\setlength{\parsep}{0pt plus1pt minus0pt}
\item\verb|function [ f_k, F_nu ] = fenster( N, M, A_0, A_1, z_0 )|\label{P3Z0}\begin{npb}
\item\verb|F = N * M|\label{P3Z1}\end{npb}
\item\verb|c = 2 / ( 1 + (N/2)^(M/3/(1-M)) / tan(pi/2/M) )|\label{P3Z2}
\item\verb|Ms = -log(eps)/log(2) * 2^log(N/3) * 3.6^(1/M)|\label{P3Z3}\begin{npb}
\item\verb|Ms = 2^ceil(log(Ms)/log(2))|\label{P3Z4}
\item\verb|Ms = max(Ms,16)|\label{P3Z5}\end{npb}\vadjust{\penalty-100}
\item\verb|Ms_OK = 0|\label{P3Z6}\begin{npb}
\item\verb|while ~Ms_OK|\label{P3Z7}\end{npb}
\item\verb|  eta = 0:Ms/2|\label{P3Z8}\begin{npb}
\item\verb|  F_eta = zeros(1,Ms/2+1)|\label{P3Z9}\end{npb}
\item\verb|  NF_1 = 1 / ( c^2 + 4 * (1-c) * sin( pi/F * (N-1-A_1) )^2 )|\label{P3Z10}\begin{npb}
\item\verb|  NF_0 = max( [ ( abs( z_0*exp( j*pi/F*(1-N) )-(1-c) ) + ...|\label{P3Z11}
\item[]\verb|                  abs( 1-(1-c)*z_0*exp( j*pi/F*(1-N) ) ) ) ; ...|
\item[]\verb|                ( abs( z_0*exp( j*pi/F*(N-1) )-(1-c) ) + ...|
\item[]\verb|                  abs( 1-(1-c)*z_0*exp( j*pi/F*(N-1) ) ) ) ] ).^(-2)|\end{npb}
\item\verb|  for nu_1 = (1-N)/2:(N-1)/2|\label{P3Z12}\begin{npb}
\item\verb|    F_eta_1 = ones(1,Ms/2+1)|\label{P3Z13}\end{npb}
\item\verb|    rho = 0|\label{P3Z14}
\item\verb|    for nu_2 = [(1-N+A_1):(nu_1+A_1-(N+1)/2), ...|\label{P3Z15}\begin{npb}
\item[]\verb|                (nu_1-A_1+(N+1)/2):(N-1-A_1)]|
\item\verb|      K_1 = NF_1 * ( c^2  + 4 * (1-c) * sin( pi/F * nu_2 )^2 )|\label{P3Z16}
\item\verb|      Psi_1 = atan( ( (1-c) * sin(2*pi/F*nu_2) ) / ...|\label{P3Z17}
\item[]\verb|                    ( c + 2 * (1-c) * sin(pi/F*nu_2)^2 ) )|
\item\verb|      F_eta_1 = F_eta_1 .* K_1 .* sin( pi/Ms*eta-pi/F*nu_2-Psi_1 ).^2|\makebox[-1pt]{}\label{P3Z18}\end{npb}
\item\verb|      rho = rho + 1|\label{P3Z19}\begin{npb}
\item\verb|      if rho < N-0.5-2*A_1-A_0|\label{P3Z20}
\item\verb|        Z_Z = z_0(rho) * exp( j*2*pi/F*nu_1 ) - (1-c)|\label{P3Z21}
\item\verb|        Z_N = 1 - (1-c) * z_0(rho) * exp(j*2*pi/F*nu_1)|\label{P3Z22}
\item\verb|        abs_Z_Z = abs( Z_Z )|\label{P3Z23}
\item\verb|        abs_Z_N = abs( Z_N )|\label{P3Z24}
\item\verb|        F_eta_1 = F_eta_1 .* NF_0(rho) .* ...|\label{P3Z25}
\item[]\verb|                  ( (abs_Z_N-abs_Z_Z)^2 + 4 * abs_Z_N * abs_Z_Z * ...|\makebox[-1pt]{}
\item[]\verb|                    sin( pi/Ms*eta-(angle(Z_Z)-angle(Z_N))/2 ).^2 )|
\item\verb|      end|\label{P3Z26}
\item\verb|    end|\label{P3Z27}\end{npb}
\item\verb|    F_eta = F_eta + F_eta_1|\label{P3Z28}\begin{npb}
\item\verb|  end|\label{P3Z29}\end{npb}\vadjust{\penalty-100}
\item\verb|  NF = 1 / sqrt( max(F_eta) * min(F_eta) )|\label{P3Z30}\begin{npb}
\item\verb|  F_eta = NF * F_eta|\label{P3Z31}\end{npb}
\item\verb|  L_eta = log(F_eta)|\label{P3Z32}
\item\verb|  Ceps_2 = ifft( [L_eta,L_eta(Ms/2:-1:2)] )|\label{P3Z33}\begin{npb}
\item\verb|  Ceps_2 = real( Ceps_2 )|\label{P3Z34}
\item\verb|  Ceps_2 = ( Ceps_2 + Ceps_2([1,Ms:-1:2]) ) / 2|\label{P3Z35}\end{npb}\vadjust{\penalty-100}
\item\verb|  sockel = eps / Ms / sqrt(48) * ...|\label{P3Z36}\begin{npb}
\item[]\verb|           sqrt( sum( max( [L_eta,L_eta(Ms/2:-1:2)].^2, 1 ) ) )|\end{npb}
\item\verb|  grenze = ( 2 * (2*N-2-2*A_1-A_0) ) ./ [1:Ms/2] .* ...|\label{P3Z37}\begin{npb}
\item[]\verb|           ( Ms * sockel/2/(2*N-2-2*A_1-A_0) ).^( 2*[1:Ms/2]/Ms )|\end{npb}
\item\verb|  max_fehl = sum( max( abs([L_eta,L_eta(Ms/2:-1:2)]), 1 ) ) * ...|\label{P3Z38}\begin{npb}
\item[]\verb|             eps * ( 2 + log(Ms)/log(2) ) / Ms|\end{npb}
\item\verb|  grenze = max( grenze, max_fehl )|\label{P3Z39}
\item\verb|  if any( abs(Ceps_2(2:Ms/2+1)) > grenze )|\label{P3Z40}\begin{npb}
\item\verb|    kriterium = abs( fft( Ceps_2(Ms/4+1:Ms/2) .* [Ms/4:Ms/2-1] ) )|\label{P3Z41}
\item\verb|    [dummy,womax] = max( kriterium(1:Ms/8+1) )|\label{P3Z42}
\item\verb|    delta_c = 1-(womax-1)*16/Ms|\label{P3Z43}
\item\verb|    delta_c = sin( delta_c*pi*(1-delta_c^2/2) )/3|\label{P3Z44}\end{npb}
\item\verb|    c = c * (1-delta_c) / ( 1 + (1-c) * delta_c )|\label{P3Z45}
\item\verb|    Ms = 2*Ms|\label{P3Z46}
\item\verb|  else|\label{P3Z47}
\item\verb|    Ms_OK = 1|\label{P3Z48}\begin{npb}
\item\verb|  end|\label{P3Z49}
\item\verb|end|\label{P3Z50}\end{npb}\vadjust{\penalty-200}
\item\verb|nu = 1:N-A_1-1|\label{P3Z51}
\item\verb|Omega = 2*pi/F*nu|\label{P3Z52}
\item\verb|Omega_s = Omega + 2 * atan( ( (1-c) * sin(Omega) ) ./ ...|\label{P3Z53}\begin{npb}
\item[]\verb|                            ( c + 2 * (1-c) * sin(Omega/2).^2 ) )|\end{npb}
\item\verb|phi = zeros(1,N-A_1-1)|\label{P3Z54}
\item\verb|for nu_i = nu|\label{P3Z55}\begin{npb}
\item\verb|  phi(nu_i) = sin(Omega_s(nu_i)*[Ms/2-1:-1:1]) * Ceps_2(Ms/2:-1:2).'|\label{P3Z56}
\item\verb|end|\label{P3Z57}\end{npb}
\item\verb|phi = phi - (N-1-A_1-A_0/2) * Omega_s + (F-1-A_0) * Omega/2|\label{P3Z58}\vadjust{\penalty-200}
\item\verb|if N == 2*A_1+1|\label{P3Z59}\begin{npb}
\item\verb|  F_nu = ones(1,N-A_1)/N|\label{P3Z60}\end{npb}
\item\verb|else|\label{P3Z61}\begin{npb}
\item\verb|  nu = 0:N-A_1-1|\label{P3Z62}
\item\verb|  F_nu = zeros(1,N-A_1)|\label{P3Z63}\end{npb}
\item\verb|  NF_1 = exp(2*log(8/pi*F)-sum(log([5:2:4*N-8*A_1]))/(N-1-2*A_1))|\label{P3Z64}\begin{npb}
\item\verb|  NF_0 = 1 ./ max( [ abs( z_0 * exp(j*pi/F*(N-1) ) - 1 ).^2 ; ...|\label{P3Z65}
\item[]\verb|                     abs( z_0 * exp(j*pi/F*(1-N) ) - 1 ).^2 ] )|\end{npb}
\item\verb|  for nu_1 = (1-N)/2:(N-1)/2|\label{P3Z66}\begin{npb}
\item\verb|    F_nu_1 = ( ( nu < nu_1+N/2-A_1 ) & ( nu > nu_1-N/2+A_1) ) / eps^2|\makebox[-1pt]{}\label{P3Z67}\end{npb}
\item\verb|    rho = 0|\label{P3Z68}\begin{npb}
\item\verb|    for nu_2 = (1-N)/2+A_1:(N-3)/2-A_1|\label{P3Z69}
\item\verb|      rho = rho + 1|\label{P3Z70}
\item\verb|      if rho < N-0.5-2*A_1-A_0|\label{P3Z71}
\item\verb|        abs_z_0 = abs(z_0(rho))|\label{P3Z72}
\item\verb|        F_nu_1 = F_nu_1 .* NF_0(rho) .* ( ( 1-abs_z_0 )^2 + ...|\label{P3Z73}
\item[]\verb|          4 * abs_z_0 * sin( pi/F*(nu-nu_1)-angle(z_0(rho))/2 ).^2 )|
\item\verb|      end|\label{P3Z74}\end{npb}
\item\verb|      nu_3 = nu - nu_1 - nu_2 - ( nu <= nu_1 + nu_2 )|\label{P3Z75}\begin{npb}
\item\verb|      F_nu_1 = F_nu_1 ./ ( NF_1 .* sin( pi/F*nu_3 ).^2 )|\label{P3Z76}
\item\verb|    end|\label{P3Z77}\end{npb}
\item\verb|    F_nu = F_nu + F_nu_1|\label{P3Z78}\begin{npb}
\item\verb|  end|\label{P3Z79}\end{npb}
\item\verb|  F_nu = sqrt(F_nu)|\label{P3Z80}\begin{npb}
\item\verb|  F_nu = F_nu / ( N * F_nu(1) )|\label{P3Z81}
\item\verb|end|\label{P3Z82}\end{npb}
\item\verb|F_nu(2:N-A_1) =  F_nu(2:N-A_1) .* exp(-j*phi)|\label{P3Z83}\vadjust{\penalty-200}
\item\verb|f_k = zeros(1,F)|\label{P3Z84}\begin{npb}
\item\verb|si_k = [ 0:2:(F+0.5)/2,...|\label{P3Z85}
\item[]\verb|         F-[2*floor((F+0.5)/4)+2:2:(3*F+0.5)/2], ...|
\item[]\verb|         [2*floor((3*F+0.5)/4)+2:2:2*F-1]-2*F        ]|
\item\verb|si_k = sin( (pi/F) * si_k )|\label{P3Z86}\end{npb}
\item\verb|co_k = [ F-[0:4:2*F-1], [4*floor((2*F-1)/4)+4:4:4*F-2]-3*F ]|\label{P3Z87}\begin{npb}
\item\verb|co_k = sin( (pi/(2*F)) * co_k )|\label{P3Z88}\end{npb}
\item\verb|for nu = N-A_1-1:-1:1|\label{P3Z89}\begin{npb}
\item\verb|  k_nu = rem( [0:F-1]*nu, F ) + 1|\label{P3Z90}
\item\verb|  f_k = f_k + 2 * real( F_nu(nu+1) ) * co_k(k_nu) - ...|\label{P3Z91}
\item[]\verb|              2 * imag( F_nu(nu+1) ) * si_k(k_nu)|
\item\verb|end|\label{P3Z92}
\item\verb|f_k = f_k + 1/N|\label{P3Z93}\end{npb}
\end{enumerate}}

Durch die bei diesem Programm freigestellte Wahl der nicht festgelegten
Nullstellen der Z-Transformierten der Basisfensterfolge ergeben sich
zwei wesentliche Modifikationen gegen"uber dem im Anhang von \cite{Diss}
angegebenen Programm. Zum einen kann nun die f"ur die Berechnung des
Cepstrums ben"otigte FFT-L"ange $\widetilde{M}$ genauso wie der optimale
Wert des Bilineartransformationsparameters $c$ nicht zu Beginn
festgelegt werden. Die hier notwendige iterative Bestimmung dieser
Parameter wird im Anschluss an die zweite wesentliche Modifikation
kommentiert. Diese besteht darin, dass sich der Betrag des Spektrums
sowohl bei der Berechnung der Betr"age der Fourierreihenkoeffizienten
der Fensterfolge $f(k)$, als auch bei der Berechnung des Betragsquadrats
der bilinear Z-Transformierten bei der Berechnung des Cepstrums
nun umfangreicher berechnet.

In der \verb|for|-Schleife zur Berechnung der kumulativen Produkte
in Gleichung~(\ref{E.10.12}), die in Zeile~\ref{P3Z15} beginnt und
in Zeile~\ref{P3Z27} endet, sind die Elemente des Vektors \verb|F_eta_1|,
mit deren Hilfe die kumulativen Produkte f"ur alle Frequenzpunkte
\mbox{$\widetilde{\Omega}=\eta\CdoT2\pi/\widetilde{M}$} mit
\mbox{$\eta=0\;(1)\;\widetilde{M}/2$} nach und nach berechnet werden,
noch mit den Betragsquadraten der Abst"ande zu den neu hinzugekommenen
Nullstellen zu multiplizieren. In Gleichung~(\ref{E.10.12}) ist dies
das kumulative Produkt mit dem Lauf"|index $\rho$. Die Berechnung der
Nullstellenabst"ande  erfolgt abwechselnd, d.~h. es wird immer
zuerst eine der Nullstellen verarbeitet, die zur Erweiterung auf
den Hauptnenner erforderlich war, und dann eine der Nullstellen
des Vektors \verb|z_0|. Dazu wird in Zeile~\ref{P3Z14} \verb|rho|
zun"achst auf null gesetzt, um dann bei jedem Schleifendurchlauf
mit einem neuen Wert von \verb|nu_2| in Zeile~\ref{P3Z19} um eins
erh"oht zu werden. In Zeile~\ref{P3Z20} wird festgestellt,
ob schon alle Nullstellen des Vektors \verb|z_0| verarbeitet
wurden. Wenn nicht, werden in den Zeilen~\ref{P3Z21} bis \ref{P3Z25}
die mit einem geeignet gew"ahlten Normierungsfaktor \verb|NF_0(rho)|
multiplizierten Betragsquadrate der Abst"ande der n"achsten um
\mbox{$e^{j\cdot\nu_1\CdoT2\pi/F}$} rotierten Nullstelle zu den Punkten
\mbox{$\Tilde{z}\!=\!e^{j\cdot\eta\cdot2\pi/\widetilde{M}}$}
f"ur alle $\eta$-Werte zugleich berechnet, wie dies in
Gleichung~(\ref{E.10.12}) angegeben ist. Da zur Erweiterung
auf den Hauptnenner \mbox{genau} \mbox{$N\!-\!1$} Nullstellen
ben"otigt werden, wird die Schleife mit dem Index \verb|nu_2|,
der alle im Anschluss an Gleichung~(\ref{E.10.9}) genannten
Werte annehmen kann, genau \mbox{$N\!-\!1$} mal durchlaufen.
Da der Vektor \verb|z_0| genau \mbox{$N\!-\!1\!-\!A_0\!-\!2\CdoT\!A_1$}
Nullstellen enth"alt, kann es nicht vorkommen, dass die Schleife
beendet wird, bevor nicht alle Nullstellen des Vektors \verb|z_0|
verarbeitet worden sind. Bei dieser Programmvariante wurde
f"ur jede der frei w"ahlbaren Nullstellen genau ein Normierungsfaktor
neu eingef"uhrt. Dies sind die Elemente des Vektors \verb|NF_0|, der
in Zeile~\ref{P3Z11} berechnet wird. Da die Normierungsfaktoren bei
allen Frequenzen \mbox{$\widetilde{\Omega}=\eta\CdoT2\pi/\widetilde{M}$}
und bei allen Summanden der Summe "uber alle verschobenen
Betragsquadratspektren der Basisfensterfolge mit dem Index \verb|nu_1|
dieselben sind, entspricht diese Art der Normierung lediglich
einer Skalierung aller Spektralwerte des Vektors \verb|F_eta| mit
einer gemeinsamen Konstante, n"amlich mit dem Produkt aller Elemente
des Vektors \verb|NF_0|. Wenn man sich in Gleichung~(\ref{E.10.12})
in der von $\widetilde{\Omega}$ abh"angigen Form einen Faktor des
kumulativen Produkts mit dem Index $\rho$ betrachtet, so stellt man fest,
dass der Faktor den Maximalwert \mbox{$\big(\,|\Tilde{z}_{N,\rho,\nu_1}|\!+\!
|\Tilde{z}_{Z,\rho,\nu_1}|\,\big)^{\!2}$} nicht "ubersteigt. F"ur
alle m"oglichen Werte von $\nu_1$ liegen die Nullstellen, die durch Rotation
mit dem Drehfaktor \mbox{$e^{j\cdot\nu_1\CdoT2\pi/F}$} aus der Nullstelle
$z_{0,\rho}$ hervorgehen, entweder auf dem Einheitskreis, oder auf
einem Kreis, der innerhalb des Einheitskreises, liegt. Au"serhalb des
Einheitskreises k"onnen diese Nullstellen nicht liegen, da wir
vorausgesetzt haben, dass die Nullstellen $z_{0,\rho}$ durch die
am Einheitskreis gespiegelten Nullstellen zu ersetzen sind, wenn sie
sich zuvor au"serhalb des Einheitskreises befinden. Auch nach der
Bilineartransformation liegen die Nullstellen entweder auf dem
Einheitskreis, oder auf einem Kreis innerhalb des Einheitskreises.
Man kann sich nun "uberlegen, dass in den F"allen, bei denen die richtige
Normierung besonders wichtig ist ---\,in der Regel ist das der Fall, wenn
$M$ besonders gro"s ist\,---,  die rotierten und bilineartransformierten
Nullstellen mit dem betraglichen Maximalwert von $\nu_1$ besonders nahe
am Einheitskreis liegen. Daher wird in Zeile~\ref{P3Z11} f"ur jede
Nullstelle des Vektors \verb|z_0| f"ur die beiden Werte 
\mbox{$\nu_1=N\!-\!1$} und \mbox{$\nu_1=1\!-\!N$} der maximal m"ogliche
Wert des Faktors des kumulativen Produkts in Gleichung~(\ref{E.10.12})
bestimmt, und von diesen beiden Werten jeweils der
gr"o"sere genommen, um dessen Reziprokwert als Normierungsfaktor
\enlargethispage{3pt}zu verwenden.

Bei dieser Programmvariante sind nun um $A_1$ weniger
Fourierreihenkoeffizienten der Fensterfolge $f(k)$ zu berechnen,
als in der in \cite{Diss} vorgestellten Variante. Deshalb hat der Vektor
\verb|nu|, der in Zeile~\ref{P3Z62} festgelegt wird, nun entsprechend
weniger Elemente. Die Berechnung der Fourierreihenkoeffizienten erfolgt
nach Gleichung~(\ref{E.10.9}) in den Zeilen~\ref{P3Z59} bis \ref{P3Z82}.
Hier ist nun eine Fallunterscheidung angebracht. Wenn alle frei
w"ahlbaren Nullstellen auf den Einheitskreis gelegt werden, so dass
$A_1$ seinen Maximalwert \mbox{$(N\!-\!1)/2$} annimmt, bleibt in 
Gleichung~(\ref{E.10.9}) von der Summe mit dem
Lauf"|index~$\nu_1$ nur ein Summand mit dem Wert Eins "ubrig. Dieser
Summand entspricht dem Summanden, bei dem die doppelte Polstelle bei
\mbox{$z\!=\!e^{j\cdot\nu\cdot2\pi/F}$} in Gleichung~(\ref{E.10.8})
mit der doppelten Nullstelle des Z"ahlers gek"urzt wurde,
und sich somit der Wert ${\D F^2}$ der quadrierten si-Funktion an der
Stelle Null ergab, der in Gleichung~(\ref{E.10.9}) vor der Summe steht.
Daher ist, wenn dieser Fall eintritt, was in Zeile~\ref{P3Z59} abgefragt
wird, der Betrag aller Fourierreihenkoeffizienten von $f(k)$ gleich
$1/N$, wie dies in Zeile~\ref{P3Z60} eingesetzt wird. Anderenfalls
werden die Betr"age aller Fourierreihenkoeffizienten von $f(k)$
nach Gleichung~(\ref{E.10.9}) berechnet, wobei auch hier die Nullstellen
und die Polstellen abwechselnd verarbeitet werden. Die Reihenfolge der
Bearbeitung wird ---\,wie bei der Berechnung des Cepstrums\,--- mit Hilfe der
Programmzeilen~\ref{P3Z68}, \ref{P3Z70}, \ref{P3Z71} und \ref{P3Z74}
realisiert. Warum diese Bearbeitungsreihenfolge gew"ahlt wurde wird
im Kommentar des Programms erl"autert, mit dessen Hilfe das Spektrum
der Fensterfolge f"ur beliebige Frequenzen $\Omega$ berechnet werden kann,
weil dort Einhaltung dieser Reihenfolge von besonderer Bedeutung ist.
Auch die in Zeile~\ref{P3Z65} berechneten Normierungsfaktoren
\verb|NF_0(rho)| werden dort n"aher erl"autert. Die Normierungskonstante
\verb|NF_1| wird in Zeile~\ref{P3Z64} so gew"ahlt, dass sich ohne
Ber"ucksichtigung der Nullstellen des Vektors \verb|z_0| {\em vor}\/ der
abschlie"senden Normierung auf \mbox{$\big(N\CdoT{\tt F\_nu}(0)\big)^2\!$}
bei \mbox{$\nu\!=\!0$} ein Wert in der Gr"o"senordnung von eins ergibt.
Dass dem so ist, wird im Kommentar zum Programm im Anhang von \cite{Diss}
erl"autert, wobei zu beachten ist, dass nun um \mbox{$2\CdoT\!A_1$}
weniger doppelte Polstellen vorliegen. Da nun um $A_1$ weniger
Fourierreihenkoeffizienten zu berechnen sind, wird auch der Vektor
\verb|F_nu|, mit dessen Hilfe die Summe in Gleichung~(\ref{E.10.9})
realisiert wird, in Zeile~\ref{P3Z63} mit entsprechend weniger Elementen
bereitgestellt. Der Vektor \verb|F_nu_1| zur Berechnung der kumulativen
Produkte ist ebensolang, und ber"ucksichtigt bei seiner Initialisierung
in Zeile~\ref{P3Z67}, die modifizierten Summengrenzen in
Gleichung~(\ref{E.10.9}), indem die Elemente zu null initialisiert
werden, die Summanden entsprechen w"urden, die nicht in der Summe
in Gleichung~(\ref{E.10.9}) vertretenen sind. Dass die anderen Elemente
dieses Vektors nicht auf eins sondern auf $\varepsilon^{-2}$ initialisiert
werden, ist Teil der Normierung, die bei dem Programm zur Berechnung
des Spektrums der Fensterfolge kommentiert wird. Die Berechnung der
Nullstellenabstandsquadrate in den Zeilen~\ref{P3Z72} und \ref{P3Z73}
erfolgt bis auf die Normierung mit dem in Gleichung~(\ref{E.10.9})
angegebenen Term. Die Berechnung der Polstellenabstandsquadrate erfolgt
in den Zeilen~\ref{P3Z75} und \ref{P3Z76} wie dies in \cite{Diss}
ausf"uhrlich erl"autert ist. Die vorl"aufige Normierung wird in
Zeile~\ref{P3Z81} endg"ultig korrigiert, nachdem in Zeile~\ref{P3Z80}
aus den Betragsquadraten der Fourierreihenkoeffizienten durch
Wurzelziehen die Betr"age berechnet worden sind. Danach werden diese in
Zeile~\ref{P3Z83} noch mit der Exponentialfunktion der Phase multipliziert,
und es kann daraus die Fensterfolge wie in dem in \cite{Diss} angegebenen
Programm berechnet werden. Es sind hier lediglich um $A_1$ weniger
Fourierreihenkoeffizienten vorhanden, so dass die \verb|for|-Schleife
in Zeile~\ref{P3Z89} nun mit einem entsprechend niedrigeren Wert
\enlargethispage{3pt}von \verb|nu| startet.

Bei der Berechnung der Phase der Fourierreihenkoeffizienten in den
Zeilen~\ref{P3Z51} bis \ref{P3Z58} ist zum einen ebenfalls zu
ber"ucksichtigen, dass nun um $A_1$ weniger Fourierreihenkoeffizienten
berechnet werden m"ussen, so dass die Vektoren \verb|nu| und \verb|phi|
in den Zeilen~\ref{P3Z51} und \ref{P3Z54} \mbox{entsprechend} k"urzer
ausfallen. Zum anderen ist bei der Addition des Phasenanteils des Teils
\mbox{$\widetilde{D}_P(\Tilde{z})$} der Bilineartransformierten, der
die bekannten Polstellen enth"alt, nun ebenso wie bei dem
linearphasigen Anteil \mbox{$D_E(z)$} die ge"anderte Vielfachheit
der Pol- bzw. Nullstellen zu beachten, so dass sich in Zeile~\ref{P3Z58}
modifizierte Vorfaktoren bei den Termen in \verb|Omega_s| und
\verb|Omega/2| ergeben.

Nun kommen wir zur dynamischen Anpassung der L"ange $\widetilde{M}$
des Cepstrums und des Bilineartransformationsparameters $c$.
Mit der Variable \verb|Ms_OK|, die in Zeile~\ref{P3Z6} auf null
initialisiert wird, wird angezeigt, dass das Cepstrum bisher
noch nicht erfolgreich berechnet werden konnte. Daher wird die
\verb|while|-Schleife, die in Zeile~\ref{P3Z7} beginnt und in
Zeile~\ref{P3Z50} endet, beim ersten mal auf jeden Fall durchlaufen.
Sollte mit den in den Zeilen~\ref{P3Z2} bis \ref{P3Z5} gew"ahlten Werten
$\widetilde{M}$ und $c$ festgestellt worden sein, dass das Cepstrum
mit einer gen"ugend gro"sen L"ange berechnet worden ist, so wird
in Zeile~\ref{P3Z48} die Variable \verb|Ms_OK| auf eins gesetzt,
so dass es nicht zu einem erneuten Schleifendurchlauf kommt. Um
\enlargethispage{3pt}festzustellen, ob die FFT-L"ange $\widetilde{M}$
bei der Berechnung des Cepstrums gro"s genug war, ist der Betrag des
Cepstrums  mit der in Gleichung~(\ref{E.10.13}) genannten Grenze
zu vergleichen. In Zeile~\ref{P3Z40} wird dieser Vergleich durchgef"uhrt,
wobei hier die zweifache Grenze herangezogen wird, weil im Programm
auch das doppelte Cepstrum \verb|Ceps_2| verwendet wird. Um die
doppelte Grenze in Zeile~\ref{P3Z37} berechnen zu k"onnen, muss
zun"achst der Mindestwert des Rauschsockels nach Gleichung~(\ref{E.10.14})
in Zeile~\ref{P3Z36} bestimmt werden. Um zu vermeiden, dass
der Rauschsockel im Cepstrum verhindert, dass eine ausreichende
FFT-L"ange $\widetilde{M}$ erkannt werden kann, wird diese
Grenze noch auf das doppelte des in Gleichung~(\ref{E.10.15})
genannten Sch"atzwertes f"ur den maximal zu erwartenden Berechnungsfehler
im Cepstrum nach unten begrenzt. Diese Begrenzung erfolgt in den
Programmzeilen~\ref{P3Z38} und \ref{P3Z39}. Nur falls eine zu
kurze FFT-L"ange $\widetilde{M}$ detektiert worden ist, wird
in Zeile~\ref{P3Z46} $\widetilde{M}$ verdoppelt, und in Zeile~\ref{P3Z45}
ein neuer Wert des Bilineartransformationsparameters $c$
nach Gleichung~(\ref{E.10.16}) berechnet. Dazu muss der Wert
\mbox{$\Delta c$} mit Gleichung~(\ref{E.10.17}) aus dem Sch"atzwert des
Winkels der Nullstelle, die dem Einheitskreis am n"achsten liegt,
berechnet werden. Diese Absch"atzung erfolgt, indem man bei der
Folge, die man durch eine FFT der L"ange \mbox{$\widetilde{M}/4$}
aus einem Abschnitt des mit $k$ multiplizierten Cepstrums gewinnt,
die Lage des betraglichen Maximums bestimmt, wie dies im Anschluss an
Gleichung~(\ref{E.10.17}) beschrieben ist, und in den Zeilen~\ref{P3Z41}
und \ref{P3Z42} ausgef"uhrt wird. Da der Wert des Maximums selbst nicht
interessiert, sondern nur die Lage des Maximums, ist daf"ur die im
weiteren nicht mehr verwendete Variable \verb|dummy| vorgesehen.
Die Variable \verb|womax| liefert uns in Zeile~\ref{P3Z42} den Index
des Elementes des Vektors \verb|kriterium|, bei dem das Maximum erreicht
wird. Da die Elemente dieses Vektors die diskreten Fouriertransformierten
des Abschnitts des mit $k$ multiplizierten Cepstrums bei den Frequenzen
\mbox{$\Tilde{\eta}\CdoT8\pi/\widetilde{M}$} mit
\mbox{$\Tilde{\eta}=0\;(1)\;\widetilde{M}/4\!-\!1$} enthalten, ist der
um eins reduzierte Index \verb|womax| proportional zum Winkel des
Punktes am Einheitskreis, wo das Spektrum sein betragliches Maximum
erreicht. Wenn man in Zeile~\ref{P3Z44} auf der rechten Seite
\mbox{${\tt delta\_c}=1\!-\!2\CdoT\psiu/\pi$} einsetzt, erh"alt man die
in Gleichung~(\ref{E.10.17}) dargestellte Form f"ur \mbox{$\Delta c$}, so dass
die beiden Zeilen~\ref{P3Z43} und \ref{P3Z44} lediglich eine zweistufige
Berechnung der Gleichung~(\ref{E.10.17}) darstellen, die vom numerischen
Standpunkt g"unstiger erscheint. Die Absch"atzung des Winkels $\psiu$
ist nur dann sinnvoll, wenn man auch schon bei dem Startwert von
$\widetilde{M}$ entscheiden kann, ob der Parameter \verb|c| verringert,
vergr"o"sert oder beibehalten werden soll. Dazu muss es bei dem
Startwert von $\widetilde{M}$ wenigstens die drei m"oglichen Sch"atzwerte
\mbox{$\psiu\in\{0;\pi/2;\pi\}$} f"ur den unbekannten Nullstellenwinkel
geben. \mbox{$\widetilde{M}/4$} sollte daher wenigstens $4$ sein. Dies
wird erzwungen, indem der Startwert von $\widetilde{M}$ in Zeile~\ref{P3Z5}
auf den Mindestwert $16$ gesetzt wird, falls sich bei der Berechnung in
den Zeilen~\ref{P3Z3} und \ref{P3Z4} ein kleinerer Wert ergibt.

Wie bei der in \cite{Diss} vorgestellten Variante kann man auch hier die 
Fensterfolge durch Auswertung der Fourierreihe nach Gleichung~(\myref{6.16}) 
berechnen. Dies erfolgt in den Zeilen \ref{P3Z84} bis \ref{P3Z93}.
Prinzipiell kann man die Fourierreihe nach Gleichung (\myref{6.16})
auch f"ur beliebiges reelles $k$ auswerten. Wenn man an nichtganzzahligen
Werten $k$, die auf einem Vektor \verb|k| abgespeichert seien, interessiert
ist, so ersetzt man den Programmaufruf in Zeile~\ref{P3Z0} sowie die
Programmzeilen ab einschlie"slich Zeile~\ref{P3Z84} durch die Programmzeilen:
{\renewcommand{\labelenumi}{{\footnotesize\arabic{enumi}:}}
\begin{enumerate}
\setlength{\itemsep}{-7pt}
\setlength{\parsep}{0pt}
\item\verb|function [ f_k, F_nu ] = fenster( N, M, A_0, A_1, z_0, k )|\label{P2Z0}\begin{npb}
\item[$\vdots\;\;$]{ }
\setcounter{enumi}{84}
\item\verb|f_k = zeros(1,length(k))|\label{P2Z84}
\item\verb|O_k = 2*pi/F*k|\label{P2Z85}\end{npb}
\item\verb|for nu = N-A_1-1:-1:1|\label{P2Z86}\begin{npb}
\item\verb|  f_k = f_k + 2*real(F_nu(nu+1))*cos(O_k*nu) - ...|\label{P2Z87}
\item[]\verb|              2*imag(F_nu(nu+1))*sin(O_k*nu)|
\item\verb|end|\label{P2Z88}
\item\verb|f_k = f_k + 1/N|\label{P2Z89}\end{npb}
\end{enumerate}}
Diese Variante ist nat"urlich nicht so exakt, wie die Variante f"ur
ganzzahliges $k$.

\section{Die Spektren der diskreten Fensterfolgen}\label{E.Kap.11.2}

Neben den im vorigen Unterkapitel genannten Parametern $N$, $M$,
$A_0$, $A_1$ und dem Vektor der Nullstellen $z_{0,\rho}$, f"ur 
die die dort genannten Restriktionen gelten sollen,  ben"otigen
wir bei diesem Programm noch einen Vektor \verb|Omega|, der die
Frequenzen $\Omega$ als Elemente enth"alt, f"ur die das Spektrum der
Fensterfolge berechnet werden soll. Von diesen wird angenommen,
dass sie im Bereich zwischen $-\pi$ und $\pi$ liegen, weil
dann die Berechnung mit bestm"oglicher Genauigkeit erfolgen kann.
Als Ergebnis werden von diesem Programm die Werte des Spektrums der
Fensterfolge f"ur die Frequenzen des Vektors \verb|Omega| auf dem Vektor
\verb|F_Omega| zur"uckgegeben.
{\renewcommand{\labelenumi}{{\footnotesize\arabic{enumi}:}}
\begin{enumerate}
\setlength{\itemsep}{-7pt plus1pt minus0pt}
\setlength{\parsep}{0pt plus1pt minus0pt}
\item\verb|function F_Omega = spektrum( N, M, A_0, A_1, z_0, Omega )|\label{P4Z0}\begin{npb}
\item\verb|F = N * M|\label{P4Z1}
\item\verb|c = 2 / ( 1 + (N/2)^(M/3/(1-M)) / tan(pi/2/M) )|\label{P4Z2}\end{npb}
\item\verb|Ms = -log(eps)/log(2) * 2^log(N/3) * 3.6^(1/M)|\label{P4Z3}\begin{npb}
\item\verb|Ms = 2^ceil(log(Ms)/log(2))|\label{P4Z4}
\item\verb|Ms = max(Ms,16)|\label{P4Z5}\end{npb}\vadjust{\penalty-100}
\item\verb|Ms_OK = 0|\label{P4Z6}\begin{npb}
\item\verb|while ~Ms_OK|\label{P4Z7}\end{npb}
\item\verb|  eta = 0:Ms/2|\label{P4Z8}\begin{npb}
\item\verb|  F_eta = zeros(1,Ms/2+1)|\label{P4Z9}\end{npb}
\item\verb|  NF_1 = 1 / ( c^2 + 4 * (1-c) * sin( pi/F * (N-1-A_1) )^2 )|\label{P4Z10}\begin{npb}
\item\verb|  NF_0 = max( [ ( abs( z_0*exp( j*pi/F*(1-N) )-(1-c) ) + ...|\label{P4Z11}
\item[]\verb|                  abs( 1-(1-c)*z_0*exp( j*pi/F*(1-N) ) ) ) ; ...|
\item[]\verb|                ( abs( z_0*exp( j*pi/F*(N-1) )-(1-c) ) + ...|
\item[]\verb|                  abs( 1-(1-c)*z_0*exp( j*pi/F*(N-1) ) ) ) ] ).^(-2)|\end{npb}
\item\verb|  for nu_1 = (1-N)/2:(N-1)/2|\label{P4Z12}\begin{npb}
\item\verb|    F_eta_1 = ones(1,Ms/2+1)|\label{P4Z13}\end{npb}
\item\verb|    rho = 0|\label{P4Z14}
\item\verb|    for nu_2 = [(1-N+A_1):(nu_1+A_1-(N+1)/2), ...|\label{P4Z15}\begin{npb}
\item[]\verb|                (nu_1-A_1+(N+1)/2):(N-1-A_1)]|
\item\verb|      K_1 = NF_1 * ( c^2  + 4 * (1-c) * sin( pi/F * nu_2 )^2 )|\label{P4Z16}
\item\verb|      Psi_1 = atan( ( (1-c) * sin(2*pi/F*nu_2) ) / ...|\label{P4Z17}
\item[]\verb|                    ( c + 2 * (1-c) * sin(pi/F*nu_2)^2 ) )|
\item\verb|      F_eta_1 = F_eta_1 .* K_1 .* sin( pi/Ms*eta-pi/F*nu_2-Psi_1 ).^2|\makebox[-1pt]{}\label{P4Z18}\end{npb}
\item\verb|      rho = rho + 1|\label{P4Z19}\begin{npb}
\item\verb|      if rho < N-0.5-2*A_1-A_0|\label{P4Z20}
\item\verb|        Z_Z = z_0(rho) * exp( j*2*pi/F*nu_1 ) - (1-c)|\label{P4Z21}
\item\verb|        Z_N = 1 - (1-c) * z_0(rho) * exp(j*2*pi/F*nu_1)|\label{P4Z22}
\item\verb|        abs_Z_Z = abs( Z_Z )|\label{P4Z23}
\item\verb|        abs_Z_N = abs( Z_N )|\label{P4Z24}
\item\verb|        F_eta_1 = F_eta_1 .* NF_0(rho) .* ...|\label{P4Z25}
\item[]\verb|                  ( (abs_Z_N-abs_Z_Z)^2 + 4 * abs_Z_N * abs_Z_Z * ...|\makebox[-1pt]{}
\item[]\verb|                     sin( pi/Ms*eta-(angle(Z_Z)-angle(Z_N))/2 ).^2 )|
\item\verb|      end|\label{P4Z26}
\item\verb|    end|\label{P4Z27}\end{npb}
\item\verb|    F_eta = F_eta + F_eta_1|\label{P4Z28}\begin{npb}
\item\verb|  end|\label{P4Z29}\end{npb}\vadjust{\penalty-200}
\item\verb|  NF = 1 / sqrt( max(F_eta) * min(F_eta) )|\label{P4Z30}\begin{npb}
\item\verb|  F_eta = NF * F_eta|\label{P4Z31}\end{npb}
\item\verb|  L_eta = log(F_eta)|\label{P4Z32}
\item\verb|  Ceps_2 = ifft( [L_eta,L_eta(Ms/2:-1:2)] )|\begin{npb}\label{P4Z33}
\item\verb|  Ceps_2 = real( Ceps_2 )|\label{P4Z34}
\item\verb|  Ceps_2 = ( Ceps_2 + Ceps_2([1,Ms:-1:2]) ) / 2|\label{P4Z35}\end{npb}\vadjust{\penalty-100}
\item\verb|  sockel = eps / Ms / sqrt(48) * ...|\label{P4Z36}\begin{npb}
\item[]\verb|           sqrt( sum( max( [L_eta,L_eta(Ms/2:-1:2)].^2, 1 ) ) )|\end{npb}
\item\verb|  grenze = ( 2 * (2*N-2-2*A_1-A_0) ) ./ [1:Ms/2] .* ...|\label{P4Z37}\begin{npb}
\item[]\verb|           ( Ms * sockel/2/(2*N-2-2*A_1-A_0) ).^( 2*[1:Ms/2]/Ms )|\end{npb}
\item\verb|  max_fehl = sum( max( abs([L_eta,L_eta(Ms/2:-1:2)]), 1 ) ) * ...|\label{P4Z38}\begin{npb}
\item[]\verb|             eps * ( 2 + log(Ms)/log(2) ) / Ms|\end{npb}
\item\verb|  grenze = max( grenze, max_fehl )|\label{P4Z39}
\item\verb|  if any( abs(Ceps_2(2:Ms/2+1)) > grenze )|\label{P4Z40}\begin{npb}
\item\verb|    kriterium = abs( fft( Ceps_2(Ms/4+1:Ms/2) .* [Ms/4:Ms/2-1] ) )|\label{P4Z41}
\item\verb|    [dummy,womax] = max( kriterium(1:Ms/8+1) )|\label{P4Z42}
\item\verb|    delta_c = 1-(womax-1)*16/Ms|\label{P4Z43}
\item\verb|    delta_c = sin( delta_c*pi*(1-delta_c^2/2) )/3|\label{P4Z44}\end{npb}
\item\verb|    c = c * (1-delta_c) / ( 1 + (1-c) * delta_c )|\label{P4Z45}
\item\verb|    Ms = 2*Ms|\label{P4Z46}
\item\verb|  else|\label{P4Z47}
\item\verb|    Ms_OK = 1|\label{P4Z48}\begin{npb}
\item\verb|  end|\label{P4Z49}
\item\verb|end|\label{P4Z50}\end{npb}\vadjust{\penalty-200}
\item\verb|len_O = length(Omega)|\label{P4Z51}
\item\verb|Omega_s = Omega + 2 * atan( ( (1-c) * sin(Omega) ) ./ ...|\label{P4Z52}\begin{npb}
\item[]\verb|                            ( c + 2 * (1-c) * sin(Omega/2).^2 ) )|\end{npb}
\item\verb|phi = zeros(1,len_O)|\label{P4Z53}
\item\verb|for nu_i = 1:len_O|\label{P4Z54}\begin{npb}
\item\verb|  phi(nu_i) = sin(Omega_s(nu_i)*[Ms/2-1:-1:1]) * Ceps_2(Ms/2:-1:2).'|\label{P4Z55}
\item\verb|end|\label{P4Z56}\end{npb}
\item\verb|phi = phi - (N-1-A_1-A_0/2) * Omega_s + (F-1-A_0) * Omega/2|\label{P4Z57}\vadjust{\penalty-200}
\item\verb|F_Omega = zeros(1,len_O+1)|\label{P4Z58}
\item\verb|if N == 2*A_1+1|\label{P4Z59}\begin{npb}
\item\verb|  NF_1 = 1|\label{P4Z60}
\item\verb|else|\label{P4Z61}
\item\verb|  NF_1 = exp(2*log(8/pi*F)-sum(log([5:2:4*N-8*A_1]))/(N-1-2*A_1))|\label{P4Z62}
\item\verb|end|\label{P4Z63}\end{npb}
\item\verb|NF_0 = 1 ./ max( [ abs( z_0 * exp(j*pi/F*(N-1) ) - 1 ).^2 ; ...|\label{P4Z64}\begin{npb}
\item[]\verb|                   abs( z_0 * exp(j*pi/F*(1-N) ) - 1 ).^2 ] )|\end{npb}
\item\verb|for nu_1 = (1-N)/2:(N-1)/2|\label{P4Z65}\begin{npb}
\item\verb|  Omega_nu_1 = [Omega,0]/(2*pi) - nu_1/F|\label{P4Z66}
\item\verb|  Omega_nu_1 = (2*pi) * ( Omega_nu_1 - round(Omega_nu_1) )|\label{P4Z67}\end{npb}
\item\verb|  nu_4 = 2 * round( F/(2*pi) * Omega_nu_1 + (N-1)/2 ) - N + 1|\label{P4Z68}\begin{npb}
\item\verb|  nu_4 = ( 1-N+2*A_1 ) .* ( nu_4 < 2*A_1-N ) + ...|\label{P4Z69}
\item[]\verb|         nu_4 .* ( abs(nu_4) < N-2*A_1 ) + ...|
\item[]\verb|         ( N-1-2*A_1 ) .* ( nu_4 > N-2*A_1 )|
\item\verb|  nu_4 = nu_4/2|\label{P4Z70}\end{npb}
\item\verb|  d_Omega = Omega_nu_1 - 2*pi/F * nu_4|\label{P4Z71}
\item\verb|  F_Omega_1 = ( abs(d_Omega) < eps/F )|\label{P4Z72}\begin{npb}
\item\verb|  F_Omega_1 = F_Omega_1 / eps^2 + (~F_Omega_1) .* ...|\label{P4Z73}
\item[]\verb|              (         sin(F/2*d_Omega) ./ ...|
\item[]\verb|                ( F*eps*sin(d_Omega/2) + F_Omega_1 ) ).^2|\end{npb}\vadjust{\penalty-100}
\item\verb|  rho = 0|\label{P4Z74}\begin{npb}
\item\verb|  for nu_2 = (1-N)/2+A_1:(N-3)/2-A_1|\label{P4Z75}
\item\verb|    rho = rho + 1|\label{P4Z76}
\item\verb|    if rho < N-0.5-2*A_1-A_0|\label{P4Z77}
\item\verb|      abs_z_0 = abs(z_0(rho))|\label{P4Z78}
\item\verb|      F_Omega_1 = F_Omega_1 .* NF_0(rho) .* ( ( 1-abs_z_0 )^2 + ...|\label{P4Z79}
\item[]\verb|                  4*abs_z_0*sin((Omega_nu_1-angle(z_0(rho)))/2).^2 )|
\item\verb|    end|\label{P4Z80}\end{npb}
\item\verb|    nu_3 = nu_2 + ( nu_2 >= nu_4 )|\label{P4Z81}\begin{npb}
\item\verb|    F_Omega_1 = F_Omega_1 ./ (NF_1.*sin( Omega_nu_1/2-pi/F*nu_3).^2 )|\makebox[-1pt]{}\label{P4Z82}
\item\verb|  end|\label{P4Z83}\end{npb}
\item\verb|  F_Omega = F_Omega + F_Omega_1|\label{P4Z84}\begin{npb}
\item\verb|end|\label{P4Z85}\end{npb}\vadjust{\penalty-100}
\item\verb|Omega_nu_1 = abs( F / (2*pi) * [Omega,0] ) - N + A_1 + 0.5|\label{P4Z86}\begin{npb}
\item\verb|Omega_nu_1 = 0.5 * ( Omega_nu_1 > 0 ) .* Omega_nu_1 + 0.25|\label{P4Z87}
\item\verb|Omega_nu_1 = 1 - 2 * ( ( Omega_nu_1 - floor(Omega_nu_1) ) > 0.5 )|\label{P4Z88}
\item\verb|F_Omega = Omega_nu_1 .* sqrt(F_Omega)|\label{P4Z89}
\item\verb|F_Omega = ( M / F_Omega(len_O+1) ) * F_Omega(1:len_O)|\label{P4Z90}
\item\verb|F_Omega =  F_Omega .* exp(-j*phi)|\label{P4Z91}\end{npb}
\end{enumerate}}
Auch wenn man die Werte des Spektrums der Fensterfolge $f(k)$ bei dieser
Programmvariante nicht mehr nur f"ur die Frequenzen 
\mbox{$\Omega=\nu\CdoT2\pi/F$} mit \mbox{$\nu=0\;(1)\;N\!-\!1$}
berechnet, weil man hier nicht an den Koeffizienten der
Fourierreihenentwicklung der Fensterfolge interessiert ist,
muss man das Cepstrum genauso berechnen, wie dies im vorigen
Unterkapitel beschrieben ist. Die Zeilen~\ref{P4Z1} bis \ref{P4Z50}
sind daher identisch mit den Zeilen des Programms in Unterkapitel~\ref{E.Kap.11.1}.
In den Zeilen~\ref{P4Z52} bis \ref{P4Z57} wird wieder die
Phase der zu bestimmenden Spektralwerte durch Auswertung der
Sinusreihe bestimmt, deren Koeffizienten die Cepstralwerte sind. Die
Auswertung erfolgt nun f"ur alle Frequenzen des Vektors \verb|Omega|,
dessen Elementeanzahl in Zeile~\ref{P4Z51} bestimmt wird. Daher ist der
Vektor \verb|phi| nun in Zeile~\ref{P4Z53} mit einer ver"anderten L"ange
bereitzustellen, und die \verb|for|-Schleife der Zeilen~\ref{P4Z54} bis
\ref{P4Z56} ist nun f"ur jede der Frequenzen des Vektors \verb|Omega|
einmal zu durchlaufen.

Ab Zeile~\ref{P4Z58} werden die Betr"age der zu bestimmenden Spektralwerte
berechnet. Da dies nun nicht mehr nur die Spektralwerte f"ur niedrige
Frequenzen im Raster der Nullstellen des Spektrums der Fensterfolge
auf dem Einheitskreis sind, und da das Spektrum f"ur gro"se Werte von
$M$ und $N$ im Bereich hoher Frequenzen \mbox{$\Omega\!\approx\!\pi$}
wegen des potenzm"a"sigen Anstiegs der Sperrd"ampfung extrem klein werden
kann, ist nun besonderes Augenmerk auf eine geeignete Normierung und
eine geeignete Reihenfolge der Verarbeitung der Pol- und Nullstellen
zu legen, wenn man die Berechnung der Spektralwerte so durchf"uhren will,
dass alle Spektralwerte, die betraglich gr"o"ser als {\tt realmin} sind,
mit dem kleinstm"oglichen relativen Fehler berechnet werden k"onnen.
Das Betragsquadrat der Fensterfolge erreicht ihr absolutes Maximum
${\D M^2}$ immer bei der Frequenz \mbox{$\Omega\!=\!0$}, weil bei dieser
Frequenz bei der "Uberlagerung der um Vielfache von \mbox{$2\pi/M$}
verschobenen Betragsquadrate der Fensterspektren in Gleichung~(\myref{2.20}),
die von der hier zu berechnenden Fensterfolge erf"ullt
wird, nur ein einziger der stets positiven Summanden "ubrigbleibt.
Wenn man die Normierungskonstanten \verb|NF_1| und \verb|NF_0(rho)|
nun so w"ahlt, dass das Betragsquadrat des Spektralwertes bei der
Frequenz \mbox{$\Omega\!=\!0$} vor der endg"ultigen Normierung gr"o"ser als
${\D M^2}$ ist, wird sich der relative Fehler der bis dahin berechneten
Spektralwerte durch die abschlie"sende Normierung nur f"ur die Werte
erh"ohen, die nach der abschlie"senden Normierung kleiner als
{\tt realmin} sind. Um dieses Ziel zu erreichen, wird in Zeile~\ref{P4Z73}
bei der Initialisierung des Vektors, mit dessen
Hilfe die kumulativen Produkte berechnet werden, im Nenner jeweils der
Faktor \mbox{${\tt eps}^2$} verwendet. Von diesem Wert kann erwartet
werden, dass er wesentlich gr"o"ser als ${\D M^2}$, und doch deutlich
kleiner als die gr"o"ste am Rechner darstellbare Zahl ist, so dass es
f"ur beliebige Werte $\Omega$ nicht zu einem "Uberlauf kommen kann,
wenn man bei der Berechnung der Zwischenergebnisse darauf achtet, dass
diese ebenfalls nicht gr"o"ser als die gr"o"ste am Rechner darstellbare
Zahl werden. Um die endg"ultige Normierung einfach realisieren zu k"onnen,
wird wieder auch der Spektralwert bei der Frequenz Null ermittelt, da man
von diesem wei"s, dass er nach der endg"ultigen Normierung $M$ sein
muss.

Bevor die konkrete Wahl der Normierungskonstanten \verb|NF_1| und
\enlargethispage{3pt}\verb|NF_0(rho)| in den Zeilen~\ref{P4Z59} bis
\ref{P4Z64} n"aher erl"autert wird, wird die im Programm realisierte
Reihenfolge der Verarbeitung der Pol- und Nullstellen analysiert und
begr"undet. Wenn man in Gleichung~(\ref{E.10.8}) bei der mit $(\ast)$
gekennzeichneten Form \mbox{$z=e^{j\cdot\Omega}$} einsetzt, kann man
f"ur jede der Frequenzen des Vektors \verb|Omega| und jeden Summanden
die um \mbox{$\nu_1\CdoT2\pi/F$} verringerte Frequenz berechnen, und 
bei beiden Nennern jeweils die Polstellen suchen, die dem Punkt
\mbox{$e^{j\cdot(\Omega-\nu_1\CdoT2\pi/F)}$} am n"achsten liegen.
Da die beiden Nenner gleiche Nullstellen aufweisen, sind diese
beiden Polstellen identisch. Diese doppelte Polstelle ist diejenige,
die jeweils bei jedem Summanden und bei jeder Frequenz des Vektors
\verb|Omega| als erste verarbeitet wird. In den Zeilen~\ref{P4Z66} und
\ref{P4Z67} werden f"ur jeden Summanden die um \mbox{$\nu_1\CdoT2\pi/F$}
verringerten Frequenzen \verb|Omega_nu_1| modulo $2\pi$ berechnet.
In den Zeilen~\ref{P4Z68} bis \ref{P4Z70} werden die auf \mbox{$2\pi/F$}
normierten Winkel der jeweils n"achstgelegenen doppelten Polstelle
bestimmt. Die Winkeldifferenz wird in Zeile~\ref{P4Z71} als Vektor
\verb|d_Omega| berechnet. Nun kann man wieder die beiden Z"ahler
in Gleichung~(\ref{E.10.8}) der Form $(\ast)$ mit der doppelten Polstelle
zu einer quadrierten, periodisch fortgesetzten si-Funktion zusammenfassen.
Das Hauptmaximum der periodisch fortgesetzten si-Funktion liegt
jeweils bei dem Polstellenwinkel, der der gerade betrachteten,
verschobenen Frequenz \mbox{$\Omega-\nu_1\CdoT2\pi/F$} am n"achsten
liegt. Wenn die verschobene Frequenz sehr nahe bei dem Hauptmaximum
liegt, was in Zeile~\ref{P4Z72} festgestellt wird, kann man im Rahmen
der Rechengenauigkeit die periodisch fortgesetzte si-Funktion durch
den Konstanten Wert des Hauptmaximums ersetzten, anderenfalls kann die
periodisch fortgesetzte si-Funktion als Quotient zweier Sinusfunktionen
sowieso mit der gew"unschten Genauigkeit berechnet werden. In der
Zeile~\ref{P4Z73} werden die Werte der quadrierten, periodisch fortgesetzten
\mbox{si-Funktion} abh"angig von den Werten des Arguments auf eine der beiden
Weisen berechnet, um den Vektor \verb|F_Omega_1|, mit dessen Hilfe die
kumulativen Produkte berechnet werden, zu initialisieren. Dabei wird
zugleich die bereits angesprochene Normierung auf \mbox{${\D\varepsilon^{2}}$}
vorgenommen. Im weiteren werden wir uns auf den kritischen Fall
beschr"anken, dass $M$ sehr gro"s ist, und eine reale Chance besteht,
dass der am Rechner darstellbare Zahlenbereich verlassen wird. In diesem
Fall sinken nun die Werte auf dem Vektor \verb|F_Omega_1| mit steigendem
$\Omega$ mit \mbox{${\D\sin(\Omega/2)^{\!-2}}$} ab. Nach der Initialisierung
des Vektors \verb|F_Omega_1| werden die Nullstellen des Vektors \verb|z_0| und
die nicht gek"urzten Polstellen der Nenner in Gleichung~(\ref{E.10.8})$(\ast)$
abwechselnd verarbeitet. In der Zeile~\ref{P4Z79} werden zuerst
die normierten Quadrate der Nullstellenabst"ande f"ur alle Frequenzen
des Vektors \verb|Omega| zugleich mit den bisher berechneten
kumulativen Produkten des Vektors \verb|F_Omega_1| multipliziert.
Danach werden jeweils in Zeile~\ref{P4Z82} die Elemente des Vektors
\verb|F_Omega_1| durch die normierten Quadrate der Polstellenabst"ande
dividiert. Dabei wird mit Hilfe des Vektors \verb|nu_3| wieder
die gek"urzte doppelte Polstelle ausgelassen, ganz "ahnlich, wie dies
bei dem Programm im Anhang~\myref{MatFen1} beschrieben worden ist. Wenn die
gerade verarbeitete Nullstelle $z_{0,\rho}$ sehr nahe bei dem Punkt
\mbox{$z\!=\!1$} liegt, werden die Quadrate der Nullstellenabst"ande
zu den Punkten \mbox{$e^{j\cdot\Omega}$} mit steigendem $\Omega$ mit
\mbox{${\D\sin(\Omega/2)^2}$} zunehmen. Da f"ur gro"se Werte von $M$
auch alle Polstellen sehr nahe bei \mbox{$z\!=\!1$} liegen, wird
dadurch der Einfluss der danach verarbeiteten Polstelle auf den
Vektor \verb|F_Omega_1| f"ur hohe Frequenzen \mbox{$\Omega\!\approx\!\pi$}
insofern weitgehend kompensiert, als dass sich als Quotient der beiden
Abstandsquadrate zu den Pol- und Nullstellen n"aherungsweise der
Konstante Quotient \mbox{$4\CdoT{\tt NF\_0(rho)}/{\tt NF\_1}$}
der Normierungskonstanten ergibt. Wenn die Nullstelle $z_{0,\rho}$
{\em nicht}\/ so nahe bei \mbox{$z\!=\!1$} liegt, wird sich bei dem Quotienten
der Abstandsquadrate zu zwei nacheinander verarbeiteten Null- und
Polstellen ein Frequenzverlauf ergeben, der bei hohen Frequenzen
insgesamt kleiner als der ebengenannte konstante Quotient ist.
Da wir vorausgesetzt hatten, dass die Nullstellen des Vektors
\verb|z_0| alle innerhalb des Einheitskreises liegen sollen,
wenn sie an dieses Programm "ubergeben werden, wird der in Zeile~\ref{P4Z64}
berechnete Normierungsfaktor \verb|NF_0(rho)| maximal,
wenn die Nullstelle $z_{0,\rho}$ genau bei eins liegt.
Die Normierungskonstante \verb|NF_1| wird in Zeile~\ref{P4Z62}
in der Art berechnet, wie dies bei dem Programm in Kapitel~\myref{MatFen1}
beschrieben ist. Es wurde nun untersucht, wie gro"s der Quotient
\mbox{$4\CdoT{\tt NF\_0(rho)}/{\tt NF\_1}$} in Abh"angigkeit von
$N$ und $A_1$ f"ur gro"ses $M$ maximal ---\,also f"ur 
\mbox{$z_{0,\rho}\!=\!1$}\,--- werden kann. Dabei stellte sich heraus,
dass er au"ser f"ur \mbox{$N\!=\!2$} und \mbox{$A_1\!=\!0$} immer
kleiner $1$ ist. Wenn $M$ hinreichend gro"s ist (\,\mbox{$M>10$}\,),
ist der Quotient bei steigendem $M$ praktisch konstant.
Desweiteren stellte sich heraus, dass Wert des Quotienten bei
einem festen Wert von $N$ und variabler Anzahl $A_1$ um so n"aher
bei eins liegt, je h"aufiger die Schleife mit dem Index \verb|nu_2|
durchlaufen wird. Im eben ausgeschlossenen Fall ergibt sich
ein maximaler Wert f"ur den Quotienten, der f"ur steigendes
$M$ gegen einen Wert von etwa $2,\!1875$ konvergiert.
Da es in diesem Fall nur maximal eine frei w"ahlbare Nullstelle
$z_{0,\rho}$ gibt, und nur ein Null-Polstellenpaar zu verarbeiten ist,
so dass die Schleife mit dem Index \verb|nu_2| nur einmal durchlaufen
wird, besteht hier selbst f"ur extrem gro"se Werte vom $M$ kaum die
Gefahr, dass der mit dem kleinstm"oglichen relativen Fehler darstellbare
Zahlenbereich verlassen wird, und somit ist dieser Maximalwert des
Quotienten akzeptabel. In allen anderen F"allen, ist es nicht
m"oglich, dass die Anhebung der Werte des Vektors \verb|F_Omega_1|
f"ur \mbox{$\Omega\!\approx\!\pi$} durch die Nullstelle $z_{0,\rho}$
nicht durch die danach verarbeitete Polstelle mehr als kompensiert wird.
Nach einem Schleifendurchlauf bei der Berechnung der kumulativen
Produkte in Gleichung~(\ref{E.10.8})$(\ast)$ werden die Werte des Vektors
\verb|F_Omega_1|, bei denen die gr"o"ste Gefahr besteht, dass der
mit dem kleinstm"oglichen relativen Fehler darstellbare Zahlenbereich
verlassen wird, kleiner sein als vor diesem Schleifendurchlauf.
Man kann sich anhand der Anzahl der zu verarbeitenden Pol- und Nullstellen
"uberlegen, dass zuletzt immer eine Polstelle behandelt wird. Es kann
also nicht vorkommen, dass ein Wert, der vor einem Schleifendurchlauf
zu klein ist, um mit voller Genauigkeit dargestellt werden zu k"onnen,
oder der gar zu null quantisiert wurde, nach dem Schleifendurchlauf
bei ideal fehlerfreier Berechnung in einen Bereich angehoben wird,
der wieder mit voller Genauigkeit darstellbar w"are.
Damit solches bei der abschlie"senden Normierung nicht passiert,
darf der letzte Wert des Vektors \verb|F_Omega_1|, der nach
Zeile~\ref{P4Z66} den Wert bei der Frequenz Null enth"alt, der f"ur die
endg"ultige Normierung ben"otigt wird, nicht kleiner als
${\D M^2}$ werden. Die Normierungskonstante \verb|NF_1| wurde
so gew"ahlt, dass der bei der Initialisierung eingestellte Wert
\mbox{${\D\varepsilon^{-2}}$} auch nach der vollst"andigen Berechnung
etwa erhalten bleibt, wenn keine Nullstelle $z_{0,\rho}$
angegeben wird. Die Normierung der Abstandsquadrate der
Nullstellen wird in Zeile~\ref{P4Z64} so gew"ahlt, dass sich
f"ur \mbox{$\Omega\!=\!0$} bei allen m"oglichen  Rotationen der
Nullstellen mit \mbox{$e^{j\cdot\nu_1\cdot2\pi/F}$} ein Abstand zum
Punkt \mbox{$z\!=\!1$} ergibt, der den Maximalwert Eins wenigstens
bei einem Summanden der Summe in Gleichung~(\ref{E.10.8})$(\ast)$
auch annimmt. Daher, und weil es immer mehr Summanden als Nullstellen 
gibt, kann man annehmen, dass in der Summe wenigstens einige Summanden
auftreten, die trotz des Einflusses der Nullstellen nicht so extrem
viel kleiner als \mbox{${\D\varepsilon^{-2}}$} sind, dass der Wert bei 
\mbox{$\Omega\!=\!0$} vor der abschlie"senden Normierung gleich unter 
${\D M^2}$ abf"allt.

\section{Die diskreten Fensterautokorrelationsfolgen}\label{E.Kap.11.3}

Dieses Programm ben"otigt die Zeitpunkte $k$, f"ur die die
Fensterautokorrelationsfolge berechnet werden soll, als Vektor \verb|k|.
Desweiteren muss der Zeilenvektor \verb|F_nu| die Fourierreihenkoeffizienten
\mbox{$F\big({\T\nu\CdoT\frac{2\pi}{F}}\big)/F$} der Fensterfolge f"ur
\mbox{$\nu=0\;(1)\;N\!-\!A_1\!-\!1$} enthalten. Diesen Vektor erh"alt man
entweder bei dem Programm im Anhang von \cite{Diss} oder bei dem
Programm in Unterkapitel~\ref{E.Kap.10.1} als Ergebnis. Als drittes
Eingabeargument ist noch die L"ange $F$ der Fensterfolge anzugeben.
{\renewcommand{\labelenumi}{{\footnotesize\arabic{enumi}:}}
\begin{enumerate}
\setlength{\itemsep}{-7pt plus1pt minus0pt}
\setlength{\parsep}{0pt plus1pt minus0pt}
\item\verb|function d_k = fenster_akf( k, F_nu, F )|\label{P5Z0}\begin{npb}
\item\verb|N_A = length(F_nu)|\label{P5Z1}
\item\verb|M = round( F * F_nu(1) )|\label{P5Z2}\end{npb}
\item\verb|k = abs(k)|\label{P5Z3}
\item\verb|O_k = 2*pi/F*k|\label{P5Z4}
\item\verb|d_k = zeros(1,length(k))|\label{P5Z5}
\item\verb|for nu = N_A:-1:2|\label{P5Z6}\begin{npb}
\item\verb|  F_sin = real( F_nu(nu)' * ...|\label{P5Z7}
\item[]\verb|          sum( [F_nu(N_A:-1:1)';F_nu([2:nu-1,nu+1:N_A]).'] ./ ...|
\item[]\verb|               tan( pi/F*[2-N_A-nu:-1,1:N_A-nu] ).' ) ) / M + ...|
\item[]\verb|          imag( F_nu(nu)' * ...|
\item[]\verb|          sum( [F_nu(N_A:-1:1)';F_nu([2:nu-1,nu+1:N_A]).'] ) ) / M|
\item\verb|  d_k = d_k + 2 * F_sin * sin(O_k*(nu-1)) + ...|\label{P5Z8}
\item[]\verb|              2 * abs(F_nu(nu))^2 / M * cos(O_k*(nu-1)) .* (F-k)|
\item\verb|end|\label{P5Z9}
\item\verb|d_k = d_k + abs(F_nu(1))^2 / M * (F-k)|\label{P5Z10}\end{npb}
\end{enumerate}}
In den Zeilen~\ref{P5Z1} und \ref{P5Z2} werden die ben"otigten Parameter
bestimmt. Die Variable \verb|N_A| enth"alt die Differenz \mbox{$N\!-\!A_1$}.
Da die Fourierreihe der Fensterfolge nur Anteile bis zum
\mbox{$(N\!-\!A_1\!-\!1)$}-fachen der Grundkreisfrequenz enth"alt, kann
der Wert f"ur \verb|N_A| als die L"ange\linebreak des Vektors \verb|F_nu| bestimmt
werden. Weil der Fourierreihenkoeffizient mit \mbox{$\nu\!=\!0$} immer
\mbox{$1/N$} ist, l"asst sich \verb|M| daraus durch Multiplikation mit der
Fensterl"ange \verb|F| berechnen. I.~allg. l"asst sich der
Fourierreihenkoeffizient des Gleichanteils \verb|F_nu(1)| am Rechner
nicht exakt darstellen. Daher wird in Zeile~\ref{P5Z2} eine Rundung
auf die n"achste am Rechner exakt darstellbare ganze Zahl vorgenommen,
um so die im Produkt \verb|F*F_nu(1)| evtl. vorhandenen Rundungsfehler
abzuschneiden. Da es sich bei der Fensterautokorrelationsfolge um eine
geradesymmetrische Folge handelt, kann man f"ur negatives $k$
ebensogut die Werte f"ur $-k$ berechnen. Durch Zeile~\ref{P5Z3}
erspart man sich so eine Fallunterscheidung bei der sp"ateren
Berechnung der Fensterautokorrelationsfolge. Diese Berechnung wird
mit Hilfe der Formeln~(\ref{E.10.19}) und (\ref{E.10.20}) als "Uberlagerung
einer Sinusreihe und einer mit der Dreiecksfunktion multiplizierten
Kosinusreihe vorgenommen, wie dies in Kapitel~\ref{E.Kap.10.3} beschrieben
ist. Dabei wird bei jedem Reihenglied die mit $k$ multiplizierte
Grundkreisfrequenz ben"otigt, die daher in Zeile~\ref{P5Z4} f"ur
alle Werte des Vektors \verb|k| auf dem Vektor \verb|O_k| bereitgestellt
wird. Die Summe in Gleichung~(\ref{E.10.20}) wird als \verb|for|-Schleife,
die in Zeile~\ref{P5Z6} beginnt und in Zeile~\ref{P5Z9} endet, realisiert,
indem zun"achst in Zeile~\ref{P5Z5} der Vektor \verb|d_k| mit null
initialisiert wird, zu dem in Zeile~\ref{P5Z8} bei jedem Schleifendurchlauf
ein Anteil der Sinusreihe und ein Anteil der mit der Dreiecksfunktion
multiplizierten Kosinusreihe addiert wird. In Zeile~\ref{P5Z7} wird
der dazu ben"otigte Sinusreihenkoeffizient nach Gleichung~(\ref{E.10.19})
berechnet. Die dabei auftretenden Summen werden mit dem \verb|MATLAB|-Befehl
\verb|sum| berechnet.

Mit der hier vorgestellten Variante kann man nicht nur die
Fensterautokorrelationsfolge f"ur ganzzahliges $k$ berechnen, sondern
auch die Werte der Funktion f"ur beliebiges reelles $k$. Ist man nur
an der Fensterautokorrelationsfolge f"ur ganzzahliges $k$ interessiert,
und will man diese Werte aber mit einem geringeren Fehler berechnen,
so empfiehlt es sich, bei der Berechnung der Reihe die dann mehrfach
auftretenden Werte der Sinus- und Kosinusfunktion wieder au"serhalb der
\verb|for|-Schleife hochgenau zu berechnen, und sich die jeweils
bei einem Schleifendurchlauf ben"otigten Werte daraus herauszugreifen.
Die dazu im eben vorgestellten Programm notwendigen Modifikationen
werden hier nicht explizit dargestellt. Man kann sie dem Programm
in Unterkapitel~\ref{E.Kap.11.1} zur Bestimmung der Fensterfolge $f(k)$ aus
den Fourierreihenkoeffizienten (\,Zeilen~\ref{P3Z84} bis \ref{P3Z93}\,)
sinngem"a"s entnehmen.

\section[Die Fourierreihenkoeffizienten der kontinuierlichen Fensterfunktionen]{{}\!Die Fourierreihenkoeffizienten der kontinuierlichen\!{} \mbox{{}\!Fensterfunktionen}}\label{E.Kap.11.4}

Die in Kapitel~\ref{E.Kap.10.7} vorgestellte Fensterfunktion stellt den
Spezialfall einer Fensterfunktion dar, bei der in Gleichung~(\ref{E.10.37})
die Laplace-Transformierte der zugrundeliegenden Basisfensterfunktion 
im Z"ahler kein Polynom in $s$ aufweist. Das hier vorgestellte Programm 
berechnet die Fourierreihenkoeffizienten der allgemeineren Fensterfunktion, 
deren zugrundeliegende Basisfensterfunktion eine Laplace-Transformierte 
aufweist, bei der in Gleichung~(\ref{E.10.37}) im Z"ahler ein Polynom 
in $s$ vom Grad \mbox{$N\!-\!A_0\!-\!1$} eingestellt werden kann. Da 
der Grad des Z"ahlerpolynoms wenigstens um eins kleiner als der Grad 
des Nennerpolynoms sein muss, wenn die Basisfensterfunktion keine 
Delta-Distributionen enthalten soll, muss der Graddefekt $A_0$ 
ebenfalls zwischen zwischen $0$ und \mbox{$N\!-\!1$} gew"ahlt werden.
Desweiteren wird das Z"ahlerpolynom durch seine Nullstellen bis auf
eine Konstante festgelegt. Gleichung~(\ref{E.10.37}) zeigt, dass die
Laplace-Transformierte der zugrundeliegenden Basisfensterfunktion
auf der imagin"aren Achse immer "aquidistante Nullstellen bei
\mbox{$s\!=\!j\omega$} mit \mbox{$\omega=(2\CdoT\nu\!-\!N\!+\!1)\CdoT\pi$}
und mit \mbox{$\nu\in{}\mathbb{Z}$} au"ser f"ur \mbox{$\nu=0\;(1)\;N\!-\!1$}
aufweist. Auch wenn nun ein zus"atzliches Z"ahlerpolynom eingef"ugt wird,
"andert sich daran nichts. Die Nullstellen des hinzugekommenen
Z"ahlerpolynoms m"ussen zur reellen Achse spiegelsymmetrisch liegen, damit
sich eine reelle Basisfensterfunktion ergibt. Von den frei einstellbaren
Nullstellen des Z"ahlerpolynoms k"onnen einige so gelegt werden, dass sie
sich im Raster $2\pi$ an die ebengenannten fest vorgegebenen "aquidistanten
Nullstellen auf der imagin"aren Achse anschlie"sen. Somit gilt hier
\mbox{$G_{\infty}\big(j\CdoT(N\!-\!1\!-\!2\CdoT\!A_1\!+\!2\CdoT\nu)\CdoT\pi\big)=0$}
f"ur \mbox{$\nu\!\in\!\mathbb{N}$} und 
\mbox{$G_{\infty}\big(j\CdoT(N\!-\!1\!-\!2\CdoT\!A_1)\CdoT\pi\big)\neq0$}.
Durch die Angabe des Parameters $A_1$ werden diese Nullstellen mit einfacher
Vielfachheit im Z"ahlerpolynom ber"ucksichtigt. Alle weiteren
\mbox{$N\!-\!A_0\!-\!1\!-\!2\CdoT\!A_1$} frei einstellbaren Nullstellen
des Z"ahlerpolynoms werden explizit als $s_{0,\rho}$ angegeben.
Die Festlegung der frei w"ahlbaren Nullstellen des hinzugekommenen
Z"ahlerpolynoms erfolgt also genau analog zu der Festlegung
der frei w"ahlbaren Nullstellen bei der Z-Transformierten der
Basisfensterfolge der in Kapitel~\ref{E.Kap.10.1} beschriebenen
diskreten Fensterfolge. Lediglich liegen nun die Nullstellen in der
$s$-Ebenen statt in der $z$-Ebene und die Nullstellen, die dort
im Raster \mbox{$2\pi/F$} auf dem Einheitskreis liegen, entsprechen nun
den Nullstellen auf der imagin"aren Achse im Raster $2\pi$. Es sei noch 
erw"ahnt, dass die in Kapitel \ref{E.Kap.10.7} beschriebene Fensterfunktion 
den Spezialfall mit \mbox{$A_0=N\!-\!1$} darstellt.

Neben dem Parameter $N$, werden die Anzahl $A_1$ der 
zus"atzlichen Nullstellen der Laplace-Transformierten \mbox{$G_{\infty}(s)$} 
der Basisfensterfunktion \mbox{$g(k)$} auf der imagin"aren Achse im Raster $2\pi$, 
die Graddifferenz $A_0$ des Nenners und des Z"ahlers von \mbox{$G_{\infty}(s)$}, 
sowie die frei w"ahlbaren Nullstellen $s_{0,\rho}$ von \mbox{$G_{\infty}(s)$} 
ben"otigt. Die ganzzahligen Parameter $N$, $A_1$ und $A_0$ m"ussen als 
skalare Gr"o"sen  (\,also als \mbox{$1\!\times\!1$} Matrizen\,) \verb|N|, 
\verb|A_1| und \verb|A_0| beim Programmaufruf angegeben werden. F"ur diese 
Parameter m"ussen die Bedingungen \mbox{$N\!>\!1$}, \mbox{$0\!\le\!A_1\!<\!N/2$} 
und \mbox{$0\!\le\!A_0\!<\!N\!-\!1\!-\!2\CdoT\!A_1$} erf"ullt sein. 
Die Nullstellen $s_{0,\rho}$ bilden die Elemente des
Vektors \verb|s_0|, der ebenfalls beim Programmaufruf zu "ubergeben ist.
Dieser Vektor muss genau \mbox{$N\!-\!1\!-\!A_0\!-2\CdoT\!A_1$}
Elemente enthalten. Nullstellen, die nicht reell sind, m"ussen
als zueinander konjugiert komplexe Paare in \verb|s_0| vorhanden sein.
Um eine m"oglichst gute Genauigkeit zu erzielen, sollten Nullstellen
in der rechten Halbebene durch die an der imagin"aren Achse gespiegelten
ersetzt worden sein, und die Nullstellen sollten nach aufsteigendem Abstand
zum Punkt \mbox{$s\!=\!0$} sortiert sein. Als Ergebnis werden von diesem
Programm die Fourierreihenkoeffizienten der Fensterfunktion f"ur
\mbox{$\nu=0\;(1)\;N\!-\!A_1\!-\!1$} auf dem Vektor \verb|F_nu|
zur"uckgegeben.
{\renewcommand{\labelenumi}{{\footnotesize\arabic{enumi}:}}
\begin{enumerate}
\setlength{\itemsep}{-7pt plus1pt minus0pt}
\setlength{\parsep}{0pt plus1pt minus0pt}
\item\verb|function F_nu = fenster_koeff( N, A_0, A_1, s_0 )|\label{P6Z0}\begin{npb}
\item\verb|c = (N/2)^(4/3)|\label{P6Z1}
\item\verb|Ms = -log(eps)/log(2) * 2^log(N/3)|\label{P6Z2}
\item\verb|Ms = 2^ceil(log(Ms)/log(2))|\label{P6Z3}
\item\verb|Ms = max(Ms,16)|\label{P6Z4}\end{npb}
\item\verb|Ms_OK = 0|\label{P6Z5}\begin{npb}
\item\verb|while ~Ms_OK|\label{P6Z6}\end{npb}
\item\verb|  eta = 0:Ms/2|\label{P6Z7}\begin{npb}
\item\verb|  F_eta = zeros(1,Ms/2+1)|\label{P6Z8}\end{npb}
\item\verb|  NF_1 = 1 / ( 1 + ( (N-1-A_1)/c )^2 )|\label{P6Z9}\begin{npb}
\item\verb|  NF_0 = 1 ./ max( [ ( abs( s_0 + j*pi*(1-N) + 2*pi*c ) + ...|\label{P6Z10}
\item[]\verb|                       abs( s_0 + j*pi*(1-N) - 2*pi*c ) ) ; ...|
\item[]\verb|                     ( abs( s_0 + j*pi*(N-1) + 2*pi*c ) + ...|
\item[]\verb|                       abs( s_0 + j*pi*(N-1) - 2*pi*c ) ) ] ).^2|\end{npb}
\item\verb|  for nu_1 = (1-N)/2:(N-1)/2|\label{P6Z11}\begin{npb}
\item\verb|    F_eta_1 = ones(1,Ms/2+1)|\label{P6Z12}\end{npb}
\item\verb|    rho = 0|\label{P6Z13}
\item\verb|    for nu_2 = [(1-N+A_1):(nu_1+A_1-(N+1)/2), ...|\label{P6Z14}\begin{npb}
\item[]\verb|                (nu_1-A_1+(N+1)/2):(N-1-A_1)]|
\item\verb|      K_1 = NF_1 * ( 1 + (  nu_2 / c )^2 )|\label{P6Z15}
\item\verb|      Psi_1 = atan( nu_2 / c )|\label{P6Z16}
\item\verb|      F_eta_1 = F_eta_1 .* K_1 .* sin( pi/Ms*eta - Psi_1 ).^2|\label{P6Z17}\end{npb}
\item\verb|      rho = rho + 1|\label{P6Z18}\begin{npb}
\item\verb|      if rho < N-0.5-2*A_1-A_0|\label{P6Z19}
\item\verb|        Z_Z = 2*pi*c + s_0(rho) + j*2*pi*nu_1|\label{P6Z20}
\item\verb|        Z_N = 2*pi*c - s_0(rho) - j*2*pi*nu_1|\label{P6Z21}
\item\verb|        abs_Z_Z = abs( Z_Z )|\label{P6Z22}
\item\verb|        abs_Z_N = abs( Z_N )|\label{P6Z23}
\item\verb|        F_eta_1 = F_eta_1 .* NF_0(rho) .* ...|\label{P6Z24}
\item[]\verb|                  ( (abs_Z_N-abs_Z_Z)^2 + 4 * abs_Z_N * abs_Z_Z * ...|\makebox[-1pt]{}
\item[]\verb|                    sin( pi/Ms*eta-(angle(Z_Z)-angle(Z_N))/2 ).^2 )|
\item\verb|      end|\label{P6Z25}
\item\verb|    end|\label{P6Z26}\end{npb}
\item\verb|    F_eta = F_eta + F_eta_1|\label{P6Z27}\begin{npb}
\item\verb|  end|\label{P6Z28}\end{npb}\vadjust{\penalty-100}
\item\verb|  NF = 1 / sqrt( max(F_eta) * min(F_eta) )|\label{P6Z29}\begin{npb}
\item\verb|  F_eta = NF * F_eta|\label{P6Z30}\end{npb}
\item\verb|  L_eta = log(F_eta)|\label{P6Z31}
\item\verb|  Ceps_2 = ifft( [L_eta,L_eta(Ms/2:-1:2)] )|\label{P6Z32}\begin{npb}
\item\verb|  Ceps_2 = real( Ceps_2 )|\label{P6Z33}
\item\verb|  Ceps_2 = ( Ceps_2 + Ceps_2([1,Ms:-1:2]) ) / 2|\label{P6Z34}\end{npb}\vadjust{\penalty-100}
\item\verb|  sockel = eps / Ms / sqrt(48) * ...|\label{P6Z35}\begin{npb}
\item[]\verb|           sqrt( sum( max( [L_eta,L_eta(Ms/2:-1:2)].^2, 1 ) ) )|\end{npb}
\item\verb|  grenze = ( 2 * (2*N-2-2*A_1-A_0) ) ./ [1:Ms/2] .* ...|\label{P6Z36}\begin{npb}
\item[]\verb|           ( Ms * sockel/2/(2*N-2-2*A_1-A_0) ).^( 2*[1:Ms/2]/Ms )|\end{npb}
\item\verb|  max_fehl = sum( max( abs([L_eta,L_eta(Ms/2:-1:2)]), 1 ) ) * ...|\label{P6Z37}\begin{npb}
\item[]\verb|             eps * ( 2 + log(Ms)/log(2) ) / Ms|\end{npb}
\item\verb|  grenze = max( grenze, max_fehl )|\label{P6Z38}
\item\verb|  if any( abs(Ceps_2(2:Ms/2+1)) > grenze )|\label{P6Z39}\begin{npb}
\item\verb|    kriterium = abs( fft( Ceps_2(Ms/4+1:Ms/2) .* [Ms/4:Ms/2-1] ) )|\label{P6Z40}
\item\verb|    [dummy,womax] = max( kriterium(1:Ms/8+1) )|\label{P6Z41}
\item\verb|    delta_c = 1-(womax-1)*16/Ms|\label{P6Z42}
\item\verb|    delta_c = sin( delta_c*pi*(1-delta_c^2/2) )/3|\label{P6Z43}\end{npb}
\item\verb|    c = c * (1-delta_c) / (1+delta_c)|\label{P6Z44}
\item\verb|    Ms = 2*Ms|\label{P6Z45}
\item\verb|  else|\label{P6Z46}
\item\verb|    Ms_OK = 1|\label{P6Z47}\begin{npb}
\item\verb|  end|\label{P6Z48}
\item\verb|end|\label{P6Z49}\end{npb}\vadjust{\penalty-200}
\item\verb|nu = 1:N-A_1-1|\label{P6Z50}
\item\verb|Omega_s = 2 * atan(nu./c)|\label{P6Z51}
\item\verb|phi = zeros(1,N-A_1-1)|\label{P6Z52}
\item\verb|for nu_i = nu|\label{P6Z53}\begin{npb}
\item\verb|  phi(nu_i) = sin(Omega_s(nu_i)*[Ms/2-1:-1:1]) * Ceps_2(Ms/2:-1:2).'|\label{P6Z54}
\item\verb|end|\label{P6Z55}\end{npb}
\item\verb|phi = phi - ( N - 1 - A_1 - A_0/2 ) * Omega_s + pi * nu|\label{P6Z56}\vadjust{\penalty-200}
\item\verb|if N == 2*A_1+1|\label{P6Z57}\begin{npb}
\item\verb|  F_nu = ones(1,N-A_1)|\label{P6Z58}\end{npb}
\item\verb|else|\label{P6Z59}\begin{npb}
\item\verb|  nu = 0:N-A_1-1|\label{P6Z60}
\item\verb|  F_nu = zeros(1,N-A_1)|\label{P6Z61}\end{npb}
\item\verb|  NF_1 = exp(2*log(8)-sum(log([5:2:4*N-8*A_1]))/(N-1-2*A_1))|\label{P6Z62}\begin{npb}
\item\verb|  NF_0 = 1 ./ max( [ abs( s_0 + j*pi*(N-1) ).^2 ; ...|\label{P6Z63}
\item[]\verb|                     abs( s_0 - j*pi*(N-1) ).^2 ] )|\end{npb}
\item\verb|  for nu_1 = (1-N)/2:(N-1)/2|\label{P6Z64}\begin{npb}
\item\verb|    F_nu_1 = ( ( nu < nu_1+N/2-A_1 ) & ( nu > nu_1-N/2+A_1) ) / eps^2|\makebox[-1pt]{}\label{P6Z65}\end{npb}
\item\verb|    rho = 0|\label{P6Z66}\begin{npb}
\item\verb|    for nu_2 = (1-N)/2+A_1:(N-3)/2-A_1|\label{P6Z67}
\item\verb|      rho = rho + 1|\label{P6Z68}
\item\verb|      if rho < N-0.5-2*A_1-A_0|\label{P6Z69}
\item\verb|        F_nu_1 = F_nu_1 .* NF_0(rho) .* ...|\label{P6Z70}
\item[]\verb|                 abs( j*2*pi*(nu-nu_1) - s_0(rho) ).^2|
\item\verb|      end|\label{P6Z71}\end{npb}
\item\verb|      nu_3 = nu - nu_1 - nu_2 - ( nu <= nu_1 + nu_2 )|\label{P6Z72}\begin{npb}
\item\verb|      F_nu_1 = F_nu_1 ./ ( NF_1 .* nu_3 .^ 2 )|\label{P6Z73}
\item\verb|    end|\label{P6Z74}\end{npb}
\item\verb|    F_nu = F_nu + F_nu_1|\label{P6Z75}\begin{npb}
\item\verb|  end|\label{P6Z76}\end{npb}
\item\verb|  F_nu = sqrt(F_nu)|\label{P6Z77}\begin{npb}
\item\verb|  F_nu = F_nu / F_nu(1)|\label{P6Z78}
\item\verb|end|\label{P6Z79}
\item\verb|F_nu(2:N-A_1) =  F_nu(2:N-A_1) .* exp(-j*phi)|\label{P6Z80}\end{npb}\vadjust{\penalty-200}
\end{enumerate}}
Gegen"uber der Berechnung der Fourierreihenkoeffizienten der diskreten
Fensterfolge nach Kapitel~\ref{E.Kap.10.1}, die im Programm des
Unterkapitels~\ref{E.Kap.11.1} enthalten ist, ergeben sich hier einige "Anderungen,
die nun kommentiert werden. Die Bilineartransformation ist nun nach 
Gleichung~(\ref{E.10.41}) und (\ref{E.10.42}) definiert.
Daher ergeben sich als Startwerte des Parameters \verb|c| und
der f"ur die Berechnung des Cepstrums ben"otigten FFT-L"ange \verb|Ms|
die in den Gleichungen~(\ref{E.10.51}) und (\ref{E.10.52}) angegebenen
Werte, die in den Zeilen~\ref{P6Z1} bis \ref{P6Z4} eingestellt werden.

Die Quadrate der Abst"ande der Punkte des Einheitskreises
\mbox{$e^{j\cdot\widetilde{\Omega}}$} von den bilinear transformierten
Nullstellen, die zur Erweiterung auf den Hauptnenner eingef"ugt wurden,
werden mit Hilfe des konstanten Normierungsfaktors \verb|NF_1| normiert.
Dieser wird in Zeile~\ref{P6Z9} wieder so
gew"ahlt, dass der maximal auftretende mit \verb|NF_1| normierte Faktor
\verb|K_1|, der in Zeile~\ref{P6Z15} berechnet wird, zu eins wird.
Dieser Faktor \verb|K_1| ist abgesehen von der Normierung gleich der
in Gleichung~(\ref{E.10.46}) angegebenen Konstante \mbox{$K_{\infty,\nu_2}$}.
In Zeile~\ref{P6Z16} wird der in Gleichung~(\ref{E.10.45}) auftretende
Winkel der bilineartransformierten Nullstelle berechnet, mit dessen Hilfe 
in Zeile~\ref{P6Z17} bei jedem Schleifendurchlauf ein Faktor mit dem
bisher berechneten kumulativen Produkt multipliziert werden kann.
Auch die Normierungsfaktoren \verb|NF_0(rho)| werden wieder als das
Reziproke des Maximalwertes \mbox{$\big(\,|\Tilde{z}_{N,\rho,\nu_1}|\!+\!
|\Tilde{z}_{Z,\rho,\nu_1}|\,\big)^{\!2}$} des normierten Quadrats
des Nullstellenabstands zum Punkt \mbox{$e^{j\cdot\widetilde{\Omega}}$}
gew"ahlt. Da hier die um \mbox{$\nu_1\CdoT2\pi$} verschobenen 
Nullstellen $s_{0,\rho}$ anders bilineartransformiert werden,
ergeben sich nach der Bilineartransformation die Nullstellen
\mbox{$\Tilde{z}_{Z,\rho,\nu_1}/\Tilde{z}_{N,\rho,\nu_1}$}, deren Z"ahler
und Nenner sich so berechnen lassen, wie dies in den Zeilen~\ref{P6Z20}
und \ref{P6Z21} durchgef"uhrt wird. Setzt man dies und die maximal
m"oglichen Verschiebungen \mbox{$\nu_1=N\!-\!1$} und \mbox{$\nu_1=1\!-\!N$}
in die maximalen Abstandsquadrate \mbox{$\big(\,|\Tilde{z}_{N,\rho,\nu_1}|\!+\!
|\Tilde{z}_{Z,\rho,\nu_1}|\,\big)^{\!2}$} ein, so erh"alt man die in
Zeile~\ref{P6Z10} berechneten Normierungsfaktoren \verb|NF_0(rho)|,
die den Vektor \verb|NF_0| bilden. Damit lassen sich die normierten
Abstandsquadrate in Zeile~\ref{P6Z22} bis \ref{P6Z24} exakt genau so
berechnen, wie bei der diskreten Fensterfolge.

Bei der evtl. notwendigen Neuberechnung des Bilineartransformationsparameters
\verb|c| in Zeile~\ref{P6Z44} ist nun eine leicht modifizierte
Formel zu verwenden, die ber"ucksichtigt, dass hier eine andere Art der
Bilineartransformation stattfindet. Aus demselben Grund berechnen sich nun
die Frequenzen \mbox{$\widetilde{\Omega}$} in Zeile~\ref{P6Z51}
nach Gleichung~(\ref{E.10.43}). Abgesehen davon, bleibt die Berechnung
des "uber das Cepstrum berechneten Phasenanteils in den Zeilen~\ref{P6Z52}
bis \ref{P6Z55} unver"andert. Bei dem linearen Phasenanteil in
Zeile~\ref{P6Z56} ergibt sich eine geringf"ugige Modifikation, da
der Anteil, der im Programm des Unterkapitels~\ref{E.Kap.11.1} mit
\mbox{$(F\!-\!1\!-\!A_0)\CdoT\Omega/2$} ansteigt, beim Grenz"ubergang
\mbox{$M\!\to\!\infty$} mit der Substitution nach Gleichung~(\ref{E.10.26})
den Term \mbox{$\omega/2$} liefert, der bei den zu bestimmenden Frequenzpunkten
gleich \mbox{$\nu\CdoT\pi$} ist. 

Die Normierung bei der Berechnung
des Betragsquadrates der Fourierreihenkoeffizienten in den
Zeilen~\ref{P6Z62} und \ref{P6Z63} weist nun zwei Ver"anderungen auf.
Wenn man ber"ucksichtigt, dass bei dem Quadrat des Polstellenabstands
immer das Quadrat der Sinusfunktion des {\em halben}\/ Differenzwinkels
auftritt, so ergibt sich mit der Substitution nach Gleichung~(\ref{E.10.26})
bei der im Unterkapitel~\myref{MatFen1} beschrieben Wahl des
Normierungsfaktors \verb|NF_1| der in Zeile~\ref{P6Z62} angegebene Wert,
bei dem nun statt \verb|log(8/pi*F)| der Term \verb|log(8)| im Exponenten
auftritt. Bei den Normierungsfaktoren des Vektors \verb|NF_0| ist 
in Zeile~\ref{P6Z63} nun nicht mehr das Quadrat des Abstands der rotierten
Nullstellen $z_{0,\rho}$ zum Punkt \mbox{$z\!=\!1$} zu verwenden, sondern
das Quadrat des Abstands der verschobenen Nullstellen $s_{0,\rho}$ zum
Punkt \mbox{$s\!=\!0$}. Dementsprechend wurden bei der Berechnung
der Faktoren des kumulativen Produkts in den Zeilen~\ref{P6Z70} und
\ref{P6Z73} die Abstandsquadrate, die sich bei der diskreten Fensterfolge
mit Hilfe der Quadrate der Sinusfunktion angeben lassen, durch die
entsprechenden Abstandsquadrate ersetzt, die sich ohne die Sinusfunktion
berechnen. Es sei noch angemerkt, dass sich im Spezialfall mit
\mbox{$A_0=N\!-\!1$} die im Nenner auftretenden Produkte der
Polstellenabst"ande durch Erweiterung mit \mbox{${\D(N-1)!^2}$}
und bei einer anderen Normierung als mit \verb|NF_1| als die Quadrate
von Binomialkoeffizienten schreiben lassen. Dies erkl"art die in
Gleichung~(\ref{E.10.39}) angegebene Formel f"ur die Betragsquadrate
der Fourierreihenkoeffizienten. Die abschlie"sende Normierung in
Zeile~\ref{P6Z78} liefert die Fourierreihenkoeffizienten in der Art,
dass sich f"ur den Gleichanteil \verb|F_nu(1)| der Wert Eins ergibt.

\section{Die kontinuierlichen Fensterfunktionen}\label{E.Kap.11.5}

Mit Hilfe der im vorigen Unterkapitel berechneten Fourierreihenkoeffizienten
\verb|F_nu| l"asst sich die kontinuierliche Fensterfunktion $f(t)$
mit dem folgenden Programm f"ur beliebige Zeitpunkte $t$ des Intervalls
\mbox{$[0;1)$}, die als Vektor \verb|t| zu "ubergeben sind, berechnen.
{\renewcommand{\labelenumi}{{\footnotesize\arabic{enumi}:}}
\begin{enumerate}
\setlength{\itemsep}{-7pt plus1pt minus0pt}
\setlength{\parsep}{0pt plus1pt minus0pt}
\item\verb|function f_t = fenster( t, F_nu )|\label{P7Z0}\begin{npb}
\item\verb|f_t = zeros(1,length(t))|\label{P7Z1}
\item\verb|N_A = length(F_nu)|\label{P7Z2}
\item\verb|O_t = 2*pi*t|\label{P7Z3}\end{npb}
\item\verb|for nu = N_A-1:-1:1|\label{P7Z4}\begin{npb}
\item\verb|  f_t = f_t + 2 * real(F_nu(nu+1)) * cos(O_t*nu) - ...|\label{P7Z5}
\item[]\verb|              2 * imag(F_nu(nu+1)) * sin(O_t*nu)|
\item\verb|end|\label{P7Z6}
\item\verb|f_t = f_t + F_nu(1)|\label{P7Z7}\end{npb}
\end{enumerate}}
Da hier die Berechnung im wesentlichen so abl"auft, wie dies 
bei der zweiten Version des in Unterkapitel~\ref{E.Kap.11.1}
angegebenen Programms beschriebenen ist, bedarf es bei diesem
Programm keines weiteren Kommentars.

\section[Die Spektren der kontinuierlichen Fensterfunktionen]{Die Spektren der kontinuierlichen \\Fensterfunktionen}\label{E.Kap.11.6}

Mit dem folgenden Programm l"asst sich das Spektrum der kontinuierlichen
Fensterfunktion bis zu sehr hohen Frequenzen mit fast der maximal m"oglichen
Genauigkeit berechnen. Neben den im Unterkapitel~\ref{E.Kap.11.4} beschriebenen
Parametern ben"otigt dieses Programm die Frequenzen $\omega$, f"ur
die das Spektrum zu berechnen ist, als Zeilenvektor \verb|omega|.
{\renewcommand{\labelenumi}{{\footnotesize\arabic{enumi}:}}
\begin{enumerate}
\setlength{\itemsep}{-7pt plus1pt minus0pt}
\setlength{\parsep}{0pt plus1pt minus0pt}
\item\verb|function F_omega = spektrum( N, A_0, A_1, s_0, omega )|\label{P8Z0}\begin{npb}
\item\verb|c = (N/2)^(4/3)|\label{P8Z1}
\item\verb|Ms = -log(eps)/log(2) * 2^log(N/3)|\label{P8Z2}
\item\verb|Ms = 2^ceil(log(Ms)/log(2))|\label{P8Z3}
\item\verb|Ms = max(Ms,16)|\label{P8Z4}\end{npb}
\item\verb|Ms_OK = 0|\label{P8Z5}\begin{npb}
\item\verb|while ~Ms_OK|\label{P8Z6}\end{npb}
\item\verb|  eta = 0:Ms/2|\label{P8Z7}\begin{npb}
\item\verb|  F_eta = zeros(1,Ms/2+1)|\label{P8Z8}\end{npb}
\item\verb|  NF_1 = 1 / ( 1 + ( (N-1-A_1)/c )^2 )|\label{P8Z9}\begin{npb}
\item\verb|  NF_0 = 1 ./ max( [ ( abs( s_0 + j*pi*(1-N) + 2*pi*c ) + ...|\label{P8Z10}
\item[]\verb|                       abs( s_0 + j*pi*(1-N) - 2*pi*c ) ) ; ...|
\item[]\verb|                     ( abs( s_0 + j*pi*(N-1) + 2*pi*c ) + ...|
\item[]\verb|                       abs( s_0 + j*pi*(N-1) - 2*pi*c ) ) ] ).^2|\end{npb}
\item\verb|  for nu_1 = (1-N)/2:(N-1)/2|\label{P8Z11}\begin{npb}
\item\verb|    F_eta_1 = ones(1,Ms/2+1)|\label{P8Z12}\end{npb}
\item\verb|    rho = 0|\label{P8Z13}
\item\verb|    for nu_2 = [(1-N+A_1):(nu_1+A_1-(N+1)/2), ...|\label{P8Z14}\begin{npb}
\item[]\verb|                (nu_1-A_1+(N+1)/2):(N-1-A_1)]|
\item\verb|      K_1 = NF_1 * ( 1 + (  nu_2 / c )^2 )|\label{P8Z15}
\item\verb|      Psi_1 = atan( nu_2 / c )|\label{P8Z16}
\item\verb|      F_eta_1 = F_eta_1 .* K_1 .* sin( pi/Ms*eta - Psi_1 ).^2|\label{P8Z17}\end{npb}
\item\verb|      rho = rho + 1|\label{P8Z18}\begin{npb}
\item\verb|      if rho < N-0.5-2*A_1-A_0|\label{P8Z19}
\item\verb|        Z_Z = 2*pi*c + s_0(rho) + j*2*pi*nu_1|\label{P8Z20}
\item\verb|        Z_N = 2*pi*c - s_0(rho) - j*2*pi*nu_1|\label{P8Z21}
\item\verb|        abs_Z_Z = abs( Z_Z )|\label{P8Z22}
\item\verb|        abs_Z_N = abs( Z_N )|\label{P8Z23}
\item\verb|        F_eta_1 = F_eta_1 .* NF_0(rho) .* ...|\label{P8Z24}
\item[]\verb|                  ( (abs_Z_N-abs_Z_Z)^2 + 4 * abs_Z_N * abs_Z_Z * ...|\makebox[-1pt]{}
\item[]\verb|                    sin( pi/Ms*eta-(angle(Z_Z)-angle(Z_N))/2 ).^2 )|
\item\verb|      end|\label{P8Z25}
\item\verb|    end|\label{P8Z26}\end{npb}
\item\verb|    F_eta = F_eta + F_eta_1|\label{P8Z27}\begin{npb}
\item\verb|  end|\label{P8Z28}\end{npb}\vadjust{\penalty-100}
\item\verb|  NF = 1 / sqrt( max(F_eta) * min(F_eta) )|\label{P8Z29}\begin{npb}
\item\verb|  F_eta = NF * F_eta|\label{P8Z30}\end{npb}
\item\verb|  L_eta = log(F_eta)|\label{P8Z31}
\item\verb|  Ceps_2 = ifft( [L_eta,L_eta(Ms/2:-1:2)] )|\label{P8Z32}\begin{npb}
\item\verb|  Ceps_2 = real( Ceps_2 )|\label{P8Z33}
\item\verb|  Ceps_2 = ( Ceps_2 + Ceps_2([1,Ms:-1:2]) ) / 2|\label{P8Z34}\end{npb}\vadjust{\penalty-100}
\item\verb|  sockel = eps / Ms / sqrt(48) * ...|\label{P8Z35}\begin{npb}
\item[]\verb|           sqrt( sum( max( [L_eta,L_eta(Ms/2:-1:2)].^2, 1 ) ) )|\end{npb}
\item\verb|  grenze = ( 2 * (2*N-2-2*A_1-A_0) ) ./ [1:Ms/2] .* ...|\label{P8Z36}\begin{npb}
\item[]\verb|           ( Ms * sockel/2/(2*N-2-2*A_1-A_0) ).^( 2*[1:Ms/2]/Ms )|\end{npb}
\item\verb|  max_fehl = sum( max( abs([L_eta,L_eta(Ms/2:-1:2)]), 1 ) ) * ...|\label{P8Z37}\begin{npb}
\item[]\verb|             eps * ( 2 + log(Ms)/log(2) ) / Ms|\end{npb}
\item\verb|  grenze = max( grenze, max_fehl )|\label{P8Z38}
\item\verb|  if any( abs(Ceps_2(2:Ms/2+1)) > grenze )|\label{P8Z39}\begin{npb}
\item\verb|    kriterium = abs( fft( Ceps_2(Ms/4+1:Ms/2) .* [Ms/4:Ms/2-1] ) )|\label{P8Z40}
\item\verb|    [dummy,womax] = max( kriterium(1:Ms/8+1) )|\label{P8Z41}
\item\verb|    delta_c = 1-(womax-1)*16/Ms|\label{P8Z42}
\item\verb|    delta_c = sin( delta_c*pi*(1-delta_c^2/2) )/3|\label{P8Z43}\end{npb}
\item\verb|    c = c * (1-delta_c) / (1+delta_c)|\label{P8Z44}
\item\verb|    Ms = 2*Ms|\label{P8Z45}
\item\verb|  else|\label{P8Z46}
\item\verb|    Ms_OK = 1|\label{P8Z47}\begin{npb}
\item\verb|  end|\label{P8Z48}
\item\verb|end|\label{P8Z49}\end{npb}\vadjust{\penalty-200}
\item\verb|len_o = length(omega)|\label{P8Z50}
\item\verb|Omega_s = 2 * atan(omega/(2*pi*c))|\label{P8Z51}
\item\verb|phi = zeros(1,len_o)|\label{P8Z52}
\item\verb|for nu_i = 1:len_o|\label{P8Z53}\begin{npb}
\item\verb|  phi(nu_i) = sin(Omega_s(nu_i)*[Ms/2-1:-1:1]) * Ceps_2(Ms/2:-1:2).'|\label{P8Z54}
\item\verb|end|\label{P8Z55}\end{npb}
\item\verb|phi = phi - ( N - 1 - A_1 - A_0/2 ) * Omega_s + omega/2|\label{P8Z56}\vadjust{\penalty-200}
\item\verb|F_omega = zeros(1,len_o+1)|\label{P8Z57}
\item\verb|if N == 2*A_1+1|\label{P8Z58}\begin{npb}
\item\verb|  NF_1 = 1|\label{P8Z59}
\item\verb|else|\label{P8Z60}
\item\verb|  NF_1 = exp(2*log(8)-sum(log([5:2:4*N-8*A_1]))/(N-1-2*A_1))|\label{P8Z61}
\item\verb|end|\label{P8Z62}\end{npb}
\item\verb|NF_0 = 1 ./ max( [ abs( s_0 + j*pi*(N-1) ).^2 ; ...|\label{P8Z63}\begin{npb}
\item[]\verb|                   abs( s_0 - j*pi*(N-1) ).^2 ] )|\end{npb}\vadjust{\penalty-200}
\item\verb|for nu_1 = (1-N)/2:(N-1)/2|\label{P8Z64}\begin{npb}
\item\verb|  omega_nu_1 = [omega,0]/(2*pi) - nu_1|\label{P8Z65}\end{npb}
\item\verb|  nu_4 = 2 * round( omega_nu_1 + (N-1)/2 ) - N + 1|\label{P8Z66}\begin{npb}
\item\verb|  nu_4 = ( 1-N+2*A_1 ) .* ( nu_4 < 2*A_1-N ) + ...|\label{P8Z67}
\item[]\verb|         nu_4 .* ( abs(nu_4) < N-2*A_1 ) + ...|
\item[]\verb|         ( N-1-2*A_1 ) .* ( nu_4 > N-2*A_1 )|
\item\verb|  nu_4 = nu_4/2|\label{P8Z68}\end{npb}
\item\verb|  d_omega = omega_nu_1 - nu_4|\label{P8Z69}
\item\verb|  F_omega_1 = ( abs(d_omega) < eps )|\label{P8Z70}\begin{npb}
\item\verb|  F_omega_1 = F_omega_1 / eps^2 + (~F_omega_1) .* ...|\label{P8Z71}
\item[]\verb|              (         sin(pi*d_omega) ./ ...|
\item[]\verb|                ( pi*eps*d_omega + F_omega_1 ) ).^2|\end{npb}\vadjust{\penalty-100}
\item\verb|  rho = 0|\label{P8Z72}\begin{npb}
\item\verb|  for nu_2 = (1-N)/2+A_1:(N-3)/2-A_1|\label{P8Z73}
\item\verb|    rho = rho + 1|\label{P8Z74}
\item\verb|    if rho < N-0.5-2*A_1-A_0|\label{P8Z75}
\item\verb|      F_omega_1 = F_omega_1 .* NF_0(rho) .* ...|\label{P8Z76}
\item[]\verb|                  abs( j*2*pi*omega_nu_1 - s_0(rho) ).^2|
\item\verb|    end|\label{P8Z77}\end{npb}
\item\verb|    nu_3 = nu_2 + ( nu_2 >= nu_4 )|\label{P8Z78}\begin{npb}
\item\verb|    F_omega_1 = F_omega_1 ./ ( NF_1 .* ( omega_nu_1 - nu_3 ) .^ 2 )|\label{P8Z79}
\item\verb|  end|\label{P8Z80}\end{npb}
\item\verb|  F_omega = F_omega + F_omega_1|\label{P8Z81}\begin{npb}
\item\verb|end|\label{P8Z82}\end{npb}
\item\verb|omega_nu_1 = abs( [omega,0] / (2*pi) ) - N + A_1 + 0.5|\label{P8Z83}\begin{npb}
\item\verb|omega_nu_1 = 0.5 * ( omega_nu_1 > 0 ) .* omega_nu_1 + 0.25|\label{P8Z84}
\item\verb|omega_nu_1 = 1 - 2 * ( ( omega_nu_1 - floor(omega_nu_1) ) > 0.5 )|\label{P8Z85}
\item\verb|F_omega = omega_nu_1 .* sqrt(F_omega)|\label{P8Z86}
\item\verb|F_omega = F_omega(1:len_o) / F_omega(len_o+1)|\label{P8Z87}
\item\verb|F_omega =  F_omega .* exp(-j*phi)|\label{P8Z88}\end{npb}
\end{enumerate}}
Die Berechnung des Cepstrums in den Zeilen~\ref{P8Z1} bis \ref{P8Z49}
erfolgt identisch, wie bei der in Unterkapitel~\ref{E.Kap.11.4} beschriebenen
Berechnung der Fourierreihenkoeffizienten des kontinuierlichen Fensters.
Wenn man mit der Substitution nach Gleichung~(\ref{E.10.26}) den
Grenz"ubergang \mbox{$M\!\to\!\infty$} betrachtet, erh"alt man aus
dem in Unterkapitel~\ref{E.Kap.11.2} beschriebenen Programm das hier
vorliegende Programm zur Berechnung des Spektrums des kontinuierlichen
Fensters f"ur beliebige Frequenzen. Die Kommentare der beiden
ebengenannten Unterkapitel lassen sich daher auf dieses Programm
sinngem"a"s "ubertragen. Eine Besonderheit ist lediglich dadurch gegeben,
dass teilweise die auf $2\pi$ normierte Kreisfrequenz ---\,also das, was
man "ublicherweise als Frequenz bezeichnet\,--- verwendet wird.

\section[Die kontinuierlichen Fensterautokorrelationsfunktionen]{Die kontinuierlichen Fensterautokorrelations-\\funktionen}\label{E.Kap.11.7}

Dieses Programm ben"otigt die Zeitpunkte $t$, f"ur die die
Fensterautokorrelationsfunktion berechnet werden soll, als Vektor \verb|t|.
Desweiteren muss der Zeilenvektor \verb|F_nu| die Fourierreihenkoeffizienten
\mbox{$F_{\infty}(\nu\CdoT2\pi)$} der Fensterfunktion f"ur
\mbox{$\nu=0\;(1)\;N\!-\!A_1\!-\!1$} enthalten. Diesen Vektor erh"alt man
bei dem Programm in Unterkapitel~\ref{E.Kap.11.4} als Ergebnis.
{\renewcommand{\labelenumi}{{\footnotesize\arabic{enumi}:}}
\begin{enumerate}
\setlength{\itemsep}{-7pt plus1pt minus0pt}
\setlength{\parsep}{0pt plus1pt minus0pt}
\item\verb|function d_t = fenster_akf( t, F_nu )|\label{P9Z0}\begin{npb}
\item\verb|N_A = length(F_nu)|\label{P9Z1}
\item\verb|t = abs(t)|\label{P9Z2}
\item\verb|O_t = 2*pi*t|\label{P9Z3}\end{npb}
\item\verb|d_t = zeros(1,length(t))|\label{P9Z4}
\item\verb|for nu = N_A:-1:2|\label{P9Z5}\begin{npb}
\item\verb|  F_sin = real( F_nu(nu)' * ...|\label{P9Z6}
\item[]\verb|          sum( [F_nu(N_A:-1:1)';F_nu([2:nu-1,nu+1:N_A]).'] ./ ...|
\item[]\verb|               [2-N_A-nu:-1,1:N_A-nu].' ) ) / pi|
\item\verb|  d_t = d_t + 2 * F_sin * sin(O_t*(nu-1)) + ...|\label{P9Z7}
\item[]\verb|              2 * abs(F_nu(nu))^2 * cos(O_t*(nu-1)) .* (1-t)|
\item\verb|end|\label{P9Z8}
\item\verb|d_t = d_t + abs(F_nu(1))^2 * (1-t)|\label{P9Z9}\end{npb}
\end{enumerate}}
In der Zeile~\ref{P5Z1} wird der ben"otigte Parameter \verb|N_A|
als die Differenz \mbox{$N\!-\!A_1$} bestimmt. Da die Fourierreihe der
Fensterfunktion nur Anteile bis zum \mbox{$(N\!-\!A_1\!-\!1)$}-fachen der
Grundkreisfrequenz enth"alt, kann der Wert f"ur \verb|N_A| als die L"ange
des Vektors \verb|F_nu| bestimmt werden. Da es sich bei der
Fensterautokorrelationsfunktion um eine geradesymmetrische Funktion handelt,
kann man f"ur negatives $t$ ebensogut die Werte f"ur $-t$ berechnen.
Durch Zeile~\ref{P9Z2} erspart man sich so eine Fallunterscheidung bei der
sp"ateren Berechnung der Fensterautokorrelationsfunktion. Diese Berechnung
wird mit Hilfe der Formeln~(\ref{E.10.54}) und (\ref{E.10.55}) als "Uberlagerung
einer Sinusreihe und einer mit der Dreiecksfunktion multiplizierten
Kosinusreihe vorgenommen, wie dies am Ende von Kapitel~\ref{E.Kap.10.7}
beschrieben ist. Dabei wird bei jedem Reihenglied die mit $t$ multiplizierte
Grundkreisfrequenz ben"otigt, die daher in Zeile~\ref{P9Z3} f"ur
alle Werte des Vektors \verb|t| auf dem Vektor \verb|O_t| bereitgestellt
wird. Die Summe in Gleichung~(\ref{E.10.55}) wird als \verb|for|-Schleife,
die in Zeile~\ref{P9Z5} beginnt und in Zeile~\ref{P9Z8} endet, realisiert,
indem zun"achst in Zeile~\ref{P9Z4} der Vektor \verb|d_t| mit null
initialisiert wird, zu dem in Zeile~\ref{P9Z7} bei jedem Schleifendurchlauf
ein Anteil der Sinusreihe und ein Anteil der mit der Dreiecksfunktion
multiplizierten Kosinusreihe addiert wird. In Zeile~\ref{P9Z6} wird
der dazu ben"otigte Sinusreihenkoeffizient nach Gleichung~(\ref{E.10.54})
berechnet. Die dabei auftretende Summe wird mit dem \verb|MATLAB|-Befehl
\verb|sum| berechnet.

\renewcommand{\thechapter}{}
\chapter{Literaturverzeichnis}


\chapter{Anhang}\label{E.Kap.A}
\renewcommand{\thechapter}{A}
\section[Identit"at der L"osungsr"aume der Gleichungssysteme~(\ref{E.2.21})
und (\ref{E.2.22})]{Identit"at der L"osungsr"aume der Gleichungs-\\
systeme~(\ref{E.2.21}) und (\ref{E.2.22})}\label{E.Kap.A.1}
Die rechten Seiten aller Gleichungen des Gleichungssystems~(\ref{E.2.21})
kann man umformen zu:
\begin{gather}
\text{E}\Big\{\big(\boldsymbol{y}(k)\!-\!
\text{E}\{\boldsymbol{y}(k)\}\big)\CdoT
\big(\Tilde{\Vec{\boldsymbol{V}}}(\Tilde{\mu})\!-\!
\text{E}\{\Tilde{\Vec{\boldsymbol{V}}}(\Tilde{\mu})\}\big)^{\HH}\Big\}\;={}
\label{E.A.1}\\*[4pt]
{}=\;\text{E}\Big\{
\boldsymbol{y}(k)\CdoT
\Tilde{\Vec{\boldsymbol{V}}}(\Tilde{\mu})^{\Hh}\!\!-\!
\boldsymbol{y}(k)\CdoT
\text{E}\{\Tilde{\Vec{\boldsymbol{V}}}(\Tilde{\mu})\}^{\Hh}\!\!-\!
\text{E}\{\boldsymbol{y}(k)\}\CdoT
\Tilde{\Vec{\boldsymbol{V}}}(\Tilde{\mu})^{\Hh}\!\!+\!
\text{E}\{\boldsymbol{y}(k)\}\CdoT
\text{E}\{\Tilde{\Vec{\boldsymbol{V}}}(\Tilde{\mu})\}^{\Hh}\Big\}\;={}
\notag\\[4pt]\begin{flalign*}
&{}=\;\text{E}\Big\{\boldsymbol{y}(k)\CdoT
\Tilde{\Vec{\boldsymbol{V}}}(\Tilde{\mu})^{\Hh}\Big\}\!-\!
\text{E}\Big\{\boldsymbol{y}(k)\CdoT
\text{E}\{\Tilde{\Vec{\boldsymbol{V}}}(\Tilde{\mu})\}^{\Hh}\Big\}-{}&&
\end{flalign*}\notag\\*\begin{flalign*}
&&{}-\text{E}\Big\{\text{E}\{\boldsymbol{y}(k)\}\CdoT
\Tilde{\Vec{\boldsymbol{V}}}(\Tilde{\mu})^{\Hh}\Big\}\!+\!
\text{E}\Big\{\text{E}\{\boldsymbol{y}(k)\}\CdoT
\text{E}\{\Tilde{\Vec{\boldsymbol{V}}}(\Tilde{\mu})\}^{\Hh}\Big\}\;={}&
\end{flalign*}\notag\\[4pt]
{}=\text{E}\Big\{\boldsymbol{y}(k)\CdoT
\Tilde{\Vec{\boldsymbol{V}}}(\Tilde{\mu})^{\Hh}\Big\}\!-\!
\text{E}\{\boldsymbol{y}(k)\}\CdoT
\text{E}\{\Tilde{\Vec{\boldsymbol{V}}}(\Tilde{\mu})\}^{\Hh}\!\!-\!
\text{E}\{\boldsymbol{y}(k)\}\CdoT
\text{E}\{\Tilde{\Vec{\boldsymbol{V}}}(\Tilde{\mu})\}^{\Hh}\!\!+\!
\text{E}\{\boldsymbol{y}(k)\}\CdoT
\text{E}\{\Tilde{\Vec{\boldsymbol{V}}}(\Tilde{\mu})\}^{\Hh}={}
\notag\\[4pt]
{}=\text{E}\Big\{\boldsymbol{y}(k)\CdoT
\Tilde{\Vec{\boldsymbol{V}}}(\Tilde{\mu})^{\Hh}\Big\}\!-\!
\text{E}\{\boldsymbol{y}(k)\}\CdoT
\text{E}\{\Tilde{\Vec{\boldsymbol{V}}}(\Tilde{\mu})\}^{\Hh}=\,
\text{E}\Big\{\boldsymbol{y}(k)\CdoT
\Tilde{\Vec{\boldsymbol{V}}}(\Tilde{\mu})^{\Hh}\Big\}\!-\!
\text{E}\Big\{\boldsymbol{y}(k)\CdoT
\text{E}\{\Tilde{\Vec{\boldsymbol{V}}}(\Tilde{\mu})\}^{\Hh}\Big\}={}
\notag\\[4pt]
{}=\;\text{E}\Big\{\boldsymbol{y}(k)\CdoT
\Tilde{\Vec{\boldsymbol{V}}}(\Tilde{\mu})^{\Hh}\!\!-\!
\boldsymbol{y}(k)\CdoT
\text{E}\{\Tilde{\Vec{\boldsymbol{V}}}(\Tilde{\mu})\}^{\Hh}\Big\}\;=\;
\text{E}\Big\{\boldsymbol{y}(k)\CdoT
\big(\Tilde{\Vec{\boldsymbol{V}}}(\Tilde{\mu})\!-\!
\text{E}\{\Tilde{\Vec{\boldsymbol{V}}}(\Tilde{\mu})\}\big)^{\HH}\Big\}
\notag\\*[4pt]
\forall\qquad\qquad k=0\;(1)\;F\!-\!1
\qquad\text{ und }\qquad\Tilde{\mu}=0\;(1)\;M\!-\!1.
\notag
\end{gather}
Da auf den linken Seiten aller Gleichungen der Gleichungssysteme~(\ref{E.2.21}) 
der Term \mbox{$e^{j\cdot\frac{2\pi}{M}\cdot\mu\cdot k}$} der einzige ist, 
der von $k$ abh"angt, steht auf der linken Seite f"ur jeden festen Wert von 
\mbox{$\Tilde{\mu}$} eine in $k$ mit $M$ periodische Folge von Zeilenvektoren. 
Die Anzahl der Elemente dieser Zeilenvektoren ist gleich der Anzahl der 
Elemente des Zufallsvektors \mbox{$\Tilde{\Vec{\boldsymbol{V}}}(\Tilde{\mu})$} 
und sei mit $R$ bezeichnet. Wenn nun f"ur diesen  festen Wert \mbox{$\Tilde{\mu}$}
auf der rechten Seite der Gleichungssysteme~(\ref{E.2.21}) keine in $k$ mit $M$ 
periodische Folge von Zeilenvektoren steht, existiert keine L"osung f"ur die Werte 
der beiden bifrequenten "Ubertragungsfunktionen. In diesem Fall macht es keinen 
Sinn, das reale System durch die beiden linearen und periodisch zeitvarianten 
Modellsysteme zu modellieren, und die "Ubertragungsfunktionen mit Hilfe des RKM 
messtechnisch abzusch"atzen. Daher wollen wir uns auf den Fall beschr"anken, 
dass eine L"osung existiert, was nur sein kann, wenn f"ur diesen festen Wert 
\mbox{$\Tilde{\mu}$} auch auf den rechten Seiten der Gleichungssysteme~(\ref{E.2.21})
eine in $k$ mit $M$ periodische Folge von Zeilenvektoren steht. Alle Gleichungen,
 bei denen sich $k$ um ein ganzzahliges Vielfaches von $M$ unterscheidet, 
w"ahrend \mbox{$\Tilde{\mu}$} gleich ist, sind somit identisch. Wenn wir nun 
eine Linearkombination dieser identischen Gleichungen bilden, wobei die Summe 
der Koeffizienten dieser Linearkombination von Null verschieden ist, "andert 
sich der L"osungsraum des Gleichungssystems nicht. Bevor wir diese
Linearkombination durchf"uhren, ersetzen wir zun"achst die
diskrete Zeitvariable $k$ durch \mbox{$k=\Tilde{k}\!+\!\kappa\CdoT M$}
mit \mbox{$\Tilde{k}=0\;(1)\;M\!-\!1$}, so dass alle Gleichungen
mit gleichem $\Tilde{k}$ f"ur alle Werte von $\kappa$ identisch sind, sofern 
die Zeitpunkte $k$ im Intervall \mbox{$[0,F\!-\!1]$} liegen.
Die Linearkombination bilden wir, indem wir die Gleichungen
mit den Werten der Fensterfolge \mbox{$f(\Tilde{k}\!+\!\kappa\CdoT M)$}
gewichten, und "uber alle $\kappa$ aufsummieren. Da die Fensterfolge f"ur
\mbox{$k\notin[0,F\!-\!1]$} nach Gleichung~(\myref{2.15}) Null ist, sind
die Gleichungen mit \mbox{$k\notin[0,F\!-\!1]$} nach der Gewichtung durch die
Fensterfolge auch immer erf"ullt, so dass \mbox{$\kappa\in\mathbb{Z}$}
gew"ahlt werden kann. Aufgrund der nach Gleichung~(\myref{2.27}) geforderten
Nullstellenlage des Spektrums der Fensterfolge ist die Summe
\mbox{$\sum_{\kappa=-\infty}^{\infty}\!f(\Tilde{k}\!+\!\kappa\CdoT M)$} 
immer gleich $1$, so dass der L"osungsraum des Gleichungssystems durch 
diese Modifikation nicht ver"andert wird. Wir erhalten $M^2$ Gleichungen,
die jeweils auf beiden Seiten einen Zeilenvektor der Dimension
\mbox{$1\!\times\!R$} stehen haben.
\begin{gather}
\begin{flalign*}
&\frac{1}{M}\cdoT\Sum{\mu=0}{M-1}\Vec{H}(\mu)\cdot
\text{E}\Big\{
\big(\Tilde{\Vec{\boldsymbol{V}}}(\mu)\!-\!
\text{E}\{\Tilde{\Vec{\boldsymbol{V}}}(\mu)\}\big)\CdoT
\big(\Tilde{\Vec{\boldsymbol{V}}}(\Tilde{\mu})\!-\!
\text{E}\{\Tilde{\Vec{\boldsymbol{V}}}(\Tilde{\mu})\}\big)^{\HH}\Big\}\cdot{}&&
\end{flalign*}\notag\\*[-6pt]\begin{flalign*}
&&{}\cdoT\underbrace{
\Sum{\kappa=-\infty}{\infty}\!f(\Tilde{k}\!+\!\kappa\CdoT M)\cdot
e^{j\cdot\frac{2\pi}{M}\cdot\mu\cdot\kappa\cdot M}}_{=1}\cdot\,
e^{j\cdot\frac{2\pi}{M}\cdot\mu\cdot\Tilde{k}}\;=&
\end{flalign*}\notag\\[4pt]
=\;\text{E}\bigg\{\;\Sum{\kappa=-\infty}{\infty}\!
f(\Tilde{k}\!+\!\kappa\CdoT M)\cdot
\boldsymbol{y}(\Tilde{k}\!+\!\kappa\CdoT M)\cdot
\big(\Tilde{\Vec{\boldsymbol{V}}}(\Tilde{\mu})-
     \text{E}\{\Tilde{\Vec{\boldsymbol{V}}}(\Tilde{\mu})\}\big)^{\HH}\,\bigg\}
\notag\\*[4pt]
\label{E.A.2}
\forall\qquad\qquad \Tilde{k}=0\;(1)\;M\!-\!1
\qquad\text{ und }\qquad \Tilde{\mu}=0\;(1)\;M\!-\!1.
\end{gather}
F"ur jeden festen Wert von $\Tilde{\mu}$ haben wir ein Gleichungssystem
mit je $M$ Gleichungen, die wir nun diskret fouriertransformieren
indem wir jeweils die $\Tilde{k}$-te Gleichung mit dem Drehfaktor 
\mbox{$e^{\!-j\cdot\frac{2\pi}{M}\cdot\Bar{\mu}\cdot\Tilde{k}}$}
multiplizieren und alle Gleichungen mit gleichem $\Tilde{\mu}$
"uber $\Tilde{k}$ aufsummieren. Wenn wir f"ur die diskrete Frequenzvariable 
$\Bar{\mu}$ im Exponenten des Drehfaktors \mbox{$\Bar{\mu}=0\;(1)\;M\!-\!1$}
w"ahlen, entspricht das ---\,abgesehen von dem konstanten und von Null
verschiedenen Faktor $\sqrt{M\,}\!$\,--- jeweils einer unit"aren
Transformation der $M$ Gleichungen mit gleichem $\Tilde{\mu}$,
so dass sich die L"osungsr"aume der Gleichungssysteme mit jeweils
festem $\Tilde{\mu}$ nicht "andern, und somit auch nicht der
L"osungsraum aller $M^2$ Gleichungssysteme. Wenn man ber"ucksichtigt,
dass die dabei auf den linken Seiten der Gleichungen
entstehenden Summen "uber die Exponentialfunktionen nur f"ur
\mbox{$\mu=\Bar{\mu}$} von Null verschieden ---\,n"amlich $M$\,--- sind,
erkennt man, dass von der Summe "uber $\mu$ nur jeweils nur der Summand
mit \mbox{$\mu=\Bar{\mu}$} "ubrig bleibt. Wenn man anschlie"send noch den
Index \mbox{$\Bar{\mu}$} durch den Index \mbox{$\mu$}, der dann ja nicht
mehr in den Gleichungen auftritt, substituiert ergeben sich wieder $M^2$
Gleichungssysteme, bei denen jeweils auf beiden Seiten ein Zeilenvektor
der Dimension \mbox{$1\!\times\!R$} steht.
\begin{gather}
\Vec{H}(\mu)\CdoT\text{E}\Big\{
\big(\Tilde{\Vec{\boldsymbol{V}}}(\mu)\!-\!
\text{E}\{\Tilde{\Vec{\boldsymbol{V}}}(\mu)\}\big)\CdoT
\big(\Tilde{\Vec{\boldsymbol{V}}}(\Tilde{\mu})\!-\!
\text{E}\{\Tilde{\Vec{\boldsymbol{V}}}(\Tilde{\mu})\}\big)^{\HH}\Big\}\;=\;
\text{E}\Big\{\boldsymbol{Y}_{\!\!\!f}(\mu)\CdoT
\big(\Tilde{\Vec{\boldsymbol{V}}}(\Tilde{\mu})\!-\!
\text{E}\{\Tilde{\Vec{\boldsymbol{V}}}(\Tilde{\mu})\}\big)^{\HH}\Big\}
\notag\\*[2pt]
\forall\qquad\qquad\mu=0\;(1)\;M\!-\!1
\qquad\text{ und }\qquad\Tilde{\mu}=0\;(1)\;M\!-\!1.
\label{E.A.3}
\end{gather}
Diese Gleichungssysteme enthalten f"ur jeden festen Wert vom $\mu$ 
genau $M$ Gleichungen mit unterschiedlichen Werten von $\Tilde{\mu}$. 
Im Gegensatz dazu weisen die Gleichungssysteme~(\ref{E.2.22}) 
f"ur jedes $\mu$ genau je {\em eine}\/ Gleichung mit einem 
\mbox{$1\!\times\!R$} Zeilenvektor auf jeder Seite auf. Diese 
Gleichung, ist eine der $M$ Gleichungen~(\ref{E.A.3}), n"amlich die, 
die man mit \mbox{$\Tilde{\mu}=\mu$} erh"alt. Daher kann der L"osungsraum 
des Gleichungssystems~(\ref{E.2.22}) allenfalls von einer h"oheren 
Dimension sein, wie der L"osungsraum des Gleichungssystems~(\ref{E.A.3}). 
Die Dimension des L"osungsraums der Vektorgleichung, die sich im 
Gleichungssystem~(\ref{E.2.22}) f"ur den festen Wert vom $\mu$ ergibt, 
ist gleich dem Rang der \mbox{$R\!\times\!R$} Kovarianzmatrix
\mbox{$\underline{C}_{\Tilde{\Vec{\boldsymbol{V}}}(\mu),\Tilde{\Vec{\boldsymbol{V}}}(\mu)}$}
des Zufallsvektors \mbox{$\Tilde{\Vec{\boldsymbol{V}}}(\mu)$} gem"a"s 
Gleichung (\ref{E.2.23}). Nach \cite{Cramer} liegt der Zufallsvektor 
\mbox{$\Tilde{\Vec{\boldsymbol{V}}}(\mu)\!-\!\text{E}\{\Tilde{\Vec{\boldsymbol{V}}}(\mu)\}$}
mit der Wahrscheinlichkeit Eins in einem Unterraum des $R$-dimensionalen
komplexen Raums, dessen Dimension gleich dem Rang der Kovarianzmatrix
\mbox{$\underline{C}_{\Tilde{\Vec{\boldsymbol{V}}}(\mu),\Tilde{\Vec{\boldsymbol{V}}}(\mu)}$} 
ist. Da sich jede Kovarianzmatrix mit einer
unit"aren Matrix $\underline{U}$ auf Diagonalform transformieren l"asst
(\,Hauptachsentransformation einer hermiteschen quadratischen Form\,),
kann man aus dem Zufallsvektor \mbox{$\Tilde{\Vec{\boldsymbol{V}}}(\mu)$}
durch eine unit"are Transformation einen neuen Zufallsvektor
\mbox{$\Acute{\Vec{\boldsymbol{V}}}(\mu)\;=\;
\underline{U}\CdoT\Tilde{\Vec{\boldsymbol{V}}}(\mu)$}
bilden, dessen Kovarianzmatrix
\begin{gather}
\underline{C}_{\Acute{\Vec{\boldsymbol{V}}}(\mu),\Acute{\Vec{\boldsymbol{V}}}(\mu)}
\;=\;\text{E}\Big\{
\big(\Acute{\Vec{\boldsymbol{V}}}(\mu)\!-\!
\text{E}\{\Acute{\Vec{\boldsymbol{V}}}(\mu)\}\big)\CdoT
\big(\Acute{\Vec{\boldsymbol{V}}}(\mu)\!-\!
\text{E}\{\Acute{\Vec{\boldsymbol{V}}}(\mu)\}\big)^{\HH}\Big\}\;=
\label{E.A.4}\\*[10pt]
=\;\text{E}\Big\{\big(\underline{U}\CdoT\Tilde{\Vec{\boldsymbol{V}}}(\mu)\!-\!
\text{E}\{\underline{U}\CdoT\Tilde{\Vec{\boldsymbol{V}}}(\mu)\}\big)\CdoT
\big(\underline{U}\CdoT\Tilde{\Vec{\boldsymbol{V}}}(\mu)\!-\!
\text{E}\{\underline{U}\CdoT\Tilde{\Vec{\boldsymbol{V}}}(\mu)\}\big)^{\HH}\Big\}\;=
\notag\\[10pt]
=\;\text{E}\Big\{\,\underline{U}\cdot
\big(\Tilde{\Vec{\boldsymbol{V}}}(\mu)\!-\!
\text{E}\{\Tilde{\Vec{\boldsymbol{V}}}(\mu)\}\big)\cdot
\big(\Tilde{\Vec{\boldsymbol{V}}}(\mu)\!-\!
\text{E}\{\Tilde{\Vec{\boldsymbol{V}}}(\mu)\}\big)^{\HH}\Cdot
\underline{U}^{\Hh}\Big\}\;=
\notag\\*[10pt]
=\;\underline{U}\cdot\text{E}\Big\{
\big(\Tilde{\Vec{\boldsymbol{V}}}(\mu)\!-\!
\text{E}\{\Tilde{\Vec{\boldsymbol{V}}}(\mu)\}\big)\CdoT
\big(\Tilde{\Vec{\boldsymbol{V}}}(\mu)\!-\!
\text{E}\{\Tilde{\Vec{\boldsymbol{V}}}(\mu)\}\big)^{\HH}
\Big\}\cdot\underline{U}^{\Hh}=\;
\underline{U}\CdoT\underline{C}_{\Tilde{\Vec{\boldsymbol{V}}}(\mu),\Tilde{\Vec{\boldsymbol{V}}}(\mu)}\CdoT\underline{U}^{\Hh}
\notag
\end{gather}
eine Diagonalmatrix ist, und deren Hauptdiagonalelemente die Varianzen
der neuen Zufallsgr"o"sen des transformierten Zufallsvektors
\mbox{$\Acute{\Vec{\boldsymbol{V}}}(\mu)$} sind. Da der Rang dieser
Diagonalmatrix gleich dem Rang der urspr"unglichen Kovarianzmatrix ist, ist
die Anzahl der transformierten Zufallsgr"o"sen, die die Varianz Null
aufweisen, gleich dem Rangdefekt der urspr"unglichen Kovarianzmatrix
\mbox{$\underline{C}_{\Tilde{\Vec{\boldsymbol{V}}}(\mu),\Tilde{\Vec{\boldsymbol{V}}}(\mu)}$}. 
Die Kovarianz jeder beliebigen Zufallsgr"o"se mit einer der Zufallsgr"o"sen 
\mbox{$\Acute{\Vec{\boldsymbol{V}}}(\mu)$} mit verschwindender Varianz 
ist immer Null. Nun fassen wir alle Zufallsgr"o"sen des Spektrums der 
Erregung, sowie die konjugierten Zufallsgr"o"sen des Spektrums der Erregung, 
die in den Zufallsspaltenvektoren \mbox{$\Vec{\boldsymbol{V}}(\Tilde{\mu})$} 
f"ur einen festen Wert vom $\mu$ f"ur alle \mbox{$\Tilde{\mu}=0\;(1)\;M\!-\!1$} 
in den Gleichungssystem~(\ref{E.A.3}) vorhanden sind, zu einem Zufallsspaltenvektor 
\mbox{$\Grave{\Vec{\boldsymbol{V}}}(\mu)$} zusammen. Die Anzahl der Elemente 
dieses Zufallsvektors bezeichnen wir mit $r$. Da die Gleichung des 
Gleichungssystems~(\ref{E.2.22}) f"ur den festen Wert den $\mu$ gerade 
eine der Gleichungen~(\ref{E.A.3}) ---\,n"amlich die mit \mbox{$\Tilde{\mu}=\mu$}\,--- 
ist, sind in dem Zufallsvektor \mbox{$\Grave{\Vec{\boldsymbol{V}}}(\mu)$} auch
die Zufallsgr"o"sen des Zufallsvektors \mbox{$\Tilde{\Vec{\boldsymbol{V}}}(\mu)$}
vorhanden. Wenn wir nun die \mbox{$R\!\times\!r$} Kovarianzmatrix
\begin{equation}
\underline{C}_{\Acute{\Vec{\boldsymbol{V}}}(\mu),\Grave{\Vec{\boldsymbol{V}}}(\mu)}
\;=\;\text{E}\Big\{
\big(\Acute{\Vec{\boldsymbol{V}}}(\mu)\!-\!
\text{E}\{\Acute{\Vec{\boldsymbol{V}}}(\mu)\}\big)\CdoT
\big(\Grave{\Vec{\boldsymbol{V}}}(\mu)\!-\!
\text{E}\{\Grave{\Vec{\boldsymbol{V}}}(\mu)\}\big)^{\HH}\Big\}
\label{E.A.5}
\end{equation}
betrachten, erkennen wir, dass alle Kovarianzen, die mit einer der Zufallsgr"o"sen 
\mbox{$\Acute{\Vec{\boldsymbol{V}}}(\mu)$} mit verschwindender Varianz gebildet 
werden, immer Null sind, so dass in den entsprechenden Zeilen von 
\mbox{$\underline{C}_{\Acute{\Vec{\boldsymbol{V}}}(\mu),\Grave{\Vec{\boldsymbol{V}}}(\mu)}$}
Nullvektoren mit $r$ Elementen zu finden sind. Weil die 
Anzahl der Nullvektoren gleich dem Rangdefekt der Kovarianzmatrix 
\mbox{$\underline{C}_{\Tilde{\Vec{\boldsymbol{V}}}(\mu),\Tilde{\Vec{\boldsymbol{V}}}(\mu)}$}
ist, und weil die Zufallsgr"o"sen des Zufallsvektors 
\mbox{$\Tilde{\Vec{\boldsymbol{V}}}(\mu)$} im Zufallsvektor 
\mbox{$\Grave{\Vec{\boldsymbol{V}}}(\mu)$} enthalten sind, 
ist der Rang dieser \mbox{$R\!\times\!r$} Kovarianzmatrix 
\mbox{$\underline{C}_{\Acute{\Vec{\boldsymbol{V}}}(\mu),\Grave{\Vec{\boldsymbol{V}}}(\mu)}$} 
gleich dem Rang der \mbox{$R\!\times\!R$} Kovarianzmatrix 
\mbox{$\underline{C}_{\Tilde{\Vec{\boldsymbol{V}}}(\mu),\Tilde{\Vec{\boldsymbol{V}}}(\mu)}$}
des Zufallsvektors $\Tilde{\Vec{\boldsymbol{V}}}(\mu)$.
Wenn man die \mbox{$R\!\times\!r$} Kovarianzmatrix
\mbox{$\underline{C}_{\Acute{\Vec{\boldsymbol{V}}}(\mu),\Grave{\Vec{\boldsymbol{V}}}(\mu)}$} 
von links mit dem konjugiert Transponierten $\underline{U}^{\HH}$ der 
unit"aren Transformationsmatrix multipliziert, erh"alt man eine 
\mbox{$R\!\times\!r$} Kovarianzmatrix desselben Rangs:
\begin{gather}
\underline{U}^{\Hh}\Cdot
\underline{C}_{\Acute{\Vec{\boldsymbol{V}}}(\mu),\Grave{\Vec{\boldsymbol{V}}}(\mu)}
\;=\;\underline{U}^{\Hh}\Cdot\text{E}\Big\{
\big(\Acute{\Vec{\boldsymbol{V}}}(\mu)\!-\!
\text{E}\{\Acute{\Vec{\boldsymbol{V}}}(\mu)\}\big)\CdoT
\big(\Grave{\Vec{\boldsymbol{V}}}(\mu)\!-\!
\text{E}\{\Grave{\Vec{\boldsymbol{V}}}(\mu)\}\big)^{\HH}\Big\}\;=
\notag\\*[10pt]
=\;\text{E}\Big\{
\big(\underline{U}^{\Hh}\!\CdoT
\Acute{\Vec{\boldsymbol{V}}}(\mu)\!-\!
\text{E}\{\underline{U}^{\Hh}\!\CdoT
\Acute{\Vec{\boldsymbol{V}}}(\mu)\}\big)\CdoT
\big(\Grave{\Vec{\boldsymbol{V}}}(\mu)\!-\!
\text{E}\{\Grave{\Vec{\boldsymbol{V}}}(\mu)\}\big)^{\HH}\Big\}\;=
\notag\\*[10pt]
=\;\text{E}\Big\{
\big(\Tilde{\Vec{\boldsymbol{V}}}(\mu)\!-\!
\text{E}\{\Tilde{\Vec{\boldsymbol{V}}}(\mu)\}\big)\CdoT
\big(\Grave{\Vec{\boldsymbol{V}}}(\mu)\!-\!
\text{E}\{\Grave{\Vec{\boldsymbol{V}}}(\mu)\}\big)^{\HH}\Big\}\;=\;
\underline{C}_{\Tilde{\Vec{\boldsymbol{V}}}(\mu),\Grave{\Vec{\boldsymbol{V}}}(\mu)}.
\label{E.A.6}
\end{gather}
Diese Kovarianzmatrix tritt im Gleichungssystem~(\ref{E.A.3}) auf, 
wenn man alle $M$ Gleichungen ---\,mit jeweils einem Zeilenvektor 
der Dimension \mbox{$1\!\times\!R$} auf beiden Seiten\,--- f"ur einen 
festen Wert vom $\mu$ und f"ur unterschiedliche Werte von $\Tilde{\mu}$ 
jeweils zu einer Gleichung mit einem Zeilenvektor der Dimension 
\mbox{$1\!\times\!r$} auf beiden Seiten zusammenfasst, wobei man 
eventuell mehrfach vorhandene, identische Gleichungen f"ur die Vektorelemente 
nur einfach ber"ucksichtigt, und die Reihenfolge der Elemente der 
Zeilenvektoren auf beiden Seiten der Gleichung in geeigneter Weise permutiert, 
so dass sich dieselbe Reihenfolge ergibt wie bei den Elementen 
des Zufallsvektors \mbox{$\Grave{\Vec{\boldsymbol{V}}}(\mu)^{\Hh}$}.
\begin{gather}
\Vec{H}(\mu)\cdot\text{E}\Big\{
\big(\Tilde{\Vec{\boldsymbol{V}}}(\mu)\!-\!
\text{E}\{\Tilde{\Vec{\boldsymbol{V}}}(\mu)\}\big)\CdoT
\big(\Grave{\Vec{\boldsymbol{V}}}(\mu)\!-\!
\text{E}\{\Grave{\Vec{\boldsymbol{V}}}(\mu)\}\big)^{\HH}\Big\}\;=\;
\text{E}\Big\{\boldsymbol{Y}_{\!\!\!f}(\mu)\CdoT
\big(\Grave{\Vec{\boldsymbol{V}}}(\mu)\!-\!
\text{E}\{\Grave{\Vec{\boldsymbol{V}}}(\mu)\}\big)^{\HH}\Big\}
\notag\\*[8pt]
\forall\qquad\mu=0\;(1)\;M\!-\!1
\label{E.A.7}
\end{gather}
Jede dieser $M$ Gleichungen mit dem Parameter $\mu$
besitzt daher denselben L"osungsraum wie die entsprechende Gleichung des 
Gleichungssystems~(\ref{E.2.22}) mit demselben Wert $\mu$. Da die Herleitung
f"ur alle Frequenzen $\mu$ gilt, ist die Identit"at der L"osungsr"aume der
Gleichungssysteme~(\ref{E.2.21}) und (\ref{E.2.22}) gezeigt.

\section{Zur Konditionierung der empirischen Kovarianzmatrix}\label{E.Kap.A.2}

Bei der Berechnung der Messwerte des RKM ist die empirisch gewonnene
Kovarianzmatrix einiger Werte des Spektrums der Erregung zu invertieren.
Da solch eine Matrix aus einer Stichprobe vom Umfang $L$ der Spektralwerte
der Erregung gewonnen wird, h"angt deren Konditionierung davon ab, welche
konkrete Stichprobe man gerade gezogen hat. Wir wollen daher eine obere
Grenze f"ur die Wahrscheinlichkeit herleiten, eine schlecht konditionierte 
Matrix zu erhalten, und zeigen, dass diese Grenze mit steigendem Umfang
$L$ der Stichprobe mindestens indirekt proportional abf"allt, wenn man die
zuf"alligen Spektralwerte so w"ahlt, dass deren theoretische
Kovarianzmatrix gut konditioniert ist. Es sei darauf hingewiesen,
dass die nun folgende Betrachtung auch f"ur die Varianz einer einzigen
Zufallsgr"o"se gilt, da es sich dabei um den Fall einer \mbox{$1\!\times\!1$}
Kovarianzmatrix handelt. In Gegensatz zu \cite{Diss} wird hier der Fall 
mittelwertbehafteter Zufallsgr"o"sen behandelt.

Gegeben seien $R$ komplexe Zufallsgr"o"sen, die den Zufallsspaltenvektor
$\Vec{\boldsymbol{V}}$ bilden, und deren Momente bis zur vierten Ordnung alle
existieren sollen. Ein Teil der zweiten Momente sind die Elemente der
theoretischen \mbox{$R\!\times\!R$} Kovarianzmatrix
\begin{equation}
\underline{C}_{\Vec{\boldsymbol{V}},\Vec{\boldsymbol{V}}}\;=\;
\text{E}\Big\{\big(\Vec{\boldsymbol{V}}\!-\!
\text{E}\{\Vec{\boldsymbol{V}}\}\big)\CdoT
\big(\Vec{\boldsymbol{V}}\!-\!
\text{E}\{\Vec{\boldsymbol{V}}\}\big)^{\HH}\Big\}\;=\;
\underline{C}_{\Vec{\boldsymbol{V}},\Vec{\boldsymbol{V}}}^{\Hh},
\label{E.A.8}
\end{equation}
die immer hermitesch und positiv semidefinit ist. Sie l"asst sich daher 
unit"ar kongruent auf Diagonalform transformieren, wobei die Diagonalelemente 
alle nichtnegativ reell sind, und mit dem Zeilenindex (\,=~Spaltenindex\,) 
des Diagonalelements monoton fallen. Die Diagonalelemente sind die 
Singul"arwerte $s_i$ mit \mbox{$i=1\;(1)\;R$} der Kovarianzmatrix, 
die zugleich deren Eigenwerte sind. Es gilt \mbox{$s_i\ge s_{i+1}$}. 
Je n"aher die Konditionszahl bei eins liegt, desto besser ist die
Kovarianzmatrix konditioniert, und desto genauer l"asst sich deren
Inverse berechnen. Jeder beliebige Vektor wird durch die Multiplikation
mit der Kovarianzmatrix auf einen Vektor abgebildet, dessen euklidische
Norm \mbox{$\snorm{\ldots}$} die Bedingung
\begin{equation}
s_R \cdot \snorm{\Vec{x}} \;\le\;
\snorm{\,\underline{C}_{\Vec{\boldsymbol{V}},\Vec{\boldsymbol{V}}} \CdoT \Vec{x}\,} \;\le\;
s_1 \cdot \snorm{\Vec{x}}
\label{E.A.9}
\end{equation}
erf"ullt. Da es sich bei der Spektralnorm um eine mit der euklidischen
Vektornorm kompatible Matrixnorm handelt, gilt au"serdem f"ur jeden
beliebigen Vektor:
\begin{equation}
\snorm{\,\underline{C}_{\Vec{\boldsymbol{V}},\Vec{\boldsymbol{V}}} \CdoT \Vec{x}\,} \;\le\;
\snorm{\underline{C}_{\Vec{\boldsymbol{V}},\Vec{\boldsymbol{V}}}} \cdot \snorm{\Vec{x}}.
\label{E.A.10}
\end{equation}

Die \mbox{$R\!\times\!L$} Matrix $\underline{V}$ enthalte eine
konkrete Stichprobe vom Umfang $L$ des Zufallsvektors $\Vec{\boldsymbol{V}}$,
d.~h. jede Spalte dieser Matrix ist ein Element der Stichprobe, also
eine konkrete Realisierung des Zufallsvektors $\Vec{\boldsymbol{V}}$, und es
wurden insgesamt $L$ konkrete Realisierungen dieses Zufallsvektors
zu einer Matrix zusammengefasst. Bei der Matrix handelt es sich um eine
konkrete Realisierung der Matrix $\underline{\boldsymbol{V}}$ der
mathematischen Stichprobe vom Umfang $L$ des Zufallsvektors
$\Vec{\boldsymbol{V}}$.
Dies ist der Fall, wenn die Stichprobenentnahme in der Art erfolgt ist,
dass jedes Element der Stichprobe --- also jede Spalte der Matrix
$\underline{\boldsymbol{V}}$ --- von jedem anderen Element unabh"angig
ist, und die gleiche Verbundverteilung besitzt, wie der Zufallsvektor
$\Vec{\boldsymbol{V}}$. Aus jeder konkreten Stichprobenmatrix
$\underline{V}$ kann man eine empirische Kovarianzmatrix berechnen.
\begin{equation}
\Hat{\underline{C}}_{\Vec{\boldsymbol{V}},\Vec{\boldsymbol{V}}}\;=\;
\frac{\;\underline{V}\cdot\underline{1}_{\bot}\Cdot
\underline{V}^{\HH}\,}{L-1}\;=\;
\underline{V}\cdot
\frac{\;L\cdot\underline{E}-\Vec{1}^{\,\Hh}\Cdot\Vec{1}\;}{L\cdot(L\!-\!1)}
\cdot\underline{V}^{\HH}
\label{E.A.11}
\end{equation}
Nun ben"otigen wir noch eine weitere Matrixnorm \mbox{$\fnorm{\ldots}$},
n"amlich die euklidische Matrixnorm, die auch Frobeniusnorm genannt wird,
und die die Wurzel aus der Summe der Betragsquadrate aller Matrixelemente
ist. Da auch diese Matrixnorm mit der euklidischen Vektornorm
kompatibel ist, gilt auch mit dieser Matrixnorm f"ur jeden
beliebigen Vektor:
\begin{equation}
\Snorm{\,\big(\underline{C}_{\Vec{\boldsymbol{V}},\Vec{\boldsymbol{V}}}\!-\!
\Hat{\underline{C}}_{\Vec{\boldsymbol{V}},\Vec{\boldsymbol{V}}}\big)\CdoT\Vec{x}\,} \,=\;
\Snorm{\,\big(\Hat{\underline{C}}_{\Vec{\boldsymbol{V}},\Vec{\boldsymbol{V}}}\!-\!
\underline{C}_{\Vec{\boldsymbol{V}},\Vec{\boldsymbol{V}}}\big)\CdoT\Vec{x}\,}\,\le\;
\fnorm{\,\Hat{\underline{C}}_{\Vec{\boldsymbol{V}},\Vec{\boldsymbol{V}}}\!-\!
\underline{C}_{\Vec{\boldsymbol{V}},\Vec{\boldsymbol{V}}}}
\cdot\snorm{\Vec{x}}.
\label{E.A.12}
\end{equation}
Zus"atzlich ben"otigen wir noch die Dreiecksungleichung
\begin{equation}
\snorm{\,\Vec{x}\!+\!\Vec{y}\,}\,\le\;\snorm{\Vec{x}}+\snorm{\Vec{y}}.
\label{E.A.13}
\end{equation}
Damit k"onnen wir absch"atzen, dass die Norm des Produkts eines
beliebigen Vektors mit der empirischen Kovarianzmatrix innerhalb eines
bestimmten Intervalls liegen muss.
\begin{gather}
\label{E.A.14}
\snorm{\,\Hat{\underline{C}}_{\Vec{\boldsymbol{V}},\Vec{\boldsymbol{V}}}\CdoT\Vec{x}\,}\,=\;
\Snorm{\,\underline{C}_{\Vec{\boldsymbol{V}},\Vec{\boldsymbol{V}}}\CdoT\Vec{x}+
\big(\Hat{\underline{C}}_{\Vec{\boldsymbol{V}},\Vec{\boldsymbol{V}}}\!-\!
\underline{C}_{\Vec{\boldsymbol{V}},\Vec{\boldsymbol{V}}}\big)
\CdoT\Vec{x}\,}\,\le\\*[4pt]
\le\;\snorm{\,\underline{C}_{\Vec{\boldsymbol{V}},\Vec{\boldsymbol{V}}}\CdoT\Vec{x}\,},+\;
\Snorm{\,\big(\Hat{\underline{C}}_{\Vec{\boldsymbol{V}},\Vec{\boldsymbol{V}}}\!-\!
\underline{C}_{\Vec{\boldsymbol{V}},\Vec{\boldsymbol{V}}}\big)
\CdoT\Vec{x}\,}\,\le\notag\\*[4pt]
\le\;\snorm{\underline{C}_{\Vec{\boldsymbol{V}},\Vec{\boldsymbol{V}}}} \Cdot \snorm{\Vec{x}}\,+\;
\fnorm{\,\Hat{\underline{C}}_{\Vec{\boldsymbol{V}},\Vec{\boldsymbol{V}}}\!-\!
\underline{C}_{\Vec{\boldsymbol{V}},\Vec{\boldsymbol{V}}}}\Cdot\snorm{\Vec{x}}\,=\;
\big(\,s_1 + \fnorm{\,\Hat{\underline{C}}_{\Vec{\boldsymbol{V}},\Vec{\boldsymbol{V}}}\!-\!
\underline{C}_{\Vec{\boldsymbol{V}},\Vec{\boldsymbol{V}}}}\,\big)
\cdot\snorm{\Vec{x}}\notag
\end{gather}
\begin{gather}
\label{E.A.15}
\snorm{\,\Hat{\underline{C}}_{\Vec{\boldsymbol{V}},\Vec{\boldsymbol{V}}}\CdoT\Vec{x}\,}\,=\;
\Snorm{\,\underline{C}_{\Vec{\boldsymbol{V}},\Vec{\boldsymbol{V}}}\CdoT\Vec{x}-
\big(\underline{C}_{\Vec{\boldsymbol{V}},\Vec{\boldsymbol{V}}}\!-\!
\Hat{\underline{C}}_{\Vec{\boldsymbol{V}},\Vec{\boldsymbol{V}}}\big)
\CdoT\Vec{x}\,}\,\ge\\*[4pt]
\ge\;\snorm{\,\underline{C}_{\Vec{\boldsymbol{V}},\Vec{\boldsymbol{V}}}\CdoT\Vec{x}\,}\,-\;
\Snorm{\,\big(\underline{C}_{\Vec{\boldsymbol{V}},\Vec{\boldsymbol{V}}}\!-\!
\Hat{\underline{C}}_{\Vec{\boldsymbol{V}},\Vec{\boldsymbol{V}}}\big)
\CdoT\Vec{x}\,}\,\ge\notag\\*[4pt]
\ge\;s_R\cdot\snorm{\Vec{x}}\,-\;
\fnorm{\,\Hat{\underline{C}}_{\Vec{\boldsymbol{V}},\Vec{\boldsymbol{V}}}\!-\!
\underline{C}_{\Vec{\boldsymbol{V}},\Vec{\boldsymbol{V}}}}\Cdot\snorm{\Vec{x}}\,=\;
\big(\,s_R - \fnorm{\,\Hat{\underline{C}}_{\Vec{\boldsymbol{V}},\Vec{\boldsymbol{V}}}\!-\!
\underline{C}_{\Vec{\boldsymbol{V}},\Vec{\boldsymbol{V}}}}\,\big)
\cdot\snorm{\Vec{x}}\notag
\end{gather}
Diese beiden Grenzen gelten f"ur beliebige Vektoren, also
auch f"ur Vektoren, die auf ihre euklidische Norm normiert worden sind
und daher die L"ange Eins haben. Der kleinste Singul"arwert 
\mbox{$\Hat{s}_R$} der empirischen Kovarianzmatrix ist die minimale
L"ange aller Vektoren, die durch eine Multiplikation mit der
empirischen Kovarianzmatrix aus allen Vektoren der L"ange Eins entstanden
sind. Die L"ange des l"angsten Bildvektors ist entsprechend der gr"o"ste
Singul"arwert \mbox{$\Hat{s}_1$} der empirischen
Kovarianzmatrix. Falls die Frobeniusnorm der Abweichung der empirischen
von der theoretischen Kovarianzmatrix kleiner als $s_R$ ist, kann
die Konditionszahl $\Hat{K}_{\Vec{\boldsymbol{V}}}$ der empirischen
Kovarianzmatrix mit den beiden letzten Ungleichungen abgesch"atzt
werden.
\begin{equation}
\Hat{K}_{\Vec{\boldsymbol{V}}}\;=\;\frac{\;\Hat{s}_1}{\;\Hat{s}_R}\;\le\;
\frac{\;s_1+\fnorm{\,\Hat{\underline{C}}_{\Vec{\boldsymbol{V}},\Vec{\boldsymbol{V}}}\!-\!
\underline{C}_{\Vec{\boldsymbol{V}},\Vec{\boldsymbol{V}}}}}
{\;s_R-\fnorm{\,\Hat{\underline{C}}_{\Vec{\boldsymbol{V}},\Vec{\boldsymbol{V}}}\!-\!
\underline{C}_{\Vec{\boldsymbol{V}},\Vec{\boldsymbol{V}}}}}
\label{E.A.16}
\end{equation}
Wenn die Frobeniusnorm der Abweichung der empirischen von der theoretischen
Kovarianzmatrix au"serdem noch kleiner als
\begin{equation}
\fnorm{\,\Hat{\underline{C}}_{\Vec{\boldsymbol{V}},\Vec{\boldsymbol{V}}}\!-\!
\underline{C}_{\Vec{\boldsymbol{V}},\Vec{\boldsymbol{V}}}}\,<\;
\frac{n-1}{\;n\!+\!K_{\Vec{\boldsymbol{V}}}^{-1}}\cdot s_R\;<\;s_R
\qquad\text{ mit }\quad n\in\mathbb{R}\quad\text{ und }\quad n>1
\label{E.A.17}
\end{equation}
ist, kann man sicher sein, dass die Konditionszahl der empirischen
Kovarianzmatrix h"ochstens $n$ mal so gro"s wie die Konditionszahl der
theoretischen Kovarianzmatrix ist:
\begin{equation}
\Hat{K}_{\Vec{\boldsymbol{V}}}\;\le\;
\frac{\;s_1+\fnorm{\,\Hat{\underline{C}}_{\Vec{\boldsymbol{V}},\Vec{\boldsymbol{V}}}\!-\!
\underline{C}_{\Vec{\boldsymbol{V}},\Vec{\boldsymbol{V}}}}}
{\;s_R-\fnorm{\,\Hat{\underline{C}}_{\Vec{\boldsymbol{V}},\Vec{\boldsymbol{V}}}\!-\!
\underline{C}_{\Vec{\boldsymbol{V}},\Vec{\boldsymbol{V}}}}}\;\le\;
\frac{\;s_1+\frac{n-1}{\;n+K_{\Vec{\boldsymbol{V}}}^{-1}}\cdot s_R}
{\;s_R-\frac{n-1}{\;n+K_{\Vec{\boldsymbol{V}}}^{-1}}\cdot s_R}\;=\;
n\CdoT K_{\Vec{\boldsymbol{V}}}.
\label{E.A.18}
\end{equation}
Falls nun die Wahrscheinlichkeit, dass die Frobeniusnorm der Matrixdifferenz 
gr"o"ser als die Schranke in der Ungleichung (\ref{E.A.17}) ist, mit steigendem 
$L$ gegen Null konvergiert, kann man durch eine Erh"ohung von $L$ erreichen, 
dass die Wahrscheinlichkeit, eine empirische Kovarianzmatrix zu erhalten, 
die $n$ mal schlechter als die theoretische Kovarianzmatrix konditioniert ist, 
unter einer beliebig kleinen tolerierbaren Schwelle bleibt. Wir wollen daher 
eine obere Schranke f"ur diese Wahrscheinlichkeit herleiten.

Dazu ben"otigen wir einen Satz, aus dem sich auch die Tschebyscheffsche
Ungleichung  ableiten l"asst, und den wir aus \cite{Fisz} entnehmen.\\[6pt]
{\sl Satz: Nimmt eine zuf"allige Ver"anderliche $\boldsymbol{Y}$
nur nichtnegative Werte an, und besitzt sie einen endlichen Mittelwert
\text{E$\{\boldsymbol{Y}\}$}, so ist f"ur jede positive Zahl $K$
die Ungleichung
\begin{gather*}
P(\,\boldsymbol{Y}\!\ge\!K\,)\;\le\;\frac{\;\text{E}\{\boldsymbol{Y}\}\;}{K}
\\[-28pt]
\end{gather*}
erf"ullt.}\\[6pt]
Da das Quadrat der Frobeniusnorm einer zuf"alligen Matrix solch eine
zuf"allige Ver"anderliche ist, erhalten wir mit
\[
\boldsymbol{Y}\;=\;
\fnorm{\,\Hat{\underline{\boldsymbol{C}}}_{\Vec{\boldsymbol{V}},\Vec{\boldsymbol{V}}}\!-\!
\underline{C}_{\Vec{\boldsymbol{V}},\Vec{\boldsymbol{V}}}}^2
\qquad\qquad\text{ und mit }\qquad
K = \Big(\frac{n-1}{\;n\!+\!K_{\Vec{\boldsymbol{V}}}^{-1}}\cdot s_R\Big)^{\!2}
\]
die gesuchte obere Grenze\vspace{-8pt}
\begin{gather}
\label{E.A.19}
P\Big(\Hat{\boldsymbol{K}}_{\Vec{\boldsymbol{V}}}\ge
n\CdoT K_{\Vec{\boldsymbol{V}}}\Big)\;\le\\[8pt]
\le\;P\bigg(\,
\fnorm{\,\Hat{\underline{\boldsymbol{C}}}_{\Vec{\boldsymbol{V}},\Vec{\boldsymbol{V}}}\!-\!
\underline{C}_{\Vec{\boldsymbol{V}},\Vec{\boldsymbol{V}}}} \ge
\frac{n-1}{\;n\!+\!K_{\Vec{\boldsymbol{V}}}^{-1}}\cdot s_R \bigg)\;={}
\notag\\[8pt]
{}=\;P\bigg(\,\fnorm{\,\Hat{\underline{\boldsymbol{C}}}_{\Vec{\boldsymbol{V}},\Vec{\boldsymbol{V}}}\!-\!
\underline{C}_{\Vec{\boldsymbol{V}},\Vec{\boldsymbol{V}}}}^2\ge
\Big(\frac{n-1}{\;n\!+\!K_{\Vec{\boldsymbol{V}}}^{-1}}\cdot s_R\Big)^{\!2}
\,\bigg)\;\le
\notag\\[8pt]
\le\;\bigg(\!\frac{n\!+\!K_{\Vec{\boldsymbol{V}}}^{-1}}
{\;(n\!-\!1)\CdoT s_R\;}\!\bigg)^{\!\!2}\Cdot\text{E}\big\{\,
\fnorm{\,\Hat{\underline{\boldsymbol{C}}}_{\Vec{\boldsymbol{V}},\Vec{\boldsymbol{V}}}\!-\!
\underline{C}_{\Vec{\boldsymbol{V}},\Vec{\boldsymbol{V}}}}^2\,\big\}
\notag
\end{gather}
f"ur die Wahrscheinlichkeit, dass die Frobeniusnorm der
Differenz der empirischen und der theoretischen Kovarianzmatrix
oberhalb der zul"assigen Schwelle liegt, die gleichzeitig eine obere Grenze
f"ur die Wahrscheinlichkeit ist, dass die Konditionszahl der empirischen
Kovarianzmatrix h"ochstens $n$ mal so gro"s ist, wie die Konditionszahl
der theoretischen Kovarianzmatrix.  Um diese Grenze angeben zu k"onnen,
m"ussen wir den Erwartungswert des Quadrats der Frobeniusnorm der
Differenz der empirischen und der theoretischen Kovarianzmatrix berechnen.
Nach der Definition der Frobeniusnorm ergibt sich diese als die Wurzel
aus der Summe der Betragsquadrate aller Matrixelemente. Der Erwartungswert
des Quadrats der Frobeniusnorm ist daher die Summe der Erwartungswerte
der Betragsquadrate aller zuf"alligen Elemente der Matrixdifferenz.

Mit \mbox{$\Vec{\boldsymbol{V}}_{\!\!i}$} sei die mathematische
Stichprobe vom Umfang $L$ der $i$-ten Zufallsgr"o"se des Zufallsvektors
$\Vec{\boldsymbol{V}}$, also die $i$-te Zeile der Zufallsmatrix
$\underline{\boldsymbol{V}}$ bezeichnet. Das Element
in der $i$-ten Zeile und $j$-ten Spalte der empirischen
Kovarianzmatrix ist
\begin{equation}
\Hat{\boldsymbol{C}}_{\boldsymbol{V}_{\!i},\boldsymbol{V}_{\!j}}\;=\;
\Vec{\boldsymbol{V}}_{\!\!i}\cdot
\frac{\underline{1}_{\bot}}{L\!-\!1}\cdot
\Vec{\boldsymbol{V}}_{\!\!j}^{\hH}.
\label{E.A.20}
\end{equation}
Der Erwartungswert dieses Elements ist
\begin{gather}
\label{E.A.21}
\text{E}\big\{\,
\Hat{\boldsymbol{C}}_{\boldsymbol{V}_{\!i},\boldsymbol{V}_{\!j}}\big\}\;=\;
\text{E}\Big\{\;
\Vec{\boldsymbol{V}}_{\!\!i}\cdot
\frac{\underline{1}_{\bot}}{L\!-\!1}\cdot
\Vec{\boldsymbol{V}}_{\!\!j}^{\hH}
\,\Big\}\;=\\[8pt]
=\;\frac{\text{spur}(\underline{1}_{\bot})}{L\!-\!1} \cdot
\text{E}\big\{\boldsymbol{V}_{\!\!i}\CdoT
\boldsymbol{V}_{\!\!j}^*\big\}\;+\;
\frac{\Vec{1}\cdot\underline{1}_{\bot}\Cdot\Vec{1}^{\,\Hh} -
      \text{spur}(\underline{1}_{\bot})}{L-1} \cdot
\text{E}\big\{\boldsymbol{V}_{\!\!i}\big\}\cdot
\text{E}\big\{\boldsymbol{V}_{\!\!j}^*\big\}\;=\notag\\*[8pt]
=\;\text{E}\big\{\boldsymbol{V}_{\!\!i}\CdoT
\boldsymbol{V}_{\!\!j}^*\big\} -
\text{E}\big\{\boldsymbol{V}_{\!\!i}\big\}\cdot
\text{E}\big\{\boldsymbol{V}_{\!\!j}^*\big\}\;=\;
C_{\boldsymbol{V}_{\!i},\boldsymbol{V}_{\!j}},\notag
\end{gather}
und somit gleich dem entsprechenden Element der theoretischen
Kovarianzmatrix \mbox{$\underline{C}_{\Vec{\boldsymbol{V}},\Vec{\boldsymbol{V}}}$}.
Die Varianz des Elements in der $i$-te Zeile und $j$-ten Spalte 
der empirischen Kovarianzmatrix ist daher gleich dem Erwartungswert 
des Betragsquadrats des Elements in der $i$-te Zeile und $j$-ten 
Spalte der Differenz der empirischen und der theoretischen 
Kovarianzmatrix. Die Varianz dieses Matrixelements berechnet sich zu
\begin{gather}
\text{E}\Big\{\,\Big|
\Hat{\boldsymbol{C}}_{\boldsymbol{V}_{\!i},\boldsymbol{V}_{\!j}}-
\text{E}\{\Hat{\boldsymbol{C}}_{\boldsymbol{V}_{\!i},\boldsymbol{V}_{\!j}}\}
\Big|^2\Big\}\;=\;
\text{E}\Big\{\,\Big|
\Hat{\boldsymbol{C}}_{\boldsymbol{V}_{\!i},\boldsymbol{V}_{\!j}}-
C_{\boldsymbol{V}_{\!i},\boldsymbol{V}_{\!j}}\Big|^2\Big\}\;={}
\label{E.A.22}\displaybreak[1]\\[8pt]
{}=\;\text{E}\bigg\{\,\bigg|\,
\Vec{\boldsymbol{V}}_{\!\!i}\cdot
\frac{\underline{1}_{\bot}}{L\!-\!1}\cdot
\Vec{\boldsymbol{V}}_{\!\!j}^{\HH} -\,
\text{E}\Big\{\big(\boldsymbol{V}_{\!\!i}\!-\!
\text{E}\{\boldsymbol{V}_{\!\!i}\}\big)\CdoT
\big(\boldsymbol{V}_{\!\!j}\!-\!
\text{E}\{\boldsymbol{V}_{\!\!j}\}\big)^{\!*}\Big\}
\bigg|^{\uP{-0.3}{2}}\bigg\}\;=
\notag\\[8pt]\begin{flalign*}
&=\,\frac{1}{L}\cdot\text{E}\bigg\{\bigg|
\big(\boldsymbol{V}_{\!\!i}\!-\!
\text{E}\{\boldsymbol{V}_{\!\!i}\}\big)\CdoT
\big(\boldsymbol{V}_{\!\!j}\!-\!
\text{E}\{\boldsymbol{V}_{\!\!j}\}\big)^{\!*}\!-
\text{E}\Big\{\big(\boldsymbol{V}_{\!\!i}\!-\!
\text{E}\{\boldsymbol{V}_{\!\!i}\}\big)\CdoT
\big(\boldsymbol{V}_{\!\!j}\!-\!
\text{E}\{\boldsymbol{V}_{\!\!j}\}\big)^{\!*}\Big\}
\bigg|^{\uP{-0.3}{2}}\bigg\}\,+{}&&
\end{flalign*}\notag\\*
{}+\,\frac{1}{L\CdoT(L\!-\!1)}\cdot\bigg(\,\Big|
\text{E}\Big\{\big(\boldsymbol{V}_{\!\!i}\!-\!
\text{E}\{\boldsymbol{V}_{\!\!i}\}\big)\CdoT
\big(\boldsymbol{V}_{\!\!j}\!-\!
\text{E}\{\boldsymbol{V}_{\!\!j}\}\big)\Big\}\Big|^2+{}
\notag\\*\begin{flalign*}
&&{}+\text{E}\Big\{\big|\boldsymbol{V}_{\!\!i}\!-\!
\text{E}\{\boldsymbol{V}_{\!\!i}\}\big|^2\Big\}\cdot
\text{E}\Big\{\big|\boldsymbol{V}_{\!\!j}\!-\!
\text{E}\{\boldsymbol{V}_{\!\!j}\}\big|^2\Big\}\bigg).&
\end{flalign*}\notag
\end{gather}
Bei dieser Berechnung wurde zun"achst die bilineare Form
\mbox{$\Vec{\boldsymbol{V}}_i\cdot\underline{1}_{\bot}\Cdot
\Vec{\boldsymbol{V}}_j^{\HH}$}
als Doppelsumme geschrieben, so dass sich durch die Bildung des
Betragsquadrats unter anderem eine Vierfachsumme ergibt. Der Erwartungswert
dieser Vierfachsumme l"asst sich durch eine Fallunterscheidung
bez"uglich der Indizes der Vierfachsumme --- wie in Tabelle \myref{TA.1}
des Kapitels \myref{4Mom} --- einigerma"sen umfangreich,
aber nicht allzu anspruchsvoll, berechnen.  Die Varianz
der Elemente der empirischen Kovarianzmatrix nimmt also mit steigendem
$L$ wenigstens mit $\alpha/L$ ab, wobei $\alpha$ eine positive
Konstante ist, die nur von den zweiten und vierten zentralen Momenten
des erregenden Zufallsprozessspektrums abh"angt. 
Da die Summe der Varianzen aller Matrixelemente der Erwartungswert 
des Quadrats der Frobeniusnorm der Differenz der empirischen und der 
theoretischen Kovarianzmatrix ist, und die Dimension dieser Matrix
mit \mbox{$R\!\times\!R$} von $L$ unabh"angig ist, nimmt auch
der Erwartungswert des Quadrats der Frobeniusnorm mindestens
mit der Ordnung $1/L$ ab. 
\begin{equation}
\text{E}\big\{\,
\fnorm{\,\Hat{\underline{\boldsymbol{C}}}_{\Vec{\boldsymbol{V}},\Vec{\boldsymbol{V}}}\!-\!
\underline{C}_{\Vec{\boldsymbol{V}},\Vec{\boldsymbol{V}}}}^2\,\big\}\;=\;
\Sum{i=1}{L}\Sum{j=1}{L}\text{E}\Big\{\,\Big|
\Hat{\boldsymbol{C}}_{\boldsymbol{V}_{\!i},\boldsymbol{V}_{\!j}}-
C_{\boldsymbol{V}_{\!i},\boldsymbol{V}_{\!j}}\Big|^2\Big\}\;\sim\;
\frac{1}{L}
\label{E.A.23}
\end{equation}
Setzt man dies in die Ungleichung (\ref{E.A.19}) ein, so sieht man,
dass auch die Wahrscheinlichkeit, eine Konditionszahl der empirischen
Kovarianzmatrix zu erhalten, die mehr als $n$ mal so gro"s ist wie die Konditionszahl
der theoretischen Kovarianzmatrix, wenigstens mit $1/L$ abnimmt.

Es sei noch angemerkt, dass dies nur eine {\em obere Schranke}\/
f"ur die Wahrscheinlichkeit ist, dass die empirische
Kovarianzmatrix eine mehr als $n$-fache Konditionszahl
der theoretischen Kovarianzmatrix aufweist. Da dieser oberen
Schranke eine Vielzahl von Ungleichungen zugrundeliegen,
ist anzunehmen, dass die wahre Wahrscheinlichkeit, deren
Berechnung ---\,wenn "uberhaupt\,--- nur bei Kenntnis der gemeinsamen
Verbundverteilung aller Elemente des Zufallsvektors $\Vec{\boldsymbol{V}}$
m"oglich w"are, wesentlich kleiner ist als die angegebene obere
Schranke. Die hier gemachte Herleitung hat den Vorteil, dass keine
Aussage "uber die genaue Verbundverteilung ben"otigt wird.
Es gen"ugt, wenn die theoretische Kovarianzmatrix gut
konditioniert ist, um zu gew"ahrleisten, dass auch die
empirische Kovarianzmatrix mit gro"ser Wahrscheinlichkeit brauchbar
konditioniert ist, wenn man $L$ nur gro"s genug w"ahlt.

\section{Zur Wahrscheinlichkeit der empirischen Varianz Null}\label{E.Kap.A.3}

F"ur den Fall, dass man wie in \cite{Diss} in Bild \ref{E.b1h} nur ein lineares 
zeitinvariantes Modellsystem mit der "Ubertragungsfunktion \mbox{$H(\Omega)$} ansetzt, 
aber dennoch die deterministische Modellst"orung \mbox{$u(k)$} beibeh"alt, 
entartet bei der Berechnung der Messwerte der "Ubertragungsfunktion nach 
Gleichung~(\ref{3.8}) die Kovarianzmatrix
\mbox{$\Hat{\underline{C}}_{\Tilde{\Vec{\boldsymbol{V}}}(\mu),\Tilde{\Vec{\boldsymbol{V}}}(\mu)}$} 
zu einer \mbox{$1\times1$} Matrix
\mbox{$\Hat{C}_{\boldsymbol{V}(\mu),\boldsymbol{V}(\mu)}$}, 
deren einziges Element die empirische Varianz des 
Spektralwertes \mbox{$\boldsymbol{V}(\mu)$} der Erregung ist. 
Hier soll nun der Fall untersucht werden, dass diese Varianz zu null wird.

Jedes Element des konkreten Stichprobenvektors
\mbox{$\Vec{V}(\mu)$} wurde zuf"allig aus der Zufallsgr"o"se
\mbox{$\boldsymbol{V}(\mu)$} gewonnen, und daher kann
\mbox{$\Vec{V}(\mu)$} als eine konkrete Realisierung des 
Zufallsvektors \mbox{$\Vec{\boldsymbol{V}}(\mu)$}
der mathematischen Stichprobe von Umgang $L$ betrachtet werden.
Wir betrachten also nicht eine konkret durchgef"uhrte Messung mit
ihren Messergebnissen, die aus den konkreten Stichprobenvektoren
gewonnen wurden, sondern die stochastischen Eigenschaften des
Messverfahrens und der Messergebnisse, die man aus den mathematischen
Stichprobenvektoren bestimmt. Es ist nicht unm"oglich, n"amlich dann,
wenn alle Elemente der konkreten Stichprobe vom Umfang $L$
der Zufallsgr"o"se \mbox{$\boldsymbol{V}\!(\mu)$} gleich sind,
dass ein oder mehrere der dann ebenfalls zuf"alligen $M$ empirischen Varianzen
\mbox{$\Hat{\boldsymbol{C}}_{\boldsymbol{V}(\mu),\boldsymbol{V}(\mu)}$} den
konkreten Wert Null annehmen. In diesem Fall werden die Messwerte
\mbox{$\Hat{H}(\mu)$} und damit auch die Messwerte \mbox{$\Hat{U}_{\!f}(\mu)$}
undefiniert. Es existiert dann ein komplexer ein- oder mehrdimensionaler
L"osungsraum f"ur die Messwerte \mbox{$\Hat{H}(\mu)$} und
\mbox{$\Hat{U}_{\!f}(\mu)$}. Um auch in diesem Fall eindeutige Messwerte
zu erhalten gehen wir folgenderma"sen vor. Wenn die empirischen Varianz
kleiner als eine beliebig kleine aber echt positive
Konstante $\Upsilon^{-1}$ ist, verwenden wir statt der Inversen der
empirischen Varianz den konstanten endlichen Wert $\Upsilon$. Die Inverse
der empirischen Varianz ist eine Funktion des Zufallsvektors
\mbox{$\Vec{\boldsymbol{V}}(\mu)$} der mathematischen Stichprobe
des Spektrums der Erregung, deren Definitionsbereich den Raum, in dem
die empirischen Varianz Null wird, nicht mit einschlie"st. Durch die
mit $\Upsilon$ eingef"uhrte Modifikation ist diese Funktion des
Zufallsvektors \mbox{$\Vec{\boldsymbol{V}}(\mu)$} dann auf
dem gesamten Raum definiert und endlich. Die Wahrscheinlichkeit,
dass wir die Inverse der empirischen Varianz begrenzen m"ussen, h"angt
neben der Wahl von $\Upsilon$ auch von der Art --- genauer gesagt von
der Verteilung --- der Zufallsgr"o"se  \mbox{$\boldsymbol{V}(\mu)$} ab.
Im Fall einer diskreten Zufallsgr"o"se \mbox{$\boldsymbol{V}(\mu)$},
ergibt sich --- wie wir gleich sehen werden --- die unerfreuliche
Tatsache, dass die Wahrscheinlichkeit, dass die empirische Varianz
so klein wird, dass eine Begrenzung ihrer Inversen erforderlich wird,
bei einer Vergr"o"serung von $\Upsilon$ nicht unter einen Minimalwert
sinkt, der echt gr"o"ser als Null ist. Dieser Fall ist der f"ur
praktische Anwendungen interessante Fall, weil man "ublicherweise
die Zufallsgr"o"se \mbox{$\boldsymbol{V}(\mu)$} an einem Rechner generiert.
Da an Rechnern nur abz"ahlbar endlich viele Zahlen dargestellt werden
k"onnen, ist der Ergebnisraum der Zufallsgr"o"se des Spektrums der
Erregung immer ein abz"ahlbarer endlicher Unterraum der am Rechner
darstellbaren Zahlen. Somit ist die Zufallsgr"o"se
\mbox{$\boldsymbol{V}(\mu)$} diskret. Unabh"angig davon, wie gro"s
$\Upsilon$ gew"ahlt worden ist, ist die Wahrscheinlichkeit,
dass alle Werte der konkreten Stichprobe vom Umfang $L$ der
Zufallsgr"o"se \mbox{$\boldsymbol{V}(\mu)$} gleich sind, und somit
die empirische Varianz Null wird, so dass ihre Inverse
zu begrenzen ist, immer echt positiv. Wir wollen nun die
Wahrscheinlichkeit des Eintretens dieses Falls berechnen,
wobei wir annehmen wollen, dass $\Upsilon$ so gro"s gew"ahlt
worden ist, dass in keinem anderen Fall, als in dem, dass alle
Werte der konkreten Stichprobe vom Umfang $L$ der Zufallsgr"o"se
\mbox{$\boldsymbol{V}(\mu)$} gleich sind, eine Begrenzung erfolgt.
Die Wahrscheinlichkeit \mbox{$P(\boldsymbol{V}(\mu)\!=\!V)$}, dass
die Zufallsgr"o"se einen bestimmten Wert des Ergebnisraums annimmt,
l"asst sich als Grenzwert aus der Verteilung der Zufallsgr"o"se
\mbox{$\boldsymbol{V}(\mu)$} berechnen.
\begin{gather}
P\big(\boldsymbol{V}(\mu)\!=\!V\big)\;=
\label{E.A.24}\\[3pt]\begin{flalign*}
&=\;\lim_{\Delta V\to0}\Big(
P\big(\boldsymbol{V}(\mu)<V\!+\!(1\!+\!j)\CdoT\Delta V\big)\,-\,
P\big(\boldsymbol{V}(\mu)<V\!+\!\Delta V\big)\,-{}&&
\end{flalign*}\notag\\*\begin{flalign*}
&&{}-\,P\big(\boldsymbol{V}(\mu)<V\!+\!j\CdoT\Delta V\big)\,+\,
P\big(\boldsymbol{V}(\mu)<V\big)\,\Big)&
\end{flalign*}\notag
\end{gather}
Da es sich bei dem Zufallsvektor \mbox{$\Vec{\boldsymbol{V}}(\mu)$} um
eine mathematische Stichprobe handelt, kann dessen Verbundverteilung
\mbox{$P\big(\Vec{\boldsymbol{V}}(\mu)<\Vec{V}(\mu)\big)$}
als die $L$-te Potenz der Verteilung der Zufallsgr"o"se
\mbox{$\boldsymbol{V}(\mu)$} geschrieben werden. Die Wahrscheinlichkeit
\mbox{$P_{L}\big(\Vec{\boldsymbol{V}}(\mu)=V\CdoT\Vec{1}\,\big)$},
dass alle Elemente des Zufallsvektors \mbox{$\Vec{\boldsymbol{V}}(\mu)$}
der mathematischen Stichprobe vom Umfang $L$ zugleich einen bestimmten
Wert $V$ annehmen, ist daher die $L$-te Potenz der Wahrscheinlichkeit
\mbox{$P\big(\boldsymbol{V}(\mu)\!=\!V\big)$}. Die Wahrscheinlichkeit
\mbox{$P_L\big(\Hat{\boldsymbol{C}}_{\boldsymbol{V}(\mu),\boldsymbol{V}(\mu)}
\!=\!0\big)$}, dass die empirische Varianz bei einer Anzahl von $L$
Einzelmessungen Null wird, kann durch Summation aus den Wahrscheinlichkeiten
\mbox{$P_{L}\big(\Vec{\boldsymbol{V}}(\mu)=V\CdoT\Vec{1}\,\big)$}
berechnet und aus der Wahrscheinlichkeit, dass die
empirische Varianz bei \mbox{$L\!-\!1$} Einzelmessungen Null
wird, abgesch"atzt werden.
\begin{gather}
P_L\big(\Hat{\boldsymbol{C}}_{\boldsymbol{V}(\mu),\boldsymbol{V}(\mu)}\!=\!0\big)
\;=\,\Sum{V}{}P\big(\boldsymbol{V}(\mu)\!=\!V\,\big)^L\;\le
\label{E.A.25}\\[4pt]
\le\;\max_{V}\Big\{P\big(\boldsymbol{V}(\mu)\!=\!V\big)\Big\}\cdoT
\Sum{V}{}P\big(\boldsymbol{V}(\mu)\!=\!V\big)^{(L-1)}\;=
\notag\\[4pt]
=\;\max_{V}\Big\{P\big(\boldsymbol{V}(\mu)\!=\!V\big)\Big\}\cdot
P_{L-1}\!\big(\Hat{\boldsymbol{C}}_{\boldsymbol{V}(\mu),\boldsymbol{V}(\mu)}
\!=\!0\big)\notag
\end{gather}
Die Summe ist ebenso wie die Maximalwertbestimmung "uber alle m"oglichen
Werte von $V\!$, also "uber alle Elemente des Ergebnisraums
von \mbox{$\boldsymbol{V}(\mu)$} zu erstrecken. Durch geeignete Wahl des
erregenden Zufallsvektors $\Vec{\boldsymbol{V}}$ kann f"ur \mbox{$L\ge2$}
erreicht werden, dass die Wahrscheinlichkeit die Varianz Null zu erhalten,
mit steigender Mittelungsanzahl $L$ rasch absinkt, weil der Faktor
zwischen der Wahrscheinlichkeit f"ur die Varianz Null f"ur $L$ und f"ur
\mbox{$L\!-\!1$} gem"a"s der letzten Gleichung mit
\mbox{$\max_{V}\!\big\{P\big(\boldsymbol{V}(\mu)\!=\!V\big)\big\}$}
nach oben abgesch"atzt werden kann. Diese obere Grenze f"ur den Faktor
wird dann besonders klein, wenn die Maximalwahrscheinlichkeit
minimal wird. Dann wird die Wahrscheinlichkeit f"ur die empirische
Varianz Null besonders rasch abklingen. Wenn man sich die
Zufallsgr"o"se \mbox{$\boldsymbol{V}(\mu)$} durch Quantisierung aus
einer wertkontinuierlichen Zufallsgr"o"se entstanden denkt,
die durch ihre Verteilungsdichtefunktion beschrieben wird,
kann man eine m"oglichst kleine Maximalwahrscheinlichkeit
besonders einfach erreichen, wenn man die Quantisierung
dort besonders fein macht, wo die Verteilungsdichtefunktion
der zugrundeliegenden wertkontinuierlichen Zufallsgr"o"se
gro"s ist, und wenn man insgesamt m"oglichst fein quantisiert.

An einem einfachen Beispiel soll gezeigt werden in welcher
Gr"o"senordnung die Wahrscheinlichkeit, die Varianz Null
zu erhalten, bei einigerma"sen realistischer Wahl des Zufallsvektors
$\Vec{\boldsymbol{V}}$ liegt. Zur Gewinnung der Stichprobe
wird ein Zufallszahlengenerator verwendet, der alle m"oglichen
4-Byte Festkommazahlen mit gleicher Wahrscheinlichkeit
unabh"angig liefert. "Ahnliche Zufallszahlengeneratoren ---
die allerdings deterministische Zahlenfolgen mit pseudozuf"alligen
Eigenschaften liefern --- sind oft in Prozessoren oder Programmen fest
eingebaut, oder sind Teil der Programmbibliothek, die man beim Erwerb
eines Compilers einer Programmiersprache erh"alt.

Wir wollen nun annehmen, dass der Real- und der Imagin"arteil
der Zufallsgr"o"sen \mbox{$V_{\lambda}(\mu)$}, die man
aus je zwei der zuf"alligen Ausgaben des Zufallszahlengenerators
 --- z.~B. mit einer nichtlinearen Kennlinie --- berechnet,
unabh"angig sind, und dass ein eineindeutiger Zusammenhang zwischen
den Zufallsgr"o"sen \mbox{$\boldsymbol{V}_{\!\lambda}(\mu)$} und
den Paaren von Ausgaben des Zufallszahlengenerators besteht, so dass
sich f"ur die M"achtigkeit des Ergebnisraums jeder der Zufallsgr"o"sen
\mbox{$\boldsymbol{V}_{\!\lambda}(\mu)$} der Wert \mbox{$2^{64}$} ergibt.
Da alle m"oglichen Werte der Zufallsgr"o"se
\mbox{$\boldsymbol{V}_{\!\lambda}(\mu)$}
gleichwahrscheinlich sind, ergibt sich f"ur die Wahrscheinlichkeit,
\mbox{$P\big(\boldsymbol{V}(\mu)\!=\!V\big)$} der von $V$ unabh"angige
konstante Wert \mbox{$2^{-64}$}, und f"ur die
Wahrscheinlichkeit, dass die Varianz bei {\em einer}\/ Frequenz $\mu$
Null wird, nach der letzten Formel der mit $L$ exponentiell fallende
Wert \mbox{$2^{64\cdot(1\!-\!L)}$}. F"ur alle Frequenzen
\mbox{$\Omega=\mu\CdoT2\pi/M$} werden unabh"angige Zufallszahlen
dieses Zufallszahlengenerators verwendet. Man kann nun die
Wahrscheinlichkeit, dass eine oder mehrere der $M$ Varianzen f"ur
die $M$ Frequenzpunkte Null werden, als Funktion von $L$ und $M$
zu \mbox{$1-\big(\,1-2^{64\cdot(1-L)}\,\big)^M$} berechnen.
Da die Wahrscheinlichkeit, dass die Varianz bei einer Frequenz $\mu$
Null wird, hinreichend klein ist, kann die Wahrscheinlichkeit, dass
eine oder mehrere der $M$ Varianzen Null werden, mit
\mbox{$M\cdot2^{64\cdot(1-L)}$} abgesch"atzt werden. Gibt
man sich die maximal zul"assige Wahrscheinlichkeit vor, dass eine
oder mehrere der $M$ Varianzen Null werden, so kann man die
notwendige Mittelungsanzahl $L$, oder die maximal m"ogliche Anzahl $M$
der Frequenzpunkte, f"ur die die Messwerte berechnet werden sollen,
als Funk\-tion von der jeweils anderen Gr"o"se absch"atzen.
Bei unserem \mbox{4-Byte} Zufallszahlengenerator ergibt sich mit dieser
Absch"atzung bereits bei einer Messung mit der k"urzest m"oglichen
Mittelungsanzahl von \mbox{$L=2$}, dass die Wahrscheinlichkeit
{\small${\T\binom{49}{6}^{-2}}$} bei zwei Spielen im Lotto
"`6 aus 49"' beide male einen "`Sechser"' zu haben gr"o"ser ist als die
Wahrscheinlichkeit, dass eine oder mehrere der $M$ Varianzen Null werden,
wenn man \mbox{$M\le94334$} w"ahlt. Daher wird man die oben beschriebene
Begrenzung der Inversen der empirischen Varianzen bei einer praktischen
Realisierung des RKM in Form eines Messprogramms nicht wirklich vornehmen,
zumal die Messwerte in diesem Fall sowieso keine Aussagekraft besitzen.
Sollte man bei einer realen Messung wirklich einmal das au"serordentliche
"`Gl"uck"' haben, auf diesen Fall zu sto"sen, so wird man die Messung einfach
wiederholen, oder die Mittelungsanzahl $L$ erh"ohen. Man wird dies auch
dann tun, wenn die Messwertvarianz zu gro"s wird. Dennoch ist Begrenzung der
Inversen der empirischen Varianz f"ur die Berechnung der stochastischen 
Eigenschaften der Messwerte unumg"anglich.

Oft verwendet man Zufallszahlengeneratoren die eine determinierte
Abfolge von Zahlen liefern, die z.~B. mit Hilfe r"uckgekoppelter
Schieberegister, oder mit Modulo-Arithmetik gewonnen werden, und die
sich mit einer extrem langen Periode wiederholen. In diesem Fall kann
der singul"are Fall nie eintreten, so dass man bei der Berechnung
der Messwerte den singul"aren Fall nicht extra abpr"ufen und getrennt
behandeln muss. Streng genommen ist die im Hauptteil beschriebene Theorie
bei Verwendung eines solchen Zufallszahlengenerators selbst dann nicht
mehr g"ultig, wenn man den Startwert des Zufallszahlengenerators zuf"allig
w"ahlt, weil dann die Unabh"angigkeit der einzelnen Elemente der Stichproben
nicht mehr gegeben ist. Die Praxis zeigt jedoch, dass die bisher
hergeleiteten Ergebnisse auch bei Verwendung solcher Zufallszahlengeneratoren
verwendet werden k"onnen.

\section[Grenzen f"ur die Hauptdiagonalelemente der inversen
Kovarianzmatrix]{Grenzen f"ur die Hauptdiagonalelemente der\\inversen
Kovarianzmatrix}\label{E.Kap.A.4}

Gegeben sei die \mbox{$R\!\times\!R$} Kovarianzmatrix \mbox{$\underline{C}$}.
Da diese hermitesch ist, l"asst sie sich
unit"ar kongruent auf Diagonalform transformieren.
\begin{equation}
\underline{S}\,=\,\underline{U}\CdoT\underline{C}\CdoT\underline{U}^{\HH}
\qquad\qquad\text{ mit }\qquad
\underline{U}\CdoT\underline{U}^{\HH} =\,\underline{E}
\quad\Leftrightarrow\quad\underline{U}^{-1} =\,\underline{U}^{\HH}
\label{E.A.26}
\end{equation}
Die Diagonalelemente der Diagonalmatrix $\underline{S}$ seien mit
$s_i$ bezeichnet. Sie sind die Eigenwerte der Kovarianzmatrix
\mbox{$\underline{C}$} und somit zugleich die Inversen der Eigenwerte
der inversen Kovarianzmatrix \mbox{$\underline{C}^{-1}$}. Da bei einer
hermiteschen Matrix die Eigenwerte gleich ihren Singul"arwerten
sind, sind $s_i$ zugleich die Singul"arwerte von \mbox{$\underline{C}$}
und die Inversen der Singul"arwerte von \mbox{$\underline{C}^{-1}$}.
Sie sind reell und nichtnegativ. Mit dem \mbox{$1\!\times\!R$}
Einheitszeilenvektor \mbox{$\Vec{E}_n$}, dessen $n$-tes Element eins ist,
und der sonst nur Nullen enth"alt, ergibt die quadratische Form
\begin{equation}
\Vec{E}_n\Cdot\underline{C}^{-1}\CdoT\Vec{E}_n^{\,\Hh}
\label{E.A.27}
\end{equation}
das $n$-te Hauptdiagonalelement der inversen Kovarianzmatrix. 
Wenn mit $\Vec{U}_n$ der $n$-te Spaltenvektor und mit $U_{j,n}$
das Element in der $j$-ten Zeile und in der $n$-ten Spalte der
Matrix $\underline{U}$ bezeichnet ist, kann man f"ur das
$n$-te Hauptdiagonalelement der inversen Kovarianzmatrix auch
\begin{gather}
\Vec{E}_n\Cdot\underline{C}^{-1}\CdoT\Vec{E}_n^{\,\Hh}\,=\;
\Vec{E}_n\Cdot\big(\,\underline{U}^{\HH}\Cdot\underline{S}\cdot\underline{U}\,
\big)^{-1}\!\Cdot\Vec{E}_n^{\,\Hh}\,=\;
\Vec{E}_n\Cdot\underline{U}^{-1}\Cdot\underline{S}^{-1}\Cdot
\big(\underline{U}^{\HH}\big)^{\!-1}\Cdot\Vec{E}_n^{\,\Hh}\,=
\notag\\*
=\;\Vec{E}_n\Cdot\underline{U}^{\HH}\Cdot\underline{S}^{-1}\Cdot
\underline{U}\cdot\Vec{E}_n^{\,\Hh}\,=\;
\Vec{U}_n^{\Hh}\Cdot\underline{S}^{-1}\Cdot\Vec{U}_n\;=\;
\Sum{j=1}{R}\,|U_{j,n}|^2\Cdot s_j^{-1}
\label{E.A.28}
\end{gather}
schreiben. F"ur diese Summe kann man eine obere und unterer Schranke
angeben.
\begin{gather}
\max_i(s_i)^{-1}\,=\; \min_i(s_i^{-1})\;=\;
\min_i(s_i^{-1})\cdoT \Sum{j=1}{R} |U_{j,n}|^2\,=\;
\Sum{j=1}{R}\,|U_{j,n}|^2\Cdot \min_i(s_i^{-1})\;\le
\notag\\*
\le\;\Sum{j=1}{R}\,|U_{j,n}|^2\Cdot s_j^{-1}\,\le
\label{E.A.29}\\*
\le\;\Sum{j=1}{R}\,|U_{j,n}|^2\Cdot \max_i(s_i^{-1})\;=\;
\max_i(s_i^{-1})\cdoT \Sum{j=1}{R}|U_{j,n}|^2\,=\;
\max_i(s_i^{-1})\;=\;\min_i(s_i)^{-1}\notag
\end{gather}
Die Hauptdiagonalelemente der inversen Kovarianzmatrix sind also alle
reell und liegen innerhalb des Intervalls
\mbox{${\D\big[\max_i(s_i)^{-1};\,\min_i(s_i)^{-1}\big]}$}.

\section[Zur Berechnung des Erwartungswertes einer bilinearen Form]{Zur Berechnung des Erwartungswertes einer \\bilinearen Form}\label{E.Kap.A.5}

Bei der Herleitung der Messwerte
\mbox{$\Hat{\Phi}_{\boldsymbol{n}}\big({\T\mu,\mu\!+\!\Tilde{\mu}\CdoT\frac{M}{K_{\Phi}}}\big)$} und 
\mbox{$\Hat{\Psi}_{\boldsymbol{n}}\big({\T\mu,\mu\!+\!\Tilde{\mu}\CdoT\frac{M}{K_{\Phi}}}\big)$} 
sowie der Varianzen der Messwerte
\mbox{$\Hat{H}\big({\T\mu,\mu\!+\!\Tilde{\mu}\CdoT\frac{M}{K_H}}\big)$}, 
\mbox{$\Hat{H}_*\big({\T\mu,\mu\!+\!\Tilde{\mu}\CdoT\frac{M}{K_H}}\big)$}, 
\mbox{$\Hat{u}(k)$} und 
\mbox{$\Hat{U}_{\!f}(\mu)$} 
tritt immer der Erwartungswert einer zuf"alligen bilinearen
Form auf. Dabei sind die Elemente der zuf"alligen Matrix der bilinearen 
Form Funktionen, die nur von dem zuf"alligen Spektrum der Erregung abh"angen.
Die Vektoren der bilinearen Form sind ebenfalls zuf"allig und h"angen nur von 
den Zeit- oder Spektralwerten des gefensterten Approximationsfehlerprozesses ab. 
Wegen der Gemeinsamkeit, die alle diese Erwartungswerte aufweisen,
werde ich in diesem Unterkapitel die Erwartungswerte 
\begin{equation}
\text{E}\Big\{\;
\Vec{\boldsymbol{N}}_{\!1}\Cdot
\underline{\boldsymbol{A}}\cdot
\Vec{\boldsymbol{N}}_{\!2}^{\hH}\,\Big\}
\label{E.A.30}
\end{equation}
einer bilinearen Form mit der allgemeineren Matrix \mbox{$\underline{\boldsymbol{A}}$}
und den allgemeineren Zufallsvektoren
\mbox{$\Vec{\boldsymbol{N}}_{\!1}$} und
\mbox{$\Vec{\boldsymbol{N}}_{\!2}$}
berechnen, um so eine Vielzahl von Berechnungen mit den im Einzelfall verwendeten 
Matrizen und Vektoren zu vermeiden. Die Zufallsvektoren sind jedoch nicht 
beliebig w"ahlbar. Es muss sich bei den $L$ Spalten der Matrix, 
die die zwei Zufallsvektoren
\mbox{$\Vec{\boldsymbol{N}}_{\!1}$} und 
\mbox{$\Vec{\boldsymbol{N}}_{\!2}$} als Zeilenvektoren enth"alt, 
um die Elemente einer mathematischen Stichprobe vom Umfang $L$ 
des Zufallsspaltenvektors \mbox{$\big[\boldsymbol{N}_{\!1},\,
\boldsymbol{N}_{\!2}\big]^{\TT}$} handeln. 
Somit sind alle $L$ Spaltenvektoren voneinander unabh"angig 
und haben dieselbe Verbundverteilung, n"amlich die Verbundverteilung 
des Zufallsspaltenvektors, aus dem die Stichprobe entnommen wurde. 
Wenn mit \mbox{$\boldsymbol{N}_{\!i,\lambda}$} das $\lambda$-te 
Element des Stichprobenvektors \mbox{$\Vec{\boldsymbol{N}}_{\!i}$} 
bezeichnet ist, gilt somit 
\begin{gather}
\text{E}\Big\{\boldsymbol{N}_{\!i,\lambda_1}\Cdot
\boldsymbol{N}_{\!j,\lambda_2}^*\Big\}\;=\;
\begin{cases}
\;\text{E}\big\{\boldsymbol{N}_{\!i}\CdoT\boldsymbol{N}_{\!j}^*\big\}&
\text{ f"ur }\lambda_1 = \lambda_2 \\
\;\text{E}\big\{\boldsymbol{N}_{\!i}\big\}\cdot
\text{E}\big\{\boldsymbol{N}_{\!j}\big\}^*\quad&
\text{ f"ur }\lambda_1 \neq \lambda_2
\end{cases}\label{E.A.31}\\*[8pt]
\forall\qquad\qquad i,j = 1\;(1)\;2
\;;\qquad\lambda,\lambda_1,\lambda_2 = 1\;(1)\;L.\notag
\end{gather}
Es sei darauf hingewiesen, dass die hier geforderte Unabh"angigkeit f"ur 
unterschiedliche Werte von $\lambda$ {\em nicht}\/ bedeutet, dass die 
Zufallsgr"o"sen \mbox{$\boldsymbol{N}_{\!1}$} und \mbox{$\boldsymbol{N}_{\!2}$} 
voneinander unabh"angig sein m"ussen. Diese k"onnen z.~B. zueinander konjugiert 
oder sogar identisch sein. Damit die weitere Herleitung ihre G"ultigkeit beh"alt, 
m"ussen die Elemente der Matrix \mbox{$\underline{\boldsymbol{A}}$} f"ur alle 
Werte von $\lambda$ von den Zufallsspaltenvektoren 
\mbox{$\big[\boldsymbol{N}_{\!1,\lambda},\,\boldsymbol{N}_{\!2,\lambda}\big]^{\TT}$}
unabh"angig sein. 

Wenn mit $\underline{\boldsymbol{A}}_{\lambda_1,\lambda_2}$ das Element 
in der $\lambda_1$-ten Zeile und der $\lambda_2$-ten Spalte bezeichnet ist, 
erhalten wir: 
\begin{subequations}\label{E.A.32}
\begin{equation}
\text{E}\Big\{\;
\Vec{\boldsymbol{N}}_{\!1}\Cdot
\underline{\boldsymbol{A}}\cdot
\Vec{\boldsymbol{N}}_{\!2}^{\hH}\,\Big\}\;=\;
\text{E}\bigg\{\,\Sum{\lambda_1=1}{L}\;\Sum{\lambda_2=1}{L}
\boldsymbol{N}_{\!1,\lambda_1}\Cdot
\underline{\boldsymbol{A}}_{\lambda_1,\lambda_2}\Cdot
\boldsymbol{N}_{\!2,\lambda_2}^*\bigg\}=\ldots
\label{E.A.32.a}
\end{equation}
Der Erwartungswert wird als Summe der Erwartungswerte berechnet.
\begin{equation}
\ldots\;=\;\Sum{\lambda_1=1}{L}\;\Sum{\lambda_2=1}{L}\text{E}\Big\{\,
\boldsymbol{N}_{\!1,\lambda_1}\Cdot
\underline{\boldsymbol{A}}_{\lambda_1,\lambda_2}\Cdot
\boldsymbol{N}_{\!2,\lambda_2}^*\,\Big\}\;=\;\ldots
\label{E.A.32.b}
\end{equation}
Die Doppelsumme wird in zwei Teilsummen aufgespalten. Die erste Teilsumme
enth"alt die Hauptdiagonalelemente, w"ahrend in der zweiten Teilsumme
die Nebendiagonalelemente auftreten.
\begin{equation}
\ldots\;=\;\Sum{\lambda=1}{L}
\text{E}\Big\{\,\boldsymbol{N}_{\!1,\lambda}\cdot
\underline{\boldsymbol{A}}_{\lambda,\lambda}\cdot\boldsymbol{N}_{\!2,\lambda}^*\,\Big\}\;+
\Sum{\lambda_1=1}{L}\;
\Sum{\substack{\lambda_2=1\;\;\\\lambda_2\neq\lambda_1}}{L}
\text{E}\Big\{\,\boldsymbol{N}_{\!1,\lambda_1}\Cdot
\underline{\boldsymbol{A}}_{\lambda_1,\lambda_2}\Cdot
\boldsymbol{N}_{\!2,\lambda_2}^*\,\Big\}\;=\;\ldots
\label{E.A.32.c}
\end{equation}
Nun wird die vorausgesetzte Unabh"angigkeit der Zufallselemente der
Matrix \mbox{$\underline{\boldsymbol{A}}$} von den Zufallsspaltenvektoren 
\mbox{$\big[\boldsymbol{N}_{\!1,\lambda},\,\boldsymbol{N}_{\!2,\lambda}\big]^{\TT}$}
dazu verwendet, den Erwartungswert des Produktes als das Produkt der
Erwartungswerte der einzelnen Faktoren zu schreiben.
\begin{equation}
\ldots\,=\Sum{\lambda=1}{L}
\text{E}\big\{\boldsymbol{N}_{\!1,\lambda}\CdoT\boldsymbol{N}_{\!2,\lambda}^*\big\}\CdoT
\text{E}\big\{\underline{\boldsymbol{A}}_{\lambda,\lambda}\big\}+\!
\Sum{\lambda_1=1}{L}\;
\Sum{\substack{\lambda_2=1\;\;\\\lambda_2\neq\lambda_1}}{L}\!
\text{E}\big\{\boldsymbol{N}_{\!1,\lambda_1}\CdoT
\boldsymbol{N}_{\!2,\lambda_2}^*\big\}\CdoT
\text{E}\big\{\underline{\boldsymbol{A}}_{\lambda_1,\lambda_2}\big\}=\ldots
\label{E.A.32.d}
\end{equation}
Desweiteren ber"ucksichtigen wir nun die Unabh"angigkeit der Stichprobenelemente, 
die in unterschiedlichen Einzelmessungen mit \mbox{$\lambda_1\!\neq\!\lambda_2$} 
gewonnen wurden. 
\begin{equation}
\ldots\,=\Sum{\lambda=1}{L}
\text{E}\big\{\boldsymbol{N}_{\!1,\lambda}\CdoT\boldsymbol{N}_{\!2,\lambda}^*\big\}\CdoT
\text{E}\big\{\underline{\boldsymbol{A}}_{\lambda,\lambda}\big\}+\!
\Sum{\lambda_1=1}{L}\;
\Sum{\substack{\lambda_2=1\;\;\\\lambda_2\neq\lambda_1}}{L}\!
\text{E}\big\{\boldsymbol{N}_{\!1,\lambda_1}\big\}\CdoT
\text{E}\big\{\boldsymbol{N}_{\!2,\lambda_2}^*\big\}\CdoT
\text{E}\big\{\underline{\boldsymbol{A}}_{\lambda_1,\lambda_2}\big\}=\ldots
\label{E.A.32.e}
\end{equation}
Au"serdem ist bei einer mathematischen Stichprobe die Verteilung 
der Stichprobenelemente gleich der Verteilung der Zufallsgr"o"se, aus der die 
Stichprobe gewonnen worden ist. Somit ist auch der Erwartungswert der 
Stichprobenelemente gleich dem Erwartungswert der Zufallsgr"o"se. 
\begin{equation}
\ldots\,=\Sum{\lambda=1}{L}
\text{E}\big\{\boldsymbol{N}_{\!1}\CdoT\boldsymbol{N}_{\!2}^*\big\}\CdoT
\text{E}\big\{\underline{\boldsymbol{A}}_{\lambda,\lambda}\big\}+\!
\Sum{\lambda_1=1}{L}\;
\Sum{\substack{\lambda_2=1\;\;\\\lambda_2\neq\lambda_1}}{L}\!
\text{E}\big\{\boldsymbol{N}_{\!1}\big\}\CdoT
\text{E}\big\{\boldsymbol{N}_{\!2}^*\big\}\CdoT
\text{E}\big\{\underline{\boldsymbol{A}}_{\lambda_1,\lambda_2}\big\}=\ldots
\label{E.A.32.f}
\end{equation}
Die von $\lambda$ bzw. von $\lambda_1$ und $\lambda_2$ unabh"angigen
Erwartungswerte werden vor die Summen gezogen.
\begin{equation}
\ldots\,=\,\text{E}\big\{\boldsymbol{N}_{\!1}\CdoT\boldsymbol{N}_{\!2}^*\big\}\CdoT\!
\Sum{\lambda=1}{L}
\text{E}\big\{\underline{\boldsymbol{A}}_{\lambda,\lambda}\big\}+
\text{E}\big\{\boldsymbol{N}_{\!1}\big\}\CdoT
\text{E}\big\{\boldsymbol{N}_{\!2}^*\big\}\CdoT\!
\Sum{\lambda_1=1}{L}\;\Sum{\substack{\lambda_2=1\;\;\\\lambda_2\neq\lambda_1}}{L}\!
\text{E}\big\{\underline{\boldsymbol{A}}_{\lambda_1,\lambda_2}\big\}=\ldots
\label{E.A.32.g}
\end{equation}
Nun erg"anzen wir die in der Doppelsumme fehlenden Hauptdiagonalelemente.
\begin{gather*}\label{E.A.32.h}
\begin{flalign}
&\ldots\,=\,\Big(\text{E}\big\{\boldsymbol{N}_{\!1}\CdoT\boldsymbol{N}_{\!2}^*\big\}-
\text{E}\big\{\boldsymbol{N}_{\!1}\big\}\CdoT
\text{E}\big\{\boldsymbol{N}_{\!2}^*\big\}\Big)\CdoT\!
\Sum{\lambda=1}{L}
\text{E}\big\{\underline{\boldsymbol{A}}_{\lambda,\lambda}\big\}\;+{}&&
\end{flalign}\\*[-2pt]\begin{flalign*}
&&{}+\text{E}\big\{\boldsymbol{N}_{\!1}\big\}\CdoT
\text{E}\big\{\boldsymbol{N}_{\!2}^*\big\}\CdoT\!
\Sum{\lambda_1=1}{L}\;\Sum{\lambda_2=1}{L}\!
\text{E}\big\{\underline{\boldsymbol{A}}_{\lambda_1,\lambda_2}\big\}=\ldots&
\end{flalign*}
\end{gather*}
Wieder vertauschen wir die Reihenfolge der Summation und der Erwartungswertbildung.
\begin{gather*}\label{E.A.32.i}
\begin{flalign}
&\ldots\,=\,\Big(\text{E}\big\{\boldsymbol{N}_{\!1}\CdoT\boldsymbol{N}_{\!2}^*\big\}-
\text{E}\big\{\boldsymbol{N}_{\!1}\big\}\CdoT
\text{E}\big\{\boldsymbol{N}_{\!2}^*\big\}\Big)\CdoT\
\text{E}\bigg\{\,\Sum{\lambda=1}{L}\underline{\boldsymbol{A}}_{\lambda,\lambda}\bigg\}\;+{}&&
\end{flalign}\\*[0pt]\begin{flalign*}
&&{}+\text{E}\big\{\boldsymbol{N}_{\!1}\big\}\CdoT
\text{E}\big\{\boldsymbol{N}_{\!2}^*\big\}\CdoT
\text{E}\bigg\{\,\Sum{\lambda_1=1}{L}\;\Sum{\lambda_2=1}{L}\!
\underline{\boldsymbol{A}}_{\lambda_1,\lambda_2}\bigg\}=\ldots&
\end{flalign*}
\end{gather*}
Mit dem \mbox{$1\times L$} Zeilenvektor \mbox{$\Vec{1}$}, der nur Einsen enth"alt, 
erhalten wir in Matrixschreibweise:
\begin{gather*}\label{E.A.32.j}
\begin{flalign}
&\ldots\,=\,\Big(\text{E}\big\{\boldsymbol{N}_{\!1}\CdoT\boldsymbol{N}_{\!2}^*\big\}-
\text{E}\big\{\boldsymbol{N}_{\!1}\big\}\CdoT
\text{E}\big\{\boldsymbol{N}_{\!2}^*\big\}\Big)\CdoT\
\text{E}\Big\{\,\text{spur}\big(\underline{\boldsymbol{A}}\big)\Big\}\;+{}&&
\end{flalign}\\*[6pt]\begin{flalign*}
&&{}+\text{E}\big\{\boldsymbol{N}_{\!1}\big\}\CdoT
\text{E}\big\{\boldsymbol{N}_{\!2}^*\big\}\CdoT
\text{E}\Big\{\Vec{1}\cdot\underline{\boldsymbol{A}}\cdot\Vec{1}^{\,\Hh}\Big\}.&
\end{flalign*}
\end{gather*}
\end{subequations}
In dieser Form k"onnen wir den gesuchten Erwartungswert der bilinearen Form im 
Hauptteil weiterverwenden.

\section{Zur Berechnung der Messwerte des LDS und des KLDS}\label{E.Kap.A.6}

Wenn man f"ur die Berechnung der Messwerte \mbox{$\Hat{\Phi}_{\boldsymbol{n}}
\big({\T\mu,\mu\!+\!\Tilde{\mu}\CdoT\frac{M}{K_{\Phi}}}\big)$}
und \mbox{$\Hat{\Psi}_{\boldsymbol{n}}
\big({\T\mu,\mu\!+\!\Tilde{\mu}\CdoT\frac{M}{K_{\Phi}}}\big)$}
die gem"a"s der Gleichungen~(\ref{E.3.27}) und (\ref{E.3.28}) definierten
Matrizen \mbox{$\underline{V}_{\bot}\!(\mu)$} verwendet, lassen
sich alle Messwerte immer durch Akkumulation aus den bei den 
Einzelmessungen $\lambda$ verwendeten Spektralwerten der Erregung 
und des gefensterten Ausgangssignals berechnen, ohne dass dazu die 
Spektralwerte aller Einzelmessungen abgespeichert werden m"ussen. 
Das gilt unabh"angig davon, welche Stichprobenvektoren
\mbox{$\Vec{V}(\Hat{\mu})$}, deren Anzahl mit $R$ bezeichnet sei,
man in die Matrizen \mbox{$\Breve{\underline{V}}(\mu)$} aufnimmt.
Hier soll nun erkl"art werden, wie sich die Messwerte in einer Form 
darstellen lassen, die diese Art der Berechnung erm"oglicht.

Nach den Gleichungen~(\ref{E.3.29}) berechnen sich die Messwerte jeweils 
als die Quotienten einer bilinearen Form und einer $M$-fachen Matrixspur. 
Nun dividieren wir Z"ahler und Nenner durch \mbox{$L\!-\!1$} und beginnen
mit der Berechnung des Z"ahlers, wobei wir jeweils die Darstellung des
Messwertes w"ahlen, die die Vektoren \mbox{$\Vec{Y}_{\!f}(\mu)$}
enth"alt. Die Matrix der bilinearen Form ist jeweils selbst ein
Produkt zweier Matrizen \mbox{$\underline{V}_{\bot}\!(\mu)$}
bei unterschiedlichen Frequenzen, wobei eine der beiden Matrizen noch
transponiert und ggf. konjugiert auftritt. Mit Gleichung~(\ref{E.3.28})
ersetzt man nun die Matrizen dieses Matrixproduktes 
jeweils durch die Differenz der Matrix \mbox{$\underline{1}_{\bot}$} 
und eines anderen Matrixproduktes. Das sich ergebende Produkt der
Matrixdifferenzen l"asst sich mit Hilfe des Distributivgesetzes der
Matrizenmultiplikation unter Beibehaltung der Reihenfolge der Matrizen
als vorzeichenbehaftete Summe von Matrixprodukten schreiben. Diese Summe
wird mit den Vektoren \mbox{$\Vec{Y}_{\!f}(\mu)$} bei
unterschiedlichen Frequenzen von links und transponiert
\vadjust{\penalty-100}und ggf. konjugiert von rechts multipliziert, was der
entsprechenden Multiplikation der einzelnen Summanden entspricht.
Wenn man nun ber"ucksichtigt, dass die Matrizen 
\mbox{$\underline{V}_{\bot}\!(\mu)$} hermitesch sind, dass
die Matrix $\underline{1}_{\bot}$ idempotent ist, und dass der
Z"ahler durch \mbox{$L\!-\!1$} dividiert wurde, erh"alt man
so eine Summe von Produkten von empirischen Kovarianzvektoren
und -matrizen und deren Inverser, wobei deren Dimensionen entweder
$1$ oder $R$ sind. Die Gr"o"se der Vektoren und Matrizen ist somit
von der Mittelungsanzahl $L$ unabh"angig. Die Elemente der
empirischen Kovarianzvektoren und -matrizen sind die empirischen
Kovarianzen der zuf"alligen Spektralwerte der Erregung und des
gefensterten Ausgangssignals. Diese kann man, wie zum Beispiel in den 
Gleichungen~(\ref{E.3.10}) und (\ref{E.3.11}) gezeigt durch Akkumulation 
aus den bei den Einzelmessungen $\lambda$ verwendeten Spektralwerten der
Erregung und der gemessenen, gefensterten Ausgangssignale
berechnen, ohne dass dazu die Spektralwerte aller Einzelmessungen
abgespeichert werden m"ussen. Am Beispiel des Messwertes
\mbox{$\Hat{\Phi}_{\boldsymbol{n}}
\big({\T\mu,\mu\!+\!\Tilde{\mu}\CdoT\frac{M}{K_{\Phi}}}\big)$} soll nun
die eben verbal beschriebene Vorgehensweise zur Berechnung des Z"ahlers des
Messwertes demonstriert werden.\vspace{-8pt}
\begin{gather}
\frac{\Vec{Y}_{\!f}(\mu)\cdot
\underline{V}_{\bot}\!(\mu)\cdot
\underline{V}_{\bot}\!\big({\T\mu\!+\!\Tilde{\mu}\CdoT\frac{M}{K_{\Phi}}}\big)^{\HH}\Cdot
\Vec{Y}_{\!f}\big({\T\mu\!+\!\Tilde{\mu}\CdoT\frac{M}{K_{\Phi}}}\big)^{\HH}}
{L\!-\!1}\;=
\label{E.A.33}\\*[8pt]\begin{flalign*}
&=\;\frac{1}{L\!-\!1}\cdot\Vec{Y}_{\!f}(\mu)\cdot\bigg(\,
\underline{1}_{\bot}\!-\underline{1}_{\bot}\Cdot\Breve{\underline{V}}(\mu)^{\Hh}
\Cdot\frac{
\Hat{\underline{C}}_{\Breve{\Vec{\boldsymbol{V}}}(\mu),\Breve{\Vec{\boldsymbol{V}}}(\mu)}^{\uP{0.4}{\!-1}}}
{L\!-\!1}\cdot
\Breve{\underline{V}}(\mu)\cdot\underline{1}_{\bot}\bigg)\cdot{}&&
\end{flalign*}\notag\\*[2pt]\begin{flalign*}
&&{}\!\CdoT\bigg(\underline{1}_{\bot}\!-\underline{1}_{\bot}\!\CdoT
\Breve{\underline{V}}\big({\T\mu\!+\!\Tilde{\mu}\CdoT\frac{M}{K_{\Phi}}}\big)^{\HH}\Cdot\frac{
\Hat{\underline{C}}_{\Breve{\Vec{\boldsymbol{V}}}(\mu+\Tilde{\mu}\cdot\frac{M}{K_{\Phi}}),\Breve{\Vec{\boldsymbol{V}}}(\mu+\Tilde{\mu}\cdot\frac{M}{K_{\Phi}})}^{\uP{0.4}{\!-1}}
}{L\!-\!1}\cdot\Breve{\underline{V}}\big({\T\mu\!+\!\Tilde{\mu}\CdoT\frac{M}{K_{\Phi}}}\big)\CdoT
\underline{1}_{\bot}\!\bigg)\CdoT\Vec{Y}_{\!f}\big({\T\mu\!+\!\Tilde{\mu}\CdoT\frac{M}{K_{\Phi}}}\big)^{\HH}\!=&
\end{flalign*}\notag\\[8pt]\begin{flalign*}
&=\;\underbrace{\frac{\Vec{Y}_{\!f}(\mu)\CdoT\underline{1}_{\bot}\!\CdoT
\Vec{Y}_{\!f}\big({\T\mu\!+\!\Tilde{\mu}\CdoT\frac{M}{K_{\Phi}}}\big)^{\HH}\!\!}{L\!-\!1}
}_{=\,\text{Kovarianz}}\;-{}&&
\end{flalign*}\notag\\*[2pt]
{}-\,\underbrace{\frac{\Vec{Y}_{\!f}(\mu)\CdoT
\underline{1}_{\bot}\!\CdoT\Breve{\underline{V}}\big({\T\mu\!+\!\Tilde{\mu}\CdoT\frac{M}{K_{\Phi}}}\big)^{\HH}\!\!}{L\!-\!1}
}_{=\,\text{Kovarianzvektor}}{}\cdot
\Hat{\underline{C}}_{\Breve{\Vec{\boldsymbol{V}}}(\mu+\Tilde{\mu}\cdot\frac{M}{K_{\Phi}}),\Breve{\Vec{\boldsymbol{V}}}(\mu+\Tilde{\mu}\cdot\frac{M}{K_{\Phi}})}^{\uP{0.4}{\!-1}}
\Cdot\underbrace{\frac{\Breve{\underline{V}}\big({\T\mu\!+\!\Tilde{\mu}\CdoT\frac{M}{K_{\Phi}}}\big)\CdoT
\underline{1}_{\bot}\!\CdoT\Vec{Y}_{\!f}\big({\T\mu\!+\!\Tilde{\mu}\CdoT\frac{M}{K_{\Phi}}}\big)^{\HH}\!\!}{L\!-\!1}
}_{=\,\text{Kovarianzvektor}}\,-{}
\notag\\*[2pt]
{}-\;\underbrace{\frac{\Vec{Y}_{\!f}(\mu)\CdoT
\underline{1}_{\bot}\!\CdoT
\Breve{\underline{V}}(\mu)^{\Hh}\!}{L\!-\!1}
}_{=\,\text{Kovarianzvektor}}{}\cdot
\Hat{\underline{C}}_{\Breve{\Vec{\boldsymbol{V}}}(\mu),\Breve{\Vec{\boldsymbol{V}}}(\mu)}^{\uP{0.4}{\!-1}}\Cdot
\underbrace{\frac{\Breve{\underline{V}}(\mu)\CdoT\underline{1}_{\bot}\!\CdoT
\Vec{Y}_{\!f}\big({\T\mu\!+\!\Tilde{\mu}\CdoT\frac{M}{K_{\Phi}}}\big)^{\HH}\!\!}{L\!-\!1}
}_{=\,\text{Kovarianzvektor}}\;+{}
\notag\\*[2pt]
{}+\;\underbrace{\frac{\Vec{Y}_{\!f}(\mu)\CdoT
\underline{1}_{\bot}\!\CdoT\Breve{\underline{V}}(\mu)^{\Hh}\!}{L\!-\!1}
}_{=\,\text{Kovarianzvektor}}{}\cdot
\Hat{\underline{C}}_{\Breve{\Vec{\boldsymbol{V}}}(\mu),\Breve{\Vec{\boldsymbol{V}}}(\mu)}^{\uP{0.4}{\!-1}}
\Cdot\underbrace{\frac{\Breve{\underline{V}}(\mu)\CdoT\underline{1}_{\bot}\!\CdoT
\Breve{\underline{V}}\big({\T\mu\!+\!\Tilde{\mu}\CdoT\frac{M}{K_{\Phi}}}\big)^{\HH}\!\!}{L\!-\!1}
}_{=\,\text{Kovarianzmatrix}}\cdot{}
\notag\\*[2pt]\begin{flalign*}
&&{}\cdot
\Hat{\underline{C}}_{\Breve{\Vec{\boldsymbol{V}}}(\mu+\Tilde{\mu}\cdot\frac{M}{K_{\Phi}}),\Breve{\Vec{\boldsymbol{V}}}(\mu+\Tilde{\mu}\cdot\frac{M}{K_{\Phi}})}^{\uP{0.4}{\!-1}}
\Cdot\underbrace{\frac{\Breve{\underline{V}}\big({\T\mu\!+\!\Tilde{\mu}\CdoT\frac{M}{K_{\Phi}}}\big)\CdoT
\underline{1}_{\bot}\!\CdoT\Vec{Y}_{\!f}\big({\T\mu\!+\!\Tilde{\mu}\CdoT\frac{M}{K_{\Phi}}}\big)^{\HH}\!\!}{L\!-\!1}
}_{=\,\text{Kovarianzvektor}}&
\end{flalign*}\notag
\end{gather}

Die Nenner der Messwerte der Gleichungen~(\ref{E.3.29}) sind jeweils die
$M$-fachen Spuren der Matrixprodukte der bilinearen Formen des Z"ahlers. 
Die dabei auftretenden Matrixspuren sind zwei der in der Tabelle~\ref{T.3.2} 
aufgef"uhrten Matrixspuren, n"amlich die, die mit \mb{d} und \mb{g} bezeichnet 
wurden. Wie man diese und auch alle anderen in der Tabelle~\ref{T.3.2} 
aufgelisteten Terme \mb{a} bis \mb{p} durch eine Akkumulation der Produkte 
der Spektralwerte der bei den Einzelmessungen verwendeten Erregungen
berechnen kann, soll nun erl"autert werden.

Zun"achst ersetzt man dazu wieder die Matrizen des Matrixproduktes,
dessen Spur berechnet werden soll, mit Hilfe der Gleichung~(\ref{E.3.28})
jeweils durch eine Differenz der Matrix \mbox{$\underline{1}_{\bot}$} 
und eines weiteren Matrixproduktes. Wieder l"asst sich das Produkt der
Matrixdifferenzen als vorzeichenbehaftete Summe von Matrixprodukten
schreiben. Die Spur ist dann die vorzeichenbehaftete Summe der Spuren
der einzelnen Summanden, also der Matrixprodukte. Jeder einzelne
Summand ist entweder die Matrix \mbox{$\underline{1}_{\bot}$} oder beginnt
mit \mbox{$\underline{1}_{\bot}\Cdot\Breve{\underline{V}}(\ldots)^{\Hh}$}
oder \mbox{$\underline{1}_{\bot}\Cdot\Breve{\underline{V}}(\ldots)^{\Tt}$}
und endet mit \mbox{$\Breve{\underline{V}}(\ldots)\cdot\underline{1}_{\bot}$}
oder \mbox{$\Breve{\underline{V}}(\ldots)^{\Kk}\Cdot\underline{1}_{\bot}$}.
Die Spur der \mbox{$L\!\times\!L$} Matrix \mbox{$\underline{1}_{\bot}$} ist 
immer \mbox{$L\!-\!1$} und braucht daher nicht durch Akkumulation der 
Stichprobenelemente berechnet zu werden. Da die Reihenfolge der Faktoren 
bei einem Produkt zweier Matrizen auf dessen Spur keinen Einfluss hat, 
kann man die ersten beiden Faktoren jedes einzelnen Matrixproduktes 
jeweils an das Ende des Matrixproduktes verschieben. Nun n"utzt man wieder 
die Tatsache, dass die Matrix $\underline{1}_{\bot}$ idempotent ist, und
erh"alt dadurch jeweils ein Matrixprodukt, bei dem sich Kovarianzmatrizen
und inverse Kovarianzmatrizen abwechseln. 

Wie Gleichung~(\ref{E.3.10}) zeigt, 
k"onnen die Kovarianzmatrizen und somit deren Spuren durch Akkumulation 
ohne Zwischenspeicherung der Spektralwerte aller Einzelmessungen berechnet 
werden. Am Beispiel der Matrixspur \mb{d} in Tabelle~\ref{T.3.2} sei dies
demonstriert.
\begin{gather}
\mb{d}\;=\;\text{spur}\Big(\,\underline{V}_{\bot}\!(\mu)\CdoT
\underline{V}_{\bot}\big({\T\mu\!+\!\Tilde{\mu}\CdoT\frac{M}{K_{\Phi}}}\big)\Big)\;=
\label{E.A.34}\\*[10pt]\begin{flalign*}
&\!=\;\text{spur}\Bigg(\bigg(\,
\underline{1}_{\bot}\!-\underline{1}_{\bot}\Cdot\Breve{\underline{V}}(\mu)^{\Hh}\Cdot
\frac{\Hat{\underline{C}}_{\Breve{\Vec{\boldsymbol{V}}}(\mu),\Breve{\Vec{\boldsymbol{V}}}(\mu)}^{\uP{0.4}{\!-1}}\!}
{L\!-\!1}\cdot\Breve{\underline{V}}(\mu)\cdot\underline{1}_{\bot}\bigg)\cdoT{}&&
\end{flalign*}\notag\\*[0pt]\begin{flalign*}
&&{}\Cdot\bigg(\,\underline{1}_{\bot}\!-\underline{1}_{\bot}\Cdot
\Breve{\underline{V}}\big({\T\mu\!+\!\Tilde{\mu}\CdoT\frac{M}{K_{\Phi}}}\big)^{\HH}\Cdot
\frac{\Hat{\underline{C}}_{\Breve{\Vec{\boldsymbol{V}}}(\mu+\Tilde{\mu}\cdot\frac{M}{K_{\Phi}}),\Breve{\Vec{\boldsymbol{V}}}(\mu+\Tilde{\mu}\cdot\frac{M}{K_{\Phi}})}^{\uP{0.4}{\!-1}}\!}
{L\!-\!1}\cdot\Breve{\underline{V}}\big({\T\mu\!+\!\Tilde{\mu}\CdoT\frac{M}{K_{\Phi}}}\big)\cdot
\underline{1}_{\bot}\bigg)\Bigg)\;=&
\end{flalign*}\notag\\[10pt]\begin{flalign*}
&\!=\;\text{spur}\Bigg(\,\underline{1}_{\bot}\Cdot\underline{1}_{\bot}-\;
\underline{1}_{\bot}\Cdot\underline{1}_{\bot}\Cdot
\Breve{\underline{V}}\big({\T\mu\!+\!\Tilde{\mu}\CdoT\frac{M}{K_{\Phi}}}\big)^{\HH}\Cdot
\frac{\Hat{\underline{C}}_{\Breve{\Vec{\boldsymbol{V}}}(\mu+\Tilde{\mu}\cdot\frac{M}{K_{\Phi}}),\Breve{\Vec{\boldsymbol{V}}}(\mu+\Tilde{\mu}\cdot\frac{M}{K_{\Phi}})}^{\uP{0.4}{\!-1}}\!}
{L\!-\!1}\cdot\Breve{\underline{V}}\big({\T\mu\!+\!\Tilde{\mu}\CdoT\frac{M}{K_{\Phi}}}\big)\cdot
\underline{1}_{\bot}-\!\!\!\!\!\!\!{}&&
\end{flalign*}\notag\\*[0pt]
{}\!-\;\underline{1}_{\bot}\Cdot\Breve{\underline{V}}(\mu)^{\Hh}
\Cdot\frac{\Hat{\underline{C}}_{\Breve{\Vec{\boldsymbol{V}}}(\mu),\Breve{\Vec{\boldsymbol{V}}}(\mu)}^{\uP{0.4}{\!-1}}\!}
{L\!-\!1}\cdot\Breve{\underline{V}}(\mu)\cdot
\underline{1}_{\bot}\Cdot\underline{1}_{\bot}\;+\;\underline{1}_{\bot}\Cdot\Breve{\underline{V}}(\mu)^{\Hh}\Cdot
\frac{\Hat{\underline{C}}_{\Breve{\Vec{\boldsymbol{V}}}(\mu),\Breve{\Vec{\boldsymbol{V}}}(\mu)}^{\uP{0.4}{\!-1}}}
{L\!-\!1}\cdoT{}
\notag\\*[4pt]\begin{flalign*}
&&{}\Cdot\Breve{\underline{V}}(\mu)\cdoT\underline{1}_{\bot}\!\CdoT
\underline{1}_{\bot}\!\CdoT\Breve{\underline{V}}\big({\T\mu\!+\!\Tilde{\mu}\CdoT\frac{M}{K_{\Phi}}}\big)^{\HH}\Cdot
\frac{\Hat{\underline{C}}_{\Breve{\Vec{\boldsymbol{V}}}(\mu+\Tilde{\mu}\cdot\frac{M}{K_{\Phi}}),\Breve{\Vec{\boldsymbol{V}}}(\mu+\Tilde{\mu}\cdot\frac{M}{K_{\Phi}})}^{\uP{0.4}{\!-1}}}
{L\!-\!1}\cdoT\Breve{\underline{V}}\big({\T\mu\!+\!\Tilde{\mu}\CdoT\frac{M}{K_{\Phi}}}\big)\CdoT
\underline{1}_{\bot}\!\Bigg)=&
\end{flalign*}\notag\displaybreak[2]\\[0pt]\begin{flalign*}
&\!=\text{spur}\big(\underline{1}_{\bot}\big)-
\text{spur}\bigg(\underline{1}_{\bot}\Cdot
\Breve{\underline{V}}\big({\T\mu\!+\!\Tilde{\mu}\CdoT\frac{M}{K_{\Phi}}}\big)^{\HH}\Cdot
\frac{\Hat{\underline{C}}_{\Breve{\Vec{\boldsymbol{V}}}(\mu+\Tilde{\mu}\cdot\frac{M}{K_{\Phi}}),\Breve{\Vec{\boldsymbol{V}}}(\mu+\Tilde{\mu}\cdot\frac{M}{K_{\Phi}})}^{\uP{0.4}{\!-1}}\!}
{L\!-\!1}\cdot\Breve{\underline{V}}\big({\T\mu\!+\!\Tilde{\mu}\CdoT\frac{M}{K_{\Phi}}}\big)
\cdot\underline{1}_{\bot}\bigg)-\!\!\!\!{}&&
\end{flalign*}\notag\\*[0pt]
{}\!-\;\text{spur}\bigg(\,\underline{1}_{\bot}\Cdot
\Breve{\underline{V}}(\mu)^{\Hh}\Cdot
\frac{\Hat{\underline{C}}_{\Breve{\Vec{\boldsymbol{V}}}(\mu),\Breve{\Vec{\boldsymbol{V}}}(\mu)}^{\uP{0.4}{\!-1}}\!}
{L\!-\!1}\cdot\Breve{\underline{V}}(\mu)\cdot\underline{1}_{\bot}\bigg)\;+\!{}
\notag\\*[4pt]\begin{flalign*}
&{}\!+\text{spur}\bigg(\underline{1}_{\bot}\Cdot
\Breve{\underline{V}}(\mu)^{\Hh}\Cdot
\frac{\Hat{\underline{C}}_{\Breve{\Vec{\boldsymbol{V}}}(\mu),\Breve{\Vec{\boldsymbol{V}}}(\mu)}^{\uP{0.4}{\!-1}}}
{L\!-\!1}\cdot\Breve{\underline{V}}(\mu)\cdoT\underline{1}_{\bot}\Cdot
\Breve{\underline{V}}\big({\T\mu\!+\!\Tilde{\mu}\CdoT\frac{M}{K_{\Phi}}}\!\big)^{\HH}\Cdot
\frac{\Hat{\underline{C}}_{\Breve{\Vec{\boldsymbol{V}}}(\mu+\Tilde{\mu}\cdot\frac{M}{K_{\Phi}}),\Breve{\Vec{\boldsymbol{V}}}(\mu+\Tilde{\mu}\cdot\frac{M}{K_{\Phi}})}^{\uP{0.4}{\!-1}}}
{L\!-\!1}\cdoT{}&&
\end{flalign*}\notag\\*[0pt]\begin{flalign*}
&&{}\Cdot\Breve{\underline{V}}\big({\T\mu\!+\!\Tilde{\mu}\CdoT\frac{M}{K_{\Phi}}}\!\big)\CdoT
\underline{1}_{\bot}\!\bigg)=&
\end{flalign*}\notag\\[4pt]\begin{flalign*}
&\!=\text{spur}\big(\underline{1}_{\bot}\big)-
\text{spur}\bigg(\,\frac{\Hat{\underline{C}}_{\Breve{\Vec{\boldsymbol{V}}}(\mu+\Tilde{\mu}\cdot\frac{M}{K_{\Phi}}),\Breve{\Vec{\boldsymbol{V}}}(\mu+\Tilde{\mu}\cdot\frac{M}{K_{\Phi}})}^{\uP{0.4}{\!-1}}\!}
{L\!-\!1}\cdot\Breve{\underline{V}}\big({\T\mu\!+\!\Tilde{\mu}\CdoT\frac{M}{K_{\Phi}}}\big)\cdot
\underline{1}_{\bot}\Cdot\underline{1}_{\bot}\cdot
\Breve{\underline{V}}\big({\T\mu\!+\!\Tilde{\mu}\CdoT\frac{M}{K_{\Phi}}}\big)^{\HH}\bigg)-\!\!\!{}&&
\end{flalign*}\notag\\*[4pt]
{}\!-\;\text{spur}\bigg(\,\frac{\Hat{\underline{C}}_{\Breve{\Vec{\boldsymbol{V}}}(\mu),\Breve{\Vec{\boldsymbol{V}}}(\mu)}^{\uP{0.4}{\!-1}}\!}
{L\!-\!1}\cdot\Breve{\underline{V}}(\mu)\cdot\underline{1}_{\bot}\Cdot
\underline{1}_{\bot}\cdot\Breve{\underline{V}}(\mu)^{\Hh}\bigg)\;+\!{}
\notag\\*[2pt]\begin{flalign*}
&{}\!+\text{spur}\bigg(\frac{\Hat{\underline{C}}_{\Breve{\Vec{\boldsymbol{V}}}(\mu),\Breve{\Vec{\boldsymbol{V}}}(\mu)}^{\uP{0.4}{\!-1}}}
{L\!-\!1}\cdot\Breve{\underline{V}}(\mu)\cdoT\underline{1}_{\bot}\Cdot
\Breve{\underline{V}}\big({\T\mu\!+\!\Tilde{\mu}\CdoT\frac{M}{K_{\Phi}}}\big)^{\HH}\Cdot
\frac{\Hat{\underline{C}}_{\Breve{\Vec{\boldsymbol{V}}}(\mu+\Tilde{\mu}\cdot\frac{M}{K_{\Phi}}),\Breve{\Vec{\boldsymbol{V}}}(\mu+\Tilde{\mu}\cdot\frac{M}{K_{\Phi}})}^{\uP{0.4}{\!-1}}}
{L\!-\!1}\cdoT{}&&
\end{flalign*}\notag\\*[0pt]\begin{flalign*}
&&{}\Cdot\Breve{\underline{V}}\big({\T\mu\!+\!\Tilde{\mu}\CdoT\frac{M}{K_{\Phi}}}\big)\CdoT
\underline{1}_{\bot}\Cdot\underline{1}_{\bot}\Cdot
\Breve{\underline{V}}(\mu)^{\Hh}\bigg)=&
\end{flalign*}\notag\\[14pt]\begin{flalign*}
&\!=\;\underbrace{\text{spur}\big(\underline{1}_{\bot}\big)}_{=\,L-1}\;-\;
\text{spur}\bigg(\,\Hat{\underline{C}}_{\Breve{\Vec{\boldsymbol{V}}}(\mu+\Tilde{\mu}\cdot\frac{M}{K_{\Phi}}),\Breve{\Vec{\boldsymbol{V}}}(\mu+\Tilde{\mu}\cdot\frac{M}{K_{\Phi}})}^{\uP{0.4}{\!-1}}\Cdot
\underbrace{\frac{\Breve{\underline{V}}\big({\T\mu\!+\!\Tilde{\mu}\CdoT\frac{M}{K_{\Phi}}}\big)\CdoT
\underline{1}_{\bot}\!\CdoT\Breve{\underline{V}}\big({\T\mu\!+\!\Tilde{\mu}\CdoT\frac{M}{K_{\Phi}}}\big)^{\HH}}{L\!-\!1}
}_{=\,\text{Kovarianzmatrix}}\bigg)\;-\!\!\!{}&&
\end{flalign*}\notag\\*[9pt]
{}\!-\;\text{spur}\bigg(\,\Hat{\underline{C}}_{\Breve{\Vec{\boldsymbol{V}}}(\mu),\Breve{\Vec{\boldsymbol{V}}}(\mu)}^{\uP{0.4}{\!-1}}\Cdot
\underbrace{\frac{\Breve{\underline{V}}(\mu)\CdoT\underline{1}_{\bot}\!\CdoT
\Breve{\underline{V}}(\mu)^{\Hh}}{L\!-\!1}
}_{=\,\text{Kovarianzmatrix}}\bigg)\;+\!{}
\notag\\*[9pt]\begin{flalign*}
&{}\!+\;\text{spur}\bigg(
\Hat{\underline{C}}_{\Breve{\Vec{\boldsymbol{V}}}(\mu),\Breve{\Vec{\boldsymbol{V}}}(\mu)}^{\uP{0.4}{\!-1}}\Cdot
\underbrace{\frac{\Breve{\underline{V}}(\mu)\CdoT\underline{1}_{\bot}\!\CdoT
\Breve{\underline{V}}\big({\T\mu\!+\!\Tilde{\mu}\CdoT\frac{M}{K_{\Phi}}}\big)^{\HH}\!}{L\!-\!1}
}_{=\,\text{Kovarianzmatrix}}{}\cdot
\Hat{\underline{C}}_{\Breve{\Vec{\boldsymbol{V}}}(\mu+\Tilde{\mu}\cdot\frac{M}{K_{\Phi}}),\Breve{\Vec{\boldsymbol{V}}}(\mu+\Tilde{\mu}\cdot\frac{M}{K_{\Phi}})}^{\uP{0.4}{\!-1}}\CdoT{}&&
\end{flalign*}\notag\\*[-6pt]\begin{flalign*}
&&{}\Cdot\underbrace{\frac{\Breve{\underline{V}}\big({\T\mu\!+\!\Tilde{\mu}\CdoT\frac{M}{K_{\Phi}}}\big)\CdoT
\underline{1}_{\bot}\!\CdoT\Breve{\underline{V}}(\mu)^{\Hh}\!}{L\!-\!1}
}_{=\,\text{Kovarianzmatrix}}\bigg)&
\end{flalign*}\notag
\end{gather}

F"uhrt man diese Umformungen bei allen in der Tabelle~\ref{T.3.2} 
aufgef"uhrten Termen durch, so stellt man bei den Termen 
\mb{a}, \mb{b}, \mb{c}, \mb{d}, \mb{g}, \mb{i} und \mb{l} fest, 
dass die Matrixspuren nur "uber die Produkte solcher empirischer Kovarianzmatrizen 
und deren Inverser zu berechnen sind, die bereits bei der Berechnung der Messwerte 
\mbox{$\Hat{\Phi}_{\boldsymbol{n}} (\mu,\mu\!+\!\Tilde{\mu}\CdoT M/K_{\Phi})$} und 
\mbox{$\Hat{\Psi}_{\boldsymbol{n}}(\mu,\mu\!+\!\Tilde{\mu}\CdoT M/K_{\Phi})$} 
aufgetreten sind. Die Terme \mb{e}, \mb{f}, \mb{h}, \mb{j}, \mb{k}, \mb{m}, \mb{n}, \mb{o} und 
\mb{p} sind nur bei den Frequenzen \mbox{$\mu=0\;\big(\frac{M}{2\cdot K_{\Phi}}\big)\;M\!-\!1$}
des Falles~(\ref{E.3.55.a}) in den Messwert"-(ko)"-varianzen zu finden. 
Bei diesen Frequenzen enthalten die Matrixspuren zus"atzlich noch einige 
weitere Kovarianzmatrizen. Diese hat man bei der Berechnung anderer Messwerte
\mbox{$\Hat{\Phi}_{\boldsymbol{n}}(\mu,\mu\!+\!\Breve{\mu}\CdoT M/K_{\Phi})$} und 
\mbox{$\Hat{\Psi}_{\boldsymbol{n}}(\mu,\mu\!+\!\Breve{\mu}\CdoT M/K_{\Phi})$}
jedoch ebenfalls bereits bestimmt. Der Parameter $\Breve{\mu}$, 
der angibt, bei der Berechnung welcher Messwerte die Kovarianzmatrizen 
bestimmt worden sind, ist dabei eine ganze Zahl, die sich mit den
Gleichungen~(\ref{E.3.55}) aus $\mu$ und $\Tilde{\mu}$ ermitteln l"asst.
Alle Kovarianzmatrizen, die man zur Berechnung der Messwert"-(ko)"-varianzen 
ben"otigt, sind also bereits nach der Berechnung der Messwerte bekannt, 
so dass man keine zus"atzlichen Akkumulatoren vorsehen muss.

\section{Zu Spur und Rang des Produktes Idempotenter Matrizen}\label{E.Kap.A.7}

Es treten immer wieder Produkte der idempotenten Matrizen
\mbox{$\underline{\boldsymbol{V}}_{\bot}\!(\mu)$} f"ur unterschiedliche
Frequenzen und deren Transponierte auf. Der Rangdefekt aller dieser Matrizen
ist auf eine gemeinsame obere Schranke \mbox{$D_{\text{max}}$} begrenzt.
Da der Rang der Transponierten einer Matrix gleich dem Rang der Matrix ist,
ist auch deren Rangdefekt auf die gleiche Schranke begrenzt. Die
Transponierte einer idempotenten Matrix ist ebenfalls wieder idempotent.
Wir wollen uns nun mit dem Produkt
\mbox{$\underline{\boldsymbol{I}}_{\Pi}\;=\;
\underline{\boldsymbol{I}}_1\cdot
\underline{\boldsymbol{I}}_2\cdot\ldots\cdot
\underline{\boldsymbol{I}}_n$}
von $n$ idempotenten \mbox{$L\!\times\!L$}-Matrizen besch"aftigen,
die alle einen Rangdefekt aufweisen, der bei allen $n$ Matrizen
auf \mbox{$D_{\text{max}}$} begrenzt ist. Die Spur einer idempotenten
Matrix ist immer gleich ihrem Rang. Daher ist die Spur jeder Matrix
des Produkts auf den Bereich
\begin{equation}
L\!-\!D_{\text{max}}\;\le\;\text{spur}(\underline{\boldsymbol{I}}_i)\;=\;
\text{rang}(\underline{\boldsymbol{I}}_i)\;\le\;
L
\qquad\qquad\forall\qquad i=1\;(1)\;n
\label{E.A.35}
\end{equation}
beschr"ankt. Jeder Zeilenvektor $\Vec{\boldsymbol{I}}_1$, der zu allen
Eigenvektoren aller $n$ Matrizen \mbox{$\underline{\boldsymbol{I}}_i$}
zum Eigenwert Null orthogonal ist, ist ein Eigenvektor aller $n$ Matrizen
\mbox{$\underline{\boldsymbol{I}}_i$} zum Eigenwert Eins.
Wenn also ein Vektor, f"ur den
\begin{equation}
\Vec{\boldsymbol{I}}_1\cdot\underline{\boldsymbol{I}}_i\;=\;
\Vec{\boldsymbol{I}}_1.
\qquad\qquad\forall\qquad i=1\;(1)\;n
\label{E.A.36}
\end{equation}
gilt, an das Matrixprodukt von links heranmultipliziert wird, so ergibt sich:
\begin{equation}
\Vec{\boldsymbol{I}}_1\cdot\,
\underline{\boldsymbol{I}}_1\Cdot
\underline{\boldsymbol{I}}_2\Cdot\ldots\cdot
\underline{\boldsymbol{I}}_n\;=\;
\Vec{\boldsymbol{I}}_1\cdot\,
\underline{\boldsymbol{I}}_2\Cdot
\underline{\boldsymbol{I}}_3\Cdot\ldots\cdot
\underline{\boldsymbol{I}}_n\;=\;\ldots\;=\;
\Vec{\boldsymbol{I}}_1\cdot\,\underline{\boldsymbol{I}}_n\;=\;
\Vec{\boldsymbol{I}}_1.
\label{E.A.37}
\end{equation}
Daher ist jeder dieser Vektoren $\Vec{\boldsymbol{I}}_1$ ein Eigenvektor
des Matrixprodukts $\underline{\boldsymbol{I}}_{\Pi}$  zum Eigenwert Eins.
Der Raum, der durch alle Vektoren aufgespannt wird, die zu allen
Eigenvektoren aller $n$ Matrizen \mbox{$\underline{\boldsymbol{I}}_i$}
zum Eigenwert Null orthogonal sind, hat mindestens die Dimension
\mbox{$L\!-\!n\CdoT D_{\text{max}}$}, weil es insgesamt h"ochstens
\mbox{$n\CdoT D_{\text{max}}$} Eigenvektoren aller $n$ Matrizen
\mbox{$\underline{\boldsymbol{I}}_i$} zum Eigenwert Null gibt. Das
Matrixprodukt hat daher --- wenn $L$ gro"s genug ist --- den Eigenwert Eins
mit einer Vielfachheit von mindestens \mbox{$L\!-\!n\CdoT D_{\text{max}}$}.
Multipliziert man einen beliebigen Vektor von links an des Matrixprodukt
\mbox{$\underline{\boldsymbol{I}}_{\Pi}$}, so entspricht dies einer
mehrmaligen Projektion des Vektors und dessen euklidische Norm wird
dadurch niemals anwachsen. Daher sind die Betr"age der restlichen maximal
\mbox{$n\CdoT D_{\text{max}}$} Eigenwerte kleiner gleich eins. Der Betrag
der Spur des Matrixprodukts \mbox{$\underline{\boldsymbol{I}}_{\Pi}$}
--- also der Betrag der Summe der Eigenwerte --- ist daher f"ur
hinreichend gro"ses $L$ immer gr"o"ser als Null und liegt im Bereich 
\begin{equation}
L-2\CdoT n\CdoT D_{\text{max}}\;\le\;
\big|\text{spur}(\underline{\boldsymbol{I}}_{\Pi})\big|\;\le\;L.
\label{E.A.38}
\end{equation}
Wird die Anzahl $n$ der am Matrixprodukt beteiligten Matrizen und der
gemeinsame maximale Rangdefekt $D_{\text{max}}$ all dieser Matrizen
konstant gehalten, so w"achst der Betrag der Spur des Matrixprodukts
f"ur hinreichend gro"ses $L$ also n"aherungsweise linear mit steigendem $L$.

\section{Zur Berechnung der Varianzen und Kovarianzen des
LDS und KLDS}\label{E.Kap.A.8}

Mit Hilfe der Gleichung~(\myref{A.5.9}) im Anhang~\myref{4Mom}
gelang es uns bei einem station"aren Approximationsfehlerprozess
die Varianzen und Kovarianzen der Messwerte
\mbox{$\Hat{\boldsymbol{\Phi}}_{\!\boldsymbol{n}}(\mu)$} und
\mbox{$\Hat{\boldsymbol{\Psi}}_{\!\boldsymbol{n}}(\mu)$}
zu berechnen. Im Fall eines zyklostation"aren Approximationsfehlerprozesses
ben"otigen wir stattdessen vier Gleichungssysteme, die sich aus
Gleichung~(\myref{A.5.9}) ableiten lassen. Das erste Gleichungssystem
erhalten wir, indem wir die Gleichung~(\myref{A.5.9}) dreimal
untereinander schreiben, wobei wir in den einzelnen
Zeilen die in der Tabelle~\ref{T.A.1} aufgef"uhrten Substitutionen vornehmen.
Die Zufallsmatrizen $\underline{\boldsymbol{V}}_1$,
$\underline{\boldsymbol{V}}_2$, $\underline{\boldsymbol{V}}_3$ und
$\underline{\boldsymbol{V}}_4$ sind dabei wieder idempotente,
hermitesche \mbox{$L\!\times\!L$} Matrizen, deren Rangdefekt mit
steigender Mittelungsanzahl $L$ auf eine Konstante begrenzt ist.
Die Elemente dieser Matrizen sind aus mathematischen Stichproben
der Spektralwerte der zuf"alligen Erregung berechnet worden.
Die Zufallsvektoren $\Vec{\boldsymbol{N}}_{\!1}$,
$\Vec{\boldsymbol{N}}_{\!2}$, $\Vec{\boldsymbol{N}}_{\!3}$
und $\Vec{\boldsymbol{N}}_{\!4}$ enthalten die Elemente der
mathematischen Stichproben der als normalverteilt und mittelwertfrei
angenommenen Spektralwerte $\boldsymbol{N}_{\!1}$,
$\boldsymbol{N}_{\!2}$, $\boldsymbol{N}_{\!3}$ und $\boldsymbol{N}_{\!4}$
des gefensterten Approximationsfehlerprozesses. Auch diese Zufallsvektoren 
werden in der in Tabelle~\ref{T.A.1} angegebenen Art substituiert. Es wird 
wieder angenommen, dass die in die Elemente der Zufallsmatrizen eingehenden 
zuf"alligen Spektralwerte der Erregung unabh"angig von den Spektralwerten 
seien, deren mathematische Stichproben die Zufallsvektoren sind, so dass 
die im Anhang~\myref{4Mom} aufgestellten Forderungen alle erf"ullt sind. 
Die drei Gleichungen, die man mit den angegebenen Substitutionen erhalten hat, 
lassen sich auch als eine Gleichung
\begin{equation}
\text{E}\big\{\Hat{\Vec{\boldsymbol{K}}}\big\}\;=\;
\Big(\underline{E}+\text{E}\big\{
\underline{\boldsymbol{D}}^{-1}\!\CdoT\underline{\boldsymbol{S}}\big\}\Big)\cdot
\Tilde{\Vec{K}}\;=\;\text{E}\{\underline{\boldsymbol{F}}\}\cdot\Tilde{\Vec{K}}
\label{E.A.39}
\end{equation}
schreiben, bei der auf beiden Seiten ein \mbox{$3\!\times\!1$} Vektor steht.
Bei dieser Gleichung enthalten die Vektoren
\begin{subequations}
\label{E.A.40}
\begin{equation}
\Hat{\Vec{\boldsymbol{K}}}\;=\;
\begin{bmatrix}
\boldsymbol{c}_1^{\!-1}\CdoT
\Vec{\boldsymbol{N}}_{\!1}\CdoT
\underline{\boldsymbol{V}}_1\CdoT
\underline{\boldsymbol{V}}_2^{\Hh}\CdoT
\Vec{\boldsymbol{N}}_{\!2}^{\hH}\CdoT
\Vec{\boldsymbol{N}}_{\!3}\CdoT
\underline{\boldsymbol{V}}_3\CdoT
\underline{\boldsymbol{V}}_4^{\Hh}\CdoT
\Vec{\boldsymbol{N}}_{\!4}^{\hH}\\[4pt]
\boldsymbol{c}_2^{\!-1}\CdoT
\Vec{\boldsymbol{N}}_{\!1}\CdoT
\underline{\boldsymbol{V}}_1\CdoT
\underline{\boldsymbol{V}}_3^{\Tt}\CdoT
\Vec{\boldsymbol{N}}_{\!3}^{\tT}\CdoT
\Vec{\boldsymbol{N}}_{\!2}^*\CdoT
\underline{\boldsymbol{V}}_2^{\Kk}\CdoT
\underline{\boldsymbol{V}}_4^{\Hh}\CdoT
\Vec{\boldsymbol{N}}_{\!4}^{\hH}\\[4pt]
\boldsymbol{c}_3^{\!-1}\CdoT
\Vec{\boldsymbol{N}}_{\!1}\CdoT
\underline{\boldsymbol{V}}_1\CdoT
\underline{\boldsymbol{V}}_4^{\Hh}\CdoT
\Vec{\boldsymbol{N}}_{\!4}^{\hH}\CdoT
\Vec{\boldsymbol{N}}_{\!2}^*\CdoT
\underline{\boldsymbol{V}}_2^{\Kk}\CdoT
\underline{\boldsymbol{V}}_3^{\Tt}\CdoT
\Vec{\boldsymbol{N}}_{\!3}^{\tT}
\end{bmatrix}
\label{E.A.40.a}
\end{equation}
als 
\begin{table}[t!]
\[
{\renewcommand{\arraystretch}{1.5}
\begin{array}{||c||c|c|c|c|c|c|c||}
\hline
\hline
\text{in (\myref{A.5.9}):}\!&
\Vec{\boldsymbol{N}}_{\!1}&
\Vec{\boldsymbol{N}}_{\!2}&
\Vec{\boldsymbol{N}}_{\!3}&
\Vec{\boldsymbol{N}}_{\!4}&
\underline{\boldsymbol{A}\,}&
\underline{\boldsymbol{B}}&
\boldsymbol{c}\\
\hline
\text{in Zeile 1:}&
\Vec{\boldsymbol{N}}_{\!1}&
\Vec{\boldsymbol{N}}_{\!2}&
\Vec{\boldsymbol{N}}_{\!3}&
\Vec{\boldsymbol{N}}_{\!4}&
\underline{\boldsymbol{V}}_1\CdoT
\underline{\boldsymbol{V}}_2^{\Hh}&
\underline{\boldsymbol{V}}_3\CdoT
\underline{\boldsymbol{V}}_4^{\Hh}&
\boldsymbol{c}_1^{\!-1}\\
\text{in Zeile 2:}&
\Vec{\boldsymbol{N}}_{\!1}&
\Vec{\boldsymbol{N}}_{\!3}^*&
\Vec{\boldsymbol{N}}_{\!2}^*&
\Vec{\boldsymbol{N}}_{\!4}&
\underline{\boldsymbol{V}}_1\CdoT
\underline{\boldsymbol{V}}_3^{\Tt}&
\underline{\boldsymbol{V}}_2^{\Kk}\CdoT
\underline{\boldsymbol{V}}_4^{\Hh}&
\boldsymbol{c}_2^{\!-1}\\
\text{in Zeile 3:}&
\Vec{\boldsymbol{N}}_{\!1}&
\Vec{\boldsymbol{N}}_{\!4}&
\Vec{\boldsymbol{N}}_{\!2}^*&
\Vec{\boldsymbol{N}}_{\!3}^*&
\underline{\boldsymbol{V}}_1\CdoT
\underline{\boldsymbol{V}}_4^{\Hh}&
\underline{\boldsymbol{V}}_2^{\Kk}\CdoT
\underline{\boldsymbol{V}}_3^{\Tt}&
\boldsymbol{c}_3^{\!-1}\\
\hline
\hline
\multicolumn{2}{||r}{\boldsymbol{c}_1\!\!}&
\multicolumn{6}{l||}{=\;\text{spur}\big(\underline{\boldsymbol{V}}_1\CdoT
\underline{\boldsymbol{V}}_2^{\Hh}\big)\cdot
\text{spur}\big(\underline{\boldsymbol{V}}_3\CdoT
\underline{\boldsymbol{V}}_4^{\Hh}\big)}\\
\multicolumn{2}{||r}{\text{mit}\qquad\qquad\boldsymbol{c}_2\!\!}&
\multicolumn{6}{l||}{=\;\text{spur}\big(\underline{\boldsymbol{V}}_1\CdoT
\underline{\boldsymbol{V}}_3^{\Tt}\big)\cdot
\text{spur}\big(\underline{\boldsymbol{V}}_2^{\Kk}\CdoT
\underline{\boldsymbol{V}}_4^{\Hh}\big)}\\
\multicolumn{2}{||r}{\boldsymbol{c}_3\!\!}&
\multicolumn{6}{l||}{=\;\text{spur}\big(\underline{\boldsymbol{V}}_1\CdoT
\underline{\boldsymbol{V}}_4^{\Hh}\big)\cdot
\text{spur}\big(\underline{\boldsymbol{V}}_2^{\Kk}\CdoT
\underline{\boldsymbol{V}}_3^{\Tt}\big)}\\
\hline
\hline
\end{array}}
\]\vspace{-17pt}
\setlength{\belowcaptionskip}{-4pt}
\caption{Drei Substitutionen in der Gleichung~(\myref{A.5.9}).}
\rule{\textwidth}{0.5pt}\vspace{-20pt}
\label{T.A.1}
\end{table}
Elemente jeweils das Produkt zweier empirischer Kovarianzen.
Die Produkte der theoretischen Kovarianzen derselben Zufallsgr"o"sen
sind in dem Vektor
\begin{equation}
\Tilde{\Vec{K}}\;=\;
\begin{bmatrix}
\text{E}\big\{\boldsymbol{N}_{\!1}  \CdoT
          \boldsymbol{N}_{\!2}^{\Kk}     \big\}\cdot
\text{E}\big\{\boldsymbol{N}_{\!3}  \CdoT
          \boldsymbol{N}_{\!4}^{\Kk}     \big\}^{\phantom{\!\Kk}}\\[4pt]
\text{E}\big\{\boldsymbol{N}_{\!1}  \CdoT
          \boldsymbol{N}_{\!3}^{\phantom{*}}\big\}\cdot
\text{E}\big\{\boldsymbol{N}_{\!2}  \CdoT
          \boldsymbol{N}_{\!4}^{\phantom{*}}\big\}^{\!\Kk}\\[4pt]
\text{E}\big\{\boldsymbol{N}_{\!1}  \CdoT
          \boldsymbol{N}_{\!4}^{\Kk}     \big\}\cdot
\text{E}\big\{\boldsymbol{N}_{\!2}  \CdoT
          \boldsymbol{N}_{\!3}^{\Kk}     \big\}^{\!\Kk}
\end{bmatrix}
\label{E.A.40.b}
\end{equation}
zusammengefasst. Der in Gleichung~(\ref{E.A.39}) gebildete Erwartungswert
des Vektors $\Hat{\Vec{\boldsymbol{K}}}$ enth"alt als Elemente somit
die theoretischen, zweiten, {\em nichtzentralen}\/ Momente der als Faktoren
auftretenden empirischen Kovarianzen. Die Matrix, deren Erwartungswert
diese zweiten Momente mit dem Vektor $\Tilde{\Vec{K}}$ verkn"upft,
und die in diesem Unterkapitel des Anhangs als $\underline{\boldsymbol{F}}$
bezeichnet sei, enth"alt neben der Einheitsmatrix $\underline{E}$ noch die Diagonalmatrix
\begin{gather}
\underline{\boldsymbol{D}}\;=\;
\begin{bmatrix}
\boldsymbol{c}_1&0&0\\
0&\boldsymbol{c}_2&0\\
0&0&\boldsymbol{c}_3
\end{bmatrix}
\label{E.A.40.c}\\*[-28pt]\notag
\end{gather}
und eine weitere Matrix
\begin{equation}
\underline{\boldsymbol{S}}\;=\;
\begin{bmatrix}
0&\text{spur}\big(
\underline{\boldsymbol{V}}_1\CdoT
\underline{\boldsymbol{V}}_2^{\Hh}\CdoT
\underline{\boldsymbol{V}}_4^{\Kk}\CdoT
\underline{\boldsymbol{V}}_3^{\Tt}\big)&
\text{spur}\big(
\underline{\boldsymbol{V}}_1\CdoT
\underline{\boldsymbol{V}}_2^{\Hh}\CdoT
\underline{\boldsymbol{V}}_3\CdoT
\underline{\boldsymbol{V}}_4^{\Hh}\big)\\
\text{spur}\big(
\underline{\boldsymbol{V}}_1\CdoT
\underline{\boldsymbol{V}}_3^{\Tt}\CdoT
\underline{\boldsymbol{V}}_4^{\Kk}\CdoT
\underline{\boldsymbol{V}}_2^{\Hh}\big)&
0&\text{spur}\big(
\underline{\boldsymbol{V}}_1\CdoT
\underline{\boldsymbol{V}}_3^{\Tt}\CdoT
\underline{\boldsymbol{V}}_2^{\Kk}\CdoT
\underline{\boldsymbol{V}}_4^{\Hh}\big)\\
\text{spur}\big(
\underline{\boldsymbol{V}}_1\CdoT
\underline{\boldsymbol{V}}_4^{\Hh}\CdoT
\underline{\boldsymbol{V}}_3\CdoT
\underline{\boldsymbol{V}}_2^{\Hh}\big)&
\text{spur}\big(
\underline{\boldsymbol{V}}_1\CdoT
\underline{\boldsymbol{V}}_4^{\Hh}\CdoT
\underline{\boldsymbol{V}}_2^{\Kk}\CdoT
\underline{\boldsymbol{V}}_3^{\Tt}\big)&0
\end{bmatrix}\!,
\label{E.A.40.d}
\end{equation}
\end{subequations}
deren Hauptdiagonale Null ist. Mit Anhang~\ref{E.Kap.A.7} kann man zeigen, dass
sowohl die Matrix $\underline{\boldsymbol{F}}$ als auch ihr Erwartungswert
f"ur hinreichend gro"se Werte der Mittelungsanzahl $L$ immer regul"ar
ist, da die Zufallsmatrizen $\underline{\boldsymbol{V}}_1$,
$\underline{\boldsymbol{V}}_2$, $\underline{\boldsymbol{V}}_3$ und
$\underline{\boldsymbol{V}}_4$ idempotent sind und einen auf eine Konstante 
begrenzten Rangdefekt aufweisen. Die inverse Matrix sei in diesem Unterkapitel
des Anhangs als $\underline{\boldsymbol{G}}=\underline{\boldsymbol{F}}^{-1}$
bezeichnet.

Da wir an den zweiten {\em zentralen}\/ Momenten
---\,also den Kovarianzen der empirischen Kovarianzen\,---
interessiert sind, berechnen wir diese, indem wir von den
zweiten {\em nichtzentralen}\/ Momenten jeweils das Produkt
der ersten Momente ---\,also den Vektor $\Tilde{\Vec{K}}$\,---
subtrahieren:
\begin{equation}
\text{E}\big\{\Hat{\Vec{\boldsymbol{K}}}\big\}-\Tilde{\Vec{K}}\;=\;
\text{E}\big\{
\underline{\boldsymbol{D}}^{-1}\!\CdoT\underline{\boldsymbol{S}}\big\}\cdot
\Tilde{\Vec{K}}.
\label{E.A.41}
\end{equation}
Nach Gleichung~(\ref{E.2.41}) und (\ref{E.2.51}) sind die meisten der
theoretischen Kovarianzen des Spektrums des gefensterten
Approximationsfehlerprozesses, die im Hauptteil der Abhandlung
in den Vektor $\Tilde{\Vec{K}}$ eingesetzt werden, in guter
N"aherung Null, wenn man eine hoch frequenzselektive Fensterfolge
verwendet. F"ur diese theoretischen Kovarianzen wurden daher auch
keine Sch"atzwerte bestimmt. Daher brauchen wir f"ur die Zeilen
des Vektors $\Hat{\Vec{\boldsymbol{K}}}$, der als Faktor solch eine
nicht berechnete empirische Kovarianz enth"alt auch keine eigene Gleichung
aufzustellen. Wir erhalten so drei weitere Gleichungen, die sich alle
in Vektorschreibweise wie Gleichung~(\ref{E.A.39}) darstellen lassen,
die nun aber andere Vektoren und Matrizen enthalten. 
Falls eine der beiden theoretischen Varianzen des zweiten Elementes
des Vektors $\Tilde{\Vec{K}}$ nach Gleichung~(\ref{E.A.40.b}) Null ist,
erhalten wir mit den Substitutionen
nach Tabelle~\ref{T.A.1} ohne die Zeile 2 die Vektoren und Matrizen
\begin{gather}
\Hat{\Vec{\boldsymbol{K}}}\,=\,\begin{bmatrix}
\boldsymbol{c}_1^{\!-1}\CdoT
\Vec{\boldsymbol{N}}_{\!1}\CdoT
\underline{\boldsymbol{V}}_1\CdoT
\underline{\boldsymbol{V}}_2^{\Hh}\CdoT
\Vec{\boldsymbol{N}}_{\!2}^{\hH}\CdoT
\Vec{\boldsymbol{N}}_{\!3}\CdoT
\underline{\boldsymbol{V}}_3\CdoT
\underline{\boldsymbol{V}}_4^{\Hh}\CdoT
\Vec{\boldsymbol{N}}_{\!4}^{\hH}\\[4pt]
\boldsymbol{c}_3^{\!-1}\CdoT
\Vec{\boldsymbol{N}}_{\!1}\CdoT
\underline{\boldsymbol{V}}_1\CdoT
\underline{\boldsymbol{V}}_4^{\Hh}\CdoT
\Vec{\boldsymbol{N}}_{\!4}^{\hH}\CdoT
\Vec{\boldsymbol{N}}_{\!2}^*\CdoT
\underline{\boldsymbol{V}}_2^{\Kk}\CdoT
\underline{\boldsymbol{V}}_3^{\Tt}\CdoT
\Vec{\boldsymbol{N}}_{\!3}^{\tT}
\end{bmatrix}\!,\quad\;
\Tilde{\Vec{K}}\,=\,
\begin{bmatrix}
\text{E}\big\{\boldsymbol{N}_{\!1}  \CdoT
          \boldsymbol{N}_{\!2}^{\Kk}     \big\}\cdot
\text{E}\big\{\boldsymbol{N}_{\!3}  \CdoT
          \boldsymbol{N}_{\!4}^{\Kk}     \big\}\\[4pt]
\text{E}\big\{\boldsymbol{N}_{\!1}  \CdoT
          \boldsymbol{N}_{\!4}^{\Kk}     \big\}\cdot
\text{E}\big\{\boldsymbol{N}_{\!2}  \CdoT
          \boldsymbol{N}_{\!3}^{\Kk}     \big\}^{\!\Kk}
\end{bmatrix}\!,
\notag\\*[6pt]
\underline{\boldsymbol{D}}\;=\;
\begin{bmatrix}
\boldsymbol{c}_1&0\\
0&\boldsymbol{c}_3
\end{bmatrix}\quad\text{und}\quad
\underline{\boldsymbol{S}}\;=\;
\begin{bmatrix}
0&\text{spur}\big(
\underline{\boldsymbol{V}}_1\CdoT
\underline{\boldsymbol{V}}_2^{\Hh}\CdoT
\underline{\boldsymbol{V}}_3\CdoT
\underline{\boldsymbol{V}}_4^{\Hh}\big)\\
\text{spur}\big(
\underline{\boldsymbol{V}}_1\CdoT
\underline{\boldsymbol{V}}_4^{\Hh}\CdoT
\underline{\boldsymbol{V}}_3\CdoT
\underline{\boldsymbol{V}}_2^{\Hh}\big)&0
\end{bmatrix},
\notag\\*[6pt]
\text{falls }\qquad
\text{E}\{\boldsymbol{N}_{\!1}\CdoT\boldsymbol{N}_{\!3}\}=0
\quad\vee\quad
\text{E}\{\boldsymbol{N}_{\!2}\CdoT\boldsymbol{N}_{\!4}\}=0.
\label{E.A.42}
\end{gather}
Ist eine der beiden theoretischen Varianzen des dritten Elementes
des Vektors $\Tilde{\Vec{K}}$ nach Gleichung~(\ref{E.A.40.b}) Null, so
ergeben sich mit den Substitutionen nach Tabelle~\ref{T.A.1} ohne die
Zeile 3 die Vektoren und Matrizen
\begin{gather}
\Hat{\Vec{\boldsymbol{K}}}\,=\,
\begin{bmatrix}
\boldsymbol{c}_1^{\!-1}\CdoT
\Vec{\boldsymbol{N}}_{\!1}\CdoT
\underline{\boldsymbol{V}}_1\CdoT
\underline{\boldsymbol{V}}_2^{\Hh}\CdoT
\Vec{\boldsymbol{N}}_{\!2}^{\hH}\CdoT
\Vec{\boldsymbol{N}}_{\!3}\CdoT
\underline{\boldsymbol{V}}_3\CdoT
\underline{\boldsymbol{V}}_4^{\Hh}\CdoT
\Vec{\boldsymbol{N}}_{\!4}^{\hH}\\[4pt]
\boldsymbol{c}_2^{\!-1}\CdoT
\Vec{\boldsymbol{N}}_{\!1}\CdoT
\underline{\boldsymbol{V}}_1\CdoT
\underline{\boldsymbol{V}}_3^{\Tt}\CdoT
\Vec{\boldsymbol{N}}_{\!3}^{\tT}\CdoT
\Vec{\boldsymbol{N}}_{\!2}^*\CdoT
\underline{\boldsymbol{V}}_2^{\Kk}\CdoT
\underline{\boldsymbol{V}}_4^{\Hh}\CdoT
\Vec{\boldsymbol{N}}_{\!4}^{\hH}
\end{bmatrix}\!,\quad\;
\Tilde{\Vec{K}}\,=\,
\begin{bmatrix}
\text{E}\big\{\boldsymbol{N}_{\!1}  \CdoT
          \boldsymbol{N}_{\!2}^{\Kk}     \big\}\cdot
\text{E}\big\{\boldsymbol{N}_{\!3}  \CdoT
          \boldsymbol{N}_{\!4}^{\Kk}     \big\}\\[4pt]
\text{E}\big\{\boldsymbol{N}_{\!1}  \CdoT
          \boldsymbol{N}_{\!3}^{\phantom{*}}\big\}\cdot
\text{E}\big\{\boldsymbol{N}_{\!2}  \CdoT
          \boldsymbol{N}_{\!4}^{\phantom{*}}\big\}^{\!\Kk}
\end{bmatrix}\!,
\notag\\*[6pt]
\underline{\boldsymbol{D}}\;=\;
\begin{bmatrix}
\boldsymbol{c}_1&0\\
0&\boldsymbol{c}_2
\end{bmatrix}\quad\text{und}\quad
\underline{\boldsymbol{S}}\;=\;
\begin{bmatrix}
0&\text{spur}\big(
\underline{\boldsymbol{V}}_1\CdoT
\underline{\boldsymbol{V}}_2^{\Hh}\CdoT
\underline{\boldsymbol{V}}_4^{\Kk}\CdoT
\underline{\boldsymbol{V}}_3^{\Tt}\big)\\
\text{spur}\big(
\underline{\boldsymbol{V}}_1\CdoT
\underline{\boldsymbol{V}}_3^{\Tt}\CdoT
\underline{\boldsymbol{V}}_4^{\Kk}\CdoT
\underline{\boldsymbol{V}}_2^{\Hh}\big)&0
\end{bmatrix},
\notag\\*[6pt]
\text{falls }\qquad
\text{E}\big\{\boldsymbol{N}_{\!1}\CdoT\boldsymbol{N}_{\!4}^{\Kk}\big\}=0
\quad\vee\quad
\text{E}\big\{\boldsymbol{N}_{\!2}\CdoT\boldsymbol{N}_{\!3}^{\Kk}\big\}=0.
\label{E.A.43}
\end{gather}
Ist sowohl eine der beiden theoretischen Varianzen des zweiten wie auch des
dritten Elementes des Vektors $\Tilde{\Vec{K}}$  nach Gleichung~(\ref{E.A.40.b})
Null, so ergibt sich mit den Substitutionen der ersten Zeile nach
Tabelle~\ref{T.A.1} keine echte Vektorgleichung mehr:
\begin{gather}
\Hat{\Vec{\boldsymbol{K}}}\;=\;
\boldsymbol{c}_1^{\!-1}\CdoT
\Vec{\boldsymbol{N}}_{\!1}\CdoT
\underline{\boldsymbol{V}}_1\CdoT
\underline{\boldsymbol{V}}_2^{\Hh}\CdoT
\Vec{\boldsymbol{N}}_{\!2}^{\hH}\CdoT
\Vec{\boldsymbol{N}}_{\!3}\CdoT
\underline{\boldsymbol{V}}_3\CdoT
\underline{\boldsymbol{V}}_4^{\Hh}\CdoT
\Vec{\boldsymbol{N}}_{\!4}^{\hH}\!,\qquad
\Tilde{\Vec{K}}\;=\;
\text{E}\big\{\boldsymbol{N}_{\!1}  \CdoT
          \boldsymbol{N}_{\!2}^{\Kk}     \big\}\cdot
\text{E}\big\{\boldsymbol{N}_{\!3}  \CdoT
          \boldsymbol{N}_{\!4}^{\Kk}     \big\},
\notag\\*[6pt]
\underline{\boldsymbol{D}}\;=\;
\boldsymbol{c}_1\quad\text{und}\quad
\underline{\boldsymbol{S}}\;=\;0,
\label{E.A.44}\\*[6pt]
\text{falls }\quad
\Big(\,\text{E}\big\{\boldsymbol{N}_{\!1}\CdoT\boldsymbol{N}_{\!3}\big\}\!=\!0
\;\vee\;
\text{E}\big\{\boldsymbol{N}_{\!2}\CdoT\boldsymbol{N}_{\!4}\big\}\!=\!0\,\Big)
\;\wedge\;
\Big(\,\text{E}\big\{\boldsymbol{N}_{\!1}\CdoT\boldsymbol{N}_{\!4}^{\Kk}\big\}\!=\!0
\;\vee\;
\text{E}\big\{\boldsymbol{N}_{\!2}\CdoT\boldsymbol{N}_{\!3}^{\Kk}\big\}\!=\!0\,\Big).
\notag
\end{gather}
Um Sch"atzwerte f"ur die Kovarianzen der empirischen Kovarianzen
der Zufallsgr"o"sen $\boldsymbol{N}_{\!1}$ bis $\boldsymbol{N}_{\!4}$
zu erhalten berechnen wir zun"achst den Erwartungswert
des Zufallsvektors
\mbox{$\underline{\boldsymbol{G}}\CdoT\Hat{\Vec{\boldsymbol{K}}}$}.
Der Erwartungswert des $i$-ten Elementes dieses Zufallsvektors
berechnet sich als Erwartungswert der Linearkombination der
Elemente $\Hat{\boldsymbol{K}}_{\!j}$ des Vektors
$\Hat{\Vec{\boldsymbol{K}}}$ mit den Elementen $\boldsymbol{G}_{i,j}$
der $i$-ten Zeile der zu $\underline{\boldsymbol{F}}$ inversen
Matrix $\underline{\boldsymbol{G}}$ als Koeffizienten.
Diese Koeffizienten sind unabh"angig von den Zufallsgr"o"sen 
$\boldsymbol{N}_{\!1}$ bis $\boldsymbol{N}_{\!4}$, da diese Funktionen
der Elemente der Zufallsmatrizen $\underline{\boldsymbol{V}}_1$ bis
$\underline{\boldsymbol{V}}_4$, und somit Funktionen der Spektralwerte
der zuf"alligen Erregung sind. Der Erwartungswert dieser Linearkombination
ist die Summe der Erwartungswerte der einzelnen Summanden.
Mit Hilfe der Gleichung~(\myref{A.5.9}) kann man nun den 
Erwartungswert jedes einzelnen Summanden dieser Linearkombination
als eine Summe dreier Summanden berechnen, wobei es je nach Zuordnung
der Spektralwerte des  gefensterten Approximationsfehlerprozesses
zu den Zufallsgr"o"sen $\boldsymbol{N}_{\!1}$ bis $\boldsymbol{N}_{\!4}$
auch vorkommen kann, dass ein oder zwei Summanden verschwinden, wenn
sie eine theoretische Kovarianz Null als Faktor enthalten. Um die
Gleichung~(\myref{A.5.9}) anzuwenden, substituiert man die
Zufallsvektoren $\Vec{\boldsymbol{N}}_{\!1}$ bis
$\Vec{\boldsymbol{N}}_{\!4}$ und die Zufallsmatrizen
$\underline{\boldsymbol{A}}$ und $\underline{\boldsymbol{B}}$
in Gleichung~(\myref{A.5.9}) durch die in Tabelle~\ref{T.A.1} angegebenen
Zufallsvektoren und -matrizen. Der Zufallsfaktor $\boldsymbol{c}$
ist nun jedoch noch mit dem entsprechenden Element $\boldsymbol{G}_{i,j}$
der $i$-ten Zeile der Zufallsmatrix $\underline{\boldsymbol{G}}$ zu
multiplizieren. Man erh"alt so f"ur jeden einzelnen Summanden eine
Linearkombination der entsprechenden Elemente $\Tilde{K}_k$ des Vektors
$\Tilde{\Vec{K}}$, wobei die Koeffizienten dieser Linearkombination
die Erwartungswerte der $\boldsymbol{G}_{i,j}$-fachen Elemente
$\boldsymbol{F}_{\!j,k}$ der $j$-ten Zeile der Zufallsmatrix
$\underline{\boldsymbol{F}}$ sind. Als Gleichung geschrieben ergibt
sich f"ur den Erwartungswert des $i$-ten Elementes des Zufallsvektors
\mbox{$\underline{\boldsymbol{G}}\CdoT\Hat{\Vec{\boldsymbol{K}}}$}:
\begin{gather}
\Sum{j}{}\text{E}\big\{\boldsymbol{G}_{i,j}\CdoT
\Hat{\boldsymbol{K}}_{\!j}\big\}\;=\;
\Sum{j}{}\,\Sum{k}{}\text{E}\big\{\boldsymbol{G}_{i,j}\CdoT
\boldsymbol{F}_{\!j,k}\big\}\CdoT\Tilde{K}_k\;=\;
\Sum{k}{}\text{E}\Big\{\!
\underbrace{\!\Sum{j}{}\boldsymbol{G}_{i,j}\CdoT\boldsymbol{F}_{\!j,k}
}_{\T=\,\Big\{\begin{array}{ll}
{\scriptstyle 1}&{\scriptstyle \text{f"ur }k=i}\\
{\scriptstyle 0}&{\scriptstyle \text{sonst}}\end{array}}\!\!\Big\}\CdoT
\Tilde{K}_k\;=\;\Tilde{K}_i.\raisetag{25pt}
\label{E.A.45}
\end{gather}
F"ur den Erwartungswert des Zufallsvektors ergibt sich das durchaus
nicht triviale Ergebnis
\begin{equation}
\text{E}\big\{\underline{\boldsymbol{F}}^{-1}\CdoT
\Hat{\Vec{\boldsymbol{K}}}\big\}\;=\;
\text{E}\big\{\underline{\boldsymbol{G}}\CdoT
\Hat{\Vec{\boldsymbol{K}}}\big\}\;=\;
\Tilde{\Vec{K}}\;=\;
\text{E}\big\{\underline{\boldsymbol{F}}\big\}^{\!-1}\cdot
\text{E}\big\{\Hat{\Vec{\boldsymbol{K}}}\big\},
\label{E.A.46}
\end{equation}
das besagt, dass sich der Erwartungswert des mit der Zufallsmatrix
$\underline{\boldsymbol{G}}$ abgebildeten Zufallsvektors
$\Hat{\Vec{\boldsymbol{K}}}$ in ein Produkt der Inversen des
Erwartungswertes einer Zufallsmatrix und den Erwartungswert
des Vektors faktorisieren l"asst, obwohl beide
Faktoren von den Spektralwerten der zuf"alligen Erregung abh"angen.
Mit dieser Erkenntnis kann man nun zeigen, dass die Zufallsgr"o"sen
des Zufallsvektors
\begin{equation}
\underline{\boldsymbol{D}}^{-1}\!\CdoT\underline{\boldsymbol{S}}\cdot
\big(\underline{E}+\underline{\boldsymbol{D}}^{-1}\!\CdoT
\underline{\boldsymbol{S}}\,\big)^{\!-1}\Cdot\Hat{\Vec{\boldsymbol{K}}}
\label{E.A.47}
\end{equation}
als Sch"atzwerte f"ur die Kovarianzen der empirischen
Kovarianzen der Zufallsgr"o"sen $\boldsymbol{N}_{\!1}$ bis
$\boldsymbol{N}_{\!4}$ dienen k"onnen, da ihre Erwartungswerte
\begin{gather}
\text{E}\Big\{
\underline{\boldsymbol{D}}^{-1}\!\CdoT\underline{\boldsymbol{S}}\cdot
\big(\underline{E}+\underline{\boldsymbol{D}}^{-1}\!\CdoT
\underline{\boldsymbol{S}}\,\big)^{\!-1}\Cdot
\Hat{\Vec{\boldsymbol{K}}}\Big\}\;=
\notag\\*[4pt]
=\;\text{E}\bigg\{\Big(\big(\underline{E}+
\underline{\boldsymbol{D}}^{-1}\!\CdoT\underline{\boldsymbol{S}}
\,\big)-\underline{E}\Big)\cdot
\big(\underline{E}+\underline{\boldsymbol{D}}^{-1}\!\CdoT
\underline{\boldsymbol{S}}\,\big)^{\!-1}\Cdot
\Hat{\Vec{\boldsymbol{K}}}\bigg\}\;=
\notag\\*[4pt]
=\;\text{E}\big\{\Hat{\Vec{\boldsymbol{K}}}\big\}-
\text{E}\Big\{\big(\underline{E}+\underline{\boldsymbol{D}}^{-1}\!\CdoT
\underline{\boldsymbol{S}}\,\big)^{\!-1}\Cdot
\Hat{\Vec{\boldsymbol{K}}}\Big\}\;=\;
\text{E}\big\{\Hat{\Vec{\boldsymbol{K}}}\big\}-\Tilde{\Vec{K}}
\label{E.A.48}
\end{gather}
mit den in Gleichung~(\ref{E.A.41}) angegebenen theoretischen 
Kovarianzen der empirischen\linebreak Kovarianzen dieser Zufallsgr"o"sen
"ubereinstimmen.

\section[Zum Vergleich der Betr"age der empirischen Varianz
und Kovarianz]{Zum Vergleich der Betr"age der empirischen\\
Varianz und Kovarianz}\label{E.Kap.A.9}

Um zu zeigen, dass die konkreten Varianzsch"atzwerte niemals kleiner
sind als die Betr"age der entsprechenden konkreten Kovarianzsch"atzwerte,
sind an zwei Stellen etwas aufwendigere "Uberlegungen anzustellen.
Um den Ablauf des Textes im Hauptteil der Abhandlung durch dieses
nicht so wichtige und im weiteren nicht mehr verwendete Detail nicht
allzusehr zu unterbrechen, wurden diese beiden Herleitungen in den Anhang
verbannt. Zum einen handelt es sich dabei um die Messwert"-(ko)"-varianzen
des LDS und des KLDS und zum anderen um die Messwert"-(ko)"-varianzen der
deterministischen St"orung.

\subsection{Messwerte des LDS und des KLDS}\label{E.Kap.A.9.1}

Zun"achst werden zwei S"atze "uber die Betr"age von Determinanten
hergeleitet, die wir f"ur den Beweis der
Formel, die wir im Hauptteil der Abhandlung brauchen, ben"otigen.

Der erste Satz bezieht sich auf den Betrag der Determinante eines
Produktes einer Matrix mit ihrer Transponierten. Die
Matrix besteht dabei aus einer Auswahl von Zeilen einer unit"aren Matrix.
Da eine Zeilenpermutation weder den Betrag der Determinante noch die
Unitarit"atseigenschaft einer Matrix beeinflusst, kann man die unit"are 
Matrix immer so permutieren, dass die ausgew"ahlten Zeilenvektoren die 
obersten Zeilen der unit"aren Matrix bilden. Daher wird der folgende Satz 
ohne Beschr"ankung der Allgemeinheit f"ur den Fall gezeigt, dass die
ausgew"ahlten Zeilenvektoren die ersten Zeilen der unit"aren Matrix sind.

Gegeben sei eine unit"are \mbox{$L\!\times\!L$} Matrix $\underline{V}$ deren
erste $R$ Zeilen die Matrix $\underline{V}_1$ bilden, und deren restliche
Zeilen in der Matrix $\underline{V}_2$ zusammengefasst sind.
\begin{equation}
\underline{V}\;=\;
\overbrace{
\left[\begin{array}{c}
\underline{V}_1\! \\
\underline{V}_2\!
\end{array}\right]}^{L \text{ Spalten}}
\!\!\!\!
\begin{array}{l}
\vphantom{\underline{V}_1}\big\} {\scriptstyle R \text{ Zeilen}}\\
\vphantom{\underline{V}_2}
\end{array}
\qquad\text{ mit }\quad
\underline{V}^{\HH} = \underline{V}^{-1}
\label{E.A.49}
\end{equation}
{\bf Satz: } {\sl Der Betrag der Determinante des Produktes
\mbox{\rm$\underline{V}_1\Cdot\underline{V}_1^{\TT}$}
ist immer kleiner gleich $1$.}

Wenn man mit $s_i$ die $R$ Singul"arwerte der Matrix
\mbox{$\underline{V}_1\Cdot\underline{V}_1^{\TT}$} bezeichnet, kann man 
unter Ausnutzung der S"atze,
{\sl\begin{itemize}
\item "`Die Spektralnorm einer Matrix ist ihr gr"o"ster
Singul"arwert"',
\item "`Die Norm eines Matrixprodukts ist niemals gr"o"ser ist als
das Produkt der Normen der daran beteiligten Matrizen"',
\item "`Die Spektralnorm einer unit"aren Matrix ist $1$"',
\item "`Die Spektralnorm einer Matrix, deren Zeilen beliebig aus den
Zeilen einer unit"aren Matrix ausgew"ahlt worden sind, ist $1$"' {\rm und}
\item "`Das Produkt der Singul"arwerte einer Matrix ist gleich dem
Betrag der Determinante einer Matrix"'
\end{itemize}}
die G"ultigkeit dieses Satzes folgenderma"sen zeigen:\vspace{-6pt}
\begin{equation}
1\,=\;\snorm{\underline{V}_1}\Cdot\snorm{\underline{V}_1^{\TT}}\,\ge\;
\Snorm{\underline{V}_1\Cdot\underline{V}_1^{\TT}}\,=\;\max_{i}(s_i)\;\ge\;
\max_{i}(s_i)^R\,\ge\;\Prod{i=1}{R}s_i\;=\;
\big|\det(\underline{V}_1\Cdot\underline{V}_1^{\TT})\big|.
\raisetag{5pt}\label{E.A.50}
\end{equation}
Die Determinante des Produktes
\mbox{$\underline{V}_1\Cdot\underline{V}_1^{\HH}$} hingegen ist immer $1$,
da dieses Produkt wegen der Unitarit"at der Matrix $\underline{V}$,
aus der die Zeilen von $\underline{V}_1$ entnommen wurden,
immer die \mbox{$R\!\times\!R$} Einheitsmatrix ist.

Der zweite Satz vergleicht die Determinante eines
Produktes einer Matrix mit ihrer konjugiert Transponierten
mit dem Betrag der Determinante des Produktes derselben Matrix mit
ihrer Transponierten.

Gegeben sei eine \mbox{$R\!\times\!L$} Matrix $\underline{M}$.\\
{\bf Satz: } {\sl Der Betrag der Determinante des Produktes
\mbox{\rm$\underline{M}\CdoT\underline{M}^{\HH}$} ist immer gr"o"ser gleich
dem Betrag der Determinante des Produktes
\mbox{\rm$\underline{M}\CdoT\underline{M}^{\TT}$}.}

Um dies zu zeigen, betrachten wir zwei F"alle. Im ersten Fall ist
\mbox{$R>L$}. Da alle Spaltenvektoren der beiden Matrixprodukte
\mbox{$\underline{M}\CdoT\underline{M}^{\HH}$} und
\mbox{$\underline{M}\CdoT\underline{M}^{\TT}$} Linearkombinationen der
$L$ Spaltenvektoren der Matrix $\underline{M}$ sind, und diese
$L$ Spaltenvektoren schon wegen ihrer zu geringen Anzahl keine Basis
der $R$-dimensionalen Raums sein k"onnen, ist die Determinante beider
Matrixprodukte immer Null, so dass der Satz mit dem Gleichheitszeichen
erf"ullt ist.

Im zweiten Fall \mbox{$R\le L$} kann man unter Ausnutzung des
zuletzt gezeigten Satzes und der S"atze\vspace{-10pt}
{\sl\begin{itemize}
\item "`Jede Matrix l"asst sich mit einer unit"aren \mbox{$R\!\times\!R$}
Matrix $\underline{U}$, einer unit"aren \mbox{$L\!\times\!L$} Matrix
$\underline{V}$, deren ersten $R$ Zeilen die Matrix $\underline{V}_1$
bilden, der \mbox{$R\!\times\!R$} Diagonalmatrix $\underline{S}$ ihrer
reellen, nichtnegativen Singul"arwerte und der \mbox{$R\!\times\!(L\!-\!R)$}
Nullmatrix $\underline{0}$ in der Form\vspace{-12pt}
\begin{gather}
\underline{M}\;=\;
\underline{U}\cdot
\big[\;\underline{S}\;,\,\,\underline{0}\;\big]\cdot\underline{V}\;=\;
\underline{U}\cdot
\big[\;\underline{S}\CdoT\underline{V}_1\;,\,\,\underline{0}\;\big]
\label{E.A.51}\\*[-26pt]\notag
\end{gather}
darstellen"',
\item "`Die Determinante eines Produkts quadratischer Matrizen ist gleich
dem Produkt der Determinanten der einzelnen daran beteiligten Matrizen"',
\item "`Der Betrag der Determinante einer unit"aren Matrix ist $1$"' {\rm und}
\item "`Der Betrag eines Produktes komplexer Zahlen ist gleich
dem Produkt der Betr"age der daran beteiligten komplexen Zahlen"'
\end{itemize}}
die G"ultigkeit des Satzes folgenderma"sen zeigen:
\begin{gather}
\det\!\big(\underline{M}\CdoT\underline{M}^{\HH}\big)\;=
\notag\\
=\;\det\!\Big(
\underline{U}\cdot\,
\big[\;\underline{S}\CdoT\underline{V}_1\;,\,\,\underline{0}\;\big]\cdot
\big[\;\underline{S}\CdoT\underline{V}_1\;,\,\,\underline{0}\;\big]^{\HH}\Cdot
\underline{U}^{\HH}\Big)\;=
\notag\\[3pt]
=\;\det\!\Big(\underline{U}\cdot\underline{S}\cdot
\underline{V}_1\Cdot\underline{V}_1^{\HH}\Cdot
\underline{S}^{\HH}\Cdot\underline{U}^{\HH}\Big)\;=
\notag\displaybreak[2]\\[5pt]
=\;\det\!\big(\underline{U}\big)\cdot
\det\!\big(\underline{S}\big)\cdot
\det\!\big(\underline{V}_1\Cdot\underline{V}_1^{\HH}\big)\cdot
\det\!\big(\underline{S}^{\HH}\big)\cdot
\det\!\big(\underline{U}^{\HH}\big)\;=
\notag\\[5pt]
=\;\det\!\big(\underline{U}\CdoT\underline{U}^{\HH}\big)\cdot
\det\!\big(\underline{S}\big)\cdot1\cdot
\det\!\big(\underline{S}^{\HH}\big)\;=
\notag\\[5pt]
=\;\big|\det\!\big(\underline{S}\big)\big|\cdot
\big|\det\!\big(\underline{S}\big)\big|\;\ge\notag\\[5pt]
\ge\;\big|\det\!\big(\underline{U}\big)\big|\cdot
\big|\det\!\big(\underline{S}\big)\big|\cdot
\Big|\det\!\Big(\underline{V}_1\Cdot\underline{V}_1^{\TT}\big)\Big|\cdot
\big|\det\!\big(\underline{S}^{\TT}\big)\big|\cdot
\big|\det\!\big(\underline{U}^{\TT}\big)\big|\;=
\notag\\[5pt]
=\;\Big|\det\!\big(\underline{U}\big)\cdot
\det\big(\underline{S}\big)\cdot
\det\Big(\underline{V}_1\Cdot\underline{V}_1^{\TT}\Big)\cdot
\det\big(\underline{S}^{\TT}\big)\cdot
\det\big(\underline{U}^{\TT}\big)\Big|\;=
\notag\\[5pt]
=\;\Big|\det\!\Big(\underline{U}\cdot\underline{S}\cdot
\underline{V}_1\Cdot\underline{V}_1^{\TT}\Cdot
\underline{S}^{\TT}\Cdot\underline{U}^{\TT}\Big)\Big|\;=
\notag\\[5pt]
=\;\Big|\det\Big(\underline{U}\cdot
\big[\;\underline{S}\CdoT\underline{V}_1\;,\,\,\underline{0}\;\big]\cdot
\big[\;\underline{S}\CdoT\underline{V}_1\;,\,\,\underline{0}\;\big]^{\TT}\!
\Cdot\underline{U}^{\TT}\Big)\Big|\;=
\notag\\[5pt]
\label{E.A.52}
=\;\Big|\det\Big(\underline{M}\CdoT\underline{M}^{\TT}\Big)\Big|
\end{gather}
Nun w"ahlen wir \mbox{$R=2$} und bezeichnen die beiden Zeilenvektoren
der \mbox{$2\!\times\!L$} Matrix $\underline{M}$ mit $\Vec{X}$ und
$\Vec{Y}^*\!\!$.\vspace{-12pt}
\begin{equation}
\underline{M}\;=\;\begin{bmatrix}
\Vec{X}\\
\,\Vec{Y}^{\Kk}
\end{bmatrix}
\label{E.A.53}
\end{equation}

Setzen wir dies in Ungleichung~(\ref{E.A.52}) ein, und berechnen wir die
Determinanten, so erhalten wir die Ungleichung
\begin{gather}
\det\Left(
\begin{bmatrix}\Vec{X}\\\,\Vec{Y}^{\Kk}\end{bmatrix}\CdoT
\Big[\,\Vec{X}^{\Hh}\;,\,\,\Vec{Y}^{\,\Tt}\,\Big]\right)\;=\;
\Vec{X}\CdoT\Vec{X}^{\Hh}\!\CdoT
\Vec{Y}^{\Kk}\!\CdoT\Vec{Y}^{\,\Tt}\!-
\Vec{Y}^{\Kk}\!\CdoT\Vec{X}^{\Hh}\!\CdoT
\Vec{X}\CdoT\Vec{Y}^{\,\Tt}
\;\ge\notag\\[5pt]
\ge\;\det\Left(
\begin{bmatrix}\Vec{X}\\\,\Vec{Y}^{\Kk}\end{bmatrix}\CdoT
\Big[\,\Vec{X}^{\,\Tt}\;,\,\,\Vec{Y}^{\Hh}\,\Big]\right)\;=\;
\Big|\,\Vec{X}\CdoT\Vec{X}^{\,\Tt}\!\CdoT
\Vec{Y}^{\Kk}\!\CdoT\Vec{Y}^{\Hh}\!-
\Vec{Y}^{\Kk}\!\CdoT\Vec{X}^{\,\Tt}\!\CdoT
\Vec{X}\CdoT\Vec{Y}^{\Hh}\Big|
\;\ge\notag\\[5pt]
\ge\;\Big|\,\big|
\Vec{X}\CdoT\Vec{X}^{\,\Tt}\!\CdoT
\Vec{Y}^{\Kk}\!\CdoT\Vec{Y}^{\Hh}\big|-
\big|\Vec{Y}^{\Kk}\!\CdoT\Vec{X}^{\,\Tt}\!\CdoT
\Vec{X}\CdoT\Vec{Y}^{\Hh}\big|\,\Big|\;\ge
\notag\\[5pt]
\label{E.A.54}
\ge\;\big|\Vec{X}\CdoT\Vec{X}^{\,\Tt}\!\CdoT
\Vec{Y}^{\Kk}\!\CdoT\Vec{Y}^{\Hh}\big|-
\big|\Vec{Y}^{\Kk}\!\CdoT\Vec{X}^{\,\Tt}\!\CdoT
\Vec{X}\CdoT\Vec{Y}^{\Hh}\big|,
\end{gather}
wobei wir bei der Determinante auf der rechten Seite mit der
Dreiecksungleichung eine untere Grenze angegeben haben. Auf beiden Seiten
der Ungleichung addiert man nun die Terme mit dem negativen Vorzeichen.
Dann wird weiter umgeformt, wobei man ber"ucksichtigt, dass das Produkt
komplexer Zahlen kommutativ ist, dass bei einem Produkt
komplexer Zahlen die Faktoren nach belieben konjugiert werden k"onnen, ohne
dass sich der Betrag des Produkts "andert, dass bei dem Produkt einer komplexen
Zahl mit ihrer Konjugierten die Betragsbildung unterbleiben kann, dass
sich das Skalarprodukt durch Transponieren nicht "andert, und dass das
Konjugieren bei einer reellen Zahl ohne Wirkung bleibt.
Wir multiplizieren die Ungleichung dann mit der reellen Zahl $a$,
von der wir festlegen, dass f"ur sie \mbox{$a\ge 2$} gilt, und
addieren auf beiden Seiten den gleichen reellen Term.
Schlie"slich verwenden wir erneut die Dreiecksungleichung um zu
der Form zu gelangen, die im Hauptteil der Abhandlung ben"otigt wird.
\begin{gather}
\label{E.A.55}
\Vec{X}\CdoT\Vec{X}^{\Hh}\!\CdoT
\Vec{Y}^{\Kk}\!\CdoT\Vec{Y}^{\,\Tt}\!+
\big|\Vec{Y}^{\Kk}\!\CdoT\Vec{X}^{\,\Tt}\!\CdoT
\Vec{X}\CdoT\Vec{Y}^{\Hh}\big|\;\ge\;
\big|\Vec{X}\CdoT\Vec{X}^{\,\Tt}\!\CdoT
\Vec{Y}^{\Kk}\!\CdoT\Vec{Y}^{\Hh}\big|+
\Vec{Y}^{\Kk}\!\CdoT\Vec{X}^{\Hh}\!\CdoT
\Vec{X}\CdoT\Vec{Y}^{\,\Tt}
\\[10pt]
\Vec{X}\CdoT\Vec{X}^{\Hh}\!\CdoT
\Vec{Y}\CdoT\Vec{Y}^{\Hh}\!+
\big|\Vec{Y}\CdoT\Vec{X}^{\Hh}\!\CdoT
\Vec{X}\CdoT\Vec{Y}^{\Hh}\big|\;\ge\;
\big|\Vec{X}\CdoT\Vec{X}^{\,\Tt}\!\CdoT
\Vec{Y}\CdoT\Vec{Y}^{\,\Tt}\big|+
\big|\Vec{Y}^{\Kk}\!\CdoT\Vec{X}^{\Hh}\!\CdoT
\Vec{X}\CdoT\Vec{Y}^{\,\Tt}\big|
\notag\\[10pt]
\Vec{X}\CdoT\Vec{X}^{\Hh}\!\CdoT
\Vec{Y}\CdoT\Vec{Y}^{\Hh}\!+
\Vec{Y}\CdoT\Vec{X}^{\Hh}\!\CdoT
\Vec{X}\CdoT\Vec{Y}^{\Hh}\ge\;
\big|\Vec{X}\CdoT\Vec{X}^{\,\Tt}\!\CdoT
\Vec{Y}\CdoT\Vec{Y}^{\,\Tt}\big|+
\big|\Vec{Y}\CdoT\Vec{X}^{\,\Tt}\!\CdoT
\Vec{X}\CdoT\Vec{Y}^{\,\Tt}\big|
\notag\\[10pt]
a\cdoT\Vec{X}\CdoT\Vec{X}^{\Hh}\!\CdoT
\Vec{Y}\CdoT\Vec{Y}^{\Hh}\!+
a\cdoT\Vec{Y}\CdoT\Vec{X}^{\Hh}\!\CdoT
\Vec{X}\CdoT\Vec{Y}^{\Hh}\ge\;
a\cdot\big|\Vec{X}\CdoT\Vec{X}^{\,\Tt}\!\CdoT
\Vec{Y}\CdoT\Vec{Y}^{\,\Tt}\big|+
a\cdot\big|\Vec{Y}\CdoT\Vec{X}^{\,\Tt}\!\CdoT
\Vec{X}\CdoT\Vec{Y}^{\,\Tt}\big|
\notag\\[10pt]\begin{flalign*}
&a\cdoT\Vec{X}\CdoT\Vec{X}^{\Hh}\!\CdoT
\Vec{Y}\CdoT\Vec{Y}^{\Hh}\!+
a\cdoT\Vec{Y}\CdoT\Vec{X}^{\Hh}\!\CdoT
\Vec{X}\CdoT\Vec{Y}^{\Hh}\!-
2\cdot\big|\Vec{Y}\CdoT\Vec{X}^{\,\Tt}\!\CdoT
\Vec{X}\CdoT\Vec{Y}^{\,\Tt}\big|\;\ge&&
\end{flalign*}\notag\\*\begin{flalign*}
&&\ge\;a\cdot\big|\Vec{X}\CdoT\Vec{X}^{\,\Tt}\!\CdoT
\Vec{Y}\CdoT\Vec{Y}^{\,\Tt}\big|+
a\cdot\big|\Vec{Y}\CdoT\Vec{X}^{\,\Tt}\!\CdoT
\Vec{X}\CdoT\Vec{Y}^{\,\Tt}\big|-
2\cdot\big|\Vec{Y}\CdoT\Vec{X}^{\,\Tt}\!\CdoT
\Vec{X}\CdoT\Vec{Y}^{\,\Tt}\big|&
\end{flalign*}\notag\\[10pt]\begin{flalign*}
&a\cdoT\Vec{X}\CdoT\Vec{X}^{\Hh}\!\CdoT
\Vec{Y}\CdoT\Vec{Y}^{\Hh}\!+
a\cdoT\Vec{Y}\CdoT\Vec{X}^{\Hh}\!\CdoT
\Vec{X}\CdoT\Vec{Y}^{\Hh}\!-
2\cdoT\Vec{Y}^{\Kk}\!\CdoT\Vec{X}^{\Hh}\!\CdoT
\Vec{X}\CdoT\Vec{Y}^{\,\Tt}\ge&&
\end{flalign*}\notag\\*\begin{flalign*}
&&\ge\;\big|a\cdoT\Vec{X}\CdoT\Vec{X}^{\,\Tt}\!\CdoT
\Vec{Y}\CdoT\Vec{Y}^{\,\Tt}\big|+
\big|(a\!-\!2)\cdoT\Vec{Y}\CdoT\Vec{X}^{\,\Tt}\!\CdoT
\Vec{X}\CdoT\Vec{Y}^{\,\Tt}\big|&
\end{flalign*}\notag\\[10pt]\begin{flalign*}
&a\cdoT\Vec{X}\CdoT\Vec{X}^{\Hh}\!\CdoT
\Vec{Y}\CdoT\Vec{Y}^{\Hh}\!+
a\cdoT\Vec{X}\CdoT\Vec{Y}^{\Hh}\!\CdoT
\Vec{Y}\CdoT\Vec{X}^{\Hh}\!-
2\cdoT\Vec{X}\CdoT\Vec{Y}^{\,\Tt}\!\CdoT
\Vec{Y}^{\Kk}\!\CdoT\Vec{X}^{\Hh}\ge&&
\end{flalign*}\notag\\*\begin{flalign*}
&&\ge\;\big|a\cdoT\Vec{X}\CdoT\Vec{X}^{\,\Tt}\!\CdoT
\Vec{Y}\CdoT\Vec{Y}^{\,\Tt}\!+
(a\!-\!2)\cdoT\Vec{Y}\CdoT\Vec{X}^{\,\Tt}\!\CdoT
\Vec{X}\CdoT\Vec{Y}^{\,\Tt}\big|&
\end{flalign*}\notag\\[10pt]\begin{flalign*}
&-2\cdoT\Vec{X}\CdoT\Vec{Y}^{\,\Tt}\!\CdoT
\Vec{Y}^{\Kk}\!\CdoT\Vec{X}^{\Hh}\!+
a\cdoT\Vec{X}\CdoT\Vec{Y}^{\Hh}\!\CdoT
\Vec{Y}\CdoT\Vec{X}^{\Hh}\!+
a\cdoT\Vec{X}\CdoT\Vec{X}^{\Hh}\!\CdoT
\Vec{Y}\CdoT\Vec{Y}^{\Hh}\ge&&
\end{flalign*}\notag\\*\begin{flalign*}
&&\ge\;\big|(a\!-\!2)\cdoT\Vec{X}\CdoT\Vec{Y}^{\,\Tt}\!\CdoT
\Vec{Y}\CdoT\Vec{X}^{\,\Tt}\!+
a\cdoT\Vec{X}\CdoT\Vec{X}^{\,\Tt}\!\CdoT
\Vec{Y}\CdoT\Vec{Y}^{\,\Tt}\big|&
\end{flalign*}\notag
\end{gather}

\subsection{Messwerte der deterministischen St"orung}\label{E.Kap.A.9.2}

Um zu zeigen, dass die konkreten Varianzsch"atzwerte der Messwerte
\mbox{$\Hat{u}(k)$} des Approximationsfehlerprozesses niemals kleiner
sind als die Betr"age der entsprechenden konkreten Kovarianzsch"atzwerte,
beginnen wir damit, in den Gleichungen~(\ref{E.3.53}) die Grenzen der
Lauf"|indizes der Doppelsumme zu ver"andern. Statt die Summe "uber $\mu$
von $0$ bis \mbox{$M\!-\!1$} gehen zu lassen, starten wir bei
\mbox{$\Hat{\mu}\CdoT M/K_{\Phi}$} und enden bei 
\mbox{$M\!+\!\Hat{\mu}\CdoT M/K_{\Phi}\!-\!1$}. Da die
Summanden mit $M$ periodisch sind, wird dadurch lediglich die Reihenfolge
der Summanden zyklisch permutiert. Wenn wir f"ur alle Werte
\mbox{$\Hat{\mu}=0\;(1)\;K_{\Phi}\!-\!1$} die so entstandenen Doppelsummen
addieren, so erhalten wir eine Dreifachsumme, die der mit $K_{\Phi}$
multiplizierten Doppelsumme entspricht. Diese Multiplikation kompensieren
wir indem wir die Dreifachsumme auf $K_{\Phi}$ normieren. Anschlie"send
ersetzen wir noch die Lauf"|indizes $\mu$ und $\Tilde{\mu}$ durch die
neuen Lauf"|indizes \mbox{$\Breve{\mu}=\mu\!-\!\Hat{\mu}\CdoT M/K_{\Phi}$}
und \mbox{$\Bar{\mu}=\Hat{\mu}\!+\!\Tilde{\mu}$} und wir erhalten,
wenn wir wieder die $M$-Periodizit"at der Varianzen und Kovarianzen der
Messwerte des Spektrums der deterministischen St"orung ausnutzen, folgenden
Dreifachsummen.
\begin{subequations}\label{E.A.56}
\begin{gather}
\Hat{C}_{\Hat{\boldsymbol{u}}(k),\Hat{\boldsymbol{u}}(k)}\;=\;
\frac{1}{M^2\CdoT K_{\Phi}}\cdoT
\Sum{\Hat{\mu}=0}{K_{\Phi}-1}\;
\Sum{\mu=\Hat{\mu}\CdoT\frac{M}{K_{\Phi}}}{M+\Hat{\mu}\CdoT\frac{M}{K_{\Phi}}-1}\;
\Sum{\Tilde{\mu}=0}{K_{\Phi}-1}
\Hat{C}_{\Hat{\boldsymbol{U}}_{\!\!f}(\mu),\Hat{\boldsymbol{U}}_{\!\!f}(\mu+\Tilde{\mu}\cdot\frac{M}{K_{\Phi}})}\Cdot
e^{\!-j\cdot\frac{2\pi}{K_{\Phi}}\cdot\Tilde{\mu}\cdot k}\;=
\notag\\[11pt]
=\;\frac{1}{M^2\CdoT K_{\Phi}}\Cdot
\Sum{\Hat{\mu}=0}{K_{\Phi}-1}\;
\Sum{\Breve{\mu}=0}{M-1}\;
\Sum{\Tilde{\mu}=0}{K_{\Phi}-1}
\Hat{C}_{\Hat{\boldsymbol{U}}_{\!\!f}(\Breve{\mu}+\Hat{\mu}\cdot\frac{M}{K_{\Phi}}),\Hat{\boldsymbol{U}}_{\!\!f}(\Breve{\mu}+(\Hat{\mu}+\Tilde{\mu})\cdot\frac{M}{K_{\Phi}})}\Cdot
e^{\!-j\cdot\frac{2\pi}{K_{\Phi}}\cdot\Tilde{\mu}\cdot k}\;=
\notag\\[16pt]
=\;\frac{1}{M^2\CdoT K_{\Phi}}\Cdot
\Sum{\Hat{\mu}=0}{K_{\Phi}-1}\;
\Sum{\Breve{\mu}=0}{M-1}\;
\Sum{\Bar{\mu}=\Hat{\mu}}{K_{\Phi}+\Hat{\mu}-1}\!\!
\Hat{C}_{\Hat{\boldsymbol{U}}_{\!\!f}(\Breve{\mu}+\Hat{\mu}\cdot\frac{M}{K_{\Phi}}),\Hat{\boldsymbol{U}}_{\!\!f}(\Breve{\mu}+\Bar{\mu}\cdot\frac{M}{K_{\Phi}})}\Cdot
e^{j\cdot\frac{2\pi}{K_{\Phi}}\cdot(\Hat{\mu}-\Bar{\mu})\cdot k}\;=
\notag\\[16pt]
=\;\frac{1}{M^2\CdoT K_{\Phi}}\cdot
\Sum{\Breve{\mu}=0}{M-1}\;
\Sum{\Hat{\mu}=0}{K_{\Phi}-1}\;
\Sum{\Bar{\mu}=0}{K_{\Phi}-1}
\Hat{C}_{\Hat{\boldsymbol{U}}_{\!\!f}(\Breve{\mu}+\Hat{\mu}\cdot\frac{M}{K_{\Phi}}),\Hat{\boldsymbol{U}}_{\!\!f}(\Breve{\mu}+\Bar{\mu}\cdot\frac{M}{K_{\Phi}})}\Cdot
e^{j\cdot\frac{2\pi}{K_{\Phi}}\cdot(\Hat{\mu}-\Bar{\mu})\cdot k}
\notag\\*[10pt]
\label{E.A.56.a}
\forall\qquad\qquad k=0\;(1)\;F\!-\!1
\intertext{und\vspace{-12pt}}
\Hat{C}_{\Hat{\boldsymbol{u}}(k),\Hat{\boldsymbol{u}}(k)^{\Kk}}\;=\;
\frac{1}{M^2\CdoT K_{\Phi}}\Cdot
\Sum{\Hat{\mu}=0}{K_{\Phi}-1}\;
\Sum{\mu=\Hat{\mu}\CdoT\frac{M}{K_{\Phi}}}{M+\Hat{\mu}\CdoT\frac{M}{K_{\Phi}}-1}\;
\Sum{\Tilde{\mu}=0}{K_{\Phi}-1}
\Hat{C}_{\Hat{\boldsymbol{U}}_{\!\!f}(\mu),\Hat{\boldsymbol{U}}_{\!\!f}(-\mu-\Tilde{\mu}\cdot\frac{M}{K_{\Phi}})^{\Kk}}\Cdot
e^{\!-j\cdot\frac{2\pi}{K_{\Phi}}\cdot\Tilde{\mu}\cdot k}\;=
\notag\\[6pt]
=\;\frac{1}{M^2\CdoT K_{\Phi}}\Cdot
\Sum{\Hat{\mu}=0}{K_{\Phi}-1}\;
\Sum{\Breve{\mu}=0}{M-1}\;
\Sum{\Tilde{\mu}=0}{K_{\Phi}-1}
\Hat{C}_{\Hat{\boldsymbol{U}}_{\!\!f}(\Breve{\mu}+\Hat{\mu}\cdot\frac{M}{K_{\Phi}}),\Hat{\boldsymbol{U}}_{\!\!f}(-\Breve{\mu}-(\Hat{\mu}+\Tilde{\mu})\CdoT\frac{M}{K_{\Phi}})^{\Kk}}\Cdot
e^{\!-j\cdot\frac{2\pi}{K_{\Phi}}\cdot\Tilde{\mu}\cdot k}\;=
\notag\\[6pt]
=\;\frac{1}{M^2\CdoT K_{\Phi}}\Cdot
\Sum{\Hat{\mu}=0}{K_{\Phi}-1}\;
\Sum{\Breve{\mu}=0}{M-1}\;
\Sum{\Bar{\mu}=\Hat{\mu}}{K_{\Phi}+\Hat{\mu}-1}\!\!
\Hat{C}_{\Hat{\boldsymbol{U}}_{\!\!f}(\Breve{\mu}+\Hat{\mu}\cdot\frac{M}{K_{\Phi}}),\Hat{\boldsymbol{U}}_{\!\!f}(-\Breve{\mu}-\Bar{\mu}\cdot\frac{M}{K_{\Phi}})^{\Kk}}\Cdot
e^{j\cdot\frac{2\pi}{K_{\Phi}}\cdot(\Hat{\mu}-\Bar{\mu})\cdot k}\;=
\notag\\[6pt]
=\;\frac{1}{M^2\CdoT K_{\Phi}}\Cdot
\Sum{\Breve{\mu}=0}{M-1}\;
\Sum{\Hat{\mu}=0}{K_{\Phi}-1}\;
\Sum{\Bar{\mu}=0}{K_{\Phi}-1}
\Hat{C}_{\Hat{\boldsymbol{U}}_{\!\!f}(\Breve{\mu}+\Hat{\mu}\cdot\frac{M}{K_{\Phi}}),\Hat{\boldsymbol{U}}_{\!\!f}(-\Breve{\mu}-\Bar{\mu}\cdot\frac{M}{K_{\Phi}})^{\Kk}}\Cdot
e^{j\cdot\frac{2\pi}{K_{\Phi}}\cdot(\Hat{\mu}-\Bar{\mu})\cdot k}
\notag\\*[6pt]
\label{E.A.56.b}
\forall\qquad\qquad k=0\;(1)\;F\!-\!1.
\end{gather}
\end{subequations}
F"ur die Varianzen und Kovarianzen der Messwerte des Spektrums der
gefensterten deterministischen St"orung setzen wir die in den
Gleichungen~(\ref{E.3.48}) angegebenen Sch"atzwerte
\mbox{$\Hat{C}_{\Hat{\boldsymbol{U}}_{\!\!f}(\Breve{\mu}+\Hat{\mu}\cdot M/K_{\Phi}),\Hat{\boldsymbol{U}}_{\!\!f}(\Breve{\mu}+\Bar{\mu}\cdot M/K_{\Phi})}$} und
\mbox{$\Hat{C}_{\Hat{\boldsymbol{U}}_{\!\!f}(\Breve{\mu}+\Hat{\mu}\cdot M/K_{\Phi}),\Hat{\boldsymbol{U}}_{\!\!f}(-\Breve{\mu}-\Bar{\mu}\cdot M/K_{\Phi})^{\Kk}}$}
in der Form ein, die die Vektoren \mbox{$\Hat{\Vec{C}}_U(\ldots)$}
enth"alt. Bei diesen Sch"atzwerten wiederum setzen wir die Messwerte
\mbox{$\Hat{\Phi}_{\boldsymbol{n}}(\ldots,\ldots)$} und
\mbox{$\Hat{\Psi}_{\boldsymbol{n}}(\ldots,\ldots)$} gem"a"s der
Gleichungen~(\ref{E.3.34}) in der Form mit den Vektoren
\mbox{$\Hat{\Vec{N}}_{\!f}(\ldots)$} ein. Da diese Messwerte mit
Matrizen \mbox{$\underline{V}_{\bot}\!(\ldots)$} gewonnen wurden,
die die Bedingung~(\ref{E.3.33}) erf"ullen, ist dort der Vorfaktor
\mbox{$M^{\up{0.3}{-1}}\!\CdoT\big(L\!-\!1\!-\!K(\Breve{\mu})\big)^{\!-1}$}
bei allen Werten $\Hat{\mu}$ und $\Bar{\mu}$ gleich. 
Er h"angt nur von $\Breve{\mu}$ ab. Somit erhalten wir:
\begin{subequations}\label{E.A.57}
\begin{gather*}\label{E.A.57.a}
\begin{flalign}
&\Hat{C}_{\Hat{\boldsymbol{u}}(k),\Hat{\boldsymbol{u}}(k)}\;=&&
\end{flalign}\\*[14pt]\begin{flalign*}
&=\;\frac{1}{M\CdoT (L\!-\!1)^2\CdoT K_{\Phi}}\cdoT
\Sum{\Breve{\mu}=0}{M-1}\;
\Sum{\Hat{\mu}=0}{K_{\Phi}-1}\;
\Sum{\Bar{\mu}=0}{K_{\Phi}-1}
\Hat{\Vec{C}}_U\big({\T\Breve{\mu}\!+\!\Bar{\mu}\CdoT\frac{M}{K_{\Phi}}}\big)\CdoT
\Hat{\Vec{C}}_U\big({\T\Breve{\mu}+\Hat{\mu}\CdoT\frac{M}{K_{\Phi}}}\big)^{\HH}\Cdot{}&&
\end{flalign*}\\*[-2pt]\begin{flalign*}
&&{}\cdot\Hat{\Phi}_{\boldsymbol{n}}\big({\T\Breve{\mu}\!+\!\Hat{\mu}\CdoT\frac{M}{K_{\Phi}},\Breve{\mu}\!+\!\Bar{\mu}\CdoT\frac{M}{K_{\Phi}}}\big)\cdot
e^{j\cdot\frac{2\pi}{K_{\Phi}}\cdot(\Hat{\mu}-\Bar{\mu})\cdot k}\;=&
\end{flalign*}\\[18pt]\begin{flalign*}
&=\;\frac{1}{M^2\CdoT(L\!-\!1)^2\CdoT K_{\Phi}}\cdoT
\Sum{\Breve{\mu}=0}{M-1}
\frac{1}{L\!-\!1\!-\!K(\Breve{\mu})}\cdoT\!
\Sum{\Hat{\mu}=0}{K_{\Phi}-1}\;
\Sum{\Bar{\mu}=0}{K_{\Phi}-1}
\Hat{\Vec{C}}_U\big({\T\Breve{\mu}\!+\!\Bar{\mu}\CdoT\frac{M}{K_{\Phi}}}\big)\CdoT
\Hat{\Vec{C}}_U\big({\T\Breve{\mu}\!+\!\Hat{\mu}\CdoT\frac{M}{K_{\Phi}}}\big)^{\HH}\Cdot{}\!\!\!&&
\end{flalign*}\\*[6pt]\begin{flalign*}
&&{}\cdot\Hat{\Vec{N}}_{\!f}\big({\T\Breve{\mu}\!+\!\Hat{\mu}\CdoT\frac{M}{K_{\Phi}}}\big)\CdoT
\Hat{\Vec{N}}_{\!f}\big({\T\Breve{\mu}\!+\!\Bar{\mu}\CdoT\frac{M}{K_{\Phi}}}\big)^{\HH}\Cdot
e^{j\cdot\frac{2\pi}{K_{\Phi}}\cdot(\Hat{\mu}-\Bar{\mu})\cdot k}&
\end{flalign*}\\[14pt]
\forall\qquad\qquad k=0\;(1)\;F\!-\!1
\\\intertext{und}\label{E.A.57.b}\begin{flalign}
&\Hat{C}_{\Hat{\boldsymbol{u}}(k),\Hat{\boldsymbol{u}}(k)^{\Kk}}\;=&&
\end{flalign}\\*[14pt]\begin{flalign*}
&=\;\frac{1}{M\CdoT(L\!-\!1)^2\CdoT K_{\Phi}}\cdoT
\Sum{\Breve{\mu}=0}{M-1}\;
\Sum{\Hat{\mu}=0}{K_{\Phi}-1}\;
\Sum{\Bar{\mu}=0}{K_{\Phi}-1}
\Hat{\Vec{C}}_U\big({\T\!-\Breve{\mu}\!-\!\Bar{\mu}\CdoT\frac{M}{K_{\Phi}}}\big)^{\!*}\!\CdoT
\Hat{\Vec{C}}_U\big({\T\Breve{\mu}\!+\!\Hat{\mu}\CdoT\frac{M}{K_{\Phi}}}\big)^{\HH}\cdoT{}&&
\end{flalign*}\\*[-2pt]\begin{flalign*}
&&{}\cdot\Hat{\Psi}_{\boldsymbol{n}}\big({\T\Breve{\mu}\!+\!\Hat{\mu}\CdoT\frac{M}{K_{\Phi}},\Breve{\mu}\!+\!\Bar{\mu}\CdoT\frac{M}{K_{\Phi}}}\big)\cdot
e^{j\cdot\frac{2\pi}{K_{\Phi}}\cdot(\Hat{\mu}-\Bar{\mu})\cdot k}\;=&
\end{flalign*}\\[18pt]\begin{flalign*}
&=\,\frac{1}{M^2\CdoT(L\!-\!1)^2\CdoT K_{\Phi}}\cdoT\!
\Sum{\Breve{\mu}=0}{M-1}
\frac{1}{L\!-\!1\!-\!K(\Breve{\mu})}\cdoT\!
\Sum{\Hat{\mu}=0}{K_{\Phi}-1}\;
\Sum{\Bar{\mu}=0}{K_{\Phi}-1}
\Hat{\Vec{C}}_U\big({\T\!-\Breve{\mu}\!-\!\Bar{\mu}\CdoT\frac{M}{K_{\Phi}}}\big)^{\!*}\!\CdoT
\Hat{\Vec{C}}_U\big({\T\Breve{\mu}\!+\!\Hat{\mu}\CdoT\frac{M}{K_{\Phi}}}\big)^{\HH}\CdoT{}\!\!\!&&
\end{flalign*}\\*[6pt]\begin{flalign*}
&&{}\cdot\Hat{\Vec{N}}_{\!f}\big({\T\Breve{\mu}\!+\!\Hat{\mu}\CdoT\frac{M}{K_{\Phi}}}\big)\CdoT
\Hat{\Vec{N}}_{\!f}\big({\T\!-\Breve{\mu}\!-\!\Bar{\mu}\CdoT\frac{M}{K_{\Phi}}}\big)^{\TT}\Cdot
e^{j\cdot\frac{2\pi}{K_{\Phi}}\cdot(\Hat{\mu}-\Bar{\mu})\cdot k}&
\end{flalign*}\\[14pt]
\forall\qquad\qquad k=0\;(1)\;F\!-\!1.
\end{gather*}
\end{subequations}
Jeder Summand dieser Dreifachsummen ist ein Produkt von zwei
Skalarprodukten von jeweils zwei Vektoren mit je $L$ Elementen.
Ausf"uhrlich geschrieben erhalten wir die F"unf"|fachsummen\vspace{0pt}
\begin{subequations}\label{E.A.58}
\begin{gather*}\label{E.A.58.a}
\begin{flalign}
&\Hat{C}_{\Hat{\boldsymbol{u}}(k),\Hat{\boldsymbol{u}}(k)}\;=&&
\end{flalign}\\*[1pt]\begin{flalign*}
&\!=\frac{1}{M^2\CdoT(L\!-\!1)^2\CdoT K_{\Phi}}\cdoT\!\!
\Sum{\Breve{\mu}=0}{M-1}\!
\frac{1}{L\!-\!1\!-\!K(\Breve{\mu})}\cdoT\!\!
\Sum{\Hat{\mu}=0}{K_{\Phi}-1}\,
\Sum{\Bar{\mu}=0}{K_{\Phi}-1}\,
\Sum{\Hat{\lambda}=1}{L}
\Hat{C}_{U,\Hat{\lambda}}\big({\T\Breve{\mu}\!+\!\Bar{\mu}\CdoT\frac{M}{K_{\Phi}}}\big)\!\CdoT
\Hat{C}_{U,\Hat{\lambda}}\big({\T\Breve{\mu}\!+\!\Hat{\mu}\CdoT\frac{M}{K_{\Phi}}}\big)^{\!*}
\!\CdoT{}\!\!\!\!\!&&
\end{flalign*}\\*[-2pt]\begin{flalign*}
&&{}\cdot\Sum{\Breve{\lambda}=1}{L}
\Hat{N}_{\!f,\Breve{\lambda}}\big({\T\Breve{\mu}\!+\!\Hat{\mu}\CdoT\frac{M}{K_{\Phi}}}\big)\CdoT
\Hat{N}_{\!f,\Breve{\lambda}}\big({\T\Breve{\mu}\!+\!\Bar{\mu}\CdoT\frac{M}{K_{\Phi}}}\big)^{\!*}\Cdot
e^{j\cdot\frac{2\pi}{K_{\Phi}}\cdot(\Hat{\mu}-\Bar{\mu})\cdot k}\;=&
\end{flalign*}\\[7pt]\begin{flalign*}
&\!=\;\frac{1}{M^2\CdoT(L\!-\!1)^2\CdoT K_{\Phi}}\cdoT
\Sum{\Hat{\lambda}=1}{L}\;
\Sum{\Breve{\lambda}=1}{L}\,
\Sum{\Breve{\mu}=0}{M-1}
\frac{1}{L\!-\!1\!-\!K(\Breve{\mu})}\cdoT{}&&
\end{flalign*}\\*[-4pt]\begin{flalign*}
&&{}\cdoT{}&\Sum{\Hat{\mu}=0}{K_{\Phi}-1}
\Hat{C}_{U,\Hat{\lambda}}\big({\T\Breve{\mu}\!+\!\Hat{\mu}\CdoT\frac{M}{K_{\Phi}}}\big)^{\!*}\!\CdoT
\Hat{N}_{\!\!f,\Breve{\lambda}}\big({\T\Breve{\mu}\!+\!\Hat{\mu}\CdoT\frac{M}{K_{\Phi}}}\big)^{\phantom{\!*}}\Cdot
e^{\phantom{\!-}j\cdot\frac{2\pi}{K_{\Phi}}\cdot\Hat{\mu}\cdot k}\CdoT{}\\*
&&{}\cdoT{}&\Sum{\Bar{\mu}=0}{K_{\Phi}-1}
\Hat{C}_{U,\Hat{\lambda}}\big({\T\Breve{\mu}\!+\!\Bar{\mu}\CdoT\frac{M}{K_{\Phi}}}\big)^{\phantom{\!*}}\!\CdoT
\Hat{N}_{\!f,\Breve{\lambda}}\big({\T\Breve{\mu}\!+\!\Bar{\mu}\CdoT\frac{M}{K_{\Phi}}}\big)^{\!*}\Cdot
e^{\!-j\cdot\frac{2\pi}{K_{\Phi}}\cdot\Bar{\mu}\cdot k}\;={}
\end{flalign*}\\[6pt]
=\;\frac{1}{M^2\CdoT(L\!-\!1)^2\CdoT K_{\Phi}}\cdoT
\Sum{\Hat{\lambda}=1}{L}\;
\Sum{\Breve{\lambda}=1}{L}\,
\Sum{\Breve{\mu}=0}{M-1}
\frac{1}{L\!-\!1\!-\!K(\Breve{\mu})}\cdot
\big|C_{\Hat{\lambda},\Breve{\lambda}}(\Breve{\mu},k)\big|^2
\\[4pt]
\forall\qquad\qquad k=0\;(1)\;F\!-\!1
\\[16pt]\label{E.A.58.b}\begin{flalign}
&\text{und}\qquad
\Hat{C}_{\Hat{\boldsymbol{u}}(k),\Hat{\boldsymbol{u}}(k)^{\Kk}}\;=&&
\end{flalign}\\*[1pt]\begin{flalign*}
&\!=\frac{1}{M^2\CdoT(\!L\!-\!1\!)^2\CdoT K_{\Phi}}\cdoT\!\!
\Sum{\Breve{\mu}=0}{M-1}\!
\frac{1}{L\!-\!1\!-\!K(\Breve{\mu})}\cdoT\!\!
\Sum{\Hat{\mu}=0}{K_{\Phi}-1}\,
\Sum{\Bar{\mu}=0}{K_{\Phi}-1}\,
\Sum{\Hat{\lambda}=1}{L}\!
\Hat{C}_{U,\Hat{\lambda}}\big({\T\!-\Breve{\mu}\!-\!\Bar{\mu}\CdoT\frac{M}{K_{\Phi}}}\!\big)^{\!*}\!\CdoT
\Hat{C}_{U,\Hat{\lambda}}\big({\T\Breve{\mu}\!+\!\Hat{\mu}\CdoT\frac{M}{K_{\Phi}}}\!\big)^{\!*}
\!\CdoT{}\!\!\!\!\!&&
\end{flalign*}\\*[-2pt]\begin{flalign*}
&&{}\cdoT\Sum{\Breve{\lambda}=1}{L}
\Hat{N}_{\!f,\Breve{\lambda}}\big({\T\Breve{\mu}\!+\!\Hat{\mu}\CdoT\frac{M}{K_{\Phi}}}\big)\CdoT
\Hat{N}_{\!f,\Breve{\lambda}}\big({\T\!-\Breve{\mu}\!-\!\Bar{\mu}\CdoT\frac{M}{K_{\Phi}}}\big)\cdot
e^{j\cdot\frac{2\pi}{K_{\Phi}}\cdot(\Hat{\mu}-\Bar{\mu})\cdot k}\;=&
\end{flalign*}\\[7pt]\begin{flalign*}
&\!=\;\frac{1}{M^2\CdoT(L\!-\!1)^2\CdoT K_{\Phi}}\cdoT
\Sum{\Hat{\lambda}=1}{L}\;
\Sum{\Breve{\lambda}=1}{L}\,
\Sum{\Breve{\mu}=0}{M-1}
\frac{1}{L\!-\!1\!-\!K(\Breve{\mu})}\cdoT{}&&
\end{flalign*}\\*[-4pt]\begin{flalign*}
&&{}\cdoT{}&\Sum{\Hat{\mu}=0}{K_{\Phi}-1}
\Hat{C}_{U,\Hat{\lambda}}\big({\T\,\Breve{\mu}\!+\!\Hat{\mu}\CdoT\frac{M}{K_{\Phi}}}\big)^{\!*}\Cdot
\Hat{N}_{\!f,\Breve{\lambda}}\big({\T\,\Breve{\mu}\!+\!\Hat{\mu}\CdoT\frac{M}{K_{\Phi}}}\big)\cdot
e^{\phantom{\!-}j\cdot\frac{2\pi}{K_{\Phi}}\cdot\Hat{\mu}\cdot k}\!\CdoT{}\\*
&&{}\cdoT{}&\Sum{\Bar{\mu}=0}{K_{\Phi}-1}
\Hat{C}_{U,\Hat{\lambda}}\big({\T\!-\Breve{\mu}\!-\!\Bar{\mu}\CdoT\frac{M}{K_{\Phi}}}\big)^{\!*}\!\CdoT
\Hat{N}_{\!f,\Breve{\lambda}}\big({\T\!-\Breve{\mu}\!-\!\Bar{\mu}\CdoT\frac{M}{K_{\Phi}}}\big)\CdoT
e^{\!-j\cdot\frac{2\pi}{K_{\Phi}}\cdot\Bar{\mu}\cdot k}\;={}
\end{flalign*}\\[6pt]
=\;\frac{1}{M^2\CdoT(L\!-\!1)^2\CdoT K_{\Phi}}\cdoT
\Sum{\Hat{\lambda}=1}{L}\;\;
\Sum{\Breve{\lambda}=1}{L}\,
\Sum{\Breve{\mu}=0}{M-1}
\frac{1}{L\!-\!1\!-\!K(\Breve{\mu})}\cdot
C_{\Hat{\lambda},\Breve{\lambda}}(\Breve{\mu},k)\cdot
C_{\Hat{\lambda},\Breve{\lambda}}(\!-\Breve{\mu},k)
\\*[4pt]
\forall\qquad\qquad k=0\;(1)\;F\!-\!1.
\end{gather*}
\end{subequations}
Dabei wurde zuletzt jeweils die Abk"urzung
\begin{gather}
C_{\Hat{\lambda},\Breve{\lambda}}(\Breve{\mu},k)\;=\;
\Sum{\Hat{\mu}=0}{K_{\Phi}-1}
\Hat{C}_{U,\Hat{\lambda}}\big({\T\Breve{\mu}\!+\!\Hat{\mu}\CdoT\frac{M}{K_{\Phi}}}\big)^{\!*}\Cdot
\Hat{N}_{\!f,\Breve{\lambda}}\big({\T\Breve{\mu}\!+\!\Hat{\mu}\CdoT\frac{M}{K_{\Phi}}}\big)\cdot
e^{j\cdot\frac{2\pi}{K_{\Phi}}\cdot\Hat{\mu}\cdot k}
\label{E.A.59}\\*[4pt]
\forall\qquad k=0\;(1)\;F\!-\!1,
\qquad\Breve{\mu}=0\;(1)\;M\!-\!1,
\qquad\Hat{\lambda}=1\;(1)\;L
\qquad\text{ und }\qquad\Breve{\lambda}=1\;(1)\;L
\notag
\end{gather}
eingef"uhrt. In Gleichung~(\ref{E.A.58.a}) ergibt sich dieselbe Summe, wenn 
wir "uber $-\Breve{\mu}$ statt "uber $\Breve{\mu}$ summieren, da dies wegen 
der $M$-Periodizit"at der Summanden lediglich die Reihenfolge der Summanden 
umkehrt. Damit k"onnen wir diese Dreifachsumme auch als die H"alfte der 
Dreifachsumme mit $-\Breve{\mu}$ und der Dreifachsumme mit $\Breve{\mu}$,
also als die Dreifachsumme der arithmetischen Mittel der Betragsquadrate
von $C_{\Hat{\lambda},\Breve{\lambda}}(\Breve{\mu},k)$ und
$C_{\Hat{\lambda},\Breve{\lambda}}(\!-\Breve{\mu},k)$ schreiben.
Da das Arithmetische Mittel zweier reeller nichtnegativer Zahlen nie
kleiner ist als deren geometrisches Mittel gilt
\begin{gather}
\frac{\big|C_{\Hat{\lambda},\Breve{\lambda}}(\Breve{\mu},k)\big|^2+
\big|C_{\Hat{\lambda},\Breve{\lambda}}(\!-\Breve{\mu},k)\big|^2}{2}\;\ge\;
\sqrt{\big|C_{\Hat{\lambda},\Breve{\lambda}}(\Breve{\mu},k)\big|^2\Cdot
\big|C_{\Hat{\lambda},\Breve{\lambda}}(\!-\Breve{\mu},k)\big|^2}\;=
\notag\\[8pt]
=\;\big|C_{\Hat{\lambda},\Breve{\lambda}}(\Breve{\mu},k)\big|\cdot
\big|C_{\Hat{\lambda},\Breve{\lambda}}(\!-\Breve{\mu},k)\big|\;=\;
\big|C_{\Hat{\lambda},\Breve{\lambda}}(\Breve{\mu},k)\CdoT
C_{\Hat{\lambda},\Breve{\lambda}}(\!-\Breve{\mu},k)\big|
\label{E.A.60}\\*[8pt]
\forall\qquad k=0\;(1)\;F\!-\!1,
\qquad\Breve{\mu}=0\;(1)\;M\!-\!1,
\qquad\Hat{\lambda}=1\;(1)\;L
\qquad\text{ und }\qquad\Breve{\lambda}=1\;(1)\;L.
\notag
\end{gather}
Da der Betrag einer Summe niemals gr"o"ser ist als die Summe der Betr"age
der Summanden, kann man nun f"ur \mbox{$L>K(\Breve{\mu})\!+\!1$} beweisen,
dass der konkrete Varianzsch"atzwert des Messwertes
\mbox{$\Hat{u}(k)$} des Approximationsfehlerprozesses zum Zeitpunk $k$
niemals kleiner als der Betrag des konkreten Kovarianzsch"atzwertes ist.
\begin{gather}
\begin{flalign*}
&\Hat{C}_{\Hat{\boldsymbol{u}}(k),\Hat{\boldsymbol{u}}(k)}\;=&&
\end{flalign*}\notag\\*[6pt]
=\;\frac{1}{M^2\Cdot(L\!-\!1)^2\Cdot K_{\Phi}}\cdoT
\Sum{\Hat{\lambda}=1}{L}\;
\Sum{\Breve{\lambda}=1}{L}\;
\Sum{\Breve{\mu}=0}{M-1}\frac{1}{L\!-\!1\!-\!K(\Breve{\mu})}\cdot
\frac{\big|C_{\Hat{\lambda},\Breve{\lambda}}(\Breve{\mu},k)\big|^2+
\big|C_{\Hat{\lambda},\Breve{\lambda}}(\!-\Breve{\mu},k)\big|^2}{2}\;\ge
\notag\\[8pt]
\ge\;\frac{1}{M^2\Cdot(L\!-\!1)^2\Cdot K_{\Phi}}\cdoT
\Sum{\Hat{\lambda}=1}{L}\;
\Sum{\Breve{\lambda}=1}{L}\;
\Sum{\Breve{\mu}=0}{M-1}\;
\bigg|\frac{C_{\Hat{\lambda},\Breve{\lambda}}(\Breve{\mu},k)\CdoT
C_{\Hat{\lambda},\Breve{\lambda}}(\!-\Breve{\mu},k)}
{L-1-K(\Breve{\mu})}\bigg|\;\ge
\notag\\[8pt]
\ge\,\Bigg|\,\frac{1}{M^2\Cdot(L\!-\!1)^2\Cdot K_{\Phi}}\cdoT
\Sum{\Hat{\lambda}=1}{L}\;
\Sum{\Breve{\lambda}=1}{L}\;
\Sum{\Breve{\mu}=0}{M-1}\frac{1}{L\!-\!1\!-\!K(\Breve{\mu})}\cdot
C_{\Hat{\lambda},\Breve{\lambda}}(\Breve{\mu},k)\CdoT
C_{\Hat{\lambda},\Breve{\lambda}}(\!-\Breve{\mu},k)\,\Bigg|\;=
\notag\\*[6pt]
=\;\big|\Hat{C}_{\Hat{\boldsymbol{u}}(k),\Hat{\boldsymbol{u}}(k)^{\Kk}}\big|
\notag\\*[6pt]
\label{E:A.61}
\forall\quad\qquad k=0\;(1)\;F\!-\!1
\end{gather}

\section{Anmerkung zur Gau"s-Verbundverteilung}\label{E.Kap.A.10}

Bei der Herleitung der Varianzen und Kovarianzen der
Messwerte \mbox{$\Hat{\boldsymbol{\Phi}}_{\boldsymbol{n}}(\mu)$} und
\mbox{$\Hat{\boldsymbol{\Psi}}_{\boldsymbol{n}}(\mu)$} eines station"aren
Approximationsfehlerprozesses in Kapitel \myref{Kova} bzw.
\mbox{$\Hat{\boldsymbol{\Phi}}_{\boldsymbol{n}}(\mu_1,\mu_2)$} und
\mbox{$\Hat{\boldsymbol{\Psi}}_{\boldsymbol{n}}(\mu_1,\mu_2)$} eines 
zyklostation"aren Approximationsfehlerprozesses in Kapitel \ref{E.Kap.3.4} 
hatten wir vorausgesetzt, dass das Zufallsgr"o"sentupel
\mbox{$\big[\boldsymbol{N}_{\!\!f}(\mu),\boldsymbol{N}_{\!\!f}(-\mu)\big]^{\Tt}\!$} 
bzw. die Zufallsgr"o"sentupel 
\mbox{$\big[\boldsymbol{N}_{\!\!f}(\mu_1),\boldsymbol{N}_{\!\!f}(\mu_2)\big]^{\Tt}$} und
\mbox{$\big[\boldsymbol{N}_{\!\!f}(\mu_1), \boldsymbol{N}_{\!\!f}(-\mu_2)\big]^{\Tt}\!$}
eine normalverteilte mittelwertfreie Verbundverteilung aufweisen.
Die Mittelwertfreiheit ist nicht das entscheidende
Kriterium f"ur die Entscheidung, ob man annehmen kann, dass die Forderung
erf"ullt ist. Wenn die Verbundverteilung des Zufallsgr"o"sentupels
eine Normalverteilung ist, so ist sie mittelwertfrei, wenn
jede einzelne der beiden Zufallsgr"o"sen des Zufallsgr"o"sentupels
mittelwertfrei ist. In \cite{Diss} hatten wir uns darauf beschr"ankt, 
die dort vorgestellte Variante des RKM nur f"ur mittelwertfreie 
Approximationsfehlerprozesse zu verwenden. Bei der hier vorgestellten 
RKM-Variante sorgt die Modellierung der deterministischen St"orung 
daf"ur, dass die Mittelwertfreiheit gegeben ist, wie Gleichung 
(\ref{E.2.30}) zeigt. Die Forderung, dass eine Verbundnormalverteilung
vorliegen muss, besagt {\em nicht}, dass die gemeinsame Verbundverteilung 
{\em aller}\/ Spektralwerte \mbox{$\boldsymbol{N}_{\!\!f}(\mu)$}
f"ur alle \mbox{$\mu\,=\,0\;(1)\;M\!-\!1$} eine Normalverteilung sein muss.
Es gibt Zufallsvektoren, die {\em keine}\/ gemeinsame Verbundnormalverteilung
aufweisen, bei denen aber alle zweidimensionalen Randverteilungen
zweier beliebigen Zufallsgr"o"sen dieses Zufallsvektors
Verbundnormalverteilungen sind. Ich m"ochte dies an einem Beispiel
demonstrieren. Um ein anschauliches Beispiel zu erhalten, werde
ich dies nicht an einem beliebig dimensionalen Zufallsvektor und
an zweidimensionalen Randverteilungen zeigen, sondern ein Beispiel angeben,
bei dem beide eindimensionalen Randverteilungen eines zweidimensionalen
{\em nicht}\/ verbundnormalverteilten Zufallsvektors jeweils eindimensionale
Normalverteilungen sind. Es bleibt dem Leser "uberlassen, dieses Beispiel auf
einen mehrdimensionalen Zufallsvektor und dessen zweidimensionale
Randverteilungen zu "ubertragen.

Gegeben sei ein zweidimensionaler reeller Zufallsvektor

\begin{equation}
\Vec{\boldsymbol{n}}\;=\;
\begin{bmatrix}
\boldsymbol{n}_1\\
\boldsymbol{n}_2
\end{bmatrix},
\label{E.A.62}
\end{equation}

der die zweidimensionale Verbundverteilungsdichte

\begin{equation}
p_{\Vec{\boldsymbol{n}}}(\Vec{n})\;=\;
\begin{cases}
\;0\qquad\text{ f"ur }\quad
n_2\!>\!n_1\!\ge\!0\;\,\vee\;-n_1\!<\!n_2\!\le\!0\;\,\vee\;\,
n_2\!<\!n_1\!\le\!0\;\,\vee\;-n_1\!>\!n_2\!\ge\!0&\\[8pt]
{\D\frac{1}{\pi}\cdot e^{-\frac{1}{2}\cdot\|\Vec{n}\|^2} = 
\frac{1}{\pi}\cdot e^{-\frac{1}{2}\cdot(n_1^2+n_2^2)}}
\qquad\qquad\text{ sonst }&
\end{cases}
\raisetag{22pt}\label{E.A.63}
\end{equation}\vspace{0pt}

besitze, 
\begin{figure}[btp]
\begin{center}
{ 
\begin{picture}(200,209)

\input{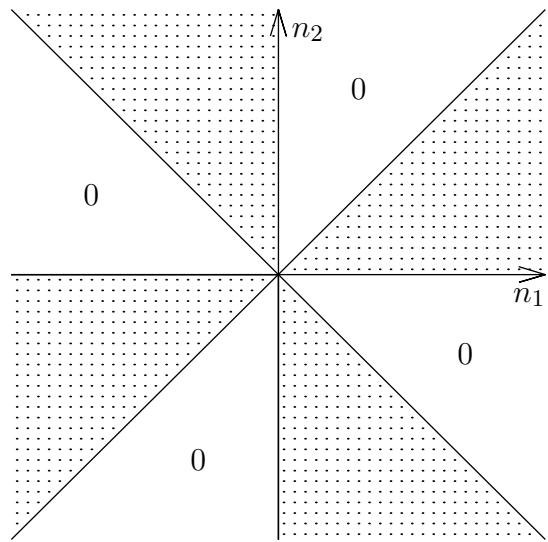}
\put(200,95){\makebox(0,0)[tr]{$n_1$}}
\put(105,195){\makebox(0,0)[tl]{$n_2$}}
\put(130,170){\makebox(0,0){$0$}}
\put(30,130){\makebox(0,0){$0$}}
\put(70,30){\makebox(0,0){$0$}}
\put(170,70){\makebox(0,0){$0$}}
\end{picture}}
\end{center}\vspace{-16pt}
\setlength{\belowcaptionskip}{-3pt}
\caption{Gebiet der \mbox{$n_1$-$n_2$-Ebene}, in dem die Verbundverteilung
zweier unkorrelierter, einzeln normalverteilter aber abh"angiger
Zufallsgr"o"sen von null verschieden ist.}
\label{E.bAa}
\rule{\textwidth}{0.5pt}\vspace{-10pt}
\end{figure}
die in der halben \mbox{$n_1$-$n_2$-Ebene} null ist,
und in der anderen H"alfte der \mbox{$n_1$-$n_2$-Ebene} doppelt so
gro"s ist, wie die Verbundverteilungsdichte eines zweidimensionalen,
unkorrelierten normierten, zentrierten und verbundnormalverteilten
Zufallsgr"o"sentupels. In Bild~\ref{E.bAa} ist das
Gebiet der \mbox{$n_1$-$n_2$-Ebene}, in dem die Verbundverteilungsdichte
\vadjust{\penalty-100}von null verschieden ist, punktiert dargestellt. 
Die Verteilungsdichte nach (\ref{E.A.63}) erf"ullt alle
Bedingungen, die bei einer zweidimensionalen Verbundverteilungsdichte erf"ullt
sein m"ussen. Die eindimensionalen Randverteilungsdichten der Zufallsgr"o"sen
$\boldsymbol{n}_1$ und $\boldsymbol{n}_2$ erh"alt man, indem man die
gemeinsame zweidimensionale Verbundverteilungsdichte "uber $n_1$ bzw. "uber
$n_2$ integriert. Diese Integrale liefern dasselbe, wie wenn man "uber
die  Verbundverteilungsdichte eines zweidimensionalen, unkorrelierten
normierten, zentrierten und normalverteilten Zufallsgr"o"sentupels
integrieren w"urde, da die Bereiche des Integrals, innerhalb derer
die Verbundverteilungsdichte (\ref{E.A.63}) null ist, gerade durch die
Bereiche kompensiert werden, in denen die Verbundverteilungsdichte doppelt
so gro"s ist wie die zentrierte normierte Verbundnormalverteilungsdichte.

Jede der beiden Zufallsgr"o"sen $\boldsymbol{n}_1$ und $\boldsymbol{n}_2$
ist daher eindimensional normalverteilt.
Da sich die Verbundverteilungsdichte nach (\ref{E.A.63}) jedoch nicht
als das Produkt der beiden Randverteilungsdichten schreiben l"asst, sind
die beiden Zufallsgr"o"sen abh"angig. Multipliziert man die
Verteilungsdichte nach (\ref{E.A.63}) mit \mbox{$n_1\CdoT n_2$} und integriert
man "uber die gesamte \mbox{$n_1$-$n_2$-Ebene}, so erh"alt man die
Korrelation der Zufallsgr"o"sen $\boldsymbol{n}_1$ und $\boldsymbol{n}_2$.
Dieses Integral ergibt dasselbe wie das entsprechende Integral
"uber die zentrierte normierte Verbundnormalverteilungsdichte,
n"amlich null. Die Zufallsgr"o"sen $\boldsymbol{n}_1$ und $\boldsymbol{n}_2$
sind also unkorreliert. Man hat es also in unserem Beispiel mit zwei
unkorrelierten normalverteilten Zufallsgr"o"sen zu tun, die
voneinander {\em abh"angig}\/ sind. Dies ist {\em kein}\/ Widerspruch
der Aussage, dass  zwei Zufallsgr"o"sen unabh"angig sind,
wenn sie unkorreliert und verbundnormalverteilt sind. Die Zufallsgr"o"sen
unseres Beispiels sind n"amlich {\em nicht}\/ verbundnormalverteilt,
sondern lediglich jede f"ur sich ist normalverteilt.

Ebenso ist es vorstellbar, dass es mehrdimensionale Verbundverteilungen
von Zufallsvektoren gibt, die sich nicht als das Produkt ihrer ein- oder
zweidimensionalen Randverteilungen schreiben lassen, obwohl jede der
Zufallsgr"o"sen bzw. Zufallsgr"o"sentupel aus dem Zufallsvektor
normalverteilt sind. Die Forderung, dass das Zufallsgr"o"sentupel
\mbox{$\big[\,\boldsymbol{N}_{\!\!f}(\mu),\,
\boldsymbol{N}_{\!\!f}(-\mu)\,\big]^{\Tt}\!\!$},
bzw. die Zufallsgr"o"sentupel
\mbox{$\big[\,\boldsymbol{N}_{\!\!f}(\mu_1),\,
\boldsymbol{N}_{\!\!f}(\mu_2)\,\big]^{\Tt}\!\!$} und
\mbox{$\big[\,\boldsymbol{N}_{\!\!f}(-\mu_1),\,
\boldsymbol{N}_{\!\!f}(\mu_2)\,\big]^{\Tt}\!\!$}
verbundnormalverteilt sein sollen,
ist daher wesentlich weniger streng, als die Forderung, dass {\em alle}\/
Spektralwerte \mbox{$\boldsymbol{N}_{\!\!f}(\mu)$} f"ur {\em alle}\/
\mbox{$\mu=0\;(1)\;M\!-\!1$} eine gemeinsame Verbundnormalverteilung
aufweisen sollen.

Es ist manchmal schwer zu entscheiden, ob man von den Zufallsgr"o"sentupeln,
die bei der Berechnung der Messwert"-(ko)"-varianzen auftreten, annehmen kann,
dass sie verbundnormalverteilt sind, weil die Zufallsgr"o"sen des Tupels
durch eine Fensterung und eine anschlie"sende DFT aus den Zufallsgr"o"sen
des Approximationsfehlerprozesses \mbox{$\boldsymbol{n}(k)$} gewonnen werden.
Es sei nun gezeigt, dass auch von den Zufallsgr"o"sen des wahren
Approximationsfehlerprozesses \mbox{$\boldsymbol{n}(k)$} {\em nicht}\/
angenommen werden muss, dass sie {\em alle}
--- f"ur {\em alle}\/ \mbox{$k\,=\,0\;(1)\;F\!-\!1$} ---
verbundnormalverteilt sind, um sicherzustellen, dass f"ur die
Zufallsgr"o"sen \mbox{$\boldsymbol{N}_{\!1}$}, \mbox{$\boldsymbol{N}_{\!2}$},
\mbox{$\boldsymbol{N}_{\!3}$} und \mbox{$\boldsymbol{N}_{\!4}$} die bei der
R"uckf"uhrung der vierten Momente auf die zweiten Momente ben"otigte
Beziehung (\myref{A.5.8}) erf"ullt ist. Dazu setzen wir dort zun"achst f"ur die
Zufallsgr"o"sen \mbox{$\boldsymbol{N}_{\!i}\;(\,i = 1\;(1)\;4\,)$} die
zuf"alligen Spektralwerte \mbox{$\boldsymbol{N}_{\!\!f}(\mu_i)$} ein,
wobei die gew"ahlten diskreten Frequenzen $\mu_i$ auch gleich sein k"onnen.

Wir beginnen damit, dass wir im Erwartungswert auf der linken Seite
der Gleichung (\myref{A.5.8}) die Zufallsgr"o"sen als Funktionen der
Zufallsgr"o"sen \mbox{$\boldsymbol{n}(k)$} des Approximationsfehlerprozesses
schreiben. Jede der vier Zufallsgr"o"sen ist entweder die diskrete
Fouriertransformierte des gefensterten Approximationsfehlerprozesses
\mbox{$\boldsymbol{n}(k)$} bei der Frequenz $\mu_i$ oder deren Konjugierte.
Wir erhalten somit den Erwartungswert "uber ein Produkt von vier Summen
(\,Lauf"|indizes $k_i$ mit \mbox{$i = 1\;(1)\;4$}\,), die jeweils
eine Linearkombination der Zufallsgr"o"sen \mbox{$\boldsymbol{n}(k)$}
oder ihrer Konjugierten enth"alt. Das Produkt der vier Summen kann
als Vierfachsumme des Produkts der Summanden geschrieben werden.
Nachdem man die Reihenfolge der Summation und der Erwartungswertbildung
vertauscht hat, kann man bei jedem Summanden einige bez"uglich der
Erwartungswertbildung konstante Faktoren vor den Erwartungswert ziehen.
Diese Faktoren sind neben den Drehfaktoren
\mbox{$e^{\pm j\cdot\frac{2\pi}{M}\cdot\mu_i\cdot k_i}$} der DFT noch die
Werte \mbox{$f(k_i)$} und \mbox{$f(k_i)^{\Kk}$} der verwendeten Fensterfolge.
Aus Gleichung (\myref{A.5.8}) erh"alt man auf diese Weise:

\begin{gather}
\text{E}\big\{\,\boldsymbol{N}_{\!1}\CdoT\boldsymbol{N}_{\!2}^*\CdoT
\boldsymbol{N}_{\!3}\CdoT\boldsymbol{N}_{\!4}^*\,\big\}\;=\;
\text{E}\big\{\,
\boldsymbol{N}_{\!\!f}(\mu_1)\cdot
\boldsymbol{N}_{\!\!f}(\mu_2)^{\Kk}\Cdot
\boldsymbol{N}_{\!\!f}(\mu_3)\cdot
\boldsymbol{N}_{\!\!f}(\mu_4)^{\Kk}\,\big\}\;=
\label{E.A.64}\\*[4pt]\begin{flalign*}
&\!\!=\!\!\Sum{k_1=-\infty}{\infty}\,
\Sum{k_2=-\infty}{\infty}\,
\Sum{k_3=-\infty}{\infty}\,
\Sum{k_4=-\infty}{\infty}\!
\text{E}\big\{\boldsymbol{n}(k_1)\CdoT\boldsymbol{n}(k_2)^{\!\Kk}\!\CdoT
              \boldsymbol{n}(k_3)\CdoT\boldsymbol{n}(k_4)^{\!\Kk}\big\}
\cdot f(k_1)\CdoT f(k_2)^{\!\Kk}\!\CdoT f(k_3)\CdoT f(k_4)^{\!\Kk}\cdot{}\!\!\!\!\!&&
\end{flalign*}\notag\\*[-2pt]\begin{flalign*}
&&{}\cdot\; e^{-j\cdot\frac{2\pi}{M}\cdot
(\mu_1\cdot k_1-\mu_2\cdot k_2+\mu_3\cdot k_3-\mu_4\cdot k_4)}.\!\!&
\end{flalign*}\notag
\end{gather}
Setzt man nun f"ur den Zufallsvektor \mbox{$\Vec{\Tilde{\boldsymbol{N}}}$} in
Gleichung (\myref{A.6.7}) des Anhangs \myref{Gauss}
\begin{equation}
\Vec{\Tilde{\boldsymbol{N}}}\;=\;\begin{bmatrix}
\boldsymbol{n}(k_1)\\
\boldsymbol{n}(k_2)\\
\boldsymbol{n}(k_3)\\
\boldsymbol{n}(k_4)
\end{bmatrix}
\qquad\qquad\text{ mit }\qquad k_1,\,k_2,\,k_3,\,k_4\,=\,0\;(1)\;F\!-\!1
\label{E.A.65}
\end{equation}
ein, so erh"alt man mit \mbox{$R\!=\!4$} und
mit dem Indexquadrupel \mbox{$[\,i,\,j,\,k,\,l\,] = [\,1,\,6,\,3,\,8\,]$}
aus Gleichung (\myref{A.6.18}) eine Gleichung, die es uns erm"oglicht das
vierte Moment in Gleichung (\ref{E.A.64}) durch die entsprechenden
zweiten Momente auszudr"ucken.
\begin{align}
\text{E}\big\{\boldsymbol{n}(k_1)\CdoT\boldsymbol{n}(k_2)^{\!\Kk}\!\CdoT
\boldsymbol{n}(k_3)\CdoT\boldsymbol{n}(k_4)^{\!\Kk}\big\}&{}\;=\,
\text{E}\big\{\boldsymbol{n}(k_1)\CdoT\boldsymbol{n}(k_2)^{\!\Kk}\big\}\cdot
\text{E}\big\{\boldsymbol{n}(k_3)\CdoT\boldsymbol{n}(k_4)^{\!\Kk}\big\}^{\phantom{\!\Kk}}+{}
\notag\\*
&{}\;+\;
\text{E}\big\{\boldsymbol{n}(k_1)\CdoT\boldsymbol{n}(k_3)^{\phantom{\!\Kk}}\big\}\cdot
\text{E}\big\{\boldsymbol{n}(k_2)\CdoT\boldsymbol{n}(k_4)^{\phantom{\!\Kk}}\big\}^{\!\Kk}+{}
\label{E.A.66}\\*
&{}\;+\;
\text{E}\big\{\boldsymbol{n}(k_1)\CdoT\boldsymbol{n}(k_4)^{\!\Kk}\big\}\cdot
\text{E}\big\{\boldsymbol{n}(k_2)\CdoT\boldsymbol{n}(k_3)^{\!\Kk}\big\}^{\!\Kk}\!.\notag
\end{align}
Dabei gen"ugt es auch beim Approximationsfehlerprozess anzunehmen,
dass die Randverteilungen jedes beliebigen Zufallsgr"o"senquadrupels
des Prozesses Verbundnormalverteilungen sind. Somit m"ussen
nicht alle Zufallsgr"o"sen \mbox{$\boldsymbol{n}(k)$} f"ur
\mbox{$k\,=\,0\;(1)\;F\!-\!1$} gemeinsam verbundnormalverteilt sein.
Desweiteren erkennt man, dass es keine Rolle spielt, welche
Fensterfolge man bei der Berechnung der Spektralwerte verwendet,
weil die Werte der Fensterfolge vor die Erwartungswertbildung
gezogen werden konnten. Ersetzt man die vierten Momente
durch die Summe dreier Produkte zweiter Momente, so kann
man die Vierfachsumme "uber je drei Summanden auch als Summe
dreier Vierfachsummen schreiben. Nun zieht man die Drehfaktoren der
DFT und die Werte der Fensterfolge wieder in die Erwartungswertbildung
hinein, wobei man auf die richtige Zuordnung der Indizes achtet.
Nach einer eventuellen Vertauschung der Reihenfolge der einzelnen
Summationen der Vierfachsummen, kann man aus den innersten
beiden Doppelsummen jeweils einen der beiden Erwartungswerte herausziehen.
Die innere Doppelsumme ist dann nicht mehr von den Indizes der
"au"seren Doppelsumme abh"angig, und kann somit vor die "au"sere
Doppelsumme gezogen werden, so dass man drei Produkte von jeweils zwei
Doppelsummen erh"alt. Nach Vertauschung der Reihenfolge der Summation
und der Erwartungswertbildung kann man die Doppelsumme innerhalb des
jeweiligen Erwartungswertes wieder als das Produkt zweier Einfachsummen
schreiben. Jede dieser Einfachsummen ist dann eine der Zufallsgr"o"sen
\mbox{$\boldsymbol{N}_{\!\!f}(\mu_i)\;(\,i = 1\;(1)\;4\,)$}.
Man erh"alt so:
\begin{gather}
\text{E}\big\{\,
\boldsymbol{N}_{\!\!f}(\mu_1)\cdot
\boldsymbol{N}_{\!\!f}(\mu_2)^{\!\Kk}\Cdot
\boldsymbol{N}_{\!\!f}(\mu_3)\cdot
\boldsymbol{N}_{\!\!f}(\mu_4)^{\!\Kk}\,\big\}\;=
\label{E.A.67}\\[3pt]\begin{flalign*}
&\!\!=\!\!
\Sum{k_1=-\infty}{\infty}
\Sum{k_2=-\infty}{\infty}
\Sum{k_3=-\infty}{\infty}
\Sum{k_4=-\infty}{\infty}\!\!\text{E}\big\{\!
\boldsymbol{n}(k_1)\CdoT
\boldsymbol{n}(k_2)^{\!\Kk}\!\big\}\CdoT\text{E}\big\{\!
\boldsymbol{n}(k_3)\CdoT
\boldsymbol{n}(k_4)^{\!\Kk}\!\big\}\CdoT
f(k_1)\CdoT\!f(k_2)^{\!\Kk}\!\CdoT\!f(k_3)\CdoT\!f(k_4)^{\!\Kk}\Cdot{}\!\!\!\!\!&&
\end{flalign*}\notag\\*[-5pt]\begin{flalign*}
&&{}\cdot e^{-j\cdot\frac{2\pi}{M}\cdot
(\mu_1\cdot k_1-\mu_2\cdot k_2+\mu_3\cdot k_3-\mu_4\cdot k_4)}+{}\!\!&
\end{flalign*}\notag\\[2pt]\begin{flalign*}
&\!\!+\!\!
\Sum{k_1=-\infty}{\infty}
\Sum{k_2=-\infty}{\infty}
\Sum{k_3=-\infty}{\infty}
\Sum{k_4=-\infty}{\infty}\!\!\text{E}\big\{\!
\boldsymbol{n}(k_1)\CdoT
\boldsymbol{n}(k_3)\!\big\}\CdoT\text{E}\big\{\!
\boldsymbol{n}(k_2)\CdoT
\boldsymbol{n}(k_4)\!\big\}^{\!\Kk}\!\CdoT
f(k_1)\CdoT\!f(k_2)^{\!\Kk}\!\CdoT\!f(k_3)\CdoT\!f(k_4)^{\!\Kk}\Cdot{}\!\!\!\!\!&&
\end{flalign*}\notag\\*[-5pt]\begin{flalign*}
&&{}\cdot e^{-j\cdot\frac{2\pi}{M}\cdot
(\mu_1\cdot k_1-\mu_2\cdot k_2+\mu_3\cdot k_3-\mu_4\cdot k_4)}+{}\!\!&
\end{flalign*}\notag\\[2pt]\begin{flalign*}
&\!\!+\!\!
\Sum{k_1=-\infty}{\infty}
\Sum{k_2=-\infty}{\infty}
\Sum{k_3=-\infty}{\infty}
\Sum{k_4=-\infty}{\infty}\!\!\text{E}\big\{\!
\boldsymbol{n}(k_1)\CdoT
\boldsymbol{n}(k_4)^{\!\Kk}\!\big\}\cdot\text{E}\big\{\!
\boldsymbol{n}(k_2)\CdoT
\boldsymbol{n}(k_3)^{\!\Kk}\!\big\}^{\!\Kk}\!\CdoT
f(k_1)\CdoT\!f(k_2)^{\!\Kk}\!\CdoT\!f(k_3)\CdoT\!f(k_4)^{\!\Kk}\Cdot{}\!\!\!\!\!&&
\end{flalign*}\notag\\*[-5pt]\begin{flalign*}
&&{}\cdot e^{-j\cdot\frac{2\pi}{M}\cdot
(\mu_1\cdot k_1-\mu_2\cdot k_2+\mu_3\cdot k_3-\mu_4\cdot k_4)}={}\!\!&
\end{flalign*}\notag\\[6pt]\begin{align*}{}=\;{}&
\text{E}\bigg\{\Sum{k_1=-\infty}{\infty}\;\Sum{k_2=-\infty}{\infty}\!
\boldsymbol{n}(k_1)^{\phantom{\!\Kk}}\!\CdoT\boldsymbol{n}(k_2)^{\!\Kk}\Cdot
f(k_1)^{\phantom{\!\Kk}}\!\CdoT f(k_2)^{\!\Kk}\Cdot
e^{-j\cdot\frac{2\pi}{M}\cdot(\mu_1\cdot k_1-\mu_2\cdot k_2)}\bigg\}\cdot{}
\\*[-4pt]{}\Cdot{}&
\text{E}\bigg\{\Sum{k_3=-\infty}{\infty}\;\Sum{k_4=-\infty}{\infty}\!
\boldsymbol{n}(k_3)^{\phantom{\!\Kk}}\!\CdoT\boldsymbol{n}(k_4)^{\!\Kk}\Cdot
f(k_3)^{\phantom{\!\Kk}}\!\CdoT f(k_4)^{\!\Kk}\Cdot
e^{-j\cdot\frac{2\pi}{M}\cdot(\mu_3\cdot k_3-\mu_4\cdot k_4)}\bigg\}\;+{}
\\[4pt]{}+\,{}&
\text{E}\bigg\{\Sum{k_1=-\infty}{\infty}\;\Sum{k_3=-\infty}{\infty}\!
\boldsymbol{n}(k_1)^{\phantom{\!\Kk}}\!\CdoT\boldsymbol{n}(k_3)^{\phantom{\!\Kk}}\Cdot
f(k_1)^{\phantom{\!\Kk}}\!\CdoT f(k_3)^{\phantom{\!\Kk}}\Cdot
e^{-j\cdot\frac{2\pi}{M}\cdot(\mu_1\cdot k_1+\mu_3\cdot k_3)}\bigg\}\cdot{}
\\*[-4pt]{}\Cdot{}&
\text{E}\bigg\{\Sum{k_2=-\infty}{\infty}\;\Sum{k_4=-\infty}{\infty}\!
\boldsymbol{n}(k_2)^{\!\Kk}\!\CdoT\boldsymbol{n}(k_4)^{\!\Kk}\Cdot
f(k_2)^{\!\Kk}\!\CdoT f(k_4)^{\!\Kk}\Cdot
e^{+j\cdot\frac{2\pi}{M}\cdot(\mu_2\cdot k_2+\mu_4\cdot k_4)}\bigg\}\;+{}
\\[4pt]{}+\,{}&
\text{E}\bigg\{\Sum{k_1=-\infty}{\infty}\;\Sum{k_4=-\infty}{\infty}\!
\boldsymbol{n}(k_1)^{\phantom{\!\Kk}}\!\CdoT\boldsymbol{n}(k_4)^{\!\Kk}\Cdot
f(k_1)^{\phantom{\!\Kk}}\!\CdoT f(k_4)^{\!\Kk}\Cdot
e^{-j\cdot\frac{2\pi}{M}\cdot(\mu_1\cdot k_1-\mu_4\cdot k_4)}\bigg\}\cdot{}
\\*[-4pt]{}\Cdot{}&
\text{E}\bigg\{\Sum{k_2=-\infty}{\infty}\;\Sum{k_3=-\infty}{\infty}\!
\boldsymbol{n}(k_2)^{\!\Kk}\!\CdoT\boldsymbol{n}(k_3)^{\phantom{\!\Kk}}\Cdot
f(k_2)^{\!\Kk}\!\CdoT f(k_3)^{\phantom{\!\Kk}}\Cdot
e^{+j\cdot\frac{2\pi}{M}\cdot(\mu_2\cdot k_2-\mu_3\cdot k_3)}\bigg\}\;=
\end{align*}\notag\\[6pt]\begin{align*}{}=\;{}&
\text{E}\big\{\,\boldsymbol{N}_{\!\!f}(\mu_1)\cdot
                \boldsymbol{N}_{\!\!f}(\mu_2)^{\!\Kk}\big\}\cdot
\text{E}\big\{\,\boldsymbol{N}_{\!\!f}(\mu_3)\cdot
                \boldsymbol{N}_{\!\!f}(\mu_4)^{\!\Kk}\big\}^{\phantom{\!\Kk}}+{}
\\*[4pt]{}+\;{}&
\text{E}\big\{\,\boldsymbol{N}_{\!\!f}(\mu_1)\cdot
                \boldsymbol{N}_{\!\!f}(\mu_3)^{\phantom{\!\Kk}}\big\}\cdot
\text{E}\big\{\,\boldsymbol{N}_{\!\!f}(\mu_2)\cdot
                \boldsymbol{N}_{\!\!f}(\mu_4)^{\phantom{\!\Kk}}\big\}^{\!\Kk}+{}
\\*[4pt]{}+\;{}&
\text{E}\big\{\,\boldsymbol{N}_{\!\!f}(\mu_1)\cdot
                \boldsymbol{N}_{\!\!f}(\mu_4)^{\!\Kk}\big\}\cdot
\text{E}\big\{\,\boldsymbol{N}_{\!\!f}(\mu_2)\cdot
                \boldsymbol{N}_{\!\!f}(\mu_3)^{\!\Kk}\big\}^{\!\Kk}
\end{align*}\notag
\end{gather}
Bei der Herleitung der Gleichung (\myref{A.5.8}) im Unterkapitel \myref{Gauss}
des Anhangs wurde gesagt, dass man die anderen vierten Momente,
die die einzelnen Zufallsgr"o"sen evtl. konjugiert enthalten,
dadurch erh"alt, indem man in Gleichung (\myref{A.5.8}) zun"achst die
gew"unschten Zufallsgr"o"sen durch ihre Konjugierten ersetzt, und dann die
Indizes entsprechend substituiert. Wenn man die Herleitung von
Gleichung (\ref{E.A.64}) bis (\ref{E.A.67}) mit den entsprechenden konjugierten
Zufallsgr"o"sen durchf"uhrt, kann man analog zeigen, dass die formal
substituierten Versionen von Gleichung (\myref{A.5.8}) ebenfalls gelten.

\renewcommand{\thechapter}{}
\renewcommand{\chaptermark}[1]{\markboth{\small\sf\thechapter\ #1}{\small\sf\thechapter\ #1}}
\renewcommand{\sectionmark}[1]{\markright{\small\sf\thesection\ #1}}
\chapter{\protect\rule{13pt}{0pt}Liste der verwendeten Abk"urzungen und Formelzeichen}
Es werden hier nur die Abk"urzungen und Formelzeichen aufgef"uhrt,
die nicht in \cite{Diss} beschrieben sind. Dies gilt auch, wenn 
die entsprechenden Folgen und Funktionen hier oftmals zweidimensional 
sind, w"ahrend sie in \cite{Diss} nur eindimensional definiert sind.

{\bf Allgemeine Formelzeichen und Funktionen}
\begin{list}{}{\setlength{\itemsep}{0.2ex plus0.2ex}
\setlength{\labelwidth}{4em}\setlength{\labelsep}{1em}
\setlength{\leftmargin}{5em}}
\item[{\tt realmin}\hfill]
     \mbox{$\D= 2^{\T2-2^{\text{Exponentenwortl"ange-$1$}}}$} beim IEEE
     Standard 754. Kleinste positive Zahl, die bei Gleitkommazahlendarstellung
     am Rechner mit voller Genauigkeit darstellbar ist \cite{Sun}.
\item[$\Vec{1}$\hfill] Zeilenvektor der nur Einsen enth"alt.
\item[$\underline{1}_{\bot}$\hfill] Idempotente Matrix, die alle Vektoren auf den Nullraum des Einservektors $\Vec{1}$ abbildet.
\item[$\Vec{E}_n$\hfill] Einheitszeilenvektor, dessen $n$-tes Element eins ist, w"ahrend alle anderen Elemente null sind.
\item[kgv$(\ldots)$\hfill] kleinstes gemeinsames Vielfaches.
\item[$\Vec{w}_{K_H,\kappa}$\hfill] Vektor der Drehfaktoren der inversen DFT
     der L"ange $K_H$ zum Zeitpunkt $\kappa$.
\end{list}

{\bf Spezielle Formelzeichen}\\
Die in \cite{Diss} beschriebenen Konventionen (\,z.~B. Fettdruck
$\Rightarrow$ Zufallsgr"o"se, -vektor, -matrix oder -prozess etc.\,)
bez"uglich der Variablennamen gelten auch hier, so dass auch in dieser
Liste im wesentlichen nur die Formelzeichen angegeben sind, die bei
der Beschreibung der Theorie der erweiterten RKM-Varianten vorkommen,
und die nicht aus den hier oder in~\cite{Diss} genannten Formelzeichen
gem"a"s dieser Konventionen abgeleitet werden k"onnen.
\begin{list}{}{\setlength{\itemsep}{0.2ex plus0.2ex}
\setlength{\labelwidth}{4em}\setlength{\labelsep}{1em}
\setlength{\leftmargin}{5em}}
\item[$A_0$\hfill] Mit diesem Parameter legt man fest, dass die letzten $A_0$ Werte der 
     Fensterfolge innerhalb des Intervalls \mbox{$k\in[0;F\!-\!1]$} null sein sollen.
\item[$A_1$\hfill] Mit diesem Parameter legt man fest, dass das Polynom \mbox{$z^{F-A_0-1}\Cdot G(z)$} 
     zus"atzlich zu den in Kapitel \myref{Fen} genannten Nullstellen weitere \mbox{$2\CdoT A_1$} 
     einfache Nullstellen auf dem Einheitskreis besitzt, die sich im Abstand
     \mbox{$2\pi/F$} anschlie"sen.
\item[$\Delta c$\hfill] "Anderung des Parameters der Bilineartransformation bei der Konstruktion
     der diskreten Fensterfolge (\,\mbox{$-1\!<\!c\!<\!1$}\,)
\item[$c_{\infty}$\hfill] Parameter der Bilineartransformation bei der
     Konstruktion der kontinuierlichen Fensterfunktion
     (\,\mbox{$0\!<\!c_{\infty}$}\,)
\item[$\Hat{\Vec{\boldsymbol{C}}}_n(\mu)$\hfill] Hilfsvektor bei der Berechnung
     der Kovarianz des $n$-ten Elementes des Mess\-wert\-vektors \mbox{$\Hat{\Vec{\boldsymbol{H}}}(\mu)$}
     der "Ubertragungsfunktionen.
\item[$\Hat{\Vec{\boldsymbol{C}}}_U(\mu)$\hfill] Hilfsvektor bei der Berechnung
     der Kovarianz des Spektrums der gefensterten deterministischen St"orung.
\item[$\Hat{\Vec{\boldsymbol{C}}}_u(\mu)$\hfill] Hilfsvektor bei der Berechnung 
     der Kovarianz der gefensterten deterministischen St"orung.
\item[$d_{\infty}(t)$\hfill]
     \mbox{$=\;f_{\infty}(t)\ast\!f_{\infty}(\!-t)\;=\;$}
     kontinuierliche Fensterautokorrelationsfunktion.
\item[$D_{\infty}(s)$\hfill] Laplacetransformierte der
     Fensterautokorrelationsfunktion \mbox{$d_{\infty}(t)$}.
\item[$D_{E,\infty}(s)$\hfill] Anteil der Laplacetransformierten
     \mbox{$D_{\infty}(s)$} der Fensterautokorrelationsfunk\-tion,
     der nur die Nullstellen auf der imagin"aren Achse enth"alt.
\item[$D_{\overline{E},\infty}(s)$\hfill] Anteil der Laplacetransformierten
     \mbox{$D_{\infty}(s)$} der Fensterautokorrelationsfunktion ohne die
     Nullstellen auf der imagin"aren Achse.
\item[$\widetilde{D}_{\overline{E},\infty}(\Tilde{z})$\hfill] Polynom in
     $\Tilde{z}$, das durch Bilineartransformation aus
     \mbox{$D_{\overline{E},\infty}(s)$} entsteht.
\item[$\widetilde{D}_{N,\infty}(\Tilde{z})$\hfill] Anteil von
     \mbox{$\widetilde{D}_{\overline{E},\infty}(\Tilde{z})$}, der die
     Nullstellen in der bilineartransformierten $z$-Ebene enth"alt, und
     dessen minimalphasiger Anteil einen nichtlinearen Phasenbeitrag liefert,
     der mit Hilfe des Cepstrums berechnet wird.
\item[$\widetilde{D}_{P,\infty}(\Tilde{z})$\hfill] Anteil von
     \mbox{$\widetilde{D}_{\overline{E},\infty}(\Tilde{z})$}, der die
     Polstellen in der bilineartransformierten $z$-Ebene enth"alt,
     und dessen minimalphasiger Anteil nach der bilinearen R"ucktransformation
     einen nichtlinearen Phasenbeitrag liefert, der geschlossen berechnet
     werden kann.
\item[$D_{P,\infty}(s)$\hfill] Bilinear R"ucktransformierte von
     \mbox{$\widetilde{D}_{P,\infty}(\Tilde{z})$}.
\item[$f_{\infty}(t)$\hfill] Kontinuierliche Fensterfunktion.
\item[$F_{\!\infty}(\omega)$\hfill] Fouriertransformierte der kontinuierlichen
     Fensterfunktion \mbox{$f_{\infty}(t)$}.
\item[$g_{\infty}(t)$\hfill] Basisfensterfunktion, die zur 
     Herleitung der Konstruktion der kontinuierlichen Fensterfunktion
     \mbox{$f_{\infty}(t)$} ben"otigt wird.
\item[$g_{Q,\infty}(t)$\hfill] \mbox{$=\;g(t)\ast g(\!-t)\;=$} kontinuierliche
     Basisfensterautokorrelationsfunktion.
\item[$G_{\infty}(j\omega)$\hfill] Fouriertransformierte der
     Basisfensterfunktion \mbox{$g_{\infty}(t)$}.
\item[$G_1(z)$\hfill] Anteil von \mbox{$G(z)$}, der die \mbox{$F\!-\!N\!+\!2\CdoT A_1$} 
     einfachen Nullstellen am Einheitskreis im Raster \mbox{$2\pi/F$} enth"alt.
\item[$G_2(z)$\hfill] Anteil von \mbox{$G(z)$}, der die \mbox{$N\!-\!1\!-\!A_0\!-\!2\CdoT\!A_1$} 
      frei w"ahlbaren Nullstellen enth"alt.
\item[$h_{\kappa}(k)$\hfill] Zeitvariante Impulsantwort des linearen, komplexwertigen Modellsystems\\
     ${\cal S}_{lin}$, das vom Eingangssignal \mbox{$\boldsymbol{v}(k)$} erregt wird.
\item[$\Tilde{h}_{\kappa}(k)$\hfill] Zeitvariante Impulsantwort, die durch periodische 
     Fortsetzung mit der Periode $M$ aus $\Tilde{h}_{\kappa}(k)$ entsteht.
\item[$H(\mu_1,\mu_2)$\hfill] Bifrequente "Ubertragungsfunktion des linearen,
     komplexwertigen Modellsystems ${\cal S}_{lin}$, das vom  
     Eingangssignal \mbox{$\boldsymbol{v}(k)$} erregt wird.
\item[$\Hat{H}\big({\T\mu,\mu\!+\!\Tilde{\mu}\CdoT\frac{M}{K_H}}\big)$\hfill]
     Messwerte der bifrequenten "Ubertragungsfunktion 
     \mbox{$H\big({\T\mu,\mu\!+\!\Tilde{\mu}\CdoT\frac{M}{K_H}}\big)$}
\item[$h_{*,\kappa}(k)$\hfill] Zeitvariante Impulsantwort des linearen, komplexwertigen Modellsystems\\
     ${\cal S}_{*,lin}$, das vom konjugierten Eingangssignal \mbox{$\boldsymbol{v}(k)^*$} erregt wird.
\item[$\Tilde{h}_{*,\kappa}(k)$\hfill] Zeitvariante Impulsantwort, die durch periodische 
     Fortsetzung mit der Periode $M$ aus $\Tilde{h}_{*,\kappa}(k)$ entsteht.
\item[$H_*(\mu_1,\mu_2)$\hfill] Bifrequente "Ubertragungsfunktion des linearen,
     komplexwertigen Modellsystems ${\cal S}_{*,lin}$, das vom konjugierten 
     Eingangssignal \mbox{$\boldsymbol{v}(k)^*$} erregt wird.
\item[$\Hat{H}_*\big({\T\mu,\mu\!+\!\Tilde{\mu}\CdoT\frac{M}{K_H}}\big)$\hfill]
     Messwerte der bifrequenten "Ubertragungsfunktion 
     \mbox{$H_*\big({\T\mu,\mu\!+\!\Tilde{\mu}\CdoT\frac{M}{K_H}}\big)$}
\item[$\Vec{H}(\mu)$\hfill] Zeilenvektor, der die Werte der sich bei der 
     Systemapproximation theoretisch ergebenden "Ubertragungsfunktionen 
     $H\!\big({\T \mu,\mu\!+\!\Hat{\mu}\CdoT\frac{M}{K_H}}\big)$ und 
     $H_*\!\big({\T \mu,\mu\!+\!\Hat{\mu}\CdoT\frac{M}{K_H}}\big)$ der beiden 
     Modellsysteme zusammenfasst.
\item[$\Hat{\Vec{H}}(\mu)$\hfill] Zeilenvektor, der die Messwerte 
     $\Hat{H}\!\big({\T \mu,\mu\!+\!\Hat{\mu}\CdoT\frac{M}{K_H}}\big)$ und 
     $\Hat{H}_*\!\big({\T \mu,\mu\!+\!\Hat{\mu}\CdoT\frac{M}{K_H}}\big)$ der 
     bifrequenten "Ubertragungsfunktionen der beiden Modellsysteme zusammenfasst.
\item[$\widetilde{L}$\hfill] Anzahl der Zeitintervalle innerhalb des
     Zeitraums der Messung, aus denen die $L$ Signalausschnitte f"ur
     eine RKM-Messung ggf. zuf"allig entnommen werden k"onnen.
\item[$K_H$\hfill] Periode der Zeitvarianz des Modellsystems.
\item[$K_{\Phi}$\hfill] Periode der Zyklostationarit"at des Approximationsfehlers.
\item[$K_S$\hfill] Kleinstes gemeinsames Vielfaches von $K_H$ und $K_{\Phi}$.
\item[$K(\mu)$\hfill] Anzahl der Zufallsgr"o"sen, die in dem Zufallsvektor 
     \mbox{$\Breve{\Vec{\boldsymbol{V}}}(\mu)$} zusammengefasst sind.
\item[$K_{\infty,\nu_2}$\hfill] konstanter Faktor der $\nu_2$-ten Nullstelle des
     bilineartransformierten Polynoms mit den Nullstellen am Einheitskreis.
\item[$P,Q$\hfill] Permutationsmatrizen
\item[${\cal S}_{*,lin}$\hfill] Eines der beiden linearen Modellsysteme. Dieses 
     Modellsystem wird von dem konjugierten Eingangssignal \mbox{$\boldsymbol{v}(k)^*$} 
     erregt, und ergibt sich bei der theoretischen Aufspaltung des gest"orten, 
     nichtlinearen realen Systems ${\cal S}$ mit Hilfe der wahren, i.~Allg.
     unbekannten Erwartungswerte der anliegenden Prozesse.
\item[$S_{\nu}$\hfill] Sinusreihenkoeffizienten der Fensterautokorrelationsfolge bzw. -funktion.
\item[$u(k)$\hfill] Deterministisches St"orsignal im Modell des realen
     Systems. Dieses zeitabh"angige St"orsignal ergibt sich
     bei der theoretischen Aufspaltung des gest"orten, nichtlinearen
     realen Systems ${\cal S}$ mit Hilfe der wahren, i.~allg.
     unbekannten Erwartungswerte der anliegenden Prozesse.
\item[$\hat{u}(k)$\hfill] Messwerte des deterministischen St"orsignals.
\item[$U_{\!f}(\mu)$\hfill] Spektrum des gefensterten deterministischen
     St"orsignals.
\item[$\Hat{U}_{\!f}(\mu)$\hfill] Messwerte des Spektrums des gefensterten 
     deterministischen St"orsignals.
\item[$\Tilde{\Vec{\boldsymbol{V}}}(\mu)$\hfill] Spaltenvektor, der die 
     Zufallswerte $\boldsymbol{V}\big({\T\mu\!+\!\Hat{\mu}\CdoT\frac{M}{K_H}}\big)$ 
     und $\boldsymbol{V}\big(\!{\T-\mu\!-\!\Hat{\mu}\CdoT\frac{M}{K_H}}\big)^{\!\Kk}$ 
     des Spektrums der Erregung zusammenfasst.
\item[$\Tilde{\underline{V}}(\mu)$\hfill] Stichprobenmatrix der Erregung. 
     Diese \mbox{$2\CdoT K_H\times L$} Matrix enth"alt eine kon\-krete Stichprobe 
     des Zufallsvektors \mbox{$\Tilde{\Vec{\boldsymbol{V}}}(\mu)$} vom Umfang $L$.
\item[$\Breve{\Vec{\boldsymbol{V}}}(\mu)$\hfill] Spaltenvektor, der einen Satz linear 
     unabh"angiger Zufallswerte $\boldsymbol{V}\big({\T\mu\!+\!\Hat{\mu}\CdoT\frac{M}{K_S}}\big)$ 
     und $\boldsymbol{V}\big(\!{\T-\mu\!-\!\Hat{\mu}\CdoT\frac{M}{K_S}}\big)^{\!\Kk}$ 
     des Spektrums der Erregung zusammenfasst.
\item[$\Breve{\underline{V}}(\mu)$\hfill] Stichprobenmatrix der Erregung. 
     Diese \mbox{$K(\mu)\times L$} Matrix enth"alt eine konkrete Stichprobe 
     des Zufallsvektors \mbox{$\Breve{\Vec{\boldsymbol{V}}}(\mu)$} vom Umfang $L$.
\item[$\boldsymbol{x}_*(k)$\hfill]Zufallsprozess am Ausgang des Modellsystems
     ${\cal S}_{*,lin}$, der bei Erregung des Modellsystems mit dem konjugierten
     Zufallsprozess \mbox{$\boldsymbol{v}(k)^*$} entsteht.
\item[$z_{0,\rho}$\hfill] $\rho$-te, frei w"ahlbare Nullstelle des Polynoms
     \mbox{$z^{(N-1-A_0-2\cdot A_1)}\cdot G_2(z)$}.
\item[$\Tilde{z}_{\infty,\nu_2}$\hfill] $\nu_2$-te Nullstelle
     des Polynoms \mbox{$\widetilde{D}_{N,\infty}(\Tilde{z})$}
     nach der Bilineartransformation am Einheitskreis.
\item[$\Tilde{z}_{N,\rho,\nu_1}$\hfill] Nenner der $\rho$-ten Nullstelle des
     bilineartransformierten Polynoms\\\mbox{$\D\Big(z\CdoT
     e^{-j\cdot\frac{2\pi}{F}\cdot\nu_1}\Big)^{\!N-1-A_0-2\cdot A_1}\Cdot
     G_2\Big(z\CdoT e^{-j\cdot\frac{2\pi}{F}\cdot\nu_1}\Big)$}.
\item[$\Tilde{z}_{Z,\rho,\nu_1}$\hfill] Z"ahler der $\rho$-ten Nullstelle des
     bilineartransformierten Polynoms\\\mbox{$\D\Big(z\CdoT
     e^{-j\cdot\frac{2\pi}{F}\cdot\nu_1}\Big)^{\!N-1-A_0-2\cdot A_1}\Cdot
     G_2\Big(z\CdoT e^{-j\cdot\frac{2\pi}{F}\cdot\nu_1}\Big)$}.
\item[$\rho$\hfill] Index der frei w"ahlbaren Nullstelle $z_{0,\rho}$ des Polynoms
     \mbox{$z^{(N-1-A_0-2\cdoT A_1)}\Cdot G_2(z)$}.
\item[$\Bar{\sigma}$\hfill] Sch"atzwert f"ur den Mindestwert der Streuung des
     Rauschsockels bei der Berechnung des Cepstrums.
\item[$\hat{\sigma}$\hfill] Sch"atzwert f"ur den maximalen Rauschwert des
     Rauschsockels bei der Berechnung des Cepstrums.
\item[$\Phi_{\boldsymbol{n}}(\Omega_1,\Omega_2)$\hfill] Bifrequentes
     Leistungsdichtespektrum des Prozesses \mbox{$\boldsymbol{n}(k)$}
     im zyklostation"aren Fall.
\item[$\Bar{\Phi}_{\boldsymbol{n}}(\mu,\mu+\Tilde{\mu}\CdoT\frac{M}{K_{\Phi}})$\hfill]
     Stufenapproximation des Leistungsdichtespektrums des Prozesses
     \mbox{$\boldsymbol{n}(\!k\!)$} im zyklostation"aren Fall.
\item[$\Tilde{\Phi}_{\boldsymbol{n}}(\mu,\mu+\Tilde{\mu}\CdoT\frac{M}{K_{\Phi}})$\hfill]
     N"aherungswerte der Stufenapproximation des Leistungsdichtespek-\linebreak
     trums des Prozesses \mbox{$\boldsymbol{n}(k)$} im zyklostation"aren Fall,
     die man mit einer endlich langen Fensterfolge gewinnt.
\item[$\Hat{\Phi}_{\boldsymbol{n}}(\mu,\mu+\Tilde{\mu}\CdoT\frac{M}{K_{\Phi}})$\hfill]
     Messwerte des Leistungsdichtespektrums des Prozesses
     \mbox{$\boldsymbol{n}(k)$} im zyklostation"aren Fall.
\item[$\Acute{\Phi}_{\boldsymbol{n}}(\Omega,k)$\hfill]
     Zeitabh"angiges Leistungsdichtespektrum des Prozesses
     \mbox{$\boldsymbol{n}(k)$} im zyklostation"aren Fall, das man aus der
     Autokorrelationsfolge gewinnt, indem man diese bez"uglich der
     Differenz der Zeitpunkte der an der Autokorrelation beteiligten
     Zufallsgr"o"sen diskret fouriertransformiert.
\item[$\Grave{\Phi}_{\boldsymbol{n}}(\Omega,\Bar{\mu})$\hfill]
     Bez"uglich $k$ invers diskret Fouriertransformierte von
     \mbox{$\Acute{\Phi}_{\boldsymbol{n}}(\Omega,k)$}.
\item[$\psi_{0,\rho}$\hfill] Winkel der $\rho$-ten, frei w"ahlbaren Nullstelle des Polynoms
     \mbox{$z^{N-1-A_0-2\cdot A_1}\CdoT G_2(z)$}.
\item[$\Tilde{\psi}_{0,\rho,\nu_1}$\hfill] Winkel der Nullstelle
     \mbox{$\Tilde{z}_{Z,\rho,\nu_1}/\Tilde{z}_{N,\rho,\nu_1}$}
     des bilineartransformierten Polynoms\\[3pt] \mbox{$\D\Big(z\CdoT
     e^{-j\cdot\frac{2\pi}{F}\cdot\nu_1}\Big)^{N-1-A_0-2\cdot A_1}\Cdot
     G_2\Big(z\CdoT e^{-j\cdot\frac{2\pi}{F}\cdot\nu_1}\Big)$}.
\item[$\psiu$\hfill] gesch"atzter Winkel der betraglich gr"o"sten,
     unbekannten Nullstelle $\zu$ des bilineartransformierten Polynoms.
\item[$\Psi_{\boldsymbol{n}}(\Omega_1,\Omega_2)$\hfill]
     Kreuzleistungsdichtespektrum des Prozesses
    \mbox{$\boldsymbol{n}(k)$} im zyklostation"aren Fall.
\item[$\Bar{\Psi}_{\boldsymbol{n}}(\mu,\mu+\Tilde{\mu}\CdoT\frac{M}{K_{\Phi}}\!)$\hfill]
     Stufenapproximation des Kreuzleistungsdichtespektrums des Prozesses
     \mbox{$\boldsymbol{n}(k)$} im zyklostation"aren Fall.
\item[$\Tilde{\Psi}_{\boldsymbol{n}}(\mu,\mu+\Tilde{\mu}\CdoT\frac{M}{K_{\Phi}})$\hfill]
     N"aherungswerte der Stufenapproximation des Kreuzleistungsdichtespektrums
     des Prozesses \mbox{$\boldsymbol{n}(k)$} im zyklostation"aren Fall,
     die man mit einer endlich langen Fenster\-folge gewinnt.
\item[$\Hat{\Psi}_{\boldsymbol{n}}(\mu,\mu+\Tilde{\mu}\CdoT\frac{M}{K_{\Phi}})$\hfill]
     Messwerte des Kreuzleistungsdichtespektrums des Prozesses
     \mbox{$\boldsymbol{n}(k)$} im zyklostation"aren Fall.
\item[$\Acute{\Psi}_{\boldsymbol{n}}(\Omega,k)$\hfill]
     Zeitabh"angiges Kreuzleistungsdichtespektrum des Prozesses
     \mbox{$\boldsymbol{n}(k)$} im zyklostation"aren Fall, das man aus der
     Kreuzkorrelationsfolge gewinnt, indem man diese bez"uglich der
     Differenz der Zeitpunkte der an der Kreuzkorrelation beteiligten
     Zufallsgr"o"sen diskret fouriertransformiert.
\item[$\Grave{\Psi}_{\boldsymbol{n}}(\Omega,\Bar{\mu})$\hfill]
     Bez"uglich $k$ invers diskret Fouriertransformierte von
     \mbox{$\Acute{\Psi}_{\boldsymbol{n}}(\Omega,k)$}.
\end{list}

{\bf Speicherplatznamen}\\
Die Namen der Variablen, f"ur die bei einer Implementierung am Rechner
Speicherbereiche bereitgestellt werden m"ussen, sind in
{\tt Schreibmaschinenschrift} dargestellt. Die Namen der
Variablen, die bei den Programmen zur der Berechnung der Fenster,
ihrer Spektren und ihrer Autokorrelationsfolgen und -funktionen
ben"otigt werden, sind jeweils im Kommentar der Programme
erl"autert. Bei der Beschreibung der RKM-Implementierung
treten folgende Variablen auf: 
\begin{list}{}{\setlength{\itemsep}{0.2ex plus0.2ex}
\setlength{\labelwidth}{4em}\setlength{\labelsep}{1em}
\setlength{\leftmargin}{5em}}
\item[$\text{\tt A\_1\_H}(\mu)$\hfill] Speicher f"ur die l"angeren
     Halbachsen der Konfidenzellipsen der Messwerte der
     "Ubertragungsfunktion des linearen Teilsystems ${\cal S}_{lin}$.
\item[$\text{\tt A\_1\_HS}(\mu)$\hfill] Speicher f"ur die l"angeren
     Halbachsen der Konfidenzellipsen der Messwerte der
     "Ubertragungsfunktion des linearen Teilsystems ${\cal S}_{*,lin}$.
\item[$\text{\tt A\_2\_H}(\mu)$\hfill] Speicher f"ur die k"urzeren
     Halbachsen der Konfidenzellipsen der Messwerte der
     "Ubertragungsfunktion des linearen Teilsystems ${\cal S}_{lin}$.
\item[$\text{\tt A\_2\_HS}(\mu)$\hfill] Speicher f"ur die k"urzeren
     Halbachsen der Konfidenzellipsen der Messwerte der
     "Ubertragungsfunktion des linearen Teilsystems ${\cal S}_{*,lin}$.
\item[$\text{\tt A\_Phi}(\mu)$\hfill] Speicher f"ur die halben
     Konfidenzintervallbreiten der LDS-Messwerte.
\item[$\text{\tt A\_1\_Psi}(\mu)$\hfill] Speicher f"ur die l"angeren
     Halbachsen der Konfidenzellipsen der Messwerte des KLDS.
\item[$\text{\tt A\_2\_Psi}(\mu)$\hfill] Speicher f"ur die k"urzeren
     Halbachsen der Konfidenzellipsen der Messwerte des KLDS.
\item[$\text{\tt A\_1\_u}$\hfill] Speicher f"ur die l"angere
     Halbachse der Konfidenzellipsen der Messwerte der
     deterministischen St"orung.
\item[$\text{\tt A\_2\_u}$\hfill] Speicher f"ur die k"urzere
     Halbachse der Konfidenzellipsen der Messwerte der
     deterministischen St"orung.
\item[$\text{\tt A\_1\_U}(\mu)$\hfill] Speicher f"ur die l"angeren
     Halbachsen der Konfidenzellipsen der Messwerte des Spektrums der
     gefensterten, deterministischen St"orung.
\item[$\text{\tt A\_2\_U}(\mu)$\hfill] Speicher f"ur die k"urzeren
     Halbachsen der Konfidenzellipsen der Messwerte des Spektrums der
     gefensterten, deterministischen St"orung.
\item[$\text{\tt Akku\_y}(k)$\hfill] Akkumulatorfeld f"ur das
     gemessene Ausgangssignal \mbox{$y_{\lambda}(k)$} des Systems.
\item[$\text{\tt Akku\_V}(\mu)$\hfill] Akkumulatorfeld f"ur die
     Spektralwerte \mbox{$V_{\lambda}(\mu)$} des Eingangssignals
     des Systems.
\item[$\text{\tt Akku\_VQ}(\mu)$\hfill] Akkumulatorfeld f"ur die
     Betragsquadrate \mbox{$\big|V_{\lambda}(\mu)\big|^2$} der
     Spektralwerte des Eingangssignals des Systems.
\item[$\text{\tt Akku\_VV}(\mu)$\hfill] Akkumulatorfeld f"ur die Produkte
     \mbox{$V_{\lambda}(-\mu)\CdoT V_{\lambda}(\mu)$} der
     Spektralwerte des Eingangssignals des Systems.
\item[$\text{\tt Akku\_VY}(\mu)$\hfill] Akkumulatorfeld f"ur die Produkte
     \mbox{$V_{\lambda}(-\mu)\CdoT Y_{\!f,\lambda}(\mu)$}
     der Spektralwerte des Eingangssignals und des gefensterten
     Ausgangssignals des Systems.
\item[$\text{\tt Akku\_YQ}(\mu)$\hfill] Akkumulatorfeld f"ur die
     Betragsquadrate \mbox{$\big|Y_{\!f,\lambda}(\mu)\big|^2$} der
     Spektralwerte des gefensterten Ausgangssignals des Systems.
\item[$\text{\tt Akku\_YV}(\mu)$\hfill] Akkumulatorfeld f"ur die Produkte
     \mbox{$Y_{\!f,\lambda}(\mu)\CdoT V_{\lambda}(\mu)^{\Kk}$}
     der Spektralwerte des Eingangssignals und des gefensterten
     Ausgangssignals des Systems.
\item[$\text{\tt Akku\_YY}(\mu)$\hfill] Akkumulatorfeld f"ur die Produkte
     \mbox{$Y_{\!f,\lambda}(-\mu)\CdoT Y_{\!f,\lambda}(\mu)$}
     der Spektralwerte des gefensterten Ausgangssignals des Systems.
\item[$\text{\tt C\_HH}(\mu)$\hfill] Speicher f"ur die Kovarianzen
     \mbox{$\Hat{C}_{\Hat{\boldsymbol{H}}(\mu),\Hat{\boldsymbol{H}}(\mu)^{\Kk}}$}
      der Messwerte der "Ubertragungsfunktion des linearen Teilsystems ${\cal S}_{lin}$.
\item[$\text{\tt C\_HQ}(\mu)$\hfill] Speicher f"ur die Varianzen
      \mbox{$\Hat{C}_{\Hat{\boldsymbol{H}}(\mu),\Hat{\boldsymbol{H}}(\mu)}$}
     der Messwerte der "Ubertragungsfunktion des linearen Teilsystems ${\cal S}_{lin}$.
\item[$\text{\tt C\_HSQ}(\mu)$\hfill] Speicher f"ur die Varianzen
      \mbox{$\Hat{C}_{\Hat{\boldsymbol{H}}_*(\mu),\Hat{\boldsymbol{H}}_*(\mu)}$}
     der Messwerte der "Ubertragungsfunktion des linearen Teilsystems ${\cal S}_{*,lin}$.
\item[$\text{\tt C\_HSHS}(\mu)$\hfill] Speicher f"ur die Kovarianzen
     \mbox{$\Hat{C}_{\Hat{\boldsymbol{H}}_*(\mu),\Hat{\boldsymbol{H}}_*(\mu)^{\Kk}}$}
      der Messwerte der "Ubertragungsfunktion des linearen Teilsystems ${\cal S}_{*,lin}$.
\item[$\text{\tt C\_PhiQ}(\mu)$\hfill] Speicher f"ur die Varianzen
     \mbox{$\Hat{C}_{\Hat{\boldsymbol{\Phi}}_{\!\boldsymbol{n}}(\mu),\Hat{\boldsymbol{\Phi}}_{\!\boldsymbol{n}}(\mu)}$}
     der LDS-Messwerte.
\item[$\text{\tt C\_PsiQ}(\mu)$\hfill] Speicher f"ur die Varianzen
     \mbox{$\Hat{C}_{\Hat{\boldsymbol{\Psi}}_{\!\boldsymbol{n}}(\mu),\Hat{\boldsymbol{\Psi}}_{\!\boldsymbol{n}}(\mu)}$}
     der KLDS-Messwerte.
\item[$\text{\tt C\_PsiPsi}(\mu)$\hfill] Speicher f"ur die Kovarianzen
     \mbox{$\Hat{C}_{\Hat{\boldsymbol{\Psi}}_{\!\boldsymbol{n}}(\mu),\Hat{\boldsymbol{\Psi}}_{\!\boldsymbol{n}}(\mu)^{\Kk}}$}
     der KLDS-Messwerte.
\item[$\text{\tt C\_uu}$\hfill] Speicher f"ur die Kovarianz
     \mbox{$\Hat{C}_{\Hat{\boldsymbol{u}}(k),\Hat{\boldsymbol{u}}(k)^{\Kk}}$}
      der Messwerte der deterministischen St"orung.
\item[$\text{\tt C\_uQ}$\hfill] Speicher f"ur die Varianz
     \mbox{$\Hat{C}_{\Hat{\boldsymbol{u}}(k),\Hat{\boldsymbol{u}}(k)}$}
      der Messwerte der deterministischen St"orung.
\item[$\text{\tt C\_UQ}(\mu)$\hfill] Speicher f"ur die Varianzen
     \mbox{$\Hat{C}_{\Hat{\boldsymbol{U}}_{\!\!f}(\mu),\Hat{\boldsymbol{U}}_{\!\!f}(\mu)}$}
      der Messwerte des Spektrums der gefensterten, deterministischen St"orung.
\item[$\text{\tt C\_UU}(\mu)$\hfill] Speicher f"ur die Kovarianzen
     \mbox{$\Hat{C}_{\Hat{\boldsymbol{U}}_{\!\!f}(\mu),\Hat{\boldsymbol{U}}_{\!\!f}(\mu)^{\Kk}}$}
      der Messwerte des Spektrums der gefensterten, deterministischen
      St"orung.
\item[$\text{\tt C\_VQ}(\mu)$\hfill] Speicher f"ur das
     \mbox{$L\CdoT(L\!-\!1)$}-fache der empirischen Varianzen
     \mbox{$\Hat{C}_{\boldsymbol{V}(\mu),\boldsymbol{V}(\mu)}$}.
\item[$\text{\tt C\_VV}(\mu)$\hfill] Speicher f"ur das
     \mbox{$L\CdoT(L\!-\!1)$}-fache der empirischen Kovarianzen
     \mbox{$\Hat{C}_{\boldsymbol{V}(-\mu),\boldsymbol{V}(\mu)^{\Kk}}$}.
\item[$\text{\tt C\_VY}(\mu)$\hfill] Speicher f"ur das
     \mbox{$L\CdoT(L\!-\!1)$}-fache der empirischen Kovarianzen
     \mbox{$\Hat{C}_{\boldsymbol{V}(-\mu),\boldsymbol{Y}_{\!\!\!f}(\mu)^{\Kk}}$}.
\item[$\text{\tt C\_YQ}(\mu)$\hfill] Speicher f"ur das
     \mbox{$L\CdoT(L\!-\!1)$}-fache der empirischen Varianzen
     \mbox{$\Hat{C}_{\boldsymbol{Y}_{\!\!\!f}(\mu),\boldsymbol{Y}_{\!\!\!f}(\mu)}$}.
\item[$\text{\tt C\_YV}(\mu)$\hfill] Speicher f"ur das
     \mbox{$L\CdoT(L\!-\!1)$}-fache der empirischen Kovarianzen
     \mbox{$\Hat{C}_{\boldsymbol{Y}_{\!\!\!f}(\mu),\boldsymbol{V}(\mu)}$}.
\item[$\text{\tt C\_YY}(\mu)$\hfill] Speicher f"ur das
     \mbox{$L\CdoT(L\!-\!1)$}-fache der empirischen Kovarianzen
     \mbox{$\Hat{C}_{\boldsymbol{Y}_{\!\!\!f}(-\mu),\boldsymbol{Y}_{\!\!\!f}(\mu)^{\Kk}}$}.
\item[$\text{\tt f}(k)$\hfill] Speicher der $F$ Werte der Fensterfolge.
\item[${\tt H}(\mu)$\hfill] Speicher der Messwerte \mbox{$\Hat{H}(\mu)$} der 
     "Ubertragungsfunktion des linearen Teilsystems ${\cal S}_{lin}$.
\item[${\tt HS}(\mu)$\hfill] Speicher der Messwerte \mbox{$\Hat{H}_*(\mu)$} der 
     "Ubertragungsfunktion des linearen Teilsystems ${\cal S}_{*,lin}$.
\item[$\text{\tt K\_VV}(\mu)$\hfill] Speicher f"ur die inversen empirischen
     Kovarianzmatrizen \mbox{$\Hat{\underline{C}}_{\Breve{\Vec{\boldsymbol{V}}}(\mu),\Breve{\Vec{\boldsymbol{V}}}(\mu)}^{\uP{0.4}{\!-1}}$}.
\item[$\text{\tt Phi}(\mu)$\hfill] Speicher der Messwerte
     \mbox{$\Hat{\Phi}_{\boldsymbol{n}}(\mu)$} des LDS.
\item[$\text{\tt Psi}(\mu)$\hfill] Speicher der Messwerte
     \mbox{$\Hat{\Psi}_{\boldsymbol{n}}(\mu)$} des KLDS.
\item[${\tt u}(k)$\hfill] Speicher der Messwerte der deterministischen St"orung \mbox{$\Hat{u}(k)$}.
\item[${\tt U}(\mu)$\hfill] Speicher der Messwerte des Spektrums
      \mbox{$\Hat{U}_{\!f}(\mu)$} der gefensterten, deterministischen St"orung.
\item[$\text{\tt v}(k)$\hfill] Speicher der Testsignalfolge
     \mbox{$v_{\lambda}(k)$}.
\item[$\text{\tt V}(\mu)$\hfill] Speicher der Spektralwerte
     \mbox{$V_{\lambda}(\mu)$} der Testsignalfolge.
\item[${\tt X\_mittel}(\mu)$\hfill] Speicher f"ur das
     \mbox{$L$}-fache der empirischen Mittelwerte der Summe der Spektren der
     Signale an den Ausg"angen der beiden linearen Modellsysteme.
\item[${\tt x\_mittel}(k)$\hfill] Speicher f"ur das
     \mbox{$L$}-fache der empirischen Mittelwerte der Summe der Signale an den Ausg"angen 
     der beiden linearen Modellsysteme.
\item[$\text{\tt y}(k)$\hfill] Speicher des gemessenen Ausgangssignals
     \mbox{$y_{\lambda}(k)$}.
\item[$\text{\tt Y}(\mu)$\hfill] Speicher der Spektralwerte
     \mbox{$Y_{\!f,\lambda}(\mu)$} des gemessenen, gefensterten
     Ausgangssignals.
\item[${\tt Y\_mittel}$\hfill] Speicher f"ur das
     \mbox{$L$}-fache der empirischen Mittelwerte
     des Spektrums des gefensterten Ausgangssignals.
\end{list}


\end{document}